\documentstyle[preprint,eqsecnum,aps,epsf,rotate,floats,sty/rmpbib,sty/times]{revtex}

\newcommand{\blankpage}{\clearpage
\thispagestyle{empty}
\
\clearpage}

\newcommand{\skipevenpage}{
\ifodd\thepage \relax
\else \blankpage \fi
}

\newcommand{\dS}{\Delta S=1}
\newcommand{\dB}{\Delta B=1}

\newcommand{\BB}{$B^0$-$\bar{B}^0$}
\newcommand{\KK}{$K^0$-$\bar{K}^0$}

\newcommand{\ord}{{\cal O}}

\newcommand{\vv}{\vec{v}}

\newcommand{\as}{\alpha_{\rm s}}
\newcommand{\aem}{\alpha}

\newcommand{\hU}{U}
\newcommand{\hR}{R}
\newcommand{\hJ}{J}
\newcommand{\hV}{V}
\newcommand{\hK}{K}
\newcommand{\hM}{M}

\newcommand{\gf}{\gamma_5}

\newcommand{\hg}{\gamma}
\newcommand{\hG}{\Gamma}

\newcommand{\gs}{\gamma_{\rm s}^{(0)}}
\newcommand{\gem}{\gamma_{\rm e}^{(0)}}

\newcommand{\gss}{\gamma_{\rm s}^{(1)}}
\newcommand{\gse}{\gamma_{\rm se}^{(1)}}

\newcommand{\gemt}{\gamma_{\rm e}^{(0) T}}

\newcommand{\gset}{\gamma_{\rm se}^{(1) T}}

\newcommand{\gssndr}{\gamma_{\rm s,NDR}^{(1)}}

\newcommand{\vC}{\vec{C}}

\newcommand{\eps}{\varepsilon}
\newcommand{\epe}{\varepsilon'/\varepsilon}

\newcommand{\vardre}{\delta r_{\rm e}}
\newcommand{\vardrs}{\delta r_{\rm s}}

\newcommand{\mt}{m_{\rm t}}
\newcommand{\mb}{m_{\rm b}}
\newcommand{\mc}{m_{\rm c}}
\newcommand{\ms}{m_{\rm s}}

\newcommand{\md}{m_{\rm d}}
\newcommand{\mw}{M_{\rm W}}
\newcommand{\mz}{M_{\rm Z}}

\newcommand{\gev}{\, {\rm GeV}}
\newcommand{\mev}{\, {\rm MeV}}

\newcommand{\Heff}{{\cal H}_{\rm eff}}

\newcommand{\Lms}{\Lambda_{\overline{\rm MS}}}

\newcommand{\V}[1]{V_{\rm #1}}

\newcommand{\RE}{{\rm Re}}
\newcommand{\IM}{{\rm Im}}

\newcommand{\Kpipi}{K \to \pi\pi}
\newcommand{\Kpiee}{K_L \to \pi^0 e^+ e^-}
\newcommand{\Kpienu}{K^+ \to \pi^0 e^+ \nu}

\newcommand{\svs}{\vbox{\vskip 5mm}}
\newcommand{\mvs}{\vbox{\vskip 8mm}}

\newcommand{\nn}{\nonumber}
\newcommand{\eqn}[1]{(\ref{#1})}

\newcommand{\kpnn}{$K^+ \to \pi^+\nu\bar\nu$\ }
\newcommand{\kpn}{K^+ \to \pi^+\nu\bar\nu}
\newcommand{\klmm}{$K_L \to \mu^+\mu^-$\ }
\newcommand{\klm}{K_L \to \mu^+\mu^-}
\newcommand{\klpnn}{$K_L \to \pi^0\nu\bar\nu$\ }
\newcommand{\klpn}{K_L \to \pi^0\nu\bar\nu}
\newcommand{\bsg}{b \to s \gamma}
\newcommand{\Bsg}{B \to X_s \gamma }
\newcommand{\Bsee}{B \to X_s e^+ e^-}
\newcommand{\bsee}{b \to s e^+ e^-}
\newcommand{\bcenu}{b \to c e \bar\nu}
\newcommand{\kpiee}{K_L \to \pi^0 e^+ e^-}
\newcommand{\Ctilde}{\widetilde C}

\newcommand{\aspi}{\frac{\as}{4 \pi}}

\newcommand{\imlt}{\IM \lambda_t}
\newcommand{\relt}{\RE \lambda_t}
\newcommand{\relc}{\RE \lambda_c}

\tighten

\begin{document}

\renewcommand{\thefootnote}{\fnsymbol{footnote}}

\preprint{\small
$
\begin{array}{r}
MPI-Ph/95-104 \\
TUM-T31-100/95 \\
FERMILAB-PUB-95/305-T \\
SLAC-PUB\ 7009
\end{array}
$
}

\title{WEAK DECAYS BEYOND LEADING LOGARITHMS}

\author{\bf
        Gerhard~{Buchalla}${}^{3}$,\,
        Andrzej~J.~{Buras}${}^{1,2}$,\,
        Markus~E.~{Lautenbacher}${}^{1,4}$
        \footnote{email:~{\tt
buchalla@fnth20.fnal.gov,\,buras@feynman.t30.physik.tu-muenchen.de, \\
\phantom{XXXXXX}lauten@feynman.t30.physik.tu-muenchen.de}}
}

\address{
\ \\
${}^{1}$ Physik Department, Technische Universit\"at M\"unchen, \\
D-85748 Garching, Germany.
\\
${}^{2}$ Max-Planck-Institut f\"ur Physik -- Werner-Heisenberg-Institut, \\
F\"ohringer Ring 6, D-80805 M\"unchen, Germany.
\\
${}^{3}$ Theoretical Physics Department,  \\
Fermi National Accelerator Laboratory, \\
P.O.\ Box 500, Batavia, IL 60510, USA.
\\
${}^{4}$ SLAC Theory Group, Stanford University, \\
P.O.\ Box 4349, Stanford, CA 94309, USA.
\\
\ 
}

\date{November 1995}

\maketitle
\thispagestyle{empty}

\centerline{to appear in Reviews of Modern Physics}

\begin{abstract}
\noindent We review the present status of QCD corrections to weak
decays beyond the leading logarithmic approximation including
particle-antiparticle mixing and rare and CP violating decays. After
presenting the basic formalism for these calculations we discuss in
detail the effective hamiltonians for all decays for which the
next-to-leading corrections are known. Subsequently, we present the
phenomenological implications of these calculations. In particular we
update the values of various parameters and we incorporate new
information on $m_t$ in view of the recent top quark discovery.  One of
the central issues in our review are the theoretical uncertainties
related to renormalization scale ambiguities which are substantially
reduced by including next-to-leading order corrections.  The impact of
this theoretical improvement on the determination of the
Cabibbo-Kobayashi-Maskawa matrix is then illustrated in various cases.
\end{abstract}

\setcounter{footnote}{0}
\renewcommand{\thefootnote}{\arabic{footnote}}

\newpage
\setcounter{page}{1}
\renewcommand{\thepage}{\roman{page}}
\tableofcontents
\newpage

\renewcommand{\thepage}{\arabic{page}}
\setcounter{page}{1}

\section{Introduction}
         \label{sec:intro}
\subsection{Preliminary Remarks}
            \label{sec:intro:rem}
Among the fundamental forces of nature the weak interactions clearly
show the most complicated and diversified pattern from the point of
view of our present day understanding represented by the
Standard Model of particle physics.
Although this theory of the strong and electroweak
forces is capable of describing very successfully a huge amount of
experimental information in a quantitative way
and a great deal of phenomena at least qualitatively, there are many
big question marks that remain. The most prominent among them like the
problem of electroweak symmetry breaking and the origin of
fermion masses and quark
mixing are closely related to the part of the Standard Model describing
weak interactions. Equally puzzling is the fact that whereas the
discrete space-time symmetries C, P, CP and T are respected by strong and
electromagnetic interactions, the weak force violates them all.
Obviously, the weak interaction is the corner of the Standard Model
that is least understood. The history of this field is full of
surprises and still more of them might be expected in the future.
\\
For these reasons big efforts have been and still are being
undertaken in order to develop our theoretical understanding of weak
interaction phenomena and to disentangle the basic mechanisms and
parameters.
An excellent laboratory for this enterprise is provided by the very
rich phenomenology of weak meson decays.

The careful investigation of these decays is mandatory for further
testing the Standard Model. Of particular importance is the
determination of all Cabibbo-Kobayashi-Maskawa (CKM) parameters as a
prerequisite for a decisive test of the consistency of the Standard
Model ansatz including the unitarity of the CKM matrix and its
compatibility with the quark masses.  Many interesting issues within
this context still remain unsettled. Let us just mention here the
question of direct CP violation in non-leptonic K decays
($\varepsilon'/\varepsilon$), the yet completely unknown pattern of CP
violation in the B system and the rare K and B decays, which are
sensitive to the effects of virtual heavy particles, most notably the
top quark, its mass and its weak couplings.  Whether the CKM
description of CP violation is correct, remains as an outstanding open
question.  It is clear that the need for a modification of the model is
conceivable and that meson decay phenomena might provide a window for
``new physics''. However, independently of this possibility it is
crucial to improve the theoretical predictions in the Standard Model
itself, either to further establish its correctness, or to be able to
make clear cut statements on its possible failure.

Now, for all attempts towards a theoretical understanding of these
issues the obvious fact that the fundamental forces do not come in
isolation is of crucial significance. Since hadrons are involved in the
decays that are of interest here, QCD unavoidably gets into the game.
In order to understand weak meson decays we have to understand the
interplay of weak interactions with the strong forces.\\
To accomplish this task it is necessary to employ the field
theoretical tools of the operator product expansion (OPE)
\cite{wilsonzim:72} and the renormalization group
\cite{stueckelberg:53}, \cite{gellmannlow:54}, \cite{ovsyannikov:56},
\cite{symanzik:70}, \cite{callan:70}, \cite{thooft:73},
\cite{weinberg:73}. The basic virtues of these two techniques may be
characterized as follows. Consider the amplitude $A$ for some weak
meson decay process. Using the OPE formalism this amplitude can be
represented as \cite{witten:77}
\begin{equation}
A=\langle {\cal H}_{eff}\rangle =\sum_i C_i(\mu, M_W)
                                        \langle Q_i(\mu)\rangle
\label{ahcq}
\end{equation}
where it is factorized into the Wilson coefficient functions $C_i$ and
the matrix elements of local operators $Q_i$. In this process the W
boson and other fields with mass bigger than the factorization scale
$\mu$
are ``integrated out'', that is removed from the theory as dynamical
degrees of freedom. The effect of their existence is however implicitly
taken into account in the Wilson coefficients. In a more intuitive
interpretation one can view the expression $\sum C_i Q_i$ as an effective
hamiltonian for the process considered, with $Q_i$ as the effective
vertices and $C_i$ the corresponding coupling constants. Usually for
weak decays only the operators of lowest dimension need to be taken into
account. Contributions of higher dimensional operators are negligible
since they are typically suppressed by powers of $p^2/M^2_W$, where $p$
is the momentum scale relevant for the decaying meson in question.
\\
The essential point about the OPE is that it achieves a separation of
the full problem into two distinct parts, the long-distance contributions
contained in the operator matrix elements and the short-distance physics
described by the Wilson coefficients. The renormalization scale $\mu$
separating the two regimes is typically chosen to be
of the order $\ord(\geq 1\gev)$ for
kaon decays and a few $GeV$ for the decays of D and B mesons.
The physical amplitude $A$ however cannot depend on
$\mu$. The $\mu$ dependence of the Wilson coefficients has
to cancel the $\mu$ dependence present in $\langle Q_i(\mu)\rangle$.
In other words it is a matter of choice what exactly
belongs to the matrix elements
and what to the coefficient functions. This cancellation of
$\mu$ dependence involves generally several terms in the expansion
in (\ref{ahcq}).\\
The long-distance part in (\ref{ahcq}) deals
with low energy strong interactions
and therefore poses a very difficult problem. Many approaches, like
lattice gauge theory, $1/N$- expansion, QCD- and hadronic sum rules or
chiral perturbation theory, have been used in the past to obtain
qualitative insight and some quantitative estimates of relevant
hadronic matrix elements. In addition heavy quark effective theory (HQET)
and heavy quark expansions (HQE) have been widely used for $B$ decays.
Despite these efforts the problem is not yet solved satisfactorily.
\\
In general in weak decays of mesons the hadronic matrix elements
constitute the most important source of theoretical uncertainty.  There
are however a few special examples of semileptonic rare decays ($\kpn$,
$K_L\to\pi^0\nu\bar\nu$, $B\to X_s\nu\bar\nu$) where the matrix
elements needed can be extracted from well measured leading decays or
calculated perturbatively or as in the case of $B_s \to \mu \bar \mu$
expressed fully in terms of meson decay constants. Thus practically the
problem of long-distance QCD can be completely avoided. This makes
these decay modes very attractive from a theoretical point of view,
although due to very small branching ratios they are quite difficult to
access experimentally today.
\\
Contrary to the long-distance contributions the short-distance part can
be analyzed systematically using well established field theoretical
methods. Due to the asymptotic freedom property of QCD the strong
interaction effects at short-distances are calculable in perturbation
theory in the strong coupling $\as(\mu)$. In fact $\as(\mu)$ is small
enough in the full range of relevant short distance scales of
$\ord(M_W)$ down to $\ord(1\gev)$ to serve as a reasonable expansion
parameter. However the presence of large logarithms $\ln(M_W/\mu)$
multiplying $\as(\mu)$ (where $\mu=\ord(1\gev)$) in the calculation of
the coefficients $C_i(\mu, M_W)$ spoils the validity of the usual
perturbation series. This is a characteristic feature of renormalizable
quantum field theories when vastly different scales are present.  It is
therefore necessary to perform a renormalization group analysis which
allows an efficient summation of logarithmic terms to all orders in
perturbation theory. In this way the usual perturbation theory is
replaced by the renormalization group improved perturbation theory in
which the leading order (LO) corresponds to summing the leading
logarithmic terms $\sim (\as \ln(M_W/\mu))^n$. Then at next-to-leading
order (NLO), all terms of the form $\sim\as (\as \ln(M_W/\mu))^n$ are
summed in addition, and so on.
\\
The evaluation of the short-distance coefficients in
renormalization group improved perturbation theory is only a part of
the entire problem, but one should stress that still it is
indispensible to analyze this part systematically; the effective
hamiltonians resulting from the short-distance analysis provide the
necessary basis for any further computation of weak decay amplitudes.
The long-distance matrix elements needed in addition
can be treated separately and will hopefully be known with
desirable accuracy one day.
\\
The rather formal expression for the decay amplitudes given in
\eqn{ahcq} can always be cast in a form \cite{buchallaetal:91}
\begin{equation}
A(M \to F)=\sum_i B_i \, V_{CKM}^{i} \, \eta^{i}_{QCD} \, F_i(m_t,m_c)
\label{PBEE}
\end{equation}
which is more useful for phenomenology.  In writing \eqn{PBEE} we
have generalized \eqn{ahcq} to include several CKM factors
$V_{CKM}^{i}$. The functions $F_i(m_t,m_c)$ result from the evaluation of
loop diagrams with internal top and charm exchanges  and may also depend
solely on $\mt$ or $\mc$. In certain cases $F_i$ are mass independent.
The factors $\eta^{i}_{QCD}$ summarize short distance QCD corrections
which can be calculated by the formal methods mentioned above. Finally
$B_i$ stand for nonperturbative factors related to the hadronic matrix
elements of the contributing operators: the main theoretical
uncertainty in the whole enterprise. A well known example of a
$B_i$-factor is the renormalization group invariant parameter $B_K$
relevant for $K^0-\bar K^0$ mixing and the indirect CP violation in
$K \to \pi\pi$.
\\
It is worth noting that the short-distance QCD contributions by
themselves have already an important impact on weak decay processes. In
non-leptonic K-decays, for example, they help to explain the famous
$\Delta I=1/2$ rule and they generate penguin operators which are
relevant for $\varepsilon' /\varepsilon$. They suppress the
semileptonic branching ratio in heavy quark decays and produce a
significant enhancement of the weak radiative process $B\to X_s\gamma$.

Starting with the pioneering work of \cite{gaillard:74} and
\cite{altarelli:74}, who calculated the first leading logarithmic QCD
effects in weak decays, considerable efforts have been devoted to the
calculation of short-distance QCD corrections to weak meson decay
processes. The analysis has been extended to a large variety of
particular modes. Of great interest are especially processes sensitive
to the virtual contribution of heavy quarks, like the top. A classic
example of this type is the 1974 analysis of \cite{gaillard:74b} of
$K^0 - \bar{K}^0$ mixing and their estimate of the charm quark mass
prior to its discovery, based on the dependence of the $\Delta S=2$
transition on virtual charm. This calculation constitutes the prototype
application for present day analyses of virtual top contributions in
$B^0 - \bar B^0$ mixing, rare decays and CP violation, which are
similar in spirit.
\\
Until 1989 most of the calculations were done in LO, i.e.\ in the
leading logarithmic approximation \cite{vainshtein:77},
\cite{gilman:79}, \cite{gilman:80}, \cite{guberina:80}. An exception
was the important work of \cite{altarelli:81} where the first NLO
calculation in the theory of weak decays has been presented.
\\
Today the effective hamiltonians for weak processes are available at
the next-to-leading level for the most important and interesting cases
due to a series of publications devoted to this enterprise beginning
with the work of \cite{burasweisz:90}. In table~\ref{tab:processes}
we give a list of decays for which NLO QCD corrections are known at
present.  With the next-to-leading short-distance effects included,
weak decays have in a sense now achieved the status that the
conceptually similar field of deep inelastic lepton nucleon scattering
had attained more then a decade ago \cite{buras:80}.

\begin{table}[htb]
\caption[]{Processes for which NLO QCD corrections have been calculated by
now.
\label{tab:processes}}
\begin{center}
\begin{tabular}{|cl|}
\bf Decay & \bf \phantom{XXXXXX} Reference \\
\hline
\hline
 \multicolumn{2}{|c|}{$\Delta F=1$ Decays} \\
\hline
current-current operators     & \cite{altarelli:81}, \cite{burasweisz:90} \\
QCD penguin operators         & \cite{burasetal:92b}, \cite{burasetal:92c}, \\
                              & \cite{ciuchini:93} \\
electroweak penguin operators & \cite{burasetal:92b}, \cite{burasetal:92c}, \\
                              & \cite{ciuchini:93} \\
magnetic penguin operators    & \cite{misiakmuenz:95} \\
$B(B \to X e\nu)$             & \cite{altarelli:81}, \cite{buchalla:93}, \\
                              & \cite{baganetal:94a}, \cite{baganetal:95} \\
Inclusive $\dS$               & \cite{jaminpich:94} \\
\hline
\multicolumn{2}{|c|}{Particle-Antiparticle Mixing} \\
\hline
$\eta_1$                   & \cite{herrlichnierste:93} \\
$\eta_2,~\eta_B$           & \cite{burasjaminweisz:90} \\
$\eta_3$                   & \cite{herrlichnierste:95} \\
\hline
\multicolumn{2}{|c|}{Rare K- and B-Meson Decays} \\
\hline
$K^0_L \rightarrow \pi^0\nu\bar{\nu}$, $B \rightarrow l^+l^-$,
$B \rightarrow X_{\rm s}\nu\bar{\nu}$ & \cite{buchallaburas:93b} \\
$K^+   \rightarrow \pi^+\nu\bar{\nu}$, $K_L \rightarrow \mu^+\mu^-$
                                      & \cite{buchallaburas:94} \\
$K^+\to\pi^+\mu\bar\mu$               & \cite{buchallaburas:94b} \\
$K_L \rightarrow \pi^0e^+e^-$         & \cite{burasetal:94a} \\
$B\rightarrow X_s e^+e^-$           & \cite{misiak:94}, \cite{burasmuenz:95} \\
\end{tabular}
\end{center}
\end{table}

Let us recall why NLO calculations are important for weak decays and why
it is worthwile to perform the very involved and complicated
computations.
\begin{itemize}
\item The NLO is first of all necessary to test the validity of
perturbation theory. In LO all the $(\as \ln(M_W/\mu))^n$ terms
are summed, yielding a result of $\ord(1)$; it is only at NLO
where one obtains a truly perturbative $\ord(\as)$
correction relative to the LO and one can check whether it is small
enough to justify the perturbative approach.
\item Without going to NLO the scheme specific
QCD scale $\Lambda_{\overline{MS}}$
extracted from various high energy processes cannot be used
meaningfully in weak decays.
\item Due to renormalization group (RG) invariance
the physical amplitudes do not depend on the exact scales $\mu_i$
at which quark masses (top) are defined or heavy particles are
integrated out. However in perturbation theory RG invariance is broken
through the truncation of the series by terms of the neglected order.
Numerically the resulting scale ambiguities, representing the
theoretical uncertainty of the short-distance part, are a serious
problem for the LO which can be reduced considerably by going to NLO.
\item The Wilson coefficients are renormalization
scheme dependent quantities. The
scheme dependence is first ``felt'' at NLO whereas the LO is completely
insensitive to this important feature. In particular this issue is
essential for a proper matching of the short distance contributions to
the long distance matrix elements as obtained from lattice calculations.
\item In some cases, particularly for $\epe$, $K_L\to\pi^0e^+e^-$ and
$B \to X_s e^+ e^-$, the central issue of the top quark mass dependence
is strictly speaking a NLO effect.
\end{itemize}
We would like to stress that short-distance QCD should be contrasted
with an ``intrinsically perturbative'' theory like QED, where
perturbation theory is almost the whole story since $\alpha_{QED}$ is
exceedingly small. In QCD the coupling is much larger at
interesting scales so that the conceptual questions like residual
scale or scheme dependences, which are formally of the neglected
higher order, become important numerically. Thus in this sense the
question of higher order corrections is not only one of a
quantitative improvement (of making precise predictions even more
accurate, like in QED), but of a qualitative improvement as well.

We think that the time is appropriate to review the subject of QCD
corrections to weak meson decays at the next-to-leading order level
and to collect the most important results obtained in this field.

\subsection{Outline}
            \label{sec:intro:outline}
This review is divided into three parts, roughly speaking ``basic
concepts'', ``technicalities'' and ``phenomenological applications''.
The division is made especially for pedagogical reasons hoping to make
the review as readable as possible to a wide audience of physicists.

In the first part we discuss the basic formalism necessary to obtain
the effective hamiltonians for weak decays from the underlying full
$SU(3) \otimes SU(2)_L \otimes U(1)_Y$ gauge theory of the Standard
Model.

The second part constitutes a compendium of effective hamiltonians for
all weak decays for which NLO corrections have been calculated in the
literature and whose list is given in table \ref{tab:processes}. We
include also the discussion of the important decay $B \to X_s \gamma$
which is known only at the LO level.

The third part of our review then presents the phenomenological picture
of weak decays beyond the leading logarithmic approximation using the
results obtained in parts one and two.

We end our review of this exciting field with a brief summary of
results and an outlook.

We are aware of the fact that some sections in this review are
necessarily rather technical which is connected to the very nature of
the subject of this review. We have however made efforts to present the
material in a pedagogical fashion. Thus part one can be regarded as an
elementary introduction to the formalism of QCD calculations which
include renormalization group methods  and the operator product
expansion. Even if our compendium in part two looks rather technical at
first sight, the guidelines to the effective hamiltonians presented in
section \ref{sec:Heffguide} should be helpful in following and using
this important part of our review. In any case the phenomenological part
three is almost self-contained and its material can be easily followed
with the help of the guidelines in section  \ref{sec:Heffguide} without
the necessity of fully understanding the details of NLO calculations.

\skipevenpage

{\Huge\bf
\noindent
Part One --

\bigskip
\bigskip
\bigskip

\noindent
The Basic Formalism
}

\vfil

\noindent
In this first part we will discuss the basic formalism behind radiative
corrections to weak decays.

In section \ref{sec:sewm} we recall those ingredients of the standard
$SU(3)\otimes SU(2)\otimes  U(1)$ model, which play an important role
in subsequent sections. In particular we recall the
Cabibbo-Kobayashi-Maskawa matrix in two useful parametrizations and we
briefly describe the unitarity triangle.

In section \ref{sec:basicform} we outline the basic formalism for the
calculation of QCD effects in weak decays.  Beginning with the idea of
effective field theories we introduce subsequently the techniques of
the operator product expansion and the renormalization group. These
important concepts are illustrated explicitly using the simple, but
phenomenologically relevant example of current-current operators, which
allows to demonstrate the procedure in a transparent way. The central
issue in this formalism is the computation of the Wilson coefficients
$C_i$ of local operators in the LO and NLO approximation. This
calculation involves the proper computation of $C_i$ at $\mu =
\ord(\mw)$ and the renormalization group evolution down to low energy
scales $\mu\ll \mw$ relevant for the weak decays considered. The latter
requires the evaluation of one-loop and two-loop anomalous dimensions
of $Q_i$ or more generally the anomalous dimension matrices, which
describe the mixing of these operators under renormalization.  We
outline the steps for a consistent calculation of the Wilson
coefficients $C_i$ and formulate recipes for the determination of the
anomalous dimensions of local operators.  In section
\ref{sec:basicform:wc} we give ``master formulae'' for the Wilson
coefficients $C_i$ , including NLO corrections.  Since these formulae
will be central for our review, we discuss their various properties in
some detail. In particular we address the $\mu$- and renormalization
scheme dependences and we show on general grounds how these dependences
are canceled by those present in the hadronic matrix elements.

\section{Standard Electroweak Model}
         \label{sec:sewm}
\subsection{Particles and Interactions}
            \label{sec:sewm:particles}
Throughout this review we will work in the context of the three
generation model of quarks and leptons based on the gauge group
$SU(3)\otimes SU(2)_L\otimes U(1)_Y $ spontaneously broken down
 to $SU(3)\otimes U(1)_Q$.
Here $Y$ and $Q$ denote the weak hypercharge and the electric charge
generators, respectively. $SU(3)$ stands for $QCD$ which will be discussed
in more detail in the following section. Here we would like to recall certain
features of the electroweak part of the Standard Model which will be
important for our considerations.

The left-handed leptons and quarks are put in $ SU(2)_L $ doublets
\begin{equation}\label{2.31}
\left(\begin{array}{c}
\nu_e \\
e^-
\end{array}\right)_L\qquad
\left(\begin{array}{c}
\nu_\mu \\
\mu^-
\end{array}\right)_L\qquad
\left(\begin{array}{c}
\nu_\tau \\
\tau^-
\end{array}\right)_L
\end{equation}
\begin{equation}\label{2.66}
\left(\begin{array}{c}
u \\
d^\prime
\end{array}\right)_L\qquad
\left(\begin{array}{c}
c \\
s^\prime
\end{array}\right)_L\qquad
\left(\begin{array}{c}
t \\
b^\prime
\end{array}\right)_L       
\end{equation}
with the corresponding right-handed fields transforming as singlets
under $ SU(2)_L $. The primes are discussed below.
The relevant electroweak charges $Q$, $Y$ and the third component of
the weak isospin $T_3$ are collected in table~\ref{tab:ewcharges}.

\begin{table}[htb]
\caption[]{Electroweak charges $Q$, $Y$ and the third component of
the weak isospin $T_3$ for quarks and leptons in the Standard Model.
\label{tab:ewcharges}}
\begin{center}
\begin{tabular}{|c||c|c|c|c|c|c|c|}
 & $\nu^e_L$ & $ e^-_L$ & $ e^-_R$ & $u_L$ & $d_L$ & $u_R$ & $d_R$ \\
\hline
$Q$ & 0 & $-1$ & $-1$ & 2/3 & $-1/3$ & 2/3 & $-1/3$ \\
\hline
$T_3$ &1/2 & $-1/2$ & 0 & 1/2 & $-1/2$ & 0 & 0 \\
\hline
$Y$ & $-1$ & $-1$ & $-2$ & 1/3 & 1/3 & 4/3 & $-2/3$
\end{tabular}
\end{center}
\end{table}

The electroweak interactions of quarks and leptons are mediated by
the massive weak gauge bosons $W^\pm$ and $Z^o$ and by the photon $A$.
These interactions are summarized by the Lagrangian
\begin{equation}\label{3}
{\cal L}_{int}={\cal L}_{\rm CC}+{\cal L}_{\rm NC}
\end{equation}
where
\begin{equation}\label{4}
{\cal L}_{\rm CC}=\frac{g_2}{2 \sqrt{2}}(J^+_\mu W^{+\mu}+J^-_\mu W^{-\mu})
\end{equation}
describes the {\it charged current} interactions and
\begin{equation}\label{5}
{\cal L}_{\rm NC}=
 e J^{em}_\mu A^{\mu}+ \frac{g_2}{2 \cos \Theta_W} J^o_\mu Z^\mu
\end{equation}
the {\it neutral current} interactions. Here $e$ is the QED coupling constant,
$g_2$ is the $SU(2)_L$ coupling constant and $\Theta_W$ is the Weinberg
angle. The currents are given as follows
\begin{equation}\label{6}
J^+_\mu=
(\bar{u} d')_{V-A} +
(\bar{c} s')_{V-A} +
(\bar{t} b')_{V-A} +
(\bar{\nu}_e e)_{V-A} +
(\bar{\nu}_\mu \mu)_{V-A} +
(\bar{\nu}_\tau \tau)_{V-A}
\end{equation}

\begin{equation}\label{7}
J^{em}_\mu=\sum_f {Q_f \bar f \gamma_\mu f}
\end{equation}
\begin{equation}\label{8}
J^o_\mu=\sum_f \bar f \gamma_\mu (v_f-a_f\gamma_5) f
\end{equation}
\begin{equation}\label{9}
v_f=T^f_3-2 Q_f \sin^2\Theta_W
\qquad
a_f=T^f_3
\end{equation}
where $Q_f$ and $T^f_3$ denote the charge and the third component of the
weak isospin of the left-handed fermion $f_L$.

In our discussion of weak decays an important role is played by
the Fermi constant:
\begin{equation}\label{2.100}
\frac{G_F}{\sqrt{2}}=\frac{g^2_2}{8 M^2_W}
\end{equation}
which has the value
\begin{equation}\label{2.98}
G_F=1.16639\cdot 10^{-5}\gev^{-2}
\end{equation}
Other values of the relevant parameters will be collected in appendix
\ref{app:numinput}. 

The interactions between the gauge bosons are standard and can be
found in any textbook on gauge theories.

The primes in (\ref{2.66}) indicate that the weak eigenstates 
$(d^\prime,s^\prime,b^\prime)$ are not equal to the corresponding
mass eigenstates $(d,s,b)$, but are rather linear combinations of
the latter. This is expressed through the relation
\begin{equation}\label{2.67}
\left(\begin{array}{c}
d^\prime \\ s^\prime \\ b^\prime
\end{array}\right)=
\left(\begin{array}{ccc}
V_{ud}&V_{us}&V_{ub}\\
V_{cd}&V_{cs}&V_{cb}\\
V_{td}&V_{ts}&V_{tb}
\end{array}\right)
\left(\begin{array}{c}
d \\ s \\ b
\end{array}\right)
\end{equation}
where the unitary matrix connecting theses two sets of states is the
Cabibbo-Kobayashi-Maskawa (CKM) matrix.  Many parametrizations of this
matrix have been proposed in the literature.  We will use in this
review two parametrizations:  the standard parametrization
recommended by the particle data group and the Wolfenstein
parametrization.

\subsection{Standard Parametrization}
            \label{sec:sewm:stdparam}
Let us introduce the notation
$c_{ij}=cos\theta_{ij}$ and $s_{ij}=sin\theta_{ij}$ with $i$ and $j$
being generation labels ($i,j=1,2,3$). The standard parametrization is
then given as follows \cite{particledata:94}
\begin{equation}\label{2.72}
V=
\left(\begin{array}{ccc}
c_{12}c_{13}&s_{12}c_{13}&s_{13}e^{-i\delta}\\ -s_{12}c_{23}
-c_{12}s_{23}s_{13}e^{i\delta}&c_{12}c_{23}-s_{12}s_{23}s_{13}e^{i\delta}&
s_{23}c_{13}\\ s_{12}s_{23}-c_{12}c_{23}s_{13}e^{i\delta}&-s_{23}c_{12}
-s_{12}c_{23}s_{13}e^{i\delta}&c_{23}c_{13}
\end{array}\right)
\end{equation}
where $\delta$ is the phase necessary for CP violation. $c_{ij}$ and
$s_{ij}$ can all be chosen to be positive and $\delta$ may vary in the
range $0\le\delta\le 2\pi$. However the measurements
of CP violation in K decays force $\delta$ to be in the range
 $0<\delta<\pi$. 

The extensive phenomenology of the last years 
has shown that
$s_{13}$ and $s_{23}$ are small numbers: $\ord(10^{-3})$ and ${\cal
O}(10^{-2})$,
respectively. Consequently to an excellent accuracy $c_{13}=c_{23}=1$
and the four independent parameters are given as follows
\begin{equation}\label{2.73}
s_{12}=| V_{us}|, \quad s_{13}=| V_{ub}|, \quad s_{23}=|
V_{cb}|, \quad \delta
\end{equation}
with the phase $\delta$ extracted from CP violating transitions or 
loop processes sensitive to $| V_{td}|$. The latter fact is based
on the observation that
 for $0\le\delta\le\pi$, as required by the analysis of CP violation,
there is a one--to--one correspondence between $\delta$ and $|V_{td}|$
given by
\begin{equation}\label{10}
| V_{td}|=\sqrt{a^2+b^2-2 a b \cos\delta},
\qquad
a=| V_{cd} V_{cb}|,
\qquad
b=| V_{ud} V_{ub}|
\end{equation} 

\subsection{Wolfenstein Parameterization Beyond Leading Order}
            \label{sec:sewm:wolfparam}
We will also use the Wolfenstein parametrization
\cite{wolfenstein:83}.
It is an approximate parametrization of the CKM matrix in which
each element is expanded as a power series in the small parameter
$\lambda=| V_{us}|=0.22$
\begin{equation}\label{2.75} 
V=
\left(\begin{array}{ccc}
1-{\lambda^2\over 2}&\lambda&A\lambda^3(\varrho-i\eta)\\ -\lambda&
1-{\lambda^2\over 2}&A\lambda^2\\ A\lambda^3(1-\varrho-i\eta)&-A\lambda^2&
1\end{array}\right)
+\ord(\lambda^4)
\end{equation}
and the set (\ref{2.73}) is replaced by
\begin{equation}\label{2.76}
\lambda, \qquad A, \qquad \varrho, \qquad \eta \, .
\end{equation}
The Wolfenstein parameterization
has several nice features. In particular it offers in conjunction with the
unitarity triangle a very transparent geometrical
representation of the structure of the CKM matrix and allows to derive
several analytic results to be discussed below. This turns out to be very
useful in the phenomenology of rare decays and of CP violation.

When using the Wolfenstein parametrization one should remember that it
is an approximation and that in certain situations neglecting
$\ord(\lambda^4)$ terms may give wrong results. The question then
arises how to find $\ord(\lambda^4)$ and higher order terms ?  The
point is that since \eqn{2.75} is only an approximation the {\em exact}
definiton of $\lambda$ is not unique by terms of the neglected order
$\ord(\lambda^4)$. This is the reason why in different papers in the
literature different $\ord(\lambda^4)$ terms can be found. They simply
correspond to different definitons of the expansion parameter
$\lambda$.  Obviously the physics does not depend on this choice.  Here
it suffices to find an expansion in $\lambda$ which allows for simple
relations between the parameters (\ref{2.73}) and (\ref{2.76}).  This
will also restore the unitarity of the CKM matrix which in the
Wolfenstein parametrization as given in (\ref{2.75}) is not satisfied
exactly.

To this end
we go back to (\ref{2.72}) and we impose the relations \cite{burasetal:94b}
\begin{equation}\label{2.77} 
s_{12}=\lambda
\qquad
s_{23}=A \lambda^2
\qquad
s_{13} e^{-i\delta}=A \lambda^3 (\varrho-i \eta)
\end{equation}
to {\it  all orders} in $\lambda$. In view of the comments made above
this can certainly be done. It follows that
\begin{equation}\label{2.84} 
\varrho=\frac{s_{13}}{s_{12}s_{23}}\cos\delta
\qquad
\eta=\frac{s_{13}}{s_{12}s_{23}}\sin\delta
\end{equation}
We observe that (\ref{2.77}) and (\ref{2.84}) represent simply
the change of variables from (\ref{2.73}) to (\ref{2.76}).
Making this change of variables in the standard parametrization 
(\ref{2.72}) we find the CKM matrix as a function of 
$(\lambda,A,\varrho,\eta)$ which satisfies unitarity exactly!
We also note that in view of $c_{13}=1-\ord(\lambda^6)$ the relations
between $s_{ij}$ and $| V_{ij}|$ in (\ref{2.73}) are 
satisfied to high accuracy. The relations in (\ref{2.84}) have
been first used in \cite{schmidtlerschubert:92}.
However, the improved treatment of the unitarity
triangle presented below goes beyond the analysis of these authors.

The procedure outlined above gives automatically the corrections to the
Wolfenstein parametrization in (\ref{2.75}).  Indeed expressing
(\ref{2.72}) in terms of Wolfenstein parameters using (\ref{2.77})
and then expanding in powers of $\lambda$ we recover the
matrix in (\ref{2.75}) and in addition find explicit corrections of
$\ord(\lambda^4)$ and higher order terms. $V_{ub}$ remains unchanged. The
corrections to $V_{us}$ and $V_{cb}$ appear only at $\ord(\lambda^7)$ and
$\ord(\lambda^8)$, respectively.  For many practical purposes the
corrections to the real parts can also be neglected.
The essential corrections to the imaginary parts are:
\begin{equation}\label{2.83g}
\Delta V_{cd}=-iA^2 \lambda^5\eta
\qquad
\Delta V_{ts}=-iA\lambda^4\eta 
\end{equation}
These two corrections have to be
taken into account in the discussion of CP violation.
On the other hand the imaginary part of $V_{cs}$ which in our expansion
in $\lambda$ appears only at $\ord(\lambda^6)$ can be fully neglected. 

In order to improve the accuracy of the unitarity triangle discussed
below we will also include the $\ord(\lambda^5)$ correction to $V_{td}$
which gives
\begin{equation}\label{2.83d}
 V_{td}= A\lambda^3 (1-\bar\varrho-i\bar\eta) 
\end{equation}
with
\begin{equation}\label{2.88d}
\bar\varrho=\varrho (1-\frac{\lambda^2}{2})
\qquad
\bar\eta=\eta (1-\frac{\lambda^2}{2}).
\end{equation}
In order to derive analytic results we need accurate explicit expressions
for $\lambda_i=V_{id}^{}V_{is}^*$ where $i=c,t$. We have
\begin{equation}\label{2.51}
 Im\lambda_t= -Im\lambda_c=\eta A^2\lambda^5 
\end{equation}
\begin{equation}\label{2.52}
 Re\lambda_c=-\lambda (1-\frac{\lambda^2}{2})
\end{equation}
\begin{equation}\label{2.53}
 Re\lambda_t= -(1-\frac{\lambda^2}{2}) A^2\lambda^5 (1-\bar\varrho) 
\end{equation}
Expressions (\ref{2.51}) and (\ref{2.52}) represent to an accuracy of
0.2\% the exact formulae obtained using (\ref{2.72}). The expression
(\ref{2.53}) deviates by at most 2\% from the exact formula in the
full range of parameters considered. 
In order to keep the analytic
expressions in the phenomenological applications in a transparent form
we have dropped a small $\ord(\lambda^7)$ term in deriving (\ref{2.53}).
After inserting the expressions (\ref{2.51})--(\ref{2.53}) in exact
formulae for quantities of interest, further expansion in $\lambda$
should not be made. 

\subsection{Unitarity Triangle Beyond Leading Order}
            \label{sec:sewm:utriag}
The unitarity of the CKM matrix provides us with several relations
of which
\begin{equation}\label{2.87h}
V_{ud}^{}V_{ub}^* + V_{cd}^{}V_{cb}^* + V_{td}^{}V_{tb}^* =0
\end{equation}
is the most useful one.
In the complex plane the relation (\ref{2.87h}) can be represented as
a triangle, the so-called ``unitarity--triangle'' (UT).
Phenomenologically this triangle is very interesting as it involves
simultaneously the elements $V_{ub}$, $V_{cb}$ and $V_{td}$ which are
under extensive discussion at present.

\begin{figure}[htb]
\vspace{0.05in}
\centerline{
\epsfysize=1.5in
\epsffile{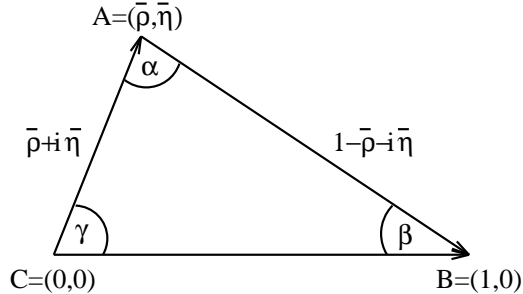}
}
\vspace{0.05in}
\caption[]{\small\sl
Unitarity triangle in the complex $(\bar\varrho,\bar\eta)$ plane.
\label{fig:triangle}}
\end{figure}

In the usual analyses of the unitarity triangle only terms ${\cal
O}(\lambda^3)$ are kept in (\ref{2.87h}) \cite{burasharlander:92},
\cite{nir:74}, \cite{harrisrosner:92}, \cite{schmidtlerschubert:92},
\cite{dibdunietzgilman:90}, \cite{alilondon:95}. It is however
straightforward to include the next-to-leading $\ord(\lambda^5)$
terms \cite{burasetal:94b}. We note first that
\begin{equation}\label{2.88a}
V_{cd}^{}V_{cb}^*=-A\lambda^3+\ord(\lambda^7).
\end{equation}
Thus to an excellent accuracy $V_{cd}^{}V_{cb}^*$ is real with
$| V_{cd}^{}V_{cb}^*|=A\lambda^3$.
Keeping $\ord(\lambda^5)$ corrections and rescaling all terms in
(\ref{2.87h})
by $A \lambda^3$ 
we find
\begin{equation}\label{2.88b}
 \frac{1}{A\lambda^3}V_{ud}^{}V_{ub}^*
=\bar\varrho+i\bar\eta
\qquad,
\qquad
 \frac{1}{A\lambda^3}V_{td}^{}V_{tb}^*
=1-(\bar\varrho+i\bar\eta)
\end{equation}
with $\bar\varrho$ and $\bar\eta$ defined in (\ref{2.88d}). 
Thus we can represent (\ref{2.87h}) as the unitarity triangle 
in the complex $(\bar\varrho,\bar\eta)$
plane. This is  shown in fig.~\ref{fig:triangle}.
 The length of the side $CB$ which lies
on the real axis equals unity when eq.~(\ref{2.87h}) is rescaled by
$V_{cd}^{}V_{cb}^*$. We observe that beyond the leading order
in $\lambda$ the point A {\it does not} correspond to  $(\varrho,\eta)$ but to
 $(\bar\varrho,\bar\eta)$.
Clearly within 3\% accuracy $\bar\varrho=\varrho$ and $\bar\eta=\eta$.
Yet in the distant future the accuracy of experimental results and
theoretical calculations may improve considerably so that the more
accurate formulation given here will be appropriate.

Using simple trigonometry one can calculate $\sin(2\phi_i$), $\phi_i=
\alpha, \beta, \gamma$, in terms of $(\bar\varrho,\bar\eta)$ with the result:
\begin{equation}\label{2.89}
\sin(2\alpha)=\frac{2\bar\eta(\bar\eta^2+\bar\varrho^2-\bar\varrho)}
  {(\bar\varrho^2+\bar\eta^2)((1-\bar\varrho)^2
  +\bar\eta^2)}  
\end{equation}
\begin{equation}\label{2.90}
\sin(2\beta)=\frac{2\bar\eta(1-\bar\varrho)}{(1-\bar\varrho)^2 + \bar\eta^2}
\end{equation}
 \begin{equation}\label{2.91}
\sin(2\gamma)=\frac{2\bar\varrho\bar\eta}{\bar\varrho^2+\bar\eta^2}=
\frac{2\varrho\eta}{\varrho^2+\eta^2}
\end{equation}
The lengths $CA$ and $BA$ in the
rescaled triangle of fig.~\ref{fig:triangle} to be denoted by $R_b$ and $R_t$,
respectively, are given by
\begin{equation}\label{2.94}
R_b \equiv \frac{| V_{ud}^{}V^*_{ub}|}{| V_{cd}^{}V^*_{cb}|}
= \sqrt{\bar\varrho^2 +\bar\eta^2}
= (1-\frac{\lambda^2}{2})\frac{1}{\lambda}
\left| \frac{V_{ub}}{V_{cb}} \right|
\end{equation}
\begin{equation}\label{2.95}
R_t \equiv \frac{| V_{td}^{}V^*_{tb}|}{| V_{cd}^{}V^*_{cb}|} =
 \sqrt{(1-\bar\varrho)^2 +\bar\eta^2}
=\frac{1}{\lambda} \left| \frac{V_{td}}{V_{cb}} \right|
\end{equation}
The expressions for $R_b$ and $R_t$ given here in terms of
$(\bar\varrho, \bar\eta)$ 
are excellent approximations. Clearly $R_b$ and $R_t$
can also be determined by measuring two of the angles $\phi_i$:
\begin{equation}\label{2.96}
R_b=\frac{\sin(\beta)}{\sin(\alpha)}=
\frac{\sin(\alpha+\gamma)}{\sin(\alpha)}=
\frac{\sin(\beta)}{\sin(\gamma+\beta)}
\end{equation}
\begin{equation}\label{2.97}
R_t=\frac{\sin(\gamma)}{\sin(\alpha)}=
\frac{\sin(\alpha+\beta)}{\sin(\alpha)}=
\frac{\sin(\gamma)}{\sin(\gamma+\beta)}
\end{equation}

\section{Basic Formalism}
         \label{sec:basicform}
\subsection{Renormalization of QCD}
            \label{sec:basicform:renorm}
As already emphasized in the introduction, the effects of QCD play an
important role in the phenomenology of weak decays of hadrons. In fact
in the theoretical analysis of these decays the investigation of
QCD corrections is the most difficult and extensive part. In the
present subsection we shall therefore briefly recall basic features
of perturbative QCD and its renormalization. Thereby we will concentrate
on those aspects, that will be needed for the present review. We
will also take the opportunity to introduce for later reference the
expressions for the running coupling, the running mass and the
corresponding renormalization group functions.\\
The Lagrangian density of QCD reads
\begin{eqnarray}\label{lqcd}
{\cal L}_{QCD} & = &
-{1\over 4}(\partial_\mu A^a_\nu-\partial_\nu A^a_\mu)
(\partial^\mu A^{a\nu}-\partial^\nu A^{a\mu})-{1\over{2\xi}}
(\partial^\mu A^a_\mu)^2   \nonumber \\
&&{}+ \bar q(i\not\!\partial-m_q)q+\chi^{a\ast}\partial^\mu
        \partial_\mu \chi^a  \nonumber \\
&&{}-{g\over 2}f^{abc}(\partial_\mu A^a_\nu-\partial_\nu A^a_\mu)
 A^{b\mu}A^{c\nu} - {g^2\over 4}f^{abe}f^{cde}A^a_\mu A^b_\nu
 A^{c\mu}A^{d\nu} \nonumber \\
&&{}+ g \bar q_i T^a_{ij}\gamma^\mu q_j A^a_\mu +
 g f^{abc} (\partial^\mu\chi^{a\ast})\chi^b A^c_\mu
\end{eqnarray}
Here $q=(q_1, q_2, q_3)$ is the color triplet of quark flavor $q$,
$q$ = $u$, $d$, $s$, $c$, $b$, $t$. $g$ is the QCD coupling, $A^a_\mu$
the gluon field, $\chi^a$ the ghost field and $\xi$ the gauge
parameter. $T^a$, $f^{abc}$ ($a$, $b$, $c$ = 1,\ldots,8) are the
generators and structure constants of $SU(3)$, respectively. From
this Lagrangian one may read off the Feynman rules for QCD, e.g.
$i g T^a_{ij}\gamma^\mu$ for the quark-gluon vertex.\\
In order to deal with divergences that appear in quantum (loop)
corrections to Green functions, the theory has to be regularized
to have an explicit parametrization
of the singularities and subsequently renormalized to render
the Green functions finite. For these purposes we will employ:
\begin{itemize}
\item Dimensional regularization (DR) by continuation to $D=4-2\,\eps$
space-time dimensions \cite{bollini:72a}, \cite{bollini:72b},
\cite{thooft:72b}, \cite{ashmore:72}, \cite{cicutamontaldi:72}.
\item Subtraction of divergences in the minimal subtraction scheme $MS$
\cite{thooft:73} or the modified minimal subtraction scheme
($\overline{MS}$) \cite{bardeen:78}.
\end{itemize}
To eliminate the divergences one has to renormalize the fields and
parameters in the Lagrangian, in general through
\begin{equation}\label{zqcd}
\begin{array}{lclcl}
A^a_{0\mu}=Z^{1/2}_3 A^a_\mu &\qquad& q_0=Z^{1/2}_q q &\qquad&
\chi^a_0=\tilde Z^{1/2}_3 \chi^a \\
g_0=Z_g g\mu^\eps &\qquad& \xi_0=Z_3 \xi &\qquad& m_0=Z_m m
\end{array}
\end{equation}
The index ``0'' indicates unrenormalized quantities. The factors $Z$
are the renormalization constants. The scale $\mu$ has been introduced
to make $g$ dimensionless in $D=4-2\,\eps$ dimensions. Since we will
not consider Green functions with external ghosts, we will not need
the ghost field renormalization. We also do not need the gauge
parameter renormalization if we are dealing with gauge independent
quantities, as e.g. Wilson coefficient functions.\\
A straightforward way to implement renormalization is provided by
the coun\-ter\-term method. Thereby the parameters and fields in the
original Lagrangian, which are to be considered as unrenormalized
(bare) quantities, are reexpressed through renormalized ones by
means of \eqn{zqcd} from the very beginning. For instance, the quark
kinetic term becomes
\begin{equation}\label{ctm}
{\cal L}_F=\bar q_0 i\not\!\partial q_0-m_0\bar q_0 q_0\equiv
\bar q i\not\!\partial q-m\bar q q+(Z_q-1)\bar q i\not\!\partial q-
(Z_q Z_m-1)m\bar q q     \end{equation}
The advantage then is, that only renormalized quantities are present
in the Lagrangian. The counterterms ($\sim(Z-1)$), appearing in
addition, can be formally treated as interaction terms that contribute
to Green functions calculated in perturbation theory. The Feynman
rule for the counterterms in \eqn{ctm}, for example, reads
($p$ is the quark momentum)
\begin{equation}\label{ctex}
i(Z_q-1)\not\! p - i(Z_q Z_m-1) m   \end{equation}
The constants $Z_i$ are then determined such that they cancel the
divergences in the Green functions according to the chosen
renormalization scheme. In an analogous way all renormalization constants
can be fixed by considering the appropriate Green functions.
\\
Of central importance for the study of perturbative QCD effects are
the renormalization group equations, which govern the dependence of
renormalized parameters and Green functions on the renormalization
scale $\mu$. These differential equations are easily derived from the
definition \eqn{zqcd} by using the fact that bare quantities are
$\mu$-independent. In this way one finds that the renormalized
coupling $g(\mu)$ obeys \cite{gross:76}
\begin{equation}\label{rgbe}
{d\over d\ln\mu}g(\mu)=\beta(\eps, g(\mu))  \end{equation}
where
\begin{equation}\label{bete}
\beta(\eps, g)=-\eps g-g{1\over Z_g}{d Z_g\over d\ln\mu}\equiv
-\eps g+\beta(g)  \end{equation}
which defines the $\beta$ function. \eqn{rgbe} is valid in arbitrary
dimensions. In four dimensions $\beta(\eps, g)$ reduces to $\beta(g)$.
Similarly, the anomalous dimension of the mass, $\gamma_m$,
defined through
\begin{equation}\label{rggm}
{d m(\mu)\over d\ln\mu}=-\gamma_m(g) m(\mu)  \end{equation}
is given by
\begin{equation}\label{gamz} \gamma_m(g)={1\over Z_m}{d Z_m\over d\ln\mu}  \end{equation}
In the $MS$ $(\overline{MS})$-scheme, where just the pole terms in $\eps$ are
present in the renorma\-lization constants $Z_i$, these can be expanded
as follows
\begin{equation}\label{ziep}
Z_i=1+\sum^\infty_{k=1} {1\over \eps^k} Z_{i, k}(g)  \end{equation}
Using \eqn{rgbe}, \eqn{bete} one finds
\begin{equation}\label{zi1}
{1\over Z_i}{d Z_i\over d\ln\mu}=-2 g^2{\partial Z_{i, 1}(g)\over\partial g^2}\end{equation}
which allows a direct calculation of the renormalization group
functions from the $1/\eps$-pole part of the renormalization
constants. Along these lines one obtains at the two loop level,
required for next-to-leading order calculations,
\begin{equation}\label{bg01}
\beta(g)=-\beta_0{g^3\over 16\pi^2}-\beta_1{g^5\over (16\pi^2)^2}  \end{equation}
In terms of
\begin{equation}\label{aasg} \as\equiv{g^2\over 4\pi}  \end{equation}
we have
\begin{equation}\label{rga}
{d\as\over d\ln\mu}=-2\beta_0 {\as^2\over 4\pi}-2\beta_1
  {\as^3\over(4\pi)^2}  \end{equation}
Similarly, the two-loop expression for the quark mass anomalous
dimension can be written as
\begin{equation}\label{gama}
\gamma_m(\as)=\gamma_{m0}\aspi + \gamma_{m1}\left(\aspi\right)^2
\end{equation}
We also give the $1/\eps$-pole part $Z_{q, 1}$ of the quark field
renormalization constant $Z_1$ to $\ord(\as^2)$, which we will need
later on
\begin{equation}\label{zq1a} Z_{q, 1}=a_1\aspi + a_2\left(\aspi\right)^2
\end{equation}
The coefficients in eqs. \eqn{rga} -- \eqn{zq1a} read
\begin{equation}\label{b0b1}
\beta_0={{11N-2f}\over 3}\qquad
\beta_1={34\over 3}N^2-{10\over 3}Nf-2C_F f\qquad
C_F={{N^2-1}\over{2N}}\end{equation}
\begin{equation}\label{gm01} \gamma_{m0}=6C_F\qquad \gamma_{m1}=C_F\left(
     3C_F+{97\over 3}N-{10\over 3}f\right)  \end{equation}
\begin{equation}\label{a1a2} a_1=-C_F\qquad a_2=C_F\left(
     {3\over 4}C_F-{17\over 4}N+{1\over 2}f\right)  \end{equation}
$N$ is the number of colors, $f$ the number of quark flavors. The
coefficients are given in the $MS$ ($\overline{MS}$) scheme.
However, $\beta_0$, $\beta_1$, $\gamma_{m0}$ and $a_1$ are scheme
independent. The expressions for $a_1$ and $a_2$ in \eqn{a1a2} are
valid in Feynman gauge, $\xi=1$.\\
At two-loop order the solution of the renormalization group equation
\eqn{rga} for $\as(\mu)$ can always be written in the form
\begin{equation}\label{amu}
\as(\mu)={4\pi\over{\beta_0\ln{{\mu^2}\over{\Lambda^2}}}} \left[1-
 {{\beta_1}\over{\beta_0^2}}{{\ln\ln{{\mu^2}\over{\Lambda^2}}}\over
    {\ln{{\mu^2}\over{\Lambda^2}}}}\right] \end{equation}
Eq. \eqn{amu} gives the running coupling constant at NLO. $\as(\mu)$
vanishes as $\mu/\Lambda\to\infty$ due to asymptotic freedom. We
remark that, in accordance with the two-loop accuracy, \eqn{amu} is
valid up to terms of the order $\ord(1/\ln^3\mu^2/\Lambda^2)$.
For the purpose of counting orders in $1/\ln\mu^2/\Lambda^2$ the
double logarithmic expression $\ln\ln\mu^2/\Lambda^2$ may formally
be viewed as a constant. Note that an additional term
${\rm const.}/\ln^2\mu^2/\Lambda^2$, which is of the same order as
the next-to-leading correction term in \eqn{amu}, can always be
absorbed into a multiplicative redefinition of $\Lambda$. Hence the
choice of the form \eqn{amu} is possible without restriction, but one
should keep in mind that the definition of $\Lambda$ is related to this
particular choice. The introduction of the $\overline{MS}$ scheme and
the corresponding definition of $\Lambda_{\overline{MS}}$ and its
relation to $\Lambda_{MS}$ is discussed in
section~\ref{sec:basicform:wc:disc}.
\\
Finally we write down the two-loop expression for the running quark
mass in the $MS$ ($\overline{MS}$) scheme, which results from
integrating \eqn{rggm}
\begin{equation}\label{mmu}
m(\mu)=m(m)\left[{\as(\mu)\over\as(m)}\right]^{\gamma_{m0}\over 2\beta_0}
\left[1+\left({\gamma_{m1}\over 2\beta_0}-{\beta_1\gamma_{m0}\over
  2\beta^2_0}\right){\as(\mu)-\as(m)\over 4\pi}\right]  \end{equation}

\subsection{Operator Product Expansion in Weak Decays -- Preliminaries}
            \label{sec:basicform:prelim}
Weak decays of hadrons are mediated through the weak interactions of
their quark constituents, whose strong interactions, binding the
constituents into hadrons, are characterized by a typical hadronic
energy scale of the order of $1\gev$. Our goal is therefore to derive
an effective low energy theory describing the weak interactions of
quarks. The formal framework to achieve this is provided by the
operator product expansion (OPE). In order to introduce the main ideas
behind it, let us consider the simple example of
the quark level transition $c\to su\bar d$,
which is relevant for Cabibbo-allowed decays of D mesons.
Disregarding QCD effects for the moment, the tree-level W-exchange
amplitude for $c\to su\bar d$ is simply given by
\begin{eqnarray}\label{aope}
A&=&i{G_F\over\sqrt{2}}V^*_{cs}V_{ud}^{}{M^2_W\over k^2-M^2_W}
  (\bar sc)_{V-A}(\bar ud)_{V-A} \nonumber\\
 &=& -i{G_F\over\sqrt{2}}V^*_{cs}V_{ud}^{}
  (\bar sc)_{V-A}(\bar ud)_{V-A} + \ord({k^2\over M^2_W})
\end{eqnarray}
where $(V-A)$ refers to the Lorentz structure $\gamma_{\mu} (1-\gf)$.

Since $k$, the momentum transfer through the W propagator, is very
small as compared to the W mass $M_W$, terms of the order
$\ord(k^2/M^2_W)$ can safely be neg\-lected and the full amplitude $A$
can be approximated by the first term on the r.h.s. of \eqn{aope}.
Now this term may obviously be also obtained from an effective
hamiltonian defined by
\begin{equation}\label{hc0}
{\cal H}_{eff}={G_F\over\sqrt{2}}V^*_{cs}V_{ud}^{}
  (\bar sc)_{V-A}(\bar ud)_{V-A} + \ldots  \end{equation}
where the ellipsis denotes operators of higher dimensions, typically
involving derivative terms, which can in principle be chosen so
as to reproduce the terms of higher order in $k^2/M^2_W$ of the
full amplitude in \eqn{aope}. This exercise already provides us with
a simple example of an OPE. The product of two charged current
operators is expanded into a series of local operators,
whose contributions are weighted by effective coupling constants,
the Wilson coefficients.\\
A more formal basis for this procedure may be given by considering
the genera\-ting functional for Green functions in the path integral
formalism. The part of the generating functional relevant for the
present discussion is, up to an overall normalizing factor, given by
\begin{equation}\label{zw1}
Z_W\sim\int [dW^+][dW^-] \exp(i\int d^4x {\cal L}_W)  \end{equation}
where ${\cal L}_W$ is the Lagrangian density containing the kinetic
terms of the W boson field and its interaction with charged currents
\begin{eqnarray}\label{lwjw}
\lefteqn{{\cal L}_W=
-{1\over 2}(\partial_\mu W^+_\nu-\partial_\nu W^+_\mu)
 (\partial^\mu W^{-\nu}-\partial^\nu W^{-\mu})+M^2_W W^+_\mu W^{-\mu}}
\hspace{3cm} \nonumber\\
& & +{g_2\over 2\sqrt{2}}(J^+_\mu W^{+\mu}+J^-_\mu W^{-\mu})
\end{eqnarray}
\begin{equation}\label{jpn}
J^+_\mu=V_{pn} \bar p\gamma_\mu(1-\gf)n\qquad p=(u, c, t)
\quad n=(d, s, b)\qquad J^-_\mu=(J^+_\mu)^\dagger  \end{equation}
Since we are not interested in Green functions with external W lines,
we have not introduced external source terms for the W fields. In the
present argument we will furthermore choose the unitary gauge for
the W field for definiteness, however physical results do not depend
on this choice.\\
Introducing the operator
\begin{equation}\label{kxy}
K_{\mu\nu}(x, y)=\delta^{(4)}(x-y)\left(g_{\mu\nu}(\partial^2+
  M^2_W)-\partial_\mu\partial_\nu\right)   \end{equation}
we may, after discarding a total derivative in the W kinetic term,
rewrite \eqn{zw1} as
\begin{eqnarray}\label{zw2}    \lefteqn{
Z_W\sim\int [dW^+][dW^-] \exp\biggl[ i\int d^4x d^4y W^+_\mu(x)
K^{\mu\nu}(x, y) W^-_\nu(y)  } \hspace{3cm}\nonumber\\
& & {}+i{g_2\over 2\sqrt{2}}\int d^4x
 J^+_\mu W^{+\mu}+J^-_\mu W^{-\mu} \biggr]
\end{eqnarray}
The inverse of $K_{\mu\nu}$, denoted by $\Delta_{\mu\nu}$, and
defined through
\begin{equation}\label{kde1}
\int d^4y K_{\mu\nu}(x, y) \Delta^{\nu\lambda}(y, z)=
 g^{\ \lambda}_\mu \delta^{(4)}(x-z)  \end{equation}
is just the W propagator in the unitary gauge
\begin{equation}\label{dexy}
\Delta_{\mu\nu}(x, y)=\int{d^4k\over (2\pi)^4}\Delta_{\mu\nu}(k)
  e^{-i k(x-y)}  \end{equation}
\begin{equation}\label{dek}
\Delta_{\mu\nu}(k)={-1\over k^2-M^2_W}\left(g_{\mu\nu}-
  {k_\mu k_\nu\over M^2_W}\right)   \end{equation}
Performing the gaussian functional integration over $W^\pm(x)$ in
\eqn{zw2} explicitly, this expression simplifies to
\begin{equation}\label{zetw}
Z_W\sim\exp\left[ -i\int{g^2_2\over 8}J^-_\mu(x)\Delta^{\mu\nu}(x, y)
J^+_\nu(y) d^4x d^4y \right]   \end{equation}
This result implies a nonlocal action functional for the quarks
\begin{equation}\label{snl}
{\cal S}_{nl}=\int d^4x {\cal L}_{kin}-
{g^2_2\over 8}\int d^4x d^4y J^-_\mu(x)\Delta^{\mu\nu}(x, y)
J^+_\nu(y)    \end{equation}
where the first piece represents the quark kinetic terms and the second
their charged current interactions.\\
We can now formally expand this second, nonlocal term in powers of
$1/M^2_W$ to yield a series of local interaction operators of
dimensions that increase with the order in $1/M^2_W$. To lowest order
\begin{equation}\label{dloc}
\Delta^{\mu\nu}(x, y)\approx{g^{\mu\nu}\over M^2_W}\delta^{(4)}(x-y) \end{equation}
and the second term in \eqn{snl} becomes
\begin{equation}\label{jjx}
-{g^2_2\over 8M^2_W}\int d^4x J^-_\mu(x) J^{+\mu}(x)  \end{equation}
corresponding to the usual effective charged current interaction
Lagrangian
\begin{equation}\label{leff}
{\cal L}_{int,eff}=-{G_F\over\sqrt{2}} J^-_\mu J^{+\mu}(x)=-
{G_F\over\sqrt{2}}V^*_{pn}V_{p'n'}^{}(\bar np)_{V-A}
  (\bar p'n')_{V-A}
\end{equation}
which contains, among other terms, the leading contribution to
\eqn{hc0}.

The simple considerations we have presented so far already illustrate
several of the basic aspects of the general approach.
\begin{itemize}
\item Formally, the procedure to approximate the interaction term in
\eqn{snl} by \eqn{jjx} is an example of a short-distance OPE. The product
of the local operators $J^-_\mu(x)$ and $J^+_\nu(y)$, to be taken at
short-distances due to the convolution with the massive, short-range
W propagator $\Delta^{\mu\nu}(x, y)$ (compare \eqn{dloc}), is expanded
into a series of composite local operators, of which the leading term
is shown in \eqn{jjx}.
\item The dominant contributions in the short-distance expansion come
from the operators of lowest dimension. In our case these are
four-fermion operators of dimension six, whereas operators of higher
dimensions can usually be neglected in weak decays.
\item Note that, as far as the charged current weak interaction is
concerned, no approximation is involved yet in the nonlocal interaction
term in \eqn{snl}, except that we do not consider higher order weak
corrections or processes with external W boson states. Correspondingly,
the OPE series into which the nonlocal interaction is expanded, is
equivalent to the original theory, when considered to all orders in
$1/M^2_W$. In other words, the full series will reproduce the complete
Green functions for the charged current weak interactions of quarks.
The truncation of the operator series then yields a systematic
approximation scheme for low energy processes, neglecting contributions
suppressed by powers of $k^2/M^2_W$. In this way one is able to
construct low energy effective theories for weak decays.
\item In going from the full to the effective theory the W boson is
removed as an explicit, dynamical degree of freedom. This step is
often refered to as ``integrating out'' the W boson, a terminology
which is very obvious in the path integral language discussed above.
Alternatively one could of course use the canonical operator
formalism, where the W field instead of being intergrated out, gets
``contracted out'' through the application of Wick's theorem.
\item The effective local four-fermion interaction terms are a
modern version of the classic Fermi-theory of weak interactions.
\item An intuitive interpretation of the OPE formalism discussed so
far is, that from the point of view of low energy dynamics, the effects
of a short-range exchange force mediated by a heavy boson
approximately corresponds to a point interaction.
\item The presentation we have given illustrates furthermore, that
the approach of evaluating the relevant Green functions (or amplitudes)
directly in order to construct the OPE, as in \eqn{aope}, actually gives
the same result as the more formal technique employing path integrals.
While the latter can give some useful insight into the general aspects
of the method, the former is more convenient for practical calculations
and we will make use of it throughout the discussion to follow.
\item Up to now we have not talked about the strong interactions
among quarks, which have of course to be taken into account. They are
described by QCD and can at short-distances be calculated in
perturbation theory, due to the property of asymptotic freedom of QCD.
The corresponding gluon exchange contributions constitute quantum
corrections to the simplified picture sketched above, which can in this
sense be viewed as a classical approximation. We will describe the
incorporation of QCD corrections and related additional features
they imply for the OPE in the following section.
\end{itemize}

\subsection{OPE and Short Distance QCD Effects}
            \label{sec:basicform:ope}
\begin{figure}[htb]
\vspace{0.10in}
\centerline{
\epsfysize=3in
\epsffile{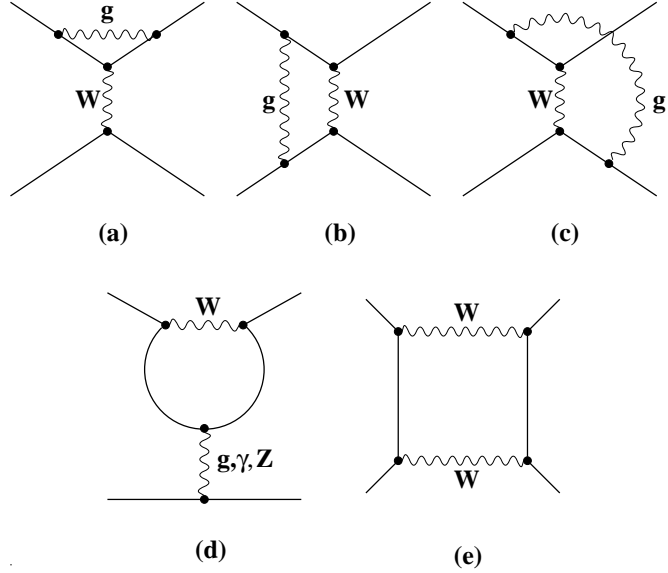}
}
\vspace{0.08in}
\caption[]{One-loop current-current (a)--(c), penguin (d) and box (e)
diagrams in the full theory. For pure QCD corrections as considered in
this section and e.g.\ in \ref{sec:HeffdF1:66} the $\gamma$- and
$Z$-contributions in diagram (d) and the diagram (e) are absent.
Possible left-right or up-down reflected diagrams are not shown.
\label{fig:1loopful}}
\end{figure}

\begin{figure}[htb]
\vspace{0.10in}
\centerline{
\epsfysize=3in
\epsffile{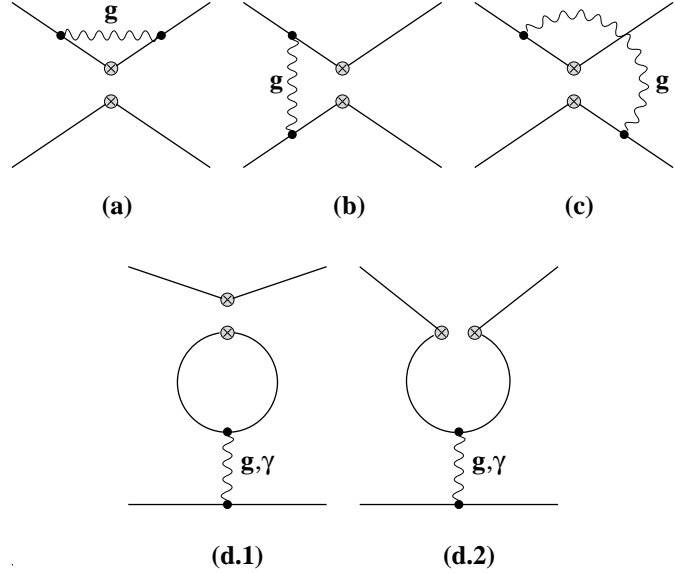}
}
\vspace{0.08in}
\caption[]{One loop current-current (a)--(c) and penguin (d) diagrams
contributing to the LO anomalous dimensions and matching conditions in
the effective theory. The 4-vertex ``$\otimes\,\,\otimes$'' denotes the
insertion of a 4-fermion operator $Q_i$. For pure QCD corrections as
considered in this section and e.g.\ in \ref{sec:HeffdF1:66} the
contributions from $\gamma$ in diagrams (d.1) and (d.2) are absent.
Again, possible left-right or up-down reflected diagrams are not
shown.
\label{fig:1loopeff}}
\end{figure}

We will now take up the discussion of QCD quantum corrections at
short-distances to the OPE for weak decays. A crucial point for this
enterprise is the property of asymptotic freedom of QCD. This allows
one to treat the short-distance corrections, that is the contribution
of hard gluons at energies of the order $\ord(M_W)$ down to hadronic
scales $\ge 1\gev$, in perturbation theory. In the following, we will
always restrict ourselves to the leading dimension six operators in
the OPE and omit the negligible contributions of higher dimensional
operators. Staying with our example of $c\to su\bar d$ transitions,
recall that we had for the amplitude without QCD
\begin{equation}\label{amp0}
A_0=-i{G_F\over\sqrt{2}}V^*_{cs}V_{ud}^{}
  (\bar s_ic_i)_{V-A}(\bar u_jd_j)_{V-A}
\end{equation}
where the summation over repeated color indices is understood.  This
result leads directly to the effective hamiltonian of \eqn{hc0} where the
color indices have been suppressed. If we now include QCD effects, the
effective hamiltonian, constructed to reproduce the low energy
approximation of the exact theory, is generalized to

\begin{equation}\label{hq12}
{\cal H}_{eff}={G_F\over\sqrt{2}}V^\ast_{cs}V_{ud}(C_1 Q_1+C_2 Q_2) \end{equation}
where
\begin{equation}\label{q1c} Q_1=(\bar s_ic_j)_{V-A}(\bar u_jd_i)_{V-A}   \end{equation}
\begin{equation}\label{q2c} Q_2=(\bar s_ic_i)_{V-A}(\bar u_jd_j)_{V-A}   \end{equation}
The essential features of this hamiltonian are:
\begin{itemize}
\item In addition to the original operator $Q_2$ (with index 2 for
historical reasons) a new operator $Q_1$ with the same flavor form
but different color structure is generated. This is because a gluon
linking the two color singlet weak current lines can ``mix'' the color
indices due to the following relation for the color charges $T^a_{ij}$
\begin{equation}\label{tata}
T^a_{ik}T^a_{jl}=-{1\over 2N}\delta_{ik}\delta_{jl}+{1\over 2}
\delta_{il}\delta_{jk}   \end{equation}
\item The Wilson coefficients $C_1$ and $C_2$, the coupling constants
for the interaction terms $Q_1$ and $Q_2$, become calculable nontrivial
functions of $\as$, $M_W$ and the renormalization scale $\mu$.
If QCD is neglected they have the trivial form $C_1=0$, $C_2=1$ and
\eqn{hq12} reduces to \eqn{hc0}.
\end{itemize}
In order to obtain the final result for the hamiltonian \eqn{hq12}, we
have to calculate the coefficients $C_{1, 2}$. These are determined
by the requirement that the amplitude $A$ in the full theory be
reproduced by the corresponding amplitude in the effective theory
\eqn{hq12}, thus
\begin{equation}\label{acq}
A=-i{G_F\over\sqrt{2}}V^*_{cs}V_{ud}^{}(C_1\langle Q_1\rangle +
C_2\langle Q_2\rangle)   \end{equation}
If we calculate the amplitude $A$ and, to the same order in $\as$,
the matrix elements of operators $\langle Q_1\rangle$,
$\langle Q_2\rangle$, we can obtain $C_1$ and $C_2$ via \eqn{acq}.
This procedure is called {\it matching} the full theory onto the
effective theory \eqn{hq12}.
\\
Here we use the term ``amplitude'' in the meaning of ``amputated Green
function''. Correspondingly operator matrix elements are -- within this
perturbative context -- amputated Green functions with operator
insertion. In a diagrammatic language these amputated Green functions
are given by Feynman graphs, but without gluonic self energy
corrections in external legs, like e.g. in figs.~\ref{fig:1loopful} and
~\ref{fig:1loopeff} for the full and effective theory, respectively.
In the present example penguin diagrams do not contribute due to the
flavor structure of the $c \to s u \bar d$ transition.

Evaluating the current-current diagrams of fig.~\ref{fig:1loopful}\,(a)--(c),
we find for the full amplitude $A$ to $\ord(\as)$
\begin{equation}\label{amp}
A=-i{G_F\over\sqrt{2}}V^\ast_{cs}V_{ud}\left[\left(1+2C_F \aspi
\ln{\mu^2\over -p^2}\right)S_2+{3\over N}\aspi\ln{M^2_W\over -p^2} S_2-
3\aspi\ln{M^2_W\over -p^2} S_1\right]   \end{equation}
Here we have introduced the spinor amplitudes
\begin{equation}\label{s1c} S_1=(\bar s_ic_j)_{V-A}(\bar u_jd_i)_{V-A}
\end{equation}
\begin{equation}\label{s2c} S_2=(\bar s_ic_i)_{V-A}(\bar u_jd_j)_{V-A}
\end{equation}
which are just the tree level matrix elements of $Q_1$ and $Q_2$.
We have employed the Feynman gauge ($\xi=1$) and taken all external
quark lines massless and carrying the off-shell momentum $p$.
Furthermore we have kept only logarithmic corrections
$\sim\as\cdot\log$ and discarded constant contributions of order
$\ord(\as)$, which corresponds to the leading log approximation.
The necessary renormalization of the quark fields in the $MS$-scheme
is already incorporated into \eqn{amp}. It has removed a $1/\eps$
singularity in the first term of \eqn{amp}, which therefore carries
an explicit $\mu$-dependence.

Under the same conditions, the unrenormalized current-current matrix elements
of the operators $Q_1$ and $Q_2$ are from fig.~\ref{fig:1loopeff}\,(a)-(c)
found to be
\begin{eqnarray}\label{q10}
\lefteqn{\langle Q_1\rangle^{(0)}=} & & \\
& & \left(1+2C_F \aspi\left({1\over\eps}+\ln{\mu^2\over -p^2}
\right)\right)S_1+{3\over N}\aspi\left({1\over\eps}+\ln{\mu^2\over -p^2}\right)S_1-
3\aspi\left({1\over\eps}+\ln{\mu^2\over -p^2}\right) S_2  \nonumber
\end{eqnarray}
\begin{eqnarray}\label{q20}
\lefteqn{\langle Q_2\rangle^{(0)}=} & & \\
& & \left(1+2C_F \aspi\left({1\over\eps}+\ln{\mu^2\over -p^2}
\right)\right)S_2+{3\over N}\aspi\left({1\over\eps}+\ln{\mu^2\over -p^2}\right)S_2-
3\aspi\left({1\over\eps}+\ln{\mu^2\over -p^2}\right) S_1  \nonumber
\end{eqnarray}
Again, the divergences in the first terms are eliminated through field
renormalization. However, in contrast to the full amplitude, the
resulting expressions are still divergent. Therefore an additional
multiplicative renormalization, refered to as {\em operator renormalization},
is necessary:
\begin{equation}
Q_i^{(0)} = Z_{ij} Q_j
\label{AL}
\end{equation}
Since \eqn{q10} and \eqn{q20} each involve both $S_1$ and
$S_2$, the renormalization constant is in this case a $2\times 2$
matrix $Z$. The relation between the unrenormalized
($\langle Q_i\rangle^{(0)}$) and the renormalized amputated Green
functions ($\langle Q_i\rangle$) is then
\begin{equation}\label{q0zq}
\langle Q_i\rangle^{(0)}=Z^{-2}_q Z_{ij}\langle Q_j\rangle
\end{equation}
From \eqn{q10}, \eqn{q20} and \eqn{zq1a} we read off ($MS$-scheme)
\begin{equation}\label{zll} Z = 1+ \aspi {1\over\eps}
 \left(\begin{array}{cc}  3/N & -3 \\
                          -3 & 3/N
    \end{array}\right)   \end{equation}
It follows that the renormalized matrix elements $\langle Q_i\rangle$
are given by
\begin{equation}\label{q1re}
\langle Q_1\rangle=\left(1+2C_F \aspi\ln{\mu^2\over -p^2}
\right)S_1+{3\over N}\aspi\ln{\mu^2\over -p^2}S_1-
3\aspi\ln{\mu^2\over -p^2} S_2   \end{equation}
\begin{equation}\label{q2re}
\langle Q_2\rangle=\left(1+2C_F \aspi\ln{\mu^2\over -p^2}
\right)S_2+{3\over N}\aspi\ln{\mu^2\over -p^2}S_2-
3\aspi\ln{\mu^2\over -p^2} S_1   \end{equation}
Inserting $\langle Q_i\rangle$ into \eqn{acq} and comparing with \eqn{amp}
we derive
\begin{equation}\label{c12}
C_1=-3\aspi\ln{M^2_W\over\mu^2}   \qquad
C_2=1+{3\over N}\aspi\ln{M^2_W\over\mu^2}   \end{equation}
We would like to digress and add a comment on the renormalization of
the interaction terms in the effective theory. The commonly used
convention is to introduce via \eqn{q0zq} the renormalization constants
$Z_{ij}$, defined to absorb the divergences of the operator matrix
elements. It is however instructive to view this renormalization in
a slightly different, but of course equivalent way, corresponding to
the standard counter\-term method in perturbative renormalization.
Consider, as usual, the hamiltonian of the effective theory as the
starting point with fields and coupling constants as bare quantities,
which are renormalized according to ($q$=$s$, $c$, $u$, $d$)
\begin{equation}\label{q0z2}
q^{(0)}=Z^{1/2}_q q  \end{equation}
\begin{equation}\label{c0zc}  C^{(0)}_i=Z^c_{ij} C_j \end{equation}
Then the hamiltonian \eqn{hq12} is essentially (omitting the factor
${G_F\over\sqrt{2}}V^\ast_{cs}V_{ud}$)
\begin{equation}\label{ctcq}
C^{(0)}_iQ_i(q^{(0)})\equiv Z^2_qZ^c_{ij} C_jQ_i\equiv
C_iQ_i+(Z^2_q Z^c_{ij}-\delta_{ij})C_jQ_i  \end{equation}
that is, it can be written in terms of renormalized couplings and
fields ($C_iQ_i$), plus coun\-ter\-terms. The argument $q^{(0)}$ in the
first term in \eqn{ctcq} indicates that the interaction term $Q_i$ is
composed of bare fields. Calculating the amplitude with the
hamiltonian \eqn{ctcq}, which includes the counterterms, we get the
finite renormalized result
\begin{equation}\label{zqzc}
Z^2_qZ^c_{ij} C_j\langle Q_i\rangle^{(0)}=
C_j\langle Q_j\rangle     \end{equation}
Hence (compare \eqn{q0zq})
\begin{equation}\label{zcz}
Z^c_{ij}=Z^{-1}_{ji}  \end{equation}
In short, it is sometimes useful to keep in mind that one can think
of the ``operator renormalization'', which sounds like a new concept,
in terms of the completely equivalent, but customary, renormalization
of the coupling constants $C_i$, as in any field theory.

Now that we have presented in quite some detail the derivation of
the Wilson coefficients in \eqn{c12}, we shall discuss and interpret
the most important aspects of the short-distance expansion for
weak decays, which can be studied very transparently on the explicit
example we have given.
\begin{itemize}
\item First of all a further remark about the phenomenon of
operator mixing that we encountered in our example. This occurs
because gluonic corrections to the matrix element of the original
operator $Q_2$ are not just proportional to $Q_2$ itself, but involve
the additional structure $Q_1$ (and vice versa). Therefore, besides
a $Q_2$-counterterm, a counterterm $\sim Q_1$ is needed to renormalize
this matrix element -- the operators in question are said to mix
under renormalization. This however is nothing new in principle. It
is just an algebraic generalization of the usual concepts. Indeed,
if we introduce a different operator basis $Q_\pm=(Q_2\pm Q_1)/2$
(with coefficients $C_\pm=C_2\pm C_1$) the renormalization becomes
diagonal and matrix elements of $Q_+$ and $Q_-$ are renormalized
multiplicatively. In this new basis the OPE reads
\begin{equation}\label{apam}  A\equiv A_++A_- =
-i{G_F\over\sqrt{2}}V^\ast_{cs}V_{ud}(C_+\langle Q_+\rangle +
C_-\langle Q_-\rangle)   \end{equation}
where ($S_\pm=(S_2\pm S_1)/2$)
\begin{equation}\label{aspm}
A_\pm=-i{G_F\over\sqrt{2}}V^\ast_{cs}V_{ud}\left[\left(1+2C_F \aspi
\ln{\mu^2\over -p^2}\right)S_\pm+({3\over N}\mp 3)\aspi\ln{M^2_W\over -p^2} S_\pm
\right]   \end{equation}
and
\begin{equation}\label{qmpm}
\langle Q_\pm\rangle=\left(1+2C_F \aspi\ln{\mu^2\over -p^2}
\right)S_\pm+({3\over N}\mp 3)\aspi\ln{\mu^2\over -p^2} S_\pm  \end{equation}
\begin{equation}\label{cpm}
C_\pm=1+({3\over N}\mp 3)\aspi \ln{M^2_W\over\mu^2}   \end{equation}
\item In the calculation of the amplitude $A$ in \eqn{amp} and of the
matrix elements in \eqn{q10} and \eqn{q20} the off-shell momentum $p$ of
the external quark legs represents an infrared regulator. The
logarithmic infrared divergence of the gluon correction diagrams
(figs.~\ref{fig:1loopful}\,(a)--(c) and \ref{fig:1loopeff}\,(a)--(c))
as $p^2\to 0$ is evident from \eqn{amp}, \eqn{q10}
and \eqn{q20}.  A similar observation can be made for the $M_W$
dependence of the full amplitude $A$. We see that \eqn{amp} is
logarithmically divergent in the limit $M_W\to\infty$. This behaviour
is reflected in the ultraviolet divergences (persisting after field
renormalization) of the matrix elements \eqn{q10}, \eqn{q20} in the
effective theory, whose local interaction terms correspond to the weak
interactions in the infinite $M_W$ limit as they are just the leading
contribution of the $1/M_W$ operator product expansion.  This also
implies, that the characteristic logarithmic functional dependence of
the leading $\ord(\as)$ corrections is closely related to the
divergence structure of the effective theory, that is to the
renormalization constants $Z_{ij}$.
\item The most important feature of the OPE is that it provides a
factorization of short-distance (coefficients) and long-distance
(operator matrix elements) contributions. This is clearly exhibited
in our example. The dependence of the amplitude \eqn{amp} on $p^2$,
representing the long-distance structure of $A$, is fully contained
in the matrix elements of the local operators $Q_i$ \eqn{q1re}, \eqn{q2re},
whereas the Wilson coefficients $C_i$ in \eqn{c12} are free from this
dependence. Essentially, this factorization has the form
(see \eqn{aspm} -- \eqn{cpm})
\begin{equation}\label{fact}
(1+\as G \ln{M^2_W\over -p^2})\doteq
(1+\as G \ln{M^2_W\over\mu^2})\cdot
(1+\as G \ln{\mu^2\over -p^2})        \end{equation}
that is, amplitude = coefficient function $\times$ operator matrix
element. Hereby the logarithm on the l.h.s. is split according to
\begin{equation}\label{splt}
\ln{M^2_W\over -p^2}=\ln{M^2_W\over\mu^2}+ \ln{\mu^2\over -p^2}  \end{equation}
Since the logarithmic behaviour results from the integration over
some virtual loop momentum, we may -- roughly speaking -- rewrite
this as
\begin{equation}\label{pmuw}
\int^{M^2_W}_{-p^2}{d k^2\over k^2}=
\int^{M^2_W}_{\mu^2}{d k^2\over k^2} +
\int^{\mu^2}_{-p^2}{d k^2\over k^2}   \end{equation}
which illustrates that the coefficient contains the contributions
from large virtual momenta of the loop correction from scales
$\mu\approx 1\gev$ to $M_W$, whereas the low energy contributions are
separated into the matrix elements.\\
Of course, the latter can not be calculated in perturbation
theory for transitions between physical meson states. The point is,
that we have calculated the OPE for unphysical off-shell quark
external states only to extract the Wilson coefficients, which we
need to construct the effective hamiltonian \eqn{hq12}. For this
purpose the fact that we have considered an unphysical amplitude is
irrelevant since the coefficient functions do not depend on the
external states, but represent the short-distance structure of the
theory. Once we have extracted the coefficients and written down
the effective hamiltonian, the latter can be used -- at least in
principle -- to evaluate the physically interesting decay amplitudes
by means of some nonperturbative approach.
\item In interpreting the role of the scale $\mu$ we may distinguish
two different aspects. From the point of view of the effective
theory $\mu$ is just a renormalization scale, introduced in the
process of renormalizing the effective local interaction terms by
the dimensional method. On the other hand, from the point of view of
the full theory, $\mu$ acts as the scale at which the full
contribution is separated into a low energy and a high energy part,
as is evident from the above discussion. For this reason $\mu$ is
sometimes also called the {\it factorization scale}.
\item In our case the infrared structure of the amplitude is
characterized by the off-shell momentum $p$.
In general one could work with any other arbitrary momentum
configuration, on-shell or off-shell, with or without external
quark mass, with infrared divergences regulated by off-shell
momenta, quark masses, a fictitious gluon mass or by dimensional
regularization. In the case of off-shell momenta the amplitude is
furthermore dependent on the gauge parameter of the gluon field.
All these things belong to the infrared or long-distance structure of
the amplitude. Therefore the dependence on these choices is the same
for the full amplitude and for the operator matrix elements and
drops out in the coefficient functions. To check that this is really
the case for a particular choice is of crucial importance for
practical calculations. On the other hand one may use this freedom and
choose the treatment of external lines according to convenience or
taste. Sometimes it may however seem preferable to keep a slightly more
inconvenient dependence on external masses and/or gluon gauge in
order to have a useful check that this dependence does indeed cancel
out for the Wilson coefficients one is calculating.
\end{itemize}

\subsection{The Renormalization Group}
            \label{sec:basicform:rg}
\subsubsection{Basic Concepts}
               \label{sec:basicform:rg:basic}
So far we have computed the Wilson coefficient functions \eqn{cpm} in ordinary
perturbation theory. This, however, is not sufficient for the problem
at hand. The appropriate scale at which to normalize the hadronic
matrix elements of local operators is a low energy scale -- low
compared to the weak scale $M_W$ -- of a few $GeV$ typically.
In our example of charm decay $\mu=\ord(m_c)$. For such a low scale
$\mu$ the logarithm $\ln(M^2_W/\mu^2)$ multiplying $\as(\mu)$
in the expression \eqn{cpm} becomes large. Although $\as(\mu)$
by itself is a valid expansion parameter down to scales of
$\ord(1\gev)$, say, this is not longer true for the combination
$\as(\mu) \ln(M^2_W/\mu^2)$. In fact, for our example \eqn{cpm}
the first order correction term amounts for $\mu=1\gev$ to
65 -- 130\% although $\as/4\pi\approx 4\%$. The reason for
this breakdown of the naive perturbative expansion lies
ultimately in the appearance of largely disparate scales $M_W$ and
$\mu$ in the problem at hand. \\
This situation can be considerably improved by employing the
method of the renormalization group (RG). The renormalization group
is the group of transformations between different choices of the
renormalization scale $\mu$. The renormalization group equations
describe the change of renormalized quantities, Green functions
and parameters, with $\mu$ in a differential form. As we shall
illustrate below, solving these differential equations allows,
in the leading logarithmic approximation (LLA),
to sum up the terms $(\as\ln(M_W/\mu))^n$ to all orders $n$ ($n=0,
\ldots, \infty$) in perturbation theory. This leads to the RG improved
perturbation theory. Going one step beyond in this modified expansion,
to the next-to-leading logarithmic approximation (NLLA), the summation
is extended to all terms $\as(\as\ln(M_W/\mu))^n$, and so on.
In this context it is useful to consider $\as\ln(M_W/\mu)$ with
a large logarithm $\ln(M_W/\mu)$ as a quantity of order $\ord(1)$
\begin{equation}\label{alog}
\as \ln{M_W\over\mu}=\ord(1)\qquad  \mu\ll M_W   \end{equation}
Therefore the series in powers of $\as\ln(M_W/\mu)$ cannot be
truncated. Summed to all orders it yields again a contribution of
order $\ord(1)$. Correspondingly the next-to-leading logs
$\as(\as\ln(M_W/\mu)^n$ represent an $\ord(\as)$
perturbative correction to the leading term.\\
The renormalization group equation for the Wilson coefficient functions
follows from the fact, that the unrenormalized Wilson coefficients
$\vec C^{(0)}=Z_c \vec C$ ($\vec C^T=(C_1, C_2)$)
are $\mu$-independent. Defining the matrix
of anomalous dimensions $\gamma$ by
\begin{equation}\label{gazz} \gamma=Z^{-1}{d\over d\ln\mu}Z  \end{equation}
and recalling that $Z^T_c=Z^{-1}$, we obtain the renormalization group
equation
\begin{equation}\label{rgc}
{d\over d\ln\mu}\vec C(\mu)=\gamma^T(\as) \vec C(\mu)  \end{equation}
The solution of \eqn{rgc} may formally be written in terms of a
$\mu$-evolution matrix $U$ as
\begin{equation}\label{rgcu}
\vec C(\mu)=U(\mu, M_W) \vec C(M_W)  \end{equation}
From \eqn{zll} and \eqn{gazz} we have to first order in $\as$
\begin{equation}\label{g120} \gamma(\as)=\aspi \gamma^{(0)}=\aspi
 \left(\begin{array}{cc} -6/N & 6 \\
                          6 & -6/N
    \end{array}\right)   \end{equation}
or in the diagonal basis
\begin{equation}\label{gpm0}
\gamma_\pm(\as)=\aspi\gamma^{(0)}_\pm \qquad  \gamma^{(0)}_\pm=
   \pm 6{N\mp1\over N}  \end{equation}
Note that if we neglect QCD loop corrections completely, the
couplings $\vec C$ are independent of $\mu$. The nontrivial
$\mu$-dependence of $\vec C$ expressed in \eqn{rgc} is a genuine
quantum effect. It implies an anomalous scaling behaviour for the
dimensionless coefficients, i.e. one that is different from the
classical theory. For this reason the factor $\gamma$ is called
anomalous (scale) dimension (compare \eqn{rgc} with
${d\over d\ln\mu}\mu^n=n \mu^n$ for an n-dimensional $\mu$-dependent
term $\mu^n$).\\
Using \eqn{rga} the RG equation \eqn{rgc} is easily solved with the
result
\begin{equation}\label{cpmrg}
C_\pm(\mu)=\left[{\as(M_W)\over\as(\mu)}\right]^{\gamma^{(0)}_\pm\over
  2\beta_0} C_\pm(M_W)  \end{equation}
At a scale $\mu_W=M_W$ no large logarithms are present and $C_\pm(M_W)$
can therefore be calculated in ordinary perturbation theory.
From \eqn{cpm} we have to the order needed for the LLA
\begin{equation}\label{cmw1}  C_\pm(M_W)=1  \end{equation}
\eqn{cpmrg} and \eqn{cmw1} give the final result for the coefficients
in the leading log approximation of RG improved perturbation theory.
\\
At this point one should emphasize, that the choice of the high energy
matching scale $\mu_W=M_W$ is of course not unique. The only
requirement is that the choice of $\mu_W$ must not introduce large
logs $\ln(M_W/\mu_W)$ in order not to spoil the applicability of
the usual perturbation theory. Therefore $\mu_W$ should be of
order $\ord(M_W)$. The logarithmic correction in \eqn{cpm} is then
$\ord(\as)$ and is neglected in LLA. Then, still,
$C_\pm(\mu_W)=1$ and
\begin{equation}\label{cpmr2}
C_\pm(\mu)=\left[{\as(\mu_W)\over\as(\mu)}\right]^{\gamma^{(0)}_\pm\over
  2\beta_0} =
\left[{\as(M_W)\over \as(\mu)}\right]^{\gamma^{(0)}_\pm\over
  2\beta_0} (1+\ord(\as)) \end{equation}
A change of $\mu_W$ around the value of $M_W$ causes an ambiguity
of $\ord(\as)$ in the coefficient. This ambiguity represents a
theoretical uncertainty in the determination of $C_\pm(\mu)$. In
order to reduce it, it is necessary to go beyond the leading order.
At NLO the scale ambiguity is then reduced from $\ord(\as)$ to
$\ord(\as^2)$. We will come back to this point below. Presently, we
will set $\mu_W=M_W$, but it is important to keep the related
uncertainty in mind.\\
Taking into account the leading order solution of the RG equation
\eqn{rga} for the coupling, which can be expressed in the form
\begin{equation}\label{alls}
\as(m)={\as(\mu)\over 1+\beta_0{\as(\mu)\over 4\pi}\ln{m^2\over\mu^2}}  \end{equation}
we may rewrite \eqn{cpmrg} as
\begin{equation}\label{clls} C_\pm(\mu)=
\left({1\over 1+\beta_0{\as(\mu)\over 4\pi}\ln{M^2_W\over\mu^2}}\right)^{
   \gamma^{(0)}_\pm\over 2\beta_0}  \end{equation}
\eqn{clls} contains the logarithmic corrections
$\sim \as\ln(M^2_W/\mu^2)$ to all orders in $\as$. This shows very
clearly that the leading log corrections have been summed up to
all orders in perturbation theory by solving the RG equation. In
particular, if we again expand \eqn{clls} in powers of $\as$, keeping
the first term only we recover \eqn{cpm}.
This observation demonstrates, that the RG method allows to obtain
solutions, which go beyond the conventional perturbation theory.

Before concluding this subsection, we would like to introduce still
two generalizations of the approach developed so far, which will
appear in the general discussion below.

\subsubsection{Threshold Effects in LLA}
               \label{sec:basicform:rg:thold}
First we may generalize the renormalization group evolution from
$M_W$ down to $\mu\approx m_c$ to include the threshold effect of
heavy quarks like $b$ or $t$ as follows
\begin{equation}\label{cmub}
\vec C(\mu)=U^{(f=4)}(\mu, \mu_b)U^{(f=5)}(\mu_b, \mu_W)\vec C(\mu_W)\end{equation}
which is valid for the LLA. In our example of the $c\to su\bar d$
transition the top quark gives no contribution at all. Being
heavier (but comparable) in mass than the W, it is simply
removed from the theory along with the W-boson. In a first step the
coefficients at the initial scale $\mu_W\approx M_W$ are evolved down to
$\mu_b\approx m_b$
in an effective theory with five quark flavors ($f=5$). Then,
again in the spirit of the effective field theory technique, for
scales below $\mu_b$ also the bottom quark is removed as an explicit
degree of freedom from the effective theory, yielding a new effective
theory with only four ``active'' quark flavors left. The matching
corrections between both theories can be calculated in ordinary
perturbation theory at the scale $\mu_b$, since due to $\mu_b\approx m_b$
no large logs can occur in this procedure. For the same reason
matching corrections $(\ord(\as))$ can be neglected in LLA and the
coefficients at $\mu_b$, $\vec C(\mu_b)$, simply serve as the initial
values for the RG evolution in the four quark theory down to
$\mu\approx m_c$. In addition, continuity of the running coupling
across the threshold $\mu_b$ is imposed by the requirement
\begin{equation}\label{af45}
\alpha_{s, f=4}(\mu_b, \Lambda^{(4)}) =
\alpha_{s, f=5}(\mu_b, \Lambda^{(5)})
\end{equation}
which defines different QCD scales $\Lambda^{(f)}$ for each effective
theory.\\
Neglecting the b-threshold, as we did before \eqn{rgcu}, one may just
perform the full evolution from $\mu_W$ to $\mu$ in an effective
four flavor theory. It turns out that in some cases the difference
of these two approaches is even negligible.\\
We would like to add a comment on this effective field theory
technique. At the first sight the idea to ``remove by hand''
heavy degrees of freedom may look somewhat artificial. However it
appears quite natural when not viewed from the evolution from high
towards low energies but vice versa (which actually corrsponds to
the historical way). Suppose only the ``light'' quarks $u$, $d$, $s$,
$c$ were known. Then in the attempt to formulate a theory of their
weak interactions one would be lead to a generalized Fermi theory
with (effective) four quark coupling constants to be determined
somehow. Of course, we are in the lucky position to know the
underlying theory in the form of the Standard Model. Therefore we
can actually derive the coupling constants of the low energy effective
theory from ``first principles''. This is exactly what is achieved
technically by going through a series of effective theories, removing
heavy degrees of freedom successively, by means of a step-by-step
procedure.

\subsubsection{Penguin Operators}
               \label{sec:basicform:rg:pop}
A second, but very important issue is the generation of QCD penguin
operators \cite{vainshtein:77}. Consider for example the local
operator $(\bar s_iu_i)_{V-A}(\bar u_jd_j)_{V-A}$, which is directly
induced by W-boson exchange. In this case, additional QCD correction
diagrams, the penguin diagrams (d.1) and (d.2 ) with a gluon in
fig.\ \ref{fig:1loopeff}, contribute and as a consequence six operators
are involved in the mixing under renormalization instead of two. These
read
\begin{equation}\label{q16}
\begin{array}{rcl}
Q_1=(\bar s_iu_j)_{V-A}(\bar u_jd_i)_{V-A} \\
Q_2=(\bar s_iu_i)_{V-A}(\bar u_jd_j)_{V-A} \\
Q_3=(\bar s_id_i)_{V-A}\sum_q(\bar q_jq_j)_{V-A} \\
Q_4=(\bar s_id_j)_{V-A}\sum_q(\bar q_jq_i)_{V-A} \\
Q_5=(\bar s_id_i)_{V-A}\sum_q(\bar q_jq_j)_{V+A} \\
Q_6=(\bar s_id_j)_{V-A}\sum_q(\bar q_jq_i)_{V+A}
\end{array}
\end{equation}
The sum over $q$ runs over all quark flavors that exist in the
effective theory in question. The operators $Q_1$ and $Q_2$ are just
the ones we have encountered in subsection~\ref{sec:basicform:ope}, but with
the c-quark replaced by $u$. This modified flavor structure gives rise to
the gluon penguin type diagrams shown in fig.\ \ref{fig:1loopeff}\,(d).
Since the gluon coupling is of course flavor conserving, it is clear
that penguins cannot be generated from the operator $(\bar
sc)_{V-A}(\bar ud)_{V-A}$. The penguin graphs induce the new local
interaction vertices $Q_3,\ldots, Q_6$, which have the same quantum
numbers. Their structure is easily understood. The flavor content is
determined by the $(\bar sd)_{V-A}$ current in the upper part and by a
$\sum_q(\bar qq)_V$ vector current due to the gluon coupling in the
lower. This vector structure is for convenience decomposed into a
$(V-A)$ and a $(V+A)$ part. For each of these, two different color
forms arise due to the color structure of the exchanged gluon (see
\eqn{tata}). Together this yields the four operators $Q_3,\ldots,
Q_6$.\\
For all operators $Q_1,\ldots, Q_6$ all possible QCD corrections (that
is all amputated Green functions with insertion of $Q_i$) of the
current-current (fig.\ \ref{fig:1loopeff}\,(a)--(c)) as well as of the
penguin type (fig.\ \ref{fig:1loopeff}\,(d.1) and (d.2)) have to be
evaluated.  In this process no new operators are generated, so that
$Q_1,\ldots, Q_6$ form a complete set. They ``close under
renormalization''. In analogy to the case of
subsection~\ref{sec:basicform:ope} the divergent parts of these Green
functions determine, after field renormalization, the operator
renormalization constants, which in the present case form a $6\times 6$
matrix. The calculation of the corresponding anomalous dimension matrix
and the renormalization group analysis then proceeds in the usual way.
We will see that the inclusion of higher order electroweak interactions
requires the introduction of still more operators.

\subsection{Summary of Basic Formalism}
            \label{sec:basicform:summary}
We think that after this rather detailed discussion of the methods
required for the short-distance calculations in weak decays, it is
useful to give at this point a concise summary of the material
covered so far. At the same time this may serve as an outline of the
necessary procedure for practical calculations. Furthermore it will
also provide a starting point for the extension of the formalism
from the LLA considered until now to the NLLA to be presented
in the next subsection.

Ultimately our goal is the evaluation of weak decay amplitudes
involving hadrons in the framework of a low energy effective theory,
of the form
\begin{displaymath}
\langle {\cal H}_{eff}\rangle={G_F\over\sqrt{2}}V_{CKM}
\langle \vec Q^T(\mu)\rangle \vec C(\mu)
\end{displaymath}
The procedure for this calculation can be divided into the
following three steps.

\bigskip
\noindent
{\bf Step 1: Perturbation Theory}
\\
Calculation of Wilson coefficients $\vec C(\mu_W)$ at
$\mu_W\approx M_W$ to the desired order in $\as$. Since
logarithms of the form $\ln(\mu_W/M_W)$ are not large, this can be
performed in ordinary perturbation theory. It amounts to matching
the full theory onto a five quark effective theory.\\

\medskip
\noindent
{\bf Step 2: RG Improved Perturbation Theory}
\\
\begin{itemize}
\item Calculation of the anomalous dimensions of the operators
\item Solution of the renormalization group equation for $\vec{C}(\mu)$
\item Evolution of the coefficients from $\mu_W$ down to the
appropriate low energy scale $\mu$
\begin{displaymath}
\vec C(\mu)=U(\mu, \mu_W)\vec C(\mu_W)
\end{displaymath}
\end{itemize}

\medskip
\noindent
{\bf Step 3: Non-Perturbative Regime}
\\
Calculation of hadronic matrix elements $\langle\vec Q(\mu)\rangle$,
normalized at the appropriate low energy scale $\mu$, by means of
some non-perturbative method.

\bigskip
\noindent
Important issues in this procedure are:
\begin{itemize}
\item The OPE achieves a {\it factorization\/} of short- and long
distance contributions.
\begin{itemize}
\item
Correspondingly, in order to disentangle the short-distance from the
long-distance part and to extract $\vec C(\mu_W)$ in actual
calculations, a proper {\it matching\/} of the full onto the
{\it effective theory\/} has to be performed.
\item Similar comments apply to the matching of an effective theory
with $f$ quark flavors to a theory with $(f-1)$ flavors during the
RG evolution to lower scales.
\item Furthermore, factorization implies, that the $\mu$-dependence
and also the dependence on the renormalization scheme, which
appears beyond the leading order, cancel between $C_i$ and
$\langle Q_i\rangle$.
\item Since the top quark is integrated out along with the W, the
coefficients $\vec C(\mu_W)$ in general contain also the full dependence
on the top quark mass $m_t$.
\end{itemize}
\item A {\it summation of large logs\/} by means of the RG method is
necessary. More specifically, in the $n$-th order of
renormalization group improved perturbation theory the terms of the
form
\begin{displaymath}
\as^n(\mu)\left(\as(\mu)\ln{M_W\over\mu}\right)^k
\end{displaymath}
are summed to all orders in $k$ ($k$=0, 1, 2,$\ldots$). This approach
is justified as long as $\as(\mu)$ is small enough, which
requires that $\mu$ not be too low, typically not less than $1\gev$.
\end{itemize}

\subsection{Wilson Coefficients Beyond Leading Order}
            \label{sec:basicform:wc}
\subsubsection{The RG Formalism}
               \label{sec:basicform:wc:rgf}
We are now going to extend the renormalization group formalism for
the coefficient functions to the next-to-leading order level.
Subsequently we will discuss important aspects of the resulting
formulae, in particular the scale- and scheme dependences and their
cancellation.\\
To have something specific in mind, we may consider the calculation
for the $\dS$ effective hamiltonian for nonleptonic decays, which
without QCD effects and for low energy is given by
\begin{equation}\label{hds1}
{\cal H}^{\dS}_{eff}={G_F\over\sqrt{2}}V^\ast_{us}V_{ud}
  (\bar su)_{V-A}(\bar ud)_{V-A}   \end{equation}
At higher energies of course also the charm, bottom and top quark
have to be taken into account. The Feynman diagrams contributing to
$\ord(\as)$ corrections to this hamiltonian are shown in
fig.~\ref{fig:1loopful} and \ref{fig:1loopeff}.
Including current-current- as well as penguin type corrections, the
relevant operator basis consists of the six operators in \eqn{q16}.
\\
On the one hand, this particular case is very important by itself since
it provides the theoretical basis for a large variety of different
decay modes. On the other hand we will at this stage keep the
discussion fairly general, so that all important features of a general
validity are exhibited. In addition, the central formulae of this
subsection will be used at several places later on, if at times
extended or modified to match the specific cases in question. In
part two of this report we will give a more detailed discussion of
the hamiltonians relevant for various decays. Here, we would rather like
to concentrate on the presentation of the OPE and renormalization group
formalism.
\\
The effective hamiltonian for nonleptonic decays may be written in
general as
\begin{equation}\label{hqtc}
{\cal H}_{eff}={G_F\over\sqrt{2}}\sum_i C_i(\mu)Q_i(\mu)\equiv
  {G_F\over\sqrt{2}}\vec Q^T(\mu) \vec C(\mu)   \end{equation}
where the index $i$ runs over all contributing operators, in our
example $Q_1,\ldots, Q_6$ of \eqn{q16}. It is straightforward to
apply ${\cal H}_{eff}$ to D- and B-meson decays as well by changing
the quark flavors appropriately. For the time being we omit CKM
parameters, which can be reinserted later on. $\mu$ is some low
energy scale of the order $\ord(1\gev)$, $\ord(m_c)$ and $\ord(m_b)$
for K-, D- and B-meson decays, respectively. The argument $\mu$ of the
operators $Q_i(\mu)$ means, that their matrix elements are to be
normalized at scale $\mu$.\\
The Wilson coefficient functions are given by
\begin{equation}\label{cucw}
\vec C(\mu)=U(\mu, \mu_W)\vec C(\mu_W)   \end{equation}
The coefficients at the scale $\mu_W=\ord(M_W)$ can be evaluated in
perturbation theory. The evolution matrix $U$ then includes the
renormalization group improved perturbative contributions from the
scale $\mu_W$ down to $\mu$.\\
In the first step we determine $\vec C(\mu_W)$ from a comparison of
the amputated Green function with appropriate external lines in the
full theory with the corresponding amplitude in the effective theory.
At NLO we have to calculate to $\ord(\as)$, including
non-logarithmic, constant terms. The full amplitude results from
the current-current- and penguin type diagrams in fig.~\ref{fig:1loopful},
is finite after field renormalization and can be written as
\begin{equation}\label{aa01} A=
{G_F\over\sqrt{2}}\vec S^T(\vec A^{(0)}+{\as(\mu_W)\over 4\pi}\vec A^{(1)})\end{equation}
Here $\vec S$ denotes the tree level matrix elements of the
operators $\vec Q$.
In the effective theory \eqn{hqtc} the current-current- and penguin
corrections of fig.~\ref{fig:1loopeff} have to be calculated for all
the operators $Q_i$. In this case, besides the field renormalization, a
renormalization of operators is necessary
\begin{equation}\label{q0z3}
Z^2_q\langle\vec Q\rangle^{(0)}=Z\langle\vec Q\rangle  \end{equation}
where the matrix Z absorbes those divergences of the Green functions
with operator $\vec Q$ insertion, that are not removed by the field
renormalization. The renormalized matrix elements of the operators
can then to $\ord(\as)$ be written as
\begin{equation}\label{qars}
\langle\vec Q(\mu_W)\rangle=(1+{\as(\mu_W)\over 4\pi} r)\vec S  \end{equation}
and the amplitude in the effective theory to the same order becomes
\begin{equation}\label{aeff}  A_{eff}=
{G_F\over\sqrt{2}}\vec S^T(1+{\as(\mu_W)\over 4\pi} r^T) \vec C(\mu_W)\end{equation}
Equating \eqn{aa01} and \eqn{aeff} we obtain
\begin{equation}\label{cmuw} \vec C(\mu_W)=
\vec A^{(0)}+{\as(\mu_W)\over 4\pi}(\vec A^{(1)}-r^T\vec A^{(0)})  \end{equation}
In general $\vec A^{(1)}$ in \eqn{aa01} involves logarithms
$\ln(M^2_W/-p^2)$ where p denotes some global external momentum for the
amplitudes in fig.~\ref{fig:1loopful}.  On the other hand, the matrix
$r$ in \eqn{qars}, characterizing the radiative corrections to
$\langle\vec Q(\mu_W)\rangle$, includes $\ln(-p^2/\mu^2_W)$. As we have
seen in subsection \ref{sec:basicform:ope}, these logarithms combine to
$\ln(M^2_W/\mu^2_W)$ in the Wilson coefficient \eqn{cmuw}. For
$\mu_W=M_W$ this logarithm vanishes altogether. For $\mu_W=\ord(M_W)$
the expression $\ln(M^2_W/\mu^2_W)$ is a ``small logarithm'' and the
correction $\sim\as \ln(M^2_W/\mu^2_W)$, which could be neglected in
LLA, has to be kept in the perturbative calculation at NLO together
with constant pieces of order $\ord(\as)$.
\\
In the second step, the renormalization group equation for $\vec C$
\begin{equation}\label{rgcv}
{d\over d\ln\mu}\vec C(\mu)=\gamma^T(g)\vec C(\mu)   \end{equation}
has to be solved with boundary condition \eqn{cmuw}. The solution is
written with the help of the U-matrix as in \eqn{cucw}, where
$U(\mu, \mu_W)$ obeys the same equation as $\vec C(\mu)$ in \eqn{rgcv}.
The general solution is easily written down iteratively
\begin{equation}\label{uit}
U(\mu, m)=1+\int^{g(\mu)}_{g(m)}dg_1{\gamma^T(g_1)\over\beta(g_1)}+
\int^{g(\mu)}_{g(m)}dg_1\int^{g_1}_{g(m)}dg_2
{\gamma^T(g_1)\over\beta(g_1)}{\gamma^T(g_2)\over\beta(g_2)}+\ldots \end{equation}
which using $dg/d\ln\mu=\beta(g)$ is readily seen to solve the
renormalization group equation
\begin{equation}\label{rgu}
{d\over d\ln\mu}U(\mu, m)=\gamma^T(g)U(\mu, m)   \end{equation}
The series in \eqn{uit} can be more compactly expressed by introducing
the notion of g-ordering
\begin{equation}\label{utge}
U(\mu, m)=T_g \exp\int^{g(\mu)}_{g(m)}dg'{\gamma^T(g')\over\beta(g')}\end{equation}
where in the case $g(\mu)>g(m)$ the g-ordering operator $T_g$ is
defined through
\begin{equation}\label{tgdf}
T_g f(g_1)\ldots f(g_n)=\sum_{perm}
\Theta(g_{i_1}-g_{i_2})\Theta(g_{i_2}-g_{i_3})\ldots
\Theta(g_{i_{n-1}}-g_{i_n})f(g_{i_1})\ldots f(g_{i_n}) \end{equation}
and brings about an ordering of the factors $f(g_i)$ such that the
coupling constants increase from right to left. The sum in \eqn{tgdf} runs
over all permutations $\{i_1,\ldots, i_n\}$ of $\{1, 2,\ldots, n\}$.
The $T_g$ ordering is necessary since in general the anomalous
dimension matrices at different couplings do not commute beyond the
leading order, $[\gamma(g_1), \gamma(g_2)]\not= 0$.\\
At next-to-leading order we have to keep the first two terms in the
perturbative expansions for $\beta(g)$ (see \eqn{bg01}) and $\gamma(g)$
\begin{equation}\label{gg01}
\gamma(\as)=\gamma^{(0)}\aspi + \gamma^{(1)}\left(\aspi\right)^2
\end{equation}
To this order the evolution matrix $U(\mu, m)$ is given by
\cite{burasetal:92a}
\begin{equation}\label{u0jj}
U(\mu,m)=
(1+{\as(\mu)\over 4\pi} J) U^{(0)}(\mu,m) (1-{\as(m)\over 4\pi} J)
\end{equation}
$U^{(0)}$ is the evolution matrix in leading logarithmic approximation
and the matrix $J$ expresses the next-to-leading corrections to this
evolution. We have
\begin{equation}\label{u0vd} U^{(0)}(\mu,m)= V
\left({\left[{\as(m)\over\as(\mu)}\right]}^{{\vec\gamma^{(0)}\over 2\beta_0}}
   \right)_D V^{-1}   \end{equation}
where $V$ diagonalizes ${\gamma^{(0)T}}$
\begin{equation}\label{ga0d} \gamma^{(0)}_D=V^{-1} {\gamma^{(0)T}} V  \end{equation}
and $\vec\gamma^{(0)}$ is the vector containing the diagonal elements of
the diagonal matrix $\gamma^{(0)}_D$.\\
If we define
\begin{equation}\label{gvg1} G=V^{-1} {\gamma^{(1)T}} V   \end{equation}
and a matrix $H$ whose elements are
\begin{equation}\label{sij} H_{ij}=\delta_{ij}\gamma^{(0)}_i{\beta_1\over 2\beta^2_0}-
    {G_{ij}\over 2\beta_0+\gamma^{(0)}_i-\gamma^{(0)}_j}  \end{equation}
the matrix $J$ is given by
\begin{equation}\label{jvs} J=V H V^{-1}   \end{equation}
The fact that \eqn{u0jj} is indeed a solution of the RG equation \eqn{rgu}
to the order considered is straightforwardly verified by differentiation
with respect to $\ln\mu$. Combining now the initial values \eqn{cmuw}
with the evolution matrix \eqn{u0jj} we obtain
\begin{equation}\label{cjua} \vec C(\mu)=(1+{\as(\mu)\over 4\pi} J)U^{(0)}(\mu,\mu_W)
(\vec A^{(0)}+{\as(\mu_W)\over 4\pi}[\vec A^{(1)}-(r^T+J)\vec A^{(0)}])\end{equation}
Using \eqn{cjua} we can calculate for example the coefficients at a scale
$\mu=\mu_b=\ord(m_b)$, working in an effective five flavor theory, $f=5$.
If we have to evolve the coefficients to still lower values, we would
like to formulate a new effective theory for $\mu<\mu_b$ where now
also the b-quark is removed as an explicit degree of freedom. To
calculate the coefficients in this new four flavor theory at the
scale $\mu_b$, we have to determine the matching corrections at
this scale.\\
We follow the same principles as in the case of integrating out the
W-boson and require
\begin{equation}\label{qfcf}
\langle\vec Q_f(m)\rangle^T\vec C_f(m)=
\langle\vec Q_{f-1}(m)\rangle^T\vec C_{f-1}(m)  \end{equation}
in the general case of a change from an f-flavor to a (f--1)-flavor
theory at a scale $m$. The ``full amplitude'' on the l.h.s., which is
now in an f-flavor effective theory, is expanded into matrix elements
of the new (f--1)-flavor theory, multiplied by new Wilson coefficients
$\vec C_{f-1}$. From \eqn{qars}, determining the matrix elements of
operators to $\ord(\as)$, one finds
\begin{equation}\label{qfdr}
\langle\vec Q_f(m)\rangle=(1+{\as(m)\over 4\pi} \delta r)
\langle\vec Q_{f-1}(m)\rangle    \end{equation}
where
\begin{equation}\label{drrf} \delta r=r^{(f)}-r^{(f-1)}
\end{equation}
In \eqn{drrf} we have made explicit the dependence of the matrix $r$
on the number of quark flavors which enters in our example via the
penguin contributions. From \eqn{qfcf} and \eqn{qfdr} we find
\begin{equation}\label{cmcf}
\vec C_{f-1}(m)=M(m)\vec C_f(m)  \end{equation}
with
\begin{equation}\label{mdrt}  M(m)=1+{\as(m)\over 4\pi} \delta r^T  \end{equation}
The general renormalization group matrix $U$ in \eqn{u0jj}, now
evaluated for (f--1) flavors, can be used to evolve $\vec C_{f-1}(m)$
to lower values of the renormalization scale. It is clear that no
large logarithms can appear in \eqn{mdrt} and that therefore the
matching corrections, expressed in the matrix $M(m)$ can be
computed in usual perturbation theory. We note that this type of
matching corrections enters in a nontrivial way for the first time
at the NLO level. In the LLA $M\equiv 1$ and one can simply omit the
b-flavor components in the penguin operators when crossing the
b-threshold.\\
We also remark that the correction matrix $M$ introduces a small
discontinuity of the coefficients, regarded as functions of $\mu$,
at the matching scale $m$. This is however not surprising. In any
case the $\vec C(\mu)$ are not physical quantities and their
discontinuity precisely cancels the effect of removing the heavy
quark flavor from the operators, which evidently is a
``discontinuous'' step. Hence, physical amplitudes are not
affected and indeed the behaviour of $\vec C$ at the matching scale
ensures that the same physical result will be obtained,
whether we choose to calculate in the f-flavor- or in the (f--1)-flavor
theory for scales around the matching scale $m$.\\
To conclude we will write down how the typical final result for
the coefficient functions at $\mu\approx 1\gev$, appropriate for
K-decays, looks like, if we combine all the contributions discussed
above. Then we can write
\begin{equation}\label{cthr}
\vec C(\mu)=U_3(\mu,\mu_c)M(\mu_c)U_4(\mu_c,\mu_b)M(\mu_b)
U_5(\mu_b,\mu_W)\vec C(\mu_W)  \end{equation}
where $U_f$ is the evolution matrix for $f$ active flavors.
In the following discussion we will not always include the flavor
thresholds when writing the expression for the RG evolution. It is
clear, that they can be added in a straightforward fashion.

\subsubsection{The Calculation of the Anomalous Dimensions}
               \label{sec:basicform:wc:adm}
The matrix of anomalous dimensions is the most important ingredient
for the renormalization group calculation of the Wilson coefficient
functions. In the following we will summarize the essential steps of
its calculation.\\
Recall that the evaluation of the amputated Green functions with
insertion of the operators $\vec Q$ gives the relation
\begin{equation}\label{qzgf}
\langle\vec Q\rangle^{(0)}=Z^{-2}_q Z\langle\vec Q\rangle
\equiv Z_{GF}\langle\vec Q\rangle   \end{equation}
$\langle\vec Q\rangle^{(0)}$, $\langle\vec Q\rangle$ denote the
unrenormalized and renormalized Green functions, respectively.
$Z_q$ is the quark field renormalization constant and $Z$ is the
renormalization constant matrix of the operators $\vec Q$.\\
The anomalous dimensions are given by
\begin{equation}\label{gaz2}
\gamma(g)=Z^{-1}{d\over d\ln\mu}Z   \end{equation}
In the $MS$ (or $\overline{MS}$) scheme the renormalization constants
are chosen to absorb the pure pole divergences $1/\eps^k$
($D=4-2\eps$), but no finite parts. $Z$ can then be expanded in
inverse powers of $\eps$ as follows
\begin{equation}\label{zeps}
Z=1+\sum^\infty_{k=1}{1\over\eps^k}Z_k(g)  \end{equation}
Using the expression for the $\beta$-function \eqn{bete}, valid for
arbitrary $\eps$ we derive the useful formula \cite{floratosetal:77}
\begin{equation}\label{ggz1}
\gamma(g)=-2g^2{\partial Z_1(g)\over\partial g^2}
=-2\as{\partial Z_1(\as)\over\partial \as}  \end{equation}
Similarly to \eqn{zeps} we expand
\begin{equation}\label{zqep}
Z_q=1+\sum^\infty_{k=1}{1\over\eps^k}Z_{q,k}(g)  \end{equation}
\begin{equation}\label{zgfe}
Z_{GF}=1+\sum^\infty_{k=1}{1\over\eps^k}Z_{GF,k}(g)  \end{equation}
From the calculation of the unrenormalized Green functions \eqn{qzgf}
we immediately obtain $Z_{GF}$. What we need to compute $\gamma(g)$
is $Z_1(g)$ \eqn{ggz1}. From \eqn{qzgf}, \eqn{zeps}, \eqn{zqep}, \eqn{zgfe}
we find
\begin{equation}\label{zqgf}
Z_1=2Z_{q,1}+Z_{GF,1}  \end{equation}
At next-to-leading order we have from the $1/\eps$ poles of the
unrenormalized Green functions
\begin{equation}\label{zgf1}
Z_{GF,1}=b_1 \aspi + b_2 \left(\aspi\right)^2  \end{equation}
The corresponding expression for the well known factor $Z_{q,1}$
has been quoted in \eqn{zq1a}. Using \eqn{zq1a}, \eqn{ggz1},
\eqn{zqgf}, \eqn{zgf1} we finally obtain for the one- and two-loop
anomalous dimension matrices $\gamma^{(0)}$ and $\gamma^{(1)}$ in
\eqn{gg01}
\begin{equation}\label{a1b1}
\gamma^{(0)}_{ij}=-2[2a_1 \delta_{ij}+(b_1)_{ij}]  \end{equation}
\begin{equation}\label{a2b2}
\gamma^{(1)}_{ij}=-4[2a_2 \delta_{ij}+(b_2)_{ij}]  \end{equation}
\eqn{a1b1} and \eqn{a2b2} may be used as recipes to immediately extract
the anomalous dimensions from the divergent parts of the
unrenormalized Green functions.

\subsubsection{Renormalization Scheme Dependence}
               \label{sec:basicform:wc:rgdep}
A further issue, which becomes important at next-to-leading order is
the dependence of unphysical quantities, like the Wilson coefficients
and the anomalous dimensions, on the choice of the renormalization
scheme. This scheme dependence arises because the renormalization
prescription involves an arbitrariness in the finite parts to be
subtracted along with the ultraviolet singularities.  Two different
schemes are then related by a finite renormalization.  Considering the
quantities, which we encountered in
subsection~\ref{sec:basicform:wc:rgf}, the following of them are
independent of the renormalization scheme
\begin{equation}\label{rsi}
\beta_0,\quad\beta_1,\quad\gamma^{(0)},\quad\vec A^{(0)},\quad
\vec A^{(1)},\quad r^T+J,\quad\langle\vec Q\rangle^T\vec C  \end{equation}
whereas
\begin{equation}\label{rsd}
r,\quad\gamma^{(1)},\quad J,\quad\vec C,\quad\langle\vec Q\rangle \end{equation}
are scheme dependent.\\
In the framework of dimensional regularization one example of how
such a scheme dependence can occur is the treatment of $\gf$ in $D$
dimensions.
Possible choices are the ``naive dimensional regularization'' (NDR)
scheme with $\gf$ taken to be
anticommuting or the 't-Hooft--Veltman (HV) scheme
\cite{thooft:72}, \cite{breitenlohner:77} with
non-anticommuting $\gf$. Another example is the use of operators in
a color singlet or a non-singlet form, such as
\begin{equation}\label{qtwi}
Q_2=(\bar s_iu_i)_{V-A}(\bar u_jd_j)_{V-A}
\quad \hbox{\rm or}\quad
\tilde Q_2=(\bar s_id_j)_{V-A}(\bar u_ju_i)_{V-A} \end{equation}
where $i$, $j$ are color indices. In $D=4$ dimensions these operators
are equivalent since they are related by a Fierz transformation. In
the NDR scheme however these two choices yield different results
for $r$, $\gamma^{(1)}$ and $J$ and thus constitute two different
schemes, related by a nontrivial finite renormalization. On the
other hand, both choices give the same $r$, $\gamma^{(1)}$ and $J$
if the HV scheme is employed.\\
Let us now discuss the question of renormalization scheme dependences
in explicit terms in order to obtain an overview on how the
scheme dependences arise, how various quantities transform under a
change of the renormalization scheme and how the cancellation of
scheme dependences is guaranteed for physically relevant
quantities.\\
First of all, it is clear that the product
\begin{equation}\label{qtc}
\langle\vec Q(\mu)\rangle^T\vec C(\mu)  \end{equation}
representing the full amplitude, is independent of the
renormalization scheme chosen. This is simply due to the fact, that
it is precisely the factorization of the amplitude into Wilson coefficients
and matrix elements of operators by means of the operator product
expansion, which introduces the scheme dependence of $\vec C$ and
$\langle\vec Q\rangle$. In other words, the scheme dependence of
$\vec C$ and $\langle\vec Q\rangle$ represents the arbitrariness one
has in splitting the full amplitude into coefficients and matrix
elements and the scheme independence of the combined product \eqn{qtc}
is manifest in the construction of the operator product expansion.
\\
More explicitly, these quantities are in different schemes
(primed and unprimed) related by
\begin{equation}\label{rsqc}\langle\vec Q\rangle'=(1+\aspi s)\langle\vec Q\rangle\qquad
 \vec C'=(1-\aspi s^T)\vec C  \end{equation}
 where $s$ is a constant matrix. \eqn{rsqc} represents a finite
renormalization of $\vec C$ and $\langle\vec Q\rangle$.
From \eqn{qars} we immediately obtain
\begin{equation}\label{rprs}  r'=r+s  \end{equation}
Furthermore from
\begin{equation}\label{qtuc} \langle\vec Q(\mu)\rangle^T\vec C(\mu)\equiv
  \langle\vec Q(\mu)\rangle^T U(\mu,M_W) \vec C(M_W)  \end{equation}
we have
\begin{equation}\label{upus} U'(\mu,M_W)=
(1-{\as(\mu)\over 4\pi}s^T)U(\mu,M_W)(1+{\as(M_W)\over 4\pi}s^T)  \end{equation}
A comparison with \eqn{u0jj} yields
\begin{equation}\label{jpjs} J'=J-s^T \end{equation}
The renormalization constant matrix in the primed scheme, $Z'$,
follows from \eqn{rsqc} and \eqn{qzgf}
\begin{equation}\label{zpzs}  Z'=Z (1- \aspi s)  \end{equation}
Recalling the definition of the matrix of anomalous dimensions,
\eqn{gaz2} and \eqn{gg01}, we derive
\begin{equation}\label{gpgs}\gamma^{(0)\prime}=\gamma^{(0)} \qquad
 \gamma^{(1)\prime}=\gamma^{(1)}+[s,\gamma^{(0)}]+2\beta_0 s \end{equation}
With these general formulae at hand it is straightforward to
clarify the cancellation of scheme dependences in all particular
cases. Alternatively, they may be used to transform scheme
dependent quantities from one scheme to another, if desired, or to
check the compatibility of results obtained in different schemes.
\\
In particular we immediately verify from \eqn{rprs} and \eqn{jpjs} the
scheme independence of the matrix $r^T+J$. This means that in the
expression for $\vec C$ in \eqn{cjua} the factor on the right hand
side of $U^{(0)}$, related to the ``upper end'' of the evolution,
is independent of the renormalization scheme, as it must be. The
same is true for $U^{(0)}$. On the other hand $\vec C$ still
depends on the renormalization scheme through the matrix $J$ to the
left of $U^{(0)}$. As is evident from \eqn{rsqc},
this dependence is compensated for by the
corresponding scheme dependence of the matrix elements of operators
so that a physically meaningful result for the decay amplitudes
is obtained.
To ensure a proper cancellation of the scheme dependence the matrix
elements have to be evaluated in the same scheme
(renormalization, $\gf$, form of operators) as the coefficient
functions, which is a nontrivial task for the necessary
non-perturbative computations. In other words, beyond the leading
order the matching between short- and long-distance contributions has
to be performed properly not only with respect to the scale $\mu$,
but also with respect to the renormalization scheme employed.

\subsubsection{Discussion}
               \label{sec:basicform:wc:disc}
We will now specialize the presentation of the general formalism to the
case of a single operator (that is without mixing).  This situation is
e.g.\ relevant for the operators $Q_+$ and $Q_-$ with four different quark
flavors, which we encountered in section~\ref{sec:basicform:ope}. The
resulting simplifications are useful in order to display some more
details of the structure of the calculation and to discuss the most
salient features of the NLO analysis in a transparent way.
\\
In the case where only one single operator contributes, the
amplitude in the full theory (dynamical W-boson) may be written as
(see \eqn{aa01})
\begin{equation}\label{amp2}
A={G_F\over\sqrt{2}}(1+{\as(\mu_W)\over 4\pi}\left[-{\gamma^{(0)}\over 2}
\ln{M^2_W\over -p^2}+\tilde A^{(1)}\right]) S  \end{equation}
where we have made the logarithmic dependence on the W mass explicit.
In the effective theory the amplitude reads
\begin{eqnarray}\label{aef2}
A_{eff}&=&{G_F\over\sqrt{2}}C(\mu_W)\langle Q(\mu_W)\rangle \\
&=&{G_F\over\sqrt{2}}C(\mu_W)(1+{\as(\mu_W)\over 4\pi}\left[{\gamma^{(0)}\over 2}
 \left(\ln{-p^2\over\mu^2_W}+\gamma_E-\ln 4\pi\right)+\tilde r
 \right])S   \nonumber
\end{eqnarray}
The divergent pole term $1/\eps$ has been subtracted minimally. A
comparison of \eqn{amp2} and \eqn{aef2} yields the Wilson coefficient
\begin{equation}\label{cmw2}
 C(\mu_W)=1+{\as(\mu_W)\over 4\pi}\left[-{\gamma^{(0)}\over 2}
 \left(\ln{M^2_W\over\mu^2_W}+\gamma_E-\ln 4\pi\right)+B\right] \end{equation}
where
\begin{equation}\label{ba1r} B=\tilde A^{(1)}-\tilde r  \end{equation}
In the leading log approximation we had simply $C(\mu_W)=1$. By
contrast at NLO the $\ord(\as)$ correction has to be taken into
account in addition. This correction term exhibits the following
new features:
\begin{itemize}
\item The expression $\gamma_E-\ln 4\pi$, which is characteristic to
dimensional regularization appears. It is proportional to $\gamma^{(0)}$.
\item A constant term $B$ is present. $B$ depends on the
factorization scheme chosen.
\item An explicit logarithmic dependence on the matching scale
$\mu_W$ shows up.
\end{itemize}
We discuss these points one by one.\\
First, the term $\gamma_E-\ln 4\pi$ is characteristic for the $MS$
scheme. It can be eliminated by going from the $MS$- to the
$\overline{MS}$ scheme. This issue is well known in the literature.  We
find it however useful to briefly repeat the definition of the
$\overline{MS}$ scheme in the present context, since this is an
important point for NLO analyses.
\\
Consider the RG solution for the coefficient
\begin{eqnarray}\label{cja2}
 \lefteqn{C(\mu)=} & & \\
 & &(1+{\as(\mu)\over 4\pi}J)\left[{\as(\mu_W)\over \as(\mu)}\right]^{
  \gamma^{(0)}\over 2\beta_0}
 (1+{\as(\mu_W)\over 4\pi}\left[-{\gamma^{(0)}\over 2}
 \left(\ln{M^2_W\over\mu^2_W}+\gamma_E-\ln 4\pi\right)+B-J\right])
 \nonumber
\end{eqnarray}
This represents the solution for the $MS$ scheme. Therefore in
\eqn{cja2} $\as=\alpha_{s,MS}$. The redefinition of $\alpha_{s,MS}$
through
\begin{equation}\label{amsb}
\alpha_{s,MS}=\alpha_{s,\overline{MS}}\left(1+\beta_0(\gamma_E-\ln 4\pi)
 {\alpha_{s,\overline{MS}}\over 4\pi}\right)  \end{equation}
is a finite renormalization of the coupling,
which defines the $\overline{MS}$ scheme. Since
\begin{equation}\label{msbr}
[\alpha_{s,MS}(\mu_W)]^{\gamma^{(0)}\over 2\beta_0}\doteq
[\alpha_{s,\overline{MS}}(\mu_W)]^{\gamma^{(0)}\over 2\beta_0}
\left(1+{\gamma^{(0)}\over 2}(\gamma_E-\ln 4\pi)
{\alpha_{s,\overline{MS}}(\mu_W)\over 4\pi}\right)   \end{equation}
we see, that this transformation eliminates, to the order considered,
the $\gamma_E-\ln 4\pi$ term in \eqn{cja2}. At the lower end of the
evolution the same redefinition yields a factor
\begin{equation}\label{msbf}
1-{\gamma^{(0)}\over 2}(\gamma_E-\ln 4\pi)
{\alpha_{s,\overline{MS}}(\mu)\over 4\pi}   \end{equation}
which removes the corresponding factor from the matrix element
(see \eqn{aef2})
\begin{equation}\label{msbq}
\langle Q(\mu)\rangle_{MS}\equiv
\left(1+{\gamma^{(0)}\over 2}(\gamma_E-\ln 4\pi)
{\alpha_{s,\overline{MS}}(\mu)\over 4\pi}\right)
\langle Q(\mu)\rangle_{\overline{MS}}   \end{equation}
At the next-to-leading log level we are working, the transformation
\eqn{amsb} is equivalent to a redefinition of the scale $\Lambda$
according to
\begin{equation}\label{msbl}
\Lambda^2_{\overline{MS}}=4\pi e^{-\gamma_E}\Lambda^2_{MS}  \end{equation}
as one can verify with the help of \eqn{amu}. In practice, one can
just drop the ($\gamma_E-\ln 4\pi$) terms in \eqn{cja2}. Then $\as(\mu)$
and $\Lambda$ are to be taken in the $\overline{MS}$ scheme.
Throughout the present report it is always understood that the
transformation to $\overline{MS}$ has been performed. Then
\begin{equation}\label{msbc}
 C(\mu)=(1+{\as(\mu)\over 4\pi}J)\left[{\as(\mu_W)\over\as(\mu)}\right]^{
  \gamma^{(0)}\over 2\beta_0}
 (1+{\as(\mu_W)\over 4\pi}\left[-{\gamma^{(0)}\over 2}
 \ln{M^2_W\over\mu^2_W}+B-J\right]) \end{equation}
Second, from the issue of the $MS$ -- $\overline{MS}$ transformation,
or more generally an arbitrary redefinition of $\as$ (or $\Lambda$) one
should distinguish the renormalization scheme dependence due to the
ambiguity in the renormalization of the operator.  It suggests itself
to use for the latter the term ``factorization scheme dependence''.
This is the scheme dependence we have discussed in
section~\ref{sec:basicform:wc:rgdep}. A change in the factorization
scheme transforms $\gamma^{(1)}$, $B$ and $J$ as
\begin{equation}\label{gbjs}
{\gamma^{(1)}}'=\gamma^{(1)} + 2\beta_0 s
\qquad B'=B - s
\qquad J'=J - s
\end{equation}
where $s$ is a constant number. This follows from the formulae in
section~\ref{sec:basicform:wc:rgdep} and from the definition of $B$ in
\eqn{ba1r}. Note that in the case of a single operator the relation
between $\gamma^{(1)}$ and $J$ simplifies to
\begin{equation}\label{jg01}
J={1\over 2\beta_0}\left({\beta_1\over\beta_0}\gamma^{(0)}
  -\gamma^{(1)}\right)   \end{equation}
Obviously the scheme dependence cancels in the difference
$B-J$ in \eqn{msbc}.\\
Third, due to the explicit $\mu_W$ dependence in the $\ord(\as)$
correction term the coefficient function is, to the order considered,
independent of the precise value of the matching scale $\mu_W$, as it
must be. Indeed
\begin{equation}\label{cmud}
{d\over d\ln\mu_W}C(\mu)=\ord(\as^2)  \end{equation}
since
\begin{equation}\label{amud}
{d\over d\ln\mu_W}\as(\mu_W)=-2\beta_0{\as(\mu_W)^2\over 4\pi}+\ord(\as^3)\end{equation}
In the same way one can also convince oneself that the coefficient
function is independent of the heavy quark threshold scales, up to
terms of the neglected order.\\
Of course the dependence on the low energy scale $\mu$ remains
and has to be matched with the corresponding dependence of the
operator matrix element.\\
All the points we have mentioned here apply in an analogous
manner also to the case with operator mixing, only the algebra is
then slightly more complicated.

We would like to stress once again that it is only at the NLO level,
that these features enter the analysis in a nontrivial way, as should
be evident from the presentation we have given above.

\subsubsection{Evanescent Operators}
               \label{sec:basicform:wc:evanop}
Finally, we would like to mention the so called {\it evanescent}
operators. These are operators which exist in $D\not=4$ dimensions but
vanish in $D=4$. It has been stressed in \cite{burasweisz:90} that a
correct calculation of two-loop anomalous dimensions requires a proper
treatment of these operators. This discussion has been extended in
\cite{dugan:91} and further generalized in \cite{herrlichnierste:94}.
In view of a rather technical nature of this aspect we refer the
interested reader to the papers quoted above.

\skipevenpage

{\Huge\bf
\noindent
Part Two --

\bigskip
\bigskip
\bigskip

\noindent
The Effective Hamiltonians
}

\bigskip
\bigskip
\bigskip

\noindent
The second part constitutes a compendium of effective hamiltonians for
weak decays.  We will deal with all decays for which NLO corrections
have been calculated in the literature and whose list is given in table
\ref{tab:processes}. This includes a listing of the initial conditions
$C_i (\mw)$, a listing of all one-loop and two-loop anomalous dimension
matrices and finally tables of numerical values of the relevant Wilson
coefficients as functions of $\Lms$, $\mt$ and the renormalization
schemes considered.  In certain cases we are able to give analytic
formulae for $C_i$.

We will discuss all effective hamiltonians one by one. With the help of the
master formulae and the procedure of section \ref{sec:basicform} it is
easy to see similarities and differences between various cases.  Our
compendium includes also the $b \to s\gamma$ and $b\to s g$ transitions
which although known only in the leading logarithmic approximation
deserve special attention.

Finally, as a preparation for the third part we give a brief
description of the ``Penguin-Box Expansion'' (PBE), which can be
regarded as a version of OPE particularly suited for the study of the
$m_t$ dependence in weak decays.

In addition we have also included a section on NLO QCD calculations in
the context of HQET. This chapter lies somewhat outside our main line of
presentation. Also a comprehensive discussion of HQET is clearly beyond
the scope of the present paper. However, we would like to illustrate the
application of the general formalism for short distance QCD corrections
within this framework and summarize a few important NLO results that
have been obtained in HQET.

\section{Guide to Effective Hamiltonians}
         \label{sec:Heffguide}
In order to facilitate the presentation of effective
hamiltonians in weak decays we give a complete compilation of the
relevant operators below. Divided into six classes, these 
operators play a dominant role in the phenomenology of weak decays.
The six classes are given as follows

\medskip

\begin{figure}[htb]
\vspace{0.10in}
\centerline{
\epsfysize=5in
\epsffile{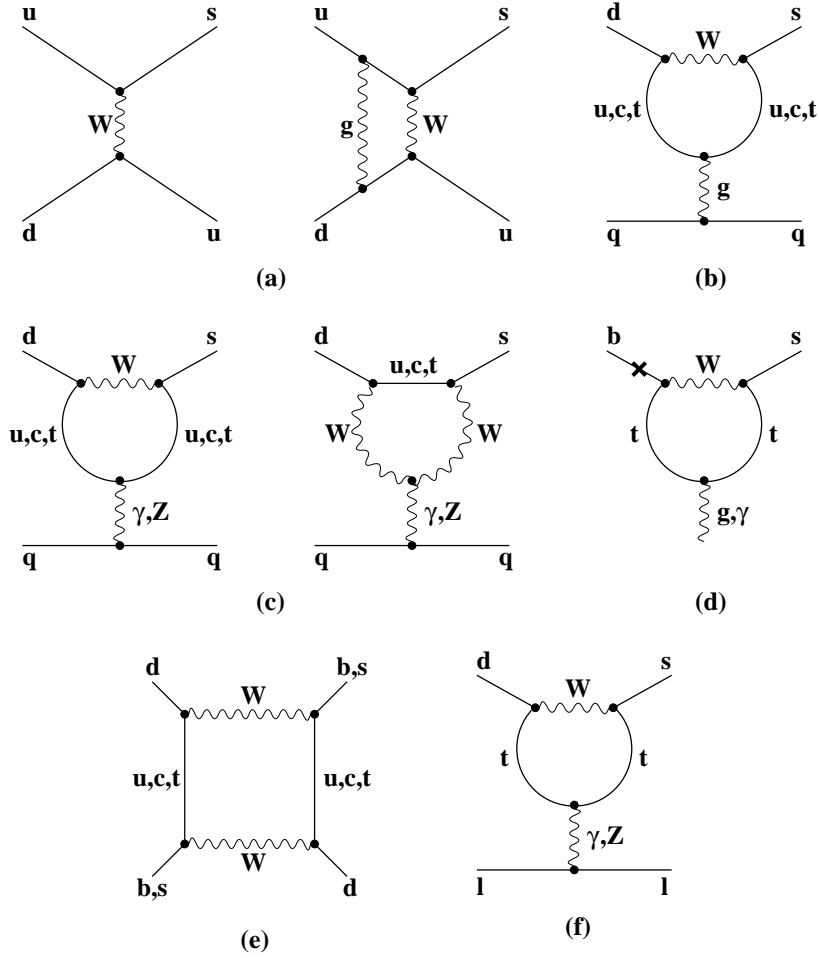}
}
\vspace{0.08in}
\caption[]{Typical diagrams in the full theory from which the operators
\eqn{eq:Q12}--\eqn{eq:Qnnmm} originate.  The cross in diagram (d) means
a mass-insertion. It indicates that magnetic penguins originate from
the mass-term on the external line in the usual QCD or QED penguin
diagrams.
\label{fig:oporig}}
\end{figure}

\leftline{\bf Current-Current Operators (fig.\ \ref{fig:oporig}\,(a)):}

\begin{eqnarray}
Q_{1} = \left( \bar s_{i} u_{j}  \right)_{\rm V-A}
            \left( \bar u_{j}  d_{i} \right)_{\rm V-A}
&\qquad&
Q_{2} = \left( \bar s u \right)_{\rm V-A}
            \left( \bar u d \right)_{\rm V-A}
\label{eq:Q12}
\end{eqnarray}

\leftline{\bf QCD-Penguins Operators (fig.\ \ref{fig:oporig}\,(b)):}

\begin{eqnarray}
Q_{3} = \left( \bar s d \right)_{\rm V-A}
   \sum_{q} \left( \bar q q \right)_{\rm V-A}
&\qquad&
Q_{4} = \left( \bar s_{i} d_{j}  \right)_{\rm V-A}
   \sum_{q} \left( \bar q_{j}  q_{i} \right)_{\rm V-A}
\label{eq:Q34} \\
Q_{5} = \left( \bar s d \right)_{\rm V-A}
   \sum_{q} \left( \bar q q \right)_{\rm V+A}
&\qquad&
Q_{6} = \left( \bar s_{i} d_{j}  \right)_{\rm V-A}
   \sum_{q} \left( \bar q_{j}  q_{i} \right)_{\rm V+A}
\label{eq:Q56}
\end{eqnarray}

\leftline{\bf Electroweak-Penguins Operators (fig.\ \ref{fig:oporig}\,(c)):}

\begin{eqnarray}
Q_{7} = \frac{3}{2} \left( \bar s d \right)_{\rm V-A}
         \sum_{q} e_{q} \left( \bar q q \right)_{\rm V+A}
&\qquad&
Q_{8} = \frac{3}{2} \left( \bar s_{i} d_{j} \right)_{\rm V-A}
         \sum_{q} e_{q} \left( \bar q_{j}  q_{i}\right)_{\rm V+A}
\label{eq:Q78} \\
Q_{9} = \frac{3}{2} \left( \bar s d \right)_{\rm V-A}
         \sum_{q} e_{q} \left( \bar q q \right)_{\rm V-A}
&\qquad&
Q_{10} = \frac{3}{2} \left( \bar s_{i} d_{j} \right)_{\rm V-A}
         \sum_{q} e_{q} \left( \bar q_{j}  q_{i}\right)_{\rm V-A}
\label{eq:Q910}
\end{eqnarray}

\leftline{\bf Magnetic-Penguins Operators (fig.\ \ref{fig:oporig}\,(d)):}

\begin{eqnarray}
Q_{7\gamma} = \frac{e}{8\pi^2} \mb \bar{s}_i \sigma^{\mu\nu}
              (1+\gamma_5) b_i F_{\mu\nu}
&\qquad&
Q_{8G} = \frac{g}{8\pi^2} \mb \bar{s}_i \sigma^{\mu\nu}
        (1+\gamma_5)T^a_{ij} b_j G^a_{\mu\nu}
\label{eq:Q78mag}
\end{eqnarray}

\leftline{\bf $\Delta S=2$ and $\Delta B=2$ Operators
(fig.\ \ref{fig:oporig}\,(e)):}
\begin{eqnarray}
Q(\Delta S=2) = (\bar s d)_{V-A} (\bar s d)_{V-A} 
&\qquad&
Q(\Delta B=2) = (\bar b d)_{V-A} (\bar b d)_{V-A} 
\label{eq:QdSB2}
\end{eqnarray}

\leftline{\bf Semi-Leptonic Operators (fig.\ \ref{fig:oporig}\,(f)):}

\begin{eqnarray}
Q_{7V}  = (\bar s d)_{V-A} (\bar e e)_{V} 
&\qquad&
Q_{7A} = (\bar s d)_{V-A} (\bar e e)_{A}
\label{eq:Q7V7A} \\
Q_{9V}  = (\bar b s)_{V-A} (\bar e e)_{V} 
&\qquad&
Q_{10A} = (\bar b s)_{V-A} (\bar e e)_{A}
\label{eq:Q9V10A} \\
Q(\bar\nu \nu) = (\bar s d)_{V-A} (\bar \nu \nu)_{V-A} 
&\qquad&
Q(\bar\mu \mu) = (\bar s d)_{V-A} (\bar \mu \mu)_{V-A} 
\label{eq:Qnnmm}
\end{eqnarray}
where indices in color singlet currents have been suppressd for
simplicity.

For illustrative purposes, typical diagrams in the full theory from
which the operators \eqn{eq:Q12}--\eqn{eq:Qnnmm} originate are shown in
fig.\ \ref{fig:oporig}.

The operators listed above will enter our review in a systematic
fashion. We begin in section \ref{sec:HeffdF1:22} with the presentation
of the effective hamiltonians involving the current-current operators
$Q_1$ and $Q_2$ only.  These effective hamiltonians are given in
(\ref{B4}), (\ref{B5}) and (\ref{B6}) for $\Delta B=1$, $\Delta C=1$
and $\Delta S=1$ non-leptonic decays, respectively.

In section \ref{sec:HeffdF1:66} we will generalize the hamiltonians
(\ref{B4}) and (\ref{B6}) to include the QCD-penguin operators
$Q_3-Q_6$.
The corresponding expressions are given in (\ref{eq:HeffdB1:66}) and
(\ref{eq:HeffKpp}), respectively.  This generalization does not affect
the Wilson coefficients of $Q_1$ and $Q_2$.

Next in section \ref{sec:HeffdF1:1010} the $\Delta S=1$ and $\Delta
B=1$ hamiltonians of section \ref{sec:HeffdF1:66} will be generalized
to include the electroweak penguin operators $Q_7-Q_{10}$. These
generalized hamiltonians are given in (\ref{eq:HeffdF1:1010}) and
(\ref{eq:HeffdB1:1010}) for $\Delta S=1$ and $\Delta B=1$ non-leptonic
decays, respectively.  The inclusion of the electroweak penguin
operators implies the inclusion of $QED$ effects. Consequently the
coefficients of the operators $Q_1-Q_6$ given in this section will
differ slightly from the ones presented in the previous sections.

In section \ref{sec:HeffKpe} the effective hamiltonian for $K_L\to
\pi^0 e^+e^- $ will be presented.  It is given in (\ref{eq:HeffKpe}).
This hamiltonian can be considered as a generalization of the $\Delta
S=1$ hamiltonian (\ref{eq:HeffKpp}) presented in section
\ref{sec:HeffdF1:66} to include the semi-leptonic operators $Q_{7V}$
and $Q_{7A}$. This generalization does not modify the numerical values
of the $\Delta S=1$ coefficients $C_i$ ($i=1,\ldots,6$) given in
section \ref{sec:HeffdF1:66}.

In section \ref{sec:Heff:BXsgamma} we will discuss the effective
hamiltonian for $B\to X_s\gamma$. It is written down in
(\ref{eq:HeffBXsgamma}). This hamiltonian can be considered as a
generalization of the $\Delta B=1$ hamiltonian (\ref{eq:HeffdB1:66})
to include the magnetic penguin operators $Q_{7\gamma}$ and $Q_{8G}$.
This generalization does not modify the numerical values of the $\Delta
B=1$ coefficients $C_i$ ($i=1,\ldots,6$) from section \ref{sec:HeffdF1:66}.

In section \ref{sec:Heff:BXsee} we present the effective hamiltonian
for $B\to X_s e^+ e^- $. It is to be found in (\ref{eq:Heff2atmu}) and
can be considered as the generalization of the $B\to X_s\gamma$
hamiltonian to include the semi-leptonic operators $Q_{9V}$ and
$Q_{10A}$.  The coefficients $C_i$ ($i=1,\ldots,6,7\gamma,8G$) given in
section \ref{sec:Heff:BXsgamma} are not affected by this
generalization.

In section \ref{sec:HeffRareKB} the effective hamiltonians for
$K^+\to\pi^+\nu\bar\nu$, $K_L\to\mu^+\mu^-$, $K_L\to \pi^0\nu\bar\nu$
($B\to X_{s,d}\nu\bar\nu$) and $B\to l^+l^-$ will be discussed.  They
are given in (\ref{hkpn}), (\ref{hklm}), (\ref{hxnu}) and (\ref{hyll})
respectively. Each of these hamiltonians involves only a single
operator:  $Q(\nu\bar\nu)$ or $Q(\mu\bar\mu)$ for
$K^+\to\pi^+\nu\bar\nu$ $(K_L\to \pi^0\nu\bar\nu)$ and
$K_L\to\mu^+\mu^-$ with analogous operators for $B\to
X_{s,d}\nu\bar\nu$ and $B\to l^+l^-$.

Finally, sections \ref{sec:HeffKKbar} and \ref{sec:HeffBBbar} present
the effective hamiltonians for $\Delta S=2$ and $\Delta B=2$
transitions, respectively. These hamiltonians involve the operators
$Q(\Delta S=2)$ and $Q(\Delta B=2)$ and can be found in (\ref{hds2})
and (\ref{hdb2}).

In table \ref{tab:heffguide} we give the list of effective hamiltonians
to be presented below, the equations in which they can be found and the
list of operators entering different hamiltonians.

\begin{table}[htb]
\caption[]{Compilation of various processes, equation no.\ of the
corresponding effective hamiltonians and contributing operators.}
\label{tab:heffguide}
\begin{center}
\begin{tabular}{|l|l|l|}
\multicolumn{1}{|c|}{\bf Process} &
\multicolumn{1}{c|}{\bf Cf.\ Equation} &
\multicolumn{1}{c|}{\bf Contributing Operators} \\
\hline
$\Delta F=1$, $F=B,C,S$ current-current &
   \eqn{B4}--\eqn{B6} & $Q_1, Q_2$       \\
$\Delta F=1$ pure QCD &
   \eqn{eq:HeffKpp}, \eqn{eq:HeffdB1:66} & $Q_1,\ldots,Q_6$ \\
$\Delta F=1$ QCD and electroweak &
   \eqn{eq:HeffdF1:1010}, \eqn{eq:HeffdB1:1010} & $Q_1,\ldots,Q_{10}$ \\
\hline
$\Kpiee$ & \eqn{eq:HeffKpe} & $Q_1,\ldots,Q_6, Q_{7V}, Q_{7A}$ \\
\hline
$B\to X_s\gamma$ & \eqn{eq:HeffBXsgamma} &
   $Q_1,\ldots,Q_6, Q_{7\gamma}, Q_{8G}$ \\
$B\to X_s e^+e^-$ & \eqn{eq:Heff2atmu} & 
   $Q_1,\ldots,Q_6, Q_{7\gamma}, Q_{8G}, Q_{9V}, Q_{10A}$ \\
\hline
$\kpn$, $(\klm)_{SD}$, $K_L\to\pi^0\nu\bar\nu$, &
   \eqn{hkpn}, \eqn{hklm}, \eqn{hxnu} & $Q(\bar\nu \nu), Q(\bar\mu \mu)$ \\
$B\to X_{s, d}\nu\bar\nu$, $B\to l^+l^-$ & \eqn{hyll} & \\
\hline
$K^0-\bar K^0$ mixing & \eqn{hds2} & $Q(\Delta S=2)$ \\
$B^0-\bar B^0$ mixing & \eqn{hdb2} & $Q(\Delta B=2)$ \\
\end{tabular}
\end{center}
\end{table}

\section{The Effective $\Delta F=1$ Hamiltonian: Current-Current Operators}
   \label{sec:HeffdF1:22}

\subsection{Operators}
   \label{sec:HeffdF1:22:op}
We begin our compendium by presenting the parts of effective
hamiltonians involving the current-current operators only.
These operators will be generally denoted by $Q_1$ and $Q_2$,
although their flavour structure depends on the decay considered.
To be specific we will consider
\begin{equation}\label{B1}
Q_1=(\bar b_i c_j)_{V-A} (\bar u_j d_i)_{V-A}
\qquad 
Q_2=(\bar b_i c_i)_{V-A} (\bar u_j d_j)_{V-A}
\end{equation}
\begin{equation}\label{B2}
Q_1=(\bar s_i c_j)_{V-A} (\bar u_j d_i)_{V-A}
\qquad 
Q_2=(\bar s_i c_i)_{V-A} (\bar u_j d_j)_{V-A}
\end{equation}
\begin{equation}\label{B3}
Q_1=(\bar s_i u_j)_{V-A} (\bar u_j d_i)_{V-A}
\qquad 
Q_2=(\bar s_i u_i)_{V-A} (\bar u_j d_j)_{V-A}
\end{equation}
for $\Delta B=1$, $\Delta C=1$ and $\Delta S=1$ decays respectively.
Then the corresponding effective hamiltonians are given by
\begin{equation}\label{B4}
H_{eff}(\Delta B=1)=\frac{G_F}{\sqrt{2}}V_{cb}^{*}V_{ud}
\lbrack C_1(\mu) Q_1+C_2(\mu)Q_2 \rbrack
\qquad
(\mu=O(m_b))
\end{equation}
\begin{equation}\label{B5}
H_{eff}(\Delta C=1)=\frac{G_F}{\sqrt{2}}V_{cs}^{*}V_{ud}
\lbrack C_1(\mu) Q_1+C_2(\mu)Q_2 \rbrack
\qquad
(\mu=O(m_c))
\end{equation}
\begin{equation}\label{B6}
H_{eff}(\Delta S=1)=\frac{G_F}{\sqrt{2}}V_{us}^{*}V_{ud}
\lbrack C_1(\mu) Q_1+C_2(\mu)Q_2 \rbrack
\qquad
(\mu=O(1\gev))
\end{equation}
As we will see in subsequent sections these hamiltonians have to be
generalized to include also penguin operators. This however will not
change the Wilson coefficients $C_1(\mu)$ and $C_2(\mu)$ except for
small $O(\aem)$ corrections in a complete analysis which includes
also electroweak penguin operators. For this reason it is useful to
present the results for $C_{1,2}$ separately as they can be used in a
large class of decays.

When analyzing $Q_1$ and $Q_2$ in isolation, it is useful to work with
the operators $Q_{\pm}$ and their coefficients $z_{\pm}$ defined by
\begin{equation}\label{B7}
Q_{\pm}=\frac{1}{2} (Q_2\pm Q_1)
\qquad
\qquad
z_\pm=C_2\pm C_1 \, .
\end{equation}
$Q_+$ and $Q_-$ do not mix under renormalization and the expression for
$z_\pm(\mu)$ is very simple.

\subsection{Wilson Coefficients and RG Evolution}
   \label{sec:HeffdF1:22:wcrg}
The initial conditions for $z_\pm$ at $\mu=M_W$ are obtained using the
matching procedure between the full (fig.\ \ref{fig:1loopful}\,(a)--(c))
and effective (fig.\ \ref{fig:1loopeff}\,(a)--(c)) theory summarized in
section \ref{sec:basicform:wc:rgf}. Given the initial conditions for
$z_\pm$ at scale $\mu=M_W$
\begin{equation}\label{B8}
z_\pm(M_W)=1+\frac{\as(M_W)}{4\pi}B_\pm
\end{equation}
and using the NLO RG evolution formula \eqn{cjua} for the case without
mixing one finds for the Wilson coefficients of $Q_\pm$ at some scale $\mu$
\begin{equation}\label{B9}
z_\pm(\mu)=\left[1+\frac{\as(\mu)}{4\pi}J_\pm\right]
      \left[\frac{\as(M_W)}{\as(\mu)}\right]^{d_\pm}
\left[1+\frac{\as(M_W)}{4\pi}(B_\pm-J_\pm)\right]
\end{equation}
with
\begin{equation}\label{B10}
J_\pm=\frac{d_\pm}{\beta_0}\beta_1-\frac{\gamma^{(1)}_\pm}{2\beta_0}
\qquad\qquad
d_\pm=\frac{\gamma^{(0)}_\pm}{2\beta_0} \, ,
\end{equation}
where the coefficients $\beta_0$ and $\beta_1$ of the QCD
$\beta$-function are given in \eqn{b0b1}. Furthermore the LO and NLO
expansion coefficients for the anomalous dimensions $\gamma_\pm$ of
$Q_\pm$ in \eqn{B10} and the coefficients $B_\pm$ in \eqn{B8} are given by
\begin{equation}\label{B11}
\gamma^{(0)}_\pm=\pm 12 \frac{N\mp 1}{2N}
\end{equation}
\begin{equation}\label{B12}
\gamma^{(1)}_{\pm}=\frac{N\mp 1}{2N}
\left[-21\pm\frac{57}{N}\mp\frac{19}{3}N \pm
\frac{4}{3}f-2\beta_0\kappa_\pm\right]
\end{equation}
\begin{equation}\label{B13}
B_\pm=\frac{N\mp 1}{2N}\left[\pm 11+\kappa_\pm\right]
\end{equation}
with $N$ being the number of colors.  Here we have introduced the
parameter $\kappa_\pm$ which conveniently distinguishes between various
renormalization schemes
\begin{equation}
\kappa_\pm = \left\{ \begin{array}{rl}
    0 & \qquad {\rm NDR}  \\
\mp 4 & \qquad {\rm HV}
\end{array}\right. \, .
\label{B14}
\end{equation}

Thus, using $N=3$ in the following, $J_\pm$ in (\ref{B10}) can also be
written as
\begin{equation}\label{B15}
J_\pm=(J_\pm)_{\rm NDR}+\frac{3\mp 1}{6}\kappa_\pm
=(J_\pm)_{\rm NDR}\pm\frac{\gamma^{(0)}_\pm}{12}\kappa_\pm
\end{equation}
Setting $\gamma_\pm^{(1)}$, $B_\pm$ and $\beta_1$ to zero gives the
leading logarithmic approximation \cite{altarelli:74}, \cite{gaillard:74}.

The NLO calculations in the NDR scheme and in the HV scheme have been
presented in \cite{burasweisz:90}.  In writing (\ref{B12}) we have
incorporated the $-2 \gamma^{(1)}_J$ correction in the HV scheme
resulting from the non-vanishing two--loop anomalous dimension of the
weak current.
\begin{equation}
\gamma^{(1)}_J = \left\{
\begin{array}{lcl}
                           0 & \qquad & \mbox{NDR} \\
\frac{N^2 -1}{N} \, 2\beta_0 & \qquad & \mbox{HV}
\end{array}
\right.
\label{eq:wcanom}
\end{equation}
The NLO corrections $\gamma^{(1)}_\pm$ in the
dimensional reduction scheme (DRED) have been first considered in
\cite{altarelli:81} and later confirmed in \cite{burasweisz:90}. Here
one has $\kappa_\pm = \mp 6 - N$. This value for $\kappa_\pm$ in DRED
incorporates also a finite renormalization of $\as$ in order to work in
all schemes with the usual $\overline{MS}$ coupling.

As already discussed in section \ref{sec:basicform:wc:rgdep}, the
expression $(B_\pm-J_\pm)$ is scheme independent.  The scheme
dependence of the Wilson coeffcients $z_\pm(\mu)$ originates then
entirely from the scheme dependence of $J_\pm$ at the lower end of the
evolution which can be seen explicitly in (\ref{B15}).

In order to exhibit the $\mu$ dependence on the same footing as the
scheme dependence, it is useful to rewrite (\ref{B9}) in the case of
B--decays as follows:
\begin{equation}\label{B16}
z_\pm(\mu)=\left[1+\frac{\as(m_b)}{4\pi} \tilde J_\pm(\mu)\right]
      \left[\frac{\as(M_W)}{\as(m_b)}\right]^{d_\pm}
\left[1+\frac{\as(M_W)}{4\pi}(B_\pm-J_\pm)\right]
\end{equation}
with 

\begin{equation}\label{B17}
\tilde J_\pm(\mu)=(J_\pm)_{NDR}\pm 
\frac{\gamma^{(0)}_\pm}{12}\kappa_\pm
+\frac{\gamma^{(0)}_\pm}{2}\ln(\frac{\mu^2}{m^2_b})
\end{equation}
summarizing both the renormalization scheme dependence and the
$\mu$--dependence. Note that in the first parenthesis in (\ref{B16}) we
have set $\as(\mu)=\as(m_b)$ as the difference in the scales
in this correction is still of higher order.  We also note that a
change of the renormalization scheme can be compensated by a change in
$\mu$. From (\ref{B17}) we find generally
\begin{equation}\label{B18}
\mu_i^\pm=\mu_{NDR} \, \exp\left(\mp\frac{\kappa_\pm^{(i)}}{12}\right)
\end{equation}
where $i$ denotes a given scheme. From (\ref{B14}) we then have
\begin{equation}\label{B19}
\mu_{HV}=\mu_{NDR} \, \exp\left(\frac{1}{3}\right)
\end{equation}
Evidently the change in $\mu$ relating HV and NDR\footnote{ The
relation $\mu_{\rm DRED}^{\pm} = \mu_{\rm NDR} \, \exp\left(\frac{2\pm
1}{4}\right)$ between NDR and DRED is more involved. In any case
$\mu_{\rm HV}$ and $\mu_{\rm DRED}^\pm$ are larger than $\mu_{\rm
NDR}$.} is the same for $z_+$ and $z_-$ and consequently for
$C_i(\mu)$.

This discussion shows that a meaningful analysis of the $\mu$
dependence of $C_i(\mu)$ can only be made simultaneously with the
analysis of the scheme dependence.

The coefficients $C_i(\mu)$ for B-decays can now be calculated using
\begin{equation}\label{B20}
C_1(\mu)=\frac{z_+(\mu)-z_-(\mu)}{2}
\qquad\qquad
C_2(\mu)=\frac{z_+(\mu)+z_-(\mu)}{2}
\end{equation}
To this end we set $f=5$ in the formulae above and use the two-loop
$\as(\mu)$ of eq.\ \eqn{amu} with $\Lambda \equiv \Lms^{(5)}$. The
actual numerical values used for $\as(M_Z)$ or equivalently
$\Lms^{(5)}$ are collected in appendix \ref{app:numinput} together with
other numerical input parameters.

In the case of D-decays and K-decays the relevant scales are
$\mu=\ord(m_c)$ and $\mu=\ord(1\gev)$, respectively. In order to
calculate $C_i(\mu)$ for these cases one has to evolve these
coefficients first from $\mu=\ord(m_b)$ further down to $\mu=\ord(m_c)$ in
an effective theory with $f=4$. Matching $\as^{(5)}(m_b) = \as^{(4)}(m_b)$
we find to a very good approximation $\Lms^{(4)}=(325\pm110)\mev$.
Unfortunately, the necessity to evolve $C_i(\mu)$ from $\mu=M_W$ down to
$\mu=m_c$ in two different effective theories ($f=5$ and $f=4$) and
eventually in the case of K-decays with $f=3$ for $\mu< m_c$ makes the
formulae for $C_i(\mu)$ in D--decays and K--decays rather complicated.
They can be found in \cite{burasetal:92d}. Fortunately all these
complications can be avoided by a simple trick, which reproduces the
results of \cite{burasetal:92d} to better than $1.5\%$.  In order to
find $C_i(\mu)$ for $1\gev\leq\mu\leq 2\gev$ one can simply use the
master formulae given above with $\Lms^{(5)}$ replaced by $\Lms^{(4)}$
and an ``{effective}'' number of active flavours $f=4.15$. The latter
effective value for $f$ allows to obtain a very good agreement with
\cite{burasetal:92d}.  This can be verified by comparing the results
presented here with those in tables \ref{tab:wc6smu1} and
\ref{tab:wc6smu2} where no ``tricks'' have been used.  The nice feature
of this method is that the $\mu$ and renormalization scheme dependences
of $C_i(\mu)$ can still be studied in simple terms.

The numerical coefficients $C_i(\mu)$ for B--decays are shown in tables
\ref{tab:c1B} and \ref{tab:c2B} for different $\mu$ and  $\Lms^{(5)}$.
In addition to the results for the NDR and HV renormalization schemes
we show the LO values\footnote{The results for the DRED scheme can be
found in \cite{buras:94}.}. The corresponding results for K--decays and
D--decays are given in tables \ref{tab:c1KD} and \ref{tab:c2KD}.

\begin{table}[htb]
\caption[]{The coefficient $C_1(\mu)$ for  B-decays.}
\label{tab:c1B}
\begin{center}
\begin{tabular}{|c|c|c|c||c|c|c||c|c|c|}
& \multicolumn{3}{c||}{$\Lms^{(5)}=140\mev$} &
  \multicolumn{3}{c||}{$\Lms^{(5)}=225\mev$} &
  \multicolumn{3}{c| }{$\Lms^{(5)}=310\mev$} \\
\hline
$\mu [{\rm GeV}]$ &LO & NDR & HV & LO & NDR & HV & LO & NDR & HV  \\
\hline
\hline
4.0 & --0.274 & --0.175 & --0.211 & --0.310 & --0.197 & --0.239 & --0.341 &
--0.216 & --0.264  \\
\hline
5.0 & --0.244 & --0.151 & --0.184 & --0.274 & --0.169 & --0.208 & --0.300 &
--0.185 & --0.228  \\
\hline
6.0 & --0.221 & --0.133 & --0.164 & --0.248 & --0.148 & --0.184 & --0.269 &
--0.161 & --0.201  \\
\hline
7.0 & --0.203 & --0.118 & --0.148 & --0.226 & --0.132 & --0.166 & --0.246 &
--0.143 & --0.181  \\
\hline
8.0 & --0.188 & --0.106 & --0.135 & --0.209 & --0.118 & --0.151 & --0.226 &
--0.128 & --0.164  \\
\end{tabular}
\end{center}
\end{table}

\begin{table}[htb]
\caption[]{The coefficient $C_2(\mu)$ for B-decays.}
\label{tab:c2B}
\begin{center}
\begin{tabular}{|c|c|c|c||c|c|c||c|c|c|}
& \multicolumn{3}{c||}{$\Lms^{(5)}=140\mev$} &
  \multicolumn{3}{c||}{$\Lms^{(5)}=225\mev$} &
  \multicolumn{3}{c| }{$\Lms^{(5)}=310\mev$} \\
\hline
$\mu [{\rm GeV}]$ & LO& NDR & HV & LO & NDR & HV & LO & NDR & HV  \\
\hline
\hline
4.0 & 1.121 & 1.074 & 1.092 & 1.141 & 1.086 & 1.107 & 1.158 &
1.096 & 1.120  \\
\hline
5.0 & 1.105 & 1.062 & 1.078 & 1.121 & 1.072 & 1.090 & 1.135 &
1.080 & 1.101  \\
\hline
6.0 & 1.093 & 1.054 & 1.069 & 1.107 & 1.062 & 1.079 & 1.118 &
1.068 & 1.087  \\
\hline
7.0 & 1.084 & 1.047 & 1.061 & 1.096 & 1.054 & 1.069 & 1.106 &
1.059 & 1.077  \\
\hline
8.0 & 1.077 & 1.042 & 1.055 & 1.087 & 1.047 & 1.062 & 1.096 &
1.052 & 1.069  \\
\end{tabular}
\end{center}
\end{table}

\noindent
From tables \ref{tab:c1B}--\ref{tab:c2Bnlo} we observe:
\begin{itemize}
\item
The scheme dependence of the Wilson coefficients is sizable.  This is
in particular the case of $C_1$ which vanishes in the absence of QCD
corrections.
\item
The differences between LO and NLO results in the case of $C_1$ are
large showing the importance of next--to--leading corrections. In fact
in the NDR scheme the corrections may be as large as $70\%$.  This
comparison of LO and NLO coefficients can however be questioned because
for the chosen values of $\Lms$ one has $\as^{(LO)}(M_Z)=0.135 \pm
0.009$ to be compared with $\as(M_Z)=0.117 \pm 0.007$ \cite{bethke:94},
\cite{webber:94}. Consequently the difference in LO and NLO results for
$C_i$ originates partly in the change in the value of the QCD
coupling.
\item
In view of the latter fact it is instructive to show also the LO
results in which the next-to-leading expression for $\as$ is used.
We give some examples in tables \ref{tab:c1Bnlo} and \ref{tab:c2Bnlo}.
Now the differences between LO and NLO results is considerably smaller
although still as large as $30-40\%$ in the case of $C_1$ and the NDR
scheme.
\item
In any case the inclusion of NLO corrections in NDR and HV schemes
weakens the impact of QCD on the Wilson coefficients of
current--current operators.  It is however important to keep in mind
that such a behavior is specific to the scheme chosen and will in
general be different in other schemes, reflecting the unphysical nature
of the Wilson coefficient functions.
\end{itemize}

\begin{table}[htb]
\caption[]{The coefficient $C_1(\mu)$ for K-decays and D-decays.}
\label{tab:c1KD}
\begin{center}
\begin{tabular}{|c|c|c|c||c|c|c||c|c|c|}
& \multicolumn{3}{c||}{$\Lms^{(4)}=215\mev$} &
  \multicolumn{3}{c||}{$\Lms^{(4)}=325\mev$} &
  \multicolumn{3}{c| }{$\Lms^{(4)}=435\mev$} \\
\hline
$\mu [{\rm GeV}]$ &LO & NDR & HV & LO & NDR & HV & LO & NDR & HV  \\
\hline
\hline
1.00 & --0.602 & --0.410 & --0.491 & --0.742 & --0.510 & --0.631 & --0.899 &
--0.632 & --0.825  \\
\hline
1.25 & --0.529 & --0.356 & --0.424 & --0.636 & --0.430 & --0.523 & --0.747 &
--0.512 & --0.642  \\
\hline
1.50 & --0.478 & --0.319 & --0.379 & --0.565 & --0.378 & --0.457 & --0.653 &
--0.439 & --0.543  \\
\hline
1.75 & --0.439 & --0.291 & --0.346 & --0.514 & --0.340 & --0.410 & --0.587 & 
--0.390 & --0.478  \\
\hline
2.00 & --0.409 & --0.269 & --0.320 & --0.475 & --0.311 & --0.375 & --0.537 &
--0.353 & --0.431  \\
\end{tabular}
\end{center}
\end{table}

\begin{table}[htb]
\caption[]{The coefficient $C_2(\mu)$ for K-decays and D-decays.}
\label{tab:c2KD}
\begin{center}
\begin{tabular}{|c|c|c|c||c|c|c||c|c|c|}
& \multicolumn{3}{c||}{$\Lms^{(4)}=215\mev$} &
  \multicolumn{3}{c||}{$\Lms^{(4)}=325\mev$} &
  \multicolumn{3}{c| }{$\Lms^{(4)}=435\mev$} \\
\hline
$\mu [{\rm GeV}]$ &LO & NDR & HV & LO & NDR & HV & LO & NDR & HV  \\
\hline
\hline
1.00 & 1.323 & 1.208 & 1.259 & 1.422 & 1.275 & 1.358 & 1.539 &  
1.363 & 1.506  \\
\hline
1.25 & 1.274 & 1.174 & 1.216 & 1.346 & 1.221 & 1.282 & 1.426 & 
1.277 & 1.367  \\
\hline
1.50 & 1.241 & 1.152 & 1.187 & 1.298 & 1.188 & 1.237 & 1.358 & 
1.228 & 1.296  \\
\hline
1.75 & 1.216 & 1.136 & 1.167 & 1.264 & 1.165 & 1.207 & 1.313 & 
1.196 & 1.252  \\
\hline
2.00 & 1.198 & 1.123 & 1.152 & 1.239 & 1.148 & 1.185 & 1.279 &
1.174 & 1.221  \\
\end{tabular}
\end{center}
\end{table}

\begin{table}[htb]
\caption[]{$C_1^{\rm LO}$ and $C_2^{\rm LO}$ for B-decays with $\as$
in NLO.}
\label{tab:c1Bnlo}
\begin{center}
\begin{tabular}{|c|c|c||c|c||c|c|}
& \multicolumn{2}{c||}{$\Lms^{(5)}=140\mev$} &
  \multicolumn{2}{c||}{$\Lms^{(5)}=225\mev$} &
  \multicolumn{2}{c| }{$\Lms^{(5)}=310\mev$} \\
\hline
$\mu [{\rm GeV}]$ & $C_1$ & $C_2$ & $C_1$ & 

$C_2$ & $C_1$ & $C_2$  \\
\hline
\hline
4.0 & --0.244 & 1.105 & --0.274 & 1.121 & --0.301 & 1.135
\\
\hline
5.0 & --0.217 & 1.091 & --0.243 & 1.105 & --0.265 & 1.116
\\
\hline
6.0 & --0.197 & 1.082 & --0.220 & 1.093 & --0.239 & 1.102
\\
\end{tabular}
\end{center}
\end{table}

\begin{table}[htb]
\caption[]{$C_1^{\rm LO}$ and $C_2^{\rm LO}$ for K and D-decays with $\as$
in NLO.}
\label{tab:c2Bnlo}
\begin{center}
\begin{tabular}{|c|c|c||c|c||c|c|}
& \multicolumn{2}{c||}{$\Lms^{(4)}=215\mev$} &
  \multicolumn{2}{c||}{$\Lms^{(4)}=325\mev$} &
  \multicolumn{2}{c| }{$\Lms^{(4)}=435\mev$} \\
\hline
$\mu [{\rm GeV}]$ & $C_1$ & $C_2$ & $C_1$ & $C_2$ & $C_1$ & $C_2$  \\
\hline
\hline
1.0 & --0.524 & 1.271 & --0.664 & 1.366 & --0.851 & 1.502
\\
\hline
1.5 & --0.413 & 1.201 & --0.493 & 1.250 & --0.579 & 1.307
\\
\hline
2.0 & --0.354 & 1.165 & --0.412 & 1.200 & --0.469 & 1.235
\\
\end{tabular}
\end{center}
\end{table}

We have made the whole discussion without invoking HQET (cf.\ section
\ref{sec:HQET}). It is sometimes stated in the literature that at
$\mu=m_b$ in the case of B-decays one {\it has to } switch to HQET.  In
this case for $\mu<m_b$ the anomalous dimensions $\gamma_\pm$ differ
from those given above. We should however stress that switching to HQET
can be done at any $\mu<m_b$ provided the logarithms $\ln(m_b/\mu)$ in
$\langle Q_i \rangle$ do not become too large.  Similar comments apply
to D-decays with respect to $\mu=m_c$. Of course the coefficients $C_i$
calculated in HQET for $\mu<m_b$ are different from the coefficients
presented here. However the corresponding matrix elements $\langle Q_i
\rangle$ in HQET are also different so that the physical amplitudes
remain unchanged.

\section{The Effective $\Delta F=1$ Hamiltonian: Inclusion of QCD Penguin
         Operators}
         \label{sec:HeffdF1:66}
In section \ref{sec:HeffdF1:22} we have restricted ourselves 
to current-current operators when considering
QCD corrections to the effective $\Delta F=1$ ($F=B, C, S$)
hamiltonian for weak decays.

As already mentioned in section \ref{sec:basicform:rg:pop} e.g.\ for
the $\dS$ case the special flavour structure of $Q_{2} = \left( \bar s
u \right)_{\rm V-A} \left( \bar u d \right)_{\rm V-A}$ allows not only
for QCD corrections of the current-current type as in
fig.\ \ref{fig:1loopeff}\,(a)--(c) from which the by now well known
second current-current operator $Q_1$ is created.  For a complete
treatment of QCD corrections all possible ways of attaching a gluon to
the initial weak $\Delta F=1$ transition operator $Q_2$ have to be
taken into account. Therefore attaching gluons to $Q_2$ in the form of
diagrams (d.1) and (d.2) in fig.\ \ref{fig:1loopeff}, generates a
completely new set of four-quark operators, the so-called QCD penguin
operators, usually denoted as $Q_3, \ldots, Q_6$\footnote{Obviously,
whether or not it is possible to form a closed fermion loop as in a
type-1 insertion or to connect the two currents to yield a continuous
fermion line as required for a type-2 insertion strongly depends on the
flavour structure of the operator considered. E.g.\ for $Q_2$ only the
type-2 penguin diagram contributes. This feature can be exploited
to obtain NLO anomalous dimension matrices in the NDR scheme 
without the necessity of calculating closed fermion loops with $\gamma_5$
\cite{burasetal:92b}, \cite{burasetal:92c}.}. This procedure is often
referred to as inserting $Q_2$ into type--1 and type--2 penguin
diagrams.

The $\dS$ effective hamiltonian for $\Kpipi$ at scales $\mu < \mc$ then
reads
\begin{equation}
\Heff(\dS) = \frac{G_F}{\sqrt{2}} \V{us}^* \V{ud}^{} \sum_{i=1}^{6}
\left( z_i(\mu) + \tau \; y_i(\mu) \right) Q_i \, ,
\label{eq:HeffKpp}
\end{equation}
with
\begin{equation}
\tau = -\frac{\V{ts}^*\V{td}^{}}{\V{us}^*\V{ud}^{}} \, .
\label{eq:tauKpp}
\end{equation}
The set of four-quark operators $\vec{Q}(\mu)$ and Wilson coefficients
$\vec{z}(\mu)$ and $\vec{y}(\mu)$ will be discussed one by one in
the subsections below.

\subsection{Operators}
            \label{sec:HeffdF1:66:op}
The basis of four-quark operators for the $\dS$ effective hamiltonian
in \eqn{eq:HeffKpp} is given in explicit form by
\begin{eqnarray}
Q_{1} & = & \left( \bar s_{i} u_{j}  \right)_{\rm V-A}
            \left( \bar u_{j}  d_{i} \right)_{\rm V-A}
\, , \nn \\
Q_{2} & = & \left( \bar s u \right)_{\rm V-A}
            \left( \bar u d \right)_{\rm V-A}
\, , \nn \\
Q_{3} & = & \left( \bar s d \right)_{\rm V-A}
   \sum_{q} \left( \bar q q \right)_{\rm V-A}
\, , \label{eq:Kppbasis} \\
Q_{4} & = & \left( \bar s_{i} d_{j}  \right)_{\rm V-A}
   \sum_{q} \left( \bar q_{j}  q_{i} \right)_{\rm V-A}
\, , \nn \\
Q_{5} & = & \left( \bar s d \right)_{\rm V-A}
   \sum_{q} \left( \bar q q \right)_{\rm V+A}
\, , \nn \\
Q_{6} & = & \left( \bar s_{i} d_{j}  \right)_{\rm V-A}
   \sum_{q} \left( \bar q_{j}  q_{i} \right)_{\rm V+A}
\, . \nn
\end{eqnarray}
As already mentioned, this basis closes under QCD renormalization.  

For $\mu < \mc$ the sums over active quark flavours in \eqn{eq:Kppbasis}
run over $u$, $d$ and $s$.  However, when  $\mb > \mu > \mc$ is
considered also $q=c$ has to be included. Moreover, in this case two
additional current--current operators have to be taken into account
\begin{equation}
Q_1^c = \left(\bar s_i c_j  \right)_{\rm V-A}
        \left(\bar c_j  d_i \right)_{\rm V-A}
\, , \qquad
Q_2^c = \left(\bar s c \right)_{\rm V-A}
        \left(\bar c d \right)_{\rm V-A} \, .
\label{eq:KppQ12c}
\end{equation}
and the effective hamiltonian takes the form
\begin{equation}
\Heff(\dS) = \frac{G_F}{\sqrt{2}} \V{us}^* \V{ud}^{} 
\left[(1-\tau) \sum_{i=1}^{2} z_i(\mu) (Q_i-Q^c_i) +
\tau \; \sum_{i=1}^{6} v_i(\mu)  Q_i \right]
\label{eq:HeffKppc}
\end{equation}

\subsection{Wilson Coefficients}
            \label{sec:HeffdF1:66:wc}
For the Wilson coefficients $y_i(\mu)$ and $z_i(\mu)$ in
eq.~\eqn{eq:HeffKpp} one has
\begin{equation}
y_i(\mu) = v_i(\mu) - z_i(\mu) \, .
\label{eq:WCy}
\end{equation}
The coefficients $z_i$ and $v_i$ are the components of the six
dimensional column vectors $\vec{v}(\mu)$ and $\vec{z}(\mu)$. Their
RG evolution is given by
\begin{equation}
\vec{v}(\mu) =
\hU_3(\mu,\mc) \hM(\mc) \hU_4(\mc,\mb) \hM(\mb) \hU_5(\mb,\mw)
\vec{C}(\mw) \, ,
\label{eq:WCv}
\end{equation}
\begin{equation}
\vec{z}(\mu) = \hU_3(\mu,\mc) \vec{z}(\mc) \, .
\label{eq:WCz}
\end{equation}
Here $\hU_f(m_1,m_2)$ denotes the full NLO evolution matrix for
$f$ active flavours. $\hM(m_i)$ is the matching matrix at quark
threshold $m_i$ given in eq.\ \eqn{mdrt}.  These two matrices will be
discussed in more detail in subsections~\ref{sec:HeffdF1:66:rge} and
\ref{sec:HeffdF1:66:Mm}, respectively.

The initial values $\vC(\mw)$ necessary for the RG evolution of
$\vv(\mu)$ in eq.~\eqn{eq:WCv} can be found according to the procedure
of matching the effective (fig.\ \ref{fig:1loopeff}) onto the full
theory (fig.\ \ref{fig:1loopful}) as summarized in
section~\ref{sec:basicform:wc}.  For the NDR scheme one obtains
\cite{burasetal:92a}
\begin{eqnarray}
C_1(\mw) &=&     \frac{11}{2} \; \frac{\as(\mw)}{4\pi} \, ,
\label{eq:CMw1QCD} \\
C_2(\mw) &=& 1 - \frac{11}{6} \; \frac{\as(\mw)}{4\pi} \, ,
\label{eq:CMw2QCD} \\
C_3(\mw) &=& -\frac{\as(\mw)}{24\pi} \widetilde{E}_0(x_t) \, ,
\label{eq:CMw3QCD} \\
C_4(\mw) &=& \frac{\as(\mw)}{8\pi} \widetilde{E}_0(x_t) \, ,
\label{eq:CMw4QCD} \\
C_5(\mw) &=& -\frac{\as(\mw)}{24\pi} \widetilde{E}_0(x_t) \, ,
\label{eq:CMw5QCD} \\
C_6(\mw) &=& \frac{\as(\mw)}{8\pi} \widetilde{E}_0(x_t) \, ,
\label{eq:CMw6QCD}
\end{eqnarray}
where
\begin{eqnarray}
E_0(x) &=& -\frac{2}{3} \ln x + \frac{x (18 -11 x - x^2)}{12 (1-x)^3} +
          \frac{x^2 (15 - 16 x  + 4 x^2)}{6 (1-x)^4} \ln x \, ,
\label{eq:Ext} \\
\widetilde{E}_0(x_t) &=& E_0(x_t) - \frac{2}{3}
\label{eq:Exttilde}
\end{eqnarray}
with
\begin{equation}
x_t = \frac{m_t^2}{\mw^2} \, .
\label{eq:xt}
\end{equation}
Here $E_0(x)$ results from the evaluation of the gluon penguin diagrams.

The initial values $\vec{C}(\mw)$ in the HV scheme can be found in
\cite{burasetal:92a}.

In order to calculate the initial conditions $\vec{z}(\mc)$ for
$z_i(\mu)$ in eq.~\eqn{eq:WCz} one has to consider the difference
$Q_2^u - Q_2^c$ of $Q_2$-type current-current operators
as can be seen explicitly in \eqn{eq:HeffKppc}. Due to the GIM
mechanism the coefficients $z_i(\mu)$ of penguin operators $Q_i$,
$i\not=1,2$ are zero in 5- and 4-flavour theories. The evolution for
scales $\mu > \mc$ involves then only the current-current operators
$Q^u_i-Q^c_i$, $i=1, 2$,  with initial conditions at
scale $\mu = \mw$
\begin{equation}
z_1(\mw) = C_1(\mw) \, ,
\qquad
z_2(\mw) = C_2(\mw) \, .
\label{eq:zMw12}
\end{equation}
$Q_{1,2}^u\equiv Q_{1,2}$ and $Q_{1,2}^c$ do not mix with each other under
renormalization. We then find
\begin{equation}
\left( \begin{array}{ll} z_1(\mc) \\ z_2(\mc) \end{array} \right) =
\hU_4(\mc,\mb) \; \hM(\mb) \; \hU_5(\mb,\mw) \;
\left( \begin{array}{ll} z_1(\mw) \\ z_2(\mw) \end{array} \right) \, ,
\label{eq:zmc12}
\end{equation}
where this time the evolution matrices $\hU_{4,5}$ contain only the $2
\times 2$ anomalous dimension submatrices describing the mixing between
current-current operators. The matching matrix $\hM(\mb)$ is then also
only the corresponding $2 \times 2$ submatrix of the full $6 \times 6$
matrix in \eqn{eq:MmKpp}. For the particular case of \eqn{eq:zmc12} it
simplifies to a unit matrix. When the charm quark is integrated out the
operators $Q_{1,2}^c$ disappear from the effective hamiltonian and the
coefficients $z_i(\mu)$, $i\not=1,2$ for penguin operators become
non-zero. In order to calculate $z_i(\mc)$ for penguin operators a
proper matching between effective 4- and 3-quark theories,
that is between \eqn{eq:HeffKppc} and \eqn{eq:HeffKpp}, has to be
made. For the 3-quark theory one obtains in the NDR scheme
\cite{burasetal:92d}
\begin{equation}
\vec{z}^{\rm }(\mc) =
\left( \begin{array}{c}
z_1^{\rm }(\mc) \\ z_2^{\rm }(\mc) \\
-\as/(24\pi) F_{\rm s}^{\rm }(\mc) \\ \as/(8\pi) F_{\rm s}^{\rm }(\mc) \\ 
-\as/(24\pi) F_{\rm s}^{\rm }(\mc) \\ \as/(8\pi) F_{\rm s}^{\rm }(\mc)
\end{array} \right) \, ,
\label{eq:zmc}
\end{equation}
where 
\begin{equation}
F_{\rm s}^{\rm }(\mc) =
-\frac{2}{3} \; z_2(\mc)
\label{eq:Fsmc}
\end{equation}
In the HV scheme $z_{1,2}$ are modified and one has $F_{\rm s}^{\rm }(\mc)
= 0$ or $z_i(\mc) = 0$ for $i\not=1,2$.

\subsection{Renormalization Group Evolution and Anomalous Dimension Matrices}
            \label{sec:HeffdF1:66:rge}
The general RG evolution matrix $\hU(m_1,m_2)$ from scale $m_2$ down to
$m_1 < m_2$ reads in pure QCD
\begin{equation}
\hU(m_1,m_2) \equiv T_g \exp
\int_{g(m_2)}^{g(m_1)} \!\! dg' \; \frac{\hg_{\rm s}^T(g'^2)}{\beta(g')} \, ,
\label{eq:UgeneralQCD}
\end{equation}
with $\hg_{\rm s}(g^2)$ being the full $6\times 6$ QCD anomalous dimension
matrix for $Q_1, \ldots, Q_6$.

For the case at hand it can be expanded in terms of $\as$ as follows
\begin{equation}
\hg_{\rm s}(g^2) = \frac{\as}{4\pi} \gs + \frac{\as^2}{(4\pi)^2} \gss
                   + \cdots \, .
\label{eq:gsexpKpp}
\end{equation}
Explicit expressions for $\gs$ and $\gss$ will be given below.

Using eq.\ \eqn{eq:gsexpKpp} the general QCD evolution matrix
$\hU(m_1,m_2)$ of eq.~\eqn{eq:UgeneralQCD} can be written as in \eqn{u0jj}
\cite{burasetal:92a}.
\begin{equation}
\hU(m_1,m_2) = 
\left( 1 + \frac{\as(m_1)}{4\pi} \hJ \right)
\hU^{(0)}(m_1,m_2)
\left( 1 - \frac{\as(m_2)}{4\pi} \hJ \right) \, ,
\label{eq:UQCDKpp}
\end{equation}
where $\hU^{(0)}(m_1,m_2)$ denotes the evolution matrix in the leading
logarithmic approximation and $\hJ$ summarizes the next-to-leading
correction to this evolution. Therefore, the full matrix $\hU(m_1,m_2)$
sums logarithms $(\as t)^n$ and $\as (\as t)^n$ with $t=\ln(m_2^2/m_1^2)$.
Explicit expressions for $\hU^{(0)}(m_1,m_2)$ and $\hJ$ are given in
eqs.~\eqn{u0vd}--\eqn{jvs}.

The LO anomalous dimension matrix $\gs$ of eq.\ \eqn{eq:gsexpKpp} has
the explicit form \cite{gaillard:74}, \cite{altarelli:74},
\cite{vainshtein:77}, \cite{gilman:79}, \cite{guberina:80}
\begin{equation}
\gs = 
\left(
\begin{array}{cccccc}
{{-6}\over N} & 6 & 0 & 0 & 0 & 0 \\ \svs
6 & {{-6}\over N} & {{-2}\over {3 N}} & {2\over 3} & {{-2}\over {3 N}} &\
  {2\over 3} \\ \svs
0 & 0 & {{-22}\over {3 N}} & {{22}\over 3} & {{-4}\over {3 N}} & {4\over 3}
\\ \svs
0 & 0 & 6 - {{2 f}\over {3 N}} & {{-6}\over N} + {{2 f}\over 3} & {{-2\
  f}\over {3 N}} & {{2 f}\over 3} \\ \svs
0 & 0 & 0 & 0 & {6\over N} & -6  \\ \svs
0 & 0 & {{-2 f}\over {3 N}} & {{2 f}\over 3} & {{-2 f}\over {3 N}} & {{-6\
  \left( -1 + {N^2} \right) }\over N} + {{2 f}\over 3}
\end{array}
\right)
\label{eq:gs0Kpp}
\end{equation}
The NLO anomalous dimension matrix $\gss$ of eq.\ \eqn{eq:gsexpKpp}
reads in the NDR scheme \cite{burasetal:92a}, \cite{ciuchini:93}
\begin{equation}
\gssndr\bigl|_{N=3} =
\left(
\begin{array}{cccccc}
-{{21}\over 2} - {{2\,f}\over 9} & {7\over 2} + {{2\,f}\over 3} & {{79}\over\
  9} & -{7\over 3} & -{{65}\over 9} & -{{7}\over{3}} \\ \mvs
{7\over 2} + {{2\,f}\over 3} & -{{21}\over 2} - {{2\,f}\over 9} &\
  -{{202}\over {243}} & {{1354}\over {81}} & -{{1192}\over {243}} &
{904 \over 81} \\ \mvs
0 & 0 & -{{5911}\over {486}} + {{71\,f}\over 9} & {{5983}\over {162}} +\
  {f\over 3} & -{{2384}\over {243}} - {{71\,f}\over 9} &
{1808 \over 81} - {f \over 3} \\ \mvs
0 & 0 & {{379}\over {18}} + {{56\,f}\over {243}} & -{{91}\over 6} +\
  {{808\,f}\over {81}} & -{{130}\over 9} - {{502\,f}\over {243}} &
-{14 \over 3} + {{646\,f} \over 81} \\ \mvs
0 & 0 & {{-61\,f}\over 9} & {{-11\,f}\over 3} & {{71}\over 3} + {{61\,f}\over\
  9} & -99 + {{11\,f} \over 3} \\ \mvs
0 & 0 & {{-682\,f}\over {243}} & {{106\,f}\over {81}} & -{{225}\over 2} +\
  {{1676\,f}\over {243}} & -{1343 \over 6} + {{1348\,f} \over 81}
\end{array}
\right)
\label{eq:gs1ndrN3Kpp}
\end{equation}
In \eqn{eq:gs0Kpp} and \eqn{eq:gs1ndrN3Kpp} $f$ denotes the number of
active quark flavours at a certain scale $\mu$. The corresponding
results for $\gss$ in the HV scheme can either be obtained by direct
calculation or by using the relation \eqn{gpgs}. They can be found in
\cite{burasetal:92a}, \cite{ciuchini:93} where also the $N$
dependence of $\gss$ is given.

\subsection{Quark Threshold Matching Matrix}
            \label{sec:HeffdF1:66:Mm}
As discussed in section~\ref{sec:basicform:wc:rgf} in general a 
matching matrix $\hM(m)$ has to be included in the RG evolution at NLO when
going from a $f$-flavour effective theory to a $(f-1)$-flavour
effective theory at quark threshold $\mu = m$ \cite{burasetal:92a},
\cite{burasetal:92d}.

For the $\dS$ decay $\Kpipi$ in pure QCD one has \cite{burasetal:92a}
\begin{equation}
\hM(m) = 1 + \frac{\as(m)}{4\pi} \; \vardrs^T \, .
\label{eq:MmKpp}
\end{equation}
At the quark thresholds $m=m_b$ and $m=m_c$ the matrix $\vardrs$ reads
\begin{equation}
\vardrs^T = -\, \frac{5}{9} P \, (0, 0, 0, 1, 0, 1)
\label{eq:drsmbcKpp}
\end{equation}
with
\begin{equation}
P^T = (0,0,-\frac{1}{3},1,-\frac{1}{3},1) \, .
\label{eq:PdrsKpp}
\end{equation}

\subsection{Numerical Results for the $\Kpipi$ Wilson Coefficients
            in Pure QCD}
            \label{sec:HeffdF1:66:numres}
\begin{table}[htb]
\caption[]{$\dS$ Wilson coefficients at $\mu=1\gev$ for $\mt=170\gev$.
$y_1 = y_2 \equiv 0$.
\label{tab:wc6smu1}}
\begin{center}
\begin{tabular}{|c|c|c|c||c|c|c||c|c|c|}
& \multicolumn{3}{c||}{$\Lms^{(4)}=215\mev$} &
  \multicolumn{3}{c||}{$\Lms^{(4)}=325\mev$} &
  \multicolumn{3}{c| }{$\Lms^{(4)}=435\mev$} \\
\hline
Scheme & LO & NDR & HV & LO & 
NDR & HV & LO & NDR & HV \\
\hline
$z_1$ & --0.602 & --0.407 & --0.491 & --0.743 & 
--0.506 & --0.636 & --0.901 & --0.622 & --0.836 \\
$z_2$ & 1.323 & 1.204 & 1.260 & 1.423 & 
1.270 & 1.362 & 1.541 & 1.352 & 1.515 \\
\hline
$z_3$ & 0.003 & 0.007 & 0.004 & 0.004 & 
0.013 & 0.007 & 0.006 & 0.022 & 0.015 \\
$z_4$ & --0.008 & --0.022 & --0.010 & --0.012 & 
--0.034 & --0.016 & --0.016 & --0.058 & --0.029 \\
$z_5$ & 0.003 & 0.006 & 0.003 & 0.004 & 
0.007 & 0.004 & 0.005 & 0.009 & 0.005 \\
$z_6$ & --0.009 & --0.021 & --0.009 & --0.013 & 
--0.034 & --0.014 & --0.018 & --0.058 & --0.025 \\
\hline
$y_3$ & 0.029 & 0.023 & 0.026 & 0.036 & 
0.031 & 0.036 & 0.045 & 0.040 & 0.048 \\
$y_4$ & --0.051 & --0.046 & --0.048 & --0.060 & 
--0.056 & --0.059 & --0.069 & --0.066 & --0.072 \\
$y_5$ & 0.012 & 0.004 & 0.013 & 0.013 & 
--0.001 & 0.016 & 0.014 & --0.013 & 0.020 \\
$y_6$ & --0.084 & --0.076 & --0.070 & --0.111 & 
--0.109 & --0.096 & --0.145 & --0.166 & --0.136 \\
\end{tabular}
\end{center}
\end{table}

\begin{table}[htb]
\caption[]{$\dS$ Wilson coefficients at $\mu=\mc=1.3\gev$ for
$\mt=170\gev$ and $f=3$ effective flavours.
$|z_3|,\ldots,|z_6|$ are numerically irrelevant relative to
$|z_{1,2}|$. $y_1 = y_2 \equiv 0$.
\label{tab:wc6smu13}}
\begin{center}
\begin{tabular}{|c|c|c|c||c|c|c||c|c|c|}
& \multicolumn{3}{c||}{$\Lms^{(4)}=215\mev$} &
  \multicolumn{3}{c||}{$\Lms^{(4)}=325\mev$} &
  \multicolumn{3}{c| }{$\Lms^{(4)}=435\mev$} \\
\hline
Scheme & LO & NDR & HV & LO & 
NDR & HV & LO & NDR & HV \\
\hline
$z_1$ & --0.518 & --0.344 & --0.411 & --0.621 & 
--0.412 & --0.504 & --0.727 & --0.487 & --0.614 \\
$z_2$ & 1.266 & 1.166 & 1.207 & 1.336 & 
1.208 & 1.269 & 1.411 & 1.258 & 1.346 \\
\hline
$y_3$ & 0.026 & 0.021 & 0.024 & 0.032 & 
0.027 & 0.031 & 0.039 & 0.035 & 0.040 \\
$y_4$ & --0.050 & --0.046 & --0.048 & --0.059 & 
--0.056 & --0.058 & --0.068 & --0.067 & --0.070 \\
$y_5$ & 0.013 & 0.007 & 0.013 & 0.015 & 
0.005 & 0.016 & 0.016 & 0.001 & 0.018 \\
$y_6$ & --0.075 & --0.067 & --0.062 & --0.095 & 
--0.088 & --0.079 & --0.118 & --0.116 & --0.102 \\
\end{tabular}
\end{center}
\end{table}

\begin{table}[htb]
\caption[]{$\dS$ Wilson coefficients at $\mu=2\gev$ for
$\mt=170\gev$. For $\mu > \mc$ the GIM mechanism gives $z_i \equiv 0$,
$i=3,\ldots,6$. $y_1 = y_2 \equiv 0$.
\label{tab:wc6smu2}}
\begin{center}
\begin{tabular}{|c|c|c|c||c|c|c||c|c|c|}
& \multicolumn{3}{c||}{$\Lms^{(4)}=215\mev$} &
  \multicolumn{3}{c||}{$\Lms^{(4)}=325\mev$} &
  \multicolumn{3}{c| }{$\Lms^{(4)}=435\mev$} \\
\hline
Scheme & LO & NDR & HV & LO & 
NDR & HV & LO & NDR & HV \\
\hline
$z_1$ & --0.411 & --0.266 & --0.318 & --0.477 & 
--0.309 & --0.374 & --0.541 & --0.350 & --0.430 \\
$z_2$ & 1.199 & 1.121 & 1.151 & 1.240 & 
1.145 & 1.185 & 1.282 & 1.170 & 1.220 \\
\hline
$y_3$ & 0.019 & 0.019 & 0.018 & 0.023 & 
0.023 & 0.022 & 0.027 & 0.027 & 0.026 \\
$y_4$ & --0.040 & --0.046 & --0.039 & --0.046 & 
--0.054 & --0.045 & --0.052 & --0.062 & --0.052 \\
$y_5$ & 0.011 & 0.010 & 0.011 & 0.012 & 
0.010 & 0.013 & 0.013 & 0.010 & 0.015 \\
$y_6$ & --0.055 & --0.057 & --0.047 & --0.067 & 
--0.070 & --0.056 & --0.078 & --0.085 & --0.067 \\
\end{tabular}
\end{center}
\end{table}

Tables \ref{tab:wc6smu1}--\ref{tab:wc6smu2} give the $\dS$ Wilson
coefficients for $Q_1,\ldots,Q_6$ in pure QCD.
\\
We observe a visible scheme dependence for all NLO Wilson coefficients.
Notably we find $|y_6|$ to be smaller in the HV than in the NDR scheme.
\\
In addition all coefficients, especially $z_1$ and $y_3,\ldots,y_6$,
show a strong dependence on $\Lms$.
\\
Next, at NLO the absolute values for $z_{1,2}$ and $y_i$ are suppressed
relative to their LO results, except for $y_5$ in HV and $y_{4,6}$ in
NDR for $\mu > \mc$. The latter behaviour is related to the effect of
the matching matrix $M(\mc)$ absent for $\mu > \mc$.
\\
For $y_3,\ldots,y_5$ there is no visible $\mt$ dependence in the range
$\mt = (170 \pm 15)\gev$. For $|y_6|$ there is a relative variation of
$\ord(\pm 1.5\%)$ for in/decreasing $\mt$.
\\
Finally, a comment on the Wilson coefficients in the HV Scheme as
presented here is appropriate. As we have mentioned in section
\ref{sec:HeffdF1:22:wcrg}, the two-loop anomalous dimensions of the
weak current in the HV scheme does not vanish. This peculiar feature of
the HV scheme is also felt in $\gss$.  The diagonal terms in $\gss$
aquire additional universal large $\ord(N^2)$ terms $(44/3)N^2$
which are absent in the NDR scheme. These artificial terms can be
removed by working with $\gss -2 \gamma^{(1)}_J$ instead of $\gss$.
This procedure, adopted in this review and in \cite{burasetal:92d},
corresponds effectively to a finite renormalization of operators which
changes the coefficient of $\as/4\pi$ in $C^{HV}_2(\mw)$ from $- 13/2$
to $- 7/6$. The Rome group \cite{ciuchini:93} has chosen not to make
this additional finite renormalization and consequently their
coefficients in the HV scheme differ from the HV coefficients presented
here by a universal factor. They can be found by using
\begin{equation}
C^{\rm HV}_{\rm Rome}(\mu) =
\left[ 1 -\frac{\as(\mu)}{4\pi} 4 C_F \right]
C^{\rm HV}(\mu)
\label{eq:Crome}
\end{equation}
Clearly this difference is compensated by the corresponding difference in the
hadronic matrix elements of the operators $Q_i$.  

\subsection{The $\dB$ Effective Hamiltonian in Pure QCD}
            \label{sec:HeffdF1:66:dB1}
An important application of the formalism developed in the previous
subsections is for the case of $B$-meson decays.  The LO calculation
can be found e.g.\ in \cite{ponce:81}, \cite{grinstein:89} where the
importance of NLO calculations has already been pointed out.  This
section can be viewed as the generalization of Grinstein's analysis
beyond the LO approximation. We will focus on the $\Delta B=1$, $\Delta
C = 0$ part of the effective hamiltonian which is of particular
interest for the study of CP violation in decays to CP self-conjugate
final states.  The part of the hamiltonian inducing $\Delta B=1$,
$\Delta C = \pm 1$ transitions involves no penguin operators and has
already been discussed in \ref{sec:HeffdF1:22}.

At tree-level the effective hamiltonian of interest here is simply given by
\begin{equation} 
\Heff(\dB) = \frac{G_F}{\sqrt{2}} \sum_{q=u,c} \sum_{q'=d,s}
              V^*_{qb} V^{}_{qq'}
             \left(\bar b q  \right)_{\rm V-A}
             \left(\bar q q' \right)_{\rm V-A} \, .
\label{eq:HeffdB1:tree}
\end{equation} 

The cases $q'=d$ and $q'=s$ can be treated separately and have the same
Wilson coefficients $C_i(\mu)$. Therefore we will restrict the
discussion to $q'=d$ in the following.

Using unitarity of the CKM matrix, $\xi_u + \xi_c + \xi_t = 0$ with
$\xi_i = V^*_{ib} V^{}_{id}$, and the fact that $Q^u_{1,2}$ and
$Q^c_{1,2}$ have the same initial conditions at
$\mu=\mw$ one obtains for the effective $\dB$ hamiltonian at scales
$\mu = \ord(\mb)$
\begin{eqnarray} 
\Heff(\dB) &=& \frac{G_F}{\sqrt{2}} \bigl\{
   \xi_c \, \left[ C_1(\mu) Q_1^c(\mu) + C_2(\mu) Q_2^c(\mu) \right] +
   \xi_u \, \left[ C_1(\mu) Q_1^u(\mu) + C_2(\mu) Q_2^u(\mu) \right] 
\nn \\
 & & - \xi_t \, \sum_{i=3}^{6} C_i(\mu) Q_i(\mu)
\bigr\} \, .
\label{eq:HeffdB1:66}
\end{eqnarray} 
Here
\begin{eqnarray}
Q^q_{1} & = & \left( \bar b_{i} q_{j}  \right)_{\rm V-A}
            \left( \bar q_{j}  d_{i} \right)_{\rm V-A}
\, , \nn \\
Q^q_{2} & = & \left( \bar b q \right)_{\rm V-A}
            \left( \bar q d \right)_{\rm V-A}
\, , \nn \\
Q_{3} & = & \left( \bar b d \right)_{\rm V-A}
   \sum_{q} \left( \bar q q \right)_{\rm V-A}
\, , \label{eq:dB1basis} \\
Q_{4} & = & \left( \bar b_{i} d_{j}  \right)_{\rm V-A}
   \sum_{q} \left( \bar q_{j}  q_{i} \right)_{\rm V-A}
\, , \nn \\
Q_{5} & = & \left( \bar b d \right)_{\rm V-A}
   \sum_{q} \left( \bar q q \right)_{\rm V+A}
\, , \nn \\
Q_{6} & = & \left( \bar b_{i} d_{j}  \right)_{\rm V-A}
   \sum_{q} \left( \bar q_{j}  q_{i} \right)_{\rm V+A}
\, , \nn
\end{eqnarray}
where the summation runs over $q=u,d,s,c,b$. 

The corresponding $\dB$ Wilson coefficients at scale $\mu = \ord(\mb)$
are simply given by a truncated version of eq.\ \eqn{eq:WCv}
\begin{equation}
\vC(\mb) = \hU_5(\mb,\mw) \, \vC(\mw) \, .
\label{eq:HeffdB1:66:wc}
\end{equation}
Here $\hU_5$ is the $6 \times 6$ RG evolution matrix of eq.\
\eqn{eq:UQCDKpp} for $f=5$ active flavours. The initial conditions
$\vC(\mw)$ are identical to those of \eqn{eq:CMw1QCD}--\eqn{eq:CMw6QCD}
for the $\dS$ case.

\subsection{Numerical Results for the $\dB$ Wilson Coefficients in Pure QCD}
            \label{sec:HeffdF1:66:dB1num}
\begin{table}[htb]
\caption[]{$\dB$ Wilson coefficients at $\mu=\overline{m}_{\rm b}(\mb)=
4.40\gev$ for $\mt=170\gev$.
\label{tab:wc6b}}
\begin{center}
\begin{tabular}{|c|c|c|c||c|c|c||c|c|c|}
& \multicolumn{3}{c||}{$\Lms^{(5)}=140\mev$} &
  \multicolumn{3}{c||}{$\Lms^{(5)}=225\mev$} &
  \multicolumn{3}{c| }{$\Lms^{(5)}=310\mev$} \\
\hline
Scheme & LO & NDR & HV & LO & 
NDR & HV & LO & NDR & HV \\
\hline
$C_1$ & --0.272 & --0.164 & --0.201 & --0.307 & 
--0.184 & --0.227 & --0.337 & --0.202 & --0.250 \\
$C_2$ & 1.120 & 1.068 & 1.087 & 1.139 & 
1.078 & 1.101 & 1.155 & 1.087 & 1.113 \\
\hline
$C_3$ & 0.012 & 0.012 & 0.011 & 0.013 & 
0.013 & 0.012 & 0.015 & 0.015 & 0.014 \\
$C_4$ & --0.026 & --0.031 & --0.026 & --0.030 & 
--0.035 & --0.029 & --0.032 & --0.038 & --0.032 \\
$C_5$ & 0.008 & 0.008 & 0.008 & 0.009 & 
0.009 & 0.009 & 0.009 & 0.009 & 0.010 \\
$C_6$ & --0.033 & --0.035 & --0.029 & --0.038 & 
--0.041 & --0.033 & --0.042 & --0.046 & --0.036 \\
\end{tabular}
\end{center}
\end{table}

Table \ref{tab:wc6b} lists the $\dB$ Wilson coefficients for
$Q_1^{u,c},Q_2^{u,c},Q_3,\ldots,Q_6$ in pure QCD.
\\
$C_1$, $C_4$ and $C_6$ show a $\ord(20\%)$ scheme dependence while
this dependence is much weaker for the rest of the coefficients.
\\
Similarly to the $\dS$ case the numerical values for $\dB$ Wilson
coefficients are sensitive to the value of $\Lms$ used to determine
$\as$ for the RG evolution. The sensitivity is however less pronounced
than in the $\dS$ case due to the higher value $\mu=\overline{m}_{\rm
b}(\mb)$ of the renormalization scale.
\\
Finally, one finds no visible $\mt$ dependence in the range $\mt = (170
\pm 15)\gev$.

\section{The Effective $\Delta F=1$ Hamiltonian: Inclusion of
         Electroweak Penguin Operators}
         \label{sec:HeffdF1:1010}
Similarly to the creation of the penguin operators $Q_3,\ldots,Q_6$
through QCD corrections the inclusion of electroweak corrections, shown
in figs.\ \ref{fig:1loopful}\,(d) and (e), generates a set of new
operators, the so-called electroweak penguin operators. For the $\dS$
decay $\Kpipi$ they are usually denoted by $Q_7, \ldots,Q_{10}$. \\
This means that although now we will have to deal with technically more
involved issues like an extended operator basis or the possibility of
mixed QCD-QED contributions the underlying principles in performing the
RG evolution will closely resemble those used in section \ref{sec:HeffdF1:66}
for pure QCD. Obviously, the fundamental step has already been made
when going from current-current operators only in
section \ref{sec:HeffdF1:22}, to the inclusion of QCD penguins in
section \ref{sec:HeffdF1:66}. Hence, in this section we will wherever
possible only point out the differences between the pure $6 \times 6$
QCD and the combined $10 \times 10$ QCD-QED case.

The full $\dS$ effective hamiltonian for $\Kpipi$ at scales
$\mu < \mc$ reads including QCD and QED corrections\footnote{
In principle also operators $Q_{11} = \frac{g_{\rm s}}{16 \pi^2} \ms
\bar s \sigma_{\mu\nu} T^a G_a^{\mu\nu} (1-\gamma_5) d$ and $Q_{12} =
\frac{e e_d}{16 \pi^2} \ms \bar s \sigma_{\mu\nu} F^{\mu\nu}
(1-\gamma_5) d$ should be considered for $\Kpipi$. However, as shown in
\cite{bertolinietal:94} their numerical contribution is negligible.
Therefore $Q_{11}$ and $Q_{12}$ will not be included here for
$\Kpipi$.}
\begin{equation}
\Heff(\dS) = \frac{G_F}{\sqrt{2}} \V{us}^* \V{ud}^{} \sum_{i=1}^{10}
\left( z_i(\mu) + \tau \; y_i(\mu) \right) Q_i(\mu) \, ,
\label{eq:HeffdF1:1010}
\end{equation}
with $\tau=-\V{ts}^* \V{td}^{}/(\V{us}^* \V{ud}^{})$.
\subsection{Operators}
            \label{sec:HeffdF1:1010:op}
The basis of four-quark operators for the $\dS$ effective hamiltonian
in \eqn{eq:HeffdF1:1010} is given by $Q_1,\ldots,Q_6$ of
\eqn{eq:Kppbasis} and the electroweak penguin operators
\begin{eqnarray}
Q_{7} & = & \frac{3}{2} \left( \bar s d \right)_{\rm V-A}
         \sum_{q} e_{q} \left( \bar q q \right)_{\rm V+A}
\, , \nn \\
Q_{8} & = & \frac{3}{2} \left( \bar s_{i} d_{j} \right)_{\rm V-A}
         \sum_{q} e_{q} \left( \bar q_{j}  q_{i}\right)_{\rm V+A}
\, , \nn \\
Q_{9} & = & \frac{3}{2} \left( \bar s d \right)_{\rm V-A}
         \sum_{q} e_{q} \left( \bar q q \right)_{\rm V-A}
\, , \label{eq:dF1:1010basis} \\
Q_{10}& = & \frac{3}{2} \left( \bar s_{i} d_{j} \right)_{\rm V-A}
         \sum_{q} e_{q} \left( \bar q_{j}  q_{i}\right)_{\rm V-A}
\, . \nn
\end{eqnarray}
Here, $e_q$ denotes the quark electric charge reflecting the
electroweak origin of $Q_7,\ldots,Q_{10}$.  The basis
$Q_1,\ldots,Q_{10}$ closes under QCD and QED renormalization.  Finally,
for $\mb > \mu > \mc$ the operators $Q_1^c$ and $Q_2^c$ of
eq.\ \eqn{eq:KppQ12c} have to be included again similarly to the case of
pure QCD.

\subsection{Wilson Coefficients}
            \label{sec:HeffdF1:1010:wc}
As far as formulae for Wilson coefficients are concerned the
generalization of section \ref{sec:HeffdF1:66:wc} to the present case is
to a large extent straightforward. \\
First, due to the extended operator basis $\vec{v}(\mu)$ and
$\vec{z}(\mu)$ in eqs.\ \eqn{eq:WCv} and \eqn{eq:WCz} are now ten
dimensional column vectors. Furthermore, the substitution
\begin{displaymath}
\hU_f(m_1,m_2) \to \hU_f(m_1,m_2,\aem)
\end{displaymath}
has to be made in the RG evolution equations \eqn{eq:WCv}, \eqn{eq:WCz} and
\eqn{eq:zmc12}. Here $\hU_f(m_1,m_2,\aem)$ denotes the
full $10 \times 10$ QCD- QED RG evolution matrix for $f$ active
flavours. $\hU_f(m_1,m_2,\aem)$ will still be discussed in more detail
in subsection \ref{sec:HeffdF1:1010:rge}.

The extended initial values $\vC(\mw)$ including now $\ord(\aem)$
corrections and additional entries for $Q_7,\ldots,Q_{10}$ can be
obtained from the usual matching procedure between figs.\ \ref{fig:1loopful}
and \ref{fig:1loopeff}. They read in the NDR scheme \cite{burasetal:92d}
\begin{eqnarray}
C_1(\mw) &=&     \frac{11}{2} \; \frac{\as(\mw)}{4\pi} \, ,
\label{eq:CMw1} \\
C_2(\mw) &=& 1 - \frac{11}{6} \; \frac{\as(\mw)}{4\pi}
               - \frac{35}{18} \; \frac{\aem}{4\pi} \, ,
\label{eq:CMw2} \\
C_3(\mw) &=& -\frac{\as(\mw)}{24\pi} \widetilde{E}_0(x_t)
             +\frac{\aem}{6\pi} \frac{1}{\sin^2\theta_W}
             \left[ 2 B_0(x_t) + C_0(x_t) \right] \, , 
\label{eq:CMw3} \\
C_4(\mw) &=& \frac{\as(\mw)}{8\pi} \widetilde{E}_0(x_t) \, ,
\label{eq:CMw4} \\
C_5(\mw) &=& -\frac{\as(\mw)}{24\pi} \widetilde{E}_0(x_t) \, ,
\label{eq:CMw5} \\
C_6(\mw) &=& \frac{\as(\mw)}{8\pi} \widetilde{E}_0(x_t) \, ,
\label{eq:CMw6} \\
C_7(\mw) &=& \frac{\aem}{6\pi} \left[ 4 C_0(x_t) + \widetilde{D}_0(x_t)
\right]\, ,
\label{eq:CMw7} \\
C_8(\mw) &=& 0 \, ,
\label{eq:CMw8} \\
C_9(\mw) &=& \frac{\aem}{6\pi} \left[ 4 C_0(x_t) + \widetilde{D}_0(x_t) +
             \frac{1}{\sin^2\theta_W} (10 B_0(x_t) - 4 C_0(x_t)) \right] \, ,
\label{eq:CMw9} \\
C_{10}(\mw) &=& 0 \, ,
\label{eq:CMw10}
\end{eqnarray}
where
\begin{eqnarray}
B_0(x) &=& \frac{1}{4} \left[ \frac{x}{1-x} + \frac{x \ln x}{(x-1)^2}
\right]\, , \label{eq:Bxt} \\
C_0(x) &=& \frac{x}{8} \left[ \frac{x-6}{x-1} + \frac{3 x + 2}{(x-1)^2}
\ln x \right]\, ,
\label{eq:Cxt} \\
D_0(x) &=& -\frac{4}{9} \ln x + \frac{-19 x^3 + 25 x^2}{36 (x-1)^3} +
         \frac{x^2 (5 x^2 - 2 x - 6)}{18 (x-1)^4} \ln x \, ,
\label{eq:Dxt} \\
\widetilde{D}_0(x_t) &=& D_0(x_t) - \frac{4}{9} \, .
\label{eq:Dxttilde} 
\end{eqnarray}
$\widetilde{E}_0(x_t)$ and $x_t$ have already been defined in
eqs.\ \eqn{eq:Exttilde} and \eqn{eq:xt}, respectively.  Here $B_0(x)$
results from the evaluation of the box diagrams, $C_0(x)$ from the
$Z^0$-penguin, $D_0(x)$ from the photon penguin and $E_0(x)$ in
$\widetilde{E}_0(x_t)$ from the gluon penguin diagrams.

The initial values $\vec{C}(\mw)$ in the HV scheme can be found in
\cite{burasetal:92d}.

Finally, the generalization of \eqn{eq:zmc} to the $Q_1,\ldots,Q_{10}$
basis reads \cite{burasetal:92d}
\begin{equation}
\vec{z}^{\rm }(\mc) =
\left( \begin{array}{c}
z_1^{\rm }(\mc) \\ z_2^{\rm }(\mc) \\
-\as/(24\pi) F_{\rm s}^{\rm }(\mc) \\ \as/(8\pi) F_{\rm s}^{\rm }(\mc) \\ 
-\as/(24\pi) F_{\rm s}^{\rm }(\mc) \\ \as/(8\pi) F_{\rm s}^{\rm }(\mc) \\ 
\aem/(6\pi) F_{\rm e}^{\rm }(\mc) \\  0  \\
\aem/(6\pi) F_{\rm e}^{\rm }(\mc) \\  0
\end{array} \right) \, ,
\label{eq:zmc:1010}
\end{equation}
with $F_{\rm s}(\mc)$ given by \eqn{eq:Fsmc} and 
\begin{equation}
F_{\rm e}^{\rm }(\mc) =
-\frac{4}{9} \; \left( 3 z_1(\mc) + z_2(\mc) \right) \, .
\label{eq:Femc}
\end{equation}
In the HV scheme, in addition to $z_{1,2}$ differing from their NDR values, 
one has $F_{\rm s}^{\rm }(\mc) = F_{\rm e}^{\rm }(\mc)
= 0$ and, consequently, $z_i(\mc) = 0$ for $i\not=1,2$.

\subsection{Renormalization Group Evolution and Anomalous Dimension Matrices}
            \label{sec:HeffdF1:1010:rge}
Besides an extended operator basis the main difference between the pure
QCD case of section \ref{sec:HeffdF1:66} and the present case consists in
the additional presence of QED contributions to the RG evolution.  This
will make the actual formulae for the RG evolution matrices more
involved, however the underlying concepts developed in
sections \ref{sec:HeffdF1:22} and \ref{sec:HeffdF1:66} remain the same.

Similarly to \eqn{eq:UgeneralQCD} for pure QCD the general RG evolution
matrix $\hU(m_1,m_2,\aem)$ from scale $m_2$ down to $m_1 < m_2$ can be
written formally as \footnote{We neglect the running of the 
electromagnetic coupling $\aem$, which is a very good approximation
\cite{buchallaetal:90}.}
\begin{equation}
\hU(m_1,m_2,\aem) \equiv T_g \exp
\int_{g(m_2)}^{g(m_1)} \!\! dg' \; \frac{\hg^T(g'^2,\aem)}{\beta(g')} \, ,
\label{eq:Ugeneral}
\end{equation}
with $\hg(g^2,\aem)$ being now the full $10\times 10$ anomalous dimension
matrix including QCD and QED contributions.

For the case at hand $\hg(g^2,\aem)$ can be expanded in the following way
\begin{equation}
\hg(g^2,\aem) = \hg_{\rm s}(g^2) + \frac{\aem}{4\pi} \hG(g^2) + \ldots \, ,
\label{eq:gexpdF1:1010}
\end{equation}
with the pure $\as$-expansion of $\hg_{\rm s}(g^2)$ given in
\eqn{eq:gsexpKpp}. The term present due to QED corrections has the
expansion
\begin{equation}
\hG(g^2) = \gem + \frac{\as}{4\pi} \gse + \ldots  \, .
\label{eq:GexpdF1:1010}
\end{equation}

Using \eqn{eq:gexpdF1:1010}--\eqn{eq:GexpdF1:1010} the general RG
evolution matrix $\hU(m_1,m_2,\aem)$ of eq.~\eqn{eq:Ugeneral} may then
be decomposed as follows
\begin{equation}
\hU(m_1,m_2,\aem) =
\hU(m_1,m_2) + \frac{\aem}{4\pi} \hR(m_1,m_2) \, ,
\label{eq:UdF1:1010}
\end{equation}
Here $\hU(m_1,m_2)$ represents the pure QCD evolution already
encountered in section \ref{sec:HeffdF1:66} but now generalized to an
extended operator basis. $\hR(m_1,m_2)$ describes the additional
evolution in the presence of the electromagnetic interaction.
$\hU(m_1,m_2)$ sums the logarithms $(\as t)^n$ and $\as (\as t)^n$ with
$t=\ln(m_2^2/m_1^2)$, whereas $\hR(m_1,m_2)$ sums the logarithms $t (\as
t)^n$ and $(\as t)^n$. \\
The formula for $\hU(m_1,m_2)$ has already been given in
\eqn{eq:UQCDKpp}.  The leading order formula for $\hR(m_1,m_2)$ can be
found in \cite{buchallaetal:90} except that there a different
overall normalization (relative factor $-4\pi$ in $\hR$) has been
used.  Here we give the general expression for $\hR(m_1,m_2)$
\cite{burasetal:92d}
\begin{eqnarray}
\hR(m_1,m_2) &=& \int_{g(m_2)}^{g(m_1)} \!\! dg' \;
                 \frac{\hU(m_1,m') \, \hG^T(g') \, \hU(m',m_2)}{\beta(g')}
\label{eq:RdF1:1010} \\
        &\equiv& -\frac{2\pi}{\beta_0} \; \hV \; \left(
                 \hK^{(0)}(m_1,m_2) +
                 \frac{1}{4\pi} \sum_{i=1}^{3} \hK_i^{(1)}(m_1,m_2)
                 \right) \hV^{-1} \, , \nn
\end{eqnarray}
with $g' \equiv g(m')$.

The matrix kernels in \eqn{eq:RdF1:1010} are defined by
\begin{equation}
(\hK^{(0)}(m_1,m_2))_{ij} = \frac{\hM^{(0)}_{ij}}{a_i - a_j - 1}
\left[
\left( \frac{\as(m_2)}{\as(m_1)} \right)^{a_j} \frac{1}{\as(m_1)} -
\left( \frac{\as(m_2)}{\as(m_1)} \right)^{a_i} \frac{1}{\as(m_2)}
\right] \, ,
\label{eq:K0}
\end{equation}
\begin{equation}
\left( \hK_1^{(1)}(m_1,m_2) \right)_{ij} =
\left\{
\begin{array}{ll}
\frac{M^{(1)}_{ij}}{a_i - a_j}
\left[ \left( \frac{\as(m_2)}{\as(m_1)} \right)^{a_j} -
       \left( \frac{\as(m_2)}{\as(m_1)} \right)^{a_i} \right] & i \not= j \\
\svs
M^{(1)}_{ii} \left( \frac{\as(m_2)}{\as(m_1)} \right)^{a_i}
             \ln\frac{\as(m_1)}{\as(m_2)}             & i=j
\end{array} \, ,
\right.
\label{eq:K11}
\end{equation}
\begin{eqnarray}
\hK_2^{(1)}(m_1,m_2) & = &
-\,\as(m_2) \; \hK^{(0)}(m_1,m_2) \; H \, ,
\label{eq:K12} \\
\hK_3^{(1)}(m_1,m_2) & = &
\phantom{-}\,\as(m_1) \; H \; \hK^{(0)}(m_1,m_2)
\label{eq:K13}
\end{eqnarray}
with
\begin{eqnarray}
\hM^{(0)} &=& \hV^{-1} \; \gemt \; \hV \, ,
\nn \\
\hM^{(1)} &=&
\hV^{-1} \left( \gset - \frac{\beta_1}{\beta_0} \gemt +
                \left[ \gemt, \hJ \right] \right) \hV \, .
\label{eq:M0M1}
\end{eqnarray}
The matrix $H$ is defined in (\ref{sij}).

After this formal description we now give explicit expressions for the
$10 \times 10$ LO and NLO anomalous dimension matrices $\gs$, $\gem$,
$\gss$ and $\gse$. The values quoted for the NLO matrices are in the
NDR scheme \cite{burasetal:92b}, \cite{burasetal:92c},
\cite{ciuchini:93}.  The corresponding results for $\gss$ and $\gse$ in
the HV scheme can either be obtained by direct calculation or by using
the QCD/QED version of eq.\ \eqn{gpgs} given in \cite{burasetal:92c}.
They can be found in \cite{burasetal:92b}, \cite{burasetal:92c}
and \cite{ciuchini:92}, \cite{ciuchini:93}.

The $6 \times 6$ submatrices for $Q_1,\ldots,Q_6$ of the full LO and
NLO $10 \times 10$ QCD matrices $\gs$ and $\gss$ are identical to the
corresponding $6 \times 6$ matrices already given in
eqs.\ \eqn{eq:gs0Kpp} and \eqn{eq:gs1ndrN3Kpp}, respectively. Next,
$Q_1,\ldots,Q_6$ do not mix to $Q_7,\ldots,Q_{10}$ under QCD and hence
\begin{equation}
\left[ \gs \right]_{ij} = \left[ \gss \right]_{ij} = 0
\qquad
i=1,\ldots,6
\quad
j=7,\ldots,10 \, .
\label{eq:gs01ij}
\end{equation}
The remaining entries for rows 7--10 in $\gs$ \cite{bijnenswise:84} and
$\gss$ \cite{burasetal:92b}, \cite{ciuchini:93} are given in tables
\ref{tab:gs0} and \ref{tab:gs1}, respectively. There $u$ and $d$
($f=u+d$) denote the number of active up- and down-type quark
flavours.

\begin{table}[htb]
\caption[]{Rows 7--10 of the LO anomalous dimension matrix $\gs$.}
\begin{tabular}{|r|c|c|c|c|c|c|c|c|c|c|}
$(i,j)$ & 1 & 2 & 3 & 4 & 5 & 6 & 7 & 8 & 9 & 10 \\
\hline
$7 $&$ 0 $&$ 0 $&$ 0 $&$ 0 $&$ 0 $&$ 0 $&$ {6\over N} $&$ -6 $&$ 0 $&$ 0 $\\
\svs
$8 $&$ 0 $&$ 0 $&$ {{-2 \left( u-d/2 \right) }\over {3 N}} $&$ {{2 \left(\
  u-d/2 \right) }\over 3} $&$ {{-2 \left( u-d/2 \right)\
  }\over {3 N}} $&$ {{2 \left( u-d/2 \right) }\over 3} $&$ 0 $&$ {{-6\
  \left( -1 + {N^2} \right) }\over N} $&$ 0 $&$ 0 $\\ \svs
$9 $&$ 0 $&$ 0 $&$ {2\over {3 N}} $&$ -{2\over 3} $&$ {2\over {3 N}} $&$
-{2\over 3} $&$  0 $&$ 0 $&$\
  {{-6}\over N} $&$ 6 $\\ \svs
$10 $&$ 0 $&$ 0 $&$ {{-2 \left( u-d/2 \right) }\over {3 N}} $&$ {{2 \left(\
  u-d/2 \right) }\over 3} $&$ {{-2 \left( u-d/2 \right)\
  }\over {3 N}} $&$ {{2 \left( u-d/2 \right) }\over 3} $&$ 0 $&$ 0 $&$ 6\
  $&$ {{-6}\over N} $
\end{tabular}
\label{tab:gs0}
\end{table}

\begin{table}[htb]
\caption[]{Rows 7--10 of the NLO anomalous dimension matrix $\gss$ for $N=3$
and NDR.}
\begin{tabular}{|r|c|c|c|c|c|}
$(i,j)$ & 1 & 2 & 3 & 4 & 5 \\
\hline
$7 $&$ 0 $&$ 0 $&$ {{-61\,(u-d/2)}\over 9} $&$ {{-11\,(u-d/2)}\over 3} $&$
{{83\, (u-d/2)}\over 9} $\\ \svs
$8 $&$ 0 $&$ 0 $&$ {{-682\,(u-d/2)}\over {243}} $&$ {{106\,(u-d/2)}
\over {81}} $&$\ {{704\,(u-d/2)}\over {243}} $\\ \svs
$9 $&$ 0 $&$ 0 $&$ {{202}\over {243}} + {{73\,(u-d/2)}\over 9} $&$
 -{{1354}\over {81}} -\
  {{(u-d/2)}\over 3} $&$ {{1192}\over {243}} - {{71\,(u-d/2)}\over 9} $\\ \svs
$10 $&$ 0 $&$ 0 $&$ -{{79}\over 9} - {{106\,(u-d/2)}\over {243}} $&$
{7\over 3} +\
  {{826\,(u-d/2)}\over {81}} $&$ {{65}\over 9} - {{502\,(u-d/2)}\over\
  {243}} $
\end{tabular}

\begin{tabular}{|r|c|c|c|c|c|}
$(i,j)$ & 6 & 7 & 8 & 9 & 10 \\
\hline
$7 $&$ {{-11\,(u-d/2)}\over 3} $&$ {{71}\over 3} - {{22\,f}\over 9} $&$ -99 +\
  {{22\,f}\over 3} $&$ 0 $&$ 0 $\\ \svs
$8 $&$ {{736\,(u-d/2)}\over {81}} $&$ -{{225}\over 2} + 4\,f $&$ -{{1343}
\over 6} +\
  {{68\,f}\over 9} $&$ 0 $&$ 0 $\\ \svs
$9 $&$ -{{904}\over {81}} - {{(u-d/2)}\over 3} $&$ 0 $&$ 0
 $&$ -{{21}\over 2} -\
  {{2\,f}\over 9} $&$ {7\over 2} + {{2\,f}\over 3} $\\ \svs
$10 $&$ {7\over 3} + {{646\,(u-d/2)}\over {81}} $&$ 0 $&$ 0 $&$ {7\over 2} +
 {{2\,f}\over\
  3} $&$ -{{21}\over 2} - {{2\,f}\over 9} $
\end{tabular}
\label{tab:gs1}
\end{table}

The full $10 \times 10$ matrices $\gem$ \cite{lusignoli:89} and $\gse$
\cite{burasetal:92c}, \cite{ciuchini:93} can be found in tables \ref{tab:gem}
and \ref{tab:gse}, respectively.

\begin{table}[htb]
\caption[]{The LO anomalous dimension matrix $\gem$.}
\begin{tabular}{|r|c|c|c|c|c|c|c|c|c|c|}
$(i,j)$ & 1 & 2 & 3 & 4 & 5 & 6 & 7 & 8 & 9 & 10 \\
\hline
$1 $&$ -{8\over 3} $&$ 0 $&$ 0 $&$ 0 $&$ 0 $&$ 0 $&$ {{16\,N}\over {27}} $&$ 0 $&$  
{{16\,N}\over\
  {27}} $&$ 0 $\\ \svs
$2 $&$ 0 $&$ -{8\over 3} $&$ 0 $&$ 0 $&$ 0 $&$ 0 $&$ {{16}\over {27}} $&$ 0 $&$  
{{16}\over {27}} $&$ 0\
  $\\ \svs
$3 $&$ 0 $&$ 0 $&$ 0 $&$ 0 $&$ 0 $&$ 0 $&$ -{{16}\over {27}} + {{16\,N\,\left( u-d/2\
  \right) }\over {27}} $&$ 0 $&$ -{{88}\over {27}} + {{16\,N\,\left( 
  u-d/2 \right) }\over {27}} $&$ 0 $\\ \svs
$4 $&$ 0 $&$ 0 $&$ 0 $&$ 0 $&$ 0 $&$ 0 $&$ {{-16\,N}\over {27}} + {{16\,\left( u-d/2\
  \right) }\over {27}} $&$ 0 $&$ {{-16\,N}\over {27}} + {{16\,\left( 
  u-d/2 \right) }\over {27}} $&$ -{8\over 3} $\\ \svs
$5 $&$ 0 $&$ 0 $&$ 0 $&$ 0 $&$ 0 $&$ 0 $&$ {8\over 3} + {{16\,N\,\left( u-d/2\
  \right) }\over {27}} $&$ 0 $&$ {{16\,N\,\left( u-d/2 \right) }\over\
  {27}} $&$ 0 $\\ \svs
$6 $&$ 0 $&$ 0 $&$ 0 $&$ 0 $&$ 0 $&$ 0 $&$ {{16\,\left( u-d/2 \right) }\over {27}} $&$\
  {8\over 3} $&$ {{16\,\left( u-d/2 \right) }\over {27}} $&$ 0 $\\ \svs
$7 $&$ 0 $&$ 0 $&$ 0 $&$ 0 $&$ {4\over 3} $&$ 0 $&$ {4\over 3} + {{16\,N\,\left( u+d/4\
  \right) }\over {27}} $&$ 0 $&$ {{16\,N\,\left( u+d/4 \right) }\over\
  {27}} $&$ 0 $\\ \svs
$8 $&$ 0 $&$ 0 $&$ 0 $&$ 0 $&$ 0 $&$ {4\over 3} $&$ {{16\,\left( u+d/4 \right) }\over\
  {27}} $&$ {4\over 3} $&$ {{16\,\left( u+d/4 \right) }\over {27}} $&$ 0\
  $\\ \svs
$9 $&$ 0 $&$ 0 $&$ -{4\over 3} $&$ 0 $&$ 0 $&$ 0 $&$ {8\over {27}} + {{16\,N\,\left( 
  u+d/4 \right) }\over {27}} $&$ 0 $&$ -{{28}\over {27}} + {{16\,N\,\left( 
  u+d/4 \right) }\over {27}} $&$ 0 $\\ \svs
$10 $&$ 0 $&$ 0 $&$ 0 $&$ -{4\over 3} $&$ 0 $&$ 0 $&$ {{8\,N}\over {27}} + {{16\,\left( 
  u+d/4 \right) }\over {27}} $&$ 0 $&$ {{8\,N}\over {27}} + {{16\,\left( 
  u+d/4 \right) }\over {27}} $&$ -{4\over 3} $
\end{tabular}
\label{tab:gem}
\end{table}

\begin{table}[htb]
\caption[]{The NLO anomalous dimension matrix $\gse$ for $N=3$ and NDR.}
\begin{tabular}{|r|c|c|c|c|c|}
$(i,j)$ & 1 & 2 & 3 & 4 & 5 \\
\hline
$1 $&$ {{194}\over 9} $&$ -{2\over 3} $&$ -{{88}\over {243}} $&$ {{88}\over {81}} $&$\
  -{{88}\over {243}} $\\ \svs 
$2 $&$ {{25}\over 3} $&$ -{{49}\over 9} $&$ -{{556}\over {729}} $&$ {{556}\over {243}} $&$\
  -{{556}\over {729}} $\\ \svs 
$3 $&$ 0 $&$ 0 $&$ {{1690}\over {729}} - {{136 \left( u - d/2 \right) }\over {243}} $&$\
  -{{1690}\over {243}} + {{136 \left( u - d/2 \right) }\over {81}} $&$\
  {{232}\over {729}} - {{136 \left( u - d/2 \right) }\over {243}} $\\ \svs 
$4 $&$ 0 $&$ 0 $&$ -{{641}\over {243}} - {{388 u}\over {729}} + {{32 d}\over {729}} $&$\
  -{{655}\over {81}} + {{388 u}\over {243}} - {{32 d}\over {243}} $&$\
  {{88}\over {243}} - {{388 u}\over {729}} + {{32 d}\over {729}} $\\ \svs 
$5 $&$ 0 $&$ 0 $&$ {{-136 \left( u - d/2 \right) }\over {243}} $&$ {{136 \left( u - d/2\
  \right) }\over {81}} $&$ -2 - {{136 \left( u - d/2 \right) }\over {243}} $\\ \svs 
$6 $&$0 $&$ 0 $&$ {{-748 u}\over {729}} + {{212 d}\over {729}} $&$ {{748 u}\over {243}} -\
  {{212 d}\over {243}} $&$ 3 - {{748 u}\over {729}} + {{212 d}\over {729}} $\\ \svs 
$7 $&$ 0 $&$ 0 $&$ {{-136 \left( u + d/4 \right) }\over {243}} $&$ {{136 \left( u + d/4\
  \right) }\over {81}} $&$ -{{116}\over 9} - {{136 \left( u + d/4 \right)\
  }\over {243}} $\\ \svs 
$8 $&$ 0 $&$ 0 $&$ {{-748 u}\over {729}} - {{106 d}\over {729}} $&$ {{748 u}\over {243}} +\
  {{106 d}\over {243}} $&$ -1 - {{748 u}\over {729}} - {{106 d}\over {729}} $\\ \svs 
$9 $&$ 0 $&$ 0 $&$ {{7012}\over {729}} - {{136 \left( u + d/4 \right) }\over {243}} $&$\
  {{764}\over {243}} + {{136 \left( u + d/4 \right) }\over {81}} $&$\
  -{{116}\over {729}} - {{136 \left( u + d/4 \right) }\over {243}} $\\ \svs 
$10 $&$ 0 $&$ 0 $&$ {{1333}\over {243}} - {{388 u}\over {729}} - {{16 d}\over {729}} $&$\
  {{107}\over {81}} + {{388 u}\over {243}} + {{16 d}\over {243}} $&$\
  -{{44}\over {243}} - {{388 u}\over {729}} - {{16 d}\over {729}} $
\end{tabular}

\begin{tabular}{|r|c|c|c|c|c|}
$(i,j)$ & 6 & 7 & 8 & 9 & 10 \\
\hline
$1 $&$ {{88}\over {81}} $&$ {{152}\over {27}} $&$ {{40}\over 9} $&$ {{136}\over {27}} $&$\
  {{56}\over 9} $\\ \svs 
$2 $&$ {{556}\over {243}} $&$ -{{484}\over {729}} $&$ -{{124}\over {27}} $&$ -{{3148}\over\
  {729}} $&$ {{172}\over {27}} $\\ \svs 
$3 $&$ -{{232}\over {243}} + {{136 \left( u - d/2 \right) }\over {81}} $&$\
  {{3136}\over {729}} + {{104 \left( u - d/2 \right) }\over {27}} $&$\
  {{64}\over {27}} + {{88 \left( u - d/2 \right) }\over 9} $&$ {{20272}\over\
  {729}} + {{184 \left( u - d/2 \right) }\over {27}} $&$ -{{112}\over {27}} +\
  {{8 \left( u - d/2 \right) }\over 9} $\\ \svs 
$4 $&$ -{{88}\over {81}} + {{388 u}\over {243}} - {{32 d}\over {243}} $&$ -{{152}\over\
  {27}} + {{3140 u}\over {729}} + {{656 d}\over {729}} $&$ -{{40}\over 9} -\
  {{100 u}\over {27}} - {{16 d}\over {27}} $&$ {{170}\over {27}} + {{908\
  u}\over {729}} + {{1232 d}\over {729}} $&$ -{{14}\over 3} + {{148 u}\over\
  {27}} - {{80 d}\over {27}} $\\ \svs 
$5 $&$ 6 + {{136 \left( u - d/2 \right) }\over {81}} $&$ -{{232}\over 9} + {{104\
  \left( u - d/2 \right) }\over {27}} $&$ {{40}\over 3} + {{88 \left( u - d/2\
  \right) }\over 9} $&$ {{184 \left( u - d/2 \right) }\over {27}} $&$ {{8 \left(\
  u - d/2 \right) }\over 9} $\\ \svs 
$6 $&$ 7 + {{748 u}\over {243}} - {{212 d}\over {243}} $&$ -2 - {{5212 u}\over {729}}\
  + {{4832 d}\over {729}} $&$ {{182}\over 9} + {{188 u}\over {27}} - {{160\
  d}\over {27}} $&$ {{-2260 u}\over {729}} + {{2816 d}\over {729}} $&$ {{-140\
  u}\over {27}} + {{64 d}\over {27}} $\\ \svs 
$7 $&$ {{20}\over 3} + {{136 \left( u + d/4 \right) }\over {81}} $&$ -{{134}\over 9} +\
  {{104 \left( u + d/4 \right) }\over {27}} $&$ {{38}\over 3} + {{88 \left( u +\
  d/4 \right) }\over 9} $&$ {{184 \left( u + d/4 \right) }\over {27}} $&$ {{8\
  \left( u + d/4 \right) }\over 9} $\\ \svs 
$8 $&$ {{91}\over 9} + {{748 u}\over {243}} + {{106 d}\over {243}} $&$ 2 - {{5212\
  u}\over {729}} - {{2416 d}\over {729}} $&$ {{154}\over 9} + {{188 u}\over\
  {27}} + {{80 d}\over {27}} $&$ {{-2260 u}\over {729}} - {{1408 d}\over {729}}\
  $&$ {{-140 u}\over {27}} - {{32 d}\over {27}} $\\ \svs 
$9 $&$ {{116}\over {243}} + {{136 \left( u + d/4 \right) }\over {81}} $&$\
  -{{1568}\over {729}} + {{104 \left( u + d/4 \right) }\over {27}} $&$\
  -{{32}\over {27}} + {{88 \left( u + d/4 \right) }\over 9} $&$ {{5578}\over\
  {729}} + {{184 \left( u + d/4 \right) }\over {27}} $&$ {{38}\over {27}} + {{8\
  \left( u + d/4 \right) }\over 9} $\\ \svs 
$10 $&$ {{44}\over {81}} + {{388 u}\over {243}} + {{16 d}\over {243}} $&$ {{76}\over\
  {27}} + {{3140 u}\over {729}} - {{328 d}\over {729}} $&$ {{20}\over 9} -\
  {{100 u}\over {27}} + {{8 d}\over {27}} $&$ {{140}\over {27}} + {{908 u}\over\
  {729}} - {{616 d}\over {729}} $&$ -{{28}\over 9} + {{148 u}\over {27}} + {{40\
  d}\over {27}} $
\end{tabular}
\label{tab:gse}
\end{table}

\subsection{Quark Threshold Matching Matrix}
            \label{sec:HeffdF1:1010:Mm}
Extending the matching matrix $\hM(m)$ of \eqn{eq:MmKpp} to the
simultaneous presence of QCD and QED corrections yields
\begin{equation}
\hM(m) = 1 + \frac{\as(m)}{4\pi} \; \vardrs^T +
         \frac{\aem}{4\pi} \; \vardre^T \, .
\label{eq:MmdF1:1010}
\end{equation}
At scale $\mu=\mb$ the matrices $\vardrs$ and $\vardre$ read
\begin{equation}\nonumber
\vardrs^T=\frac{5}{18} P \, (0,0,0,-2,0,-2,0,1,0,1)
\end{equation}
\begin{equation}
\vardre^T=\frac{10}{81} \bar{P} \, (0,0,6,2,6,2,-3,-1,-3,-1)
\label{eq:drsembdF1:1010}
\end{equation}
and at $\mu=\mc$
\begin{equation}\nonumber
\vardrs^T=-\frac{5}{9} P \, (0,0,0,1,0,1,0,1,0,1)
\end{equation}
\begin{equation}
\vardre^T=-\frac{40}{81} \bar{P} \, (0,0,3,1,3,1,3,1,3,1)
\label{eq:drsemcdF1:1010}
\end{equation}
with eq.\ \eqn{eq:PdrsKpp} generalized to
\begin{eqnarray}
P^T &=& (0,0,-\frac{1}{3},1,-\frac{1}{3},1,0,0,0,0) \, ,
\label{eq:Pdrs:10} \\
\bar{P}^T &=& (0,0,0,0,0,0,1,0,1,0) \, .
\label{eq:Pbardre:10}
\end{eqnarray}

\subsection{Numerical Results for the $\Kpipi$ Wilson Coefficients}
            \label{sec:HeffdF1:1010:numres}
\begin{table}[htb]
\caption[]{$\dS$ Wilson coefficients at $\mu=1\gev$ for $\mt=170\gev$.
$y_1 = y_2 \equiv 0$.
\label{tab:wc10smu1}}
\begin{center}
\begin{tabular}{|c|c|c|c||c|c|c||c|c|c|}
& \multicolumn{3}{c||}{$\Lms^{(4)}=215\mev$} &
  \multicolumn{3}{c||}{$\Lms^{(4)}=325\mev$} &
  \multicolumn{3}{c| }{$\Lms^{(4)}=435\mev$} \\
\hline
Scheme & LO & NDR & HV & LO & 
NDR & HV & LO & NDR & HV \\
\hline
$z_1$ & --0.607 & --0.409 & --0.494 & --0.748 & 
--0.509 & --0.640 & --0.907 & --0.625 & --0.841 \\
$z_2$ & 1.333 & 1.212 & 1.267 & 1.433 & 
1.278 & 1.371 & 1.552 & 1.361 & 1.525 \\
\hline
$z_3$ & 0.003 & 0.008 & 0.004 & 0.004 & 
0.013 & 0.007 & 0.006 & 0.023 & 0.015 \\
$z_4$ & --0.008 & --0.022 & --0.010 & --0.012 & 
--0.035 & --0.017 & --0.017 & --0.058 & --0.029 \\
$z_5$ & 0.003 & 0.006 & 0.003 & 0.004 & 
0.008 & 0.004 & 0.005 & 0.009 & 0.005 \\
$z_6$ & --0.009 & --0.022 & --0.009 & --0.013 & 
--0.035 & --0.014 & --0.018 & --0.059 & --0.025 \\
\hline
$z_7/\aem$ & 0.004 & 0.003 & --0.003 & 0.008 & 
0.011 & --0.002 & 0.011 & 0.021 & --0.001 \\
$z_8/\aem$ & 0 & 0.008 & 0.006 & 0.001 & 
0.014 & 0.010 & 0.001 & 0.027 & 0.017 \\
$z_9/\aem$ & 0.005 & 0.007 & 0 & 0.008 & 
0.018 & 0.005 & 0.012 & 0.034 & 0.011 \\
$z_{10}/\aem$ & 0 & --0.005 & --0.006 & --0.001 & 
--0.008 & --0.010 & --0.001 & --0.014 & --0.017 \\
\hline
$y_3$ & 0.030 & 0.025 & 0.028 & 0.038 & 
0.032 & 0.037 & 0.047 & 0.042 & 0.050 \\
$y_4$ & --0.052 & --0.048 & --0.050 & --0.061 & 
--0.058 & --0.061 & --0.071 & --0.068 & --0.074 \\
$y_5$ & 0.012 & 0.005 & 0.013 & 0.013 & 
--0.001 & 0.016 & 0.014 & --0.013 & 0.021 \\
$y_6$ & --0.085 & --0.078 & --0.071 & --0.113 & 
--0.111 & --0.097 & --0.148 & --0.169 & --0.139 \\
\hline
$y_7/\aem$ & 0.027 & --0.033 & --0.032 & 0.036 & 
--0.032 & --0.030 & 0.043 & --0.031 & --0.027 \\
$y_8/\aem$ & 0.114 & 0.121 & 0.133 & 0.158 & 
0.173 & 0.188 & 0.216 & 0.254 & 0.275 \\
$y_9/\aem$ & --1.491 & --1.479 & --1.480 & --1.585 & 
--1.576 & --1.577 & --1.700 & --1.718 & --1.722 \\
$y_{10}/\aem$ & 0.650 & 0.540 & 0.547 & 0.800 & 
0.690 & 0.699 & 0.968 & 0.892 & 0.906 \\
\end{tabular}
\end{center}
\end{table}

\begin{table}[htb]
\caption[]{$\dS$ Wilson coefficients at $\mu=\mc=1.3\gev$ for
$\mt=170\gev$ and $f=3$ effective flavours.
$|z_3|,\ldots,|z_{10}|$ are numerically irrelevant relative to
$|z_{1,2}|$. $y_1 = y_2 \equiv 0$.
\label{tab:wc10smu13}}
\begin{center}
\begin{tabular}{|c|c|c|c||c|c|c||c|c|c|}
& \multicolumn{3}{c||}{$\Lms^{(4)}=215\mev$} &
  \multicolumn{3}{c||}{$\Lms^{(4)}=325\mev$} &
  \multicolumn{3}{c| }{$\Lms^{(4)}=435\mev$} \\
\hline
Scheme & LO & NDR & HV & LO & 
NDR & HV & LO & NDR & HV \\
\hline
$z_1$ & --0.521 & --0.346 & --0.413 & --0.625 & 
--0.415 & --0.507 & --0.732 & --0.490 & --0.617 \\
$z_2$ & 1.275 & 1.172 & 1.214 & 1.345 & 
1.216 & 1.276 & 1.420 & 1.265 & 1.354 \\
\hline
$y_3$ & 0.027 & 0.023 & 0.025 & 0.034 & 
0.029 & 0.033 & 0.041 & 0.036 & 0.042 \\
$y_4$ & --0.051 & --0.048 & --0.049 & --0.061 & 
--0.057 & --0.060 & --0.070 & --0.068 & --0.072 \\
$y_5$ & 0.013 & 0.007 & 0.014 & 0.015 & 
0.005 & 0.016 & 0.017 & 0.001 & 0.018 \\
$y_6$ & --0.076 & --0.068 & --0.063 & --0.096 & 
--0.089 & --0.081 & --0.120 & --0.118 & --0.103 \\
\hline
$y_7/\aem$ & 0.030 & --0.031 & --0.031 & 0.039 & 
--0.030 & --0.028 & 0.048 & --0.029 & --0.026 \\
$y_8/\aem$ & 0.092 & 0.103 & 0.112 & 0.121 & 
0.136 & 0.145 & 0.155 & 0.179 & 0.189 \\
$y_9/\aem$ & --1.428 & --1.423 & --1.423 & --1.490 & 
--1.479 & --1.479 & --1.559 & --1.548 & --1.549 \\
$y_{10}/\aem$ & 0.558 & 0.451 & 0.457 & 0.668 & 
0.547 & 0.553 & 0.781 & 0.656 & 0.664 \\
\end{tabular}
\end{center}
\end{table}

\begin{table}[htb]
\caption[]{$\dS$ Wilson coefficients at $\mu=2\gev$ for
$\mt=170\gev$. For $\mu > \mc$ the GIM mechanism gives $z_i \equiv 0$,
$i=3,\ldots,10$. $y_1 = y_2 \equiv 0$.
\label{tab:wc10smu2}}
\begin{center}
\begin{tabular}{|c|c|c|c||c|c|c||c|c|c|}
& \multicolumn{3}{c||}{$\Lms^{(4)}=215\mev$} &
  \multicolumn{3}{c||}{$\Lms^{(4)}=325\mev$} &
  \multicolumn{3}{c| }{$\Lms^{(4)}=435\mev$} \\
\hline
Scheme & LO & NDR & HV & LO & 
NDR & HV & LO & NDR & HV \\
\hline
$z_1$ & --0.413 & --0.268 & --0.320 & --0.480 & 
--0.310 & --0.376 & --0.544 & --0.352 & --0.432 \\
$z_2$ & 1.206 & 1.127 & 1.157 & 1.248 & 
1.151 & 1.191 & 1.290 & 1.176 & 1.227 \\
\hline
$y_3$ & 0.021 & 0.020 & 0.019 & 0.025 & 
0.024 & 0.023 & 0.028 & 0.028 & 0.027 \\
$y_4$ & --0.041 & --0.046 & --0.040 & --0.047 & 
--0.055 & --0.046 & --0.053 & --0.063 & --0.053 \\
$y_5$ & 0.011 & 0.010 & 0.012 & 0.012 & 
0.011 & 0.013 & 0.014 & 0.011 & 0.015 \\
$y_6$ & --0.056 & --0.058 & --0.047 & --0.068 & 
--0.071 & --0.057 & --0.079 & --0.086 & --0.068 \\
\hline
$y_7/\aem$ & 0.031 & --0.023 & --0.020 & 0.037 & 
--0.019 & --0.020 & 0.042 & --0.016 & --0.019 \\
$y_8/\aem$ & 0.068 & 0.076 & 0.084 & 0.084 & 
0.094 & 0.102 & 0.101 & 0.113 & 0.121 \\
$y_9/\aem$ & --1.357 & --1.361 & --1.357 & --1.393 & 
--1.389 & --1.389 & --1.430 & --1.419 & --1.423 \\
$y_{10}/\aem$ & 0.442 & 0.356 & 0.360 & 0.513 & 
0.414 & 0.419 & 0.581 & 0.472 & 0.477 \\
\end{tabular}
\end{center}
\end{table}

\begin{figure}[htb]
\vspace{0.10in}
\centerline{
\epsfysize=7in
\rotate[r]{
\epsffile{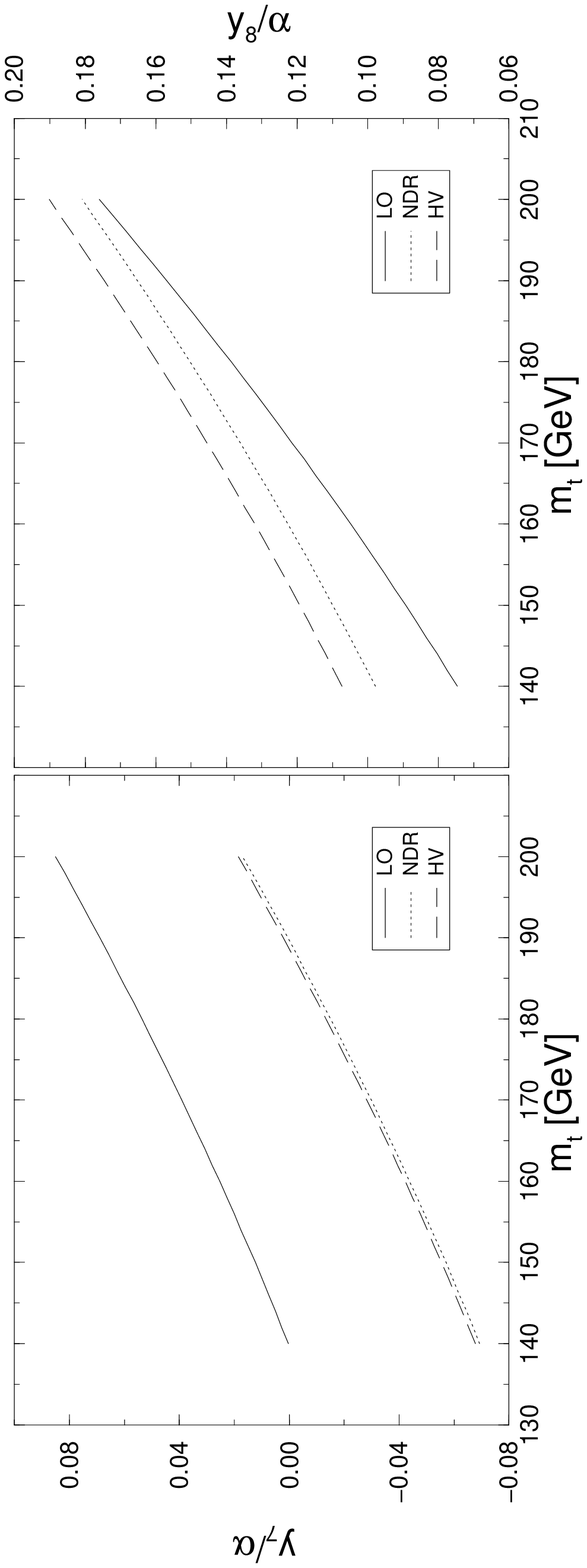}
}}
\vspace{0.08in}
\caption[]{
Wilson coefficients $y_{7}(\mc)/\aem$ and $y_{8}(\mc)/\aem$ as
functions of $\mt$ for $\Lms^{(4)}=325\mev$.
\label{fig:dF1:mty78}}
\end{figure}

\begin{figure}[htb]
\vspace{0.10in}
\centerline{
\epsfysize=7in
\rotate[r]{
\epsffile{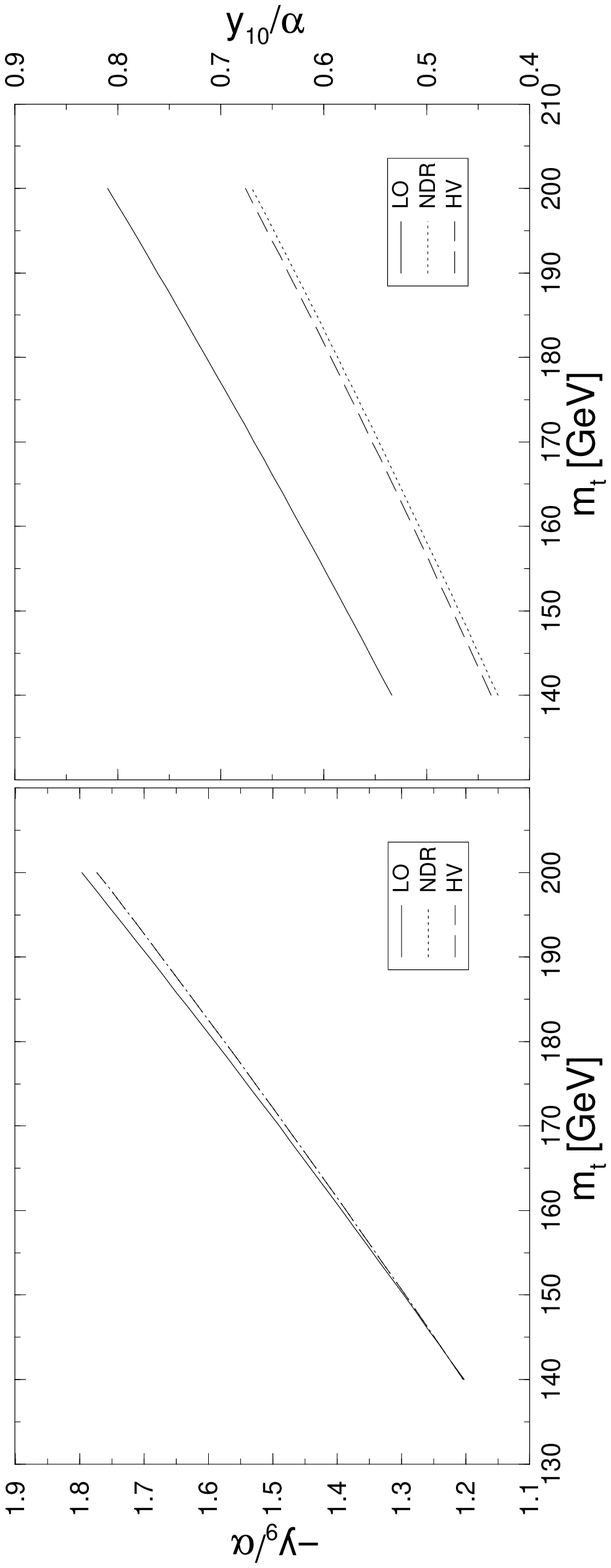}
}}
\vspace{0.08in}
\caption[]{
Wilson coefficients $y_{9}(\mc)/\aem$ and $y_{10}(\mc)/\aem$ as a
function of $\mt$ for $\Lms^{(4)}=325\mev$.
\label{fig:dF1:mty910}}
\end{figure}

Tables \ref{tab:wc10smu1}--\ref{tab:wc10smu2} give the $\dS$ Wilson
coefficients for $Q_1,\ldots,Q_{10}$ in the mixed case of QCD and QED.
\\
The coefficients for the current-current and QCD penguin operators
$Q_1,\ldots,Q_6$ are only very weakly affected by the extension of the
operator basis to the electroweak penguin operators $Q_7,\ldots,Q_{10}$.
Therefore the discussion for $Q_1,\ldots,Q_6$ given in connection with
tables \ref{tab:wc6smu1}--\ref{tab:wc6smu2} for the case of pure
QCD basically still holds and will not be repeated here.
\\
For the remaining coefficients of $Q_7,\ldots,Q_{10}$ one finds a
moderate scheme dependence for $y_7$, $y_9$ and $y_{10}$, but a $\ord(9\%)$
one for $y_8$. The notable feature of $|y_6|$ being larger in NDR than in
HV still holds, but is now confronted with an exactly opposite
dependence for the other important $\dS$ Wilson coefficient $y_8$ which
is in addition enhanced over its LO value.
\\
The particular dependence of $y_6$ and $y_8$ with respect to scheme,
LO/NLO and $\mt$ (see below) should be kept in mind for the later
discussion of $\epe$ in section \ref{sec:nloepe}.
\\
We also note that in the range of $\mt$ considered here, $y_7$ is very
small, $y_9$ is essentially unaffected by NLO QCD corrections and
$y_{10}$ is suppressed for $\mu \ge \mc$. It should also be stressed
that $|y_9|$  and $|y_{10}|$ are substantially larger than $|y_8|$
although, as we will see in the analysis of $\epe$, the operator $Q_8$
is more important than $Q_9$ and $Q_{10}$ for this ratio.
\\
Next, one infers from tables \ref{tab:wc10smu1}--\ref{tab:wc10smu2} 
that also in the mixed QCD/QED case the Wilson coefficients show a
strong dependence on $\Lms$.
\\
In contrast to the coefficients $y_3,\ldots,y_6$ for QCD penguins,
$y_7,\ldots,y_{10}$ for the electroweak penguins show a sizeable
$\mt$ dependence in the range $\mt = (170 \pm 15)\gev$. With
in/decreasing $\mt$ there is a relative variation of $\ord(\pm 19\%)$
and $\ord(\pm 10\%)$ for the absolute values of $y_8$ and $y_{9,10}$,
respectively.
This is illustrated further in figs.\ \ref{fig:dF1:mty78} and
\ref{fig:dF1:mty910} where the $\mt$ dependence of these coefficients is
shown explicitly. This strong $\mt$-dependence originates in the
$Z^0$-penguin diagrams.
The $\mt$-dependence of $y_9$ and $y_{10}$ can be conveniently
parametrized by a linear function to an accuracy better than $0.5\,\%$.
Details of this $\mt$-parametriziation can be found in table
\ref{tab:linfity910}.
\\
Finally, in tables \ref{tab:wc10smu1}--\ref{tab:wc10smu2} one
observes again the usual feature of decreasing Wilson coefficients
with increasing scale $\mu$.

\begin{table}[htb]
\caption[]{Coefficients in linear $\mt$-parametriziation $y_i/\aem = a
+ b \cdot (\mt/170\gev)$ of Wilson coefficients $y_9/\aem$ and
$y_{10}/\aem$ at scale $\mu=\mc$ for $\Lms^{(4)}=325\mev$.
\label{tab:linfity910}}
\begin{center}
\begin{tabular}{|c||c|c||c|c|}
\hline
 & \multicolumn{2}{|c||}{$y_9/\aem$} &
   \multicolumn{2}{|c| }{$y_{10}/\aem$} \\
\hline
 & a & b & a & b \\
\hline
LO  & 0.189 & --1.682 & --0.111 & 0.780 \\
NDR & 0.129 & --1.611 & --0.128 & 0.676 \\
HV  & 0.129 & --1.611 & --0.121 & 0.676 \\
\hline
\end{tabular}
\end{center}
\end{table}


\subsection{The $\dB$ Effective Hamiltonian Including Electroweak
            Penguins}
            \label{sec:HeffdF1:1010:dB1}
Finally we present in this section the Wilson coefficient functions of
the $\Delta B=1$, $\Delta C=0$ hamiltonian, including the effects of
electroweak penguin contributions \cite{burasetal:92d}.  These effects
play a role in certain penguin-induced B meson decays as discussed in
\cite{fleischer:94a}, \cite{fleischer:94b}, \cite{deshpandeetal:94},
\cite{deshpandehe:94}.

The generalization of the $\Delta B=1$, $\Delta C=0$ hamiltonian in pure
QCD (\ref{eq:HeffdB1:66}) to incorporate also electroweak penguin
operators is straightforward. One obtains
\begin{eqnarray} 
\Heff(\dB) &=& \frac{G_F}{\sqrt{2}} \bigl\{
   \xi_c \, \left[ C_1(\mu) Q_1^c(\mu) + C_2(\mu) Q_2^c(\mu) \right] +
   \xi_u \, \left[ C_1(\mu) Q_1^u(\mu) + C_2(\mu) Q_2^u(\mu) \right] 
\nn \\
 & & - \xi_t \, \sum_{i=3}^{10} C_i(\mu) Q_i(\mu)
\bigr\} \, .
\label{eq:HeffdB1:1010}
\end{eqnarray} 
where the operator basis now includes the electroweak penguin
operators
\begin{eqnarray}
Q_{7} & = & \frac{3}{2} \left( \bar b d \right)_{\rm V-A}
         \sum_{q} e_{q} \left( \bar q q \right)_{\rm V+A}
\, , \nn \\
Q_{8} & = & \frac{3}{2} \left( \bar b_{i} d_{j} \right)_{\rm V-A}
         \sum_{q} e_{q} \left( \bar q_{j}  q_{i}\right)_{\rm V+A}
\, , \nn \\
Q_{9} & = & \frac{3}{2} \left( \bar b d \right)_{\rm V-A}
         \sum_{q} e_{q} \left( \bar q q \right)_{\rm V-A}
\, , \label{eq:dB1:1010basis} \\
Q_{10}& = & \frac{3}{2} \left( \bar b_{i} d_{j} \right)_{\rm V-A}
         \sum_{q} e_{q} \left( \bar q_{j}  q_{i}\right)_{\rm V-A}
\nn
\end{eqnarray}
in addition to (\ref{eq:dB1basis}). The Wilson coefficients at $\mu=m_b$
read
\begin{equation}
\vC(\mb) = \hU_5(\mb,\mw,\aem) \, \vC(\mw) \, .
\label{eq:HeffdB1:1010:wc}
\end{equation}
where $\hU_5$ is the $10\times 10$ evolution matrix of
(\ref{eq:UdF1:1010}) for $f=5$ flavors. The $\vC(\mw)$ are given in
(\ref{eq:CMw1}) -- (\ref{eq:CMw10}) in the NDR scheme.

\subsection{Numerical Results for the $\dB$ Wilson Coefficients}
            \label{sec:HeffdF1:1010:dB1num}
\begin{table}[htb]
\caption[]{$\dB$ Wilson coefficients at $\mu=\overline{m}_{\rm b}(\mb)=
4.40\gev$ for $\mt=170\gev$.
\label{tab:wc10b}}
\begin{center}
\begin{tabular}{|c|c|c|c||c|c|c||c|c|c|}
& \multicolumn{3}{c||}{$\Lms^{(5)}=140\mev$} &
  \multicolumn{3}{c||}{$\Lms^{(5)}=225\mev$} &
  \multicolumn{3}{c| }{$\Lms^{(5)}=310\mev$} \\
\hline
Scheme & LO & NDR & HV & LO & 
NDR & HV & LO & NDR & HV \\
\hline
$C_1$ & --0.273 & --0.165 & --0.202 & --0.308 & 
--0.185 & --0.228 & --0.339 & --0.203 & --0.251 \\
$C_2$ & 1.125 & 1.072 & 1.091 & 1.144 & 
1.082 & 1.105 & 1.161 & 1.092 & 1.117 \\
\hline
$C_3$ & 0.013 & 0.013 & 0.012 & 0.014 & 
0.014 & 0.013 & 0.016 & 0.016 & 0.015 \\
$C_4$ & --0.027 & --0.031 & --0.026 & --0.030 & 
--0.035 & --0.029 & --0.033 & --0.039 & --0.033 \\
$C_5$ & 0.008 & 0.008 & 0.008 & 0.009 & 
0.009 & 0.009 & 0.009 & 0.009 & 0.010 \\
$C_6$ & --0.033 & --0.036 & --0.029 & --0.038 & 
--0.041 & --0.033 & --0.043 & --0.046 & --0.037 \\
\hline
$C_7/\aem$ & 0.042 & --0.003 & 0.006 & 0.045 & 
--0.002 & 0.005 & 0.047 & --0.001 & 0.005 \\
$C_8/\aem$ & 0.041 & 0.047 & 0.052 & 0.048 & 
0.054 & 0.060 & 0.054 & 0.061 & 0.067 \\
$C_9/\aem$ & --1.264 & --1.279 & --1.269 & --1.280 & 
--1.292 & --1.283 & --1.294 & --1.303 & --1.296 \\
$C_{10}/\aem$ & 0.291 & 0.234 & 0.237 & 0.328 & 
0.263 & 0.266 & 0.360 & 0.288 & 0.291 \\
\end{tabular}
\end{center}
\end{table}

Table \ref{tab:wc10b} lists the $\dB$ Wilson coefficients for
$Q_1^{u,c},Q_2^{u,c},Q_3,\ldots,Q_{10}$ in the mixed case of QCD and
QED.
\\
Similarly to the $\dS$ case the coefficients for the current-current and
QCD penguin operators $Q_1,\ldots,Q_6$ are only very weakly affected by
the extension of the operator basis to the electroweak penguin
operators $Q_7,\ldots,Q_{10}$. Therefore the discussion of
$C_1,\ldots,C_6$ in connection with table \ref{tab:wc6b} is also valid
for the present case.
\\
Here we therefore restrict our discussion to the coefficients
$C_7,\ldots,C_{10}$ of the operators $Q_7,\ldots,Q_{10}$ in the
extended basis.
\\
The coefficients $C_7,\ldots,C_{10}$ show a visible dependence on the
scheme, $\Lms$ and LO/NLO. However, this dependence is less pronounced
for the coefficient $C_9$ than it is for $C_{7,8,10}$. This is
noteworthy since in $B$-meson decays $C_9$ usually resides in the
dominant electroweak penguin contribution \cite{fleischer:94a},
\cite{fleischer:94b}, \cite{deshpandeetal:94}, \cite{deshpandehe:94}.
\\
In contrast to $C_1,\ldots,C_6$ the additional coefficients
$C_7,\ldots,C_{10}$ show a non negligible $\mt$ dependence in the range
$\mt = (170 \pm 15)\gev$. With in/decreasing $\mt$ there is similarly to
the $\dS$ case a relative variation of $\ord(\pm 19\%)$ and $\ord(\pm
10\%)$ for the absolute values of $C_8$ and $C_{9,10}$, respectively.

\begin{figure}[htb]
\vspace{0.10in}
\centerline{
\epsfysize=7in
\rotate[r]{
\epsffile{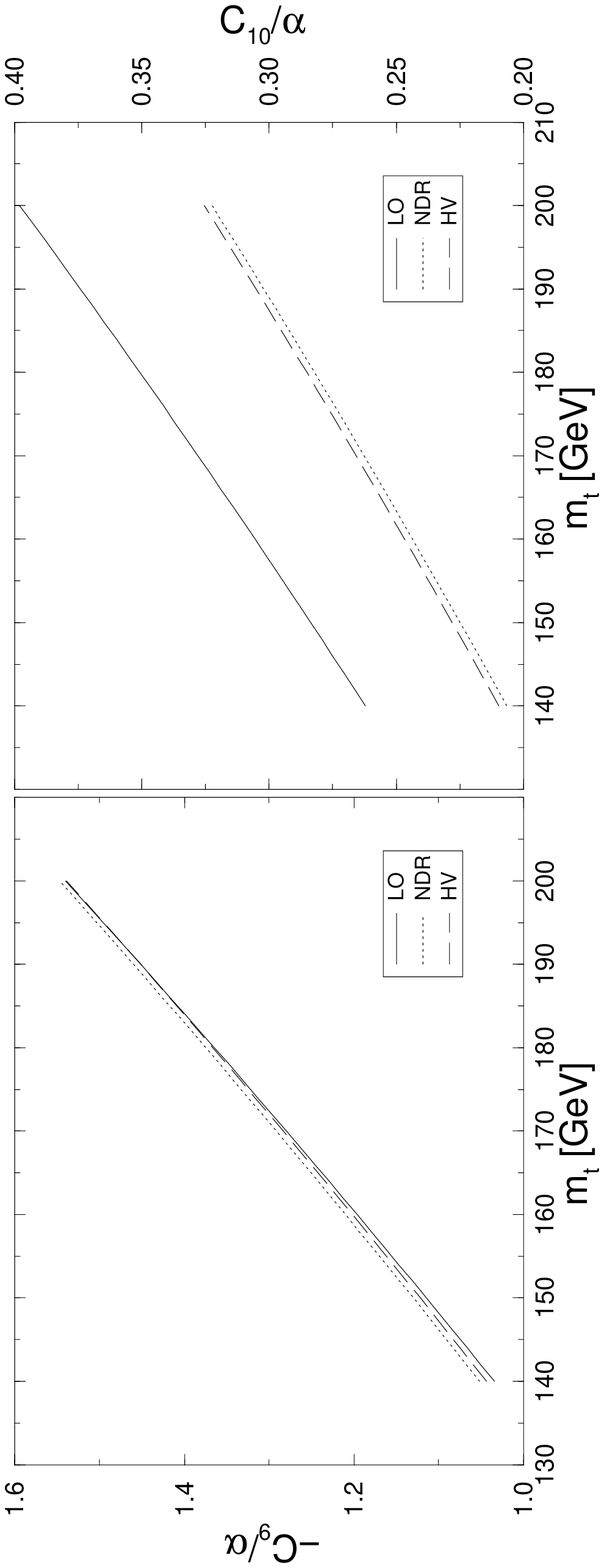}
}}
\vspace{0.08in}
\caption[]{
Wilson coefficients $C_{9}/\aem$ and $C_{10}/\aem$ at $\mu=
\overline{m}_{\rm b}(\mb)= 4.40\gev$ as a function of $\mt$ for
$\Lms^{(5)}=225\mev$.
\label{fig:dF1:mtC910}}
\end{figure}

Since the coefficients $C_9$ and $C_{10}$ play an important role in
B decays we show in fig.\ \ref{fig:dF1:mtC910} their $\mt$ dependence
explicitly. Again the $\mt$-dependence can be parametrized by a linear
function to an accuracy better than $0.5\,\%$. Details of the
$\mt$-parametriziation are given in table \ref{tab:linfitC910}.

\begin{table}[htb]
\caption[]{Coefficients in linear $\mt$-parametriziation $C_i/\aem = a
+ b \cdot (\mt/170\gev)$ of Wilson coefficients $C_9/\aem$ and
$C_{10}/\aem$ at scale $\mu=\mb=4.4\gev$ for $\Lms^{(5)}=225\mev$.
\label{tab:linfitC910}}
\begin{center}
\begin{tabular}{|c||c|c||c|c|}
\hline
 & \multicolumn{2}{|c||}{$C_9/\aem$} &
   \multicolumn{2}{|c| }{$C_{10}/\aem$} \\
\hline
 & a & b & a & b \\
\hline
LO  & 0.152 & --1.434 & --0.056 & 0.385 \\
NDR & 0.109 & --1.403 & --0.065 & 0.328 \\
HV  & 0.117 & --1.403 & --0.062 & 0.328 \\
\hline
\end{tabular}
\end{center}
\end{table}

\section{The Effective Hamiltonian for $\Kpiee$}
         \label{sec:HeffKpe}
The $\dS$ effective hamiltonian for $\Kpiee$ at scales $\mu < \mc$ is given by
\begin{equation}
\Heff(\dS) = \frac{G_F}{\sqrt{2}} \V{us}^* \V{ud}^{} \left[ \; \sum_{i=1}^{6,7V}
             \left( z_i(\mu) + \tau \; y_i(\mu) \right) Q_i(\mu)
             + \tau \; y_{7A}(\mw) \; Q_{7A}(\mw) \; \right]
\label{eq:HeffKpe}
\end{equation}
with
\begin{equation}
\tau = - \frac{ \V{ts}^* \V{td}^{} }{ \V{us}^* \V{ud}^{} } \, .
\label{eq:tauagain}
\end{equation}

\subsection{Operators}
            \label{sec:HeffKpe:op}
In \eqn{eq:HeffKpe} $Q_{1,2}$
denote the $\dS$ current-current and $Q_3,\ldots,Q_6$ the QCD penguin
operators of eq.~\eqn{eq:Kppbasis}. For scales $\mu > \mc$ again the
current-current operators $Q^c_{1,2}$ of eq.~\eqn{eq:KppQ12c} have to
be taken into account.

The new operators specific to the decay $\Kpiee$ are
\begin{eqnarray}
Q_{7V} &=& (\bar{s} d)_{\rm V-A} (\bar{e} e)_{\rm V} \, ,
\label{eq:Q7VKpe} \\
Q_{7A} &=& (\bar{s} d)_{\rm V-A} (\bar{e} e)_{\rm A} \, .
\label{eq:Q7AKpe}
\end{eqnarray}
They originate through the $\gamma$- and $Z^0$-penguin and box diagrams of
fig.\ \ref{fig:1loopful}.

It is convenient to introduce the auxiliary operator
\begin{equation}
Q'_{7V} = (\aem/\as) \; (\bar{s} d)_{\rm V-A} (\bar{e}e)_{\rm V}
\label{eq:Q7VprimeKpe}
\end{equation}
and work for the renormalization group analysis in the basis
$Q_1,\ldots,Q_6$, $Q'_{7V}$. The factor $\aem/\as$ in the definition of
$Q'_{7V}$ implies that in this new basis the anomalous
dimension matrix $\hg$ will be a function of $\as$ alone. At the end of the
renormalization group analysis, this factor will be put back into the
Wilson coefficient $C_{7V}(\mu)$ of the operator $Q_{7V}$ in
eq.~\eqn{eq:Q7VKpe}. There is no need to include a similar factor in
$Q_{7A}$ as this operator does not mix under renormalization with the
remaining operators. Since $Q_{7A}$ has no anomalous dimension its Wilson
coeffcient is $\mu$-independent.

In principle one can think of including the electroweak four-quark
penguin operators $Q_7,\ldots,Q_{10}$ of eq.~\eqn{eq:dF1:1010basis} in
$\Heff$ of \eqn{eq:HeffKpe}. However, their Wilson coefficients and
matrix elements for the decay $\Kpiee$ are both of order $\ord(\aem)$
implying that these operators eventually would enter the amplitude
$A(\Kpiee)$ at $\ord(\aem^2)$.  To the order considered here this
contribution is thus negligible. This should be distinguished from the
case of $\Kpipi$ discussed in section~\ref{sec:HeffdF1:1010}. There, in
spite of being suppressed by $\aem/\as$ relative to QCD penguin
operators, the electroweak penguin operators have to be included in the
analysis because of the additional enhancement factor $\RE A_0/\RE A_2
\simeq 22$ present in the formula for $\epe$ (see section
\ref{sec:nloepe}). Such an enhancement factor is not present in the
$\Kpiee$ case and the electroweak penguin operators can be safely
neglected. \\
Concerning the Wilson coefficients, the electroweak four-quark penguin
operators would also affect through mixing under renormalization the
coefficients $C_3,\ldots,C_6$ at $\ord(\aem)$ and $C_{7V}$ at
$\ord(\aem^2)$.  Since the corresponding matrix elements are
$\ord(\aem)$ and $\ord(1)$, respectively, we again obtain a negligible
$\ord(\aem^2)$ effect in $A(\Kpiee)$. \\
In summary, the electroweak four-quark penguin operators $Q_7, \ldots,
Q_{10}$ can safely be neglected in the following discussion of
$\Heff(\dS)$ for $\Kpiee$.

We also neglect the ``magnetic moment'' operators.These operators,
being of dimension five, do not influence the Wilson coefficients of
the operators $Q_1,\ldots,Q_6$, $Q_{7V}$ and $Q_{7A}$. Since their
contributions to $\Kpiee$ are suppressed by an additional factor $\ms$,
they appear strictly speaking at higher order in chiral perturbation
theory. On the other hand the magnetic moment type operators play a
crucial role in $b \to s \gamma$ and $b \to d \gamma$ transitions as
discussed in sections \ref{sec:Heff:BXsgamma} and
\ref{sec:Heff:Bsgamma}. They also have to be kept in the decay $B\to
X_se^+e^-$.

\subsection{Wilson Coefficients}
            \label{sec:HeffKpe:wc}
Eqs.~\eqn{eq:WCy}--\eqn{eq:WCz} remain valid in the case of $\Kpiee$ with
$\hU_{f}(m_1,m_2)$ and $\hM(m_i)$ now denoting $7 \times 7$ matrices in
the $Q_1,\ldots,Q_6$, $Q'_{7V}$ basis. The Wilson coefficients are
given by seven-dimensional column vectors $\vec{z}(\mu)$ and
$\vec{v}(\mu)$ having components $(z_1,\ldots,z_6,z'_{7V})$ and
$(v_1,\ldots,v_6,v'_{7V})$, respectively. Here
\begin{equation}
v'_{7V}(\mu) = \frac{\as(\mu)}{\aem} \; v_{7V}(\mu) \, ,
\qquad\qquad
z'_{7V}(\mu) = \frac{\as(\mu)}{\aem} \; z_{7V}(\mu)
\end{equation}
are the rescaled Wilson coefficients of the auxiliary operator $Q'_{7V}$
used in the renormalization group evolution.

The initial conditions $C_1(\mw),\ldots,C_6(\mw)$, $z_1(\mw)$, $z_2(\mw)$
and $z_1(\mc),\ldots,z_6(\mc)$ for the four-quark operators $Q_1,\ldots,Q_6$
are readily obtained from eqs.~\eqn{eq:CMw1QCD}--\eqn{eq:CMw6QCD},
\eqn{eq:zMw12} and \eqn{eq:zmc}.

The corresponding initial conditions for the remaining operators
$Q'_{7V}$ and $Q_{7A}$ specific to $\Kpiee$ are then given in the NDR
scheme by
\begin{equation}
C'_{7V}(\mw) = \frac{\as(\mw)}{2 \pi}
\left[ \frac{C_0(x_t)-B_0(x_t)}{\sin^2\theta_W} -\tilde{D}_0(x_t) -
4 C_0(x_t) \right]
\label{eq:C7VprimeMw}
\end{equation}
and
\begin{equation}
C_{7A}(\mw) = 
y_{7A}(\mw) = \frac{\aem}{2 \pi} \frac{B_0(x_t)-C_0(x_t)}{\sin^2\theta_W}.
\label{eq:C7AMw}
\end{equation}
In order to find $z'_{7V}(\mc)$ which results from the diagrams of
fig.\ \ref{fig:1loopeff}, we simply have to rescale the NDR result for
$z_7(\mc)$ in eq.~\eqn{eq:zmc:1010} by a factor of $-3 \as/\aem$. This
yields
\begin{equation}
z'_{7V}(\mc) = -\frac{\as(\mc)}{2 \pi} F_{\rm e}(\mc) \, .
\label{eq:z7primemc}
\end{equation}

\subsection{Renormalization Group Evolution and Anomalous Dimension Matrices}
            \label{sec:HeffKpe:rge}
Working in the rescaled basis $Q_1,\ldots,Q_6$, $Q'_{7V}$ the anomalous
dimension matrix $\hg$ has per construction a pure $\ord(\as)$ expansion
\begin{equation}
\hg = \frac{\as}{4\pi}       \gamma^{(0)} +
      \frac{\as^2}{(4\pi)^2} \gamma^{(1)} + \ldots
\, ,
\label{eq:gsexpKpe}
\end{equation}
where $\hg^{(0)}$ and $\hg^{(1)}$ are $7 \times 7$ matrices. The
evolution matrices $\hU_{f}(m_1,m_2)$ in eqs.~\eqn{eq:WCv} and
\eqn{eq:WCz} are for the present case simply given by \eqn{eq:UQCDKpp}
and \eqn{u0vd}--\eqn{jvs}.

The $6 \times 6$ submatrix of $\gamma^{(0)}$ involving the
operators $Q_1,\ldots,Q_6$ has already been given in eq.~\eqn{eq:gs0Kpp}.
Here we only give the remaining entries of $\gamma^{(0)}$ related to
the additional presence of the operator $Q'_{7V}$
\begin{equation}
\begin{array}{lclclcl}
\gamma^{(0)}_{17} &=& -\frac{16}{9} N
&\qquad&
\gamma^{(0)}_{27} &=& -\frac{16}{9}
\svs \\
\gamma^{(0)}_{37} &=& -\frac{16}{9} N \left(u-\frac{d}{2}-\frac{1}{N} \right)
&\qquad&
\gamma^{(0)}_{47} &=& -\frac{16}{9} \left( u -\frac{d}{2} - N \right)
\svs \\
\gamma^{(0)}_{57} &=& -\frac{16}{9} N \left( u -\frac{d}{2} \right)
&\qquad&
\gamma^{(0)}_{67} &=& -\frac{16}{9} \left( u -\frac{d}{2} \right)
\svs \\
\gamma^{(0)}_{77} &=& -2 \beta_0 = -\frac{22}{3} N + \frac{4}{3} f
&\qquad&
\gamma^{(0)}_{7i} &=& 0 \qquad i=1,\ldots,6
\svs
\end{array}
\label{eq:g0col7kpe}
\end{equation}
where $N$ denotes the number of colours. These elements have been first
calculated in \cite{gilman:80} except that $\gamma^{(0)}_{37}$ and
$\gamma^{(0)}_{47}$ have been corrected in \cite{eegpicek:88},
\cite{flynn:89b}.

The $6 \times 6$ submatrix of $\gamma^{(1)}$ involving the
operators $Q_1,\ldots,Q_6$ has already been presented as $\gss$ in
eq.~\eqn{eq:gs1ndrN3Kpp} and
the seventh column of $\gamma^{(1)}$ is given as follows in the
NDR scheme \cite{burasetal:94a}
\begin{eqnarray}
\gamma^{(1)}_{17} &=& \frac{8}{3} \left( 1 - N^2 \right) \, ,
\nn \\
\gamma^{(1)}_{27} &=& \frac{200}{81} \left( N - \frac{1}{N} \right) \, ,
\nn \\
\gamma^{(1)}_{37} &=& \frac{8}{3} \left( u - \frac{d}{2} \right) \left( 1
    - N^2 \right) + \frac{464}{81} \left( \frac{1}{N} - N \right) \, ,
\nn \\
\gamma^{(1)}_{47} &=& \left( u \frac{280}{81} + d \frac{64}{81} \right)
    \left( \frac{1}{N} - N \right) + \frac{8}{3} \left( N^2 - 1 \right) \, ,
\label{eq:g1col7kpe} \\
\gamma^{(1)}_{57} &=& \frac{8}{3} \left( u - \frac{d}{2} \right)
                      \left( 1 - N^2 \right) \, ,
\nn \\
\gamma^{(1)}_{67} &=& \left( u \frac{440}{81} - d \frac{424}{81} \right)
                      \left( N - \frac{1}{N} \right) \, ,
\nn \\
\gamma^{(1)}_{77} &=& -2 \beta_1 = -\frac{68}{3} N^2 + \frac{20}{3} N f +
                      4 C_F f
\nn \\
\gamma^{(1)}_{7i} &=& 0 \qquad i=1,\ldots,6
\nn
\end{eqnarray}
where $C_F = (N^2 -1)/(2 N)$. The corresponding results in the HV
scheme can be found in \cite{burasetal:94a}.

\subsection{Quark Threshold Matching Matrix}
            \label{sec:HeffKpe:Mm}
For the case of $\Kpiee$ the matching matrix $M(m)$ has in the
basis $Q_1,\ldots,Q_6,Q'_{7V}$ the form
\begin{equation}
\hM(m) = 1 + \frac{\as(m)}{4 \pi} \; \vardrs^T
\label{eq:MmKpe}
\end{equation}
where $1$ and $\vardrs^T$ are $7 \times 7$ matrices and $m$ is
the scale of the quark threshold.

The $6 \times 6$ submatrix of $\hM(m)$ involving $Q_1,\ldots,Q_6$ has been
given in eq.~\eqn{eq:drsmbcKpp}. The remaining entries of $\vardrs$ can
be found from the matrix $\vardre$ given in eqs.\ \eqn{eq:drsembdF1:1010}
and \eqn{eq:drsemcdF1:1010} by making a simple rescaling by $-3 \;
\as/\aem$ as in the case of $z_7(\mc)$.

In summary, for the quark threshold $m = \mb$ the matrix $\vardrs$ reads
\begin{equation}
\vardrs =
\left(
\begin{array}{ccccccc}
0 & 0 & 0 & 0 & 0 & 0 & 0 \\
0 & 0 & 0 & 0 & 0 & 0 & 0 \\
0 & 0 & 0 & 0 & 0 & 0 & -\frac{20}{9} \\
0 & 0 & \frac{5}{27} & -\frac{5}{9} & \frac{5}{27} & -\frac{5}{9} &
  -\frac{20}{27} \\
0 & 0 & 0 & 0 & 0 & 0 & -\frac{20}{9} \\
0 & 0 & \frac{5}{27} & -\frac{5}{9} & \frac{5}{27} & -\frac{5}{9} &
  -\frac{20}{27} \\
0 & 0 & 0 & 0 & 0 & 0 & 0
\end{array}
\right) \, .
\label{eq:drsKpe}
\end{equation}
For $m=\mc$ the seventh column of $\vardrs$ in \eqn{eq:drsKpe} has
to be multiplied by $-2$.

\subsection{Numerical Results for the $\Kpiee$ Wilson Coefficients}
            \label{sec:HeffKpe:numres}
\begin{table}[htb]
\caption[]{$K_{\rm L} \rightarrow \pi^0\,e^+\,e^-$ Wilson coefficients
for $Q_{7V,A}$ at $\mu=1\gev$ for $\mt=170\gev$. The corresponding
coefficients for $Q_1,\ldots,Q_6$ can be found in table \ref{tab:wc6smu1}
of section \ref{sec:HeffdF1:66}.
\label{tab:wckpemu1}}
\begin{center}
\begin{tabular}{|c|c|c|c||c|c|c||c|c|c|}
& \multicolumn{3}{c||}{$\Lms^{(4)}=215\mev$} &
  \multicolumn{3}{c||}{$\Lms^{(4)}=325\mev$} &
  \multicolumn{3}{c| }{$\Lms^{(4)}=435\mev$} \\
\hline
Scheme & LO & NDR & HV & LO & 
NDR & HV & LO & NDR & HV \\
\hline
$z_{7V}/\aem$ & --0.014 & --0.015 & 0.005 & --0.024 & 
--0.046 & --0.003 & --0.035 & --0.084 & --0.011 \\
\hline
$y_{7V}/\aem$ & 0.575 & 0.747 & 0.740 & 0.540 & 
0.735 & 0.725 & 0.509 & 0.720 & 0.710 \\
$y_{7A}/\aem$ & --0.700 & --0.700 & --0.700 & --0.700 & 
--0.700 & --0.700 & --0.700 & --0.700 & --0.700 \\
\end{tabular}
\end{center}
\end{table}

In the case of $K_L \to \pi^0 e^+ e^-$, due to
$\gamma^{(0)}_{7i}=\gamma^{(1)}_{7i}=0$, $i=1,\ldots,6$ in
eq.\ \eqn{eq:g0col7kpe} and \eqn{eq:g1col7kpe}, respectively, the RG
evolution of $Q_1,\ldots,Q_6$ is completely unaffected by the additional
presence of the operator $Q_{7V}$. The $K_L \to \pi^0 e^+ e^-$ Wilson
coefficients $z_i$ and $y_i$, $i=1,\ldots,6$ at scale $\mu=1\gev$
can therefore be found in table \ref{tab:wc6smu1} of section
\ref{sec:HeffdF1:66}.
\\
The $K_L \to \pi^0 e^+ e^-$ Wilson coefficients for the remaining operators
$Q_{7V}$ and $Q_{7A}$ are given in table \ref{tab:wckpemu1}. Some
insight in the analytic structure of $y_{7V}$ will be gained by studying
the analogous decay $B \to X_s e^+ e^-$ in section \ref{sec:Heff:BXsee}
and also in section \ref{sec:KLpee} where the phenomenology of $K_L \to
\pi^0 e^+ e^-$ will be presented.

\begin{table}[htb]
\caption[]{$K_{\rm L} \to \pi^0 e^+ e^-$ Wilson coefficients
$z_{7V}/\aem$ and  $y_{7V}/\aem$ for $\mt=170\gev$ and various values
of $\mu$.
\label{tab:kpemuzy7}}
\begin{center}
\begin{tabular}{|c|c|c|c||c|c|c||c|c|c|}
& \multicolumn{3}{c||}{$\Lms^{(4)}=215\mev$} &
  \multicolumn{3}{c||}{$\Lms=^{(4)}325\mev$} &
  \multicolumn{3}{c| }{$\Lms=^{(4)}435\mev$} \\
\hline
Scheme & LO & NDR & HV & LO & NDR & HV & LO & NDR & HV \\
\hline
$\mu\,[\gev]$ & \multicolumn{9}{c|}{$z_{7V}/\aem$} \\
\hline
0.8 & --0.031 & --0.029 & 0.004 & --0.053 &
      --0.081 & --0.012 & --0.077 & --0.149 & --0.023 \\
1.0 & --0.014 & --0.015 & 0.005 & --0.024 &
      --0.046 & --0.003 & --0.035 & --0.084 & --0.011 \\
1.2 & --0.004 & --0.009 & 0.002 & --0.006 &
      --0.029 &       0 & --0.009 & --0.051 & --0.002 \\
\hline
$\mu\,[\gev]$ & \multicolumn{9}{c|}{$y_{7V}/\aem$} \\
\hline
0.8 & 0.578 & 0.751 & 0.744 & 0.545 & 0.739 & 0.730 & 0.514 & 0.722 & 0.712 \\
1.0 & 0.575 & 0.747 & 0.740 & 0.540 & 0.735 & 0.725 & 0.509 & 0.720 & 0.710 \\
1.2 & 0.571 & 0.744 & 0.736 & 0.537 & 0.731 & 0.721 & 0.505 & 0.716 & 0.706
\end{tabular}
\end{center}
\end{table}

\begin{figure}[hbt]
\vspace{0.10in}
\centerline{
\epsfysize=5in
\rotate[r]{
\epsffile{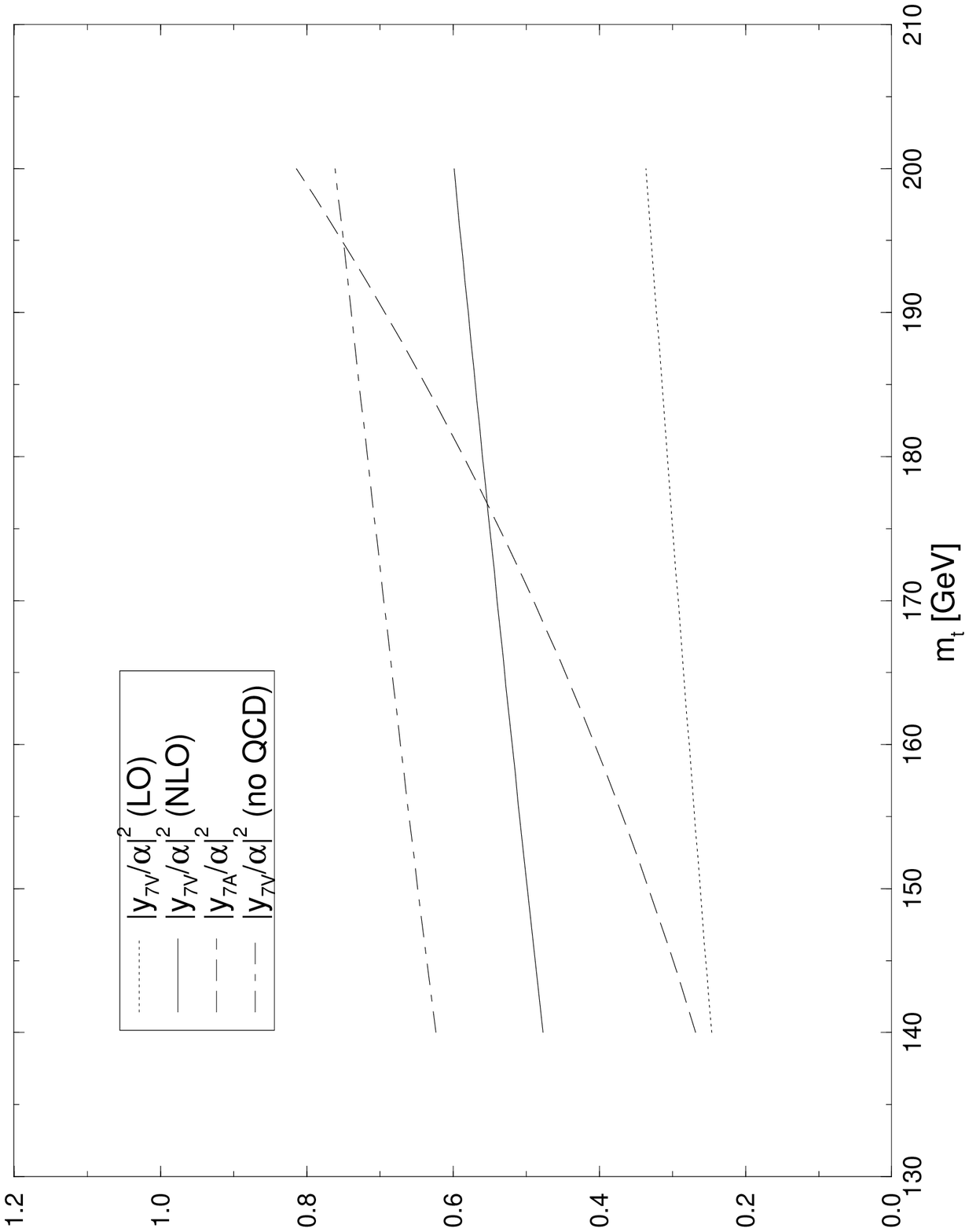}
} }
\vspace{0.08in}
\caption[]{
Wilson coefficients $|y_{7V}/\aem|^2$ and $|y_{7A}/\aem|^2$ as a
function of $\mt$ for $\Lms^{(4)}=325\mev$ at scale $\mu=1.0\gev$.
\label{fig:kpiee:mty7VA}}
\end{figure}

In table \ref{tab:kpemuzy7} we show the $\mu$-dependence of
$z_{7V}/\aem$ and $y_{7V}/\aem$.  We find a pronounced scheme and
$\mu$-dependence for $z_{7V}$.  This signals that these dependences
have to be carefully addressed in the calculation of the CP conserving
part in the $K_L \to \pi^0 e^+ e^-$ amplitude. On the other hand, the
scheme and $\mu$-dependences for $y_{7V}$ are below $\ord(1.5\%)$.
\\
Similarly, $z_{7V}$ shows a strong dependence on the choice of the QCD
scale $\Lms$ while this dependence is small or absent for $y_{7V}$ and
$y_{7A}$, respectively.
\\
Finally, as seen from eq.\ \eqn{eq:z7primemc} $z_{7V}$ is independent
of $\mt$. However, with in/decreasing $\mt$ in the range $\mt = (170
\pm 15)\gev$ there is a relative variation of $\ord(\pm 3\%)$ and
$\ord(\pm 14\%)$ for the absolute values of $y_{7V}$ and $y_{7A}$,
respectively.
This is illustrated further in fig.\ \ref{fig:kpiee:mty7VA} and table
\ref{tab:kpey7VAmt} where the $\mt$ dependence of these coefficients is
shown explicitly. Accidentally for $m_t\approx 175\gev$ one finds
$|y_{7V}| \approx |y_{7A}|$.  Most importantly the impact of NLO
corrections is to enhance the Wilson coefficient $y_{7V}$ by roughly
$25\%$. As we will see in section \ref{sec:KLpee} this implies an
enhancement of the direct CP violation in $\Kpiee$.

\begin{table}[htb]
\caption[]{$K_{\rm L} \to \pi^0 e^+ e^-$ Wilson coefficients
$y_{7V}/\aem$ and $y_{7A}/\aem$ for $\mu=1.0\gev$ and various values of
$\mt$.
\label{tab:kpey7VAmt}}
\begin{center}
\begin{tabular}{|c|c|c|c||c|c|c||c|c|c||c|}
& \multicolumn{9}{c||}{$y_{7V}/\aem$} & $y_{7A}/\aem$ \\
\hline
& \multicolumn{3}{c||}{$\Lms^{(4)}=215\mev$} &
  \multicolumn{3}{c||}{$\Lms^{(4)}=325\mev$} &
  \multicolumn{3}{c||}{$\Lms^{(4)}=435\mev$} & \\
\hline
$\mt [\gev]$ & LO & NDR & HV & LO & NDR & HV & LO & NDR & HV & \\
\hline
150 & 0.546 & 0.719 & 0.711 & 0.512 & 
0.706 & 0.697 & 0.481 & 0.692 & 0.681 & --0.576 \\
160 & 0.560 & 0.733 & 0.726 & 0.526 & 
0.721 & 0.711 & 0.495 & 0.706 & 0.696 & --0.637 \\
170 & 0.575 & 0.747 & 0.740 & 0.540 & 
0.735 & 0.725 & 0.509 & 0.720 & 0.710 & --0.700 \\
180 & 0.588 & 0.761 & 0.753 & 0.554 & 
0.748 & 0.739 & 0.523 & 0.734 & 0.723 & --0.765 \\
190 & 0.601 & 0.774 & 0.766 & 0.567 & 
0.761 & 0.752 & 0.536 & 0.747 & 0.736 & --0.833 \\
200 & 0.614 & 0.786 & 0.779 & 0.580 & 
0.774 & 0.764 & 0.549 & 0.760 & 0.749 & --0.902 \\
\end{tabular}
\end{center}
\end{table}

\section{The Effective Hamiltonian for $B\to X_{\lowercase{s}}\gamma$} 
         \label{sec:Heff:BXsgamma}
The effective hamiltonian for $B\to X_s\gamma$ at scales $\mu=O(\mb)$
is given by
\begin{equation}
\Heff(b\to s\gamma) = - \frac{G_F}{\sqrt{2}} V_{ts}^* V^{}_{tb}
\left[ \sum_{i=1}^6 C_i(\mu) Q_i(\mu) + C_{7\gamma}(\mu) Q_{7\gamma}(\mu)
+C_{8G}(\mu) Q_{8G}(\mu) \right]
\label{eq:HeffBXsgamma}
\end{equation}
where in view of $|V_{us}^*V_{ub}^{} / V_{ts}^* V_{tb}^{}| < 0.02$
we have neglected the term proportional to $V_{us}^*V_{ub}^{}$.

\subsection{Operators}
         \label{sec:Heff:BXsgamma:ops}
The complete list of operators is given as follows
\begin{eqnarray}
Q_1    & = & (\bar{s}_{i}  c_{j})_{V-A}
           (\bar{c}_{j}  b_{i})_{V-A}        \nn \\
Q_2    & = & (\bar{s} c)_{V-A}  (\bar{c} b)_{V-A}      \nn \\
Q_3    & = & (\bar{s} b)_{V-A}\sum_q(\bar{q}q)_{V-A}   \nn \\
Q_4    & = & (\bar{s}_{i}  b_{j})_{V-A} \sum_q (\bar{q}_{j}
          q_{i})_{V-A}     \nn \\
Q_5    & = & (\bar{s} b)_{V-A}\sum_q(\bar{q}q)_{V+A}
             \label{eq:BXsgamma:ops} \\
Q_6    & = & (\bar{s}_{i}  b_{j})_{V-A}
   \sum_q  (\bar{q}_{j}  q_{i})_{V+A}        \nn \\
Q_{7\gamma}    & = & \frac{e}{8\pi^2} \mb \bar{s}_i \sigma^{\mu\nu}
          (1+\gamma_5) b_i F_{\mu\nu}             \nn \\
Q_{8G}    & = & \frac{g}{8\pi^2} \mb \bar{s}_i \sigma^{\mu\nu}
   (1+\gamma_5)T^a_{ij} b_j G^a_{\mu\nu}  \nn
\end{eqnarray}
The current-current operators $Q_{1,2}$ and the QCD penguin operators
$Q_3,\ldots,Q_6$ have been already contained in the usual $\Delta B=1$
hamiltonian presented in section \ref{sec:HeffdF1:66:dB1}. The new
operators $Q_{7\gamma}$ and $Q_{8G}$ specific for $b\to s\gamma$ and
$b\to s g$ transitions carry the name of magnetic penguin operators.
They originate from the mass-insertion on the external b-quark line in
the QCD and QED penguin diagrams of fig.\ \ref{fig:oporig}\,(d),
respectively. In view of $\ms \ll \mb$ we do not include the corresponding
contributions from mass-insertions on the external s-quark line.

\subsection{Wilson Coefficients}
         \label{sec:Heff:BXsgamma:wc}
A very peculiar feature of the renormalization group analysis of the
set of operators in (\ref{eq:BXsgamma:ops}) is that the mixing under
(infinite) renormalization between the set $Q_1,\ldots,Q_6$ and the
operators $Q_{7\gamma},Q_{8G}$ vanishes at the one-loop level.
Consequently in order to calculate the coefficients $C_{7\gamma}(\mu)$
and $C_{8G}(\mu)$ in the leading logarithmic approximation, two-loop
calculations of ${\cal{O}}(e g^2_s)$ and ${\cal{O}}(g^3_s)$ are
necessary. The corresponding NLO analysis requires the evaluation of
the mixing in question at the three-loop level. In view of this new
feature it is useful to include additional couplings in the definition
of $Q_{7\gamma}$ and $Q_{8G}$ as done in (\ref{eq:BXsgamma:ops}).  In
this manner the entries in the anomalous dimension matrix representing
the mixing between $Q_1,\ldots,Q_6$ and $Q_{7\gamma},Q_{8G}$ at the
two-loop level are $O(g_s^2)$ and enter the anomalous dimension matrix
$\gamma^{(0)}_s$. Correspondingly the three-loop mixing between these
two sets of operators enters the matrix $\gamma^{(1)}_s$.

The mixing under renormalization in the sector $Q_{7\gamma},Q_{8G}$
proceeds in the usual manner and the corresponding entries in
$\gamma^{(0)}_s $ and $\gamma^{(1)}_s$ result from one-loop and 
two-loop calculations, respectively.

At present, the coefficients $C_{7\gamma}$ and $C_{8G}$ are only known
in the leading logarithmic approximation. Consequently we are in the
position to give here only their values in this approximation.
The work on NLO corrections to $C_{7\gamma}$ and $C_{8G}$ is in
progress and we will summarize below what is already known about
these corrections.

The peculiar feature of this decay mentioned above caused that the
first fully correct calculation of the leading  anomalous dimension
matrix has been obtained only in 1993 \cite{CFMRS:93}, \cite{CFRS:94}.
It is instructive to clarify this right at the beginning. We follow
here \cite{BMMP:94}.

The point is that the mixing between the sets $Q_1,\ldots,Q_6$ and
$Q_{7\gamma},Q_{8G}$ in $\gamma^{(0)}_s$ resulting from two-loop
diagrams is generally regularization scheme dependent. This is
certainly disturbing because the matrix $\gamma^{(0)}_s$, being the
first term in the expansion for $\gamma_s$, is usually scheme
independent.  There is a simple way to circumvent this difficulty
\cite{BMMP:94}.

As noticed in \cite{CFMRS:93}, \cite{CFRS:94} the regularization scheme
dependence of $\gamma^{(0)}_s$ in the case of $b\to s\gamma$ and
$b\to s g$ is signaled in the one-loop matrix elements of $Q_1,\ldots,Q_6$
for on-shell photons or gluons.  They vanish in any 4-dimensional
regularization scheme and in the HV scheme but some of them are
non-zero in the NDR scheme.  One has
\begin{equation}
\langle Q_i \rangle_{\rm one-loop}^\gamma =
y_i \, \langle Q_{7\gamma} \rangle_{\rm tree},
\qquad i=1,\ldots,6
\label{eq:defy}
\end{equation}
and
\begin{equation}
\langle Q_i\rangle_{\rm one-loop}^G =
z_i \, \langle Q_{8G} \rangle_{\rm tree},
\qquad i=1,\ldots,6.
\end{equation}

In the HV scheme all the $y_i$'s and $z_i$'s vanish, while in the NDR
scheme $\vec{y} = (0,0,0,0,-\frac{1}{3},-1)$ and $\vec{z} =
(0,0,0,0,1,0)$.  This regularization scheme dependence is canceled by a
corresponding regularization scheme dependence in $\gamma_s^{(0)}$
as first demonstrated in \cite{CFMRS:93}, \cite{CFRS:94}. It should be
stressed that the numbers $y_i$ and $z_i$ come from divergent, i.e.
purely short-distance parts of the one-loop integrals. So no reference
to the spectator-model or to any other model for the matrix elements is
necessary here.

In view of all this  it is convenient in the leading order to introduce
the so-called ``effective coefficients'' \cite{BMMP:94} for the
operators $Q_{7\gamma}$ and $Q_{8G}$ which are regularization scheme
independent. They are given as follows:
\begin{equation} \label{eq:defc7eff}
C^{(0)eff}_{7\gamma}(\mu) =
C^{(0)}_{7\gamma}(\mu) + \sum_{i=1}^6 y_i C^{(0)}_i(\mu).
\end{equation}
and 
\begin{equation}
C^{(0)eff}_{8G}(\mu) = C^{(0)}_{8G}(\mu) + \sum_{i=1}^6 z_i C^{(0)}_i(\mu)
\end{equation}
One can then introduce a scheme-independent vector
\begin{equation} 
\vec{C}^{(0)eff}(\mu) = \left( C^{(0)}_1(\mu),\ldots, C^{(0)}_6(\mu), 
C^{(0)eff}_{7\gamma}(\mu),C^{(0)eff}_{8G}(\mu) \right) \, .
\end{equation}
From the RGE for $\vec{C}^{(0)}(\mu)$ it is straightforward
to derive the RGE for $\vec{C}^{(0)eff}(\mu)$. It has the form
\begin{equation} \label{RGEeff}
\mu \frac{d}{d \mu} C^{(0)eff}_i(\mu) = 
\frac{\as}{4\pi} \gamma^{(0)eff}_{ji} C^{(0)eff}_j(\mu)
\end{equation}
where
\begin{equation} \label{def.geff}
\gamma^{(0)eff}_{ji} = \left\{ \begin{array}{ccl}
\gamma^{(0)}_{j7} +
\sum_{k=1}^6 y_k\gamma^{(0)}_{jk} -y_j\gamma^{(0)}_{77} -z_j\gamma^{(0)}_{87}
&\quad& $i=7$,\ $j=1,\ldots,6$ \\
\gamma^{(0)}_{j8} +
\sum_{k=1}^6 z_k\gamma^{(0)}_{jk} -z_j\gamma^{(0)}_{88}
&\quad& $i=8$,\ $j=1,\ldots,6$ \\
\gamma^{(0)}_{ji} &\quad& \mbox{otherwise.}
\end{array}
\right.
\end{equation}
The matrix $\gamma^{(0)eff}$ is a scheme-independent quantity.
It equals the matrix which one would directly obtain from two-loop
diagrams in the HV scheme.  In order to simplify the notation we will
omit the label ``eff'' in the expressions for the elements of this
effective one loop anomalous dimension matrix given below and keep it
only in the Wilson coefficients of the operators $Q_{7\gamma}$ and
$Q_{8G}$.

This discussion clarifies why it took so long to find the correct
leading anomalous dimension matrix for the $\bsg$ decay
\cite{CFMRS:93}, \cite{CFRS:94}. All previous calculations \cite{Grin},
\cite{cella:90a}, \cite{misiak:93}, \cite{Yao1}, \cite{Yao2} of the
leading order QCD corrections to $\bsg$ used the NDR scheme setting
unfortunately $z_i$ and $y_i$ to zero, or not including all operators
or making other mistakes.
The discrepancy between the calculation of \cite{grigjanis:88} (DRED
scheme) and \cite{Grin} (NDR scheme) has been clarified by
\cite{misiak:94b}.

\subsection{Renormalization Group Evolution and Anomalous Dimension Matrices}
         \label{sec:Heff:BXsgamma:RGE}
The coefficients $C_i(\mu)$ in (\ref{eq:HeffBXsgamma}) can be calculated
by using
\begin{equation}
\vec C(\mu)= U_5(\mu,\mw)\vec C(\mw)
\end{equation}
Here $ U_5(\mu,\mw)$ is the $8\times 8$ evolution matrix which is
given in general terms in \eqn{u0jj} with $\gamma$ being this
time an $8\times 8$ anomalous dimension matrix. In the leading order
$U_5(\mu,\mw)$ is to be replaced by $U_5^{(0)}(\mu,\mw)$ and
the initial conditions by $\vec C^{(0)}(\mw)$ given by \cite{Grin}
\begin{equation}\label{c2}
C^{(0)}_2(\mw) = 1                               
\end{equation}
\begin{equation}\label{c7}
C^{(0)}_{7\gamma} (\mw) = \frac{3 x_t^3-2 x_t^2}{4(x_t-1)^4}\ln x_t + 
   \frac{-8 x_t^3 - 5 x_t^2 + 7 x_t}{24(x_t-1)^3}
   \equiv -\frac{1}{2} D'_0(x_t)
\end{equation}
\begin{equation}\label{c8}
C^{(0)}_{8G}(\mw) = \frac{-3 x_t^2}{4(x_t-1)^4}\ln x_t +
   \frac{-x_t^3 + 5 x_t^2 + 2 x_t}{8(x_t-1)^3}                               
   \equiv -\frac{1}{2} E'_0(x_t)
\end{equation}
with all remaining coefficients being zero at $\mu=\mw$. The functions
$D'_0(x_t)$ and $E'_0(x_t)$ are sometimes used in the literature. The
$6 \times 6$ submatrix of $\gamma^{(0)}_s$ involving the operators
$Q_1,\ldots,Q_6$ is given in \eqn{eq:gs0Kpp}. Here we only give the
remaining non-vanishing entries of $\gamma^{(0)}_s$
\cite{CFMRS:93}, \cite{CFRS:94}.

Denoting for simplicity $\gamma_{ij} \equiv (\gamma_s)_{ij}$, the elements
$\gamma^{(0)}_{i7}$ with $i=1,\ldots,6$ are:
\begin{eqnarray}
\gamma^{(0)}_{17} = 0, &\qquad&  \gamma^{(0)}_{27} =
\frac{104}{27} C_F
\label{eq:g0127} \\
\gamma^{(0)}_{37} = -\frac{116}{27} C_F
 &\qquad&  \gamma^{(0)}_{47}  = \left(\frac{104}{27} u -\frac{58}{27}d
\right) C_F
\label{eq:g0347} \\
\gamma^{(0)}_{57} = \frac{8}{3} C_F &\qquad&
\gamma^{(0)}_{67} = \left( \frac{50}{27}d -\frac{112}{27}u \right) C_F
\label{eq:g0567}
\end{eqnarray}
The elements $\gamma^{(0)}_{i8}$ with $i=1,\ldots,6$ are:
\begin{eqnarray}
\gamma^{(0)}_{18} = 3, &\quad& \gamma^{(0)}_{28} =
\frac{11}{9} N-\frac{29}{9}\frac{1}{N}
\label{eq:g0128} \\
\gamma^{(0)}_{38} = \frac{22}{9} N-\frac{58}{9}\frac{1}{N}+3 f
 &\quad& \gamma^{(0)}_{48}  = 
6+\left(\frac{11}{9} N -\frac{29}{9}\frac{1}{N}\right) f
\label{eq:g0348} \\
\gamma^{(0)}_{58} = -2 N+\frac{4}{N} -3 f  &\quad&
\gamma^{(0)}_{68} = -4-\left( \frac{16}{9} N -
\frac{25}{9}\frac{1}{N}\right) f
\label{eq:g0568}
\end{eqnarray}

Finally the $2\times 2$ one-loop anomalous dimension matrix in the
sector $Q_{7\gamma},Q_{8G}$ is given by \cite{Grin}
\begin{eqnarray}
\gamma^{(0)}_{77} = 8 C_F
&\qquad&
\gamma^{(0)}_{78} = 0
\label{gammaB0} \\
\gamma^{(0)}_{87} = -\frac{8}{3} C_F
&\qquad&
\gamma^{(0)}_{88} = 16 C_F - 4 N
\nn
\end{eqnarray}

As we discussed above, the first correct calculation of the two-loop
mixing between $Q_1,\ldots,Q_6$ and $Q_{7\gamma}$, $Q_{8G}$ has been
presented in \cite{CFMRS:93}, \cite{CFRS:94} and confirmed subsequently
in \cite{CCRV:94a}, \cite{CCRV:94b}, \cite{misiak:94}.  In order to
extend these calculations beyond the leading order one would have to
calculate $\gamma_s^{(1)}$ (see (\ref{gg01})) and $O(\as)$ corrections to
the initial conditions in (\ref{c7}) and (\ref{c8}). We summarize below
the present status of this NLO calculation.

The $6\times 6$ two-loop submatrix of $\gamma^{(1)}_s$ involving
the operators $Q_1,\ldots,Q_6$ is given in eq. \eqn{eq:gs1ndrN3Kpp}.
The two-loop generalization of (\ref{gammaB0}) has been calculated only
last year \cite{misiakmuenz:95}. It is given for both NDR and HV
schemes as follows
\begin{eqnarray}
\gamma^{(1)}_{77} &=& 
   C_F \left(\frac{548}{9} N - 16 C_F - \frac{56}{9} f \right)
\nn \\
\gamma^{(1)}_{78} &=& 0
\label{gammaB1} \\
\gamma^{(1)}_{87} &=& 
   C_F \left(-\frac{404}{27} N +\frac{32}{3} C_F +\frac{56}{27} f \right)
\nn \\
\gamma^{(1)}_{88} &=& -\frac{458}{9} -\frac{12}{N^2}+ \frac{214}{9} N^2 +
   \frac{56}{9} \frac{f}{N} - \frac{13}{9} f N
\nn
\end{eqnarray}

The generalization of \eqn{eq:g0127}--\eqn{eq:g0568} to next-to-leading
order requires three loop calculations which have not been done yet.
The $O(\as)$ corrections to $C_{7\gamma}(\mw)$ and $C_{8G}(\mw)$ have
been considered in \cite{Yao1}.

\subsection{Results for the Wilson Coefficients}
         \label{sec:Heff:BXsgamma:wcres}
The leading order results for the Wilson Coefficients of all operators
entering the effective hamiltonian in (\ref{eq:HeffBXsgamma}) can be written
in an analytic form. They are \cite{BMMP:94}
\begin{eqnarray}
\label{coeffs}
C_j^{(0)}(\mu)    & = & \sum_{i=1}^8 k_{ji} \eta^{a_i}
  \qquad (j=1,\ldots,6)  \\
\label{C7eff}
C_{7\gamma}^{(0)eff}(\mu) & = & 
\eta^\frac{16}{23} C_{7\gamma}^{(0)}(\mw) + \frac{8}{3}
   \left(\eta^\frac{14}{23} - \eta^\frac{16}{23}\right) C_{8G}^{(0)}(\mw) +
    C_2^{(0)}(\mw)\sum_{i=1}^8 h_i \eta^{a_i},
\label{C7Geff}
\\
C_{8G}^{(0)eff}(\mu) & = & 
\eta^\frac{14}{23} C_{8G}^{(0)}(\mw) 
   + C_2^{(0)}(\mw) \sum_{i=1}^8 \bar h_i \eta^{a_i},
\end{eqnarray}
with
\begin{eqnarray}
\eta & = & \frac{\as(\mw)}{\as(\mu)}, 
\end{eqnarray}
and $C_{7\gamma}^{(0)}(\mw)$
and $ C_{8G}^{(0)}(\mw)$ given in (\ref{c7}) and (\ref{c8}),
respectively. The numbers $a_i$, $k_{ji}$, $h_i$ and $\bar h_i$ are
given in table \ref{tab:akh}.

\begin{table}[htb]
\caption[]{
\label{tab:akh}}
\begin{center}
\begin{tabular}{|r|r|r|r|r|r|r|r|r|}
$i$ & 1 & 2 & 3 & 4 & 5 & 6 & 7 & 8 \\
\hline
$a_i $&$ \frac{14}{23} $&$ \frac{16}{23} $&$ \frac{6}{23} $&$
-\frac{12}{23} $&$
0.4086 $&$ -0.4230 $&$ -0.8994 $&$ 0.1456 $\\
$k_{1i} $&$ 0 $&$ 0 $&$ \frac{1}{2} $&$ - \frac{1}{2} $&$
0 $&$ 0 $&$ 0 $&$ 0 $\\
$k_{2i} $&$ 0 $&$ 0 $&$ \frac{1}{2} $&$  \frac{1}{2} $&$
0 $&$ 0 $&$ 0 $&$ 0 $\\
$k_{3i} $&$ 0 $&$ 0 $&$ - \frac{1}{14} $&$  \frac{1}{6} $&$
0.0510 $&$ - 0.1403 $&$ - 0.0113 $&$ 0.0054 $\\
$k_{4i} $&$ 0 $&$ 0 $&$ - \frac{1}{14} $&$  - \frac{1}{6} $&$
0.0984 $&$ 0.1214 $&$ 0.0156 $&$ 0.0026 $\\
$k_{5i} $&$ 0 $&$ 0 $&$ 0 $&$  0 $&$
- 0.0397 $&$ 0.0117 $&$ - 0.0025 $&$ 0.0304 $\\
$k_{6i} $&$ 0 $&$ 0 $&$ 0 $&$  0 $&$
0.0335 $&$ 0.0239 $&$ - 0.0462 $&$ -0.0112 $\\
$h_i $&$ 2.2996 $&$ - 1.0880 $&$ - \frac{3}{7} $&$ -
\frac{1}{14} $&$ -0.6494 $&$ -0.0380 $&$ -0.0185 $&$ -0.0057 $\\
$\bar h_i $&$ 0.8623 $&$ 0 $&$ 0 $&$ 0
 $&$ -0.9135 $&$ 0.0873 $&$ -0.0571 $&$ 0.0209 $\\
\end{tabular}
\end{center}
\end{table}

\subsection{Numerical Analysis}
         \label{sec:Heff:BXsgamma:num}
The decay $B \to X_s \gamma$ is the only decay in our review for which
the complete NLO corrections are not available. In presenting the
numerical values for the Wilson coefficients a few remarks on the choice
of $\as$ should therefore be made. In the leading order the leading
order expression for $\as$ should be used. The question then is what to
use for $\Lambda_{\rm QCD}$ in this expression. In other decays for
which NLO corrections were available this was not important because LO
results were secondary. We have therefore simply inserted our standard
$\Lms$ values into the LO formula for $\as$. This procedure gives
$\as^{(5)}(\mz)=0.126, 0.136, 0.144$ for $\Lms^{(5)}=140\mev,
225\mev, 310\mev$, respectively. In view of these high values of
$\as^{(5)}(\mz)$ we will proceed here differently. Following
\cite{BMMP:94} we will use $\as^{(5)}(\mz)=0.110, 0.117, 0.124$ as
in our NLO calculations , but we will evolve $\as(\mu)$ to $\mu \approx
\ord(\mb)$ using the leading order expressions. In short, we will use
\begin{equation} 
\as(\mu) = \frac{\as(\mz)}{1 - \beta_0 \as(\mz)/2\pi \, \ln(\mz/\mu)} \, .
\label{eq:asmumz}
\end{equation} 
This discussion shows again the importance of the complete NLO
calculation for this decay.

Before starting the discussion of the numerical values for the
coefficients $C^{(0)eff}_{7\gamma}$ and $C^{(0)eff}_{8G}$, let us
illustrate the relative numerical importance of the three terms in
expression (\ref{C7eff}) for $C^{(0)eff}_{7\gamma}$. 

For instance, for $\mt = 170\gev$, $\mu = 5\gev$ and $\as^{(5)}(\mz)
=0.117$ one obtains
\begin{eqnarray}
C^{(0)eff}_{7\gamma}(\mu) &=&
0.698 \; C^{(0)}_{7\gamma}(\mw) +
0.086 \; C^{(0)}_{8G}(\mw) - 0.156 \; C^{(0)}_2(\mw)
\nn\\
 &=& 0.698 \; (-0.193) + 0.086 \; (-0.096) - 0.156 = -0.299 \, .
\label{eq:C7geffnum}
\end{eqnarray}

In the absence of QCD we would have $C^{(0)eff}_{7\gamma}(\mu) =
C^{(0)}_{7\gamma}(\mw)$ (in that case one has $\eta = 1$). Therefore, the
dominant term in the above expression (the one proportional to
$C^{(0)}_2(\mw)$) is the additive QCD correction that causes the
enormous QCD enhancement of the $\bsg$ rate \cite{Bert}, \cite{Desh}.
It originates solely from the two-loop diagrams. On the other hand, the
multiplicative QCD correction (the factor 0.698 above) tends to
suppress the rate, but fails in the competition with the additive
contributions.

In the case of $C^{(0)eff}_{8G}$ a similar enhancement is observed
\begin{eqnarray}
C^{(0)eff}_{8G}(\mu) &=&
0.730 \; C^{(0)}_{8G}(\mw) - 0.073 \; C^{(0)}_2(\mw)
\nn \\
 &=& 0.730 \; (-0.096) - 0.073 = -0.143 \, .
\label{eq:C8Geffnum}
\end{eqnarray}

In table \ref{tab:c78effnum} we give the values of
$C^{(0)eff}_{7\gamma}$ and $C^{(0)eff}_{8G}$ for different values of
$\mu$ and $\as^{(5)}(\mz)$. To this end \eqn{eq:asmumz} has been used.
A strong $\mu$-dependence of both coefficients is observed.  We will
return to this dependence in section \ref{sec:Heff:Bsgamma}.

\begin{table}[htb]
\caption[]{Wilson coefficients $C^{(0)eff}_{7\gamma}$ and $C^{(0)eff}_{8G}$
for $\mt = 170 \gev$ and various values of $\as^{(5)}(\mz)$ and $\mu$.
\label{tab:c78effnum}}
\begin{center}
\begin{tabular}{|c||c|c||c|c||c|c|}
& \multicolumn{2}{c||}{$\as^{(5)}(\mz) = 0.110$} &
  \multicolumn{2}{c||}{$\as^{(5)}(\mz) = 0.117$} &
  \multicolumn{2}{c| }{$\as^{(5)}(\mz) = 0.124$} \\
\hline
$\mu [\gev]$ & 
$C^{(0)eff}_{7\gamma}$ & $C^{(0)eff}_{8G}$ &
$C^{(0)eff}_{7\gamma}$ & $C^{(0)eff}_{8G}$ &
$C^{(0)eff}_{7\gamma}$ & $C^{(0)eff}_{8G}$ \\
\hline
 2.5 & --0.323 & --0.153 & --0.334 & --0.157 & --0.346 & --0.162 \\
 5.0 & --0.291 & --0.140 & --0.299 & --0.143 & --0.307 & --0.147 \\
 7.5 & --0.275 & --0.133 & --0.281 & --0.136 & --0.287 & --0.139 \\
10.0 & --0.263 & --0.129 & --0.268 & --0.131 & --0.274 & --0.133
\end{tabular}
\end{center}
\end{table}

\section{The Effective Hamiltonian for $B\to X_{\lowercase{s}}
         \lowercase{e}^+\lowercase{e}^-$}
         \label{sec:Heff:BXsee}
The effective hamiltonian for $B\to X_s e^+e^-$ at scales $\mu=O(\mb)$
is given by
\begin{eqnarray} \label{eq:Heff2atmu}
\Heff(b\to s e^+e^-) &=&
\Heff(b\to s\gamma)  - \\
& & \frac{G_F}{\sqrt{2}} V_{ts}^* V_{tb}^{} \left[ C_{9V}(\mu) Q_{9V}(\mu)+
C_{10A}(\mu) Q_{10A}(\mu) \right]
\nn
\end{eqnarray}
where again we have neglected the term proportional to $V_{us}^*V_{ub}^{}$
and $\Heff(b\to s\gamma)$ is given in (\ref{eq:HeffBXsgamma}).

\subsection{Operators}
         \label{sec:Heff:BXsee:ops}
In addition to the operators relevant for $B\to X_s\gamma$,
there are two new operators
\begin{equation}\label{Q9V}
Q_{9V}    = (\bar{s} b)_{V-A}  (\bar{e}e)_V         
\qquad
Q_{10A}  =  (\bar{s} b)_{V-A}  (\bar{e}e)_A
\end{equation}
where $V$ and $A$ refer to $\gamma_{\mu}$ and $ \gamma_{\mu}\gamma_5$,
respectively.

They originate in the $Z^0$- and $\gamma$-penguin diagrams
with external $\bar{e}e$ of fig.\ \ref{fig:oporig}\,(f) and from the
corresponding box diagrams.
\subsection{Wilson Coefficients}
         \label{sec:Heff:BXsee:wc}
The coefficient $C_{10A}(\mu)$ is given by
\begin{equation} \label{C10}
C_{10A}(\mw) =  \frac{\aem}{2\pi} \Ctilde_{10}(\mw), \qquad
\Ctilde_{10}(\mw) = - \frac{Y_0(x_t)}{\sin^2\Theta_W}
\end{equation}
with $Y_0(x)$ given in \eqn{eq:yz0}. Since $Q_{10A}$ does not renormalize
under QCD, its coefficient does not depend on $\mu\approx {\cal
O}(\mb)$. The only renormalization scale dependence in (\ref{C10})
enters through the definition of the top quark mass. We will return to
this issue in section \ref{sec:Heff:BXsee:nlo:num}.

The coefficient $C_{9V}(\mu)$ has been calculated over the last years
with increasing precision by several groups \cite{grinstein:89a},
\cite{GDSN:89}, \cite{cellaetal:91}, \cite{misiak:93} culminating in two
complete next-to-leading QCD calculations \cite{misiak:94},
\cite{burasmuenz:95} which agree with each other.

In order to calculate the coefficient $C_{9V}$ including
next-to-leading order corrections we have to perform in principle a
two-loop renormalization group analysis for the full set of operators
contributing to (\ref{eq:Heff2atmu}). However, $Q_{10A}$ is not
renormalized and the dimension five operators $Q_{7\gamma}$ and
$Q_{8G}$ have no impact on $C_{9V}$. Consequently only a set of seven
operators, $Q_1,\ldots,Q_6$ and $Q_{9V}$, has to be considered. This is
precisely the case of the decay $\kpiee$ discussed in \cite{burasetal:94a}
and in section \ref{sec:HeffKpe}, except for an appropriate change of quark
flavours and the fact that now $\mu\approx {\cal O}(\mb)$ instead of
$\mu\approx {\cal O}(1\gev)$. Since the NLO analysis of $\kpiee$ has
already been presented in section \ref{sec:HeffKpe} we will only give the
final result for $C_{9V}(\mu)$. Because of the one step evolution from
$\mu=\mw$ down to $\mu=\mb$ without any thresholds in between it is
possible to find an analytic formula for $C_{9V}(\mu)$. Defining
$\tilde C_{9}$ by
\begin{equation} \label{C9}
C_{9V}(\mu) = \frac{\aem}{2\pi} \Ctilde_9(\mu) 
\end{equation}
one finds \cite{burasmuenz:95} in the NDR scheme
\begin{equation}\label{C9tilde}
\Ctilde_9^{NDR}(\mu)  =  
P_0^{NDR} + \frac{Y_0(x_t)}{\sin^2\Theta_W} -4 Z_0(x_t) +
P_E E_0(x_t)
\end{equation}
with
\begin{eqnarray}
\label{P0NDR}
P_0^{NDR} & = & \frac{\pi}{\as(\mw)} (-0.1875+ \sum_{i=1}^8 p_i
\eta^{a_i+1}) \nn \\ 
          &   & + 1.2468 +  \sum_{i=1}^8 \eta^{a_i} \lbrack
r^{NDR}_i+s_i \eta \rbrack \\ 
\label{PE}
P_E & = & 0.1405 +\sum_{i=1}^8 q_i\eta^{a_i+1}  \, .
\end{eqnarray}
The functions $Y_0(x)$ and $Z_0(x)$ are defined by
\begin{equation}
Y_0(x) = C_0(x) - B_0(x)
\qquad
Z_0(x) = C_0(x) + \frac{1}{4} D_0(x)
\label{eq:yz0}
\end{equation}
with $B_0(x)$, $C_0(x)$ and $D_0(x)$ given in \eqn{eq:Bxt},
\eqn{eq:Cxt} and \eqn{eq:Dxt}, respectively. $E_0(x)$ is given in
eq.\ \eqn{eq:Ext}. The powers $a_i$ are the same as in table
\ref{tab:akh}.  The coefficients $p_i$, $r^{NDR}_i$, $s_i$, and $q_i$
can be found in table \ref{tab:prsq}.  $P_E$ is ${\cal O}(10^{-2})$ and
consequently the last term in (\ref{C9tilde}) can be neglected. We keep
it however in our numerical analysis. These results agree with
\cite{misiak:94}.

\begin{table}[htb]
\caption[]{
\label{tab:prsq}}
\begin{center}
\begin{tabular}{|r|r|r|r|r|r|r|r|r|}
$i$ & 1 & 2 & 3 & 4 & 5 & 6 & 7 & 8 \\
\hline
$p_i $&$ 0, $&$ 0, $&$ -\frac{80}{203}, $&$  \frac{8}{33}, $&$
0.0433 $&$  0.1384 $&$ 0.1648 $&$ - 0.0073 $\\
$r^{NDR}_{i} $&$ 0 $&$ 0 $&$ 0.8966 $&$ - 0.1960 $&$
- 0.2011 $&$ 0.1328 $&$ - 0.0292 $&$ - 0.1858 $\\
$s_i $&$ 0 $&$ 0 $&$ - 0.2009 $&$  -0.3579 $&$
0.0490 $&$ - 0.3616 $&$ -0.3554 $&$ 0.0072 $\\
$q_i $&$ 0 $&$ 0 $&$ 0 $&$  0 $&$
0.0318 $&$ 0.0918 $&$ - 0.2700 $&$ 0.0059 $\\
\svs
$r^{HV}_{i} $&$ 0 $&$ 0 $&$ -0.1193 $&$ 0.1003 $&$
- 0.0473 $&$ 0.2323 $&$ - 0.0133 $&$ - 0.1799 $
\end{tabular}
\end{center}
\end{table}

In the HV scheme only the coefficients $r_i$ are changed. 
They are given on the last line of table \ref{tab:prsq}.
Equivalently we can write
\begin{equation} \label{P0HV}
P_0^{k} = P_0^{NDR} + \xi_{k} \frac{4}{9} \left( 3 C_1^{(0)} +
C_2^{(0)} - C_3^{(0)} -3 C_4^{(0)} \right)
\end{equation}
with
\begin{equation} \label{xi}
\xi_k = \left\{
\begin{array}{rl}
0  &\quad k=\mbox{NDR} \\
-1 &\quad k=\mbox{HV}
\end{array} \, .
\right.
\end{equation}
We note that
\begin{eqnarray}
\label{sums1}
\sum_{i=1}^8 p_i = 0.1875, &\quad& \sum_{i=1}^8 q_i = -
0.1405, \\
\label{sums2}
\sum_{i=1}^8 (r_i^k + s_i) = - 1.2468 + \frac{4}{9} (1 +
\xi_k), &\quad& \sum_{i=1}^8 p_i (a_i + 1) = - \frac{16}{69}.
\end{eqnarray}
In this way for $\eta=1$ one finds $P_E=0$, $P_0^{NDR} = 4/9$ and
$P_0^{HV} = 0$ in accordance with the initial conditions at $\mu=\mw$.
 Moreover, the second relation in (\ref{sums2})
assures the correct large logarithm in $P_0^{NDR}$, i.e.\ $8/9\, \ln
(\mw/\mu)$. 

The special feature of $C_{9V}(\mu)$ compared to the coefficients
of the remaining operators contributing to $B\to X_s e^+e^-$ is the
large logarithm represented by $1/\as$ in $P_0$ in
(\ref{P0NDR}). Consequently the renormalization group improved
perturbation theory for $C_{9V}$ has the structure $ {\cal O}(1/\as) +
{\cal O}(1) + {\cal O}(\as)+ \ldots$, whereas the corresponding series
for the remaining coefficients is $ {\cal O}(1) + {\cal O}(\as)+
\ldots$\,. Therefore in order to find the next-to-leading ${\cal O}(1)$
term in the branching ratio for $B\to X_s e^+e^-$, the full two-loop
renormalization group analysis has to be performed in order to find
$C_{9V}$, but the coefficients of the remaining operators should be
taken in the leading logarithmic approximation. This is gratifying
because the coefficient of the magnetic operator $Q_{7\gamma}$ is known
only in the leading logarithmic approximation.

\subsection{Numerical Results}
         \label{sec:Heff:BXsee:num}
In our numerical analysis we will use the two-loop expression for
$\as$ and the parameters collected in the appendix. Our
presentation follows closely the one given in \cite{burasmuenz:95}. 

In table \ref{tab:p0C9} we show the constant $P_0$ in (\ref{P0NDR}) for
different $\mu$ and $\Lms$, in the leading order corresponding to the
first term in (\ref{P0NDR}) and for the NDR and HV schemes as given by
(\ref{P0NDR}) and (\ref{P0HV}), respectively. In table \ref{tab:BXsee:C9} we
show the corresponding values for $\Ctilde_9(\mu)$. To this end
we set $\mt= 170 \gev$. 

\begin{table}[htb]
\caption[]{The coefficient $P_0$ of $\widetilde C_9$ for various values
of $\Lms^{(5)}$ and $\mu$.
\label{tab:p0C9}}
\begin{center}
\begin{tabular}{|c||c|c|c||c|c|c||c|c|c|}
& \multicolumn{3}{c||}{$\Lms^{(5)} = 140 \mev$} &
  \multicolumn{3}{c||}{$\Lms^{(5)} = 225 \mev$} &
  \multicolumn{3}{c| }{$\Lms^{(5)} = 310 \mev$} \\
\hline
$\mu [\gev]$ & LO & NDR & HV & LO & NDR & HV & LO & NDR & HV \\
\hline
2.5 & 2.053 & 2.928 & 2.797 & 1.933 & 2.846 & 2.759 & 1.835 & 2.775 &
2.727 \\
5.0 & 1.852 & 2.625 & 2.404 & 1.788 & 2.591 & 2.395 & 1.736 & 2.562 &
2.388 \\
7.5 & 1.675 & 2.391 & 2.127 & 1.632 & 2.373 & 2.127 & 1.597 & 2.358 &
2.128 \\
10.0 & 1.526 & 2.204 & 1.912 & 1.494 & 2.194 & 1.917 & 1.469 & 2.185 &
1.921
\end{tabular}
\end{center}
\end{table}

\begin{table}[htb]
\caption[]{Wilson coefficient $\widetilde C_9$ for $\mt = 170 \gev$ and
various values of $\Lms^{(5)}$ and $\mu$.
\label{tab:BXsee:C9}}
\begin{center}
\begin{tabular}{|c||c|c|c||c|c|c||c|c|c|}
& \multicolumn{3}{c||}{$\Lms^{(5)} = 140 \mev$} &
  \multicolumn{3}{c||}{$\Lms^{(5)} = 225 \mev$} &
  \multicolumn{3}{c| }{$\Lms^{(5)} = 310 \mev$} \\
\hline
$\mu [\gev]$ & LO & NDR & HV & LO & NDR & HV & LO & NDR & HV \\
\hline
2.5 & 2.053 & 4.493 & 4.361 & 1.933 & 4.410 & 4.323 & 1.835 & 4.338 &
4.290 \\
5.0 & 1.852 & 4.191 & 3.970 & 1.788 & 4.156 & 3.961 & 1.736 & 4.127 &
3.954 \\
7.5 & 1.675 & 3.958 & 3.694 & 1.632 & 3.940 & 3.694 & 1.597 & 3.924 &
3.695 \\
10.0 & 1.526 & 3.772 & 3.480 & 1.494 & 3.761 & 3.485 & 1.469 & 3.752 &
3.488
\end{tabular}
\end{center}
\end{table}

\noindent
We observe:
\begin{itemize}
\item
The NLO corrections to $P_0$ enhance this constant relatively to the
LO result by roughly 45\% and 35\% in the NDR and HV schemes,
respectively. This enhancement is analogous to the one found in the
case of $\kpiee$.
\item
In calculating $P_0$ in the LO we have used $\as(\mu)$ at one-loop
level. Had we used the two-loop expression for $\as(\mu)$ we
would find for $\mu=5 \gev$ and $\Lms^{(5)} = 225 \mev$ the value $P_0^{LO}
\approx 1.98$. Consequently the NLO corrections would have smaller
impact. Ref.~\cite{grinstein:89a} including the next-to-leading term $4/9$
would find $P_0$ values roughly 20\% smaller than $P_0^{NDR}$ given in
table \ref{tab:p0C9}.
\item
It is tempting to compare $P_0$ in table \ref{tab:p0C9} with that found in
the absence of QCD corrections. In the limit $\as \to 0$ we
find $P_0^{NDR} = 8/9 \, \ln(\mw/\mu) + 4/9$ and $P_0^{HV} = 8/9\,
\ln(\mw/\mu)$ which for $\mu = 5 \gev$ give $P_0^{NDR} = 2.91$ and
$P_0^{HV} = 2.46$. Comparing these values with table~\ref{tab:p0C9} we
conclude that the QCD suppression of $P_0$ present in the leading order
approximation is considerably weakened in the NDR treatment of
$\gamma_5$ after the inclusion of NLO corrections. It is essentially
removed for $\mu > 5 \gev$ in the HV scheme.
\item
The NLO corrections to $\Ctilde_9$ which include also the
$\mt$-dependent contributions are large as seen in table
\ref{tab:BXsee:C9}. The results in HV and NDR schemes are by more than
a factor of two larger than the leading order result $\Ctilde_9 =
P_0^{LO}$ which consistently should not include $\mt$-contributions.
This demonstrates very clearly the necessity of NLO calculations which
allow a consistent inclusion of the important $\mt$-contributions. For
the same set of parameters the authors of \cite{grinstein:89a}
would find $\Ctilde_9$ to be smaller than $\Ctilde_9^{NDR}$ by
10--15\%.
\item
The $\Lms$ dependence of $\Ctilde_9$ is rather weak.  On the other
hand its $\mu$ dependence is sizable ($\sim 15\%$ in the range of $\mu$
considered) although smaller than in the case of the coefficients
$C_{7\gamma}$ and $C_{8G}$ given in table \ref{tab:c78effnum}.  We also
find that the $\mt$ dependence of $\Ctilde_9$ is rather weak. Varying
$\mt$ between $150 \gev$ and $190 \gev$ changes $\Ctilde_9$ by at most
10\%. This weak $\mt$ dependence of $\Ctilde_9$ originates in the
partial cancelation of $\mt$ dependences between $Y_0(x_t)$ and
$Z_0(x_t)$ in (\ref{C9tilde}) as already seen in the case of $\kpiee$
in fig.\ \ref{fig:kpiee:mty7VA}. Finally, the difference between
$\Ctilde_9^{NDR}$ and $\Ctilde_9^{HV}$ is small and amounts to roughly 5\%.
\item
The dominant $\mt$-dependence in this decay originates in the $\mt$
dependence of $\tilde C_{10}(\mw)$. In fact, as can be seen in section
\ref{sec:HeffKpe} $\tilde C_{10}(\mw)=2\pi y_{7A}/\aem$ with $y_{7A}$
present in $K_L \to \pi^0 e^+ e^-$. The $\mt$ dependence of $y_{7A}$ is
shown in fig.\ \ref{fig:kpiee:mty7VA}.
\end{itemize}

\section{Effective Hamiltonians for Rare $K$- and $B$-Decays}
         \label{sec:HeffRareKB}
\subsection{Overview}
            \label{sec:HeffRareKB:overview}
In the present section we will summarize the effective hamiltonians
valid at next-to-leading logarithmic accuracy in QCD, which describe
the semileptonic rare Flavour Changing Neutral Current (FCNC)
transitions $\kpn$, $(\klm)_{SD}$, $K_L\to\pi^0\nu\bar\nu$, $B\to X_{s,
d}\nu\bar\nu$ and $B\to l^+l^-$.  These decay modes all are very
similar in their structure and it is natural to discuss them together.
On the other hand they differ from the decays $\Kpipi$, $K \to \pi e^+
e^-$, $B \to X_s \gamma$ and $B \to X_s e^+ e^-$ discussed in previous
sections. Before giving the detailed formulae, it will be useful to
recall the most important general features of this class of processes
first. In addition, characteristic differences between the specific
modes will also become apparent from our presentation.
\begin{itemize}
\item
Within the Standard Model all the decays listed above are loop-induced
semileptonic FCNC processes determined by $Z^0$-penguin and box diagrams
(fig.~\ref{fig:1loopful}\,(d) and (e)).\\
In particular, a distinguishing feature of the present class of decays
is the absence of a photon penguin contribution. For the decay modes
with neutrinos in the final state this is obvious, since the photon
does not couple to neutrinos. For the mesons decaying into a charged
lepton pair the photon penguin amplitude vanishes due to vector current
conservation.\\
An important consequence is, that the decays considered here exhibit a
hard GIM suppression, quadratic in (small) internal quark masses, which
is a property of the $Z^0$-penguin  and box graphs. By contrast, the
GIM suppression resulting from photon penguin contributions is
logarithmic. Decays where the photon penguin contributes are for
example $K_L\to\pi^0e^+e^-$ and $B \to X_s e^+ e^-$. The differences in
the basic structure of these processes, resulting from the different
pattern of GIM suppression, are the reason why we have discussed
$K_L\to\pi^0e^+e^-$ and $B \to X_s e^+ e^-$ in a separate context.
\item
The investigation of low energy rare decay processes allows to probe,
albeit indirectly, high energy scales of the theory. Of particular
interest is the sensitivity to properties of the top quark,
its mass $m_t$ and its CKM couplings $V_{ts}$ and $V_{td}$.
\item
A particular and very important advantage of the processes under
discussion is, that theoretically clean predictions can be obtained.
The reasons for this are:
\begin{itemize}
\item The low energy hadronic
matrix elements required are just the matrix elements of quark currents
between hadron states, which can be extracted from the leading
(non-rare) semileptonic decays. Other long-distance contributions
are negligibly small.\\
An exception is the decay $\klm$ receiving important contributions from
the two-photon intermediate state, which are difficult to calculate
reliably. However, the short-distance part $(\klm)_{SD}$ alone, which we
shall discuss here, is on the same footing as the other modes. The
essential difficulty for phenomenological applications then is to
separate the short-distance from the long-distance piece in the
measured rate.
\item According to the comments just made, the processes at hand are
short-distance processes, calculable within a perturbative framework,
possibly including renormalization group improvement. The necessary
separation of the short-distance dynamics from the low energy matrix
elements  is achieved by means of an operator product expansion.
The scale ambiguities, inherent to perturbative QCD, essentially
constitute the only theoretical uncertainties present in the analysis.
These uncertainties are well under control as they may be
systematically reduced through calculations beyond leading order.
\end{itemize}
\item
The points made above emphasize, that the short-distance dominated
loop-induced FCNC decays provide highly promising possibilities to
investigate flavourdynamics at the quantum level. However, the very fact
that these processes are based on higher order electroweak effects,
which makes them interesting theoretically, at the same time implies,
that the branching ratios will be very small and not easy to
access experimentally.
\end{itemize}
The effective hamiltonians governing the decays
$\kpn$, $(\klm)_{SD}$, $K_L\to\pi^0\nu\bar\nu$,
$B\to X_{s, d}\nu\bar\nu$, $B\to l^+l^-$,
resulting from the $Z^0$-penguin and box-type contributions, can all be
written down in the following general form
\begin{equation}\label{hnr} {\cal H}_{eff}={G_F \over{\sqrt 2}}{\alpha\over 2\pi \sin^2\Theta_W}
 \left( \lambda_c F(x_c) + \lambda_t F(x_t)\right)
 (\bar nn^\prime)_{V-A}(\bar rr)_{V-A}  \end{equation}
where $n$, $n^\prime$ denote down-type quarks
($n, n^\prime=d, s, b$ but $n\not= n^\prime$) and $r$ leptons,
$r=l, \nu_l$ ($l=e, \mu, \tau$). The $\lambda_i$ are products of CKM elements,
in the general case $\lambda_i=V^*_{in}V_{in^\prime}^{}$. Furthermore
$x_i=m^2_i/M^2_W$.\\
The functions $F(x_i)$ describe the dependence on the internal
up-type quark masses $m_i$ (and on lepton masses if necessary)
and are understood to include QCD corrections.
They are increasing functions of the quark masses, a property that is
particularly important for the top contribution.\\
Crucial features of the structure of the hamiltonian
are furthermore determined
by the hard GIM suppression characteristic for this class of decays.
First we note that the dependence of the hamiltonian on the internal
quarks comes in the form
\begin{equation}\label{lifx}
\sum_{i=u, c, t}\lambda_i F(x_i)=
\lambda_c(F(x_c)-F(x_u))+\lambda_t(F(x_t)-F(x_u))  \end{equation}
where we have used the unitarity of the CKM matrix. Now, hard GIM
suppression means that for $x\ll 1$ $F$ behaves quadratically in
the quark masses. In the present case we have
\begin{equation}\label{fxln}
F(x)\sim x\ln x\qquad\qquad  {\rm for}\quad x\ll 1   \end{equation}
The first important consequence is, that $F(x_u)\approx 0$ can be
neglected. The hamiltonian acquires the form anticipated in \eqn{hnr}.
It effectively consists of a charm and a top contribution. Therefore
the relevant energy scales are $M_W$ or $m_t$ and, at least, $m_c$,
which are large compared to $\Lambda_{QCD}$. This fact indicates the
short-distance nature of these processes.\\
A second consequence of \eqn{fxln} is that $F(x_c)/F(x_t)\approx
\ord(10^{-3})\ll 1$. Together with the weighting introduced by the
CKM factors this relation determines the relative importance of
the charm versus the top contribution in \eqn{hnr}. As seen in
table~\ref{tab:lambdaexp} a simple pattern emerges if one writes down
the order of magnitude of $\lambda_c$, $\lambda_t$ in terms of powers
of the Wolfenstein expansion parameter $\lambda$.

\begin{table}[htb]
\caption[]{
Order of magnitude of CKM parameters relevant for the various decays,
expressed in powers of the Wolfenstein parameter $\lambda=0.22$. In the
case of $K_L\to\pi^0\nu\bar\nu$, which is CP-violating, only the
imaginary parts of $\lambda_{c, t}$ contribute.
\label{tab:lambdaexp}}
\begin{center}
\begin{tabular}{|r|c|c|c|c|}
&$\kpn$&$K_L\to\pi^0\nu\bar\nu$&$B\to X_s\nu\bar\nu$&
$B\to X_d\nu\bar\nu$\\
&$(\klm)_{SD}$&&$B_s\to l^+l^-$&$B_d\to l^+l^-$\\  \hline\hline
$\lambda_c$&$\sim\lambda$&(${\rm Im}\lambda_c\sim\lambda^5$)&
$\sim\lambda^2$&$\sim\lambda^3$\\  \hline
$\lambda_t$&$\sim\lambda^5$&(${\rm Im}\lambda_t\sim\lambda^5$)&
$\sim\lambda^2$&$\sim\lambda^3$
\end{tabular}
\end{center}
\end{table}

For the CP-violating decay $K_L\to\pi^0\nu\bar\nu$ and the B-decays
the CKM factors $\lambda_c$ and $\lambda_t$ have the same order of
magnitude. In view of $F(x_c)\ll F(x_t)$ the charm contribution is
therefore negligible and these decays are entirely determined by the
top sector.\\
For \kpnn and $(\klm)_{SD}$ on the other hand $\lambda_t$ is
suppressed compared to $\lambda_c$ by a factor of order
$\ord(\lambda^4)\approx\ord(10^{-3})$, which roughly compensates
for the $\ord(10^3)$ enhancement of $F(x_t)$ over $F(x_c)$. Hence
the top and charm contributions have the same order of magnitude and
must both be taken into account.

In principle, as far as flavordynamics is concerned, the top and the
charm sector have the same structure. The only difference comes from
the quark masses. However, this difference has striking implications
for the detailed formalism necessary to treat the strong interaction
corrections. We have $m_t/M_W=\ord(1)$ and $m_c/M_W\ll 1$.
Correspondingly the QCD coupling $\as$ is also somewhat smaller at
$m_t$ than at $m_c$. For the charm contribution this implies that one
can work to lowest order in the mass ratio $m_c/M_W$. On the other
hand, for the same reason, logarithmic QCD corrections
$\sim\as\ln M_W/m_c$ are large and have to be resummed to all
orders in perturbation theory by renormalization group methods.
On the contrary, no large logarithms are present in the top sector, so that
ordinary perturbation theory is applicable, but all orders in $m_t/M_W$ have
to be taken into account. In fact we see that from the
point of view of QCD corrections the charm and top contributions are
quite ``complementary'' to each other, representing in a sense
opposite limiting cases.\\
We are now ready to list the explicit expressions for the effective
hamiltonians.

\subsection{The Decay \kpnn}
            \label{sec:HeffRareKB:kpnn}
\subsubsection{The Next-to-Leading Order Effective Hamiltonian}
               \label{sec:HeffRareKB:kpnn:heff}
The final result for the effective hamiltonian inducing \kpnn can
be written as
\begin{equation}\label{hkpn} {\cal H}_{eff}={G_F \over{\sqrt 2}}{\alpha\over 2\pi \sin^2\Theta_W}
 \sum_{l=e,\mu,\tau}\left( V^{\ast}_{cs}V_{cd} X^l_{NL}+
V^{\ast}_{ts}V_{td} X(x_t)\right)
 (\bar sd)_{V-A}(\bar\nu_l\nu_l)_{V-A} \, .
\end{equation}
The index $l$=$e$, $\mu$, $\tau$ denotes the lepton flavor.
The dependence on the charged lepton mass, resulting from the box-graph,
is negligible for the top contribution. In the charm sector this is the
case only for the electron and the muon, but not for the $\tau$-lepton.
\\
The function $X(x)$, relevant for the top part, reads
to $\ord(\as)$ and to all orders in $x=m^2/M^2_W$
\begin{equation}\label{xx} X(x)=X_0(x)+\aspi X_1(x) \end{equation}
with \cite{inamilim:81}
\begin{equation}\label{xx0} X_0(x)={x\over 8}\left[ -{2+x\over 1-x}+{3x-6\over (1-x)^2}\ln x\right] \end{equation}
and the QCD correction \cite{buchallaburas:93b}
\begin{eqnarray}\label{xx1}
X_1(x)=&-&{23x+5x^2-4x^3\over 3(1-x)^2}+{x-11x^2+x^3+x^4\over (1-x)^3}\ln x
\nonumber\\
&+&{8x+4x^2+x^3-x^4\over 2(1-x)^3}\ln^2 x-{4x-x^3\over (1-x)^2}L_2(1-x)
\nonumber\\
&+&8x{\partial X_0(x)\over\partial x}\ln x_\mu
\end{eqnarray}
where $x_\mu=\mu^2/M^2_W$ with $\mu=\ord(m_t)$ and
\begin{equation}\label{l2} L_2(1-x)=\int^x_1 dt {\ln t\over 1-t}   \end{equation}
The $\mu$-dependence in the last term in (\ref{xx1}) cancels to the
order considered the $\mu$-dependence of the leading term $X_0(x(\mu))$.
\\
The expression corresponding to $X(x_t)$ in the charm sector is the function
$X^l_{NL}$. It results from the RG calculation in NLLA and is given
as follows:
\begin{equation}\label{xlnl}X^l_{NL}=C_{NL}-4 B^{(1/2)}_{NL}  \end{equation}
$C_{NL}$ and $B^{(1/2)}_{NL}$ correspond to the $Z^0$-penguin and the
box-type contribution, respectively. One has \cite{buchallaburas:94}
\begin{eqnarray}\label{cnln}
\lefteqn{C_{NL}={x(m)\over 32}K^{{24\over 25}}_c\left[\left({48\over 7}K_++
 {24\over 11}K_--{696\over 77}K_{33}\right)\left({4\pi\over\as(\mu)}+
 {15212\over 1875} (1-K^{-1}_c)\right)\right.}\nonumber\\
&&+\left(1-\ln{\mu^2\over m^2}\right)(16K_+-8K_-)-{1176244\over 13125}K_+-
 {2302\over 6875}K_-+{3529184\over 48125}K_{33} \nonumber\\
&&+\left. K\left({56248\over 4375}K_+-{81448\over 6875}K_-+{4563698\over 144375}K_{33}
  \right)\right]
\end{eqnarray}
where
\begin{equation}\label{kkc} K={\as(M_W)\over\as(\mu)}\qquad
  K_c={\as(\mu)\over\as(m)}  \end{equation}
\begin{equation}\label{kkn} K_+=K^{{6\over 25}}\qquad K_-=K^{{-12\over 25}}\qquad
          K_{33}=K^{{-1\over 25}}  \end{equation}
\begin{eqnarray}\label{bnln}
\lefteqn{B^{(1/2)}_{NL}={x(m)\over 4}K^{24\over 25}_c\left[ 3(1-K_2)\left(
 {4\pi\over\as(\mu)}+{15212\over 1875}(1-K^{-1}_c)\right)\right.}\nonumber\\
&&-\left.\ln{\mu^2\over m^2}-
  {r\ln r\over 1-r}-{305\over 12}+{15212\over 625}K_2+{15581\over 7500}K K_2
  \right]
\end{eqnarray}
Here $K_2=K^{-1/25}$, $m=m_c$, $r=m^2_l/m^2_c(\mu)$ and $m_l$ is the
lepton mass.  We will at times omit the index $l$ of $X^l_{NL}$.  In
(\ref{cnln}) -- (\ref{bnln}) the scale is $\mu=\ord(m_c)$.  The
two-loop expression for $\as(\mu)$ is given in \eqn{amu}.  Again -- to
the considered order -- the explicit $\ln(\mu^2/m^2)$ terms in
(\ref{cnln}) and (\ref{bnln}) cancel the $\mu$-dependence of the
leading terms.
\\
These formulae give the complete next-to-leading effective hamiltonian
for $\kpn$. The leading order expressions \cite{novikovetal:77},
\cite{ellishagelin:83}, \cite{dibetal:91}, \cite{buchallaetal:91} are
obtained by replacing $X(x_t)\to X_0(x_t)$ and $X^l_{NL}\to X_L$ with
$X_L$ found from (\ref{cnln}) and (\ref{bnln}) by retaining there only
the $1/\as(\mu)$ terms. In LLA the one-loop expression should be used
for $\as$.  This amounts to setting $\beta_1=0$ in \eqn{amu}.  The
numerical values for $X_{NL}$ for $\mu = \mc$ and several values of
$\Lms^{(4)}$ and $\mc(\mc)$ are given in table \ref{tab:xnlnum}. The
$\mu$ dependence will be discussed in part three.

\begin{table}[htb]
\caption[]{The functions $X^e_{NL}$ and $X^\tau_{NL}$
for various $\Lms^{(4)}$ and $\mc$.
\label{tab:xnlnum}}.
\begin{center}
\begin{tabular}{|c|c|c|c|c|c|c|}
& \multicolumn{3}{c|}{$X^e_{NL}/10^{-4}$} &
  \multicolumn{3}{c|}{$X^\tau_{NL}/10^{-4}$} \\
\hline
$\Lms^{(4)}\ [\mev]\;\backslash\;\mc\ [\gev]$ &
1.25 & 1.30 & 1.35 & 1.25 & 1.30 & 1.35 \\
\hline
215 & 10.55  & 11.40  & 12.28 & 7.16 & 7.86 & 8.59 \\
325 &  9.71  & 10.55  & 11.41 & 6.32 & 7.01 & 7.72 \\
435 &  8.75  &  9.59  & 10.45 & 5.37 & 6.05 & 6.76
\end{tabular}
\end{center}
\end{table}

\subsubsection{Z-Penguin and Box Contribution in the Top Sector}
               \label{sec:HeffRareKB:kpnn:Zptop}
For completeness we give here in addition the expressions for the
$Z^0$-penguin function $C(x)$ and the box function $B(x,1/2)$ separately,
which contribute to $X(x)$ in \eqn{xx} according to
\begin{equation}\label{xc4b} X(x)=C(x) - 4 \, B(x,1/2)  \end{equation}
The functions $C$ and $B$ depend on the gauge of the $W$-boson.
In 't Hooft--Feynman-gauge ($\xi=1$) they read
\begin{equation}\label{cx01} C(x)=C_0(x)+{\as\over 4\pi} C_1(x)   \end{equation}
where \cite{inamilim:81}
\begin{equation}\label{cx0} C_0(x) = {x \over 8} \left[{6-x\over 1-x} +
   {3x+2\over (1-x)^2} \ln x\right]   \end{equation}
and \cite{buchallaburas:93a}
\begin{eqnarray}\label{cx1}
C_1(x) &=& {29x + 7x^2+4x^3 \over 3(1-x)^2}
           - {x - 35x^2 -3 x^3 -3 x^4\over 3(1-x)^3} \ln x\nonumber\\
       & &- {20x^2 - x^3+x^4\over 2(1-x)^3} \ln^2 x
           + {4x+x^3\over (1-x)^2} L_2 (1-x)\nonumber\\
       & &+ 8x {\partial C_0(x)\over \partial x} \ln x_\mu
\end{eqnarray}
Similarly
\begin{equation}\label{bx01p}
B(x, 1/2) = B_0(x) + {\as\over 4\pi} B_1 (x, 1/2)   \end{equation}
with the one-loop function \cite{inamilim:81}
\begin{equation}\label{bx0}
B_0(x) = {1\over 4}\left[{x\over 1-x} + {x\over (1-x)^2} \ln x\right] \end{equation}
and \cite{buchallaburas:93b}
\begin{eqnarray}\label{bx1p}
B_1(x, 1/2)&=&{13x + 3x^2\over 3(1-x)^2} -
{x-17x^2\over 3(1-x)^3} \ln x
-  {x+3x^2\over (1-x)^3}\ln^2x + {2x\over(1-x)^2} L_2(1-x)\nonumber\\
& &+ 8x {\partial B_0(x)\over \partial x} \ln x_\mu
\end{eqnarray}
The gauge dependence of $C$ and $B$ is canceled in the
combination $X$ \eqn{xc4b}. The second argument in $B(x,1/2)$ indicates the
weak isospin of the external leptons (the neutrinos in this case).

\subsubsection{The $Z$-Penguin Contribution in the Charm Sector}
               \label{sec:HeffRareKB:kpnn:Zpcharm}
In the next two paragraphs we would like to summarize the essential
ingredients of the RG calculation for the charm sector leading to
\eqn{cnln} and \eqn{bnln}. In particular we present the operators involved,
the initial values for the RG evolution of the Wilson coefficients and
the required two-loop anomalous dimensions. We will first treat
the $Z^0$-penguin contribution \eqn{cnln} and discuss the box part \eqn{bnln}
subsequently. Further details can be found in \cite{buchallaburas:94}.

At renormalization scales of the order $\ord(M_W)$ and after
integrating out the $W$- and $Z$-bosons the effective hamiltonian
responsible for the $Z^0$-penguin contribution of the charm sector is given
by
\begin{equation}\label{hzop}{\cal H}^{(Z)}_{eff,c}={G_F \over{\sqrt 2}}
{\alpha\over 2\pi \sin^2\Theta_W}\lambda_c{\pi^2\over 2 M^2_W}
\left( v_+ O_+ +v_- O_- +v_3 Q\right) \end{equation}
where the operator basis is
\begin{equation}\label{o1} O_1=
   -i\int d^4x\ T\left((\bar s_ic_j)_{V-A}(\bar c_jd_i)_{V-A}\right)(x)\
       \left((\bar c_kc_k)_{V-A}(\bar\nu\nu)_{V-A}\right)(0)\ -
       \{c\to u\}    \end{equation}
\begin{equation}\label{o2} O_2=
   -i\int d^4x\ T\left((\bar s_ic_i)_{V-A}(\bar c_jd_j)_{V-A}\right)(x)\
       \left((\bar c_kc_k)_{V-A}(\bar\nu\nu)_{V-A}\right)(0)\ -
       \{c\to u\}    \end{equation}
\begin{equation}\label{opm} O_\pm ={1\over 2}(O_2 \pm O_1) \end{equation}
\begin{equation}\label{qnu} Q={m^2\over g^2} (\bar sd)_{V-A}(\bar\nu\nu)_{V-A}   \end{equation}
The Wilson coefficients at $\mu=M_W$ are
($\vec v^T\equiv(v_+,v_-,v_3)$)
\begin{equation}\label{vmw} \vec v(M_W)=\vec v^{(0)}+\frac{\as(M_W)}{4 \pi}
\vec v^{(1)}
\end{equation}
\begin{equation}\label{v0} {\vec v^{(0)T}}=(1,1,0)  \end{equation}
\begin{equation}\label{v1} {\vec v^{(1)T}}=(B_+,B_-,B_3)  \end{equation}
where in the NDR scheme ($\overline{MS}$, anticommuting $\gamma_5$
in $D\not=4$ dimensions)
\begin{equation}\label{bpm3} B_\pm=\pm 11{N\mp 1\over 2N}\qquad B_3=16  \end{equation}
with $N$ denoting the number of colors.\\
In the basis of operators $\{ O_+,O_-,Q\}$ the matrix of anomalous
dimensions has the form
\begin{equation}\label{gz} \gamma =
 \left(\begin{array}{ccc} \gamma_+ & 0 & \gamma_{+3} \\
                           0 & \gamma_- & \gamma_{-3} \\
                           0 & 0 & \gamma_{33}
    \end{array}\right)   \end{equation}
with the perturbative expansion
\begin{equation}\label{ga2}
\gamma(\as)=\aspi \gamma^{(0)}+\left(\aspi\right)^2\gamma^{(1)}
\end{equation}
The nonvanishing entries of the anomalous dimension matrix read
\begin{equation}\label{gzij}
\begin{tabular}{lcl}
$\gamma^{(0)}_{33}=2(\gamma_{m0}-\beta_0)$ &\qquad\qquad&
   $\gamma^{(1)}_{33}=2(\gamma_{m1}-\beta_1)$ \\  & & \\
$\gamma^{(0)}_\pm=\pm 6{N\mp 1\over N}$ &  &
   $\gamma^{(1)}_\pm={N\mp 1\over 2N}\left(-21\pm{57\over N}\mp{19\over 3}N\pm{4\over 3}f
     \right)$ \\                                 & & \\
$\gamma^{(0)}_{\pm 3}=\pm 8(N\pm 1)$ &  &
   $\gamma^{(1)}_{\pm 3}=C_F(\pm 88 N-48)$ \\
\end{tabular}  \end{equation}
where $\gamma_{m0}$, $\gamma_{m1}$, $\beta_0$, $\beta_1$ can be found
in \eqn{gm01} and \eqn{b0b1}, respectively. The expressions
$\gamma^{(1)}$ refer to the NDR scheme, consistent with the scheme
chosen for $\vec v(M_W)$. Following the general method for the solution
of the RG equations explained in section~\ref{sec:basicform:wc:rgf}, we
can compute the Wilson coefficients $\vec v(\mu)$ at a scale
$\mu=\ord(m_c)$. It is convenient to work in an effective four-flavor
theory ($f=4$) in the full range of the RG evolution from $M_W$ down to
$\mu$. The possible inclusion of a $b$-quark threshold would change the
result for $X_{NL}$ by not more than 0.1\% and can therefore be safely
neglected.\\ After integrating out the charm quark at the scale
$\mu=\ord(m_c)$, the $Z^0$-penguin part of the charm contribution to
the effective hamiltonian becomes
\begin{equation}\label{hzc} {\cal H}^{(Z)}_{eff,c}={G_F \over{\sqrt 2}}
  {\alpha\over 2\pi \sin^2\Theta_W} \lambda_c\ C_{NL}\
   (\bar sd)_{V-A}(\bar\nu\nu)_{V-A}  \end{equation}
\begin{equation}\label{cnl}
C_{NL}={x(\mu)\over 32}\left[{1\over 2}\left(1-\ln{\mu^2\over m^2}\right)
  \left(\gamma^{(0)}_{+3} K_++\gamma^{(0)}_{-3} K_-\right)+
  {4\pi\over\as(\mu)}v_3(\mu)\right] \, .
\end{equation}
The explicit expression for $v_3(\mu)$ as obtained from solving the RG
equation is given in \cite{buchallaburas:94}. Inserting this
expression in \eqn{cnl}, expressing the charm quark mass $m(\mu)$ in
terms of $m(m)$ and setting $N=3$, $f=4$, we finally end up with
\eqn{cnln}.

\subsubsection{The Box Contribution in the Charm Sector}
               \label{sec:HeffRareKB:kpnn:boxcharm}
The RG analysis for the box contribution proceeds in analogy to the
$Z^0$-penguin case. The box part is even somewhat simpler. When the
$W$ boson is integrated out, the hamiltonian based on the box diagram
reads
\begin{equation}\label{hbop}{\cal H}^{(B)}_{eff,c}=-{G_F \over{\sqrt 2}}
{\alpha\over 2\pi \sin^2\Theta_W}\lambda_c\left(-{\pi^2\over M^2_W}\right)
\left( c_1 O +c_2 Q\right) \end{equation}
\begin{equation}\label{ob} O=
   -i\int d^4x\ T\left((\bar sc)_{V-A}(\bar \nu l)_{V-A}\right)(x)\
       \left((\bar l\nu)_{V-A}(\bar cd)_{V-A}\right)(0)\ -
       \{c\to u\}    \end{equation}
with $Q$ alread given in \eqn{qnu}.
The Wilson coefficients at $M_W$ in the NDR scheme are given by
\begin{equation}\label{cmw}
\vec c^T(M_W)\equiv (c_1(M_W),c_2(M_W))=(1,0)+\frac{\as(M_W)}{4 \pi} (0,B_2)
  \qquad B_2=-36  \end{equation}
In the operator basis $\{O,Q\}$ the anomalous dimension matrix  has the form
\begin{equation}\label{gb} \gamma =
 \left(\begin{array}{cc}   0 & \gamma_{12} \\
                           0 & \gamma_{22}
    \end{array}\right)   \end{equation}
When expanded as
\begin{equation}\label{gb2}
\gamma=\aspi \gamma^{(0)}+\left(\aspi\right)^2\gamma^{(1)}  \end{equation}
the non-zero elements read (NDR scheme for $\gamma^{(1)}$)
\begin{equation}\label{gbij}
\begin{tabular}{lcl}
$\gamma^{(0)}_{22}=2(\gamma_{m0}-\beta_0)$ &\qquad\qquad&
   $\gamma^{(1)}_{22}=2(\gamma_{m1}-\beta_1)$ \\  & & \\
$\gamma^{(0)}_{12}=-32$ &  &
   $\gamma^{(1)}_{12}=80C_F$ \\
\end{tabular}  \end{equation}
Finally, after integrating out charm at $\mu=\ord(m_c)$
\begin{equation}\label{hbcn} {\cal H}^{(B)}_{eff,c}=-{G_F \over{\sqrt 2}}
  {\alpha\over 2\pi \sin^2\Theta_W} \lambda_c\ 4 B^{(1/2)}_{NL}\
   (\bar sd)_{V-A}(\bar\nu_l\nu_l)_{V-A}  \end{equation}
\begin{equation}\label{bnl} B^{(1/2)}_{NL}=-{x(\mu)\over 64}\left[
16\left(\ln{\mu^2\over m^2}+
{5\over 4}+{r\ln r\over 1-r}\right) + \frac{4 \pi}{\as(\mu)} c_2(\mu) \right]
\end{equation}
(\ref{hbcn}) is written here for one neutrino flavor. The index $(1/2)$
refers to the weak isospin of the final state leptons.
From this result \eqn{bnln} can be derived ($N=3$, $f=4$).
The explicit expression for $c_2(\mu)$ can be found in
\cite{buchallaburas:94}.

Although Wilson coefficients and anomalous dimensions depend on the
renormalization scheme, the final results in \eqn{cnln} and \eqn{bnln}
are free from this dependence. The argument proceeds as in the
general case presented in section~\ref{sec:basicform:wc:rgdep}.

\subsubsection{Discussion}
               \label{sec:HeffRareKB:kpnn:disc}
It is instructive to consider furthermore the function $X(x)$ in the
limiting case of small masses ($x\ll 1$), keeping only terms linear
in $x$ and including $\ord(\as)$ corrections:
\begin{equation}\label{xxc} X(x)\doteq -{3\over 4}x\ln x-{1\over 4}x+
     \aspi\left(-2x\ln^2 x-7x\ln x-{23+2\pi^2\over 3}x\right)  \end{equation}
This simple and transparent expression can be regarded as a common
limiting case of the top- and the charm contribution: On the one hand
it follows from keeping only terms linear in $x$ in the top function
\eqn{xx}. On the other hand it can be obtained
(up to the last term in \eqn{xxc} which is $\ord(\as x)$ and
goes beyond the NLLA) from expanding $X_{NL}$ \eqn{xlnl} (for $m_l=0$)
to first order in $\as$.\\
This exercise provides one with a nice cross-check between the
rather different looking functions $X_{NL}$ and $X(x_t)$ of the
charm- and the top sector. Viewed the other way around, \eqn{xxc} may
serve to further illustrate the complementary character of the
calculations necessary in each of the two sectors. $X(x_t)$ is the
generalization of \eqn{xxc} that includes all the higher order mass terms.
$X_{NL}$ on the other hand generalizes \eqn{xxc} to include all the
leading logarithmic, $\ord(x \alpha^n_s \ln^{n+1}x)$, as well as
the next-to-leading logarithmic $\ord(x \alpha^n_s \ln^{n}x)$
corrections, to all orders $n$ in $\as$. Of these only the terms
with $n=0$ and $n=1$ are contained in \eqn{xxc}.\\
Applying the approximation \eqn{xxc} to the charm part directly, one can
furthermore convince oneself, that the $\ord(\as)$
correction term would amount to more than 50\% of the lowest order
result. This observation illustrates very clearly the necessity to
go beyond straightforward perturbation theory and to employ the
RG summation technique. The importance of going still to
next-to-leading order accuracy in the RG calculation is suggested by the
relatively large size of the $\ord(x \as \ln x)$ term.
Note also, that formally the non-logarithmic mass term $(-x/4)$ in
\eqn{xxc} is a next-to-leading order effect in the framework of
RG improved perturbation theory. The same is true for the dependence
on the charged lepton mass, which can be taken into account
consistently only in NLLA.

A crucial issue is the residual dependence of the functions $X_{NL}$
and $X(x_t)$ on the corresponding renormalization scales $\mu_c$ and
$\mu_t$. Since the quark current operator in \eqn{hnr} has no
anomalous dimension, its matrix elements do not depend on the
renormalization scale. The same must then hold for the coefficient
functions $X_{NL}$ and $X(x_t)$. However, in practice this is only
true up to terms of the neglected order in perturbation theory.
The resulting scale ambiguities represent the theoretical uncertainties
present in the calculation of the short-distance dominated processes
under discussion. They can be systematically reduced by going to
higher orders in the analysis. In table~\ref{tab:scaledep} we compare
the order of the residual scale dependence in LLA and in NLLA for the
top- and the charm contribution.

\begin{table}[htb]
\caption[]{Residual scale ambiguity in the top and charm sector
in LLA and NLLA.
\label{tab:scaledep}}
\begin{center}
\begin{tabular}{|r|c|c|}
&Top Sector ($\mu_t=\ord(m_t)$) & Charm Sector ($\mu_c=\ord(m_c)$)
\\  \hline
LLA & $\ord(\as)$&$\ord(x_c)$ \\  \hline
NLLA & $\ord(\alpha^2_s)$&$\ord(\as x_c)$
\end{tabular}
\end{center}
\end{table}

For numerical investigations we shall use $1\gev\leq\mu_c\leq 3\gev$
for the renormalization scale $\mu_c=\ord(m_c)$ in the charm sector.
Similarly, in the case of the top contribution we choose
$\mu_t=\ord(m_t)$ in the range $100\gev\leq\mu_t\leq 300\gev$ for
$m_t=170\gev$. Then, comparing LLA and NLLA, the theoretical
uncertainty due to scale ambiguity is typically reduced from
$\ord(10\%)$ to $\ord(1\%)$ in the top sector and from more than 50\%
to less than 20\% in the charm sector. Here the quoted percentages refer
to the total variation $(X_{max}-X_{min})/X_{central}$ of the functions
$X(x_t)$ or $X_{NL}$ within the range of scales considered.
Phenomenological implications of this gain in accuracy will be
discussed in section~\ref{sec:Kpnn}.

\subsection{The Decay $(\klm)_{SD}$}
            \label{sec:HeffRareKB:klmm}
\subsubsection{The Next-to-Leading Order Effective Hamiltonian}
               \label{sec:HeffRareKB:klmm:heff}
The analysis of $(\klm)_{SD}$ proceeds in essentially the same
manner as for $\kpn$. The only difference is introduced through the
reversed lepton line in the box contribution. In particular there is
no lepton mass dependence, since only massless neutrinos appear as
virtual leptons in the box diagram.\\
The effective hamiltonian in next-to-leading order can be written as
follows:
\begin{equation}\label{hklm}{\cal H}_{eff}=-{G_F \over{\sqrt 2}}{\alpha\over 2\pi \sin^2\Theta_W}
 \left( V^{\ast}_{cs}V_{cd} Y_{NL}+
V^{\ast}_{ts}V_{td} Y(x_t)\right)
 (\bar sd)_{V-A}(\bar\mu\mu)_{V-A} + h.c. \end{equation}
The function $Y(x)$ is given by
\begin{equation}\label{yy}
Y(x) = Y_0(x) + \aspi Y_1(x)\end{equation}
where \cite{inamilim:81}
\begin{equation}\label{yy0}
Y_0(x) = {x\over 8}\left[{4-x\over 1-x}+{3x\over (1-x)^2}\ln x\right]
\end{equation}
and \cite{buchallaburas:93b}
\begin{eqnarray}\label{yy1}
Y_1(x) = &&{4x + 16 x^2 + 4x^3 \over 3(1-x)^2} -
           {4x - 10x^2-x^3-x^4\over (1-x)^3} \ln x\nonumber\\
         &+&{2x - 14x^2 + x^3 - x^4\over 2(1-x)^3} \ln^2 x
           + {2x + x^3\over (1-x)^2} L_2(1-x)\nonumber\\
         &+&8x {\partial Y_0(x) \over \partial x} \ln x_\mu
\end{eqnarray}
The RG expression $Y_{NL}$ representing the charm contribution reads
\begin{equation}\label{ynl} Y_{NL}=C_{NL}-B^{(-1/2)}_{NL}  \end{equation}
where $C_{NL}$ is the $Z^0$-penguin part given in (\ref{cnln}) and
$B^{(-1/2)}_{NL}$ is the box contribution in the charm sector, relevant
for the case of final state leptons with weak isospin $T_3=-1/2$.
One has \cite{buchallaburas:94}
\begin{eqnarray}\label{bmnln}
\lefteqn{B^{(-1/2)}_{NL}={x(m)\over 4}K^{24\over 25}_c\left[ 3(1-K_2)\left(
 {4\pi\over\as(\mu)}+{15212\over 1875}(1-K^{-1}_c)\right)\right.}\nonumber\\
&&-\left.\ln{\mu^2\over m^2}-
  {329\over 12}+{15212\over 625}K_2+{30581\over 7500}K K_2
  \right]
\end{eqnarray}
Note the simple relation to $B^{(1/2)}_{NL}$ in (\ref{bnln}) (for
$r=0$)
\begin{equation}\label{dbnl}
B^{(-1/2)}_{NL}-B^{(1/2)}_{NL}={x(m)\over 2}K^{24\over 25}_c (K K_2-1)
\end{equation}
More details on the RG analysis in this case may be found in
\cite{buchallaburas:94}.

\begin{table}[htb]
\caption[]{The function $Y_{NL}$ for various $\Lms^{(4)}$ and $\mc$.
\label{tab:ynlnum}}.
\begin{center}
\begin{tabular}{|c|c|c|c|}
&\multicolumn{3}{c|}{$Y_{NL}/10^{-4}$}\\
\hline
$\Lms^{(4)}\ [\mev]\;\backslash\;\mc\ [\gev]$ & 1.25 & 1.30 & 1.35 \\
\hline
215 & 3.09 & 3.31 & 3.53 \\
325 & 3.27 & 3.50 & 3.73 \\
435 & 3.40 & 3.64 & 3.89
\end{tabular}
\end{center}
\end{table}

\subsubsection{Discussion}
               \label{sec:HeffRareKB:klmm:disc}
The gauge independent function $Y$ can be decomposed into the
$Z^0$-penguin- and the box contribution
\begin{equation}\label{yxcb}  Y(x)=C(x)-B(x, -1/2)    \end{equation}
In Feynman-gauge for the $W$ boson $C(x)$ is given in \eqn{cx01}.
In the same gauge the box contribution reads
\begin{equation}\label{bx01m}
B(x,-1/2)=B_0(x)+{\as\over 4\pi} B_1(x,-1/2)
\end{equation}
with $B_0(x)$ from \eqn{bx0} and
\begin{eqnarray}\label{bx1m}
B_1(x, -1/2)&=&{25x-9x^2\over 3(1-x)^2}
+ {11x + 5x^2\over 3(1-x)^3}\ln x
- {x+3x^2\over(1-x)^3} \ln^2x + {2x\over (1-x)^2}L_2(1-x)\nonumber \\
& &+ 8x {\partial B_0(x)\over \partial x} \ln x_\mu
\end{eqnarray}
The equality $B(x,1/2)=B(x,-1/2)$ at the one-loop level is a
particular property of the Feynman-gauge. It is violated by
$\ord(\as)$ corrections. There is however a very simple
relation between $B_1(x,1/2)$ and $B_1(x,-1/2)$
\begin{equation}\label{b1pm}
B_1(x, -1/2)- B_1(x, 1/2) = 16 B_0(x)   \end{equation}

We add a few comments on the most important differences between
$Y_{NL}$ and $X_{NL}$.\\
Expanding $Y_{NL}$ to first order in $\as$ we find
\begin{equation}\label{ylin}
Y_{NL}\doteq {1\over 2}x +{\as\over 4\pi} x \ln^2x +\ord(\as x)  \end{equation}
In contrast to $X_{NL}$ both the terms of $\ord(x\ln x)$ and of
$\ord(\as x\ln x)$ are absent in $Y_{NL}$. The cancellation
of the leading $\ord(x\ln x)$ terms between $Z^0$-penguin and box
contribution implies that the non-leading $\ord(x)$ term plays a
much bigger role for $Y_{NL}$. A second consequence are the increased
importance of QCD effects and the related larger sensitivity to $\mu_c$,
resulting in a bigger theoretical uncertainty for $Y_{NL}$ than it
happened to be the case for $X_{NL}$. In addition, whereas $X(x_c)$
is suppressed by $\sim 30\%$ through QCD effects, the zeroth order
expression for $Y$ is enhanced by as much as a factor of about 2.5.
Nevertheless, QCD corrections included, $X_{NL}$ still exceeds $Y_{NL}$
by a factor of four, so that $Y_{NL}$ is less important for
$(\klm)_{SD}$ than $X_{NL}$ is for $\kpn$. Although the impact of the
bigger uncertainties in $Y_{NL}$ is thus somewhat reduced in the
complete result for $(\klm)_{SD}$, the remaining theoretical uncertainty
due to scale ambiguity is still larger than for $\kpn$. It will be
investigated numerically in section \ref{sec:KLmm}. The numerical
values for $Y_{NL}$ for $\mu=\mc$ and several values of $\Lms^{(4)}$
and $\mc(\mc)$ are given in table \ref{tab:ynlnum}.

\subsection{The Decays $K_L\to\pi^0\nu\bar\nu$, $B\to X_{s,d}\nu\bar\nu$ and
            $B_{s,d}\to l^+l^-$}
            \label{sec:HeffRareKB:klpinn}
After the above discussion it is easy to write down also the effective
hamiltonians for $K_L\to\pi^0\nu\bar\nu$, $B\to X_{s,d}\nu\bar\nu$
and $B_{s,d}\to l^+l^-$. As we have seen, only the top contribution is
important in these cases and we can write
\begin{equation}\label{hxnu}
{\cal H}_{eff} = {G_F\over \sqrt 2} {\alpha \over
2\pi \sin^2 \Theta_W} V^\ast_{tn} V_{tn^\prime}
X (x_t) (\bar nn^\prime)_{V-A} (\bar\nu\nu)_{V-A} + h.c.   \end{equation}
for the decays $K_L\to\pi^0\nu\bar\nu$, $B\to X_s\nu\bar\nu$
and $B\to X_d\nu\bar\nu$, with $(\bar nn^\prime)=(\bar sd)$, $(\bar bs)$,
$(\bar bd)$ respectively. Similarly
\begin{equation}\label{hyll}
{\cal H}_{eff} = -{G_F\over \sqrt 2} {\alpha \over
2\pi \sin^2 \Theta_W} V^\ast_{tn} V_{tn^\prime}
Y (x_t) (\bar nn^\prime)_{V-A} (\bar ll)_{V-A} + h.c.   \end{equation}
for $B_s\to l^+l^-$ and $B_d\to l^+l^-$, with
$(\bar nn^\prime)=(\bar bs)$, $(\bar bd)$.
The functions $X$, $Y$ are given in \eqn{xx} and \eqn{yy}.

\section{The Effective Hamiltonian for $K^0-\bar K^0$ Mixing}
         \label{sec:HeffKKbar}

\subsection{General Structure}
\label{sec:HeffKKbar:General}
The following chapter is devoted to the presentation of the
effective hamiltonian for $\Delta S=2$ transitions. This hamiltonian
incorporates the short-distance physics contributing to
$K^0-\bar K^0$ mixing and is essential for the description of
CP violation in the neutral K-meson system.
\\
Being a FCNC process, $K^0-\bar K^0$ mixing can only occur at the
loop level within the Standard Model. To lowest order it is induced
through the box diagrams in fig.\ \ref{fig:oporig}\,(e).
Including QCD corrections the
effective low energy hamiltonian, to be derived from these diagrams,
can be written as follows $(\lambda_i = V_{is}^* V_{id}^{})$
\begin{eqnarray}\label{hds2}
{\cal H}^{\Delta S=2}_{eff}&=&\frac{G^2_F}{16\pi^2}M^2_W
 \left[\lambda^2_c\eta_1 S_0(x_c)+\lambda^2_t \eta_2 S_0(x_t)+
 2\lambda_c\lambda_t \eta_3 S_0(x_c, x_t)\right] \times
\nonumber\\
& & \times \left[\as(\mu)\right]^{-2/9}\left[
  1 + \frac{\as(\mu)}{4\pi} J_3\right]  Q + h. c.
\end{eqnarray}
This equation, together with \eqn{eta1}, \eqn{eta2}, \eqn{eta3}
for $\eta_1$, $\eta_2$ and $\eta_3$ respectively, represents the
complete next-to-leading order short-distance hamiltonian for
$\Delta S=2$ transitions. (\ref{hds2}) is valid for scales $\mu$
below the charm threshold $\mu_c=\ord(m_c)$. In this case
${\cal H}^{\Delta S=2}_{eff}$ consists of a single four-quark operator
\begin{equation}\label{qsdsd}
Q=(\bar sd)_{V-A}(\bar sd)_{V-A}
\end{equation}
which is multiplied by the corresponding coefficient function.
It is useful and customary to decompose this function into a
charm-, a top- and a mixed charm-top contribution, as displayed
in (\ref{hds2}). This form is obtained upon eliminating $\lambda_u$
by means of CKM matrix unitarity and setting $x_u=0$. The basic
electroweak loop contributions without QCD correction are then
expressed through the functions $S_0$, which read \cite{inamilim:81}
\begin{equation}\label{s0c}
S_0(x_c)\doteq x_c
\end{equation}
\begin{equation}\label{s0t}
S_0(x_t)=\frac{4x_t-11x^2_t+x^3_t}{4(1-x_t)^2}-
 \frac{3x^3_t \ln x_t}{2(1-x_t)^3}
\end{equation}
\begin{equation}\label{s0ct}
S_0(x_c, x_t)=x_c\left[\ln\frac{x_t}{x_c}-\frac{3x_t}{4(1-x_t)}-
 \frac{3 x^2_t\ln x_t}{4(1-x_t)^2}\right]
\end{equation}
Here again we keep only linear terms in $x_c\ll 1$, but of course
all orders in $x_t$.
\\
Short-distance QCD effects are described through the correction
factors $\eta_1$, $\eta_2$, $\eta_3$ and the explicitly
$\as$-dependent terms in (\ref{hds2}). The discussion of
these corrections will be the subject of the following sections.
\\
Without QCD, i.e. in the limit $\as\to 0$, one has
$\eta_i [\as]^{-2/9}\to 1$. In general, the complete
coefficient function multiplying $Q$ in (\ref{hds2}) contains the
QCD effects at high energies $\mu_W=\ord(M_W)$,
$\mu_t=\ord(m_t)$ together with their RG evolution down to the
scale $\mu=\ord(1\gev)$. A common ingredient for the three
different contributions in (\ref{hds2}) is the anomalous dimension
of the operator $Q$ and the corresponding evolution of its coefficient.
The Fierz symmetric flavor structure of $Q$ implies that it acquires
the same anomalous dimension as the Fierz symmetric operator
$Q_+=(Q_2+Q_1)/2$ (see section \ref{sec:HeffdF1:22}), explicitly
\begin{equation}\label{gamq1}
\gamma=\frac{\as}{4\pi}\gamma^{(0)}+
       \left(\frac{\as}{4\pi}\right)^2\gamma^{(1)}
\end{equation}
\begin{equation}\label{gamq2}
\gamma^{(0)}=6\frac{N-1}{N}\qquad
\gamma^{(1)}=\frac{N-1}{2N}\left[-21+\frac{57}{N}-\frac{19}{3}N+
    \frac{4}{3} f\right]\quad ({\rm NDR})
\end{equation}
The resulting evolution of the coefficient of $Q$ between general
scales $\mu_1$ and $\mu$ then reads
\begin{equation}\label{cqmu}
C_Q(\mu)=\left[1+\frac{\as(\mu)-\as(\mu_1)}{4\pi}J_f\right]
  \left[\frac{\as(\mu_1)}{\as(\mu)}\right]^{d_f}
  C_Q(\mu_1)
\end{equation}
where
\begin{equation}\label{zd0}
d_f=\frac{\gamma^{(0)}}{2\beta_0} \qquad
J_f= \frac{d_f}{\beta_0}\beta_1 - \frac{\gamma^{(1)}}{2\beta_0}
\end{equation}
depend on the number of active flavors $f$. At the lower end of the
evolution $f=3$. The terms in (\ref{cqmu}) depending on
$\as(\mu)$ are factored out explicitly in (\ref{hds2}) to
exhibit the $\mu$-dependence of the coefficient function in the
$f=3$ regime, which has to cancel the corresponding $\mu$-dependence
of the hadronic matrix element of $Q$ between meson states in
physical applications. A similar comment applies to the scheme
dependence entering $J_f$ in (\ref{zd0}) through the scheme
dependence of $\gamma^{(1)}$. Splitting off the $\mu$-dependence in
(\ref{hds2}) is of course not unique. The way it is done belongs to the
{\em definition} of the $\eta_i$-factors.
\\
Let us finally compare the structure of (\ref{hds2}) with the
effective hamiltonians for rare decays discussed in chapter
\ref{sec:HeffRareKB}. Common features of both types of processes
include:
\begin{itemize}
\item
Both are generated to lowest order via electroweak FCNC loop
transitions involving heavy quarks.
\item
They contain a charm and a top contribution.
\item
The hamiltonian consists of a single dimension-6 operator.
\end{itemize}
Besides these similarities, however, there are also a few important
differences, which have their root in the fact that the $\Delta S=2$
box diagrams involve two distinct quark lines as compared to the single
quark line in semileptonic rare decays:
\begin{itemize}
\item
The CKM parameter combinations $\lambda_i$ appear quadratically in
(\ref{hds2}) instead of only linearly.
\item
(\ref{hds2}) in addition receives contributions from a mixed
top-charm sector. This part in fact turns out to have the most
involved structure of the three contributions.
\item
The operator $Q$ has a non-vanishing QCD anomalous dimension, resulting
in a non-trivial scale and scheme dependence of the Wilson
coefficient.
\item
The hadronic matrix element of the four-quark operator $Q$ is
a considerably more complicated object than the quark current
matrix elements in semileptonic rare decays.
\end{itemize}
We will now present the complete next-to-leading order results for
$\eta_2$, $\eta_1$ and $\eta_3$ in turn and discuss their most
important theoretical features. The first leading log calculations of
$\eta_1$ have been presented in \cite{vainshteinetal:76},
\cite{novikovetal:77} and of $\eta_2$ in \cite{vysotskij:80}. The
complete leading log calculation inlcuding also $\eta_3$ has been first
performed in \cite{gilmanwise:83}.  Leading order calculations in the
presence of a heavy top can be found in \cite{kaufmanetal:89},
\cite{flynn:90}, \cite{dattaetal:90} and \cite{dattaetal:95}.

\subsection{The Top Contribution -- $\eta_2$}
\label{sec:HeffKKbar:eta2}

The basic structure of the top quark sector in
${\cal H}^{\Delta S=2}_{eff}$ is easy to understand. First the
top quark is integrated out, along with the $W$, at a matching
scale $\mu_t=\ord(m_t)$, leaving a $m_t$-dependent coefficient
normalized at $\mu_t$, multiplying the single operator $Q$.
Subsequently the coefficient is simply renormalized down to
scales $\mu=\ord(1\gev)$ by means of (\ref{cqmu}). Including
NLO corrections the resulting QCD factor $\eta_2$ from (\ref{hds2})
may be written (in $\overline{MS}$) as follows \cite{burasjaminweisz:90}
\begin{eqnarray}\label{eta2}
\eta_2&=&\left[\as(m_c)\right]^{6/27}
 \left[\frac{\as(m_b)}{\as(m_c)}\right]^{6/25}
 \left[\frac{\as(\mu_t)}{\as(m_b)}\right]^{6/23}
\\
& & \cdot\left[1+\frac{\as(m_c)}{4\pi}(J_4-J_3)+
         \frac{\as(m_b)}{4\pi}(J_5-J_4) \right.
\nonumber\\
& & + \left.\frac{\as(\mu_t)}{4\pi} \left(
 \frac{S_1(x_t)}{S_0(x_t)}+B_t-J_5+\frac{\gamma^{(0)}}{2}
 \ln\frac{\mu^2_t}{M^2_W}+\gamma_{m0}
 \frac{\partial\ln S_0(x_t)}{\partial\ln x_t}\ln\frac{\mu^2_t}{M^2_W}
\right)\right] \nonumber
\end{eqnarray}
where $\gamma_{m0} = 6 C_F$,
\begin{equation}\label{btndr}
B_t=5\frac{N-1}{2N}+3\frac{N^2-1}{2N}\qquad ({\rm NDR})
\end{equation}
and
\begin{equation}\label{s1x}
S_1(x)=\frac{N-1}{2N}S^{(8)}_1(x)+\frac{N^2-1}{2N}S^{(1)}_1(x)
\end{equation}
\begin{eqnarray}\label{s1x8}
S^{(8)}_1(x)=&-&\frac{64-68x-17x^2+11x^3}{4(1-x)^2}+
                \frac{32-68x+32x^2-28x^3+3x^4}{2(1-x)^3}\ln x
\nonumber\\
&+&\frac{x^2(4-7x+7x^2-2x^3)}{2(1-x)^4}\ln^2 x+
   \frac{2x(4-7x-7x^2+x^3)}{(1-x)^3}L_2(1-x)
\nonumber\\
&+&\frac{16}{x}\left(\frac{\pi^2}{6}-L_2(1-x)\right)
\end{eqnarray}
\begin{eqnarray}\label{s1x1}
S^{(1)}_1(x)=&-&\frac{x(4-39x+168x^2+11x^3)}{4(1-x)^3}-
                \frac{3x(4-24x+36x^2+7x^3+x^4)}{2(1-x)^4}\ln x
\nonumber\\
&+&\frac{3x^3(13+4x+x^2)}{2(1-x)^4}\ln^2 x-
   \frac{3x^3(5+x)}{(1-x)^3}L_2(1-x)
\end{eqnarray}
where the dilogarithm $L_2$ is defined in \eqn{l2}.

In the expression (\ref{eta2}) we have taken into account the heavy
quark thresholds at $m_b$ and $m_c$ in the RG evolution. As it must
be, the dependence on the threshold scales is of the neglected
order $\ord(\alpha^2_s)$. In fact the threshold ambiguity is here of
$\ord(\as^2)$ also in LLA since $\gamma^{(0)}$ is flavor independent.
It turns out that this dependence is also very weak numerically and
we therefore set $\mu_c=m_c$ and $\mu_b=m_b$. Furthermore it is
a good approximation to neglect the b-threshold completely
using an effective 4-flavor theory from $\mu_t$ down to $m_c$. This
can be achieved by simply substituting $m_b\to\mu_t$ in (\ref{eta2}).
\\
The leading order expression for $\eta_2$ is given by the first three
factors on the r.h.s. of (\ref{eta2}). The fourth factor represents
the next-to-leading order generalization. Let us discuss now the
most interesting and important features of the NLO result for $\eta_2$
exhibited in (\ref{eta2}).
\begin{itemize}
\item
$\eta_2$ is proportional to the initial value of the Wilson
coefficient function at $\mu_t=\mw$
\begin{equation}\label{s01x}
S(x)=S_0(x)+\frac{\as}{4\pi}\left(S_1(x)+B_t S_0(x)\right)
\end{equation}
which has to be extracted from the box graphs in fig.\
\ref{fig:oporig}\,(e) and the corresponding gluon correction diagrams
after a proper factorization of long- and short-distance
contributions.
\item
$S(x)$ in (\ref{s01x}) is similar to the functions $X(x)$ and
$Y(x)$ in sections \ref{sec:HeffRareKB:kpnn:heff} and
\ref{sec:HeffRareKB:klmm:heff} except that $S(x)$ is scheme
dependent due to the renormalization that is required for the
operator $Q$. This scheme dependence enters (\ref{s01x}) through the
scheme dependent constant $B_t$, given in the NDR scheme in
(\ref{btndr}). This scheme dependence is canceled in the combination
$B_t-J_5$ by the two-loop anomalous dimension contained in $J_5$.
Likewise the scheme dependence of $J_f$ cancels in the differences
$(J_{f}-J_{f-1})$ as is evident from the discussion of section
\ref{sec:basicform:wc:rgdep}.
\item
A very important point is the dependence on the high energy matching
scale $\mu_t$. This dependence enters the NLO
$\as(\mu_t)$-correction in (\ref{eta2}) in two distinct ways:
First as a term proportional to $\gamma^{(0)}$ and, secondly, in
conjunction with $\gamma_{m0}$. The first of these terms cancels to
$\ord(\as)$ the $\mu_t$-dependence present in the leading
term $[\as(\mu_t)]^{6/23}$. The second, on the other hand, leads
to an $\ord(\as)$ $\mu_t$-dependence of $\eta_2$ which is
just the one needed to cancel the $\mu_t$-ambiguity of the leading
function $S_0(x_t(\mu_t))$ in the product $\eta_2 S_0(x_t)$, such
that in total physical results become independent of $\mu_t$ to
$\ord(\as)$. From these observations it is obvious that one
may interpret $\mu_t$ in the first case as the initial scale of the
RG evolution and in the second case as the scale at which the top
quark mass is defined. These two scales need not necessarily have the
same value.
\\
The important point is, that to leading logarithmic accuracy the
$\mu_t$-de\-pen\-dence of both $\eta^{LO}_2(\mu_t)$ and
$S_0(x_t(\mu_t))$ remains uncompensated, leaving a dis\-tur\-bing\-ly
large uncertainty in the short-distance calculation.
\item
It is interesting to note that in the limit $\mt\gg \mw$ the dependence
on $\mu_t$ enters $\eta_2$ as $\ln\mu_t/\mt$, rather than
$\ln\mu_t/\mw$. This feature provides a formal justification for
choosing $\mu_t=\ord(\mt)$ instead of $\mu_t=\ord(\mw)$. An explicit
expression for the large $\mt$ limit in the similar case of $\eta_{2B}$
may be found in section \ref{sec:HeffBBbar}.
\item
Although at NLO the product $\eta_2 S_0(x_t)$ depends only very weakly
on the precise value of $\mu_t$ as long as it is of $\ord(m_t)$, the
choice $\mu_t=m_t$ is again convenient:  With this choice $\eta_2$
becomes almost independent of the top quark mass $m_t(m_t)$. By
contrast, for $\mu_t=M_W$, say, $\eta_2$ would decrease with rising
$m_t(m_t)$ in order to compensate for the increase of $S_0(x_t(M_W))$
due to the use of a -- for high $m_t$ -- ``unnaturally'' low scale
$M_W$.
\item
As mentioned above the dependence of the Wilson coefficient on the
low energy scale $\mu$ and the remaining scheme dependence ($J_3$)
has been factored out explicitly in (\ref{hds2}). Therefore the
QCD correction factor $\eta_2$ is {\it by definition\/}
scale and scheme independent on the lower end of the RG evolution.
\end{itemize}

\subsection{The Charm Contribution -- $\eta_1$}
\label{sec:HeffKKbar:eta1}

The calculation of $\eta_1$ beyond leading logs has been presented in
great detail in \cite{herrlichnierste:93}, \cite{herrlich:94}. Our task
here will be to briefly describe the basic procedure and to summarize
the main results.
\\
In principle the charm contribution is similar in structure to the
top contribution. However, since the quark mass $m_c\ll M_W$,
the charm degrees of freedom can no longer be integrated out
simultaneously with the $W$ boson, which would introduce large
logarithmic corrections $\sim\as\ln M_W/m_c$. To resum these logarithms
one first constructs an effective theory at a scale $\ord(M_W)$, where
the $W$ boson is removed. The relevant operators are subsequently
renormalized down to scales $\mu_c=\ord(m_c)$, where the charm quark is
then integrated out. After this step only the operator $Q$
(\ref{qsdsd}) remains and $\eta_1$ is finally obtained as discussed in
section \ref{sec:HeffKKbar:General}.
\\
Let us briefly outline this procedure for the case at hand.
After integrating out $W$ the effective hamiltonian to first order
in weak interactions, which is needed for the charm contribution,
can be written as
\begin{equation}\label{hc1}
{\cal H}^{(1)}_c=\frac{G_F}{\sqrt 2}\sum_{q, q^\prime=u, c}
 V^\ast_{q^\prime s}V_{qd}\left(C_+Q^{q^\prime q}_+ +
 C_-Q^{q^\prime q}_- \right)
\end{equation}
where we have introduced the familiar $\Delta S=1$ four-quark
operators in the multiplicatively renormalizable basis
\begin{equation}\label{qqpm}
Q^{q^\prime q}_\pm=\frac{1}{2}\left[
(\bar s_iq^\prime_i)_{V-A}(\bar q_jd_j)_{V-A}\pm
(\bar s_iq^\prime_j)_{V-A}(\bar q_jd_i)_{V-A} \right]
\end{equation}
We remark that no penguin operators appear in the present case
due to GIM cancellation between charm quark and up quark
contributions.
\\
$\Delta S=2$ transitions occur to second order in the effective
interaction (\ref{hc1}). The $\Delta S=2$ effective hamiltonian
is therefore given by
\begin{equation}\label{heff2c}
{\cal H}^{\Delta S=2}_{eff,c}=-\frac{i}{2}\int d^4x\
T \left( {\cal H}^{(1)}_c(x) {\cal H}^{(1)}_c(0) \right)
\end{equation}
Inserting (\ref{hc1}) into (\ref{heff2c}), keeping only pieces that can
contribute to the charm box diagrams and taking the GIM constraints
into account, one obtains
\begin{equation}\label{hds2c}
{\cal H}^{\Delta S=2}_{eff,c}=\frac{G^2_F}{2}\lambda^2_c
 \sum_{i,j=+,-}C_i C_j O_{ij}
\end{equation}
where
\begin{equation}\label{oijdef}
O_{ij}=-\frac{i}{2}\int d^4x\ T\left[
Q^{cc}_i(x)Q^{cc}_j(0)-Q^{uc}_i(x)Q^{cu}_j(0)-
Q^{cu}_i(x)Q^{uc}_j(0)+Q^{uu}_i(x)Q^{uu}_j(0) \right]
\end{equation}
From the derivation of (\ref{hds2c}) it is evident, that the
Wilson coefficients of the bilocal operators $O_{ij}$ are simply given
by the product $C_i C_j$ of the coefficients pertaining to the
local operators $Q_i$, $Q_j$. The evolution of the $C_i$ from $M_W$
down to $\mu_c$ proceeds in the standard fashion and is described
by equations of the type shown in (\ref{cqmu}) with the
appropriate anomalous dimensions inserted. In the following we list
the required ingredients.
\\
The Wilson coefficients at scale $\mu=M_W$ read
\begin{equation}\label{cpmw}
C_\pm(M_W)=1+\frac{\as(M_W)}{4\pi}B_\pm
\end{equation}
\begin{equation}\label{bpmn}
B_\pm=\pm 11\frac{N\mp 1}{2N}   \qquad {\rm (NDR)}
\end{equation}
The two-loop anomalous dimensions are
\begin{equation}\label{gapm1}
\gamma_\pm=\frac{\as}{4\pi}\gamma^{(0)}_\pm+
       \left(\frac{\as}{4\pi}\right)^2\gamma^{(1)}_\pm
\end{equation}
\begin{equation}\label{gapm2}
\gamma^{(0)}_\pm=\pm 6\frac{N\mp 1}{N}\quad
\gamma^{(1)}_\pm=\frac{N\mp 1}{2N}\left[-21\pm\frac{57}{N}\mp\frac{19}{3}N
 \pm\frac{4}{3} f\right]\quad ({\rm NDR})
\end{equation}
For $i,j=+,-$ we introduce
\begin{equation}\label{dzi}
d^{(f)}_i=\frac{\gamma^{(0)}_i}{2\beta_0} \qquad
J^{(f)}_i= \frac{d^{(f)}_i}{\beta_0}\beta_1 - \frac{\gamma^{(1)}_i}{2\beta_0}
\end{equation}
and
\begin{equation}\label{dzij}
d^{(f)}_{ij}=d^{(f)}_i+d^{(f)}_j \qquad
J^{(f)}_{ij}=J^{(f)}_i+J^{(f)}_j
\end{equation}
The essential step consists in matching (\ref{hds2c}) onto an
effective theory without charm, which will contain the single
operator $Q=(\bar sd)_{V-A}(\bar sd)_{V-A}$. In NLO this matching
has to be performed to $\ord(\as)$. At a normalization
scale $\mu_c$ it reads explicitly, expressed in terms of operator
matrix elements $(i, j=+, -)$
\begin{equation}\label{oijq}
\langle O_{ij}\rangle=\frac{m^2_c(\mu_c)}{8\pi^2}\left[ \tau_{ij}+
 \frac{\as(\mu_c)}{4\pi}\left( \kappa_{ij}
 \ln\frac{\mu^2_c}{m^2_c}+\beta_{ij}\right)\right]
 \langle Q\rangle
\end{equation}
\begin{equation}\label{tauij}
\tau_{++}=\frac{N+3}{4}
\qquad
\tau_{+-}=\tau_{-+}=-\frac{N-1}{4}
\qquad
\tau_{--}=\frac{N-1}{4}
\end{equation}
\begin{equation}\label{kapij}
\kappa_{++}=3(N-1)\tau_{++}
\qquad
\kappa_{+-}=\kappa_{-+}=3(N+1)\tau_{+-}
\qquad
\kappa_{--}=3(N+3)\tau_{--}
\end{equation}
The $\beta_{ij}$ are scheme dependent. In the NDR scheme they are
given by \cite{herrlichnierste:93}
\begin{eqnarray}\label{betij}
\beta_{++} &=& (1-N)\left(\frac{N^2-6}{12N}\pi^2+
 3\frac{-N^2+2N+13}{4N} \right)  \nonumber \\
\beta_{+-}=\beta_{-+} &=& (1-N)\left(\frac{-N^2+2N-2}{12N}\pi^2+
 \frac{3N^2+13}{4N} \right)    \\
\beta_{--} &=& (1-N)\left(\frac{N^2-4N+2}{12N}\pi^2-
 \frac{3N^2+10N+13}{4N} \right)  \nonumber
\end{eqnarray}
Now, starting from (\ref{hds2c}), evolving $C_i$ from $M_W$ down to
$\mu_c$, integrating out charm at $\mu_c$ with the help of
(\ref{oijq}), evolving the resulting coefficient according to
(\ref{cqmu}) and recalling the definition of $\eta_1$ in
(\ref{hds2}), one finally obtains
\begin{eqnarray}\label{eta1}
\eta_1=&&[\as(\mu_c)]^{d_3}\sum_{i,j=+,-}
\left(\frac{\as(m_b)}{\as(\mu_c)}\right)^{d^{(4)}_{ij}}
\left(\frac{\as(M_W)}{\as(m_b)}\right)^{d^{(5 )}_{ij}}\times
\nonumber \\
&&\times\left[\tau_{ij}+\frac{\as(\mu_c)}{4\pi}\left(\kappa_{ij}
\ln\frac{\mu^2_c}{m^2_c}+\tau_{ij}(J^{(4)}_{ij}-J_3)+\beta_{ij}
\right)+\right. \nonumber \\
&&\left. +\tau_{ij}\left(\frac{\as(m_b)}{4\pi}
(J^{(5)}_{ij}-J^{(4)}_{ij})+\frac{\as(M_W)}{4\pi}
(B_i+B_j-J^{(5)}_{ij})\right)\right]
\end{eqnarray}
We conclude this section with a discussion of a few important
issues concerning the structure of this formula.
\begin{itemize}
\item
(\ref{eta1}), as first obtained in \cite{herrlichnierste:93}, represents the
next-to-leading order generalization of the leading log expression
for $\eta_1$ given in \cite{gilmanwise:83}. The latter follows as a
special case of (\ref{eta1}) when the $\ord(\as)$ correction
terms are put to zero.
\item
The expression (\ref{eta1}) is independent of the renormalization
scheme up to terms of the neglected order $\ord(\alpha^2_s)$.
We have written $\eta_1$ in a form, in which this scheme
independence becomes manifest: While the various $J$-terms, $B_i$ and
$\beta_{ij}$ in (\ref{eta1}) all depend on the renormalization
scheme when considered separately, the combinations
$\tau_{ij}(J^{(4)}_{ij}-J_3)+\beta_{ij}$,
$J^{(5)}_{ij}-J^{(4)}_{ij}$ and $B_i+B_j-J^{(5)}_{ij}$
are scheme invariant.
\item
The product $\eta_1(\mu_c) x_c(\mu_c)$ is independent of $\mu_c$
to the considered order,
\begin{equation}
\frac{d}{d\ln\mu_c}\eta_1(\mu_c) x_c(\mu_c)=\ord(\alpha^2_s)
\end{equation}
in accordance with the requirements of renormalization group
invariance. The cancellation of the $\mu_c$-dependence to
$\ord(\as)$ is related to the presence of an explicitly
$\mu_c$-dependent term at NLO in (\ref{eta1}) and is
guaranteed through the identity
\begin{equation}\label{kapid}
\kappa_{ij}=\tau_{ij}\left(\gamma_{m0}+\frac{\gamma^{(0)}}{2}-
\frac{\gamma^{(0)}_i+\gamma^{(0)}_j}{2}\right)
\end{equation}
which is easily verified using (\ref{gm01}), (\ref{gamq2}),
(\ref{gapm2}), (\ref{tauij}) and (\ref{kapij}).
\item
Also the ambiguity in the scale $\mu_W$, at which $W$ is integrated out,
is reduced from $\ord(\as)$ to $\ord(\alpha^2_s)$
when going from leading to next-to-leading order. As mentioned
above the dependence on the $b$-threshold scale $\mu_b$ is
$\ord(\alpha^2_s)$ in NLLA as well as in LLA.
Numerically the dependence on $\mu_b$ is very small. Also the
variation of the result with the high energy matching scale $\mu_W$
is considerably weaker than the residual dependence on $\mu_c$.
Therefore we have set $\mu_b=m_b$ and $\mu_W=M_W$ in (\ref{eta1}).
In numerical analyses we will take the dominant $\mu_c$-dependence
as representative for the short-distance scale ambiguity of $\eta_1$.
The generalization to the case $\mu_W\not= M_W$ is discussed in
\cite{herrlichnierste:93}. The more general case $\mu_b\not= m_b$ is trivially
obtained by substituting $m_b\to\mu_b$ in (\ref{eta1}).
\item
Note that due to the GIM structure of $O_{ij}$ no mixing under infinite
renormalization occurs between $O_{ij}$ and the local operator $Q$.
This is related to the absence of the logarithm in the function
$S_0(x_c)$ in \eqn{s0c}.
\end{itemize}
It is instructive to compare the results obtained for $\eta_1$
and $\eta_2$. Expanding (\ref{eta1}) to first order in $\as$,
in this way ``switching off'' the RG summations, we find
\begin{eqnarray}\label{e1lim}
&&[\as(\mu)]^{-2/9}\left(1+\frac{\as(\mu)}{4\pi}J_3\right)
\eta_1\doteq  \\
&& 1+\frac{\as}{4\pi}\left[\frac{\gamma^{(0)}}{2}\left(
\ln\frac{\mu^2}{M^2_W}+\ln\frac{m^2}{M^2_W}-1+\frac{2}{9}\pi^2\right)
+\gamma_{m0}\left(\ln\frac{\mu^2}{m^2}+\frac{1}{3}\right)\right]
\nonumber
\end{eqnarray}
where we have replaced $\mu_c\to\mu$ and $m_c\to m$. In deriving
(\ref{e1lim}) besides (\ref{kapid}) the following identities are
useful
\begin{equation}\label{tauid}
\sum_{i,j=+,-}\tau_{ij}=1\qquad
\sum_{i,j=+,-}\tau_{ij}\frac{\gamma^{(0)}_i+\gamma^{(0)}_j}{2}=
\gamma^{(0)}
\end{equation}
\begin{equation}\label{bid}
\sum_{i,j=+,-}\tau_{ij}(B_i+B_j)=2 B_+
\end{equation}
The same result (\ref{e1lim}) is obtained from $\eta_2$ as well,
if we set $m_c=m_b=\mu_t=\mu$, $m_t=m$ in (\ref{eta2}) and expand
for $m\ll M_W$. This exercise yields a useful cross-check between
the calculations for $\eta_1$ and $\eta_2$. In addition it gives
some further insight into the structure of the QCD corrections to
$\Delta S=2$ box diagrams, establishing $\eta_1$ and $\eta_2$ as two
different generalizations of the same asymptotic limit (\ref{e1lim}).

\subsection{The Top-Charm Contribution -- $\eta_3$}
\label{sec:HeffKKbar:eta3}
To complete the description of the $K^0-\bar K^0$ effective hamiltonian we
now turn to the mixed top-charm component, defined by the contribution
$\sim\lambda_c\lambda_t$ in (\ref{hds2}), and the associated QCD
correction factor $\eta_3$. The short distance QCD effects have been
first obtained within the leading log approximation by
\cite{gilmanwise:83}. The calculation of $\eta_3$ at next-to-leading
order is due to the work of \cite{herrlichnierste:95},
\cite{nierste:95}. As already mentioned, the renormalization group
analysis necessary for $\eta_3$ is more involved than in the cases of
$\eta_1$ and $\eta_2$. The characteristic differences will become clear
from the following presentation.
\\
We begin by writing down the relevant $\Delta S=1$ hamiltonian,
obtained after integrating out $W$ and top, which provides the basis
for the construction of the $\Delta S=2$ effective hamiltonian we want
to derive. It reads
\begin{equation}\label{hct1}
{\cal H}^{(1)}_{ct}=\frac{G_F}{\sqrt 2}\left(\sum_{q, q^\prime=u, c}
 V^\ast_{q^\prime s}V_{qd}\sum_{i=1,2}C_iQ^{q^\prime q}_i -
 \lambda_t\sum_{i=3}^6 C_i Q_i \right)
\end{equation}
with
\begin{equation}\label{qpq12}
Q^{q^\prime q}_1=(\bar s_iq^\prime_j)_{V-A}(\bar q_jd_i)_{V-A} \qquad
Q^{q^\prime q}_2=(\bar s_iq^\prime_i)_{V-A}(\bar q_jd_j)_{V-A}
\end{equation}
and corresponds to the hamiltonian (\ref{eq:HeffKppc}), discussed in
chapter \ref{sec:HeffdF1:66}, except that we have included the
$\Delta C=1$ components $Q^{uc}_i$, $Q^{cu}_i$, which contribute in the
analysis of $\eta_3$.
By contrast to the simpler case of $\eta_1$ presented in the previous
section, now also the penguin operators $Q_i$, $i=3,\ldots, 6$
(\ref{eq:Kppbasis}) have to be considered. Being proportional to
$\lambda_t=V^\ast_{ts}V_{td}$ they will contribute to the
$\lambda_c\lambda_t$-part of (\ref{hds2}). We remark in this context
that, on the other hand, the penguin contribution to the 
$\lambda^2_t$-sector is entirely negligible. Since only light quarks
are involved in $Q_3,\ldots ,Q_6$, the double penguin diagrams, which 
would contribute to the $\lambda^2_t$-piece of the $\Delta S=2$
amplitude, are suppressed by at least a factor of $m^2_b/m^2_t$
compared with the dominant top-exchange effects discussed in section
\ref{sec:HeffKKbar:eta2}.
\\
At second order in (\ref{hct1}) $\Delta S=2$ transitions are generated.
Inserting (\ref{hct1}) in an expression similar to (\ref{heff2c}),
eliminating $\lambda_u$ by means of $\lambda_u=-\lambda_c-\lambda_t$
and collecting the terms proportional to $\lambda_c\lambda_t$, we
obtain the top-charm component of the effective $\Delta S=2$
hamiltonian in the form
\begin{equation}\label{hds2ct}
{\cal H}^{\Delta S=2}_{eff,ct}=\frac{G^2_F}{2}\lambda_c\lambda_t
 \sum_{i=\pm}\left[\sum_{j=1}^6 C_i C_j Q_{ij}
 +C_{7i} Q_7\right]
\end{equation}
where   
\begin{equation}\label{qij12}
Q_{ij}=-\frac{i}{2}\int d^4x\ T\left[
2Q^{uu}_i(x)Q^{uu}_j(0)-Q^{uc}_i(x)Q^{cu}_j(0)-
Q^{cu}_i(x)Q^{uc}_j(0) \right]
\end{equation}
for $j=1, 2$ and
\begin{equation}\label{qij36}
Q_{ij}=-\frac{i}{2}\int d^4x\ T\left[
\left(Q^{uu}_i(x)-Q^{cc}_i(x)\right)Q_j(0)+
Q_j(x)\left(Q^{uu}_i(0)-Q^{cc}_i(0)\right)\right]
\end{equation}
for $j=3, \ldots,6$.
\\
In defining these operators we have already omitted bilocal products
with flavor structure like $(\bar su\bar ud)\cdot(\bar sc\bar cd)$,
which cannot contribute to $\Delta S=2$ box diagrams. Furthermore,
for the factor entering the bilocal operators with index
$i$ we have changed the basis from $Q^{q^\prime q}_{1,2}$ to
$Q^{q^\prime q}_\pm$ given in (\ref{qqpm}).
In addition local counterterms proportional to the $\Delta S=2$
operator
\begin{equation}\label{q7def}
Q_7=\frac{m^2_c}{g^2}(\bar sd)_{V-A}(\bar sd)_{V-A}
\end{equation}
have been added to (\ref{hds2ct}). These are necessary here because
the bilocal operators can in general mix into $Q_7$ under infinite
renormalization, a fact related to the logarithm present in the
leading term $-x_c\ln x_c$ entering $S_0(x_c, x_t)$ in (\ref{s0ct}).
This behaviour is in contrast to the charm contribution, where the
corresponding function $S_0(x_c)=x_c$ does not contain a logarithmic term
and consequently no local $\Delta S=2$ counterterm is necessary in
(\ref{hds2c}). On the other hand the situation here is analogous to the
case of the charm contribution to the effective hamiltonian for
$K^+\to\pi^+\nu\bar\nu$ in section \ref{sec:HeffRareKB:kpnn} which
similarly behaves as $x_c\ln x_c$ in lowest order and correspondingly
requires a counterterm, as displayed in (\ref{hzop}) and (\ref{hbop}).

After integrating out top and $W$ at the high energy matching scale
$\mu_W={\cal O}(M_W)$, the Wilson coefficients $C_j$, $j=1, \ldots 6$
of (\ref{hct1}) and (\ref{hds2ct}) are given in the NDR-scheme by
(see section \ref{sec:HeffdF1:66})
\begin{eqnarray}\label{ctmuw}
\vec C^T(\mu_W) & = & (0, 1, 0, 0, 0, 0)+ 
  \frac{\alpha_s(\mu_W)}{4\pi}
  \left(6, -2, -\frac{2}{9}, \frac{2}{3}, -\frac{2}{9}, \frac{2}{3}
  \right)\ln\frac{\mu_W}{M_W} \nonumber\\ 
  & + & \frac{\alpha_s(\mu_W)}{4\pi}
  \left(\frac{11}{2}, -\frac{11}{6}, -\frac{1}{6}\tilde E_0(x_t),
  \frac{1}{2}\tilde E_0(x_t), -\frac{1}{6}\tilde E_0(x_t),
  \frac{1}{2}\tilde E_0(x_t)\right)
\end{eqnarray}
and $C_\pm=C_2\pm C_1$. $\tilde E_0(x_t)$ can be found in
(\ref{eq:Exttilde}).
The coefficient of $Q_7$ is obtained through matching the $\Delta S=2$
matrix element of the effective theory (\ref{hds2ct}) to the
corresponding full theory matrix element, which is in the required
approximation ($x_c\ll 1$) given by (compare (\ref{hds2}))
\begin{equation}\label{afullct}
A_{full, ct}=\frac{G^2_F}{16\pi^2}M^2_W 2\lambda_c\lambda_t
S_0(x_c,x_t)\langle Q\rangle
\end{equation}
At next-to-leading order this matching has to be done to one loop,
including finite parts. Note that here the loop effect is due to
electroweak interactions and QCD does not contribute explicitly in this
step.  The matching condition determines the sum $C_7\equiv
C_{7+}+C_{7-}$, which in the NDR scheme and with the conventional
definition of evanescent operators, \cite{burasweisz:90}, see also
\cite{herrlichnierste:95}, \cite{nierste:95}, reads
\begin{equation}\label{c7muw}
C_7(\mu_W)=\frac{\alpha_s(\mu_W)}{4\pi}\left[-8\ln\frac{\mu_W}{M_W}+
4\ln x_t-\frac{3x_t}{1-x_t}-\frac{3x^2_t\ln x_t}{(1-x_t)^2}+2\right]
\end{equation}
at next-to-leading order.
In leading log approximation one simply would have $C_7(\mu_W)=0$.
\\
The distribution of $C_7$ among $C_{7+}$ and $C_{7-}$ is arbitrary
and has no impact on the physics. For example we may choose
\begin{equation}\label{c7pm}
C_{7+}=C_7\qquad\qquad  C_{7-}=0
\end{equation}
Having determined the initial values of the Wilson coefficients
\begin{equation}\label{cvpmt}
\vec C^{(\pm)T}\equiv (C_\pm C_1, \ldots, C_\pm C_6, C_{7\pm})
\end{equation}
at a scale $\mu_W$, $\vec C^{(\pm)}(\mu_W)$, the next step consists
in solving the RG equations to determine $\vec C^{(\pm)}(\mu_c)$
at the charm mass scale $\mu_c={\cal O}(m_c)$. The renormalization
group evolution of $\vec C^{(\pm)}$ is given by
\begin{equation}\label{rgct}
\frac{d}{d\ln\mu}\vec C^{(\pm)}(\mu)=\gamma^{(\pm)T}_{ct}
  \vec C^{(\pm)}(\mu)
\end{equation}
\begin{equation}\label{gpmct}
\gamma^{(\pm)}_{ct}=
\left( \begin{array}{cc}
\gamma_s+\gamma_\pm\cdot 1 & \vec\gamma_{\pm 7}\\
\vec 0^T  & \gamma_{77}  \end{array}  \right)
\end{equation}
Here $\gamma_s$ is the standard $6\times 6$ anomalous dimension matrix
for the $\Delta S=1$ effective hamiltonian including QCD penguins
from (\ref{eq:gsexpKpp}), (\ref{eq:gs0Kpp}) and (\ref{eq:gs1ndrN3Kpp})
(NDR-scheme). Similarly $\gamma_\pm$ are the anomalous dimensions of the
current-current operators. They can be obtained as
$\gamma_\pm=\gamma_{s,11}\pm\gamma_{s,12}$ and are also given in
section \ref{sec:HeffdF1:22}.
\\
$\gamma_{77}$ represents the anomalous dimension of $Q_7$ (\ref{q7def})
and reads
\begin{equation}\label{g77}
\gamma_{77}=\gamma_++2\gamma_m+2\beta(g)/g=
\frac{\alpha_s}{4\pi}\gamma^{(0)}_{77}+
\left(\frac{\alpha_s}{4\pi}\right)^2\gamma^{(1)}_{77}
\end{equation}
For $N=3$ and in NDR
\begin{equation}\label{g7701}
\gamma^{(0)}_{77}=-2+\frac{4}{3}f\qquad
\gamma^{(1)}_{77}=\frac{175}{3}+\frac{152}{9}f
\end{equation}
Finally $\vec\gamma_{\pm 7}$, the vector of anomalous dimensions expressing
the mixing of the bilocal operators $Q_{\pm i}$ ($i=1,\ldots,6$)
into $Q_7$, is given by
\begin{equation}\label{gpm7}
\vec\gamma_{\pm 7}=
\frac{\alpha_s}{4\pi}\vec\gamma^{(0)}_{\pm 7}+
\left(\frac{\alpha_s}{4\pi}\right)^2\vec\gamma^{(1)}_{\pm 7}
\end{equation}
where
\begin{equation}\label{gp70}
\vec\gamma^{(0)T}_{+7}=
\left(-16,-8,-32,-16,32,16\right)
\end{equation}
\begin{equation}\label{gm70}
\vec\gamma^{(0)T}_{-7}=
\left(8,0,16,0,-16,0\right)
\end{equation}
\begin{equation}\label{gp71}
\vec\gamma^{(1)T}_{+7}=
\left(-212,-28,-424,-56,\frac{1064}{3},\frac{832}{3}\right)
\end{equation}
\begin{equation}\label{gm71}
\vec\gamma^{(1)T}_{-7}=
\left(276,-92,552,-184,-\frac{1288}{3},0 \right)
\end{equation}
The scheme-dependent numbers in $\vec\gamma^{(1)}_{\pm 7}$ are given
here in the NDR-scheme with the conventional treatment of evanescent
operators as described in \cite{burasweisz:90}, \cite{herrlichnierste:95},
\cite{nierste:95}.
\\
In order to solve the RG equation (\ref{rgct}) it is useful
\cite{herrlichnierste:95}, \cite{nierste:95} to define the
eight-dimensional vector ($\vec C^T=(C_1,\ldots,C_6)$)
\begin{equation}\label{d8def}
\vec D^T=(\vec C^T,C_{7+}/C_+,C_{7-}/C_-)
\end{equation}
which obeys
\begin{equation}\label{rgd8}
\frac{d}{d\ln\mu}\vec D=\gamma^T_{ct}\vec D
\end{equation}
where
\begin{equation}\label{gct8}
\gamma_{ct}=
\left(\begin{array}{ccc}
\gamma_s & \vec\gamma_{+7} & \vec\gamma_{-7} \\
\vec 0^T & \gamma_{77}-\gamma_+ & 0 \\
\vec 0^T & 0 & \gamma_{77}-\gamma_-
\end{array}\right)
\end{equation}
The solution of (\ref{rgd8}) proceeds in the standard fashion as
described in section \ref{sec:basicform:wc:rgf} and has the form
\begin{equation}\label{dcsol}
\vec D(\mu_c)=U_4(\mu_c,\mu_b)M(\mu_b)U_5(\mu_b,\mu_W)\vec D(\mu_W)
\end{equation}
similarly to (\ref{cthr}).
The b-quark-threshold matching matrix $M(\mu_b)$ is an $8\times 8$ 
matrix whose $6\times 6$ submatrix $M_{ij}$, $i,j=1,\ldots,6$
is identical to the matrix $M$ described in section
\ref{sec:HeffdF1:66:Mm}. The remaining elements are $M_{77}=M_{88}=1$
and zero otherwise.
From (\ref{dcsol}) the Wilson coefficients $C_i(\mu_c)$ are
obtained as
\begin{equation}\label{cdrel}
C_i(\mu_c)=D_i(\mu_c)\quad i=1,\ldots,6\qquad
C_7(\mu_c)=C_+(\mu_c)D_7(\mu_c)+C_-(\mu_c)D_8(\mu_c)
\end{equation}

The final step in the calculation of $\eta_3$ consists in removing
the charm degrees of freedom from the effective theory.
Without charm the effective short-distance hamiltonian corresponding
to (\ref{hds2ct}) can be written as 
\begin{equation}\label{hctq}
{\cal H}^{\Delta S=2}_{eff,ct}=\frac{G^2_F}{2}\lambda_c\lambda_t
C_{ct}  Q
\end{equation}
The matching condition is obtained by equating the matrix elements of
(\ref{hds2ct}) and (\ref{hctq}), evaluated at a scale 
$\mu_c={\cal O}(m_c)$. At next-to-leading order one needs the finite
parts of the matrix elements of $Q_{ij}$, which can be written in the
form
\begin{equation}\label{qijq}
\langle Q_{ij}(\mu_c)\rangle=\frac{m^2_c(\mu_c)}{8\pi^2}
r_{ij}(\mu_c)\langle Q\rangle
\end{equation}
where in the renormalization scheme described above after eq. (\ref{gm71})
the $r_{ij}$ are given by
\begin{equation}\label{rijt}
r_{ij}(\mu_c)=
\left\{ \begin{array}{ll}
(4\ln(\mu_c/m_c)-1)\tau_{ij} & j=1,2 \\
(8\ln(\mu_c/m_c)-4)\tau_{ij} & j=3,4 \\
(-8\ln(\mu_c/m_c)+4)\tau_{ij} & j=5,6 \end{array}\right.
\end{equation}
\begin{equation}\label{tjodd}
\tau_{\pm 1}=\tau_{\pm 3}=\tau_{\pm 5}=(1\pm 3)/2
\end{equation}
\begin{equation}\label{tjevn}
\tau_{+j}=1\qquad \tau_{-j}=0\qquad  \mbox{$j$ even}
\end{equation}
Using (\ref{qijq}), the matching condition at $\mu_c$ between
(\ref{hds2ct}) and (\ref{hctq}) implies
\begin{equation}\label{cctmuc}
C_{ct}(\mu_c)=\sum_{i=\pm}\sum_{j=1}^6 C_i(\mu_c) C_j(\mu_c)
\frac{m^2_c(\mu_c)}{8\pi^2}r_{ij}(\mu_c)+C_7(\mu_c)
\frac{m^2_c(\mu_c)}{4\pi\alpha_s(\mu_c)}
\end{equation}
Evolving $C_{ct}$ from $\mu_c$ to $\mu<\mu_c$ in a three-flavor theory
using (\ref{cqmu}) and comparing (\ref{hctq}) with (\ref{hds2}), 
we obtain the final result
\begin{displaymath}
\eta_3=\frac{x_c(\mu_c)}{S_0(x_c(\mu_c),x_t(\mu_W))}\alpha_s(\mu_c)^{2/9}
\Bigl[ \frac{\pi}{\alpha_s(\mu_c)}C_7(\mu_c)
\left(1-\frac{\alpha_s(\mu_c)}{4\pi}J_3\right)+ 
\end{displaymath}
\begin{equation}\label{eta3}
 +\frac{1}{2}\sum_{i=\pm}\sum_{j=1}^6 C_i(\mu_c) C_j(\mu_c)
r_{ij}(\mu_c)\Bigr] 
\end{equation}
One may convince oneself, that $\eta_3 S_0(x_c,x_t)$ is independent of the 
renormalization scales, in particular of $\mu_c$, up to terms of 
${\cal O}(x_c \alpha^{2/9}_s \alpha_s)$. 

Furthermore, using the formulae given in this
section, it is easy to see from the explicit expression (\ref{eta3}),
that $\eta_3 \alpha^{-2/9}_s\to 1$ in the limit $\alpha_s\to 0$,
as it should indeed be the case.
\\
The next-to-leading order formula (\ref{eta3}) for $\eta_3$, first
calculated in \cite{herrlichnierste:95}, \cite{nierste:95}, provides
the generalization of the leading log result obtained by
\cite{gilmanwise:83}. It is instructive to compare (\ref{eta3}) with
the leading order approximation, which can be written as
\begin{equation}\label{eta3lo}
\eta^{LO}_3=\alpha_s(\mu_c)^{2/9}
\frac{-\pi C^{LO}_7(\mu_c)}{\alpha_s(\mu_c)\ln x_c}
\end{equation}
using the notation of (\ref{eta3}). $C^{LO}_7$ denotes the coefficient
$C_7$, restricted to the leading logarithmic approximation.
Formula (\ref{eta3lo}), derived here as a special case of (\ref{eta3}),
is equivalent to the result obtained in \cite{gilmanwise:83}.
\\
If penguin operators and the b-quark threshold in the RG
evolution are neglected, it is possible to write down in closed form
a relatively simple, explicit expression for $\eta_3$. Using a 4-flavor
effective theory for the evolution from the $W$-scale down to the
charm scale, we find in this approximation
\begin{eqnarray}\label{eta3apx}
\eta_3 &=& \frac{x_c(\mu_c)}{S_0(x_c(\mu_c),x_t)}\alpha_s(\mu_c)^{2/9}\cdot
\nonumber \\
&\cdot& \Biggl[ \frac{\pi}{\alpha_s(\mu_c)}\left(-\frac{18}{7}K_{++}-
\frac{12}{11}K_{+-}+\frac{6}{29}K_{--}+\frac{7716}{2233}K_7\right)
\left(1-\frac{\alpha_s(\mu_c)}{4\pi}\frac{307}{162}\right)+ \nonumber \\
& & + \left(\ln\frac{\mu_c}{m_c}-\frac{1}{4}\right)
\left(3 K_{++}-2 K_{+-}+ K_{--}\right) + \nonumber \\
& & +\frac{262497}{35000}K_{++}-\frac{123}{625}K_{+-}+
\frac{1108657}{1305000}K_{--}-\frac{277133}{50750}K_7+ \nonumber \\
& & + K\left(-\frac{21093}{8750}K_{++}+\frac{13331}{13750}K_{+-}-
\frac{10181}{18125}K_{--}-\frac{1731104}{2512125}K_7\right)+ \nonumber \\
& & +\left(\ln x_t-\frac{3 x_t}{4(1-x_t)}-\frac{3x^2_t\ln x_t}{4(1-x_t)^2}
+\frac{1}{2}\right) K K_7 \Biggr]
\end{eqnarray}
where
\begin{equation}\label{kpm}
K_{++}=K^{12/25}\qquad K_{+-}=K^{-6/25}\qquad K_{--}=K^{-24/25}
\end{equation}
\begin{equation}\label{k7k}
K_7=K^{1/5}\qquad  K=\frac{\alpha_s(M_W)}{\alpha_s(\mu_c)}
\end{equation}
Here we have set $\mu_W=M_W$. (\ref{eta3apx}) represents the
next-to-leading order generalization of an approximate formula for the
leading log $\eta_3$, also omitting gluon penguins, that has been first
given in \cite{gilmanwise:83}.
The analytical expression for $\eta_3$ in \eqn{eta3apx}
provides an excellent approximation, deviating generally by less
than $1\%$ from the full result.

\subsection{Numerical Results}
\label{sec:HeffKKbar:Num}

\subsubsection{General Remarks}
\label{sec:HeffKKbar:Num:Rem}
After presenting the theoretical aspects of the short-distance
QCD factors $\eta_1$, $\eta_2$ and $\eta_3$ in the previous
sections, we shall now turn to a discussion of their
numerical values. However, before considering explicit numbers,
we would like to make a few general remarks.
\\
First of all, it is important to recall
that in the matrix element
$\langle\bar K^0|{\cal H}^{\Delta S=2}_{eff}|K^0\rangle$
(see (\ref{hds2})), only the complete products
\begin{equation}\label{setaiq}
S_{0i}\cdot \eta_i[\as(\mu)]^{-2/9}\left[
1+\frac{\as(\mu)}{4\pi}J_3\right]
\langle\bar K^0|Q(\mu)|K^0\rangle\equiv
C_i(\mu) \langle\bar K^0|Q(\mu)|K^0\rangle
\end{equation}
are physically relevant.  Here $S_{0i}$ denote the appropriate quark
mass dependent functions $S_0$ for the three contributions ($i=1$, $2$,
$3$) in (\ref{hds2}).  None of the factors in (\ref{setaiq}) is
physically meaningful by itself. In particular, there is some
arbitrariness in splitting the product (\ref{setaiq}) into the
short-distance part and the matrix element of $Q$ (\ref{qsdsd})
containing long distance contributions. This arbitrariness has of
course no impact on the physical result. However, it is essential to
employ a definition for the operator matrix element that is consistent
with the short-distance QCD factor used.
\\
Conventionally, the matrix element $\langle\bar K^0|Q|K^0\rangle$
is expressed in terms of the so-called bag parameter $B_K(\mu)$
defined through
\begin{equation}\label{bkdef}
\langle\bar K^0|Q(\mu)|K^0\rangle\equiv\frac{8}{3}F^2_K m^2_K B_K(\mu)
\end{equation}
where $m_K$ is the kaon mass and $F_K=160 MeV$ is the kaon decay
constant. In principle, one could just use the scale- and scheme
dependent bag factor $B_K(\mu)$ along with the coefficient
functions $C_i(\mu)$ as defined by (\ref{setaiq}), evaluated
at the same scale and in the same renormalization scheme.
However, it has become customary to define the short-distance
QCD correction factors $\eta_i$ by splitting off from the
Wilson coefficient $C_i(\mu)$ the factor
$[\as(\mu)]^{-2/9}[1+\as(\mu)/(4\pi)\ J_3]$, which
carries the dependence on the renormalization scheme and the scale $\mu$.
This factor is then
attributed to the matrix element of $Q$, formally cancelling its
scale and scheme dependence. Accordingly one defines a
renormalization scale and scheme invariant bag parameter $B_K$
(compare (\ref{setaiq}), (\ref{bkdef}))
\begin{equation}\label{bkbkmu}
B_K\equiv [\as(\mu)]^{-2/9}\left[
1+\frac{\as(\mu)}{4\pi}J_3\right] B_K(\mu)
\end{equation}
If the $\eta_i$ as described in this report are employed to
describe the short-distance QCD corrections, eq. (\ref{bkbkmu})
is the consistent definition to be used for the kaon bag
parameter.
\\
Eventually the quantity $B_K(\mu)$ should be calculated within
lattice QCD. At present, the analysis of \cite{sharpe:94}, for
example, gives a central value of $B_K(2 GeV)_{NDR}=0.616$, with
some still sizable uncertainty. For a recent review see also
\cite{soni:95}.
This result already incorporates the lattice-continuum
theory matching and refers to the usual NDR scheme.
It is clear that the NLO calculation of short-distance QCD effects
is essential for consistency with this matching and for a proper
treatment of the scheme dependence. Both require ${\cal O}(\as)$
corrections, which go beyond the leading log approximation.
\\
To convert to the scheme invariant parameter $B_K$ one uses
(\ref{bkbkmu}) with the NDR-scheme value for $J_3=307/162$ to
obtain $B_K=0.84$. Note that the factor involving $J_3$ in (\ref{bkbkmu}),
which appears at NLO, increases the r.h.s. of (\ref{bkbkmu}) by
$\approx 4.5\%$. The leading factor $\as^{-2/9}$ is about $1.31$.
Of course, the fact that there is presently still a rather large
uncertainty in the calculation of the hadronic matrix element is
somewhat forgiving, regarding the precise definition of $B_K$.
However, as the lattice calculations improve further and the errors
decrease, the issue of a consistent definition of the $\eta_i$
and $B_K$ will become crucial and it is important to keep
relation (\ref{bkbkmu}) in mind.

Let us next add a side remark concerning the separation of the
full amplitude into the formally RG invariant factors $\eta_i$
and $B_K$.
This separation is essentially unique, up to trivial constant factors,
if the evolution from the charm scale $\mu_c$ down to a "hadronic"
scale $\mu<\mu_c$ is written in the resummed form as shown in
(\ref{cqmu}) and one requires that all factors depending on the scale
$\mu$ are absorbed into the matrix element. On the other hand the
hadronic scale $\mu={\cal O}(1 GeV)$ is not really much different
from the charm scale $\mu_c={\cal O}(m_c)$, so that the logarithms
$\ln\mu/\mu_c$ are not very large. Therefore one could argue that
it is not necessary to resum those logarithms. In this case the
first two factors on the r.h.s. of (\ref{cqmu}) could be expanded
to first order in $\as$ and the amplitude (\ref{setaiq})
would read
\begin{equation}\label{ciqmumuc}
C_i(\mu_c)\left(1+\frac{\as}{\pi}\ln\frac{\mu}{\mu_c}\right)
\langle\bar K^0|Q(\mu)|K^0\rangle
\end{equation}
From this expression it is obvious, that the separation of the
physical amplitude into scheme invariant short-distance factors
and a scheme invariant matrix element is in general not unique.
This illustrates once more the ambiguity existing for theoretical
concepts such as operator matrix elements or QCD correction factors,
which only cancels in physical quantities.
\\
For definiteness, we will stick to the RG improved form also for the
evolution between $\mu_c$ and $\mu$ and the definitions for
$\eta_i$ and $B_K$ that we have discussed in detail above.

\subsubsection{Results for $\eta_1$, $\eta_2$ and $\eta_3$}
\label{sec:HeffKKbar:Num:Res}
We are now ready to quote numerical results for the
short-distance QCD corrections $\eta_i$ at next-to-leading order
and to compare them with the leading order approximation.
\\
The factors $\eta_1$ and $\eta_3$ have been analyzed in detail in
\cite{herrlichnierste:93} and \cite{nierste:95}.
Here we summarize briefly their main results.
Using our central parameter values $m_c(m_c)=1.3 GeV$,
$\Lms^{(4)}=0.325 GeV$, $m_t(m_t)=170 GeV$ and
fixing the scales as $\mu_c=m_c$, $\mu_W=M_W$ for $\eta_1$,
$\mu_W=130 GeV$ for $\eta_3$, one obtains at NLO
\begin{equation}\label{eta13nlo}
\eta_1=1.38\qquad\qquad \eta_3=0.47
\end{equation}
This is to be compared with the LO values corresponding to the same
input $\eta^{LO}_1=1.12$, $\eta^{LO}_3=0.35$.  We note that the
next-to-leading order corrections are sizable, typically $20\%-30\%$,
but still perturbative.  The numbers above may be compared with the
leading log values $\eta^{LO}_1=0.85$ and $\eta^{LO}_3=0.36$ that have
been previously used in the literature, based on the choice $m_c=1.4
GeV$, $\Lambda_{QCD}=0.2 GeV$ and $\mu_W=M_W$.  The considerable
difference between the two LO values for $\eta_1$ mainly reflects the
large dependence of $\eta_1$ on $\Lambda_{QCD}$.
\\
In fact, when the QCD scale is allowed to vary within
$\Lms^{(4)}=(0.325\pm 0.110)GeV$, the
value for $\eta_1$ (NLO) changes by $\sim\pm 35\%$. The leading order
result $\eta^{LO}_1$ appears to be slightly less sensitive to
$\Lambda_{QCD}$. However, in this approximation the relation of
$\Lambda_{QCD}$ to $\Lms^{(4)}$ is not well
defined, which introduces an additional source of uncertainty
when working to leading logarithmic accuracy.
\\
The situation is much more favorable in the case of $\eta_3$, where
the sensitivity to $\Lms^{(4)}$ is quite small,
$\sim\pm 3\%$.
Likewise the dependence on the charm quark mass is very small for both
$\eta_1$ and $\eta_3$. Using $m_c(m_c)=(1.3\pm 0.05)GeV$ and the
central value for $\Lms^{(4)}$ it is about $\pm 4\%$ for $\eta_1$ and
entirely negligible for $\eta_3$.
\\
Finally, there are the purely theoretical uncertainties due to the
renormalization scales. They are dominated by the ambiguity related
to $\mu_c$. The products $S_0(x_c(\mu_c))\cdot\eta_1(\mu_c)$ and
$S_0(x_c(\mu_c),x_t)\cdot\eta_3(\mu_c)$ are independent of $\mu_c$
up to terms of the neglected order in RG improved perturbation theory.
In the case of $S_0(x_c(\mu_c))\cdot\eta_1(\mu_c)$
($S_0(x_c(\mu_c),x_t)\cdot\eta_3(\mu_c)$) the remaining sensitivity to
$\mu_c$ amounts to typically $\pm 15\%$ ($\pm 7\%$) at NLO. These
scale dependences are somewhat reduced compared to the leading order
calculation, where the corresponding uncertainty is around $\pm 30\%$
($\pm 10\%$).
\\
To summarize, sizable uncertainties are still associated with the
number for the QCD factor $\eta_1$, whose central value is found to
be $\eta_1=1.38$ \cite{herrlichnierste:93}.
On the other hand, the prediction for $\eta_3$ appears to be quite
stable and can be reliably determined as $\eta_3=0.47\pm 0.03$
\cite{herrlichnierste:95}, \cite{nierste:95}.  One should emphasize
however, that these conclusions have their firm basis only within the
framework of a complete NLO analysis, as the one performed in
\cite{herrlichnierste:93}, \cite{nierste:95}.  Fortunately the quantity
$\eta_1$, for which a high precision seems difficult to achieve, plays
a less important role in the phenomenology of indirect CP violation.

Finally, we turn to a brief discussion of $\eta_2$ \cite{burasjaminweisz:90},
representing the short-distance QCD effects of the top-quark
contribution. For central parameter values, in particular
$\Lms^{(4)}=0.325 GeV$ and $m_t(m_t)=170 GeV$, and for $\mu_t=m_t(m_t)$
the numerical value is
\begin{equation}\label{eta2num}
\eta_2=0.574
\end{equation}
Varying the QCD scale within
$\Lms^{(4)}=(0.325\pm 0.110) GeV$ results in a
$\pm 0.5\%$ change in $\eta_2$. The dependence on $m_t(m_t)$ is even
smaller, only $\pm 0.3\%$ for $m_t(m_t)=(170\pm 15)GeV$.
It is worthwhile to compare the NLO results with the leading log
approximation. Using the same input as before yields a central value of
$\eta^{LO}_2=0.612$, about $7\%$ larger as the NLO result (\ref{eta2num}).
However, what is even more important than the difference in central
values is the quite striking reduction of scale uncertainty when going
from the leading log approximation to the full NLO treatment.
Recall that the $\mu_t$-dependence in $\eta_2$ has to cancel the
scale dependence of the function $S_0(x_t(\mu_t))$. Allowing for a
typical variation of the renormalization scale $\mu_t={\cal O}(m_t)$
from $100 GeV$ to $300 GeV$ results in a sizable change in
$S_0(x_t(\mu_t)) \eta^{LO}_2$ of $\pm 9\%$. In fact, in leading order
the $\mu_t$-dependence of $\eta_2$ has even the wrong sign, re-inforcing
the scale dependence present in $S_0(x_t(\mu_t))$ instead of
reducing it. The large sensitivity to the unphysical parameter $\mu_t$
is essentially eliminated (to $\pm 0.4\%$) for $\eta_2 S_0(x_t)$ at
NLO, a quite remarkable improvement of the theoretical accuracy. The
situation here is similar to the case of the top-quark dominated
rare K and B decays discussed in sections
\ref{sec:HeffRareKB}, \ref{sec:Kpnn} and \ref{sec:BXnnBmm}.
For a further illustration of the reduction in scale uncertainty see
the discussion of the analogous case of $\eta_{2B}$ in section
\ref{sec:HeffBBbar:Num}.
\\
The dependence of $\eta_2$ on the charm and bottom threshold scales
$\mu_c={\cal O}(m_c)$ and  $\mu_b={\cal O}(m_b)$ is also extremely
weak. Taking $1GeV\leq\mu_c\leq 3GeV$ and $3GeV\leq\mu_b\leq 9GeV$
results in a variation of $\eta_2$ by merely $\pm 0.26\%$ and
$\pm 0.06\%$, respectively.
\\
In summary, the NLO result for $\eta_2 S_0(x_t)$ is, by contrast to the
leading logarithmic approximation, essentially free from theoretical
uncertainties. Furthermore, $\eta_2$ is also rather insensitive to the
input parameters $\Lms$ and $m_t$. The top
contribution plays the dominant role for indirect CP violation in the
neutral kaon system. The considerable improvement in the theoretical
analysis of the short-distance QCD factor $\eta_2$ brought about by
the next-to-leading order calculation is therefore particularly
satisfying.

\section{The Effective Hamiltonian for $B^0-\bar B^0$ Mixing}
\label{sec:HeffBBbar}

\subsection{General Structure}
\label{sec:HeffBBbar:General}
Due to the particular hierarchy of the CKM matrix elements only the
top sector can contribute significantly to $B^0-\bar B^0$ mixing.
The charm sector and the mixed top-charm contributions are
entirely negligible here, in contrast to the $K^0-\bar K^0$ case,
which considerably simplifies the analysis.

Refering to  our earlier presentation of the top sector for $\Delta
S=2$ transitions in section \ref{sec:HeffKKbar:eta2} we can immediately
write down the effective $\Delta B=2$ hamiltonian. Performing the RG
evolution only down to scales $\mu_b=\ord(m_b)$ and making the
necessary replacements ($s\to b$) we get, in analogy to (\ref{hds2})
\cite{burasjaminweisz:90}
\begin{equation}\label{hdb2}
{\cal H}^{\Delta B=2}_{eff}=\frac{G^2_F}{16\pi^2}M^2_W
 \left(V^\ast_{tb}V_{td}\right)^2 \eta_{2B}
 S_0(x_t) \left[\as(\mu_b)\right]^{-6/23}\left[
  1 + \frac{\as(\mu_b)}{4\pi} J_5\right]  Q + h. c.
\end{equation}
where here
\begin{equation}\label{qbdbd}
Q=(\bar bd)_{V-A}(\bar bd)_{V-A}
\end{equation}
and
\begin{eqnarray}\label{eta2b}
&&\eta_{2B}=\left[\as(\mu_t)\right]^{6/23}\times
\\
&& \times \left[1+\frac{\as(\mu_t)}{4\pi} \left(
 \frac{S_1(x_t)}{S_0(x_t)}+B_t - J_5+\frac{\gamma^{(0)}}{2}
 \ln\frac{\mu^2_t}{M^2_W}+\gamma_{m0}
 \frac{\partial\ln S_0(x_t)}{\partial\ln x_t}\ln\frac{\mu^2_t}{M^2_W}
\right)\right] \nonumber
\end{eqnarray}
The definitions of the various quantities in (\ref{eta2b}) can
be found in section \ref{sec:HeffKKbar:eta2}.
Several important aspects of $\eta_2$ in the kaon system have also
been discussed in this section. Similar comments apply to the
present case of $\eta_{2B}$.
Here we would still like to supplement this discussion by writing 
down the formula for $\eta_{2B}$ in the limiting case $\mt\gg \mw$,
\begin{eqnarray}\label{eta2bas}
&&\eta_{2B}=\left[\alpha_s(\mu_t)\right]^{6/23}\times
\\
&& \times \left[1+\frac{\alpha_s(\mu_t)}{4\pi} \left(
\frac{\gamma^{(0)}}{2}\ln\frac{\mu^2_t}{m^2_t}+
\gamma_{m0}\ln\frac{\mu^2_t}{m^2_t}+11-\frac{20}{9}\pi^2+B_t-J_5+
\ord\left(\frac{M^2_W}{m^2_t}\right) \right)\right] \nonumber
\end{eqnarray}
This expression clarifies the structure of the RG evolution in the
limit $\mt\gg \mw$. It also suggests that the 
renormalization scale is most naturally to be taken as
$\mu_t=\ord(\mt)$ rather than $\mu_t=\ord(\mw)$,
both in the definition of the top quark mass and as the initial
scale of the RG evolution. Formula (\ref{eta2bas}) also holds,
with obvious modifications, for the $\eta_2$ factor in the
kaon system, which has been discussed in sec. \ref{sec:HeffKKbar:eta2}.

We finally mention that in the literature the $\mu_b$-dependent
factors in (\ref{hdb2}) are sometimes not attributed to the
matrix elements of $Q$, as implied by (\ref{hdb2}), but absorbed
into the definition of the QCD correction factor
\begin{equation}\label{e2bbar}
\bar\eta_{2B}=\eta_{2B} \left[\as(\mu_b)\right]^{-6/23}\left[
  1 + \frac{\as(\mu_b)}{4\pi} J_5\right]
\end{equation}
Whichever definition is employed, it is important to remember this
difference and to evaluate the hadronic matrix element consistently.
Note that, in contrast to $\eta_{2B}$, $\bar\eta_{2B}$ is scale
and scheme dependent.

\subsection{Numerical Results}
\label{sec:HeffBBbar:Num}
\begin{figure}[htb]
\vspace{0.15in}
\centerline{
\epsfysize=4.5in
\rotate[r]{
\epsffile{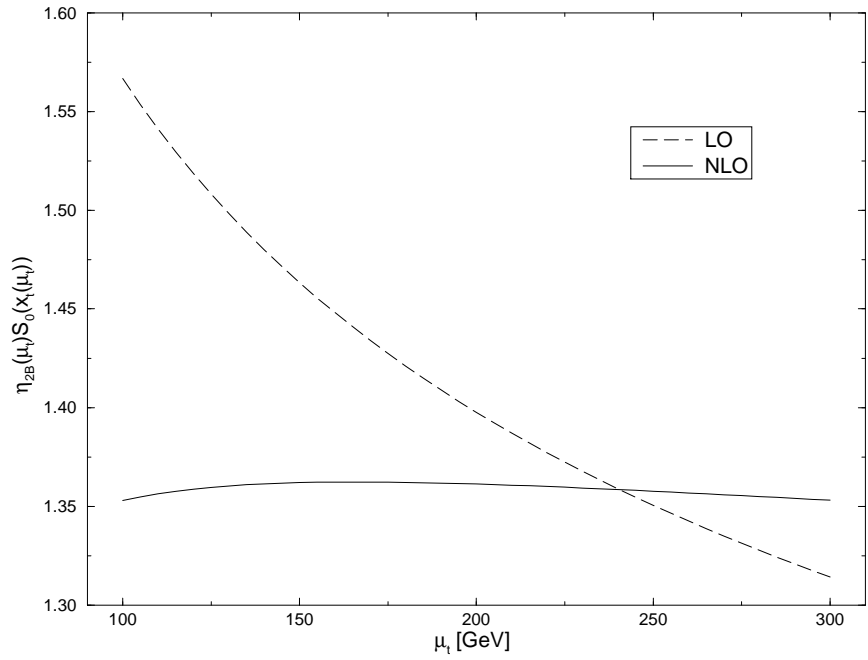}
}}
\vspace{0.15in}
\caption[]{
Scale $\mu_t$ dependence of $\eta_{2B}(\mu_t) S_0(x_t(\mu_t))$ in LO and
NLO.
The quantity $\eta_{2B}(\mu_t) S_0(x_t(\mu_t))$ enters the theoretical
expression for $\Delta m_B$, describing $B^0-\bar B^0$ mixing. It is
independent of the precise value of the renormalization scale $\mu_t$
up to terms of the neglected order in $\as$. The remaining
sensitivity represents an unavoidable theoretical uncertainty.
This ambiguity is shown here for the leading order (dashed) and
the next-to-leading order calculation (solid).
\label{fig:eta2mut}}
\end{figure}

The correction factor $\eta_{2B}$ describes the short-distance
QCD effects in the theoretical expression for $B^0-\bar B^0$
mixing.
Due to the arbitrariness that exists in dividing
the physical amplitude into short-distance contribution and
hadronic matrix element, the short-distance QCD factor is strictly
speaking an unphysical quantity and hence definition dependent.
The $B$-factor, parametrizing the hadronic matrix element, has to
match the convention used for $\eta_{2B}$. With the definition of
$\eta_{2B}$ employed in this article and given explicitly in the previous
section, the appropriate $B$-factor to be used is the so-called
scheme independent bag-parameter $B_B$ as defined in eq.
(\ref{eq:BBrenorm}), where $\mu=\mu_b={\cal O}(m_b)$.
We remark, that the factor $\eta_{2B}$ is identical for
$B_d-\bar B_d$ and $B_s-\bar B_s$ mixing. The effects of $SU(3)$
breaking enter only the hadronic matrix elements. This feature
is a consequence of the factorization of short-distance and
long-distance contributions inherent to the operator product expansion.
For further
comments see also the discussion of the analogous case of
short-distance QCD factors in the neutral kaon system in section
\ref{sec:HeffKKbar:Num:Rem}.
\\
In the following we summarize the main results of a numerical analysis
of $\eta_{2B}$. The factor $\eta_{2B}$ is analogous to $\eta_2$
entering the top contribution to $K^0-\bar K^0$ mixing and both
quantities share many important features.
\\
The value of $\eta_{2B}$ for
$\Lms^{(4)}=0.325 GeV$, $\mt(\mt)=170 GeV$ and
with $\mu_t$ set equal to $\mt(\mt)$ reads at NLO
\begin{equation}\label{eta2bnum}
\eta_{2B}=0.551
\end{equation}
This can be compared with $\eta^{LO}_{2B}=0.580$, obtained, using
the same input, in the leading logarithmic approximation.
In the latter case the product
$\eta^{LO}_{2B}(\mu_t)\cdot S(x_t(\mu_t))$ is, however, affected
by a residual scale ambiguity of $\pm 9\%$ (for
$100 GeV\leq\mu_t\leq 300 GeV$). This uncertainty is reduced to the
negligible amount of $\pm 0.3\%$ in the complete NLO expression of
$\eta_{2B}(\mu_t)\cdot S(x_t(\mu_t))$, corresponding to an increase
in accuracy by a factor of 25. The sensitivity to the unphysical
scale $\mu_t$ in leading and next-to-leading order is illustrated
in fig. \ref{fig:eta2mut}.

In addition the number shown in (\ref{eta2bnum}) is also very stable
against changes in the input parameters. Taking $\Lms^{(4)}=(0.325\pm
0.110) GeV$ and $\mt(\mt)=(170\pm 15)GeV$ results in a variation of
$\eta_{2B}$ by $\pm 1.3\%$ and $\pm 0.3\%$, respectively.

It is clear from this discussion, that the short-distance QCD effects
in $B^0-\bar B^0$ mixing are very well under control, once NLO
corrections have been properly included, and the remaining
uncertainties are extremely small. The effective hamiltonian given in
(\ref{hdb2}) therefore provides a solid foundation for the
incorporation of non-perturbative effects, to be determined from
lattice gauge theory, and for further phenomenological investigations
related to $B^0-\bar B^0$ mixing phenomena.

\section{Penguin Box Expansion for FCNC Processes}
         \label{sec:PBE}
An important virtue of OPE and RG is that with $m_t > \mw$ the
dependence of weak decays on the top quark mass is very elegantly
isolated. It resides only in the initial conditions for the Wilson
coefficients at scale $\mu \approx \mw$ i.e.~in $C_i(\mw)$. A quick look
at the initial conditions in the previous sections reveals the important
fact that the leading $m_t$-dependence in all decays considered is
represented universally by the $m_t$-dependent functions which result
from exact calculations of the relevant penguin and box diagrams with
internal top quark exchanges. These are the functions
\begin{equation}
S_0(x_t), \quad
B_0(x_t), \quad
C_0(x_t), \quad
D_0(x_t), \quad
E_0(x_t), \quad
D'_0(x_t), \quad
E'_0(x_t)
\label{eq:SBCDE}
\end{equation}
for which explicit expressions are given in \eqn{s0t},
eqs.~\eqn{eq:Bxt}--\eqn{eq:Dxt}, \eqn{eq:Ext}, \eqn{c7} and \eqn{c8},
respectively. In certain decays some of these functions do not appear
because the corresponding penguin or box diagram does not contribute to
the initial conditions. However, the function $C_0(x_t)$ resulting from
the $Z^0$-penguin diagram enters all $\Delta F=1$ decays but $B \to X_s
\gamma$. Having a quadratic dependence on $m_t$, this function
is responsible for the dominant $m_t$-dependence of these
decays. Since the non-leading $m_t$-dependence of $C_0(x_t)$ is gauge
dependent, $C_0(x_t)$ is always accompanied by $B_0(x_t)$ or $D_0(x_t)$
in such a way that this dependence cancels. For this reason it is
useful to replace the gauge dependent functions $B_0(x_t)$, $C_0(x_t)$
and $D_0(x_t)$ by the gauge independent set \cite{buchallaetal:91}
\begin{eqnarray}
X_0(x_t) &=& C_0(x_t) - 4 \, B_0(x_t)           \nn \\
Y_0(x_t) &=& C_0(x_t) - B_0(x_t)                \label{eq:XYZ} \\
Z_0(x_t) &=& C_0(x_t) + \frac{1}{4} \, D_0(x_t) \nn
\end{eqnarray}
as we have already done at various places in this review.  The
inclusion of NLO QCD corrections to \BB-, \KK-mixing and the rare $K$-
and $B$-decays of section~\ref{sec:HeffRareKB} requires the calculation
of QCD corrections to penguin and box diagrams in the full theory. This
results in the functions $\tilde{S}(x_t)=\eta_2 S_0(x_t)$, $X(x_t)$ and
$Y(x_t)$, with the latter two given in \eqn{xx} and \eqn{yy}, respectively.
\\
It turns out however that if the top quark mass is definded as $m_t
\equiv \bar{m}_t(m_t)$ one has
\begin{equation}
\tilde{S}(x_t) = \eta_2 \, S_0(x_t), \quad
X(x_t) = \eta_X \, X_0(x_t), \quad
Y(x_t) = \eta_Y \, Y_0(x_t)
\label{eq:SXY}
\end{equation}
with $\eta_2$, $\eta_X$ and $\eta_Y$ almost independent of $m_t$.
Numerical values of $\eta_X$ and $\eta_Y$ are given in part three.

Consequently with this definition of $m_t$ the basic $m_t$-dependent
functions listed in \eqn{eq:SBCDE} and \eqn{eq:XYZ} represent the
$m_t$-dependence of weak decays at the NLO level to a good
approximation.  It should be remarked that the QCD corrections to
$D_0$, $E_0$, $D'_0$ and $E'_0$ have not been calculated yet. They
would however be only required for still higher order corrections
(NNLO) in the renormalization group improved perturbation theory as
far as $D_0$ and $E_0$ are concerned. On the other hand, in the case
of $D_0'$ and $E_0'$, which are relevant for the $b \to s \gamma$ decay,
these corrections are necessary.

An inspection of the effective hamiltonians derived in the previous
sections shows that for \BB-mixing, \KK-mixing and the rare decays of
section~\ref{sec:HeffRareKB} the $m_t$ dependence of the effective
hamiltonian is explicitly given in terms of the basic functions listed
above. Due to the one step evolution from $\mu_t$ to $\mu_b$ we have
also presented the explicit $\mt$-dependence for $B \to X_s \gamma$ and
$B \to X_s e^+ e^-$ decays. On the other hand in the case of $\Kpipi$
and $K_L \to \pi^0 e^+ e^-$ where the renormalization group evolution
is very complicated the $m_t$ dependence of a given box or penguin
diagram is distributed among various Wilson coefficient functions. In
other words the $m_t$-dependence acquired at scale $\mu \approx
\ord(\mw)$ is hidden in a complicated numerical evaluation of
$\hU(\mu,\mw)$.

For phenomenological applications it is more elegant and more
convenient to have a formalism in which the final formulae for all
amplitudes are given explicitly in terms of the basic $m_t$-dependent
functions discussed above.

In \cite{buchallaetal:91} an approach has been presented which
accomplishes this task. It gives the decay amplitudes as linear
combinations of the basic, universal, process independent but
$m_t$-dependent functions $F_r(x_t)$ of eq.~\eqn{eq:SBCDE} with
corresponding coefficients $P_r$ characteristic for the decay under
consideration. This approach termed ``Penguin Box Expansion'' (PBE) has
the following general form
\begin{equation}
A({\rm decay}) = P_0({\rm decay}) + \sum_r P_r({\rm decay}) \, F_r(x_t)
\label{eq:generalPBE}
\end{equation}
where the sum runs over all possible functions contributing to a given
amplitude. In \eqn{eq:generalPBE} we have separated a $m_t$-independent
term $P_0$ which summarizes contributions stemming from internal quarks
other than the top, in particular the charm quark.

Many examples of PBE appear in this review. Several decays or
transitions depend only on a single function out of the complete set
\eqn{eq:SBCDE}. For completeness we give here the correspondence
between various processes and the basic functions

\begin{center}
\begin{tabular}{lcl}
\BB-mixing &\qquad\qquad& $S_0(x_t)$ \\
$K \to \pi \nu \bar\nu$, $B \to K \nu \bar\nu$, 
$B \to \pi \nu \bar\nu$ &\qquad\qquad& $X_0(x_t)$ \\
$K \to \mu \bar\mu$, $B \to l\bar l$ &\qquad\qquad& $Y_0(x_t)$ \\
$K_L \to \pi^0 e^+ e^-$ &\qquad\qquad& $Y_0(x_t)$, $Z_0(x_t)$, $E_0(x_t)$ \\
$\varepsilon'$ &\qquad\qquad& $X_0(x_t)$, $Y_0(x_t)$, $Z_0(x_t)$,
$E_0(x_t)$ \\
$B \to X_s \gamma$ &\qquad\qquad& $D'_0(x_t)$, $E'_0(x_t)$ \\
$B \to X_s e^+ e^-$ &\qquad\qquad&
$Y_0(x_t)$, $Z_0(x_t)$, $E_0(x_t)$, $D'_0(x_t)$, $E'_0(x_t)$
\end{tabular}
\end{center}

In \cite{buchallaetal:91} an explicit transformation from OPE to
PBE has been made. This transformation and the relation between these
two expansions can be very clearly seen on the basis of 
\begin{equation}
A(P \to F) = \sum_{i,k} \langle F | O_k(\mu) | P \rangle \,
U_{kj}(\mu,\mw) \, C_j(\mw)
\label{eq:RGTransf}
\end{equation}
where $U_{kj}(\mu,\mw)$ represents the renormalization group
transformation from $\mw$ down to $\mu$. As we have seen, OPE puts the
last two factors in this formula together, mixing this way the physics
around $\mw$ with all physical contributions down to very low energy
scales. The PBE is realized on the other hand by putting the first two
factors together and rewriting $C_j(\mw)$ in terms of the basic
functions \eqn{eq:SBCDE}. This results in the expansion of
eq.~\eqn{eq:generalPBE}. Further technical details and the methods for
the evaluation of the coefficients $P_r$ can be found in
\cite{buchallaetal:91}, where further virtues of PBE are discussed.

Finally, we give approximate formulae having power-like
dependence on $x_t$ for the basic, gauge independent functions of PBE
\begin{equation}
\begin{array}{lclclcl}
S_0(x_t)  &=& 0.784 \cdot x_t^{ 0.76}  &\qquad\qquad&
X_0(x_t)  &=& 0.660 \cdot x_t^{0.575} \\
Y_0(x_t)  &=& 0.315 \cdot x_t^{ 0.78}  &\qquad\qquad&
Z_0(x_t)  &=& 0.175 \cdot x_t^{0.93} \\
E_0(x_t)  &=& 0.564 \cdot x_t^{-0.51}  &\qquad\qquad&
D'_0(x_t) &=& 0.244 \cdot x_t^{0.30} \\
E'_0(x_t) &=& 0.145 \cdot x_t^{ 0.19} \, . &\qquad\qquad&  
\end{array}
\label{eq:approxSXYZE}
\end{equation}
In the range $150\gev \le m_t \le 200\gev$ these approximations
reproduce the exact expressions to an accuracy better than 1\%.

\section{Heavy Quark Effective Theory Beyond Leading Logs}
\label{sec:HQET}

\subsection{General Remarks}
\label{sec:HQET:General}

Since its advent in 1989 heavy quark effective theory (HQET) has
developped into an elaborate and well-established formalism, providing a
systematic framework for the treatment of hadrons containing a heavy quark.
HQET represents a static approximation for
the heavy quark, covariantly formulated in the language of an effective
field theory.
It allows to extract the dependence of hadronic matrix elements on the
heavy quark mass and to exploit the simplifications that arise in QCD in
the static limit.
\\
There are several excellent reviews on this subject \cite{neubert:94},
\cite{georgi:91}, \cite{grinstein:91}, \cite{isgurwise:92},
\cite{mannel:93} and we do not attempt here to cover the details of
this extended field. However, we would like to emphasize the close
parallels in the general formalism employed to calculate perturbative
QCD effects for the effective weak hamiltonians we have been discussing
in this review and in the context of HQET. In particular we will
concentrate on results that have been obtained in HQET beyond the
leading logarithmic approximation in QCD perturbation theory. Such
calculations have been done mainly for bilinear current operators
involving heavy quark fields, which have important applications in
semileptonic decays of heavy hadrons.  For the purpose of illustration
we will focus on the simplest case of heavy-light currents as an
important example.  Furthermore, while existing reviews concentrate on
semileptonic decays and current operators, we will also include results
obtained for nonleptonic transitions and summarize the calculation of
NLO QCD corrections to $B^0-\bar B^0$ mixing in HQET \cite{flynnetal:91},
\cite{gimenez:93}. These latter papers generalize the leading-log
results first obtained in \cite{voloshinshifman:87}, \cite{politzerwise:88a},
\cite{politzerwise:88b}.
\\
Throughout we will restrict ourselves to the leading order in HQET and
not address the question of $1/m$ corrections. For a discussion of this
topic we refer the reader to the literature, in particular the above
mentioned review articles.

\subsection{Basic Concepts}
\label{sec:HQET:Basic}

Let us briefly recall the most important basic concepts
underlying the idea of HQET. 
\\
The Lagrangian describing a quark field $\Psi$ with mass $m$ and its QCD
interactions with gluons reads
\begin{equation}\label{lful}
{\cal L}=\bar\Psi i\not\!\! D\Psi-m\bar\Psi\Psi
\end{equation}
where $D_\mu=\partial_\mu-i g T^a A^a_\mu$ is the gauge-covariant
derivative. If $\Psi$ is a heavy quark, i.e. its mass is large compared to
the QCD scale, $\Lambda_{QCD}/m\ll 1$, it acts approximately like a
static color source and its QCD interactions simplify.
A heavy quark inside a hadron moving with velocity $v$ has approximately
the same velocity. Thus its momentum can be written as $p=m v+k$, where $k$
is a small residual momentum of the order of $\Lambda_{QCD}$ and
subject to changes of the same order through soft QCD interactions.
To implement this approximation, the quark field $\Psi$ is decomposed into
\begin{equation}\label{qehv}
\Psi(x)=e^{-i m v\cdot x}\left(h_v(x)+H_v(x)\right)
\end{equation}
with $h_v$ and $H_v$ defined by
\begin{equation}\label{hvpl}
h_v(x)=e^{imv\cdot x}\frac{1+\not\! v}{2} \Psi(x)
\end{equation}
\begin{equation}\label{hvmi}
H_v(x)=e^{imv\cdot x}\frac{1-\not\! v}{2} \Psi(x)
\end{equation}
To be specific we consider here the case of a hadron containing a heavy
quark, as opposed to a heavy antiquark. In order to describe a heavy
antiquark, the definitions (\ref{hvpl}) and (\ref{hvmi}) are replaced by
\begin{equation}\label{ahvp}
h^{(-)}_v(x)=e^{-imv\cdot x}\frac{1-\not\! v}{2} \Psi(x)
\end{equation}
\begin{equation}\label{ahvm}
H^{(-)}_v(x)=e^{-imv\cdot x}\frac{1+\not\! v}{2} \Psi(x)
\end{equation}
Consequently, for a heavy antiquark, one only needs to substitute $v\to -v$ in the  
expressions given below for the case of a heavy quark.
\\
In the rest frame of the heavy quark $h_v$ and $H_v$ correspond to the
upper and lower components of $\Psi$, respectively. In general, for
$m\to\infty$ $h_v$ and $H_v$ represent the "large" and "small" components
of $\Psi$. In fact, the equations of motion of QCD imply that $H_v$ is
suppressed by a factor $\Lambda_{QCD}/m$ in comparison to $h_v$.
The inclusion of an explicit exponential factor $\exp(-i mv\cdot x)$ in
(\ref{qehv}) ensures that the momentum associated with the field $h_v$
is only a small residual momentum of order $\Lambda_{QCD}$.
\\
Now an effective theory for $h_v$ is constructed by eliminating the
small component field $H_v$ from explicitly appearing in the
description of the heavy quark. On the classical level this can be done
by using the equations of motion or, equivalently, by directly
integrating out the $H_v$ degrees of freedom in the context of a path
integral formulation \cite{manneletal:92}.  The effective Lagrangian
one obtains from (\ref{lful}) along these lines is given by
($D^\mu_\perp=D^\mu-v^\mu v\cdot D$)
\begin{equation}\label{lhqt}
{\cal L}_{eff,tot}=\bar h_v iv\cdot D h_v+\bar h_v i\not\!\! D_\perp
\frac{1}{iv\cdot D+2m-i\varepsilon}i\not\!\! D_\perp h_v
\end{equation}
The first term in (\ref{lhqt})
\begin{equation}\label{lehq}
{\cal L}_{eff}=\bar h_v (i v^\mu\partial_\mu+g v^\mu T^a A^a_\mu)h_v
\end{equation}
represents the Lagrangian of HQET to lowest order in $1/m$ and will be 
sufficient for our purposes. The second, nonlocal contribution in
(\ref{lhqt}) can be expanded into a series of local, higher dimension
operators, carrying coefficients with increasing powers of $1/m$. To first
order it yields the correction due to the residual heavy quark kinetic
energy and the chromo-magnetic interaction term, coupling the heavy
quark spin to the gluon field. 
\\
From (\ref{lehq}) one can obtain the Feynman rules of HQET, the
propagator of the effective field $h_v$
\begin{equation}\label{hqpr}
\frac{i}{v\cdot k} \frac{1+\not\! v}{2}
\end{equation}
and the $\bar h_v$-$h_v$-gluon vertex, $igv^\mu T^a$. The explicit
factor $(1+\not\! v)/2$ in (\ref{hqpr}) arises because the effective
field $h_v$ is a constrained spinor, satisfying $\not\! v h_v\equiv h_v$,
as is obvious from (\ref{hvpl}). The velocity $v_\mu$ is a constant in the
effective theory and plays the role of a label for the effective field $h_v$.
In principle, a different field $h_v$ has to be considered for every
velocity $v$.
\\
The Lagrangian in (\ref{lehq}) exhibits the crucial features of HQET:
The quark-gluon coupling is independent of the quark's spin degrees of
freedom and the Lagrangian is independent of the heavy quark flavor,
since the heavy quark mass has been eliminated. This observation forms
the basis for the spin-flavor symmetry of HQET \cite{isgurwise:89},
\cite{isgurwise:90}, which gives rise to important simplifications in
the strong interactions of heavy quarks and allows to establish
relations among the form factors of different heavy hadron matrix
elements. The heavy quark symmetries are broken by $1/m$-contributions
as well as radiative corrections.
\\
So far our discussion has been limited to the QCD interactions of the
heavy quark. Weak interactions introduce new operators into the theory,
which may be current operators, bilinear in quark fields, or four-quark
operators, relevant for semileptonic and nonleptonic transitions,
respectively. Such operators form the basic ingredients to be studied in
weak decay phenomenology. They can as well be expanded in $1/m$ and
incorporated into the framework of HQET. For example a heavy-light current
operator $\bar q\Gamma \Psi$ (evaluated at the origin, $x=0$), can be
written (\ref{qehv})
\begin{equation}\label{qgqh}
\bar q\Gamma \Psi=\bar q\Gamma h_v+{\cal O}(\frac{1}{m})
\end{equation}
to lowest order in HQET.
\\
Up to now we have restricted our discussion to the classical level. In
addition, of course, quantum radiative corrections have to be
included. They will for example modify relations such as (\ref{qgqh}).
Technically their effects are taken into account by performing the
appropriate matching calculations, relating operators in the effective
theory to the corresponding operators in the full theory to the
required order in renormalization group improved QCD perturbation theory.
The procedure is very similar to the calculation of the usual effective hamiltonians
for weak decays. The basic difference consists in the heavy degrees of
freedom that are being integrated out in the matching process. In the
general case of effective weak hamiltonians the heavy field to be 
removed as a dynamical variable is the W boson, whereas it is the lower
component heavy quark field $H_v$ in the case of HQET. This similarity
will become obvious from our presentation below.
\\
At this point some comment might be in order concerning the 
relationship of the HQET formalism to the general weak effective
hamiltonians discussed primarily in this review, in particular those
relevant for b-physics.
\\
The effective hamiltonians for $\Delta B=1, 2$ nonleptonic transitions
are the relevant hamiltonians for scales $\mu={\cal O}(m_b)$, which are
appropriate for B hadron decays, and their Wilson coefficients
incorporate the QCD short distance dynamics between scales of ${\cal
O}(M_W)$ and ${\cal O}(m_b)$. As already mentioned at the end of
section \ref{sec:HeffdF1:22} it is therefore not necessary to invoke
HQET. The physics below $\mu={\cal O}(m_b)$ is completely contained
within the relevant hadronic matrix elements.  On the other hand, HQET
may be useful in certain cases, like e.g.  $B^0-\bar B^0$ mixing, to
gain additional insight into the structure of the hadronic matrix
elements for scales below $m_b$, but still large compared to
$\Lambda_{QCD}$. These scales are still perturbative and the related
contributions can be extracted analytically within HQET.  In
particular, this procedure makes the dependence of the matrix element
on the heavy quark mass explicit, as we will see on examples below.
Furthermore, this approach can be useful e.g. in connection with
lattice calculations of hadronic matrix elements, which are easier to
perform in the static limit for b-quarks, i.e.\ employing HQET
\cite{sachrajda:92}. The simplifications obtained are however at the
expense of the approximation due to the expansion in $1/m$.
\\
The most important application of HQET has been to the analysis of
exclusive semileptonic transitions involving heavy quarks, where this
formalism allows to exploit the consequences of heavy quark symmetry to
relate formfactors and provides a basis for systematic corrections to the
$m\to\infty$ limit. This area of weak decay phenomenology has been already
reviewed in detail \cite{neubert:94}, \cite{georgi:91}, \cite{grinstein:91},
\cite{isgurwise:92}, \cite{mannel:93} and will not be covered
in the present article.

\subsection{Heavy-Light Currents}
\label{sec:HQET:Currents}
As an example of a next-to-leading QCD calculation within the context of
HQET, we will now discuss the case of a weak current, composed of one
heavy and one light quark field, to leading order in the $1/m$ expansion.
For definitness we consider the axial vector heavy-light current, whose
matrix elements determine the decay constants of pseudoscalar mesons
containing a single heavy quark, like $f_B$ and $f_D$.
\\
The axial vector current operator in the full theory is given by
\begin{equation}\label{aqq}
A=\bar q\gamma_\mu\gamma_5 \Psi
\end{equation}
where $\Psi$ is the heavy and $q$ the light quark field. In HQET this operator
can be expanded in the following way
\begin{equation}\label{ac12}
A=C_1(\mu)\tilde A_1+C_2(\mu)\tilde A_2+{\cal O}(\frac{1}{m})
\end{equation}
where the operator basis in the effective theory reads
\begin{equation}\label{at12}
\tilde A_1=\bar q\gamma_\mu\gamma_5 h_v \qquad\qquad
\tilde A_2=\bar q v_\mu\gamma_5 h_v
\end{equation}
with the heavy quark effective field $h_v$ defined in (\ref{hvpl}).
The use of the expansion (\ref{ac12}) is to make the dependence of the
matrix elements of $A$ on the heavy quark mass $m$ explicit. The dependence
on this mass is two-fold. First, there is a power dependence, which is 
manifest in the heavy quark expansion in powers of $1/m$. From this
series only the lowest order term is shown in (\ref{ac12}). Second, there is
a logarithmic dependence on $m$ due to QCD radiative corrections, which
can be calculated in perturbation theory. This dependence is factorized
into the coefficient functions $C_1$, $C_2$ in much the same way as the 
logarithmic dependence of nonleptonic weak decay amplitudes on the W boson 
mass is factorized into the Wilson coefficients of the usual weak
hamiltonians. Since the dynamics of HQET is, by construction, independent 
of $m$, no further $m$ dependence is present in the matrix
elements of the effective theory operators $\tilde A_{1,2}$, except for
trivial factors of $m$ related to the normalization of meson states.
Consequently the $m$ dependence of (\ref{ac12}) is determined explicitly. 

We remark that in general the meson states in HQET to be used for the
r.h.s. of (\ref{ac12}) differ from the meson states in the full theory
to be used to sandwich the operator $A$ on the l.h.s.. For the leading
order in $1/m$ we are working in this distinction is irrelevant, however.
\\
An important point is that the operators $\tilde A_{1,2}$ in the effective
theory have anomalous dimensions, although the operator $A$ in the full
theory, being an axial vector current operator, does not. As a 
consequence matrix elements of $\tilde A_{1,2}$ will depend on the
renormalization scale and scheme. This dependence is canceled however
through a corresponding dependence of the coefficients so that the physical
matrix elements of $A$ will be scale and scheme independent as they must be.
The existence of anomalous dimensions for the effective theory operators
merely reflects the logarithmic dependence on the heavy quark mass $m$
due to QCD effects. This dependence results in logarithmic divergences in
the limit $m\to\infty$, corresponding to the effective theory, which
require additional infinite renormalizations not present in full QCD.
Obviously the situation is completely analogous to the case of constructing
effective weak hamiltonians through integrating out the W boson, which we
have described in detail in section \ref{sec:basicform}. In fact, the
extraction of the coefficient functions by factorizing long and short
distance contributions to quark level amplitudes and the renormalization
group treatment follow exactly the same principles.
\\
The Wilson coefficients at the high matching scale $\mu_h={\cal O}(m)$,
the initial condition to the RG evolution, can be calculated in
ordinary perturbation theory with the result (NDR scheme)
\begin{equation}\label{c1mh}
C_1(\mu_h)=1+\frac{\alpha_s}{4\pi}\left(\gamma^{(0)}_{hl}\ln\frac{\mu_h}{m}
+B_1\right)
\end{equation}
\begin{equation}\label{c2mh}
C_2(\mu_h)=\frac{\alpha_s}{4\pi} B_2
\end{equation}
with
\begin{equation}\label{b12h}
B_1=-4 C_F\qquad\qquad B_2=-2 C_F
\end{equation}
and $\gamma^{(0)}_{hl}$ given in (\ref{g0hl}) below. $C_F$ is the QCD color
factor $(N^2-1)/(2N)$. We remark that the coefficient of the new
operator $\tilde A_2$, generated at ${\cal O}(\alpha_s)$, is finite without
requiring renormalization. As a consequence no explicit scale dependence
appears in (\ref{c2mh}) and $B_2$ is a scheme independent constant.
For the same reason $\tilde A_1$ and $\tilde A_2$ do not mix under
renormalization, but renormalize only multiplicatively.
The anomalous dimension of the effective heavy quark currents is independent
of the Dirac structure. It is the same for $\tilde A_1$ and $\tilde A_2$
and reads at two-loop order
\begin{equation}\label{g01h}
\gamma_{hl}=\gamma^{(0)}_{hl}\frac{\alpha_s}{4\pi}+\gamma^{(1)}_{hl}
\left(\frac{\alpha_s}{4\pi}\right)^2
\end{equation}
where
\begin{equation}\label{g0hl}
\gamma^{(0)}_{hl}=-3 C_F
\end{equation}
\begin{eqnarray}\label{g1hl}
\gamma^{(1)}_{hl} &=& \left(-\frac{49}{6}+\frac{2}{3}\pi^2\right)N C_F+
 \left(\frac{5}{2}-\frac{8}{3}\pi^2\right) C^2_F+\frac{5}{3}C_F f=
 \nonumber\\
&=& -\frac{254}{9}-\frac{56}{27}\pi^2+\frac{20}{9}f\qquad\qquad {\rm (NDR)}
\end{eqnarray}
$N$ ($f$) denotes the number of colors (flavors). The anomalous
dimension $\gamma^{(0)}_{hl}$ has been first calculated by
\cite{voloshinshifman:87} and \cite{politzerwise:88a},
\cite{politzerwise:88b}. The generalization to next-to-leading order
has been performed in \cite{jimuslof:91} and
\cite{broadhurstgrozin:91}.
\\
The RG equations are readily solved to obtain the coefficients at a lower
but still perturbative scale $\mu$, where, formally,
$\mu\ll\mu_h={\cal O}(m)$. Using the results of section
\ref{sec:basicform:wc} we have
\begin{equation}\label{c1hl}
C_1(\mu)=\left(1+\frac{\alpha_s(\mu)}{4\pi}J_{hl}\right)
\left[\frac{\alpha_s(\mu_h)}{\alpha_s(\mu)}\right]^
{d_{hl}}\left(1+\frac{\alpha_s(\mu_h)}{4\pi}
\left[\gamma^{(0)}_{hl}\ln\frac{\mu_h}{m}+B_1-J_{hl}\right]\right)
\end{equation}
\begin{equation}\label{c2hl}
C_2(\mu)=\left[\frac{\alpha_s(\mu_h)}{\alpha_s(\mu)}\right]^
{d_{hl}}\frac{\alpha_s(\mu_h)}{4\pi}B_2
\end{equation}
with
\begin{equation}\label{djhl}
d_{hl}=\frac{\gamma^{(0)}_{hl}}{2\beta_0}\qquad\qquad
J_{hl}=\frac{d_{hl}}{\beta_0}\beta_1-\frac{\gamma^{(1)}_{hl}}{2\beta_0}
\end{equation}
We remark that the corresponding formulae for the vector current can
be simply obtained from the above expressions by letting $\gamma_5\to 1$
and changing the sign of $B_2$.
\\
In addition to the case of heavy-light currents considered here, the
NLO corrections have also been calculated for flavor-conserving and
flavor-changing heavy-heavy currents of the type $\bar\Psi\Gamma\Psi$
and $\bar\Psi_1\Gamma\Psi_2$ respectively, where $\Psi$, $\Psi_{1,2}$
are heavy quark fields ($\Gamma=\gamma_\mu$, $\gamma_\mu\gamma_5$). In
these cases the anomalous dimensions become velocity dependent.
Additional complications arise in the analysis of flavor changing
heavy-heavy currents due to the presence of two distinct heavy mass
scales. For a detailed presentation see \cite{neubert:94} and
references cited therein.

\subsection{The Pseudoscalar Decay Constant in the Static Limit}
\label{sec:HQET:DecCons}
An important application of the results summarized in the last
section is the calculation of the short distance QCD effects, from
scales between $\mu_h={\cal O}(m)$ and the low scale $\mu={\cal O}(1 GeV)$,
for the decay constants $f_P$ of pseudoscalar heavy mesons. Using only
the leading term in the expansion (\ref{ac12}), omitting all $1/m$
power corrections, corresponds to the static limit for $f_P$, which plays
some role in lattice studies. As already mentioned we will restrict
ourselves here to this limit. We should remark however, that
non-negligible power corrections are known to exist for the realistic case 
of B or D meson decay constants \cite{sachrajda:92}.
\\
The decay constant $f_P$ is defined through
\begin{equation}\label{apme}
\langle 0|A|P\rangle =-i f_P m_P v_\mu
\end{equation}
where the pseudoscalar meson state $|P\rangle$ is normalized in the
conventional way ($\langle P|P\rangle=2 E V$). The matrix elements of
$\tilde A_{1,2}$ are related via heavy quark symmetry and are given by
\begin{equation}\label{atme}
\langle 0|\tilde A_1|P\rangle =-\langle 0|\tilde A_2|P\rangle =
  -i \tilde f(\mu) \sqrt{m_P} v_\mu
\end{equation}
Apart from the explicit mass factor $\sqrt{m_P}$, which is merely due to
the normalization of $|P\rangle$, these matrix elements are independent
of the heavy quark mass. The "reduced" decay constant $\tilde f(\mu)$ is
therefore $m$-independent. It does however depend on the 
renormalization scale and scheme chosen. The computation of
$\tilde f(\mu)$ is a nonperturbative problem involving strong dynamics
below scale $\mu$. Using (\ref{ac12}), (\ref{c1hl}), (\ref{c2hl}),
(\ref{apme}) and (\ref{atme}) we obtain
\begin{equation}\label{fpft}
f_P=\frac{\tilde f(\mu)}{\sqrt{m_P}}\left(1+\frac{\alpha_s(\mu)}{4\pi}J_{hl}
\right)\left[\frac{\alpha_s(\mu_h)}{\alpha_s(\mu)}\right]^{d_{hl}}
\left(1+\frac{\alpha_s(\mu_h)}{4\pi}
\left[\gamma^{(0)}_{hl}\ln\frac{\mu_h}{m}+B_1-J_{hl}-B_2\right]\right)
\end{equation}
The dependence of the coefficient function on the renormalization scheme
through $J_{hl}$ in the second factor in (\ref{fpft}), and its dependence on
$\mu$ cancel the corresponding dependences in the hadronic quantity
$\tilde f(\mu)$ to the considered order in $\alpha_s$. The last factor
in (\ref{fpft}) is scheme independent. Furthermore the cancellation of
the dependence on $\mu_h$ to the required order can be seen explicitly.
Note also the leading scaling behaviour $f_P\sim 1/\sqrt{m_P}$, which
is manifest in (\ref{fpft}).
\\
Although $\tilde f(\mu)$ cannot be calculated without nonperturbative
input, its independence of the heavy quark mass $m$ implies that $\tilde f$
will drop out in the ratio of $f_B$ over $f_D$, if charm is treated as
a heavy quark. One thus obtains
\begin{eqnarray}\label{fbfd}
\lefteqn{\frac{f_B}{f_D} = \sqrt{\frac{m_D}{m_B}}
\left[\frac{\alpha_s(\mu_b)}{\alpha_s(\mu_c)}\right]^{d_{hl}}
 \left( 1 +\frac{\alpha_s(\mu_b)-\alpha_s(\mu_c)}{4\pi}
\left(B_1-J_{hl}-B_2\right)+ \right.} \hspace{2cm} \nonumber\\
& & \hspace{3cm} + \left. \frac{\alpha_s(\mu_b)}{4\pi}
\gamma^{(0)}_{hl}\ln\frac{\mu_b}{m_b}-\frac{\alpha_s(\mu_c)}{4\pi}
\gamma^{(0)}_{hl}\ln\frac{\mu_c}{m_c}\right) 
\end{eqnarray}
The QCD factor on the right hand side of (\ref{fbfd}) amounts to $1.14$
for $m_b=4.8 GeV$, $m_c=1.4GeV$ and $\Lambda_{\overline{MS}}=0.2 GeV$
if we set $\mu_i=m_i$, $i=b,c$. If we allow for a variation of the
renormalization scales as $2/3\leq\mu_i/m_i\leq 2$, this factor lies
within a range of $1.12$ to $1.16$. This is to be compared with the
leading log approximation, where the central value reads $1.12$ with a
variation from $1.10$ to $1.15$. Note that due to cancellations in the
ratio $f_B/f_D$ the scale ambiguity is not much larger in LLA than in
NLLA. However the next-to-leading order QCD effects further enhance
$f_B/f_D$ independently of the renormalization scheme. 

\subsection{$\Delta B=2$ Transitions in the Static Limit}
\label{sec:HQET:DelB2}
In section \ref{sec:HeffBBbar} we have described the effective
hamiltonian for $B^0-\bar B^0$ mixing. The calculation of the mixing
amplitude requires in particular the evaluation of the matrix element
$\langle \bar B^0|Q|B^0\rangle\equiv\langle Q\rangle$ of the operator
\begin{equation}\label{qbd}
Q=(\bar bd)_{V-A}(\bar bd)_{V-A}
\end{equation}
in addition to the short-distance Wilson coefficient. Coefficient
function and operator matrix element are to be evaluated at a common
renormalization scale, $\mu_b={\cal O}(m_b)$, say. In contrast to the 
determination of the Wilson coefficient, the computation of the hadronic
matrix element involves nonperturbative long-distance contributions.
Ultimately this problem should be solved using lattice QCD. However,
the b quark is rather heavy and it is therefore difficult to
incorporate it as a fully dynamical field in the context of a
lattice regularization approach. On the other hand QCD effects from
scales below $\mu_b={\cal O}(m_b)$ down to $\sim 1GeV$ are still
accessible to a perturbative treatment. HQET provides the tool to
calculate these contributions. At the same time it allows one to
extract the dependence of $\langle \bar B^0|Q|B^0\rangle$ on the bottom
mass $m_b$ explicitly, albeit at the prize of the further
approximation introduced by the expansion in inverse powers of $m_b$.
\\
In a first step the operator $Q$ in (\ref{qbd}) is expressed as a
linear combination of HQET operators by matching the "full" to the
effective theory at a scale $\mu_b={\cal O}(m_b)$
\begin{equation}\label{qqtm}
\langle Q(\mu_b)\rangle=\left(1+\frac{\alpha_s(\mu_b)}{4\pi}
\left[(\tilde\gamma^{(0)}-\gamma^{(0)})\ln\frac{\mu_b}{m_b}+\tilde B-B
\right]\right)\langle\tilde Q(\mu_b)\rangle+
\frac{\alpha_s(\mu_b)}{4\pi}\tilde B_s \langle\tilde Q_s(\mu_b)\rangle
\end{equation}
Here 
\begin{equation}\label{qqst}
\tilde Q=2(\bar hd)_{V-A}(\bar h^{(-)}d)_{V-A}\qquad\qquad
\tilde Q_s=2(\bar hd)_{S-P}(\bar h^{(-)}d)_{S-P}
\end{equation}
($(\bar hd)_{S-P}\equiv\bar h(1-\gamma_5)d$)
are the necessary operators in HQET relevant for the case of a
$B^0\to\bar B^0$ transition. The field $\bar h$ creates a heavy quark,
while $\bar h^{(-)}$ annihilates a heavy antiquark. Since the
effective theory field $\bar h$ ($\bar h^{(-)}$) cannot, unlike the
full theory field $\bar b$ in $Q$, at the same time annihilate (create)
the heavy antiquark (heavy quark), explicit factors of two have to appear
in (\ref{qqst}). Similarly to the case of the heavy-light current
discussed in the previous section a new operator $\tilde Q_s$ with
scalar-pseudoscalar structure is generated. Its coefficient is finite
and hence no operator mixing under infinite renormalization occurs
between $\tilde Q$ and $\tilde Q_s$.
\\
In a second step, the matrix element $\langle\tilde Q(\mu_b)\rangle$
at the high scale $\mu_b$ has to be expressed through the matrix
element of $\tilde Q$ evaluated at a lower scale $\mu\sim 1 GeV$,
which is relevant for a nonperturbative calculation, for example
using lattice gauge theory. This relation can be obtained through the
usual renormalization group evolution and reads in NLLA
\begin{equation}\label{qtrg}
\langle\tilde Q(\mu_b)\rangle=
\left[\frac{\alpha_s(\mu_b)}{\alpha_s(\mu)}\right]^{\tilde d}
\left(1+\frac{\alpha_s(\mu)-\alpha_s(\mu_b)}{4\pi}\tilde J\right)
\langle\tilde Q(\mu)\rangle
\end{equation}
where
\begin{equation}\label{djt}
\tilde d=\frac{\tilde\gamma^{(0)}}{2\beta_0}\qquad\qquad
\tilde J=\frac{\tilde d}{\beta_0}\beta_1-\frac{\tilde\gamma^{(1)}}{2\beta_0}
\end{equation}
with the beta-function coefficients $\beta_0$ and $\beta_1$ given in
(\ref{b0b1}). The calculation of the one-loop anomalous dimension
$\tilde\gamma^{(0)}$ of the HQET operator $\tilde Q$, required for the
leading log approximation to (\ref{qtrg}), has been first performed in
\cite{voloshinshifman:87} and \cite{politzerwise:88a},
\cite{politzerwise:88b}. The computation of the two-loop anomalous
dimension $\tilde\gamma^{(1)}$ is due to \cite{gimenez:93}. Finally,
the next-to-leading order matching condition (\ref{qqtm}) has been
determined in \cite{flynnetal:91}.  In the following we summarize the
results obtained in these papers.

The scheme dependent next-to-leading order quantities $B$, $\tilde B$
and $\tilde\gamma^{(1)}$ refer to the NDR scheme with anticommuting
$\gamma_5$ and the usual subtraction of evanescent terms as defined
in \cite{burasweisz:90}. For $N=3$ colors we then have
\begin{equation}\label{ggt}
\tilde\gamma^{(0)}=-8\qquad\qquad\gamma^{(0)}=4
\end{equation}
\begin{equation}\label{bbt}
\tilde B-B=-14\qquad B=\frac{11}{3}\qquad \tilde B_s=-8
\end{equation}
\begin{equation}\label{gt1}
\tilde\gamma^{(1)}=-\frac{808}{9}-\frac{52}{27}\pi^2+\frac{64}{9}f
\end{equation}
where $f$ is the number of active flavors.  

At this point we would like to make the following comments.
\begin{itemize}
\item
The logarithmic term in (\ref{qqtm}) reflects the ${\cal O}(\alpha_s)$
scale dependence of the matrix elements of $Q$ and
$\tilde Q$. Accordingly its coefficient is given by
the difference in the one-loop anomalous dimensions of these operators,
$\gamma^{(0)}$ and $\tilde\gamma^{(0)}$.
\item
The one-loop anomalous dimension of the effective theory operator
$\tilde Q$, $\tilde\gamma^{(0)}$, is exactly twice as large as the
one-loop anomalous dimension of the heavy-light current discussed in
section \ref{sec:HQET:Currents} (see eq. (\ref{g0hl}). Therefore the
scale dependence of $\langle\tilde Q\rangle$ below $\mu_b$ is entirely
contained in the scale dependence of the decay constant squared
$\tilde f^2(\mu)$. This implies the well known result that in leading log
approximation the parameter $B_B$ has no perturbative scale
dependence in the static theory below $\mu_b$. As the result of
\cite{gimenez:93} for $\tilde\gamma^{(1)}$ shows, this somewhat accidental
cancellation is not valid beyond the one-loop level.
\item
The matching condition (\ref{qqtm}) contains besides the logarithm
a scheme dependent constant term in the relation between
$\langle Q\rangle$ and $\langle\tilde Q\rangle$. We have written this
coefficient in the form $\tilde B-B$ in order to make the
cancellation of scheme dependences, to be discussed below, more
transparent. Here $B$ is identical to $B_+$ introduced in (\ref{B8})
and characterizes the scheme dependence of $\langle Q\rangle$ 
(see also sections \ref{sec:HeffKKbar} and \ref{sec:HeffBBbar}).
\item
The quantity $\tilde\gamma^{(1)}$ has been originally calculated in
\cite{gimenez:93} using dimensional reduction (DRED) instead of NDR as
renormalization scheme. However, $\tilde B$ turns out to be the same in
DRED and NDR, implying that also $\tilde\gamma^{(1)}$ coincides in
these schemes \cite{gimenez:93}.
\end{itemize}

Finally we may put together (\ref{qqtm}) and (\ref{qtrg}), omitting
for the moment the scheme independent, constant correction due to
$\tilde Q_s$, to obtain 
\begin{equation}\label{qqtl}
\langle Q(\mu_b)\rangle=
\left[\frac{\alpha_s(\mu_b)}{\alpha_s(\mu)}\right]^{\tilde d}
\left(1+\frac{\alpha_s(\mu_b)}{4\pi}
\left[(\tilde\gamma^{(0)}-\gamma^{(0)})\ln\frac{\mu_b}{m_b}+\tilde B-B
-\tilde J\right]+\frac{\alpha_s(\mu)}{4\pi}\tilde J\right)
\langle\tilde Q(\mu)\rangle
\end{equation}
This relation serves to express the $B^0-\bar B^0$ matrix element of the
operator $Q$, evaluated at a scale $\mu_b={\cal O}(m_b)$, which is the
relevant scale for the effective hamiltonian of section 
\ref{sec:HeffBBbar}, in terms of the static theory matrix element
$\langle\tilde Q(\mu)\rangle$ normalized at a low scale $\mu\sim 1 GeV$.
The latter is more readily accessible to a nonperturbative lattice
calculation than the full theory matrix element $\langle Q(\mu_b)\rangle$.
Note that (\ref{qqtl}) as it stands is valid in the continuum theory.
In order to use lattice results one still has to perform an ${\cal
O}(\alpha_s)$ matching of $\tilde Q$ to its lattice counterpart. This
step however does not require any further renormalization group
improvement since by means of (\ref{qqtl}) $\tilde Q$ is already
normalized at the appropriate low scale $\mu$.  The continuum --
lattice theory matching was determined in \cite{flynnetal:91} and is
also discussed in \cite{gimenez:93}.
\\
Of course, the right hand side in (\ref{qqtl}) gives only the leading
contribution in the $1/m$ expansion of the full matrix element $\langle
Q(\mu_b)\rangle$ (apart from $\tilde Q_s$). Going beyond this
approximation would require the consideration of several new operators,
which appear at the next order in $1/m$.  These contributions have been
studied in \cite{kilianmannel:93} in the leading logarithmic
approximation. On the other hand (\ref{qqtl}), while restricted to the
static limit, includes and resums all leading and next-to-leading
logarithmic corrections between the scales $\mu_b={\cal O}(m_b)$ and
$\mu\sim 1GeV$ in the relation among $Q$ and $\tilde Q$. It is
interesting to consider the scale and scheme dependences present in
(\ref{qqtl}). The dependence on $\mu$ in the first factor on the r.h.s.
of (\ref{qqtl}) is canceled by the $\mu$-dependence of $\langle\tilde
Q(\mu)\rangle$. The dependence on $\mu_b$ of this factor is canceled by
the explicit $\ln\mu_b$ term proportional to $\tilde\gamma^{(0)}$.
Hence the only scale dependence remaining on the r.h.s., to the
considered order ${\cal O}(\alpha_s)$, is the one
$\sim\alpha_s(\mu_b)\gamma^{(0)}\ln\mu_b$. This is precisely the scale
dependence of the full theory matrix element on the l.h.s., which is
required to cancel the corresponding dependence of the Wilson
coefficient. Similarly the term $\sim\alpha_s(\mu_b)B$ represents the
correct scheme dependence of $\langle Q(\mu_b)\rangle$, while the
scheme dependence of $\alpha_s(\mu)\tilde J$ cancels with the scheme
dependence of $\langle\tilde Q(\mu)\rangle$ and the difference $\tilde
B-\tilde J$ is scheme independent by itself.  This discussion
demonstrates explicitly that the transition from full QCD to HQET can
be made at an arbitrary scale $\mu_b={\cal O}(m_b)$, as we have already
emphasized above.
\\
Finally we would like to remark that since the logarithm $\ln\mu_b/\mu$
is not really very large in the present case, one might take the
attitude of neglecting higher order resummations of logarithmic terms
altogether and restricting oneself to the ${\cal O}(\alpha_s)$
corrections alone. Then (\ref{qqtm}) would be already the final result,
as it was used in \cite{flynnetal:91}. This approximation is fully
consistent from a theoretical point of view. Yet it is useful to have
the more complete expression (\ref{qqtl}) at hand. Of course, as
indicated above, the finite ${\cal O}(\alpha_s)$ correction due to the
matrix element of $\tilde Q_s$ in (\ref{qqtm}) must still be added to
the r.h.s. of (\ref{qqtl}).  However, to complete the NLO
renormalization group calculation also the leading logarithmic
corrections related to the operator $\tilde Q_s$ should then be
resummed. To our knowledge this part of the analysis has not yet been
performed in the literature so far.

\skipevenpage

{\Huge\bf
\noindent
Part Three --

\bigskip
\bigskip
\bigskip

\noindent
The Phenomenology of Weak Decays
}

\vfil

\noindent
The third part of our review presents the phenomenological picture
of weak decays beyond the leading logarithmic approximation.

There is essentially a one-to-one correspondence between the sections
in the second and in the third part of this review. Part three
uses heavily the results derived in part two. In spite of this, the
third part is meant to be essentially self-contained and can be followed
without difficulties by those readers who only scanned the material
of the second part and read section \ref{sec:sewm}.

The phenomenological part of our review is organized as follows.  We
begin with a few comments on the input parameters in section
\ref{sec:inputparams}. Next, as an application of the NLO corrections
in the current-current sector, we summarize the present status of the
tree level inclusive B-decays, in particular the theoretical status of
the semi-leptonic branching ratio.  The issue of exclusive two-body
non-leptonic decays and the question of factorization will not be
discussed here. The numerical values of the related factors $a_i$ for
various renormalization schemes can be found in \cite{buras:94}.
\\
The main part of the phenomenology begins in section \ref{sec:epsBBUT}
where we update the "standard" analysis of the unitarity triangle based
on the indirect CP violation in $K\to\pi\pi$ (the parameter
$\varepsilon_K$) and the $B^0_d-\bar B^0_d$ mixing described by $x_d$.
We incorporate in this analysis the most recent values of $m_t$,
$V_{ub}/V_{cb}$, $V_{cb}$, $B_K$ and $F_B$.  In addition to the
analysis of the unitarity triangle we determine several quantities of
interest. These results will be used frequently in subsequent sections.
\\
In section \ref{sec:nloepe} we present $\varepsilon'/\epsilon$ beyond
leading logarithms, summarizing and updating the extensive analysis
presented in \cite{burasetal:92d}. $\varepsilon'$ measures the size of
the direct CP violation in $K\to \pi\pi$ and its accurate estimate is
an important but very difficult task. In section \ref{sec:mki12} we
discuss briefly the $K_L-K_S$ mass difference and the $\Delta I=1/2$
rule. Next, in section \ref{sec:KLpee} we present an update for 
$K_L\to\pi^0 e^+e^-$.
\\
Next in sections \ref{sec:Heff:Bsgamma} and \ref{sec:Heff:BXsee:nlo} we
consider $B\to X_s\gamma$ and $B\to X_s e^+e^-$, respectively.  $B\to
X_s\gamma$ is known only in the LO approximation. However, in view of
its importance we summarize the leading order formulae and show the
standard model prediction compared with the CLEO II findings. We also
summarize the present status of NLO calculations for this decay. The
NLO calculations for $B\to X_s e^+ e^-$ have been completed and we give
a brief account of these results.
\\
Sections \ref{sec:Kpnn}--\ref{sec:BXnnBmm} discuss  $K\to
\pi\nu\bar\nu$, $K_L\to \mu^+\mu^-$ and rare B-decays ($B\to
X_s\nu\bar\nu$, $B\to l^+l^-$).  Except for $K_L\to \mu^+\mu^-$, all
these decays have only very small hadronic uncertainties and the
dominant theoretical errors are related to various renormalization
scale ambiguities. We demonstrate that these uncertainties are
considerably reduced by including NLO corrections, which will improve
the determination of the CKM matrix in forthcoming experiments. Using
the results of section \ref{sec:epsBBUT}, we also give updated standard
model predictions for these decays.

\section{Comments on Input Parameters}
         \label{sec:inputparams}
The phenomenology of weak decays depends sensitively on a number of
input parameters. We have collected the numerical values of these
parameters in appendix \ref{app:numinput}. To this end we have
frequently used the values quoted by \cite{particledata:94}. The basis
for our choice of the numerical values for various non-perturbative
parameters, such as $B_K$ or $F_B$, will be given in the course of our
presentation. In certain cases, like the B-meson life-times and the
size of the $B^0_d-\bar B^0_d$ mixing, for which the experimental world
averages change constantly we have chosen values, which are in the ball
park of those presented at various conferences and workshops during the
summer of 1995. Here we would like to comment briefly on three
important parameters:  $\left| V_{cb} \right|$, $\left| V_{ub}/V_{cb}
\right|$ and $\mt$.  The importance of these parameters lies in the
fact that many branching ratios and also the CP violation in the
Standard Model depend sensitively on them.

\subsection{CKM Element $\left| V_{cb} \right|$}
         \label{subsec:inputparams:Vcb}
During the last two years there has been a considerable progress made
by experimentalists \cite{patterson:95} and theorists in the extraction of
$\left| V_{cb} \right|$ from the exclusive and inclusive B-decays.  In
these investigations the HQET in the case of exclusive decays and the
Heavy Quark Expansions for inclusive decays played a considerable
role.  In particular we would like to mention the important papers
\cite{neubert:94b}, \cite{shifmanetal:95} and \cite{braunetal:95} on
the basis of which one is entitled to use:
\begin{equation}
\label{eq:Vcberr}
\left| V_{cb} \right|=0.040\pm0.003\quad =>\quad A=0.82\pm 0.06
\end{equation}
This should be compared with an error of $\pm 0.006$  for 
$\left| V_{cb} \right|$ quoted still in 1993. The corresponding
reduction of the error in $A$ by a factor of 2 has important
consequences for the phenomenology of weak decays.

\subsection{CKM Element Ratio $\left| V_{ub}/V_{cb} \right|$}
         \label{subsec:inputparams:Vubcb}
Here the situation is much worse and the value
\begin{equation}
\label{eq:Vubcberr}
\left| \frac{V_{ub}}{V_{cb}} \right|=0.08\pm0.02
\end{equation}
quoted by \cite{particledata:94} appears to be still valid. There is a
hope that the error could be reduced by a factor of 2 to 4 in the
coming years both due to the theory \cite{braunetal:95} and the recent
CLEO measurements of the exclusive semileptonic decays $B \to
(\pi,\varrho)l\nu_l$ \cite{thorndike:95}.

\subsection{Top Quark Mass $\mt$}
         \label{subsec:inputparams:mt}
Next it is important to stress that the discovery of the top quark
\cite{abeetal:94a}, \cite{abeetal:94b}, \cite{abeetal:95}, \cite{D0:95}
and its mass measurement had an important impact on the field of rare
decays and CP violation reducing considerably one potential
uncertainty. It is however important to keep in mind that the parameter
$\mt$, the top quark mass, used in weak decays is not equal to the one
used in the electroweak precision studies at LEP or SLD. In the latter
investigations the so-called pole mass is used, whereas in all the NLO
calculations listed in table \ref{tab:processes} and used in this
review, $\mt$ refers to the running current top quark mass normalized
at $\mu=\mt$:  $\bar \mt(\mt)$. One has
\begin{equation}
\mt^{\rm (pole)}=\bar \mt(\mt)
\left[ 1+\frac{4}{3}\frac{\as(\mt)}{\pi}\right]
\label{eq:mtpolebar}
\end{equation}
so that for $\mt={\cal O}(170\gev)$, $\bar \mt(\mt)$ is typically by
$8\gev$ smaller than $\mt^{\rm (Pole)}$. This difference will matter in
a few years.

In principle any definition $\bar \mt(\mu_t)$ with $\mu_t=\ord(\mt)$
could be used. In the leading order this arbitrariness in the choice of
$\mu_t$ introduces a potential theoretical uncertainty in those
branching ratios which depend sensitively on the top quark mass. The
inclusion of NLO corrections reduces this uncertainty considerably so
that the resulting branching ratios remain essentially independent of
the choice of $\mu_t$. We have discussed this point already in previous
sections. Numerical examples will be given in this part below.  The
choice $\mu_t=\mt$ turns out to be convenient and will be adopted in
what follows.

Using the $\mt^{\rm (pole)}$ quoted by CDF \cite{abeetal:94a},
\cite{abeetal:94b}, \cite{abeetal:95} together with the relation
\eqn{eq:mtpolebar} we find roughly
\begin{equation}
\mt\equiv\bar \mt(\mt)=(170\pm15)\gev
\end{equation}
which we will use in our phenomenological applications. In principle an
error of $\pm 11\gev$ could be used but we prefer to be
conservative.

\section{Inclusive B Decays}
\label{sec:InclB}
\subsection{General Remarks}
\label{sec:InclB:General}
Inclusive decays of $B$ mesons constitute an important testing ground
for our understanding of strong interaction dynamics in its interplay
with the weak forces. At the same time inclusive semileptonic modes
provide useful information on $|V_{cb}|$.
\\
Due to quark-hadron duality inclusive decays of heavy mesons can, in
general, be calculated more reliably than corresponding exclusive
modes. During recent years a systematic formulation for the treatment
of inclusive heavy meson decays has been developed. It is based on
operator product and heavy quark expansion, which are applied to the
$B$ meson inclusive width, expressed as the absorptive part of the $B$
forward scattering amplitude
\begin{equation}\label{gabx}
\Gamma(B\to X)=\frac{1}{2m_B}{\IM}\left(i \int d^4x
 \langle B | T \, {\cal H}^{(X)}_{eff}(x){\cal H}^{(X)}_{eff}(0) | B
\rangle\right)
\end{equation}
Here ${\cal H}^{(X)}_{eff}$ is the part of the complete $\Delta B=1$
effective hamiltonian that contributes to the particular inclusive
final state $X$ under consideration. E.g.\ for inclusive semileptonic decays
\begin{equation}\label{hbsl}
{\cal H}^{(SL)}_{eff,\Delta B=1}=\frac{G_F}{\sqrt{2}} V_{cb}
(\bar cb)_{V-A} \sum_{l=e,\mu,\tau}(\bar l\nu_l)_{V-A}+ h.c.
\end{equation}
For nonleptonic modes the relevant expression is the $\Delta B=1$ short
distance effective hamiltonian given in \eqn{eq:HeffdB1:66}.  It has
been shown in \cite{Chay}, \cite{Bj}, \cite{bigietal:92},
\cite{bigietal:93}, \cite{manoharwise:94}, \cite{bloketal:94},
\cite{falketal:94}, \cite{mannel:94}, \cite{Bigi}, that the leading
term in a systematic expansion of (\ref{gabx}) in $1/m_b$ is determined
by the decay width of a free b-quark calculated in the parton picture.
Furthermore, the nonperturbative corrections to this perturbative
result start at order $(\Lambda/m_b)^2$, where $\Lambda$ is a hadronic
scale $\sim 1\gev$, and are quite small in the case of B decays.  In
the light of this formulation it becomes apparent that the
perturbative, partonic description of heavy hadron decay is thus
promoted from the status of a model calculation to the leading
contribution in a systematic expansion based on QCD.  We will still
comment on the $(\Lambda/m_b)^2$ corrections below.  In the following
we will however concentrate on the leading quark level analysis of
inclusive $B$ decays. As we shall see, the treatment of short-distance
QCD effects at the next-to-leading order level -- at least for the
dominant modes -- is of crucial importance for a proper understanding
of these processes.
\\
The calculation of b-quark decay starts from the effective 
$\Delta B=1$ hamiltonian containing the relevant four-fermion
operators multiplied by Wilson coefficients. To obtain the decay rate,
the matrix elements (squared) of these operators have to be calculated
perturbatively to the required order in $\as$. While in LLA a
zeroth order evaluation is sufficient, ${\cal O}(\as)$ virtual
gluon effects (along with real gluon bremsstrahlung contributions for
the proper cancellation of infrared divergences in the inclusive rate)
have to be taken into account at NLO. In this way the renormalization
scale and scheme dependence present in the coefficient functions is
canceled to the considered order (${\cal  O}(\as)$) in the
decay rate. Thus, by contrast to low energy decays, in the case of
inclusive heavy quark decay, a physical final result can be obtained
within perturbation theory alone.
\\
Our goal will be in particular to review the present status of the
theoretical prediction for the $B$ meson semileptonic branching ratio
$B_{SL}$. This quantity has received some attention in recent years
since theoretical calculations \cite{altarellipetrarca:91},
\cite{tanimoto:92}, \cite{palmerstech:93}, \cite{bigietal:94},
\cite{falketal:94b} tended to yield values around $12.5-13.5\%$, above
the experimental figure $B_{SL}=(10.4\pm 0.4)\%$
\cite{particledata:94}.  However, these earlier analyses have not been
complete in regard to the inclusion of final state mass effects and NLO
QCD corrections in the nonleptonic widths. More precisely, these
calculations took into account mass effects appropriate for the leading
order in QCD along with NLO QCD corrections obtained for massless final
state quarks.  Recently the most important of these -- so far lacking
-- mass effects have been properly included in the NLO QCD calculation
through the work of \cite{baganetal:94a}, \cite{baganetal:94b},
\cite{baganetal:95}. These ${\cal O}(\as)$ mass effects tend to
decrease $B_{SL}$ and, according to the analysis of these authors
essentially bring it, within theoretical uncertainties, into agreement
with the experimental number.  Before further discussing these issues,
it is appropriate to start with a short overview summarizing the
possible b-quark decay modes and classifying their relative
importance.

\subsection{b-Quark Decay Modes}
\label{sec:InclB:bdecay}
First of all, a b-quark can decay {\it semileptonically\/} to the
final states $cl\bar\nu_l$ and $ul\bar\nu_l$ with $l=e$, $\mu$, $\tau$.
\\
In the case of nonleptonic final states we may distinguish three classes:
Decays induced through current-current operators alone (class I),
decays induced by both current-current and penguin operators (class II)
and pure penguin transitions (class III). We have

\begin{center}
\begin{tabular}{|c|l|}
\hline
Class & \multicolumn{1}{c|}{Final State} \\
\hline
I    & $c\bar ud$, \quad $c\bar us$; \qquad  $u\bar cs$, \quad $u\bar cd$ \\
II   & $c\bar cs$, \quad $c\bar cd$; \qquad  $u\bar ud$, \quad $u\bar us$ \\
III  & $d\bar dd$, \quad $d\bar ds$; \qquad  $s\bar sd$, \quad $s\bar ss$ \\
\hline
\end{tabular}
\end{center}

Clearly there is a rich structure of possible decay modes even at the
quark level and a complete treatment would be quite complicated.
However, not all of these final states are equally important. In order
to perform the analysis of b-quark decay, in particular in view of the
calculation of $B_{SL}$, it is useful to identify the most important
channels and to introduce appropriate approximations in dealing
with less prominent decays. To organize the procedure, we make the
following observations:
\begin{itemize}
\item
The dominant, i.e. CKM allowed and tree-level induced, decays are
$b\to cl\nu$, $b\to c\bar ud$ and $b\to c\bar cs$. For these a complete
NLO calculation including final state mass effects is necessary.
\item
The channels $c\bar us$, $c\bar cd$, $u\bar cd$,
$u\bar us$ may be incorporated with excellent accuracy into the modes
$c\bar ud$, $c\bar cs$, $u\bar cs$, $u\bar ud$, respectively, using
the approximate CKM unitarity in the first two generations.
The error introduced thereby through the $s$-$d$ mass difference is
entirely negligible.
\item
Penguin transitions are generally suppressed by the smallness of their
Wilson coefficient functions, which are typically of the order of
a few percent. For this reason, one may neglect the pure penguin
decays of class III altogether as their decay rates involve
penguin coefficients squared.
\item 
Furthermore we may neglect the penguin contributions to the CKM
suppressed $b\to u$ transitions of class II.
\item
In addition one may treat the remaining smaller effects, namely $b\to u$
transitions and the interference of penguins with the leading
current-current contribution in $b\to c\bar cs$ within the
leading log approximation.
\item
Finally, rare, flavor-changing neutral current b-decay modes are
negligible in the present context as well.
\end{itemize}
Next we will write down expressions for the relevant decay rate
contributions we have discussed.
\\
For the dominant modes $b\to cl\nu$, $b\to c\bar ud$ and
$b\to c\bar cs$ (without penguin effects) one has at
next-to-leading order:
\begin{equation}\label{bcln}
\Gamma(b\to cl\nu)=\Gamma_0 P(x_c,x_l,0)\left[
 1+\frac{2\as(\mu)}{3\pi} g(x_c,x_l,0)\right]
\end{equation}
\begin{eqnarray}\label{bcud}
\Gamma(b\to c\bar ud)&=&\Gamma_0 P(x_c,0,0)\left[2L^2_++L^2_-+
 \frac{\as(M_W)-\as(\mu)}{2\pi}(2L^2_+ R_++L^2_- R_-)\right.
\nonumber\\
 &+&\frac{2\as(\mu)}{3\pi}\left(\frac{3}{4}(L_+-L_-)^2 
g_{11}(x_c)+\frac{3}{4}(L_++L_-)^2g_{22}(x_c)\right.
\nonumber\\
&&+\left.\left.\frac{1}{2}
(L^2_+-L^2_-)(g_{12}(x_c)-12 \ln\frac{\mu}{m_b})\right)\right]
\end{eqnarray}
\begin{eqnarray}\label{bccs}
\Gamma(b\to c\bar cs)&=&\Gamma_0 P(x_c,x_c,x_s)\left[2L^2_++L^2_-+
 \frac{\as(M_W)-\as(\mu)}{2\pi}(2L^2_+ R_++L^2_- R_-)\right.
\nonumber\\
 &+&\frac{2\as(\mu)}{3\pi}\left(\frac{3}{4}(L_+-L_-)^2 
h_{11}(x_c)+\frac{3}{4}(L_++L_-)^2h_{22}(x_c)\right.
\nonumber\\
&&+\left.\left.\frac{1}{2}
(L^2_+-L^2_-)(h_{12}(x_c)-12 \ln\frac{\mu}{m_b})\right)\right]
\end{eqnarray}
Eq.\ \eqn{bccs} neglects small strange quark mass effects in the NLO
terms, which have however been included in the numerical analysis in
\cite{baganetal:95}.
In the equations above $\Gamma_0=G^2_Fm^5_b|V_{cb}|^2/(192\pi^3)$ and
$P(x_1,x_2,x_3)$ is the leading order phase space factor given for
arbitrary masses $x_i=m_i/m_b$ by
\begin{equation}\label{px123}
P(x_1,x_2,x_3)=12\int\limits_{(x_2+x_3)^2}^{(1-x_1)^2} \frac{ds}{s}
(s-x^2_2-x^2_3)(1+x^2_1-s) w(s,x^2_2,x^2_3) w(s,x^2_1,1)
\end{equation}
\begin{equation}\label{wabc}
w(a,b,c)=(a^2+b^2+c^2-2ab-2ac-2bc)^{1/2}
\end{equation}
$P$ is a completely symmetric function of its arguments.
\\
Furthermore
\begin{equation}\label{lpmmu}
L_\pm=L_\pm(\mu)=\left[\frac{\as(M_W)}{\as(\mu)}
\right]^{d_\pm}
\end{equation}
with $d_+=6/23$, $d_-=-12/23$ (see (\ref{B10})) and $\mu={\cal
O}(m_b)$.  The scheme independent $R_\pm$ come from the NLO
renormalization group evolution and are given by $R_\pm=B_\pm-J_\pm$
(see (\ref{B9})).  For $f=5$ flavors $R_+=6473/3174$,
$R_-=-9371/1587$.  Note that the leading dependence of $L_\pm$ on the
renormalization scale $\mu$ is canceled to ${\cal O}(\as)$ by the
explicit $\mu$-dependence in the $\as$-correction terms.  Virtual
gluon and bremsstrahlung corrections to the matrix elements of four
fermion operators are contained in the mass dependent functions $g$,
$g_{ij}$ and $h_{ij}$.
\\
The function $g(x_1,x_2,x_3)$ is available for arbitrary $x_1$, $x_2$,
$x_3$ from \cite{hokimpham:83}, \cite{hokimpham:84}. The special case
$g(x_1,0,0)$ has been analysed also in \cite{CM:78}. Analytical
expressions have been given in \cite{nir:89} for $g(x_1,0,0)$ and in
\cite{baganetal:94a} for $g(0,x_2,0)$.  The functions $g_{11}(x)$,
$g_{12}(x)$ and $g_{22}(x)$ are calculated analytically in
\cite{baganetal:94a}. Furthermore, as discussed in
\cite{baganetal:94a}, $h_{11}(x)$ and $h_{22}(x)$ can be obtained from
the work of \cite{hokimpham:83}, \cite{hokimpham:84}. Finally,
$h_{12}(x)$ has been determined in \cite{baganetal:95}.
For the full mass dependence of these functions we refer the
reader to the cited literature. Here we quote the results obtained in
the massless limit.  These have been computed in \cite{altarelli:81},
\cite{buchalla:93} for $g_{ij}$, $h_{ij}$ ($g_{ij}(0)=h_{ij}(0)$)
\begin{equation}\label{ghij0}
g_{11}(0)=g_{22}(0)=\frac{31}{4}-\pi^2
\qquad g_{12}(0)=g_{11}(0)-\frac{19}{2}
\end{equation}
Furthermore
\begin{equation}\label{g000}
g(0,0,0)=\frac{25}{4}-\pi^2
\end{equation}
In table \ref{tab:bNLOnum} we have listed some typical numbers
extracted from \cite{baganetal:94b}, \cite{baganetal:95} illustrating
the impact of charm mass effects (for $x_c=0.3$) in the NLO correction
terms by giving the enhancment factor of the NLO over the LO results.
There are of course various ambiguities involved in this comparison.
The numbers in table \ref{tab:bNLOnum} are therefore merely intended to
show the general trend.  Note the sizable enhancement through NLO mass
effects in the nonleptonic channels, in particular $b\to c\bar cs$. A
large QCD enhancement in the latter case has also been reported in
\cite{voloshin:94}.

\begin{table}[htb]
\caption[]{Typical values for the ratio of NLO to LO results for dominant
b-decay channels with (I) and without (II) including finite charm mass
effects in the NLO correction terms. The leading order final state mass
effects (through the function $P$) are taken into account in all
cases.
\label{tab:bNLOnum}}
\begin{center}
\begin{tabular}{|c||c|c|c|c|}
 & $b\to ce\nu$ & $b\to c\tau\nu$ & $b\to c\bar ud$ & $b\to c\bar cs$\\
\hline
I & 0.85 & 0.88 & 1.06 & 1.32 \\
\hline
II & 0.79 & 0.80 & 1.01 & 1.02
\end{tabular}
\end{center}
\end{table}

To complete the presentation of b decay modes we next write down
expressions for the CKM suppressed channels $b \to u l \nu$, $b\to
u\bar cs$ and $b\to u\bar ud$ (without penguins) as well as the
contribution to the $b\to c\bar cs$ rate due to interference of the
leading, current-current type transitions with penguin operators.
Restricting ourselves to the LLA for these small contributions we
obtain
\begin{equation}\label{buln}
\Gamma(b\to u\sum_l l\nu)=\Gamma_0 |\frac{V_{ub}}{V_{cb}}|^2
\sum_l P(0,x_l,0)
\end{equation} 
\begin{equation}\label{bucs}
\Gamma(b\to u\bar cs)=\Gamma_0 |\frac{V_{ub}}{V_{cb}}|^2
 P(0,x_c,x_s) \left[2L^2_++L^2_-\right]
\end{equation} 
\begin{equation}\label{buud}
\Gamma(b\to u\bar ud)=\Gamma_0 |\frac{V_{ub}}{V_{cb}}|^2
  \left[2L^2_++L^2_-\right]
\end{equation} 
\begin{eqnarray}\label{pbccs}
\Delta\Gamma_{penguin}(b\to c\bar cs) &=& 6\Gamma_0P(x_c,x_c,x_s)
\left[c_1\left(c_3+\frac{1}{3}c_4+F(c_5+\frac{1}{3}c_6)\right)\right.
\nonumber \\
&&+\left.c_2\left(\frac{1}{3}c_3+c_4+F(\frac{1}{3}c_5+c_6)\right)\right]
\end{eqnarray}
where $c_1, \ldots, c_6$ are the leading order Wilson coefficients and
\begin{equation}\label{fpeng}
F=\frac{6x^2_c}{P(x_c,x_c,x_s)}\int\limits_{(x_c+x_s)^2}^{(1-x_c)^2}
\frac{ds}{s^2}(s+x^2_s-x^2_c)(1+s-x^2_c) w(s,x^2_c,x^2_s) w(1,s,x^2_c)
\end{equation}
Numerically we have for $|V_{ub}/V_{cb}|=0.1$
\begin{equation}\label{bupnum1}
\Gamma(b\to u\sum_l l\nu)\approx 0.024 \Gamma_0 \qquad
\Gamma(b\to u\bar cs)\approx 0.017 \Gamma_0
\end{equation}
\begin{equation}\label{bupnum2}
\Gamma(b\to u\bar ud)\approx 0.034 \Gamma_0 \qquad
\Delta\Gamma_{penguin}(b\to c\bar cs)\approx -0.041 \Gamma_0
\end{equation}
Note that the contribution due to the interference with penguin
transitions in $b\to c\bar cs$ is negative. Hence, in addition to 
being small the effects in (\ref{bupnum1}) and (\ref{bupnum2}) tend
to cancel each other in the total nonleptonic width.

Finally one may also incorporate nonperturbative corrections. These
have been derived in \cite{bigietal:92} and are also discussed in
\cite{baganetal:94a}. As mentioned above, nonperturbative effects are
suppressed by two powers of the heavy b-quark mass and amount typically
to a few percent. For details we refer the reader to the cited
articles.

\subsection{The B Meson Semileptonic Branching Ratio}
\label{sec:InclB:BSL}
An important application of the results described in the
previous section is the theoretical prediction for the inclusive
semileptonic branching ratio of $B$ mesons
\begin{equation}\label{bsldef}
B_{SL}=\frac{\Gamma(B\to Xe\nu)}{\Gamma_{tot}(B)}
\end{equation}
On the parton level $\Gamma(B\to Xe\nu)\simeq\Gamma(b\to ce\nu)$
and
\begin{equation}\label{gtotb}
\Gamma_{tot}(B)\simeq \sum_{l=e,\mu,\tau}\Gamma(b\to cl\nu)+
\Gamma(b\to c\bar ud)+\Gamma(b\to c\bar cs)+
\Delta\Gamma_{penguin}(b\to c\bar cs)+\Gamma(b\to u)
\end{equation}
Here we have applied the approximations discussed above. $\Gamma(b\to u)$
summarizes the $b\to u$ transitions.
\\
Based on a similar treatment of the partonic rates, including in
particular next-to-leading QCD corrections for the dominant channels
and also incorporating nonperturbative corrections, the authors of
\cite{baganetal:94b}, \cite{baganetal:95} have carried out an analysis
of $B_{SL}$ and estimated the theoretical uncertainties. They obtain
\cite{baganetal:95}
\begin{equation}\label{bslnum}
B_{SL} = (12.0 \pm 1.4)\%
\qquad \mbox{and} \qquad
B_{SL} = (11.2 \pm 1.7)\%
\end{equation}
using pole and $\overline{MS}$ masses, respectively. The error is
dominated in both cases by the renormalization scale uncertainty ($\mb/2
< \mu < 2 \mb$). Note also the sizable scheme ambiguity.
\\
Within existing uncertainties, the theoretical prediction does not disagree
significantly with the experimental value $B_{SL,exp}=(10.4\pm 0.4)\%$
\cite{particledata:94}, although it seems to lie still somewhat on the
high side.
\\
It is amusing to note, that the naive mode counting estimate for
$B_{SL}$, neglecting QCD and final state mass effects completely,
yields $B_{SL}=1/9=11.1\%$ in (almost) "perfect agreement" with
experiment. Including the final state masses, still neglecting QCD,
enhances this number to $B_{SL}=15.8\%$. Incorporating in addition QCD
effects at the leading log level {\it increases\/} the hadronic modes,
thus leading to a {\it decrease\/} in $B_{SL}$, resulting typically in
$B_{SL}=14.7\%$. A substantial further decrease is finally brought
about through the NLO QCD corrections, which both further enhance
hadronic channels, in particular $b\to c\bar cs$, and simultaneously
reduce $b\to ce\nu$. As pointed out in \cite{baganetal:94b},
\cite{baganetal:95} and illustrated in table \ref{tab:bNLOnum} final
state mass effects in the NLO correction terms play a nonnegligile role
for this enhancement of hadronic decays. The nonperturbative effects
also lead to a slight decrease of $B_{SL}$.
\\
In short, leading final state mass effects and QCD corrections, acting
in opposite directions on $B_{SL}$, tend to cancel each other, 
resulting in a number for $B_{SL}$ not too different from the
simple modecounting guess.

We finally mention that, besides a calculation of $B_{SL}$, the
partonic treatment of heavy meson decay has further important
applications, such as the determination of $|V_{cb}|$ from inclusive
semileptonic $B$ decay, $B\to X_ce\nu$. Analyses of this type have been
presented in \cite{lukesavage:94}, \cite{bigiuraltsev:94},
\cite{ballnierste:94}, \cite{shifmanetal:95}.

Exact results beyond the presently known NLO accuracy seem extremely
difficult to obtain, even for relatively simple quantities like the
semileptonic b-quark decay rate.  There exist however calculations in
the literature devoted to the investigation of these higher order
perturbative effects. Due to the severe technical difficulties, those
calculations require additional assumptions. For instance, in an
interesting study \cite{braunetal:95} have investigated the effects of
the running of $\as$ on the semileptonic b-quark decay rate to all
orders in perturbation theory. This calculation is equivalent to a
resummation of all terms of the form $\as(\beta_0\as)^n$, which are
related to one-gluon exchange diagrams containing an arbitrary number
$n$ of fermion bubbles. The work of \cite{braunetal:95} applies the
renormalon techniques developped in \cite{benekebraun:95},
\cite{balletal:95} and generalizes the ${\cal O}(\beta_0\alpha^2_s)$
results computed in \cite{lukeetal:95}. The underlying idea is similar
in spirit to the BLM approach \cite{brodskyetal:83}.  An important
application of the result is the extraction of $|V_{cb}|$
\cite{braunetal:95}.  The formalism has also been used to study higher
order QCD corrections to the $\tau$ lepton hadronic width
\cite{balletal:95}.  Irrespective of the ultimate reliability of the
approximation, these investigations are useful from a conceptual point
of view as they help to illustrate important features of the higher
order behavior of the perturbative expansion.

In principle the discussion we have given for
b-decays may of course, with appropriate modifications, be applied to
the case of charm as well.  However here the nonperturbative
corrections to the parton picture, which scale like $1/m^2_Q$ with the
heavy quark mass $m_Q$, are by an order of magnitude larger than for B
mesons and accurate theoretical predictions are much more difficult to
obtain \cite{blokshifman:93}.

\section{$\eps_K$, $B^0$-$\bar B^0$ Mixing and the Unitarity Triangle}
        \label{sec:epsBBUT}
\subsection{Basic Formula for $\eps_K$}
            \label{subsec:epsformula}
The indirect CP violation in $K \to \pi\pi$ is described by the well
known parameter $\eps_K$. The general formula for $\eps_K$ is given as
follows
\begin{equation}
\eps_K = \frac{\exp(i \pi/4)}{\sqrt{2} \Delta M_K} \,
\left( \IM M_{12} + 2 \xi \RE M_{12} \right)
\label{eq:epsdef}
\end{equation}
where
\begin{equation}
\xi = \frac{\IM A_0}{\RE A_0}
\label{eq:xi}
\end{equation}
with $A_0 \equiv A(K \to (\pi\pi)_{I=0})$ and $\Delta M_K$ being
the $K_L$-$K_S$ mass difference. The off-diagonal element $M_{12}$ in
the neutral $K$-meson mass matrix represents the $K^0$-$\bar K^0$
mixing. It is given by
\begin{equation}
2 m_K M^*_{12} = \langle \bar K^0| \Heff(\Delta S=2) |K^0\rangle
\label{eq:M12Kdef}
\end{equation}
where $\Heff(\Delta S=2)$ is the effective hamiltonian of eq.\
\eqn{hds2}. Defining the renormalization group invariant parameter
$B_K$ by
\begin{equation}
B_K = B_K(\mu) \left[ \as^{(3)}(\mu) \right]^{-2/9} \,
\left[ 1 + \frac{\as^{(3)}(\mu)}{4\pi} J_3 \right]
\label{eq:BKrenorm}
\end{equation}
\begin{equation}
\langle \bar K^0| (\bar s d)_{V-A} (\bar s d)_{V-A} |K^0\rangle
\equiv \frac{8}{3} B_K(\mu) F_K^2 m_K^2
\label{eq:KbarK}
\end{equation}
and using \eqn{hds2} we find
\begin{equation}
M_{12} = \frac{G_F^2}{12 \pi^2} F_K^2 B_K m_K \mw^2
\left[ {\lambda_c^*}^2 \eta_1 S_0(x_c) + {\lambda_t^*}^2 \eta_2 S_0(x_t) +
2 {\lambda_c^*} {\lambda_t^*} \eta_3 S_0(x_c, x_t) \right]
\label{eq:M12K}
\end{equation}
where the functions $S_0(x_i)$ and $S_0(x_i, x_j)$ are those of eq.\
\eqn{s0c}--\eqn{s0ct}. $F_K$ is the $K$-meson decay constant and $m_K$
the $K$-meson mass. The coefficient $J_3$ is given in \eqn{zd0} and the
QCD factors $\eta_i$ have been discussed in section
\ref{sec:HeffKKbar}. Their numerical values are
\begin{equation}
\eta_1 = 1.38 
\qquad
\eta_2 = 0.57
\qquad
\eta_3 = 0.47 \, .
\label{eq:etaknum}
\end{equation}
The last term in \eqn{eq:epsdef} constitutes at
most a 2\,\% correction to $\eps_K$ and consequently can be neglected
in view of other uncertainties, in particular those connected with
$B_K$.  Inserting \eqn{eq:M12K} into \eqn{eq:epsdef} we find
\begin{equation}
\eps_K = C_{\eps} B_K \IM\lambda_t \left\{
\RE\lambda_c \left[ \eta_1 S_0(x_c) - \eta_3 S_0(x_c, x_t) \right] -
\RE\lambda_t \eta_2 S_0(x_t) \right\} \exp(i \pi/4)
\label{eq:epsformula}
\end{equation}
where we have used the unitarity relation $\IM\lambda_c^* = {\rm
Im}\lambda_t$ and we have neglected $\RE\lambda_t/\RE\lambda_c
= \ord(\lambda^4)$ in evaluating $\IM(\lambda_c^* \lambda_t^*)$.
The numerical constant $C_\eps$ is given by
\begin{equation}
C_\eps = \frac{G_F^2 F_K^2 m_K \mw^2}{6 \sqrt{2} \pi^2 \Delta M_K}
       = 3.78 \cdot 10^4 \, .
\label{eq:Ceps}
\end{equation}
Using the standard parametrization of \eqn{2.72} to evaluate ${\rm
Im}\lambda_i$ and $\RE\lambda_i$, setting the values for $s_{12}$,
$s_{13}$, $s_{23}$ and $\mt$ in accordance with appendix
\ref{app:numinput} and taking a value for $B_K$ (see below) one can
determine the phase $\delta$ by comparing \eqn{eq:epsformula} with the
experimental value for $\eps_K$.

Once $\delta$ has been determined in this manner one can find the
corresponding point $(\bar\varrho,\bar\eta)$ by using \eqn{2.84} and
\eqn{2.88d}. Actually for a given set ($s_{12}$, $s_{13}$, $s_{23}$,
$\mt$, $B_K$) there are two solutions for $\delta$ and consequently two
solutions for $(\bar\varrho,\bar\eta)$. In order to see this clearly it is
useful to use the Wolfenstein parametrization in which ${\rm
Im}\lambda_t$, $\RE\lambda_c$ and $\RE\lambda_t$ are given to
a very good approximation by \eqn{2.51}--\eqn{2.53}. We then find that
\eqn{eq:epsformula} and the experimental value for $\eps_K$ specify a
hyperbola in the $(\bar\varrho,\bar\eta)$ plane given by
\begin{equation}
\bar\eta \left\{ (1 - \bar\varrho) A^2 \eta_2 S_0(x_t) + P_0(\eps)
\right\} A^2 B_K = 0.226 \, .
\label{eq:hyperbola}
\end{equation}
where
\begin{equation}
P_0(\eps) =
\left[ \eta_3 S_0(x_c, x_t) - \eta_1 x_c \right] \frac{1}{\lambda^4} \, .
\label{eq:p0eps}
\end{equation}
The hyperbola \eqn{eq:hyperbola} intersects the circle given by
\eqn{2.94} in two points which correspond to the two solutions for
$\delta$ mentioned earlier.

The position of the hyperbola \eqn{eq:hyperbola} in the
$(\bar\varrho,\bar\eta)$ plane depends on $\mt$, $|V_{cb}|=A \lambda^2$
and $B_K$. With decreasing $\mt$, $|V_{cb}|$ and $B_K$ the
$\eps_K$-hyperbola moves away from the origin of the
$(\bar\varrho,\bar\eta)$ plane. When the hyperbola and the circle
\eqn{2.94} touch each other lower bounds consistent with $\eps_K^{\rm
exp}$ for $\mt$, $|V_{cb}|$, $|V_{ub}/V_{cb}|$ and $B_K$ can be found.
The lower bound on $\mt$ is discussed in \cite{buras:93}.
Corresponding results for $|V_{ub}/V_{cb}|$ and $B_K$ are shown in
fig.\ \ref{fig:ut:vubcbmin} and \ref{fig:ut:bkmin}, respectively.
They will be discussed below.
\\
Moreover approximate analytic expressions for these bounds can be
derived. One has
\begin{eqnarray}
(\mt)_{\rm min} &=& \mw \left[ \frac{1}{2 A^2} \left( \frac{1}{A^2 B_K
R_b} - 1.4 \right) \right]^{0.658}
\label{eq:mtmin} \\
|V_{ub}/V_{cb}|_{\rm min} &=&
\frac{\lambda}{1-\lambda^2/2} \,
\left[ A^2 B_K \left( 2 x_t^{0.76} A^2 + 1.4 \right) \right]^{-1}
\label{eq:Vubcbmin} \\
(B_K)_{\rm min} &=& \left[ A^2 R_b \left( 2 x_t^{0.76} A^2 + 1.4 \right)
                    \right]^{-1}
\label{eq:BKmin}
\end{eqnarray}

Concerning the parameter $B_K$, the analyses of \cite{sharpe:94},
\cite{ishizuka:93} ($B_K=0.83\pm 0.03$) using the lattice method and of
\cite{bijnenspardes:95} using a somewhat modified form of the $1/N$
approach of \cite{bardeenetal:88}, \cite{gerard:90} give results in the
ball park of the $1/N$ result $B_K=0.70\pm 0.10$ obtained some time ago
in \cite{bardeenetal:88}, \cite{gerard:90}. In particular the analysis
of \cite{bijnenspardes:95} seems to have explained the difference
between these values for $B_K$ and the lower values obtained using the
QCD Hadronic Duality approach \cite{pichderaf:85}, \cite{pradesetal:91}
($B_K=0.39\pm 0.10$) or using SU(3) symmetry and PCAC ($B_K=1/3$)
\cite{donoghueetal:82}. These higher values of $B_K$ are also found in
the most recent lattice analysis \cite{crisafullietal:95}
($B_K=0.86 \pm 0.15$) and in the lattice calculations of Bernard and
Soni ($B_K=0.78 \pm 0.11$) and the JLQCD group ($B_K=0.67 \pm 0.07$)
with the quoted values obtained on the basis on the review by
\cite{soni:95}. In our numerical analysis we will use
\begin{equation}
B_K = 0.75 \pm 0.15 \, .
\label{eq:BKnum}
\end{equation}

\subsection{Basic Formula for $B^0$-$\bar B^0$ Mixing}
            \label{subsec:BBformula}
The $B^0$-$\bar B^0$ mixing is usually described by
\begin{equation}
x_{d,s} \equiv \frac{(\Delta M)_{B_{d,s}}}{\Gamma_{B_{d,s}}} =
\frac{2 |M_{12}|_{B_{d,s}}}{\Gamma_{B_{d,s}}}
\label{eq:xdsdef}
\end{equation}
where $(\Delta M)_{B_{d,s}}$ is the mass difference between the mass
eigenstates in the $B_d^0-\bar B_d^0$ system and the $B_s^0-\bar B_s^0$
system, respectively, and $\Gamma_{B_{d,s}} = 1/\tau_{B_{d,s}}$ with
$\tau_{B_{d,s}}$ being the corresponding lifetimes. The off-diagonal
term $M_{12}$ in \eqn{eq:xdsdef} is given by
\begin{equation}
2 m_B |M_{12}| = |\langle \bar B^0| \Heff(\Delta B=2) |B^0\rangle|
\label{eq:M12Bdef}
\end{equation}
where $\Heff(\Delta B=2)$ is the effective hamiltonian of
\eqn{hdb2}. Defining the renormalization group invariant parameter $B_B$ by
\begin{equation}
B_B = B_B(\mu) \left[ \as^{(5)}(\mu) \right]^{-6/23} \,
\left[ 1 + \frac{\as^{(5)}(\mu)}{4\pi} J_5 \right]
\label{eq:BBrenorm}
\end{equation}
\begin{equation}
\langle \bar B^0| (\bar b d)_{V-A} (\bar b d)_{V-A} |B^0\rangle
\equiv \frac{8}{3} B_B(\mu) F_B^2 m_B^2
\label{eq:BbarB}
\end{equation}
and using \eqn{hdb2} we find
\begin{equation}
x_{d,s} = \tau_{B_{d,s}} \frac{G_F^2}{6 \pi^2} \eta_B m_{B_{d,s}} (
B_{B_{d,s}} F_{B_{d,s}}^2 ) \mw^2 S_0(x_t) |V_{t(d,s)}|^2
\label{eq:xds}
\end{equation}
with the QCD factor $\eta_B$ discussed in section \ref{sec:HeffBBbar}
and given by $\eta_B=0.55$.

The measurement of $B_d^0$-$\bar B_d^0$ mixing allows then to determine
$|V_{td}|$ or $R_t$ of \eqn{2.95}
\begin{equation}
|V_{td}| = A \lambda^3 R_t
\qquad
\qquad
R_t = 1.52 \frac{R_0}{\sqrt{S_0(x_t)}}
\label{eq:VtdRt}
\end{equation}
where
\begin{equation}
R_0 = 
\left[ \frac{0.040}{|V_{cb}|} \right]
\left[ \frac{200\mev}{ \sqrt{B_{B_d}} F_{B_d} } \right]
\left[ \frac{x_d}{0.75}       \right]^{0.5} 
\left[ \frac{1.6\,ps}{\tau_B} \right]^{0.5}
\left[ \frac{0.55}{\eta_B}    \right]^{0.5} \, .
\label{eq:R0}
\end{equation}
which gives setting $\eta_B = 0.55$
\begin{equation}
|V_{td}| = 8.56 \cdot 10^{-3}
\left[ \frac{170\gev}{\bar m_t(m_t)} \right]^{0.76} 
\left[ \frac{200\mev}{ \sqrt{B_{B_d}} F_{B_d} } \right]
\left[ \frac{x_d}{0.75}             \right]^{0.5} 
\left[ \frac{1.6\,ps}{\tau_B}       \right]^{0.5} \, .
\label{eq:Vtdnum}
\end{equation}

There is a vast literature on the lattice calculations of $F_B$. The
most recent results are somewhat lower than quoted a few years ago.
Based on a review by \cite{sachrajda:94}, the recent extensive study by
\cite{duncanetal:94} and the analyses in \cite{bernardetal:94},
\cite{drapermcneile:94} we conclude that $F_{B_d}=(180\pm40)\mev$. This
together with the earlier result of the European Collaboration
\cite{abadaetal:92} for $B_B$, gives $F_{B_d}\sqrt{B_{B_d}}=194\pm
45\mev$. A reduction of the error in this important quantity is
desirable. These results for $F_B$ are compatible with the results
obtained using QCD sum rules (e.g. \cite{baganetal:92},
\cite{neubert:92}). An interesting upper bound $F_{B_d}<195\mev$ using
QCD dispersion relations has also recently been obtained
\cite{boydetal:94}. In our numerical analysis we will use
\begin{equation}
\sqrt{B_{B_d}} F_{B_d} = (200 \pm 40)\mev \, .
\label{eq:Fbnum}
\end{equation}

The accuracy of the determination of $R_t$ can be considerably improved
by measuring simultaneously the $B_s^0$-$\bar B_s^0$ mixing described by
$x_s$. We have
\begin{equation}
R_t = \frac{1}{\sqrt{R_{ds}}} \sqrt{\frac{x_d}{x_s}} \frac{1}{\lambda}
\sqrt{1 - \lambda^2 (1 - 2 \varrho)}
\qquad
R_{ds} = \frac{\tau_{B_d}}{\tau_{B_s}} \frac{m_{B_d}}{m_{B_s}}
\left[ \frac{F_{B_d} \sqrt{B_{B_d}}}{F_{B_s} \sqrt{B_{B_s}}} \right]^2 \, .
\label{eq:Rt}
\end{equation}
Note that $\mt$ and $|V_{cb}|$ have been eliminated in this way and that
$R_{ds}$ depends only on $SU(3)$-flavour breaking effects which contain
much smaller theoretical uncertainties than the hadronic matrix elements
in $x_d$ and $x_s$ separately.
Provided $x_d/x_s$ has been accurately measured a determination of
$R_t$ within $\pm 10\%$ should be possible. Indeed the most recent
lattice results \cite{duncanetal:94}, \cite{baxteretal:94} give
$F_{B_s}/F_{B_d} = 1.22\pm0.04$. A similar result $F_{B_s}/F_{B_d} =
1.16\pm0.05$ has been obtained using QCD sum rules \cite{narison:94}.
It would be useful to know $B_{B_s}/B_{B_d}$ with a similar precision.
For $B_{B_s}=B_{B_d}$ we find using the lattice result $R_{ds} =
0.66 \pm 0.07$.

\subsection{$\sin(2\beta)$ from $\eps_K$ and $B^0$-$\bar B^0$ Mixing}
           \label{subsec:sin2bepskBB}
Combining \eqn{eq:hyperbola} and \eqn{eq:xds} one can derive an analytic
formula for $\sin(2\beta)$ \cite{burasetal:94b}
\begin{equation}
\sin(2\beta) = \frac{1}{1.16 A^2 \eta_2 R_0^2}
\left[ \frac{0.226}{A^2 B_K} - \bar\eta P_0(\eps) \right] \, .
\label{eq:sin2b}
\end{equation}
$P_0(\eps)$ is weakly dependent on $\mt$ and for $155\gev \le \mt \le
185\gev$ one has $P_0(\eps) \approx 0.31 \pm 0.02$. As $\bar\eta \le
0.45$ for $|V_{ub}/V_{cb}| \le 0.1$ the first term in parenthesis is
generally by a factor of 2--3 larger than the second term. Since this
dominant term is independent of $\mt$, the values for $\sin(2\beta)$
extracted from $\eps_K$ and $B^0$-$\bar B^0$ mixing show only a weak
dependence on $\mt$ as stressed in particular in \cite{rosner:00}.

Since in addition $A^2 R_0^2$ is independent of $|V_{cb}|$, the dominant
uncertainty in this determination of $\sin(2\beta)$ resides in $A^2 B_K$
in the first term in the parenthesis and in $F_{B_d} \sqrt{B_{B_d}}$
contained in $R_0^2$.

\subsection{Phenomenological Analysis}
           \label{subsec:phenoUT}
We will now combine the analyses of $\eps_K$ and of $B^0_d-\bar B^0_d$
mixing to obtain allowed ranges for several quantities of interest. We
consider two sets of input parameters, which are collected in the
appendix. The first set represents the present situation. The second
set can be considered as a ``future vision'' in which the errors on
various input parameters have been decreased. It is plausible that such
errors will be achieved at the end of this decade, although one cannot
guarantee that the central values will remain. In table
\ref{tab:predictions} we show the results for $\delta$, $\IM\lambda_t$,
$\sin 2\alpha$, $\sin 2\beta$, $\sin \gamma$, $|V_{td}|$ and $x_s$.
They correspond to the two sets of parameters in question, with and
without the constraint from $B^0_d-\bar B^0_d$ mixing. The results for
$\IM\lambda_t$ and $|V_{td}|$ will play an important role in the
phenomenology of rare decays and CP violation.  For completeness we
also show the expectations for $\sin 2\alpha$, $\sin 2\beta$ and
$\sin\gamma$ which enter various CP asymmetries in B-decays. As already
discussed in detail in \cite{burasetal:94b}, $\sin 2\alpha$ cannot be
predicted accurately this way.  On the other hand $\sin 2\beta$ and
$\sin\gamma$ are more constrained and the resulting ranges for these
quantities indicate that large CP asymmetries should be observed in a
variety of B-decays.

\begin{table}[htb]
\caption[]{
Predictions for various quantities using present and future input
parameter ranges given in appendix \ref{app:numinput}.
${\rm Im}\lambda_t$ and $|V_{td}|$ are given
in units of $10^{-4}$ and $10^{-3}$, respectively. $\delta$ is in degrees.
\label{tab:predictions}}
\begin{center}
\begin{tabular}{|c|r|r|r|r|}
& \multicolumn{2}{c|}{no   $x_d$ constraint} &
  \multicolumn{2}{c|}{with $x_d$ constraint} \\
\hline
& \multicolumn{1}{c|}{Present} & \multicolumn{1}{c|}{Future} &
  \multicolumn{1}{c|}{Present} & \multicolumn{1}{c|}{Future} \\
\hline
$\delta$                     & 37.7 -- 160.0 & 57.4 -- 144.9
                             & 37.7 -- 140.2 & 58.5 -- 93.3 \\
${\rm Im}\lambda_t$              & 0.64 -- 1.75 & 0.82 -- 1.50
                             & 0.87 -- 1.75 & 1.12 -- 1.50 \\
$|V_{td}| $                  & 6.7 -- 13.5 & 7.7 -- 12.1
                             & 6.7 -- 11.9 & 7.8 -- 9.3 \\
$x_s$                        &  --  &  -- 
                             & 11.1 -- 47.0 & 19.6 -- 29.6 \\
$\sin 2\alpha$               & --0.86 -- 1.00 & --0.323 -- 1.00
                             & --0.86 -- 1.00 & --0.30 -- 0.73 \\
$\sin 2\beta$                & 0.21 -- 0.80 & 0.34 -- 0.73
                             & 0.34 -- 0.80 & 0.57 -- 0.73 \\
$\sin\gamma$                 & 0.34 -- 1.00 & 0.58 -- 1.00
                             & 0.61 -- 1.00 & 0.85 -- 1.00 \\
\end{tabular}
\end{center}
\end{table}

\begin{figure}[hbt]
\vspace{0.10in}
\centerline{
\epsfysize=5in
\rotate[r]{
\epsffile{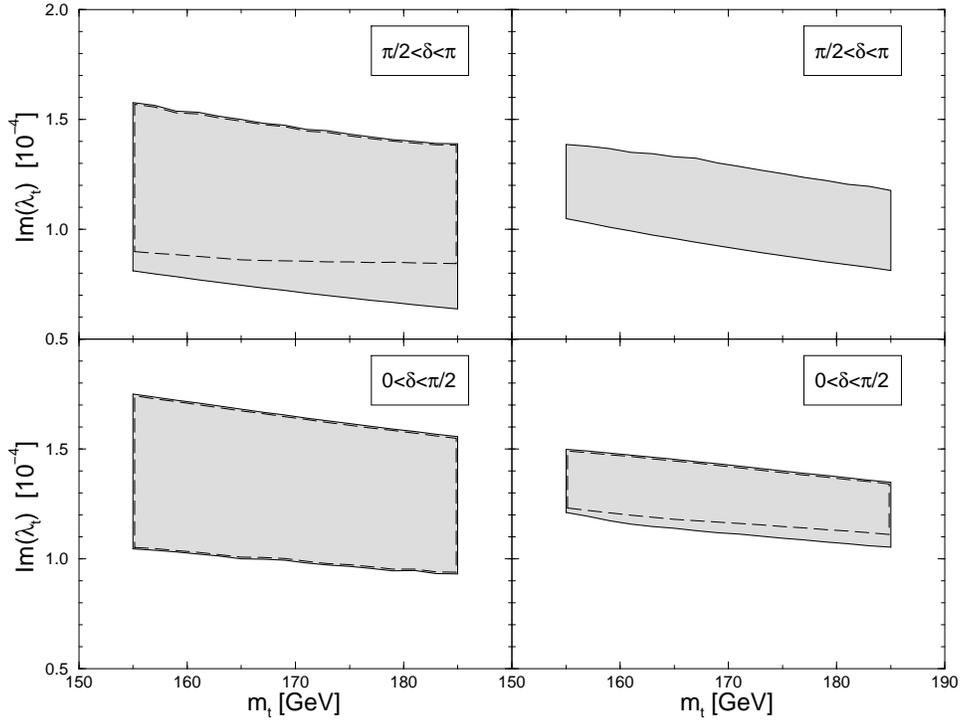}
} }
\vspace{0.08in}
\caption[]{
Present (left) and future (right) allowed ranges for $\IM(\lambda_t)$.
The ranges have been obtained by fitting $\eps_K$ in \eqn{eq:epsformula} to 
the experimental value. Input parameter ranges are given in appendix
\ref{app:numinput}. The impact of the additional constraint coming from
$x_d$ is illustrated by the dashed lines. With the $x_d$ constraint
imposed the solution $\pi/2 < \delta < \pi$ is completely eliminated
for the future scenario.
\label{fig:ut:imtimltxd}}
\end{figure}

\begin{figure}[hbt]
\vspace{0.10in}
\centerline{
\epsfysize=5in
\rotate[r]{
\epsffile{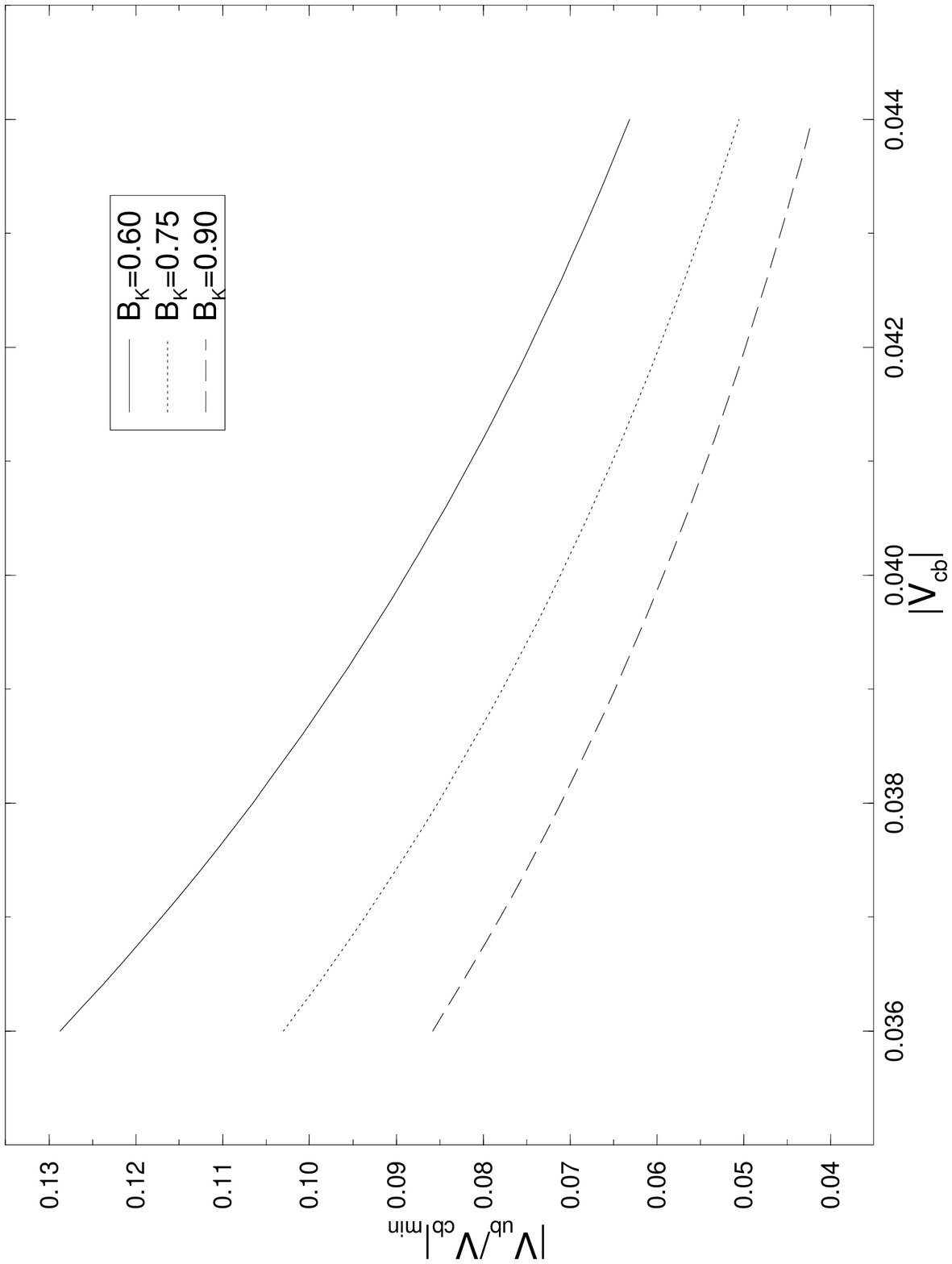}
} }
\vspace{0.08in}
\caption[]{
$|V_{ub}/V_{cb}|_{\rm min}$ for $\mt \le 185\gev$ and various choices
of $B_K$.
\label{fig:ut:vubcbmin}}
\end{figure}

\begin{figure}[hbt]
\vspace{0.10in}
\centerline{
\epsfysize=5in
\rotate[r]{
\epsffile{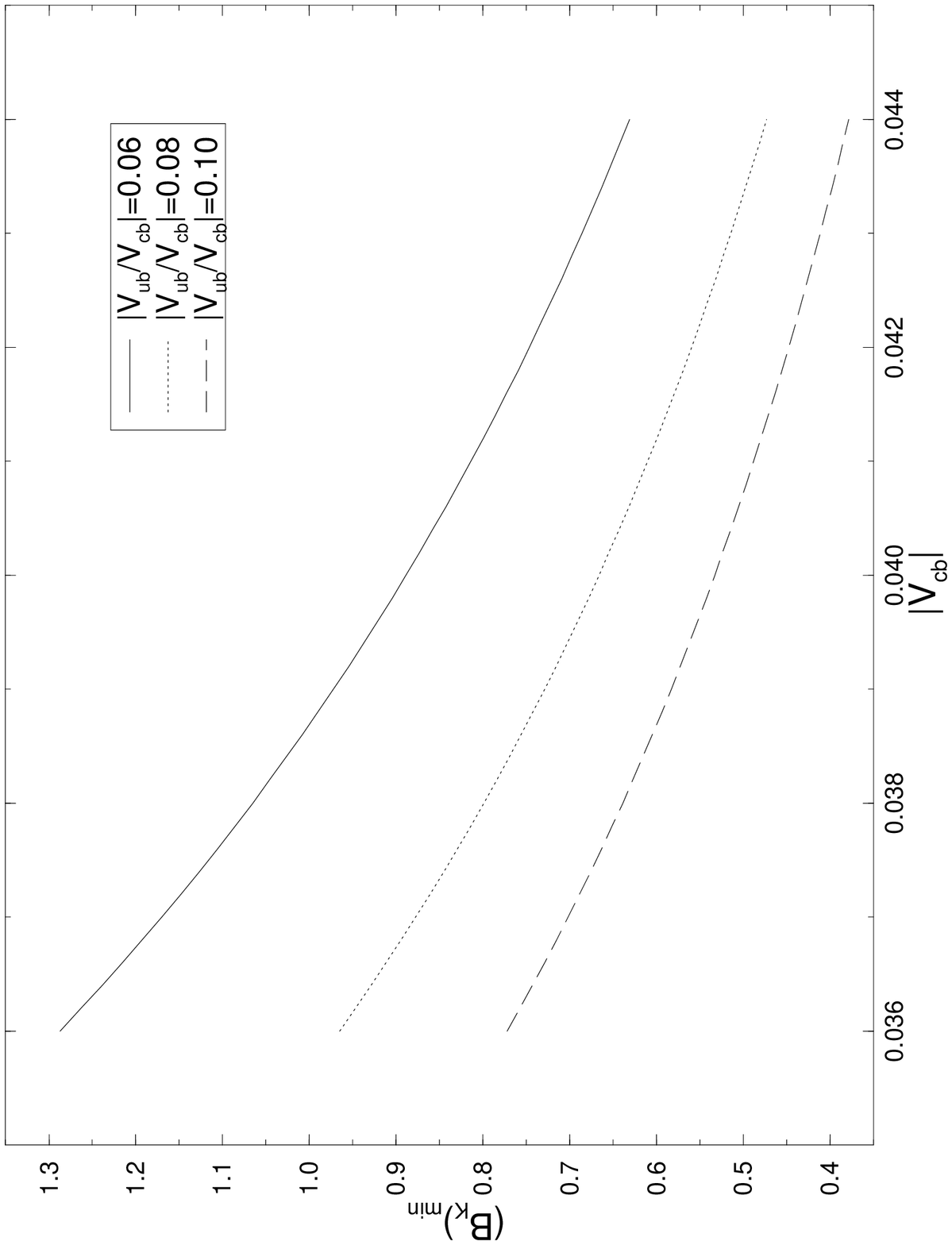}
} }
\vspace{0.08in}
\caption[]{
$(B_K)_{\rm min}$ of eq.\ \eqn{eq:BKmin} for $\mt \le 185\gev$ and
various choices of $|V_{ub}/V_{cb}|$.
\label{fig:ut:bkmin}}
\end{figure}

In fig.\ \ref{fig:ut:imtimltxd} we show $\IM\lambda_t$ as a function of
$\mt$. In fig.\ \ref{fig:ut:vubcbmin} the lower bound on
$|V_{ub}/V_{cb}|$ resulting from the $\eps_K$-constraint is shown as a
function of $|V_{cb}|$ for various values of $B_K$. To this end we have
set $\mt=185\gev$.  For lower values of $\mt$ the lower bound on
$|V_{ub}/V_{cb}|$ is stronger.  A similar analysis has been made by
\cite{herrlichnierste:95}. The latter work and the plot in
fig.\ \ref{fig:ut:vubcbmin} demonstrate clearly the impact of the
$\eps_K$ constraint on the allowed values of $|V_{ub}/V_{cb} |$ and
$|V_{cb}|$.  Simultaneously small values of $|V_{ub}/V_{cb} |$ and $|
V_{cb}|$, although still consistent with tree-level decays, are not
allowed by the size of the indirect CP violation observed in $K \to
\pi\pi$.  Another representation of this behaviour is shown in
fig.\ \ref{fig:ut:bkmin} where we plot the minimal value of $B_{K}$
consistent with the experimental value of $\eps_K$ as a function of
$V_{cb}$ for different $|V_{ub}/V_{cb} |$ and $\mt < 185\gev$.

\begin{figure}[hbt]
\vspace{0.10in}
\centerline{
\epsfysize=5in
\rotate[r]{
\epsffile{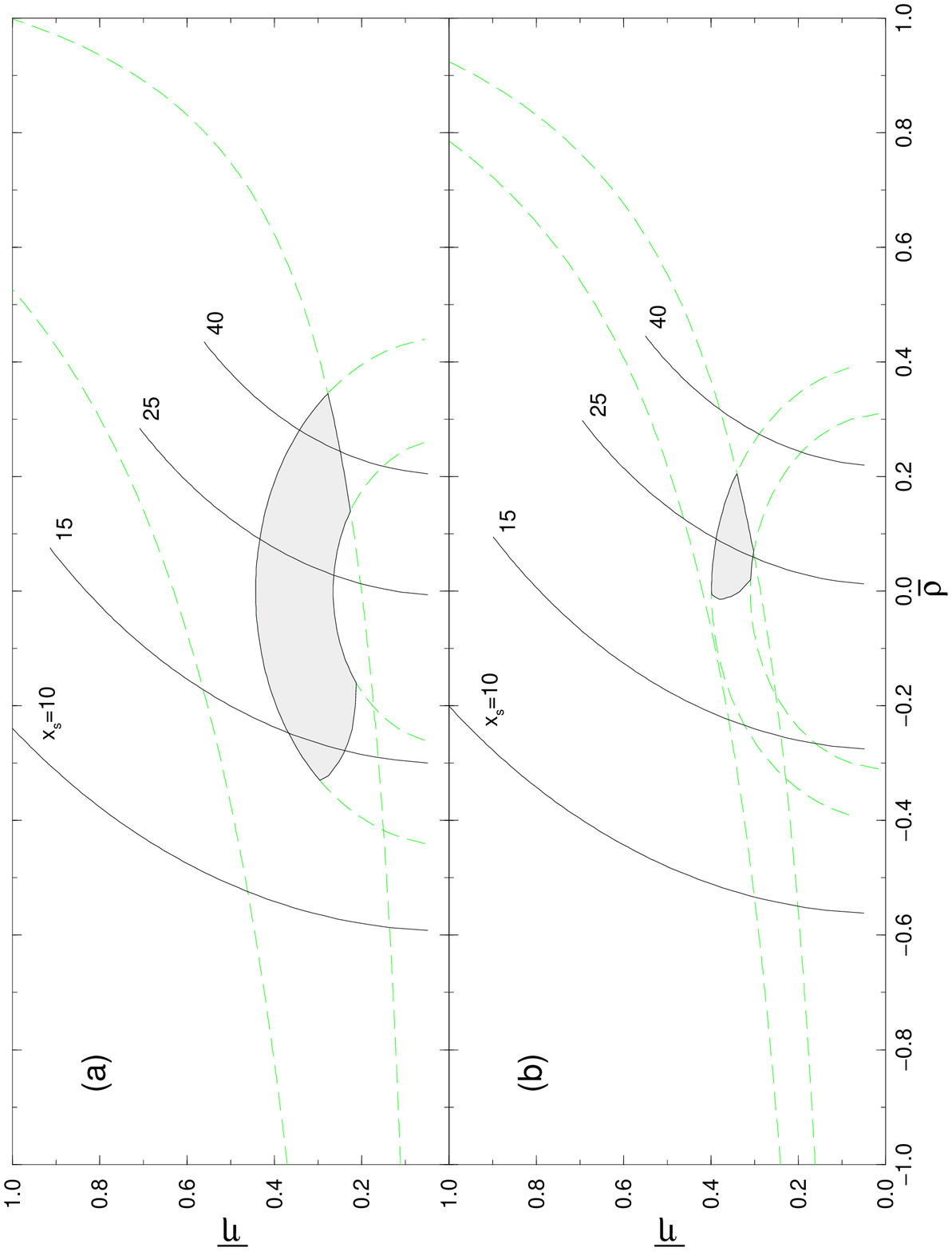}
} }
\vspace{0.08in}
\caption[]{
Present (a) and future (b) allowed ranges for the upper corner A
of the UT using data from $K^0-\bar K^0$-, $B^0-\bar B^0$-mixing
and tree-level $B$-decays. Input parameter ranges are given in appendix
\ref{app:numinput}.
The solid lines correspond to $(R_t)_{\rm max}$ from eq.\eqn{eq:Rt}
using $R_{ds}=0.66$ and $x_s \ge 10, 15, 25$ and $40$, respectively.
\label{fig:ut:rhoeta}}
\end{figure}

Finally in fig.\ \ref{fig:ut:rhoeta} we show the allowed ranges in the
$(\bar\rho,\bar\eta)$ plane obtained using the information from
$V_{cb}$, $|V_{ub}/V_{cb}|$, $\eps_K$ and $B^0_d-\bar B^0_d$ mixing.
In this plot we also show the impact of a future measurement of
$B^0_s-\bar B^0_s$ mixing with $x_s=10$, $15$, $25$, $40$, which by
means of the formula (\ref{eq:Rt}) gives an important measurement of
the side $R_t$ of the unitarity triangle. Whereas at present a broad
range in the $(\bar\rho,\bar\eta)$ plane is allowed, the situation
might change in the future allowing only the values $0 \le \bar \rho
\le 0.2$ and $0.30 \le \bar \eta \le 0.40$.  This results in smaller
ranges for various quantities of interest as explicitly seen in table
\ref{tab:predictions}.

Other analyses of the unitarity triangle can be found in
\cite{pecceiwang:95}, \cite{ciuchini:95}, \cite{herrlichnierste:95},
\cite{alilondon:95}.

\section{$\epe$ Beyond Leading Logarithms}
        \label{sec:nloepe}
\subsection{Basic Formulae}
           \label{subsec:epeformulae}
The direct CP violation in $K \to \pi\pi$ is described by $\eps'$.
The parameter $\eps'$ is given in terms of the amplitudes $A_0 \equiv
A(K \to (\pi\pi)_{I=0})$ and $A_2 \equiv
A(K \to (\pi\pi)_{I=2})$ as follows
\begin{equation}
\eps' = -\frac{\omega}{\sqrt{2}} \xi (1 - \Omega) \exp(i \Phi) \, ,
\label{eq:epsprim}
\end{equation}
where
\begin{equation}
\xi = \frac{\IM A_0}{\RE A_0} \, , \quad
\omega = \frac{\RE A_2}{\RE A_0} \, , \quad
\Omega = \frac{1}{\omega} \frac{\IM A_2}{\IM A_0}
\label{eq:xiomega}
\end{equation}
and $\Phi = \pi/2 + \delta_2 - \delta_0 \approx \pi/4$.

When using \eqn{eq:epsprim} and \eqn{eq:xiomega} in phenomenological
applications one usually takes $\RE A_0$ and $\omega$ from
experiment, i.e.
\begin{equation}
\RE A_0 = 3.33 \cdot 10^{-7}\gev
\qquad
\RE A_2 = 1.50 \cdot 10^{-8}\gev
\qquad
\omega = 0.045
\label{eq:ReA0data}
\end{equation}
where the last relation reflects the so-called $\Delta I=1/2$ rule. The
main reason for this strategy is the unpleasant fact that until today
nobody succeded in fully explaining this rule which to a large extent is
believed to originate in the long-distance QCD contributions. We will be
more specific about this in the next section. On the other hand the
imaginary parts of the amplitudes in \eqn{eq:xiomega} being related to
CP violation and the top quark physics should be dominated by
short-distance contributions. Therefore $\IM A_0$ and $\IM A_2$ are
usually calculated using the effective hamiltonian given in
\eqn{eq:HeffdF1:1010}. Using this hamiltonian and the experimental
values for $\eps$, $\RE A_0$ and $\omega$ the ratio $\epe$ can be
written as follows
\begin{equation}
\epe = \IM \lambda_t \left[ P^{(1/2)} - P^{(3/2)} \right]
\label{eq:epe}
\end{equation}
where
\begin{eqnarray}
P^{(1/2)} &=& \sum P_i^{(1/2)} = r \sum y_i \langle Q_i\rangle_0
(1-\Omega_{\eta+\eta'})
\label{eq:P12} \\
P^{(3/2)} &=& \sum P_i^{(3/2)} = \frac{r}{\omega}
\sum y_i \langle Q_i\rangle_2
\label{eq:P32}
\end{eqnarray}
with
\begin{equation}
r = \frac{G_F \omega}{2 |\eps| \RE A_0} \, .
\label{eq:repe}
\end{equation}
Here the hadronic matrix element shorthand notation is
\begin{equation}
\langle Q_i\rangle_I \equiv \langle (\pi\pi)_I | Q_i | K \rangle
\label{eq:QiKpp}
\end{equation}
and the sum in \eqn{eq:P12} and \eqn{eq:P32} runs over all contributing
operators. This means for $\mu > \mc$ also contributions from operators
$Q^c_{1,2}$ to $P^{(1/2)}$ and $P^{(3/2)}$ have to be taken into
account. These are necessary for $P^{(1/2)}$ and $P^{(3/2)}$ to be
independent of the renormalization scale $\mu$. Next,
\begin{equation}
\Omega_{\eta+\eta'} = \frac{1}{\omega} \frac{(\IM A_2)_{\rm
I.B.}}{\IM A_0}
\label{eq:Omegaeta}
\end{equation}
represents the contribution stemming from isospin breaking in the quark masses
($m_u \not= m_d$). For $\Omega_{\eta+\eta'}$ we will take
\begin{equation}
\Omega_{\eta+\eta'} = 0.25 \pm 0.05
\label{eq:Omegaetadata}
\end{equation}
which is in the ball park of the values obtained in the $1/N_c$ approach
\cite{burasgerard:87} and in chiral perturbation theory
\cite{donoghueetal:86}, \cite{lusignoli:89}. $\Omega_{\eta+\eta'}$ is
independent of $\mt$.

The numerical values of the Wilson coefficients $y_i$ have been already given
in section \ref{sec:HeffdF1:1010:numres}. We therefore turn now our
attention to the hadronic matrix elements \eqn{eq:QiKpp} which
constitute the main source of uncertainty in the calculation of
$\epe$.

\subsection{Hadronic Matrix Elements for $K \to \pi\pi$}
           \label{subsec:matelKpp}
The hadronic matrix elements $\langle Q_i \rangle_I$ depend generally
on the renormalization scale $\mu$ and on the scheme used to
renormalize the operators $Q_i$. These two dependences are canceled by
those present in the Wilson coefficients $C_i(\mu)$ so that the
resulting physical amplitudes do not depend on $\mu$ and on the
renormalization scheme of the operators.  Unfortunately the accuracy of
the present non-perturbative methods used to evalutate $\langle Q_i
\rangle_I$, like lattice methods or $1/N_c$ expansion, is not
sufficient to obtain the required $\mu$ and scheme dependences of
$\langle Q_i \rangle_I$. A review of the existing methods and their
comparison can be found in \cite{burasetal:92d}, \cite{ciuchini:95}.
In view of this situation it has been suggested \cite{burasetal:92d} to
determine as many matrix elements $\langle Q_i \rangle_I$ as possible
from the leading CP conserving $K \to \pi\pi$ decays, for which the
experimental data are summarized in \eqn{eq:ReA0data}. To this end it
turned out to be very convenient to determine $\langle Q_i \rangle_I$
at a scale $\mu = \mc$.  Using the renormalization group evolution one
can then find $\langle Q_i \rangle_I$ at any other scale $\mu \not=
\mc$. The details of this procedure can be found in
\cite{burasetal:92d}. Here we simply summarize the results of this
work.

We first express the matrix elements
$\langle Q_i \rangle_I$ in terms of the non-perturbative parameters
$B_i^{(1/2)}$ and $B_i^{(3/2)}$ for $\langle Q_i \rangle_0$ and
$\langle Q_i \rangle_2$, respectively. For $\mu \le \mc$ we have
\cite{burasetal:92d}
\begin{eqnarray}
\langle Q_1 \rangle_0 &=& -\,\frac{1}{9} X B_1^{(1/2)} \, ,
\label{eq:Q10} \\
\langle Q_2 \rangle_0 &=&  \frac{5}{9} X B_2^{(1/2)} \, ,
\label{eq:Q20} \\
\langle Q_3 \rangle_0 &=&  \frac{1}{3} X B_3^{(1/2)} \, ,
\label{eq:Q30} \\
\langle Q_4 \rangle_0 &=&  \langle Q_3 \rangle_0 + \langle Q_2 \rangle_0
                          -\langle Q_1 \rangle_0 \, ,
\label{eq:Q40} \\
\langle Q_5 \rangle_0 &=&  \frac{1}{3} B_5^{(1/2)} 
                           \langle \overline{Q_6} \rangle_0 \, ,
\label{eq:Q50} \\
\langle Q_6 \rangle_0 &=&  -\,4 \sqrt{\frac{3}{2}} 
\left[ \frac{m_{\rm K}^2}{\ms(\mu) + \md(\mu)}\right]^2
\frac{F_\pi}{\kappa} \,B_6^{(1/2)} \, ,
\label{eq:Q60} \\
\langle Q_7 \rangle_0 &=& 
- \left[ \frac{1}{6} \langle \overline{Q_6} \rangle_0 (\kappa + 1) 
         - \frac{X}{2} \right] B_7^{(1/2)} \, ,
\label{eq:Q70} \\
\langle Q_8 \rangle_0 &=& 
- \left[ \frac{1}{2} \langle \overline{Q_6} \rangle_0 (\kappa + 1) 
         - \frac{X}{6} \right] B_8^{(1/2)} \, ,
\label{eq:Q80} \\
\langle Q_9 \rangle_0 &=& 
\frac{3}{2} \langle Q_1 \rangle_0 - \frac{1}{2} \langle Q_3 \rangle_0 \, ,
\label{eq:Q90} \\
\langle Q_{10} \rangle_0 &=& 
    \langle Q_2 \rangle_0 + \frac{1}{2} \langle Q_1 \rangle_0
  - \frac{1}{2} \langle Q_3 \rangle_0 \, ,
\label{eq:Q100}
\end{eqnarray}
\begin{eqnarray}
\langle Q_1 \rangle_2 &=& 
\langle Q_2 \rangle_2 = \frac{4 \sqrt{2}}{9} X B_1^{(3/2)} \, ,
\label{eq:Q122} \\
\langle Q_i \rangle_2 &=&  0 \, , \qquad i=3,\ldots,6 \, ,
\label{eq:Q362} \\
\langle Q_7 \rangle_2 &=& 
  -\left[ \frac{\kappa}{6 \sqrt{2}} \langle \overline{Q_6} \rangle_0
          + \frac{X}{\sqrt{2}}
   \right] B_7^{(3/2)} \, ,
\label{eq:Q72} \\
\langle Q_8 \rangle_2 &=& 
  -\left[ \frac{\kappa}{2 \sqrt{2}} \langle \overline{Q_6} \rangle_0
          + \frac{\sqrt{2}}{6} X
   \right] B_8^{(3/2)} \, ,
\label{eq:Q82} \\
\langle Q_9 \rangle_2 &=& 
   \langle Q_{10} \rangle_2 = \frac{3}{2} \langle Q_1 \rangle_2 \, ,
\label{eq:Q9102}
\end{eqnarray}
where
\begin{equation}
\kappa = \frac{\Lambda_\chi^2}{m_{\rm K}^2 - m_\pi^2} =
         \frac{F_\pi}{F_{\rm K} - F_\pi} \, ,
\label{eq:kappaQi}
\end{equation}
\begin{equation}
X = \sqrt{\frac{3}{2}} F_\pi \left( m_{\rm K}^2 - m_\pi^2 \right) \, ,
\label{eq:XQi}
\end{equation}
and
\begin{equation}
\langle \overline{Q_6} \rangle_0 =
   \frac{\langle Q_6 \rangle_0}{B_6^{(1/2)}} \, .
\label{eq:Q60bar}
\end{equation}
The actual numerical values used for $m_{\rm K}$, $m_\pi$, $F_{\rm K}$,
$F_\pi$ are collected in appendix \ref{app:numinput}.

In the vacuum insertion method $B_i=1$ independent of $\mu$. In QCD,
however, the hadronic parameters $B_i$ generally depend on the
renormalizations scale $\mu$ and the renormalization scheme considered.

\subsection{$\langle Q_i(\mu) \rangle_2$ for $(V-A)\otimes (V-A)$ Operators}
           \label{subsec:Qi2VmAVmA}
The matrix elements $\langle Q_1 \rangle_2$, $\langle Q_2 \rangle_2$,
$\langle Q_9 \rangle_2$ and $\langle Q_{10} \rangle_2$ can to a very
good approximation be determined from $\RE A_2$ in
\eqn{eq:ReA0data} as functions of $\Lms$, $\mu$ and the renormalization
scheme considered. To this end it is useful to set $\aem=0$, as the
$\ord(\aem)$ effects in CP conserving amplitudes, such as the
contributions of electroweak penguins, are very small. One then finds
\begin{equation}
\langle Q_1(\mu) \rangle_2 = \langle Q_2(\mu) \rangle_2 =
\frac{10^6\gev^2}{1.77} \frac{\RE A_2}{z_+(\mu)} =
\frac{8.47 \cdot 10^{-3}\gev^3}{z_+(\mu)}
\label{eq:Q122data}
\end{equation}
and comparing with \eqn{eq:Q122}
\begin{equation}
B_1^{(3/2)}(\mu) = \frac{0.363}{z_+(\mu)}
\label{eq:B321}
\end{equation}
with $z_+ = z_1 + z_2$.
Since $z_+(\mu)$ depends on the scale $\mu$ and the renormalization
scheme used, \eqn{eq:B321} gives automatically the scheme and $\mu$
dependence of $B_1^{(3/2)}$ and of the related matrix elements $\langle
Q_1 \rangle_2$, $\langle Q_2 \rangle_2$,
$\langle Q_9 \rangle_2$ and $\langle Q_{10} \rangle_2$. The impact of
$\ord(\aem)$ corrections on this result has been analysed in
\cite{burasetal:92d}. It amounts only to a few percent as expected.
These corrections are of course included in the numerical analysis
presented in this reference and here as well. Using $\mu=\mc=1.3\gev$,
$\Lms^{(4)}=325\mev$ and $z_+(\mc)$ of table \ref{tab:wc10smu13}
we find according to \eqn{eq:B321}
\begin{equation}
B_{1,NDR}^{(3/2)}(\mc) =  0.453
\qquad
B_{1,HV}^{(3/2)}(\mc) =  0.472 \, .
\label{eq:B321mc}
\end{equation}

\noindent
The following comments should be made:
\begin{itemize}
\item
$B_1^{(3/2)}(\mu)$ decreases with increasing $\mu$.
\item
The extracted value for $B_1^{(3/2)}$ is by more than a factor of two
smaller than the vacuum insertion estimate.
\item
It is compatible with the $1/N_c$ value $B_1^{(3/2)}(1\gev) \approx
0.55$ \cite{bardeen:87b} and somewhat smaller than the lattice result
$B_1^{(3/2)}(2\gev) \approx 0.6$ \cite{ciuchini:95}.
\end{itemize}

\subsection{$\langle Q_i(\mu) \rangle_0$ for $(V-A)\otimes (V-A)$ Operators}
           \label{subsec:Qi0VmAVmA}
The determination of $\langle Q_i(\mu) \rangle_0$ matrix elements is
more involved because several operators may contribute to $\RE
A_0$. The main idea of \cite{burasetal:92d} is then to set $\mu=\mc$, as
at this scale only $Q_1$ and $Q_2$ operators contribute to $\RE
A_0$ in the HV scheme. One then finds $\langle Q_1(\mc) \rangle_0$ as a
function of  $\langle Q_2(\mc) \rangle_0$
\begin{equation}
\langle Q_1(\mc) \rangle_0 = \frac{10^6\gev^2}{1.77} \frac{\RE
A_0}{z_1(\mc)} - \frac{z_2(\mc)}{z_1(\mc)} \langle Q_2(\mc) \rangle_0
\label{eq:Q10mc}
\end{equation}
where the reference in $\langle Q_{1,2}(\mc) \rangle_0$ to the HV scheme
has been suppressed for convenience. Using next the relations
\eqn{eq:Q40}, \eqn{eq:Q90} and \eqn{eq:Q100} one is able to obtain
$\langle Q_4(\mc) \rangle_0$, $\langle Q_9(\mc) \rangle_0$ and $\langle
Q_{10}(\mc) \rangle_0$ as functions of $\langle Q_2(\mc) \rangle_0$ and
$\langle Q_3(\mc) \rangle_0$. Because $\langle Q_3(\mc) \rangle_0$ is
colour suppressed it is less essential for this analysis than $\langle
Q_2(\mc) \rangle_0$. Moreover its Wilson coefficient is small and
similarly to $\langle Q_9(\mc) \rangle_0$ and $\langle 
Q_{10}(\mc) \rangle_0$ also $\langle Q_3(\mc) \rangle_0$ has only a
small impact on $\epe$. On the other hand the coefficient $y_4$ is
substantial and consequently $\langle Q_4(\mc) \rangle_0$ plays a
considerable role in the analysis of $\epe$. The matrix element $\langle
Q_3(\mc) \rangle_0$ has then an indirect impact on $\epe$ through
relation \eqn{eq:Q40}. For numerical evaluation, $\langle Q_3(\mc)
\rangle_0$ of \eqn{eq:Q30} with $B_3^{(1/2)} = 1$ can be used keeping in
mind that this may introduce a small uncertainty in the final analysis.
This uncertainty has been investigated in \cite{burasetal:92d}.

Once the matrix elements in question have been determined as functions
of $\langle Q_2(\mc) \rangle_0$ in the HV scheme, they can be found by a
finite renormalization in any other scheme. Details can be found in
\cite{burasetal:92d}.

If one in addition makes the very plausible assumption valid in all
known non-perturbative approaches that $\langle Q_-(\mc) \rangle_0 \ge
\langle Q_+(\mc) \rangle_0 \ge 0$ the experimental value of $\RE
A_0$ in \eqn{eq:ReA0data} together with \eqn{eq:Q10mc} and table
\ref{tab:wc10smu13} implies for $\Lms^{(4)}=325\mev$
\begin{equation}
B_{2,LO}^{(1/2)}(\mc)  =  5.7 \pm 1.1
\qquad
B_{2,NDR}^{(1/2)}(\mc) =  6.6 \pm 1.0
\qquad
B_{2,HV}^{(1/2)}(\mc) =  6.2 \pm 1.0 \, .
\label{eq:B122mc}
\end{equation}
The extraction of $B_1^{(1/2)}(\mc)$ and of an analogous parameter
$B_4^{(1/2)}(\mc)$ are presented in detail in \cite{burasetal:92d}.
$B_1^{(1/2)}(\mc)$ depends very sensitively on $B_2^{(1/2)}(\mc)$ and
its central value is as high as 15. $B_4^{(1/2)}(\mc)$ is less sensitive
and typically by (10--15)\,\% lower than $B_2^{(1/2)}(\mc)$. In any case
this analysis shows very large departures from the results of the
vacuum insertion method.

\subsection{$\langle Q_i(\mu) \rangle_{0,2}$ for $(V-A)\otimes (V+A)$ Operators}
           \label{subsec:Qi0VmAVpA}
The matrix elements of the $(V-A) \otimes (V+A)$ operators $Q_5$--$Q_8$
cannot be constrained by CP conserving data and one has to rely on
existing non-perturbative methods to calculate them. Fortunately, there
are some indications that the existing non-perturbative estimates of
$\langle Q_i(\mu) \rangle_{0,2}$, $i=5,\ldots,8$ are more reliable than
the corresponding calculations for $(V-A) \otimes (V-A)$ operators.

First of all, the parameters $B_{5,6}^{(1/2)}$ \cite{kilcup:91},
\cite{sharpe:91} and $B_{7,8}^{(3/2)}$ \cite{francoetal:89},
\cite{kilcup:91}, \cite{sharpe:91}, \cite{bernardsoni:91} calculated in
the lattice approach
\begin{equation}
B_{5,6}^{(1/2)} = 1.0 \pm 0.2 
\qquad
B_{7,8}^{(3/2)} = 1.0 \pm 0.2 
\label{eq:B1258}
\end{equation}
agree well with the vacuum insertion values ($B_i=1$) and in the case
of $B_6^{(1/2)}$ and $B_8^{(3/2)}$ with the $1/N_c$ approach ($B_6^{(1/2)}
= B_8^{(3/2)} = 1$) \cite{bardeen:87a}, \cite{burasgerard:87}.

We note next that with fixed values for $B_{5,6}^{(1/2)}$ and
$B_{7,8}^{(3/2)}$ the $\mu$-dependence of $\langle Q_{5,6} \rangle_0$
and $\langle Q_{7,8} \rangle_2$ is governed by the $\mu$ dependence of
$\ms(\mu)$. For $\langle Q_6 \rangle_0$ and $\langle Q_8 \rangle_2$
this property has been first found in the $1/N_c$ approach
\cite{burasgerard:87}: in the large-$N_c$ limit the anomalous
dimensions of $Q_6$ and $Q_8$ are simply twice the anomalous dimension
of the mass operator leading to $\sim 1/\ms^2(\mu)$ for the
corresponding matrix elements. Another support comes from a
renormalization study in \cite{burasetal:92d}. In this analysis the
$B_i$-factors in \eqn{eq:B1258} have been set to unity at $\mu=\mc$.
Subsequently the evolution of the matrix elements in the range $1\gev
\le \mu \le 4\gev$ has been calculated showing that for the NDR scheme
$B_{5,6}^{(1/2)}$ and $B_{7,8}^{(3/2)}$ were $\mu$ independent within
an accuracy of (2--3)\,\%. The $\mu$ dependence in the HV scheme has
been found to be stronger but still below 10\,\%.

Concerning $B_{7,8}^{(1/2)}$ one can simply set $B_{7,8}^{(1/2)}=1$ as
the matrix elementes $\langle Q_{7,8} \rangle_0$ play only a minor role
in the $\epe$ analysis.

In summary, our treatment of $\langle Q_i \rangle_{0,2}$, $i=5,\ldots 8$
follows the one used in \cite{burasetal:92d}. We will set
\begin{equation}
B_{7,8}^{(1/2)}(\mc) = 1
\qquad
B_5^{(1/2)}(\mc) = B_6^{(1/2)}(\mc)
\qquad
B_7^{(3/2)}(\mc) = B_8^{(3/2)}(\mc)
\label{eq:B1278mc}
\end{equation}
and we will treat $B_6^{(1/2)}(\mc)$ and $B_8^{(3/2)}(\mc)$ as free
parameters in the neighbourhood of the values given in \eqn{eq:B1258}.
Then the main uncertainty in the values of $\langle Q_i \rangle_{0,2}$,
$i=5,\ldots 8$ results from the value of the strange quark mass
$\ms(\mc)$. The present estimates give
\begin{equation}
\ms(\mc) = (170 \pm 20)\mev
\label{eq:msmc}
\end{equation}
with the lower values coming from recent lattice calculations
\cite{alltonetal:94} and the higher ones from QCD sum rules
\cite{jaminmuenz:95}, \cite{chetyrkinetal:95}.

\subsection{The Four Dominant Contributions to $\epe$}
           \label{subsec:epe4dom}
$P^{(1/2)}$ and $P^{(3/2)}$ in \eqn{eq:epe} can be written as linear
combinations of two independent hadronic parameters $B_6^{(1/2)}$ and
$B_8^{(3/2)}$ \cite{burasetal:92d}. This $B_i$-expansion reads
\begin{eqnarray}
P^{(1/2)} &=& a_0^{(1/2)} +
  \left[ \frac{178\mev}{\ms(\mc)+\md(\mc)} \right]^2 a_6^{(1/2)} B_6^{(1/2)}
\label{eq:P12Bi} \\
P^{(3/2)} &=& a_0^{(3/2)} +
  \left[ \frac{178\mev}{\ms(\mc)+\md(\mc)} \right]^2 a_8^{(3/2)} B_8^{(3/2)}
  \, .
\label{eq:P32Bi}
\end{eqnarray}
Here $a_0^{(1/2)}$ and $a_0^{(3/2)}$ effectively summarize all
dependences other than $B_6^{(1/2)}$ and $B_8^{(3/2)}$, especially
$B_2^{(1/2)}$ in the case of $a_0^{(1/2)}$.
Note that in contrast to \cite{burasetal:92d} we have absorbed the
dependence on $B_2^{(1/2)}$ into $a_0^{(1/2)}$ and we have exhibited the
dependence on $\ms$ which was not shown explicitly there.
The residual $\ms$ dependence present in $a_0^{(1/2)}$ and
$a_0^{(3/2)}$ is negligible.  Setting $\mu=\mc$, and using the strategy
for hadronic matrix elements outlined above one finds the coefficients
$a_i^{(1/2)}$ and $a_i^{(3/2)}$ as functions of $\Lms$, $\mt$ and the
renormalization scheme considered. These dependences are given in
tables \ref{tab:bip12} and \ref{tab:bip32}. We should however stress
that $P^{(1/2)}$ and $P^{(3/2)}$ are independent of $\mu$ and the
renormalization scheme considered.

\begin{table}[htb]
\caption[]{$B_i$-expansion coefficients for $P^{(1/2)}$.
\label{tab:bip12}}
\begin{center}
\begin{tabular}{|c|c||c|c||c|c||c|c|}
& & \multicolumn{2}{c||}{LO} &
  \multicolumn{2}{c||}{NDR} &
  \multicolumn{2}{c|}{HV} \\
\hline
$\Lms^{(4)}\,[\mev]$ & $\mt\,[\gev]$ &
$a_0^{(1/2)}$ & $a_6^{(1/2)}$ &
$a_0^{(1/2)}$ & $a_6^{(1/2)}$ &
$a_0^{(1/2)}$ & $a_6^{(1/2)}$ \\
\hline
    & 155  &  --2.138 & 5.110  &  --2.251 & 4.676  &  --2.215 & 4.159 \\
215 & 170  &  --2.070 & 5.138  &  --2.187 & 4.698  &  --2.150 & 4.181 \\
    & 185  &  --1.996 & 5.162  &  --2.117 & 4.716  &  --2.081 & 4.200 \\
\hline 
    & 155  &  --2.231 & 6.540  &  --2.414 & 6.255  &  --2.362 & 5.389 \\
325 & 170  &  --2.161 & 6.576  &  --2.350 & 6.282  &  --2.298 & 5.416 \\
    & 185  &  --2.085 & 6.606  &  --2.281 & 6.306  &  --2.229 & 5.439 \\
\hline 
    & 155  &  --2.288 & 8.171  &  --2.549 & 8.417  &  --2.473 & 6.972 \\
435 & 170  &  --2.212 & 8.214  &  --2.482 & 8.451  &  --2.406 & 7.005 \\
    & 185  &  --2.130 & 8.251  &  --2.409 & 8.480  &  --2.333 & 7.035
\end{tabular}
\end{center}
\end{table}

\begin{table}[htb]
\caption[]{$B_i$-expansion coefficients for $P^{(3/2)}$.
\label{tab:bip32}}
\begin{center}
\begin{tabular}{|c|c||c|c||c|c||c|c|}
& & \multicolumn{2}{c||}{LO} &
  \multicolumn{2}{c||}{NDR} &
  \multicolumn{2}{c|}{HV} \\
\hline
$\Lms^{(4)}\,[\mev]$ & $\mt\,[\gev]$ &
$a_0^{(3/2)}$ & $a_8^{(3/2)}$ & $a_0^{(3/2)}$ &
$a_8^{(3/2)}$ & $a_0^{(3/2)}$ & $a_8^{(3/2)}$ \\
\hline
    & 155  &  --0.797 & 1.961  &  --0.819 & 1.887  &  --0.838 & 2.114 \\
215 & 170  &  --0.880 & 2.602  &  --0.900 & 2.438  &  --0.919 & 2.666 \\
    & 185  &  --0.965 & 3.296  &  --0.983 & 3.036  &  --1.002 & 3.263 \\
\hline 
    & 155  &  --0.788 & 2.645  &  --0.814 & 2.639  &  --0.837 & 2.894 \\
325 & 170  &  --0.870 & 3.422  &  --0.895 & 3.305  &  --0.917 & 3.560 \\
    & 185  &  --0.956 & 4.264  &  --0.978 & 4.027  &  --1.000 & 4.281 \\
\hline 
    & 155  &  --0.779 & 3.425  &  --0.809 & 3.622  &  --0.835 & 3.899 \\
435 & 170  &  --0.861 & 4.360  &  --0.889 & 4.435  &  --0.915 & 4.712 \\
    & 185  &  --0.947 & 5.372  &  --0.971 & 5.316  &  --0.998 & 5.593
\end{tabular}
\end{center}
\end{table}

Inspecting \eqn{eq:P12Bi}, \eqn{eq:P32Bi} and tables \ref{tab:bip12},
\ref{tab:bip32} we identify the following four contributions which
govern the ratio $\epe$ at scales $\mu=\ord(\mc)$:
\renewcommand{\theenumi}{\roman{enumi}}
\begin{enumerate}
\item
\label{enum:i}
The contribution of $(V-A) \otimes (V-A)$ operators to $P^{(1/2)}$ is
dominantly represented by $a_0^{(1/2)}$. This term is to a large extent
fixed by the experimental value of $A_0$ and consequently is only very
weakly dependent on $\Lms$ and the renormalization scheme considered.
The weak dependence on $\mt$ results from small contributions of
electroweak penguin operators.  Taking $\Lms^{(4)} = 325\mev$,
$\mu=\mc$ and $\mt = 170\gev$ we have $a_0^{(1/2)} \approx -2.3$ for
both schemes considered.  We observe that the contribution of $(V-A)
\otimes (V-A)$ operators, in particular $Q_4$, to $\epe$ is {\em
negative}.
\item
\label{enum:ii}
The contribution of $(V-A) \otimes (V+A)$ QCD penguin operators to
$P^{(1/2)}$ is given by the second term in \eqn{eq:P12Bi}. This
contribution is large and {\em positive}. The coefficient $a_6^{(1/2)}$
depends sensitively on $\Lms$ which results from the strong dependence
of $y_6$ on the QCD scale. The dependence on $\mt$ is very weak on the
other hand. Taking $\Lms^{(4)} = 325\mev$, $\ms(\mc)=170\mev$ and $\mt
= 170\gev$ and setting as an example $B_6^{(1/2)} = 1$ in the NDR and
HV schemes we find a positve contribution to $\epe$ amounting to 6.3
and 5.4 in the NDR and HV scheme, respectively.

\item
\label{enum:iii}
The contribution of the $(V-A) \otimes (V-A)$ electroweak penguin
operators $Q_9$ and $Q_{10}$ to $P^{(3/2)}$ is represented by
$a_0^{(3/2)}$. As in the case of the contribution \ref{enum:i}, the
matrix elements contributing to $a_0^{(3/2)}$ are fixed by the CP
conserving data, this time by the amplitude $A_2$. Consequently, the
scheme and the $\Lms$ dependence of $a_0^{(3/2)}$ is very weak. The
sizeable $\mt$ dependence of $a_0^{(3/2)}$ results from the $\mt$
dependence of $y_9 + y_{10}$. $a_0^{(3/2)}$ contributes {\em positively}
to $\epe$. For $\mt = 170\gev$ this contribution is roughly 0.9 for
both renormalization schemes and the full range of $\Lms$ considered.
\item
\label{enum:iv}
The contribution of the $(V-A) \otimes (V+A)$ electroweak penguin
operators $Q_7$ and $Q_8$ to $P^{(3/2)}$ is represented by the second
term in \eqn{eq:P32Bi}. This contribution depends sensitively on $\mt$
and $\Lms$ as could be expected on the basis of $y_7$ and $y_8$. Taking
again $B_8^{(3/2)}=1$ in both renormalization schemes we find for the
central values of $\Lms^{(4)}$, $\mt$ and $\mc$ a {\em
negative} contribution to $\epe$ equal to $-3.9$ and $-3.6$ for the NDR
and HV scheme, respecetively.
\end{enumerate}
\renewcommand{\theenumi}{\arabic{enumi}}

Before analysing $\epe$ numerically in more detail, let us just set
$\IM \lambda_t = 1.3 \cdot 10^{-4}$ and $B_6^{(1/2)} = B_8^{(3/2)} = 1$
in both schemes. Then for the central values of the remaining
parameters one obtains $\epe = 2.0 \cdot 10^{-4}$ and $\epe = 0.6 \cdot
10^{-4}$ for the NDR and HV scheme, respectively. This strong scheme
dependence can only be compensated for by having $B_6^{(1/2)}$ and
$B_8^{(3/2)}$ different in the two schemes considered. As we will see
below the strong cancellations between various contributions at $\mt
\approx 170\gev$ make the prediction for $\epe$ rather uncertain. One
should also stress that the formulation presented here does not exhibit
analytically the $\mt$ dependence. As the coefficients $a_0^{(3/2)}$
and $a_8^{(3/2)}$ depend very sensitively on $\mt$ it is useful to
display this dependence in an analytic form.

\subsection{An Analytic Formula for $\epe$}
           \label{subsec:epeanalytic}
As shown in \cite{buraslauten:93} it is possible to cast the above
discussion into an analytic formula which exhibits the $\mt$ dependence
together with the dependence on $\ms$, $B_6^{(1/2)}$ and $B_8^{(3/2)}$.
Such an analytic formula should be useful for those phenomenologists
and experimentalists who are not interested in getting involved with
the technicalities discussed in preceding sections.

In order to find an analytic expression for $\epe$ which exactly
reproduces the results discussed above one uses the PBE presented in
section \ref{sec:PBE}. The resulting analytic expression for $\epe$ is
then given as follows
\begin{equation}
\epe =
\IM \lambda_t F(x_t)
\label{eq:epePBE}
\end{equation}
where
\begin{equation}
F(x_t) = P_0 + P_X X_0(x_t) + P_Y Y_0(x_t) + P_Z Z_0(x_t) + P_E E_0(x_t)
\label{eq:Fxt}
\end{equation}
with the $\mt$ dependent functions listed in section \ref{sec:PBE}. The
coefficients $P_i$ are given in terms of $B_6^{(1/2)} \equiv
B_6^{(1/2)}(\mc)$, $B_8^{(3/2)} \equiv B_8^{(3/2)}(\mc)$ and $\ms(\mc)$
as follows
\begin{equation}
P_i = r_i^{(0)} + \left[ \frac{178\mev}{\ms(\mc)+\md(\mc)} \right]^2
\left(r_i^{(6)} B_6^{(1/2)} + r_i^{(8)} B_8^{(3/2)} \right) \, .
\label{eq:pbePi}
\end{equation}
The $P_i$ are $\mu$ and renormalization scheme independent. They depend
however on $\Lms$. In table \ref{tab:pbendr} we give the numerical
values of $r_i^{(0)}$, $r_i^{(6)}$ and $r_i^{(8)}$ for different values
of $\Lms$ at $\mu=\mc$ in the NDR renormalization scheme. Analogous
results in the HV scheme are given in table \ref{tab:pbehv}. The
coefficients $r_i^{(0)}$, $r_i^{(6)}$ and $r_i^{(8)}$do not depend on
$\ms(\mc)$ as this dependence has been factored out. $r_i^{(0)}$ does,
however, depend on the particular choice for the parameter
$B_2^{(1/2)}$ in the parametrization of the matrix element $\langle Q_2
\rangle_0$. The values given in the tables correspond to the central
values in \eqn{eq:B122mc}. Variation of $B_2^{(1/2)}$ in the full
allowed range introduces an uncertainty of at most 18\,\% in the
$r_i^{(0)}$ column of the tables.  Since the parameters $r_i^{(0)}$
give only subdominant contributions to $\epe$ keeping $B_2^{(1/2)}$ and
$r_i^{(0)}$ at their central values is a very good approximation.

For different scales $\mu$ the numerical values in the tables change
without modifying the values of the $P_i$'s as it should be. To this
end also $B_6^{(1/2)}$ and $B_8^{(3/2)}$ have to be modified as they
depend albeit weakly on $\mu$.

Concerning the scheme dependence we note that whereas $r_0$ coefficients
are scheme dependent, the coefficients $r_i$, $i=X, Y, Z, E$ do not show
any scheme dependence. This is related to the fact that the $\mt$
dependence in $\epe$ enters first at the NLO level and consequently all
coefficients $r_i$ in front of the $\mt$ dependent functions must be
scheme independent. That this turns out to be indeed the case is a nice
check of our calculations.

Consequently when changing the renormalization scheme one is only
obliged to change appropriately $B_6^{(1/2)}$ and $B_8^{(3/2)}$ in the
formula for $P_0$ in order to obtain a scheme independence of $\epe$.
In calculating $P_i$ where $i \not= 0$, $B_6^{(1/2)}$ and $B_8^{(3/2)}$
can in fact remain unchanged, because their variation in this part
corresponds to higher order contributions to $\epe$ which would have to
be taken into account in the next order of perturbation theory.

For similar reasons the NLO analysis of $\epe$ is still insensitive to
the precise definition of $\mt$. In view of the fact that the NLO
calculations of $\IM \lambda_t$ have been done with
$\mt=\overline{m}_t(\mt)$ we will also use  this definition in
calculating $F(x_t)$.

\begin{table}[htb]
\caption[]{$\Delta S=1$ PBE coefficients for various $\Lms$ in the NDR scheme.
\label{tab:pbendr}}
\begin{center}
\begin{tabular}{|c||c|c|c||c|c|c||c|c|c|}
& \multicolumn{3}{c||}{$\Lms^{(4)}=215\mev$} &
  \multicolumn{3}{c||}{$\Lms^{(4)}=325\mev$} &
  \multicolumn{3}{c| }{$\Lms^{(4)}=435\mev$} \\
\hline
$i$ & $r_i^{(0)}$ & $r_i^{(6)}$ & $r_i^{(8)}$ &
      $r_i^{(0)}$ & $r_i^{(6)}$ & $r_i^{(8)}$ &
      $r_i^{(0)}$ & $r_i^{(6)}$ & $r_i^{(8)}$ \\
\hline
 0  & --2.644 & 4.784 & 0.876 & --2.749 & 6.376 & 0.689 & --2.845 & 8.547
    & 0.436 \\
$X$ & 0.555 & 0.008 & 0 & 0.521 & 0.012 & 0 & 0.495 & 0.017 & 0 \\
$Y$ & 0.422 & 0.037 & 0 & 0.385 & 0.046 & 0 & 0.356 & 0.057 & 0 \\
$Z$ & 0.074 & --0.007 & --4.798 & 0.149 & --0.009 & --5.789 & 0.237 & --0.011
    & --7.064 \\
$E$ & 0.209 & --0.591 & 0.205 & 0.181 & --0.727 & 0.265 & 0.152 & --0.892
    & 0.342
\end{tabular}
\end{center}
\end{table}

\begin{table}[htb]
\caption[]{$\Delta S=1$ PBE coefficients for various $\Lms$ in the HV scheme.
\label{tab:pbehv}}
\begin{center}
\begin{tabular}{|c||c|c|c||c|c|c||c|c|c|}
& \multicolumn{3}{c||}{$\Lms^{(4)}=215\mev$} &
  \multicolumn{3}{c||}{$\Lms^{(4)}=325\mev$} &
  \multicolumn{3}{c| }{$\Lms^{(4)}=435\mev$} \\
\hline
$i$ & $r_i^{(0)}$ & $r_i^{(6)}$ & $r_i^{(8)}$ &
      $r_i^{(0)}$ & $r_i^{(6)}$ & $r_i^{(8)}$ &
      $r_i^{(0)}$ & $r_i^{(6)}$ & $r_i^{(8)}$ \\
\hline
 0  & --2.631 & 4.291 & 0.668 & --2.735 & 5.548 & 0.457 & --2.830 & 7.163
    & 0.185 \\
$X$ & 0.555 & 0.008 & 0 & 0.521 & 0.012 & 0 & 0.495 & 0.017 & 0 \\
$Y$ & 0.422 & 0.037 & 0 & 0.385 & 0.046 & 0 & 0.356 & 0.057 & 0 \\
$Z$ & 0.074 & --0.007 & --4.798 & 0.149 & --0.009 & --5.789 & 0.237 & --0.011
    & --7.064 \\
$E$ & 0.209 & --0.591 & 0.205 & 0.181 & --0.727 & 0.265 & 0.152 & --0.892
    & 0.342
\end{tabular}
\end{center}
\end{table}

The inspection of tables \ref{tab:pbendr} and \ref{tab:pbehv} shows
that the terms involving $r_0^{(6)}$ and $r_Z^{(8)}$ dominate the ratio
$\epe$. The function $Z_0(x_t)$ representing a gauge invariant
combination of $Z^0$- and $\gamma$-penguins grows rapidly with $\mt$
and due to $r_Z^{(8)} < 0$ these contributions suppress $\epe$ strongly
for large $\mt$ \cite{flynn:89}, \cite{buchallaetal:90}. These two
dominant terms $r_0^{(6)}$ and $r_Z^{(8)}$ correspond essentially to
the second terms in \eqn{eq:P12Bi} and \eqn{eq:P32Bi}, respectively.
The first term in \eqn{eq:P12Bi} corresponds roughly to $r_0^{(0)}$
given here, while the first term in \eqn{eq:P32Bi} is represented to a
large extent by the positve contributions of $X_0(x_t)$ and $Y_0(x_t)$.
The last term in \eqn{eq:Fxt} representing the residual $\mt$
dependence of QCD penguins plays only a minor role in the full analysis
of $\epe$.

\subsection{Numerical Results}
           \label{subsec:epenumres}
\begin{figure}[htb]
\vspace{0.15in}
\centerline{
\epsfysize=6in
\rotate[r]{
\epsffile{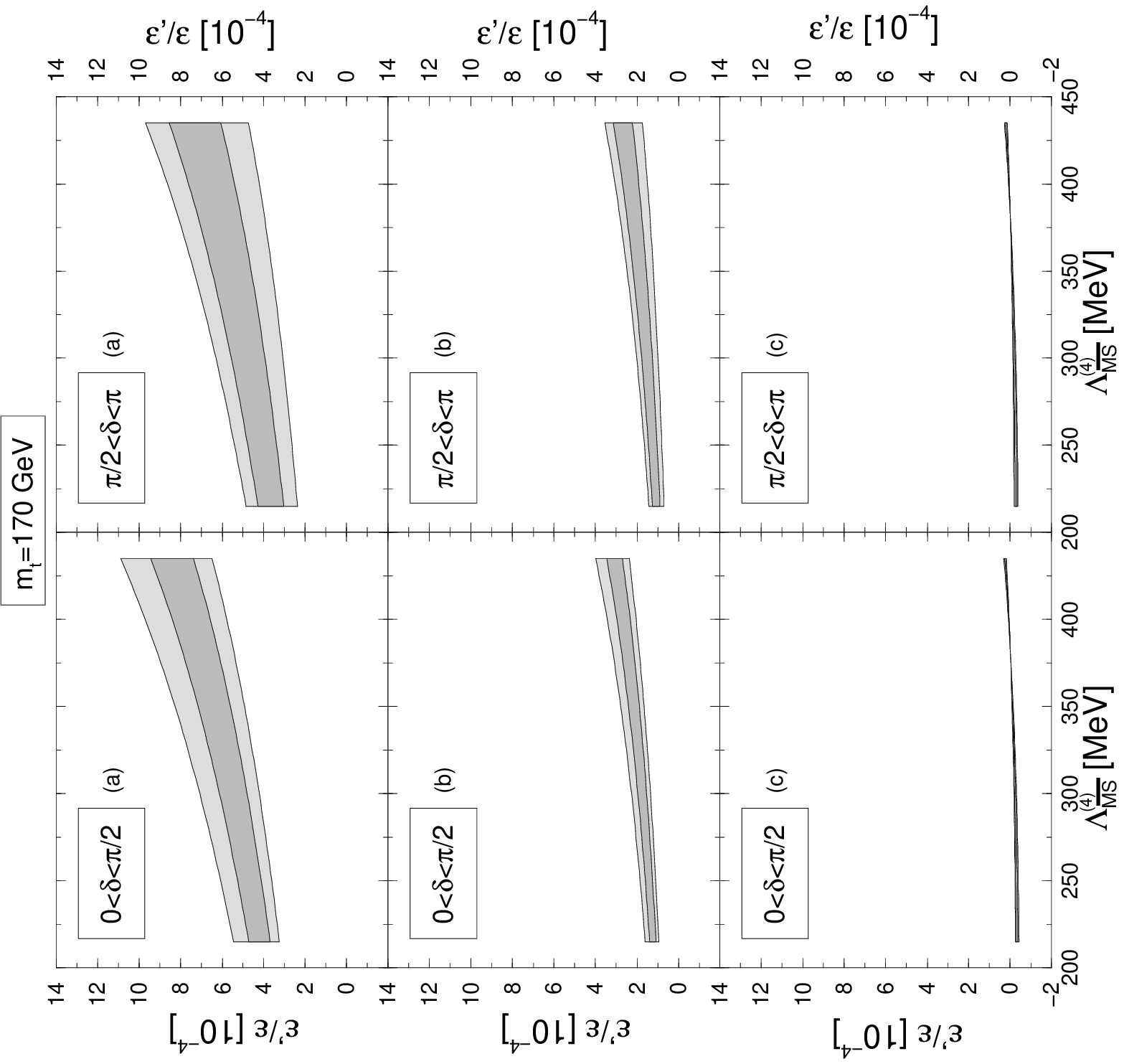}
}}
\vspace{0.15in}
\caption[]{
The ranges of $\epe$ in the NDR scheme as a function of $\Lms^{(4)}$
for $\mt=170\gev$ and present (light grey) and future (dark grey)
parameter ranges given in appendix \ref{app:numinput}. The three pairs
of $\epe$ plots correspond to hadronic parameter sets
(a) $(B_6^{(1/2)}(\mc))_{\rm eff}=1.5$, $(B_8^{(3/2)}(\mc))_{\rm eff}=1.0$,
(b) $(B_6^{(1/2)}(\mc))_{\rm eff}=1.0$, $(B_8^{(3/2)}(\mc))_{\rm eff}=1.0$,
and
(c) $(B_6^{(1/2)}(\mc))_{\rm eff}=1.0$, $(B_8^{(3/2)}(\mc))_{\rm eff}=1.5$,
respectively.
\label{fig:lepemt170}}
\end{figure}

Let us define two effective B-factors:
\begin{equation}\label{7e}
(B_i^{(j)}(m_c))_{eff}= 
\left [\frac{178\mev}{\bar m_s(m_c)+\bar m_d(m_c)} \right ]^2 B_i^{(j)}(m_c)
\end{equation}
In fig.\ \ref{fig:lepemt170} we show $\epe$ for $\mt=170\gev$ as a function
of $\Lms$ for different choices of the effective $B_i$ factors.
We show here only the results in the NDR scheme. As discussed above
$\epe$ is generally lower in the HV scheme, if the same values for
$B_6^{(1/2)}$ and $B_8^{(3/2)}$ are used in both schemes. In view of the
fact that the differences between NDR and HV schemes are smaller than
the uncertainties in $B_6^{(1/2)}$ and $B_8^{(3/2)}$ we think it is
sufficient to present only the results in the NDR scheme here. The
results in the HV scheme can be found in \cite{burasetal:92d},
\cite{ciuchini:95}.

\begin{figure}[htb]
\vspace{0.15in}
\centerline{
\epsfysize=6in
\rotate[r]{
\epsffile{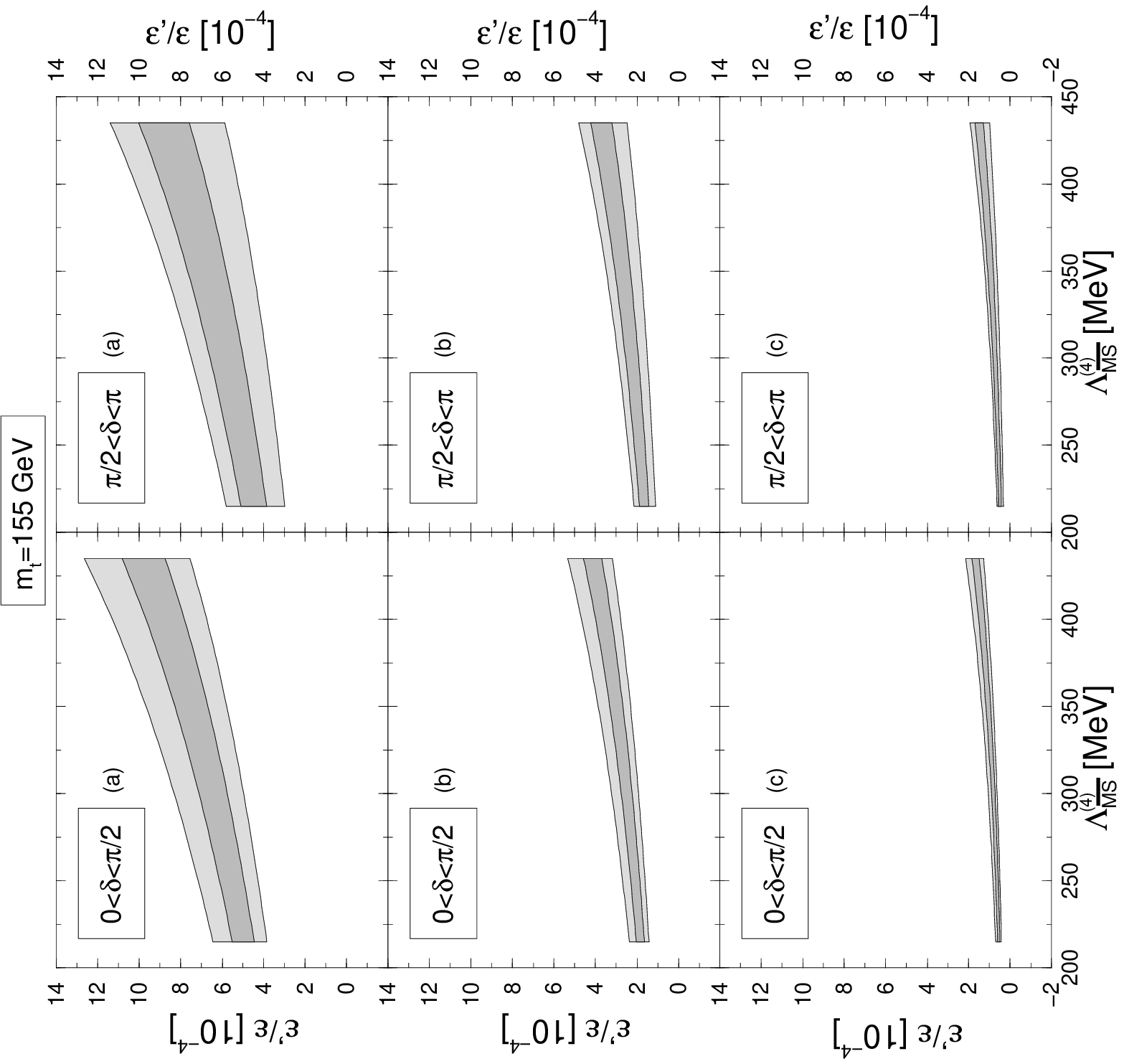}
}}
\vspace{0.15in}
\caption[]{
Same as fig.\ \ref{fig:lepemt170} but for $\mt=155\gev$.
\label{fig:lepemt155}}
\end{figure}

Fig.\ \ref{fig:lepemt170} shows strong dependence of $\epe$ on $\Lms$.
However the main uncertainty originates in the poor knowledge of
$(B_i)_{eff}$.  In case a) in which the QCD-penguin contributions
dominate, $\epe$ can reach values as high as $1 \cdot 10^{-3}$.
However, in case c) the electroweak penguin contributions are large
enough to cancel essentially the QCD-penguin contributions completely.
Consequently in this case $|\epe|< 2 \cdot 10^{-5}$ and the standard
model prediction of $\epe$ cannot be distinguished from a superweak
theory. As shown in fig.\ \ref{fig:lepemt155} higher values of $\epe$
can be obtained for $\mt = 155\gev$ although still $\epe < 13 \cdot
10^{-4}$.

\begin{figure}[htb]
\vspace{0.15in}
\centerline{
\epsfysize=6in
\rotate[r]{
\epsffile{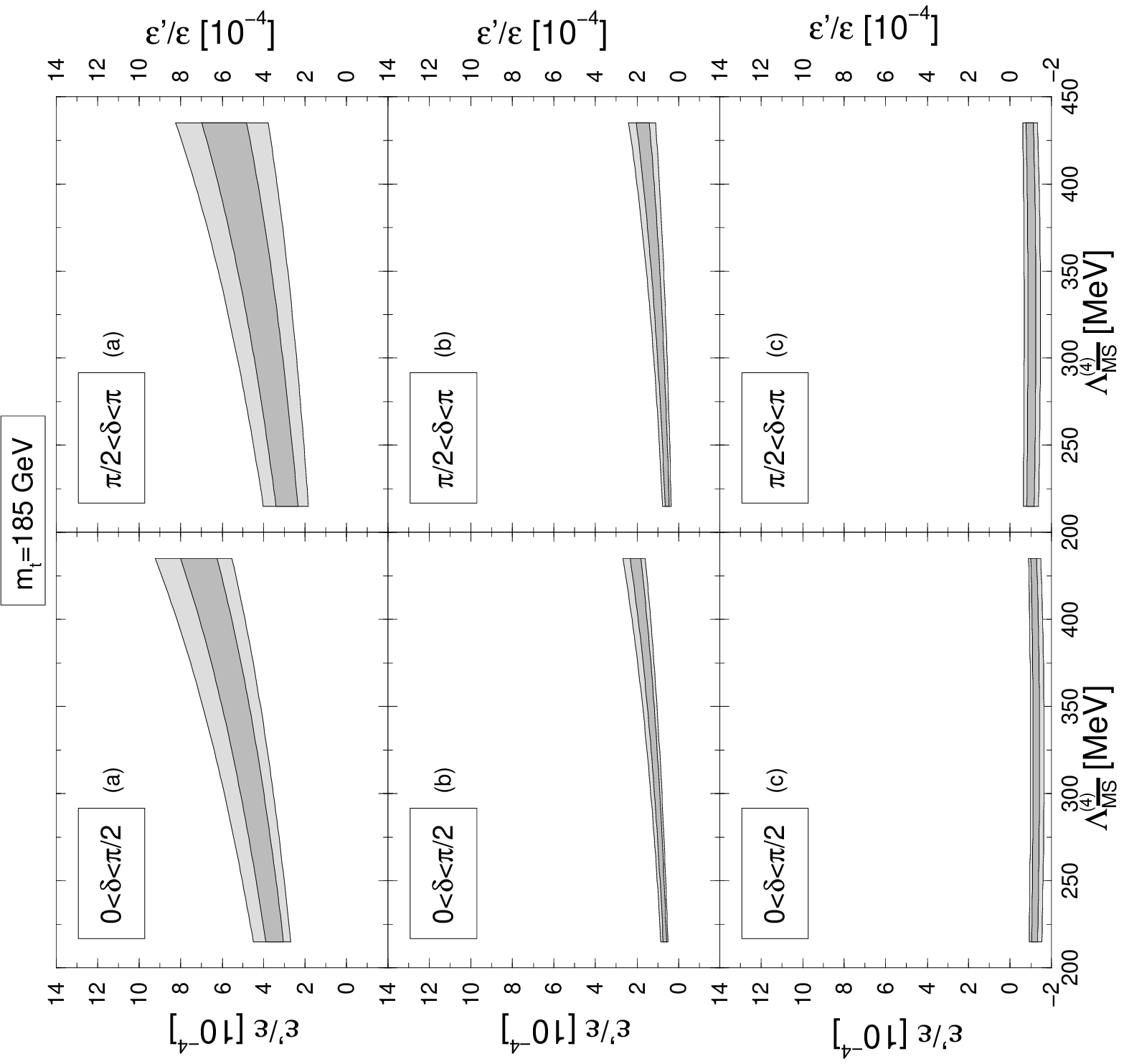}
}}
\vspace{0.15in}
\caption[]{
Same as fig.\ \ref{fig:lepemt170} but for $\mt=185\gev$.
\label{fig:lepemt185}}
\end{figure}

For $\mt = 185\gev$ the values of $\epe$ are correspondingly smaller
and in case c) small negative values are found for $\epe$.  In figs.\
\ref{fig:lepemt170}--\ref{fig:lepemt185} the dark grey regions refer to
the future ranges for $\IM\lambda_t$. Of course one should hope that
also the knowledge of $(B_i)_{eff}$ and of $\Lms^{(4)}$ will be
improved in the future so that a firmer prediction for $\epe$ can be
obtained.

\begin{figure}[htb]
\vspace{0.15in}
\centerline{
\epsfysize=6in
\rotate[r]{
\epsffile{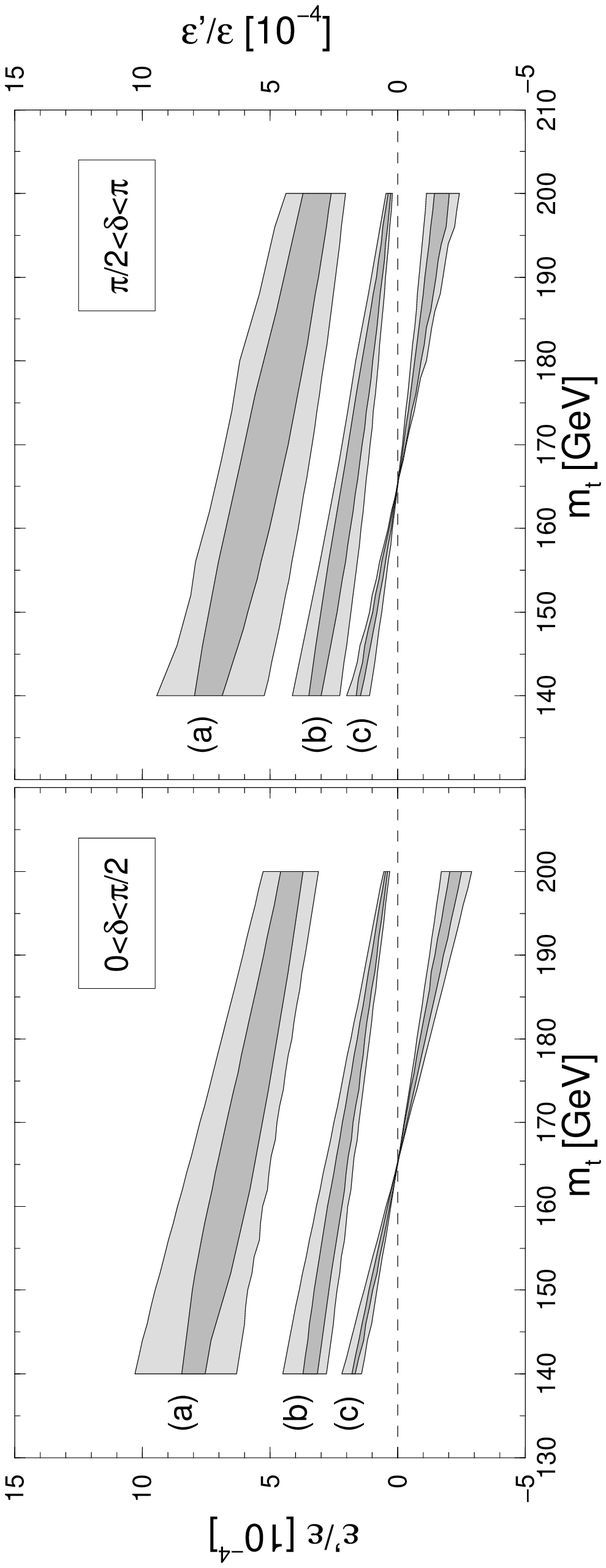}
}}
\vspace{0.15in}
\caption[]{
The ranges of $\epe$ in the NDR scheme as a function of $\mt$
for $\Lms^{(4)}=325\mev$ and present (light grey) and future (dark grey)
parameter ranges given in appendix \ref{app:numinput}. The three bands
correspond to hadronic parameter sets
(a) $(B_6^{(1/2)}(\mc))_{\rm eff}=1.5$, $(B_8^{(3/2)}(\mc))_{\rm eff}=1.0$,
(b) $(B_6^{(1/2)}(\mc))_{\rm eff}=1.0$, $(B_8^{(3/2)}(\mc))_{\rm eff}=1.0$,
and
(c) $(B_6^{(1/2)}(\mc))_{\rm eff}=1.0$, $(B_8^{(3/2)}(\mc))_{\rm eff}=1.5$,
respectively.
\label{fig:mteper1r2f}}
\end{figure}

Finally, fig.\ \ref{fig:mteper1r2f} shows the interrelated influence of
$\mt$ and the two most important hadronic matrix elements for penguin
operators on the theoretical prediction of $\epe$. For a dominant QCD
penguin matrix element $<Q_6>_0$ $\epe$ stays positive for all $\mt$
values considered. $\epe \approx 0$ becomes possible for equally
weighted matrix elements $<Q_6>_0$ and $<Q_8>_2$ around $\mt=205\gev$.
A dominant electroweak pengiun matrix element $<Q_8>_2$ shifts the
point $\epe \approx 0$ to $\mt \approx 165\gev$ and even allows for a
negative $\epe$ for higher values of $\mt$. The key issue to understand
this behaviour of $\epe$ is the observation that the $Q_6$ contribution
to $\epe$ is positive and only weakly $\mt$ dependent. On the other
hand the contribution coming from $Q_8$ is negative and shows a strong
$\mt$ dependence.

The results in fig.\ \ref{fig:lepemt170}--\ref{fig:mteper1r2f}
use only the $\eps_K$ constraint. In order to complete our analysis we
want to impose also the $x_d$-constraint and vary $m_s(m_c)$,
$B^{(1/2)}_6$ and $B^{(3/2)}_8$ in the full ranges given in
\eqn{eq:B1258} and \eqn{eq:msmc}.

This gives for the ``present'' scenario
\begin{equation}
-2.1 \cdot 10^{-4} \le \epe \le 13.2 \cdot 10^{-4}
\label{eq:eperangepresent}
\end{equation}
to be compared with
\begin{equation}
-1.1 \cdot 10^{-4} \le \epe \le 10.4 \cdot 10^{-4}
\label{eq:eperangefuture}
\end{equation}
in the case of the ``future'' scenario. In both cases the
$x_d$-constraint has essentially no impact on the predicted range for
$\epe$.
\\
Finally, extending the ``future'' scenario to $\ms(\mc)=(170 \pm
10)\mev$, $\Lms^{(4)}=(325 \pm 50)\mev$ and $B^{(1/2)}_6,
B^{(3/2)}_8=1.0 \pm 0.1$ would give
\begin{equation}
0.3 \cdot 10^{-4} \le \epe \le 5.4 \cdot 10^{-4}
\label{eq:eperangefuture2}
\end{equation}
again with no impact from imposing the $x_d$-constraint.

Allowing for the additional variation $B_{2,NDR}^{(1/2)}(\mc) = 6.6 \pm
1.0$ extends ranges \eqn{eq:eperangepresent}--\eqn{eq:eperangefuture2}
to $-2.5 \cdot 10^{-4} \le \epe \le 13.7 \cdot 10^{-4}$, $-1.5 \cdot
10^{-4} \le \epe \le 10.8 \cdot 10^{-4}$ and $0.1 \cdot 10^{-4} \le
\epe \le 5.8 \cdot 10^{-4}$, respectively.

An analysis of the Rome group \cite{ciuchini:95} gives
$\RE(\epe)=(3.1\pm 2.5)\cdot 10^{-4}$ which is compatible with our
results. Similar results are found with hadronic matrix elements
calculated in the chiral quark model \cite{bertolinietal:94},
\cite{bertolinietal:95}.

The difference in the range for $\epe$ presentend here by us and the
Rome group is related to the different treatment of theoretical and
experimental errors. Whereas we simply scan all parameters within one
standard deviation, \cite{ciuchini:95} use Gaussian distributions in
treating the experimental errors. Consequently our procedure is more
conservative. We agree however with these authors that values for
$\epe$ above $1 \cdot 10^{-3}$ although not excluded are very
improbable. This should be contrasted with the work of the Dortmund
group \cite{froehlich:91}, \cite{heinrichetal:92} which finds values
for $\epe$ in the ball park of $(2 - 3) \cdot 10^{-3}$. We do not know
any consistent framework for hadronic matrix elements which would give
such high values within the Standard Model.

The experimental situation on $\RE(\epe)$ is unclear at present.  While
the result of the NA31 collaboration at CERN with $\RE(\epe) = (23 \pm
7)\cdot 10^{-4}$ \cite{barr:93} clearly indicates direct CP violation,
the value of E731 at Fermilab, $\RE(\epe) = (7.4 \pm 5.9)\cdot 10^{-4}$
\cite{gibbons:93}, is compatible with superweak theories
\cite{wolfenstein:64} in which $\epe = 0$.  The E731 result is in the
ball park of the theoretical estimates.  The NA31 value appears a bit
high compared to the range given in \eqn{eq:eperangepresent} above.

Hopefully, in about three years the experimental situation concerning
$\epe$ will be clarified through the improved measurements by the two
collaborations at the $10^{-4}$ level and by experiments at the $\Phi$
factory in Frascati.  One should also hope that the theoretical
situation of $\epe$ will improve by then to confront the new data.

\section{$K_L-K_S$ Mass Difference and $\Delta I=1/2$ Rule}
         \label{sec:mki12}
It is probably a good moment to make a few comments on the $K_L-K_S$
mass difference given by \begin{equation}\label{delmk} \Delta
M=M(K_L)-M(K_S)=3.51\cdot 10^{-15}\gev \end{equation} and the
approximate $\Delta I=1/2$ rule in $K\to\pi\pi$ decays.  As we have
already briefly mentioned in the beginning of section
\ref{subsec:epeformulae}, this empirical rule manifests itself in the
dominance of $\Delta I=1/2$ over $\Delta I=3/2$ decay amplitudes.
It can be expressed as
\begin{equation}\label{deli12}
\frac{{\rm Re}A_0}{{\rm Re}A_2}=22.2
\end{equation}
using the notation of section \ref{subsec:epeformulae}.

\subsection{$\Delta M(K_L-K_S)$}
            \label{sec:mki12:mk}
The $K_L-K_S$ mass difference can be written as
\begin{equation}\label{2rem12}
\Delta M=2 {\rm Re} M_{12}+(\Delta M)_{LD}
\end{equation}
with $M_{12}$ given in \eqn{eq:M12K} and $(\Delta M)_{LD}$
representing long distance contributions, corresponding for instance to
the exchange of intermediate light pseudoscalar mesons ($\pi^0$,
$\eta$). The first term in \eqn{2rem12}, the so-called short distance
contribution, is dominated by the first term in \eqn{eq:M12K} so that
\begin{equation}\label{dmsd}
(\Delta M)_{SD}=\frac{G^2_F}{6\pi^2}F^2_K B_K m_K M^2_W
\left[ \lambda^2_c\eta_1\frac{m^2_c}{M^2_W}+\Delta_{top}\right]
\end{equation}
where $\Delta_{top}$ represents the two top dependent terms in
\eqn{eq:M12K}. In writing \eqn{dmsd} we are neglecting the tiny
imaginary part in $\lambda_c=V^*_{cs}V_{cd}$. A very extensive
numerical analysis of \eqn{dmsd} has been presented by
\cite{herrlichnierste:93}, who calculated the NLO corrections to
$\eta_1$ and also to $\eta_3$ \cite{herrlichnierste:95} which enters
$\Delta_{top}$.  The NLO calculation of the short distance
contributions improves the matching to the non-perturbative matrix
element parametrized by $B_K$ and clarifies the proper definition of
$B_K$ to be used along with the QCD factors $\eta_i$. In addition the
NLO study reveals an enhancement of $\eta_1$ over its LO estimate by
about 20\%. Although sizable, this enhancement can still be considered
being perturbative, as required by the consistency of the calculation.
This increase in $\eta_1$, reinforced by updates in input parameters
($\Lms$), brings $(\Delta M)_{SD}$ closer to the experimental value in
\eqn{delmk}.  With $\Lms^{(4)}=325\mev$ and $\mc=1.3\gev$, giving
$\eta^{NLO}_1=1.38$, one finds that typically $70\%$ of $\Delta M$ can
be described by the short distance component. The exact value is still
somewhat uncertain because $\eta_1$ is rather sensitive to $\Lms$.
Further uncertainties are introduced by the error in $B_K$ and due to
the renormalization scale ambiguity, which is still quite pronounced
even at NLO. Yet the result is certainly more reliable than previous LO
estimates.  Using the old value $\eta^{LO}_1=0.85$, corresponding to
$\mc=1.4\gev$ and $\Lambda_{QCD}=200\mev$, $(\Delta M)_{SD}/\Delta M$
would be below $50\%$, suggesting a dominance of long distance
contributions in $\Delta M$. As discussed in \cite{herrlichnierste:93},
such a situation would be "unnatural" since the long distance component
is formally suppressed by $\Lambda^2_{QCD}/m^2_c$. Hence the short
distance dominance indicated by the NLO analysis is also gratifying in
this respect.
\\
The long distance contributions, to which one can attribute the 
remaining $\sim 30\%$ in $\Delta M$ not explained by the short distance
part, are nicely discussed in \cite{bijnensetal:91}.
\\
In summary, the observed $K_L-K_S$ mass difference can be roughly
described within the standard model after the NLO corrections
have been taken into account. The remaining theoretical uncertainties
in the dominant part in \eqn{dmsd} and the uncertainties in
$(\Delta M)_{LD}$ do not allow however to use $\Delta M$ as a 
constraint on the CKM parameters.

\subsection{The $\Delta I=1/2$ Rule}
            \label{sec:mki12:i12}
Using the effective hamiltonian in \eqn{eq:HeffdF1:1010} and 
keeping only the dominant terms one has
\begin{equation}\label{rea0a2}
\frac{{\rm Re}A_0}{{\rm Re}A_2}\approx
\frac{z_1(\mu)\langle Q_1(\mu)\rangle_0+
      z_2(\mu)\langle Q_2(\mu)\rangle_0+z_6(\mu)\langle Q_6(\mu)\rangle_0} 
{z_1(\mu)\langle Q_1(\mu)\rangle_2+z_2(\mu)\langle Q_2(\mu)\rangle_2}
\end{equation}
where $\langle Q_i\rangle_{0,2}$ are defined in \eqn{eq:QiKpp}.  The
coefficients $z_i(\mu)$ can be found in table \ref{tab:wc10smu1}.  For
the hadronic matrix elements we use the formulae \eqn{eq:Q10},
\eqn{eq:Q20}, \eqn{eq:Q60} and \eqn{eq:Q122}, which have been
discussed in section \ref{subsec:matelKpp}.  We find then, separating
current-current and penguin contributions
\begin{equation}\label{rcrp}
\frac{{\rm Re}A_0}{{\rm Re}A_2}=R_c+R_p
\end{equation}
\begin{equation}\label{rczb}
R_c=\frac{5z_2(\mu) B^{(1/2)}_2-z_1(\mu) B^{(1/2)}_1}{4\sqrt{2}z_+(\mu)
  B^{(3/2)}_1} \qquad\qquad   z_+=z_1+z_2
\end{equation}
\begin{equation}\label{rpzb}
R_p=-11.9\frac{z_6(\mu)}{z_+(\mu)}\frac{B^{(1/2)}_6}{B^{(3/2)}_1}
\left[\frac{178\mev}{m_s(\mu)+m_d(\mu)}\right]^2
\end{equation}
The factor $11.9$ expresses the enhancement of the matrix elements of
the penguin operator $Q_6$ over $\langle Q_{1,2}\rangle$ first pointed
out in \cite{vainshtein:77}. It is instructive to calculate $R_c$ and
$R_p$ using the vacuum insertion estimate for which $B^{(1/2)}_1=$
$B^{(1/2)}_2=$ $B^{(3/2)}_1=$ $B^{(1/2)}_6=1$.  Without QCD effects one
finds then $R_c=0.9$ and $R_p=0$ in complete disagreement with the
data. In table \ref{tab:rcrp} we show the values of $R_c$ and $R_p$ at
$\mu=1\gev$ using the results of table \ref{tab:wc10smu1}. We have set
$m_s+m_d=178\mev$.

\begin{table}[htb]
\caption[]{The quantities $R_c$ and $R_p$ contributing to ${\rm
Re}A_0/{\rm Re}A_2$ as described in the text, calculated using the
vacuum  insertion estimate for the hadronic matrix elements. The Wilson
coefficient functions are evaluated for various $\Lms^{(4)}$in leading
logarithmic approximation as well as in next-to-leading order in two
different schemes (NDR and HV).
\label{tab:rcrp}}
\begin{center}
\begin{tabular}{|c|c|c|c|c|c|c|c|c|c|}
&\multicolumn{3}{c|}{$\Lms^{(4)}=215\mev$}
&\multicolumn{3}{c|}{$\Lms^{(4)}=325\mev$}
&\multicolumn{3}{c|}{$\Lms^{(4)}=435\mev$}\\
 \hline
Scheme&LO&NDR&HV&LO&NDR&HV&LO&NDR&HV\\ \hline
$R_c$&1.8&1.4&1.6&2.0&1.6&1.8&2.4&1.8&2.2\\
 \hline
$R_p$&0.1&0.3&0.1&0.2&0.5&0.2&0.3&1.0&0.4
\end{tabular}
\end{center}
\end{table}

The inclusion of QCD effects enhances both $R_c$ and $R_p$
\cite{gaillard:74}, \cite{altarelli:74}, however even for the highest
values of $\Lms^{(4)}$ the ratio ${\rm Re}A_0/{\rm Re}A_2$ is by at
least a factor of 8 smaller than the experimental value in
\eqn{deli12}. Moreover a considerable scheme dependence is observed.
Lowering $\mu$ would improve the situation, but for $\mu< 1\gev$ the
perturbative calculations of $z_i(\mu)$ can no longer be trusted.
Similarly lowering $m_s$ down to $100\mev$ would increase the penguin
contribution. In view of the most recent estimates in \eqn{eq:msmc}
such a low value of $m_s$ seems to be excluded however. We conclude
therefore, as already known since many years, that the vacuum insertion
estimate fails completely in explaining the $\Delta I=1/2$ rule. As we
have discussed in section \ref{sec:nloepe} the vacuum insertion
estimate $B^{(1/2)}_6=1$ is supported by the $1/N$ expansion approach
and by lattice calculations. Consequently the only solution to the
$\Delta I=1/2$ rule problem appears to be a change in the values of the
remaining $B_i$ factors. For instance repeating the above calculation
with $B^{(3/2)}_1=0.48$, $B^{(1/2)}_2=5$ and  $B^{(1/2)}_1=10$ would
give in the NDR scheme $R_c\approx 20$, $R_p\approx 2$ and ${\rm
Re}A_0/{\rm Re}A_2\approx 22$ in accordance with the experimental
value.

There have been several attempts to explain the $\Delta I=1/2$ rule,
which basically use the effective hamiltonian in
\eqn{eq:HeffdF1:1010} but employ different methods for the hadronic
matrix elements.  In particular we would like to mention the $1/N$
approach \cite{bardeen:87b}, the work of \cite{pichderafael:91} based
on an effective action for four-quark operators, the diquark
approach in \cite{neubertstech:91}, QCD sum rules \cite{jaminpich:94},
the chiral perturbation calculations in \cite{kamboretal:90},
\cite{kamboretal:91} and very recently an analysis \cite{antonellietal:95}
 in the framework of the chiral quark model \cite{cohenmanohar:84}.

With these methods values for ${\rm
Re}A_0/{\rm Re}A_2$ in the range 15--20 can be obtained. It is beyond
the scope of this review to discuss the weak and strong points of each
method, although at least one of us believes that the "meson evolution"
picture advocated in \cite{bardeen:87b} represents the main bulk of the
physics behind the number 22.  In view of the uncertainties present in
these approaches, we have not used them in our analysis of
$\varepsilon'/\varepsilon$, but have constrained the hadronic matrix
elements so that they satisfy the $\Delta I=1/2$ rule exactly.

\section{The Decay $K_L\to\pi^0 \lowercase{e}^+\lowercase{e}^-$}
\label{sec:KLpee}

\subsection{General Remarks}
\label{sec:KLpee:General}

Let us next move on to discuss the rare decay $K_L\to\pi^0e^+e^-$.
Whereas in $K\to\pi\pi$ decays the CP violating contribution is
only a tiny part of the full amplitude and the direct CP violation
as we have just seen is expected to be at least by three orders of
magnitude smaller than the indirect CP violation, the corresponding
hierarchies are very different for $K_L\to\pi^0e^+e^-$. At lowest
order in electroweak interactions (one-loop photon penguin,
$Z^0$-penguin and W-box diagrams), this decay takes place only if
CP symmetry is violated. The CP conserving contribution to the
amplitude comes from a two photon exchange, which although of higher
order in $\alpha$ could in principle be sizable. Extensive studies
of several groups indicate however that the CP conserving part is
likely to be smaller than the CP violating contributions. We will be
more specific about this at the end of this section.  \\ The CP
violating part can again be divided into a direct and an indirect one.
The latter is given by the $K_S\to\pi^0e^+e^-$ amplitude times the CP
violating parameter $\varepsilon_K$.  The amplitude
$A(K_S\to\pi^0e^+e^-)$ can be written as \begin{equation}\label{akspee}
A(K_S\to\pi^0e^+e^-)=\langle\pi^0e^+e^-|{\cal H}_{eff}| K_S\rangle
\end{equation} where ${\cal H}_{eff}$ can be found in
(\ref{eq:HeffKpe}) with the operators $Q_1, \ldots, Q_6$ defined in
(\ref{eq:Kppbasis}), the operators $Q_{7V}$ and $Q_{7A}$ given by
\begin{equation}\label{q7v7a} Q_{7V}=(\bar sd)_{V-A}(\bar ee)_V  \qquad
Q_{7A}=(\bar sd)_{V-A}(\bar ee)_A \end{equation} and the Wilson
coefficients $z_i$ and $y_i$ calculated in section \ref{sec:HeffKpe}.
\\ Let us next note that the coefficients of $Q_{7V}$ and $Q_{7A}$ are
$\ord(\alpha)$, but their matrix elements $\langle\pi^0e^+e^-|
Q_{7V,A}| K_S\rangle$ are $\ord(1)$. In the case of $Q_i$ ($i=1,\ldots,
6$) the situation is reversed: the Wilson coefficients are $\ord(1)$,
but the matrix elements $\langle\pi^0e^+e^-| Q_i| K_S\rangle$ are
$\ord(\alpha)$. Consequently at $\ord(\alpha)$ all operators contribute
to $A(K_S\to\pi^0e^+e^-)$. However because $K_S\to\pi^0e^+e^-$ is CP
conserving, the coefficients $y_i$ multiplied by $\tau=\ord(\lambda^4)$
can be fully neglected and the operator $Q_{7A}$ drops out in this
approximation.  Now whereas $\langle\pi^0e^+e^-| Q_{7V}| K_S\rangle$
can be trivially calculated, this is not the case for
$\langle\pi^0e^+e^-| Q_i| K_S\rangle$ with $i=1,\ldots, 6$ which can
only be evaluated using non-perturbative methods. Moreover it is clear
from the short-distance analysis of section \ref{sec:HeffKpe} that the
inclusion of $Q_i$ in the estimate of $A(K_S\to\pi^0e^+e^-)$ cannot be
avoided. Indeed, whereas $\langle\pi^0e^+e^-| Q_{7V}| K_S\rangle$ is
independent of $\mu$ and the renormalization scheme, the coefficient
$z_{7V}$ shows very strong scheme and $\mu$-dependences.  They can only
be canceled by the contributions from the four-quark operators $Q_i$.
All this demonstrates that the estimate of the indirect CP violation in
$K_L\to\pi^0e^+e^-$ cannot be done very reliably at present. Some
estimates in the framework of chiral perturbation theory will be discussed
below. On the other hand, a much better assessment of the importance of
indirect CP violation in $K_L\to\pi^0e^+e^-$ will become possible after
a measurement of $B(K_S\to\pi^0e^+e^-)$.  \\ Fortunately the directly
CP violating contribution can be fully calculated as a function of
$m_t$, CKM parameters and the QCD coupling constant $\as$. There are
practically no theoretical uncertainties related to hadronic matrix
elements because $\langle\pi^0|(\bar sd)_{V-A}| K_L\rangle$ can be
extracted using isospin symmetry from the well measured decay
$K^+\to\pi^0e^+\nu$. In what follows, we will concentrate on this
contribution.

\subsection{Analytic Formula for $B(K_L\to\pi^0e^+e^-)_{dir}$}
\label{sec:KLpee:Analytic}

The directly CP violating contribution is governed by the coefficients
$y_i$ and consequently only the penguin operators $Q_3,\ldots, Q_6$,
$Q_{7V}$ and $Q_{7A}$ have to be considered. Since
$y_i=\ord(\as)$ for $i=3,\ldots, 6$, the contribution of
QCD penguins to $B(K_L\to\pi^0e^+e^-)_{dir}$ is really
$\ord(\alpha\as)$ to be compared with the $\ord(\alpha)$
contributions of $Q_{7V}$ and $Q_{7A}$. In deriving the final
formula we will therefore neglect the contributions of the
operators $Q_3,\ldots, Q_6$, i.e. we will assume that
\begin{equation}\label{qillq7v1}
\sum_{i=3}^6 y_i(\mu)\langle\pi^0e^+e^-|Q_i|K_L\rangle\ll
y_{7V}(\mu)\langle\pi^0e^+e^-|Q_{7V}|K_L\rangle
\end{equation}
This assumption is supported by the corresponding relation for the
quark-level matrix elements
\begin{equation}\label{qillq7v2}
\sum_{i=3}^6 y_i(\mu)\langle d e^+e^-|Q_i| s \rangle\ll
y_{7V}(\mu)\langle d e^+e^-|Q_{7V}| s \rangle
\end{equation}
that can be easily verified perturbatively.
\\
The neglect of the QCD penguin operators is compatible with the scheme
and $\mu$-independence of the resulting branching ratio.  Indeed
$y_{7A}$ does not depend on $\mu$ and the renormalization scheme at all
and the corresponding dependences in $y_{7V}$ are at the level of $\pm
1\%$ as discussed in section \ref{sec:HeffKpe:numres}.  Introducing the
numerical constant
\begin{equation}
\label{kappae}
\kappa_e=\frac{1}{V_{us}^2}\frac{\tau(K_L)}{\tau(K^+)}
\left( \frac{\aem}{2 \pi} \right)^2 B(K^+\to\pi^0e^+\nu)
= 6.3 \cdot 10^{-6}
\end{equation}
one then finds
\begin{equation}\label{bklpedir}
B(K_L\to\pi^0e^+e^-)_{dir}=\kappa_e ({\rm Im}\lambda_t)^2
\left(\tilde{y}_{7V}^2 + \tilde{y}_{7A}^2\right)
\end{equation}
where
\begin{equation}
y_i = \frac{\aem}{2 \pi} \tilde{y_i} \, .
\label{eq:ytilde}
\end{equation}
Using next the method of the penguin-box expansion
(section \ref{sec:PBE}) we can write similarly to \eqn{C9tilde} and
\eqn{C10}
\begin{equation}\label{y7vpbe}
\tilde{y}_{7V} =
P_0 + \frac{Y_0(x_t)}{\sin^2\Theta_W} - 4 Z_0(x_t)+ P_E E_0(x_t)
\end{equation}
\begin{equation}\label{y7apbe}
\tilde{y}_{7A}=-\frac{1}{\sin^2\Theta_W} Y_0(x_t)
\end{equation}
with $Y_0$, $Z_0$ and $E_0$ given in (\ref{yy0}), (\ref{eq:XYZ}) and
(\ref{eq:Ext}).  $P_E$ is $\ord(10^{-2})$ and consequently the last
term in (\ref{y7vpbe}) can be neglected. $P_0$ is given for different
values of $\mu$ and $\Lambda_{\overline{MS}}$ in table
\ref{tab:P0klpee}.  There we also show the leading order results and
the case without QCD corrections.

\begin{table}[htb]
\caption[]{PBE coefficient $P_0$ of $y_{7V}$ for various values of
$\Lms^{(4)}$ and $\mu$. In the absence of QCD $P_0=8/9\;\ln(M_W/m_c)
= 3.664$ holds universally.
\label{tab:P0klpee}}
\begin{center}
\begin{tabular}{|c|c||c|c|c|}
\multicolumn{2}{|c||}{}     &
\multicolumn{3}{c|}{$P_0$} \\
\hline
$\Lms^{(4)}\,[\mev]$ & $\mu\,[\gev]$ & {\rm LO} & {\rm NDR} & {\rm HV} \\
\hline
    & 0.8 & 2.073 &  3.159 &  3.110 \\
215 & 1.0 & 2.048 &  3.133 &  3.084 \\
    & 1.2 & 2.027 &  3.112 &  3.063 \\
\hline
    & 0.8 & 1.863 &  3.080 &  3.024 \\
325 & 1.0 & 1.834 &  3.053 &  2.996 \\
    & 1.2 & 1.811 &  3.028 &  2.970 \\
\hline
    & 0.8 & 1.672 &  2.976 &  2.914 \\
435 & 1.0 & 1.640 &  2.965 &  2.899 \\
    & 1.2 & 1.613 &  2.939 &  2.872
\end{tabular}
\end{center}
\end{table}

The analytic expressions in (\ref{y7vpbe}) and (\ref{y7apbe}) are
useful as they display not only the explicit $m_t$-dependence, but also
isolate the impact of leading and next-to-leading QCD effects. These
effects modify only the constants $P_0$ and $P_E$. As anticipated from
the results of section \ref{sec:HeffKpe:numres}, $P_0$ is strongly
enhanced relatively to the LO result. This enhancement amounts roughly
to a factor of $1.6\pm 0.1$. Partially this enhancement is however due
to the fact that for $\Lambda_{LO}=\Lambda_{\overline{MS}}$ the QCD
coupling constant in the leading order is $20-30\%$ larger than its
next-to-leading order value. Calculating $P_0$ in LO but with the full
$\as$ of (\ref{amu}) we have found that the enhancement then amounts to
a factor of $1.33\pm 0.06$. In any case the inclusion of NLO QCD
effects and a meaningful use of $\Lambda_{\overline{MS}}$ show that the
next-to-leading order effects weaken the QCD suppression of $y_{7V}$.
As seen in table \ref{tab:P0klpee}, the suppression of $P_0$ by QCD
corrections amounts to about $15\%$ in the complete next-to-leading
order calculation.

\subsection{Numerical Analysis}
\label{sec:KLpee:Numerical}

In fig.\ \ref{fig:kpiee:mty7VA} of section \ref{sec:HeffKpe:numres} we
have shown $|y_{7V}/\alpha|^2$ and $|y_{7A}/\alpha|^2$ as functions of
$m_t$ together with the leading order result for $|y_{7V}/\alpha|^2$
and the case without QCD corrections. From there it is obvious that the
dominant $m_t$-dependence of $B(K_L\to\pi^0e^+e^-)_{dir}$ originates
from the coefficient of the operator $Q_{7A}$. Another noteworthy
feature was that accidentally for $m_t\approx 175\gev$ one finds
$y_{7V}\approx y_{7A}$.

\begin{figure}[hbt]
\vspace{0.10in}
\centerline{
\epsfysize=5in
\rotate[r]{
\epsffile{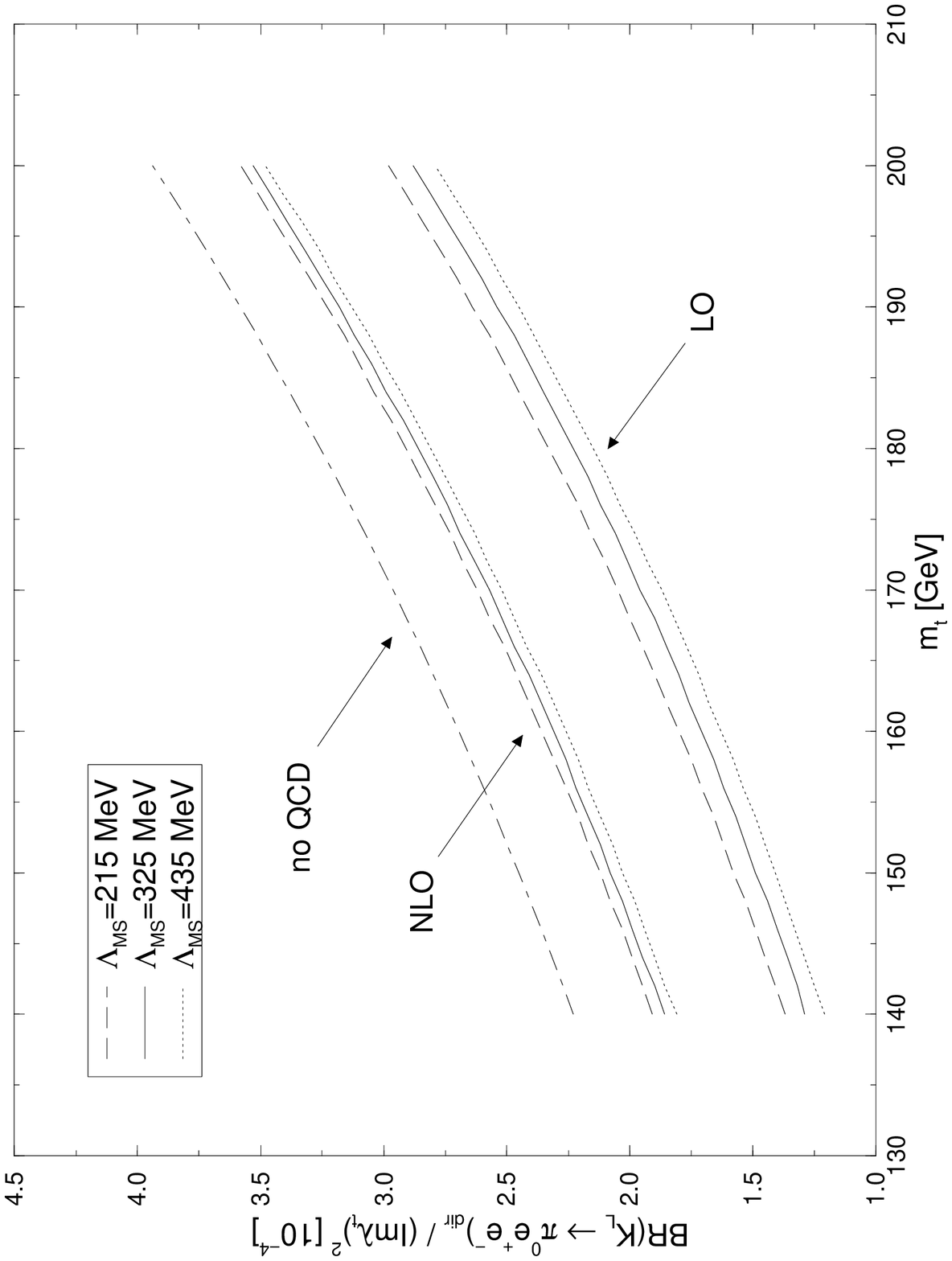}
}
}
\vspace{0.10in}
\caption[]{
$B(K_L \to \pi^0 e^+e^-)_{\rm dir}/(\IM\lambda_t)^2$
as a function of $\mt$ for various values of $\Lms^{(4)}$ at scale
$\mu =1.0\gev $.
\label{fig:mtbrimltAll}}
\end{figure}

In fig.\ \ref{fig:mtbrimltAll} the ratio
$B(K_L\to\pi^0e^+e^-)_{dir}/({\rm Im}\lambda_t)^2$
is shown as a function of $m_t$. The enhancement of the directly
CP violating contribution through NLO corrections relatively to the
LO estimate is clearly visible on this plot. As we will see below,
due to large uncertainties present in ${\rm Im}\lambda_t$ this
enhancement cannot yet be fully appreciated phenomenologically.
\\
The very weak dependence on $\Lambda_{\overline{MS}}$ should be
contrasted with the very strong dependence found in the case of
$\varepsilon^\prime/\varepsilon$. Therefore, provided the other two
contributions to $K_L\to\pi^0e^+e^-$ can be shown to be small or can
be reliably calculated one day, the measurement of
$B(K_L\to\pi^0e^+e^-)$ should offer a good determination of
${\rm Im}\lambda_t$.

Next we would like to comment on the possible uncertainties due to the
definition of $m_t$. At the level of accuracy at which we work we
cannot fully address this question yet. In order to be able to do it,
one needs to know the perturbative QCD corrections to $Y_0(x_t)$ and
$Z_0(x_t)$ and for consistency an additional order in the
renormalization group improved calculation of $P_0$.  Since the
$m_t$-dependence of $y_{7V}$ is rather moderate, the main concern in
this issue is the coefficient $y_{7A}$ whose $m_t$-dependence is fully
given by $Y(x_t)$. Fortunately the QCD corrected function $Y(x_t)$ is
known from the analysis of $K_L\to\mu^+\mu^-$ and can be directly used
here. As we will discuss in section \ref{sec:KLmm}, for $m_t=m_t(m_t)$
the QCD corrections to $Y_0(x_t)$ are around 2\%. On this basis we
believe that if $m_t=m_t(m_t)$ is chosen, the additional  QCD
corrections to $B(K_L\to\pi^0e^+e^-)_{dir}$ should be small.

Finally we give the predictions for the present and future sets of
input parameters as described in apppendix \ref{app:numinput}. It
should be  emphasized that the uncertainties in these predictions
result entirely from the CKM parameters. This situation will improve
considerably in the era of dedicated B-physics experiments in the
next decade, allowing a precise prediction for $B(K_L \to \pi^0
e^+e^-)_{\rm dir}$.

We find
\begin{equation}
B(K_L \to \pi^0 e^+ e^-)_{\rm dir}=
\left\{\begin{array}{ll}
(4.26 \pm 3.03) \cdot 10^{-12} & \mbox{no $x_d$ constraint} \\
(4.48 \pm 2.77) \cdot 10^{-12} & \mbox{with $x_d$ constraint} 
\end{array} \right.
\label{eq:BKLdirpresent}
\end{equation}
\begin{equation}
B(K_L \to \pi^0 e^+ e^-)_{\rm dir}=
\left\{\begin{array}{ll}
(3.71 \pm 1.61) \cdot 10^{-12} & \mbox{no $x_d$ constraint} \\
(4.32 \pm 0.96) \cdot 10^{-12} & \mbox{with $x_d$ constraint}
\end{array} \right.
\label{eq:BKLdirfuture}
\end{equation}
These results are compatible with those found in \cite{burasetal:94a},
\cite{donoghuegabbiani:95}, \cite{kohlerpaschos:95} with differences
originating in various choices of CKM parameters.

\subsection{The Indirectly CP Violating and CP Conserving Parts}
            \label{sec:KLpee:Comparison}
Now we want to compare the results obtained for the direct CP violating
part with the estimates made for the indirect CP-violating contribution
and the CP-conserving one. The most recent discussions have been
presented in \cite{cohenetal:93}, \cite{heiligerseghal:93},
\cite{donoghuegabbiani:95}, \cite{kohlerpaschos:95} where references to
earlier papers can be found.

The indirect CP violating amplitude is given by the
$K_S \to \pi^0 e^+ e^-$ amplitude times the CP parameter $\eps_K$.
Once $B(K_S \to \pi^0 e^+ e^-)$ has been accurately measured, it will
be possible to calculate this contribution precisely. Using chiral
perturbation theory it is however possible to get an estimate by
relating $K_S \to \pi^0 e^+ e^-$ to the $K^+ \to \pi^+ e^+ e^-$
transition \cite{eckeretal:87}, \cite{eckeretal:88}. To this end one
can write 

\begin{equation}
B(K_L \to \pi^0 e^+ e^-)_{indir}=B(K^+ \to \pi^+e^+e^-)
\frac{\tau(K_L)}{\tau(K^+)} |\eps_K|^2 r^2
\label{eq:BKLindir1}
\end{equation}
where
\begin{equation}
r^2=\frac{\Gamma(K_S \to \pi^0 e^+ e^-)}{\Gamma(K^+ \to \pi^+ e^+ e^-)}
\label{eq:r2}
\end{equation}
With $B(K^+ \to \pi^+e^+e^-)=(2.74\pm 0.23)\cdot 10^{-7}$
\cite{alliegro:92} and the most recent chiral perturbation theory
estimate $|r| \le 0.5 $ \cite{eckeretal:88}, \cite{brunoprades:93}
one has
\begin{equation}
B(K_L \to \pi^0 e^+ e^-)_{indir}=(5.9\pm 0.5)\cdot 10^{-12} r^2
 \le 1.6\cdot 10^{-12},
\label{eq:BKLindir2}
\end{equation}
i.e.\ a branching ratio more than a factor of 2 below the direct CP
violating contribution. \\
Yet as emphasized recently in \cite{donoghuegabbiani:95} and also in
\cite{heiligerseghal:93} the knowledge of $r$ is very uncertain at
present. In particular the estimate in \eqn{eq:BKLindir2} is based on
a relation between two non-perturbative parameters, which is rather
ad hoc and certainly not a consequence of chiral symmetry. As shown
in \cite{donoghuegabbiani:95} a small deviation from this relation
increases $r$ to values above unity so that $B(K_L \to \pi^0
e^+e^-)_{\rm indir}$ could be comparable or even large than 
$B(K_L \to \pi^0 e^+e^-)_{\rm dir}$. It appears then that this
enormous uncertainty in the indirectly CP violating part can only be 
removed by measuring the rate of $K_S \to \pi^0 e^+e^-$.

It should also be stressed, that in reality the CP indirect amplitude
may interfere with the vector part of the CP direct amplitude.  The
full CP violating amplitude can then be written following
\cite{dib1:89}, \cite{dib2:89} as follows
\begin{equation}
B(K_L \to \pi^0 e^+ e^-)_{CP}=| 2.43\cdot 10^{-6} r e^{i\pi/4}-
i\sqrt{\kappa_e} Im\lambda_t\tilde y_{7V}|^2+
\kappa_e (Im\lambda_t)^2\tilde y_{7A}^2
\label{eq:BKLCP}
\end{equation}

\begin{figure}[hbt]
\vspace{0.10in}
\centerline{
\epsfysize=4in
\rotate[r]{
\epsffile{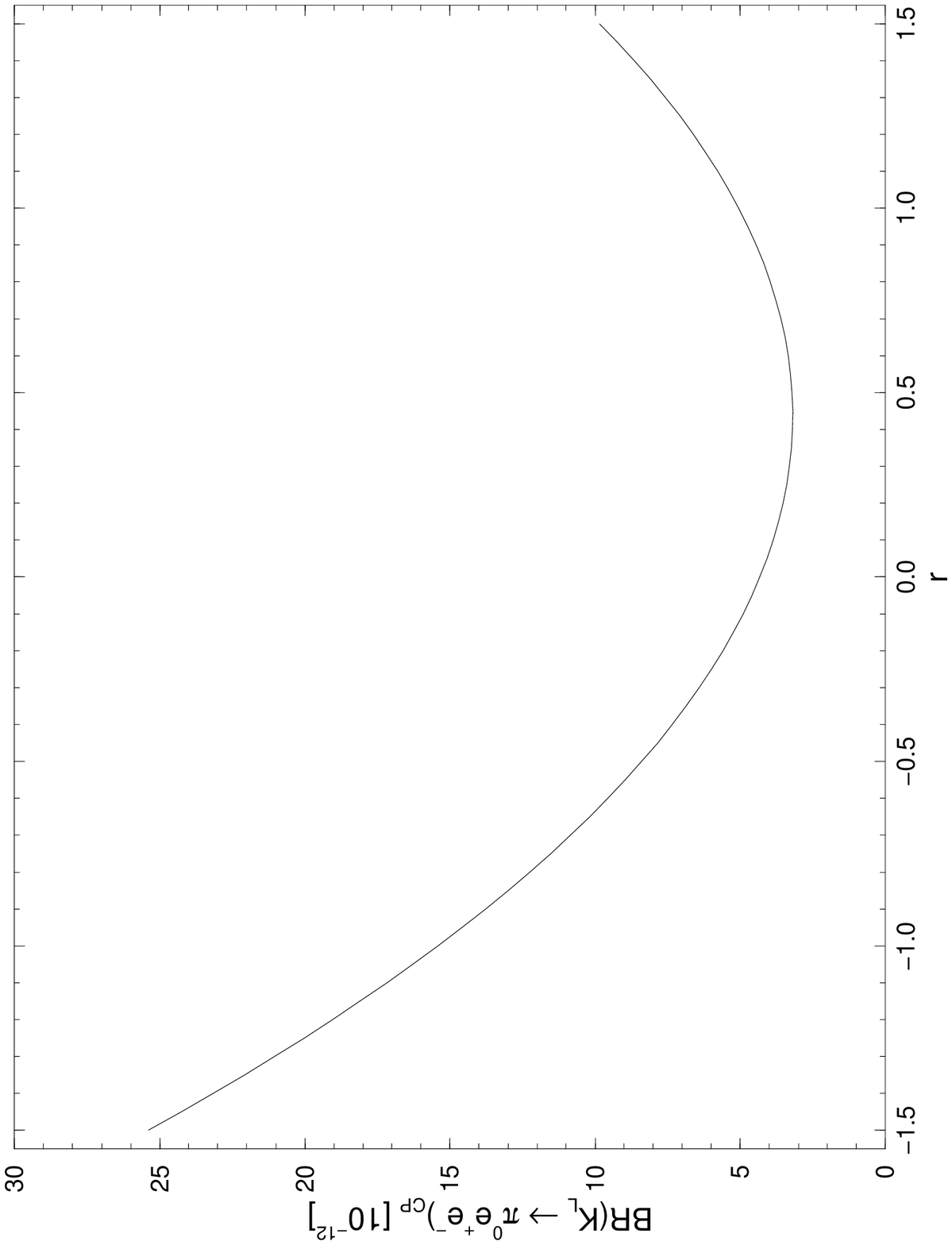}
}
}
\vspace{0.10in}
\caption[]{
$B(K_L \to \pi^0 e^+e^-)_{\rm CP}$ for $\mt=170\gev$,
$\Lms^{(4)}=325\mev$ and $\IM\lambda_t = 1.3 \cdot 10^{-4}$ as
a function of $r$.
\label{fig:BKLCP}}
\end{figure}

As an example we show in fig.\ \ref{fig:BKLCP} $B(K_L \to \pi^0
e^+e^-)_{\rm CP}$ for $\mt=170\gev$, $\Lms^{(4)}=325\mev$ and
$\IM\lambda_t = 1.3 \cdot 10^{-4}$ as a function of $r$.
We observe that whereas for $0 \le r \le 1$ the dependence of $B(K_L \to
\pi^0 e^+e^-)_{\rm CP}$ on $r$ is moderate, it is rather strong
otherwise and already for $r < -0.6$ values as high as $10^{-11}$ are
found.

The estimate of the CP conserving contribution is also difficult. We
refer the reader to \cite{cohenetal:93}, \cite{heiligerseghal:93} and
\cite{donoghuegabbiani:95} where further references to an extensive
literature on this subject can be found. The measurement of the
branching ratio
\begin{equation}
B(K_L \to \pi^0 \gamma\gamma) \leq
\left\{ \begin{array}{ll}
(1.7\pm 0.3) \cdot 10^{-6} & \hbox{\cite{barretal:92}} \\
(2.0\pm 1.0) \cdot 10^{-6} & \hbox{\cite{papadimitriou:91}}
\end{array} \right.
\label{eq:BKLexp}
\end{equation}
and of the shape of the $\gamma\gamma$ mass spectrum plays an important
role in this estimate. The most recent analyses give
\begin{equation}
B(K_L \to \pi^0 e^+ e^-)_{cons}\approx\left\{ \begin{array}{ll}
(0.3-1.8)\cdot 10^{-12} & \hbox{\cite{cohenetal:93}} \\
     4.0 \cdot 10^{-12} & \hbox{\cite{heiligerseghal:93}} \\
(5 \pm 5)\cdot 10^{-12} & \hbox{\cite{donoghuegabbiani:95}}
\end{array} \right.
\label{eq:BKLtheo}
\end{equation}
i.e.\ not necesarily below the CP violating contribution.
An improved estimate of this component is certainly desirable.
It should be noted that there is no interference in the rate between
the CP conserving and CP violating contributions so that the results
in fig.\ \ref{fig:BKLCP} and \eqn{eq:BKLtheo} can simply be added.

\subsection{Outlook} 
            \label{sec:KLpee:Outlook}
The results discussed above indicate that within the Standard Model
$B(K_L \to \pi^0 e^+ e^-)$ could be as high as $1\cdot 10^{-11}$.
Moreover the direct CP violating contribution is found to be important
and could even be dominant. Unfortunately the large uncertainties in
the remaining two contributions will probably not allow an easy
identification of the direct CP violation by measuring the branching
ratio only. The future measurements of $B(K_S \to \pi^0 e^+e^-)$ and
improvements in the estimate of the CP conserving part may of course
change this unsatisfactory situation. Alternatively the measurements
of the electron energy asymmetry \cite{heiligerseghal:93}, 
\cite{donoghuegabbiani:95} and the study of the time evolution of $K^0 \to
\pi^0 e^+e^-$ \cite{littenberg:89b}, \cite{donoghuegabbiani:95},
\cite{kohlerpaschos:95} could allow for a refined study of CP violation
in this decay.

The present experimental bounds
\begin{equation}
B(K_L \to \pi^0 e^+ e^-) \leq
\left\{ \begin{array}{ll}
4.3 \cdot 10^{-9} & \hbox{\cite{harrisetal:93}} \\
5.5 \cdot 10^{-9} & \hbox{\cite{ohletal:90}}
\end{array} \right.
\label{eq:BKLbounds}
\end{equation}
are still by three orders of magnitude away from the theoretical
expectations. Yet the prospects of getting the required sensitivity of
order $10^{-11}$--$10^{-12}$ in five years are encouraging
\cite{littenbergvalencia:93}, \cite{winsteinwolfenstein:93},
\cite{ritchiewojcicki:93}.

\section{The Decay $B \to X_{\lowercase{s}} \gamma$}
         \label{sec:Heff:Bsgamma}
\subsection{General Remarks}
         \label{sec:Heff:Bsgamma:rem}
The $B \to X_s \gamma$ decay is known to be extremely sensitive to the
structure of fundamental interactions at the electroweak scale. As any
FCNC process, it does not arise at the tree level in the Standard
Model. The one-loop W-exchange diagrams that generate this decay at the
lowest order in the Standard Model are small enough to be comparable to
possible nonstandard contributions (charged scalar exchanges, SUSY one
loop diagrams, $W_R$ exchanges in the L--R symmetric models, etc.).\\

The $B \to X_s \gamma$ decay is particularly interesting because its
rate is of order $G_F^2 \aem$, while most of the other FCNC processes
involving leptons or photons are of order $G_F^2 \aem^2$.  The
long-range strong interactions are expected to play a minor role in the
inclusive $\Bsg$ decay.  This is because the mass of the b-quark is
much larger than the QCD scale $\Lambda$. Moreover, the only relevant
intermediate hadronic states $\psi X_s$ are expected to give very small
contributions, as long as we assume no interference between short- and
long-distance terms in the inclusive rate.  Therefore, it has become
quite common to use the following approximate equality to estimate the
$\Bsg$ rate:

\begin{equation}\label{ratios}
\frac{\Gamma(B \to X_s \gamma)}
     {\Gamma(B \to X_c e \bar{\nu}_e)}
 \simeq                                                     
\frac{\Gamma(b \to s \gamma)}
     {\Gamma(b \to c e \bar{\nu}_e)} \equiv R(\mt,\as,\xi)
\end{equation}
where the quantities on the r.h.s are calculated in the spectator model
corrected for short-distance QCD effects. The normalization to the
semileptonic rate is usually introduced in order to cancel the
uncertainties due to the Cabibbo-Kobayashi-Maskawa (CKM) matrix
elements and factors of $\mb^5$ in the r.h.s. of eq. (\ref{ratios}).
Additional support for the approximation given above comes from the
heavy quark expansions.  Indeed the spectator model has been shown to
correspond to the leading order approximation of an expansion in
$1/\mb$.  The first corrections appear at the ${\cal O}(1/\mb^2)$
level. The latter terms have been studied by several authors
\cite{Chay}, \cite{Bj}, \cite{bigietal:92}, \cite{bigietal:93},
\cite{manoharwise:94}, \cite{bloketal:94}, \cite{falketal:94},
\cite{mannel:94}, \cite{Bigi} with the result that they affect
$B(\Bsg)$ and $B(B \to X_c e \bar{\nu}_e)$ by only a few percent.

As indicated above, the ratio $R$ depends only on $\mt$ and $\as$ in
the Standard Model. In extensions of the Standard Model, additional
parameters are present. They have been commonly denoted by $\xi$. The
main point to be stressed here is that $R$ is a calculable function of
its parameters in the framework of a renormalization group improved
perturbation theory. Consequently, the decay in question is
particularly suited for tests of the Standard Model and its extensions.

One of the main difficulties in analyzing the inclusive $\Bsg$ decay is
calculating the short-distance QCD effects due to hard gluon exchanges
between the quark lines of the leading one-loop electroweak diagrams.
These effects are known \cite{Bert}, \cite{Desh}, \cite{Grin},
\cite{grigjanis:88}, \cite{grigjanis:92}, \cite{misiak:91} to enhance
the $B \to X_s \gamma$ rate in the Standard Model by a factor of 2--3,
depending on the top quark mass. So the $B \to X_s \gamma$ decay
appears to be the only known short distance process in the Standard
Model that is dominated by two-loop contributions.

The $B \to X_s \gamma$ decay has already been measured.  In 1993
CLEO reported \cite{CLEO:93} the following branching ratio for the
exclusive $B \to K^* \gamma$ decay
\begin{equation}
B(B \to K^* \gamma) = (4.5 \pm 1.5 \pm 0.9) \times 10^{-5}.
\label{excl}
\end{equation}

In 1994 a first measurement of the inclusive rate has been
presented \cite{CLEO:94}
\begin{equation}
B(B \to X_s\gamma) = (2.32 \pm 0.57 \pm 0.35) \times 10^{-4}
\label{incl}
\end{equation}
where the first error is statistical and the second is systematic.

As we will see below
these experimental findings are in the ball park of the Standard Model
expectations based on the leading logarithmic approximation. 

In fact a complete leading order analysis of $B(B \to
X_s\gamma)$ in the Standard Model has been presented almost a year
before the CLEO result giving \cite{BMMP:94}
\begin{equation}
B(B \to X_s\gamma)_{TH} = (2.8 \pm 0.8) \times 10^{-4}.
\label{theo}
\end{equation}
where the error is dominated by the uncertainty in the choice of the
renormalization scale $\mb/2<\mu<2 \mb$ as first stressed by Ali and
Greub \cite{aligreub:93} and confirmed in \cite{BMMP:94}. Since $\Bsg$ is
dominated by QCD effects, it is not surprising that this
scale-uncertainty in the leading order is particularly large.  Such an
uncertainty, inherent in any finite order of perturbation theory can be
reduced by including next-to-leading order corrections.  Unfortunately,
it will take some time before the $\mu$-dependences present in $\Bsg$
can be reduced in the same manner as it was done for the other decays
\cite{burasjaminweisz:90}, \cite{buchallaburas:93b},
\cite{buchallaburas:94}, \cite{herrlichnierste:93}.  As we already
stated in section \ref{sec:Heff:BXsgamma:wc}, a full next-to-leading
order computation of $\Bsg$ would require calculation of three-loop
mixings between the operators $Q_1,\ldots,Q_6$ and the magnetic penguin
operators $Q_{7\gamma},Q_{8G}$.
Moreover, certain two-loop matrix elements of the relevant operators
should be calculated in the spectator model.  A formal analysis at the
next-to-leading level \cite{BMMP:94} is however very encouraging and
shows that the $\mu$-dependence can be considerably reduced once all
the necessary calculations have been performed.  We will return to this
issue below.

\subsection{The Decay $B\to X_s\gamma$ in the Leading Log Approximation}
         \label{sec:Heff:Bsgamma:lo}
The leading logarithmic calculations \cite{Grin}, \cite{misiak:93},
\cite{aligreub:93}, \cite{CFRS:94}, \cite{CCRV:94a}, \cite{misiak:94},
\cite{BMMP:94} can be summarized in a compact form, as follows:
\begin{equation}\label{main}
R = 
\frac{\Gamma(b \to s \gamma)}{\Gamma(b \to c e
\bar{\nu}_e)}
 =  \frac{|V_{ts}^* V_{tb}^{}|^2}{|V_{cb}|^2} 
\frac{6 \alpha}{\pi f(z)} |C^{(0)eff}_{7\gamma}(\mu)|^2
\end{equation}
where $C^{(0)eff}_{7\gamma}(\mu)$ is the effective coefficient 
given in (\ref{C7eff}) and table \ref{tab:c78effnum}, $z =
\frac{\mc}{\mb}$, and
\begin{equation}\label{g}
f(z) = 1 - 8z^2 + 8z^6 - z^8 - 24z^4 ln \; z           
\end{equation}
is the phase space factor in the semileptonic b-decay. Note, that at
this stage one should  not include the ${\cal O}(\as)$ corrections to
$\Gamma(b \to c e \bar{\nu})$ since they are part of the
next-to-leading effects. For the same reason we do not include the
${\cal O}(\as)$ QCD corrections to the matrix element of the operator
$Q_{7\gamma}$ (the QCD bremsstrahlung $b\to s\gamma+g$ and the virtual
corrections to $b\to s\gamma$) which are known \cite{aligreub:91a},
\cite{aligreub:91b}, \cite{pott:95} and will be a part of a future NLO
analysis.

Formula (\ref{main}) and the expression (\ref{C7eff}) for
$C^{(0)eff}_{7\gamma}(\mu)$ summarize the complete leading logarithmic
(LO) approximation for the $\Bsg$ rate in the Standard Model.  Their
important property is that they are exactly the same in many
interesting extensions of the Standard Model, such as the
Two-Higgs-Doublet Model (2HDM) \cite{Grin}, \cite{Hewett},
\cite{Barger}, \cite{Hayashi}, \cite{BMMP:94} or the Minimal
Supersymmetric Standard Model (MSSM) \cite{Borzum}, \cite{Barbieri},
\cite{Borzum2}.  The only quantities that change are the coefficients
$C^{(0)}_2(\mw)$, $C^{(0)}_{7\gamma}(\mw)$ and $C^{(0)}_{8G}(\mw)$ .
On the other hand in a general $SU(2)_L \times SU(2)_R \times U(1)$
model additional modifications are necessary, because new operators
enter \cite{LR}.

A critical analysis of theoretical and experimental
uncertainties present in the prediction for $B(\Bsg)$ based on the
above formulae has been made \cite{BMMP:94}. Here we just briefly 
list the main findings:

\begin{itemize}
\item
First of all, eq.\ \eqn{main} is based on the spectator model. As we
have mentioned above the heavy quark expansion gives a strong support
for this model in inclusive B-decays.  On a conservative side one can
assume the error due to the use of the spectator model in $\Bsg$ to
amount to at most $\pm 10\%$.
\item
The uncertainty coming from the ratio $z = \frac{\mc}{\mb}$ in the
phase-space factor $f(z)$ for the semileptonic decay is estimated to be
around 6\%.
\item
The error due to the ratio of the CKM parameters in
eq. (\ref{main}) is small. Assuming unitarity of the $3 \times 3$ CKM
matrix and imposing the constraints from the CP-violating parameter
$\epsilon_{\rm K}$ and $B^0-\bar B^0$ mixing one finds
\begin{equation}\label{KM}
\frac{|V_{ts}^* V_{tb}^{}|^2}{|V_{cb}|^2} = 0.95 \pm 0.03     
\end{equation}
\item
There exists an uncertainty due to the determination of $\as$. This
uncertainty is not small because of the importance of QCD corrections
in the considered decay.  For instance the difference between the
ratios $R$ of eq.  (\ref{main}) obtained with help of
$\alpha_{\overline{MS}}(\mz)=0.11$ and $0.13$, respectively, is roughly
20\%.
\item
The dominant uncertainty in eq. (\ref{main}) comes from the
unknown next-to-leading order contributions. This uncertainty is best
signaled by the strong $\mu$-dependence of the leading order
expression (\ref{main}), which is shown by the solid line in
fig.\ \ref{fig:bsg:rmu}, for the case $\mt=170\gev$. 

\begin{figure}[hbt]
\vspace{0.10in}
\centerline{
\epsfysize=5in
\rotate[r]{
\epsffile{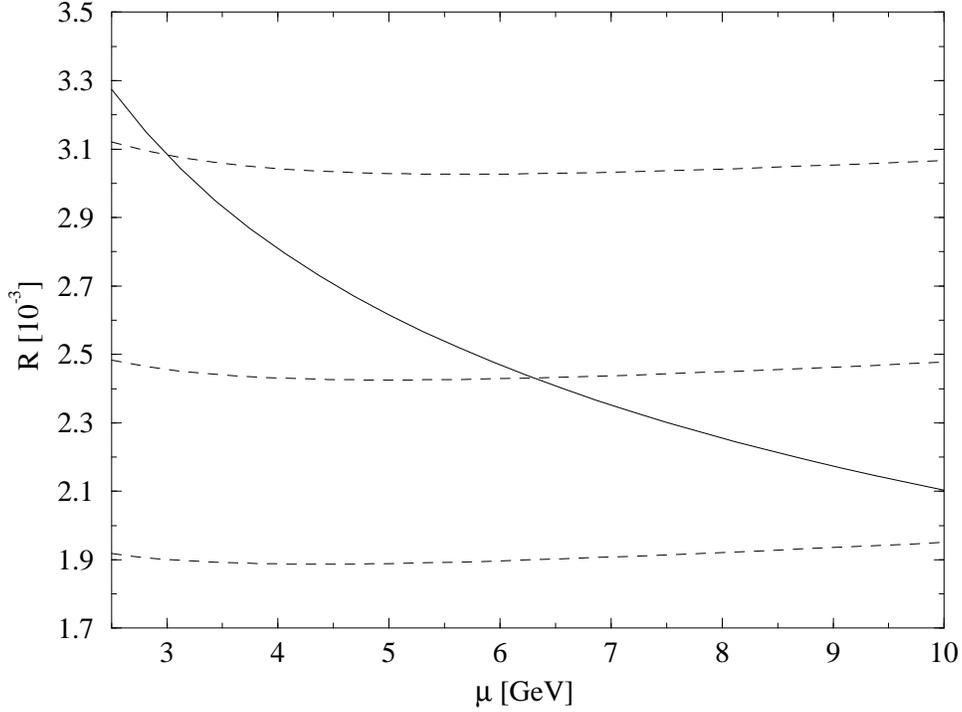}
}}
\vspace{0.08in}
\caption[]{
$\mu$-dependence of the theoretical prediction for the ratio $R$ for
$\mt=170\gev$ and $\Lms^{(5)}=225\mev$. The solid line corresponds to the
leading order prediction. The dashed lines describe possible
next-to-leading results.
\label{fig:bsg:rmu}}
\end{figure}

One can see that when $\mu$ is varied by a factor of 2 in both
directions around $\mb \simeq 5\gev$, the ratio (\ref{main}) changes
by around $\pm 25\%$, i.e. the ratios $R$ obtained for $\mu=2.5\gev$
and $\mu=10\gev$ differ by a factor of 1.6 \cite{aligreub:93}. 

The dashed lines in fig.\ \ref{fig:bsg:rmu} show the expected
$\mu$-dependence of the ratio (\ref{main}) once a complete
next-to-leading calculation is performed. The $\mu$-dependence is then
much weaker, but until one performs the calculation explicitly one
cannot say which of the dashed curves is the proper one. The way the
dashed lines are obtained is described in \cite{BMMP:94}.
\item	
Finally, there exists a $\pm 2.4\%$ error in determining $B(B \to
X_s \gamma)$ from eq. (\ref{ratios}), which is due to the error in the
experimental measurement of $B(B \to X_c e \bar{\nu}_e) = (10.43 \pm
0.24)\%$ \cite{particledata:94}.
\item
The uncertainty due to the value of $\mt$ is small as is shown
explicitly below.
\end{itemize}

\begin{figure}[hbt]
\vspace{0.10in}
\centerline{
\epsfysize=5in
\rotate[r]{
\epsffile{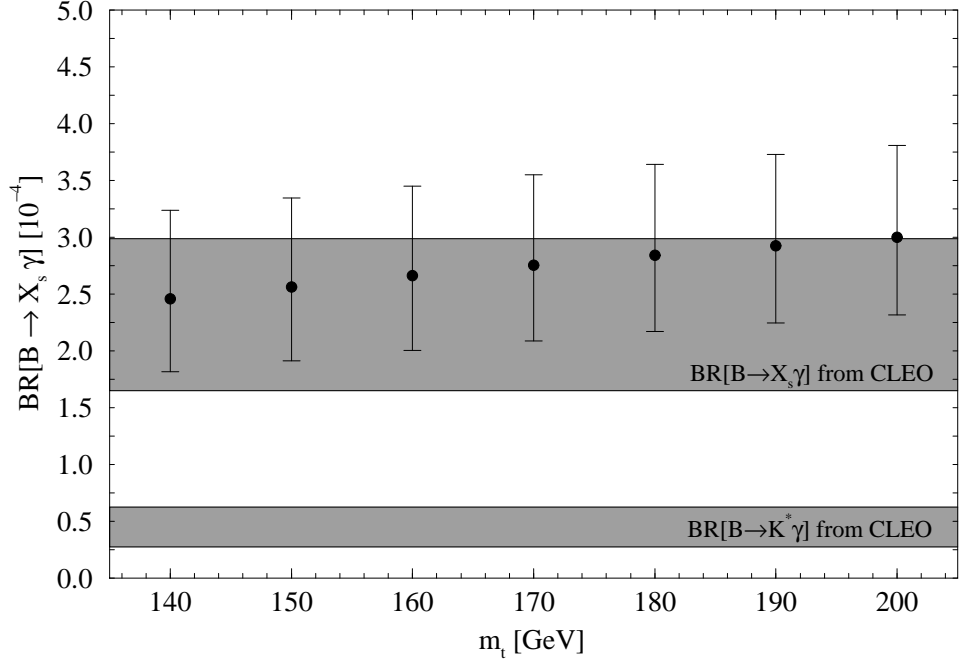}
}}
\vspace{0.08in}
\caption[]{
Predictions for $B \to X_s \gamma$ in the SM as a function of the top
quark mass with the theoretical uncertainties taken into account.
\label{fig:bsg:brmt}}
\end{figure}

Fig.\ \ref{fig:bsg:brmt} based on \cite{BMMP:94} presents the Standard
Model prediction for the inclusive $\Bsg$ branching ratio including the
errors listed above as a function of $\mt$ together with the CLEO
result.

We stress that the theoretical curves have been obtained prior to the
experimental result. Since the theoretical error is dominated by scale
ambiguities a complete NLO analysis is very desirable.

\subsection { Looking at $\Bsg$ Beyond Leading Logarithms}
         \label{sec:Heff:Bsgamma:nlo}
In this section we describe briefly a complete next-to-leading
calculation of $\Bsg$ in general terms. 
This section collects the most important findings of section 4 of
\cite{BMMP:94}.

Let us first enumerate what has been already calculated in the
literature and which calculations are still required in order to
complete the next-to-leading calculation of $B(B \to X_s \gamma)$.

\noindent
The present status is as follows:
\begin{itemize}
\item
The $6 \times 6$ submatrix of
$\gamma^{(1)}$ describing the two-loop mixing of $(Q_1,\ldots,
Q_6)$ and the corresponding ${\cal{O}}(\as)$ corrections in
$\vec{C}(\mw)$ have been already calculated. They are given in section
\ref{sec:HeffdF1:66}.
\item
The two-loop mixing in the $(Q_{7\gamma},Q_{8G})$ sector of
$\gamma^{(1)}$ is known \cite{misiakmuenz:95} and given in
section \ref{sec:Heff:BXsgamma:RGE}.
\item
The ${\cal{O}}(\as)$ corrections to the matrix element
of the operators $Q_{7\gamma}$ and $Q_{8G}$ have 
been calculated \cite{aligreub:91a}, \cite{aligreub:91b}.
They have been recently confirmed by \cite{pott:95} who also presents
the results for the matrix elements of the remaining operators.
\end{itemize}
The remaining ingredients of a next-to-leading analysis of
$B(B \to X_s \gamma)$ are:
\begin{itemize}
\item
The three-loop mixing between the sectors $(Q_1,\ldots,Q_6)$ and
$(Q_{7\gamma},Q_{8G})$ which, with our normalizations, contributes to
$\gamma^{(1)}$.\\
\item
The ${\cal O}(\as)$ corrections to $C_{7\gamma}(\mw)$ and $C_{8G}(\mw)$
in (\ref{c7}) and (\ref{c8}). This requires evaluation of two-loop
penguin diagrams with internal W and top quark masses and a proper
matching with the effective five-quark theory. An attempt to calculate
the necessary two-loop Standard Model diagrams has been made in \cite{Yao2}.
\item
The finite parts of the effective theory two-loop diagrams with the
insertions of the four-quark operators .
\end{itemize}
All these calculations are very involved, and the necessary three-loop
calculation is a truly formidable task! Yet, as stressed in
\cite{BMMP:94} all these calculations have to be done if we want to
reduce the theoretical uncertainties in $\bsg$ to around 10\%.

As demonstrated formally in \cite{BMMP:94} the cancellation of the
dominant $\mu$-dependence in the leading order can be achieved by
calculating the relevant two-loop matrix element of the dominant
four-quark operator $Q_2$.  This matrix element is however
renormalization-scheme dependent and moreover mixing with other
operators takes place.  This scheme dependence can only be canceled by
calculating $\gamma^{(1)}$ in the same renormalization scheme.
This point has been extensively discussed in this review
and  we will not repeat this discussion here. However,
it is clear from these remarks, that in order to address the
$\mu$-dependence and the renormalization-scheme dependence as well as
their cancellations, it is necessary to perform a complete
next-to-leading order analysis of $\vec{C}(\mu)$ and of the corresponding
matrix elements.

In this context we would like to comment on an analysis of
\cite{Ciu:94} in which the known two-loop mixing of $Q_1,\ldots,Q_6$
has been added to the leading order analysis of $\Bsg$.  Strong
renormalization scheme dependence of the resulting branching ratio has
been found, giving the branching ratio $(1.7 \pm 0.2)\times 10^{-4}$
and $(2.3 \pm 0.2)\times 10^{-4}$ at $\mu=5\gev$ for HV and NDR
schemes, respectively. It has also been observed that whereas in the HV
scheme the $\mu$ dependence has been weakened, it remained still strong
in the NDR scheme. In our opinion this partial cancellation of the
$\mu$-dependence in the HV scheme is rather accidental and has nothing
to do with the cancellation of the $\mu$-dependence discussed above. The
latter requires the evaluation of finite parts in two-loop matrix
elements of the four-quark operators $Q_1,\ldots,Q_6$.  On the other
hand the strong scheme dependence in the partial NLO analysis presented
in \cite{Ciu:94} demonstrates very clearly the need for a full
analysis.  In view of this discussion we think that the decrease of the
branching ratio for $\Bsg$ relative to the LO prediction, found in
\cite{Ciu:94}, and given by $B(B \to X_s \gamma) = (1.9 \pm 0.2 \pm
0.5) \cdot 10^{-4}$ is still premature and one should wait until the
full NLO analysis has been done.

\section{The Decay $B\to X_{\lowercase{s}} \lowercase{e}^+\lowercase{e}^-$}
         \label{sec:Heff:BXsee:nlo}
\subsection{General Remarks} 
         \label{sec:Heff:BXsee:nlo:rem}
The rare decay $\Bsee$ has been the subject of many theoretical studies
in the framework of the Standard Model and its extensions such as the
two Higgs doublet models and models involving supersymmetry
\cite{HWS:87}, \cite{grinstein:89a}, \cite{JW:90}, \cite{BBMR:91},
\cite{AMM:91}, \cite{DPT:93}, \cite{AGM:94}, \cite{GIW:94}.  In
particular the strong dependence of $\Bsee$ on $\mt$ has been stressed
in \cite{HWS:87}. It is clear that once $\Bsee$ has been observed, it
will offer a useful test of the Standard Model and its extensions.
To this end the relevant branching ratio, the dilepton invariant mass
distribution and other distributions of interest should be calculated
with sufficient precision. In particular the QCD effects should be
properly taken into account.

The central element in any analysis of $\Bsee$ is the effective
hamiltonian for this decay given in section \ref{sec:Heff:BXsee} where a
detailed analysis of the Wilson coefficients has been presented.
However, the actual calculation of $\Bsee$ involves not only the
evaluation of Wilson coefficients of the relevant local operators but
also the calculation of the corresponding matrix elements of these
operators relevant for $\Bsee$. The latter part of the analysis can be
done in the spectator model, which, as indicated by the heavy quark
expansion should offer a good approximation to QCD for B-decays. One can
also include the non-perturbative ${\cal O}(1/\mb^2)$ corrections to
the spectator model which enhance the rate for $\Bsee$ by roughly 10\%
\cite{falketal:94}. A realistic phenomenological analysis should also
include the long-distance contributions which are mainly due to the
$J/\psi$ and $\psi'$ resonances \cite{LMS:89}, \cite{DTP:89},
\cite{DT:91}. Since in this review we are mainly interested in the
next-to-leading short-distance QCD effects we will not include these
complications in what follows. This section closely follows
\cite{burasmuenz:95} execpt that the numerical results in figs.\
\ref{fig:bsee:rs}--\ref{fig:bsee:rs2} have been slightly changed in
accordance with the input parameters of appendix \ref{app:numinput}.

We stress again that in a consistent NLO analysis of the decay $\Bsee$,
one should on one hand calculate the Wilson coefficient of the operator
$Q_{9V} = (\bar s b)_{V-A} (\bar e e)_V$ including leading and
next-to-leading logarithms, but on the other hand only leading
logarithms should be kept in the remaining Wilson coefficients. Only
then a scheme independent amplitude can be obtained. As already
discussed in section \ref{sec:Heff:BXsee}, this special treatment of
$Q_9$ is related to the fact that strictly speaking in the leading
logarithmic approximation only this operator contributes to $\Bsee$.
The contributions of the usual current-current operators, QCD penguin
operators, magnetic penguin operators and of $Q_{10A} = (\bar s
b)_{V-A}(\bar e e)_A$ enter only at the NLO level and to be consistent
only the leading contributions to the corresponding Wilson coefficients
should be included.

\subsection{The Differential Decay Rate}
         \label{sec:Heff:BXsee:nlo:rate}
Introducing
\begin{equation} \label{invleptmass}
\hat s = \frac{(p_{e^+} + p_{e^-})^2}{\mb^2}, \qquad z =
\frac{\mc}{\mb}
\end{equation}
and calculating the one-loop matrix elements of $Q_i$ using the
spectator model in the NDR scheme one finds \cite{misiak:94},
\cite{burasmuenz:95}
\begin{eqnarray} \label{rate}
& &
R(\hat s) \equiv \frac{{d}/{d\hat s} \, \Gamma (\bsee)}{\Gamma
(\bcenu)} = \frac{\aem^2}{4\pi^2}
\left|\frac{V_{ts}}{V_{cb}}\right|^2 \frac{(1-\hat s)^2}{f(z)\kappa(z)}
\times \\ 
& &
\biggl[(1+2\hat s)\left(|\Ctilde_9^{eff}|^2 + |\Ctilde_{10}|^2\right) + 
4 \left( 1 + \frac{2}{\hat s}\right) |C_{7\gamma}^{(0)eff}|^2 + 12
C_{7\gamma}^{(0)eff} \ \RE\,\Ctilde_9^{eff}  \biggr]
\nn
\end{eqnarray}
where
\begin{eqnarray} \label{C9eff}
\Ctilde_9^{eff} & = & \Ctilde_9^{NDR} \tilde\eta(\hat s) + h(z, \hat
s)\left( 3 C_1^{(0)} + C_2^{(0)} + 3 C_3^{(0)} + C_4^{(0)} + 3
C_5^{(0)} + C_6^{(0)} \right) - \nn \\
& & \frac{1}{2} h(1, \hat s) \left( 4 C_3^{(0)} + 4 C_4^{(0)} + 3
C_5^{(0)} + C_6^{(0)} \right) - \\
& & \frac{1}{2} h(0, \hat s) \left( C_3^{(0)} + 3 C_4^{(0)} \right) +
\frac{2}{9} \left( 3 C_3^{(0)} + C_4^{(0)} + 3 C_5^{(0)} + C_6^{(0)}
\right) \, .
\nn
\end{eqnarray}
The general expression (\ref{rate}) with $\kappa(z)=1$ has been first
presented by \cite{grinstein:89a} who in their approximate leading
order renormalization group analysis kept only the operators $Q_1, Q_2,
Q_{7\gamma},Q_{9V}, Q_{10A}$.

The various entries in (\ref{rate}) are given as follows
\begin{eqnarray} \label{phasespace}
h(z, \hat s) & = & -\frac{8}{9}\ln\frac{\mb}{\mu} - \frac{8}{9}\ln z +
\frac{8}{27} + \frac{4}{9} x - \\
& & \frac{2}{9} (2+x) |1-x|^{1/2} \left\{
\begin{array}{ll}
\left( \ln\left| \frac{\sqrt{1-x} + 1}{\sqrt{1-x} - 1}\right| - i\pi \right),
 & \mbox{for } x \equiv 4 z^2/\hat s < 1 \nn \\
2 \arctan \frac{1}{\sqrt{x-1}}, & \mbox{for } x \equiv 4 z^2/\hat s
 > 1,
\end{array}
\right. \\
h(0, \hat s) & = &
\frac{8}{27} -\frac{8}{9} \ln\frac{\mb}{\mu} - \frac{4}{9} \ln
\hat s + \frac{4}{9} i\pi. \\ 
f(z) & = & 1 - 8 z^2 + 8 z^6 - z^8 - 24 z^4 \ln z, \\
\kappa(z)  & = & 1 - \frac{2 \as(\mu)}{3\pi}\left[(\pi^2 -
\frac{31}{4})(1-z)^2 + \frac{3}{2} \right]  \label{eq:kappaz} \\
\tilde\eta(\hat s) & = & 1 + \frac{\as(\mu)}{\pi}\, \omega(\hat s)
\end{eqnarray}
with
\begin{eqnarray} \label{omega}
\omega(\hat s) & = & - \frac{2}{9} \pi^2 - \frac{4}{3}\mbox{Li}_2(s) -
\frac{2}{3}
\ln s \ln(1-s) - \frac{5+4s}{3(1+2s)}\ln(1-s) - \nn \\
& &  \frac{2 s (1+s) (1-2s)}{3(1-s)^2 (1+2s)} \ln s + \frac{5+9s-6s^2}{6
(1-s) (1+2s)}.
\end{eqnarray}
Here $f(z)$ is the phase-space factor for $b \to c e \bar\nu$.
$\kappa(z)$ is the corresponding single
gluon QCD correction \cite{CM:78} in the approximation of \cite{KM:89}.
$\tilde\eta$ on the other hand represents single gluon
corrections to the matrix element of $Q_9$ with $\ms = 0$ \cite{JK:89},
\cite{misiak:94}. For consistency reasons this correction should only
multiply the leading logarithmic term in $\tilde{C}_9^{\rm NDR}$.

In the HV scheme the one-loop matrix elements are different and one
finds an additional explicit contribution to (\ref{C9eff}) given by
\cite{burasmuenz:95}
\begin{equation} \label{MEHV}
- \xi^{HV} \frac{4}{9} \left( 3 C_1^{(0)} + C_2^{(0)} - C_3^{(0)} - 3
C_4^{(0)} \right).
\end{equation}
However $\Ctilde_9^{NDR}$ has to be replaced by $\Ctilde_9^{HV}$ given in
(\ref{C9tilde}) and (\ref{P0HV}) and consequently $\Ctilde_9^{eff}$ is the
same in both schemes.

The first term in the function $h(z, \hat s)$ in (\ref{phasespace})
represents the leading $\mu$-dependence in the matrix elements. It is
canceled by the $\mu$-dependence present in the leading logarithm in
$\tilde C_{9}$. This is precisely the type of cancellation of the
$\mu$-dependence which one would like to achieve in the case of $B \to
X_s \gamma$. The $\mu$-dependence present in the coefficients of
the other operators can only be canceled by going to still higher order
in the renormalization group improved perturbation theory. To this end
the matrix elements of four-quark operators should be evaluated at
two-loop level. Also certain unknown three-loop anomalous dimensions
should be included in the evaluation of $C_{7\gamma}^{eff}$ and
$C_{9V}$. Certainly this is beyond the scope of this review and we will
only investigate the left-over $\mu$-dependence below.

\subsection{Numerical Analysis}
         \label{sec:Heff:BXsee:nlo:num}
A detailed numerical analysis of the formulae above has been presented
in \cite{burasmuenz:95}. We give here a brief account of this work.  We
set first $|V_{ts}/V_{cb}|  = 1$ which in view of (\ref{KM}) is a good
approximation.  We keep in mind that for $\hat s \approx m_\psi^2 /
\mb^2$, $\hat s \approx m_{\psi'}^2 / \mb^2$ etc.~the spectator model
cannot be the full story and additional long-distance contributions
discussed in \cite{LMS:89}, \cite{DTP:89}, \cite{DT:91} have to be
taken into account in a phenomenological analysis. Similarly we do not
include $1/\mb^2$ corrections calculated in \cite{falketal:94} which
typically enhance the differential rate by about 10\%.

\begin{figure}[hbt]
\vspace{0.10in}
\centerline{
\epsfysize=7in
\rotate[r]{
\epsffile{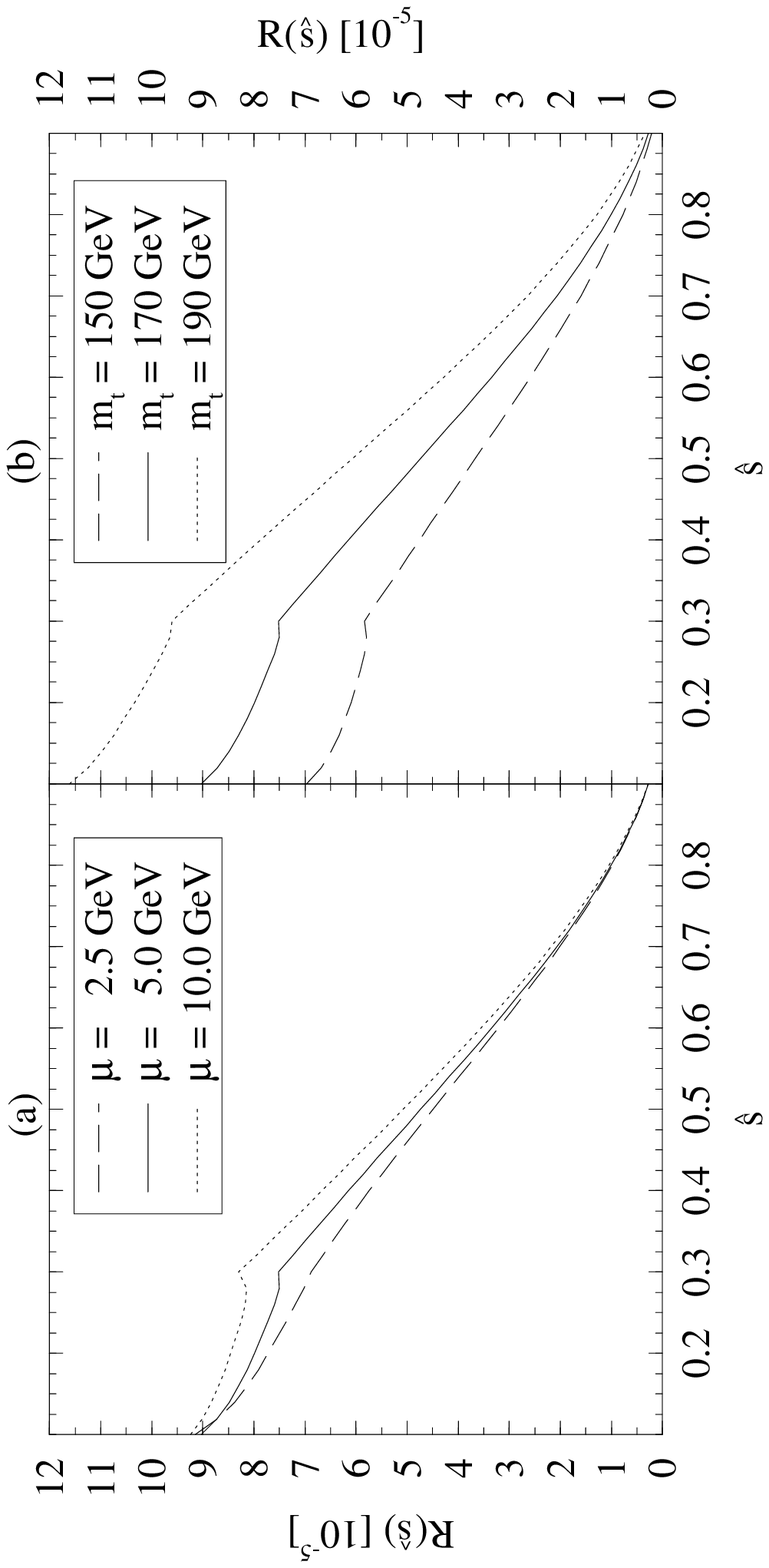}
}}
\vspace{0.08in}
\caption[]{
(a) $R(\hat{s})$ for $\mt=170\gev$, $\Lms^{(5)}=225\mev$ and differents
values of $\mu$. 
\\
\phantom{xxxxxxxxxx}
(b) $R(\hat{s})$ for $\mu=5\gev$, $\Lms^{(5)}=225\mev$ and various
values of $\mt$. 
\label{fig:bsee:rs}}
\end{figure}

In fig.\ \ref{fig:bsee:rs}\,(a) we show $R(\hat s)$ for $\mt = 170 \gev$,
$\Lms = 225 \mev$ and different values of $\mu$. In fig.\
\ref{fig:bsee:rs}\,(b) we set $\mu = 5 \gev$ and vary $\mt$ from $150 \gev$
to $190 \gev$. The remaining $\mu$ dependence is rather weak and
amounts to at most $\pm 8\%$ in the full range of parameters
considered. The $\mt$ dependence of $R(\hat s)$ is sizeable. Varying
$\mt$ between $150\gev$ and $190\gev$ changes $R(\hat s)$ by typically
60--65\% which in this range of $\mt$ corresponds to $R(\hat s) \sim
\mt^2$. It is easy to verify that this strong $\mt$ dependence
originates in the coefficient $\Ctilde_{10}$ given in (\ref{C10}) as
already stressed by several authors in the past \cite{HWS:87},
\cite{grinstein:89a}, \cite{BBMR:91}, \cite{DPT:93}, \cite{GIW:94},
\cite{AGM:94}, \cite{AMM:91}, \cite{JW:90}.

We do not show the $\Lms$ dependence as it is very weak. Typically,
changing $\Lms$ from $140\mev$ to $310\mev$ decreases $R(\hat s)$ by
about 5\%.

\begin{figure}[hbt]
\vspace{0.10in}
\centerline{
\epsfysize=4in
\rotate[r]{
\epsffile{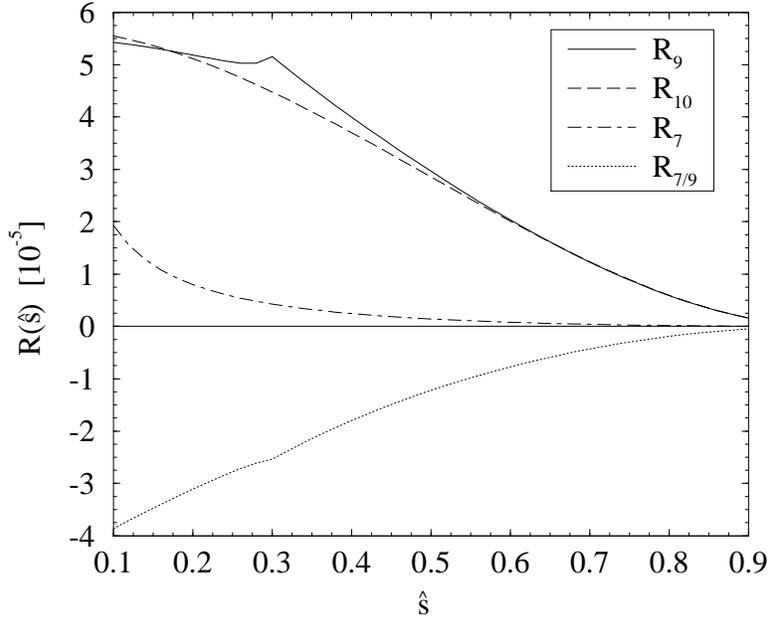}
}}
\vspace{0.08in}
\caption[]{
Comparison of the four different contributions to $R(\hat{s})$ according
to eq.\ \eqn{rate}.
\label{fig:bsee:rscmp}}
\end{figure}

$R(\hat s)$ is governed by three coefficients, $\Ctilde_9^{eff}$,
$\Ctilde_{10}$ and $C_{7\gamma}^{(0)eff}$.  The importance of various
contributions has been investigated in \cite{burasmuenz:95}. To this
end one sets $\Lms^{(5)}=225\gev$, $\mt=170\gev$ and $\mu = 5 \gev$. In
fig.\ \ref{fig:bsee:rscmp} we show $R(\hat s)$ keeping only
$\Ctilde_9^{eff}$, $\Ctilde_{10}$, $C_{7\gamma}^{(0)eff}$ and the
$C_{7\gamma}^{(0)eff}$--$\Ctilde_9^{eff}$ interference term,
respectively.  Denoting these contributions by $R_9$, $R_{10}$, $R_7$
and $R_{7/9}$ we observe that the term $R_7$ plays only a minor role in
$R(\hat s)$. On the other hand the presence of $C_{7\gamma}^{(0)eff}$
cannot be ignored because the interference term $R_{7/9}$ is
significant. In fact the presence of this large interference term could
be used to measure experimentally the relative sign of
$C_{7\gamma}^{(0)eff}$ and $\mbox{Re}\,\Ctilde_9^{eff}$
\cite{grinstein:89a}, \cite{JW:90}, \cite{AMM:91}, \cite{GIW:94},
\cite{AGM:94} which as seen in fig.\ \ref{fig:bsee:rscmp} is negative
in the Standard Model. However, the most important contributions are
$R_9$ and $R_{10}$ in the full range of $\hat s$ considered. For $\mt
\approx 170 \gev$ these two contributions are roughly of the same size.
Due to a strong $\mt$ dependence of $R_{10}$, this contribution
dominates for higher values of $\mt$ and is less important than $R_9$
for $\mt < 170 \gev$.

\begin{figure}[hbt]
\vspace{0.10in}
\centerline{
\epsfysize=4in
\rotate[r]{
\epsffile{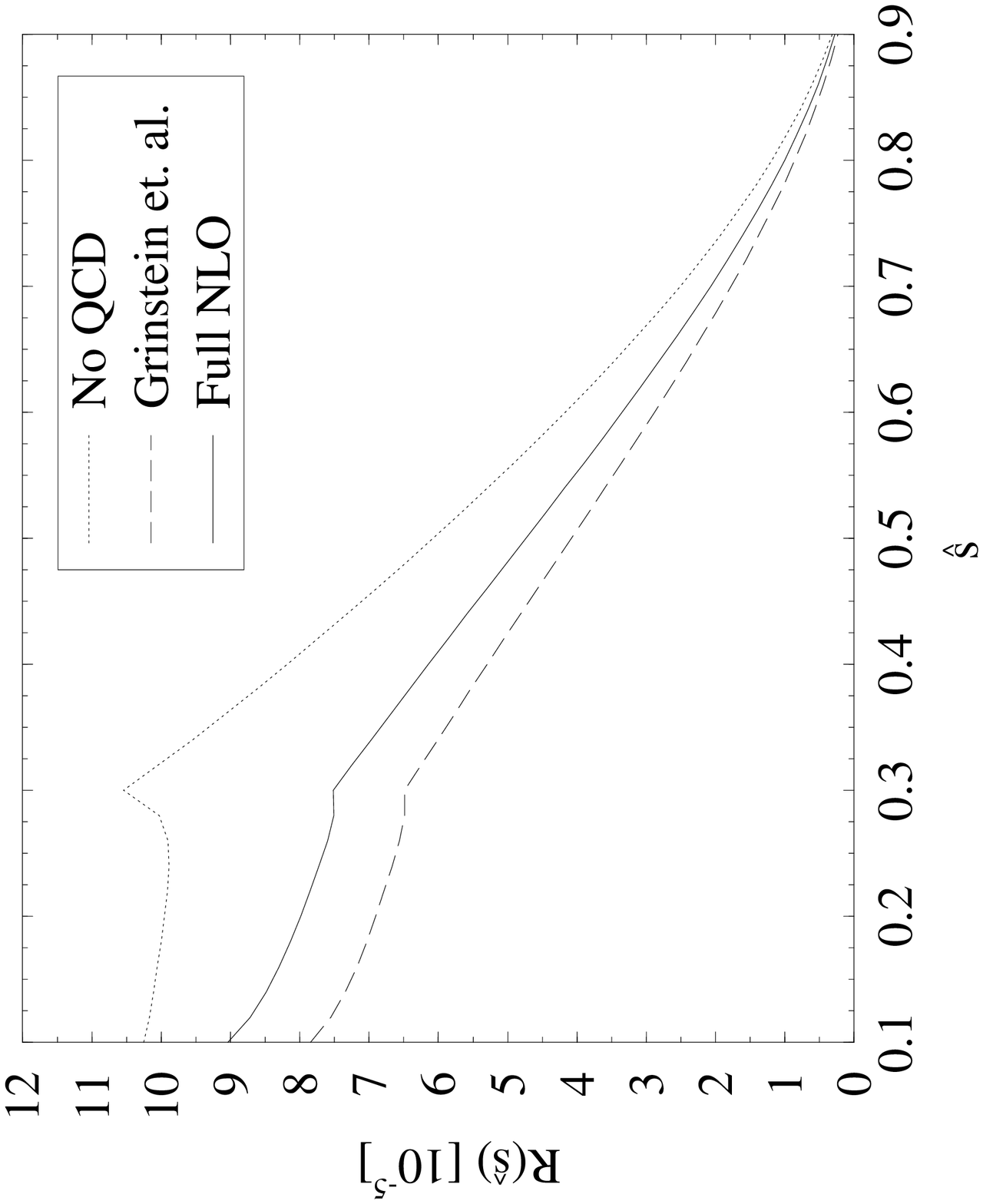}
}}
\vspace{0.08in}
\caption[]{
$R(\hat{s})$ for $\mt=170\gev$, $\Lms^{(5)}=225\mev$ and $\mu=5\gev$. 
\label{fig:bsee:rs2}}
\end{figure}

Next, in fig.\ \ref{fig:bsee:rs2} we show $R(\hat s)$ for $\mu = 5
\gev$, $\mt = 170 \gev$ and $\Lms = 225 \mev$ compared to the case of
no QCD corrections and to the results \cite{grinstein:89a} would obtain
for our set of parameters using their approximate leading order
formulae.

The discussion of the definition of $\mt$ used here is identical to the
one in the case of $K_L \to \pi^0 e^+ e^-$ and will not be repeated here.
On the basis of the arguments given there we believe that if $\mt =
\overline{m}_{\rm t}(\mt)$ is chosen, the additional short-distance QCD
corrections to $B(\Bsee)$ should be small.

Our discussion has been restricted to $B(B \to X_s \gamma)$. Also the
photon spectrum has been the subject of several papers. We just refer
to the most recent articles \cite{neubert:94c}, \cite{shifmanetal:94},
\cite{dikemanetal:94}, \cite{kapustinligeti:95}, \cite{kapustinetal:95},
\cite{aligreub:95}, \cite{pott:95} where further references can be found.

\section{The Decays \kpnn and \klpnn}
\label{sec:Kpnn}
\subsection{General Remarks on \kpnn}
\label{sec:Kpnn:GeneralKp}
The rare decay \kpnn is one of the theoretically cleanest decays.
As such it is very well suited for the determination of CKM
parameters, in particular of the element $V_{td}$. \kpnn is CP
conserving and receives contributions from both internal top and
charm exchanges. The inclusion of next-to-leading QCD corrections
incorporated in the effective hamiltonian in (\ref{hkpn}) and
discussed in detail in section \ref{sec:HeffRareKB:kpnn} reduces
considerably the theoretical uncertainties due to the choice of the
renormalization scales present in the leading order expressions.
We will illustrate this below. Since in addition the relevant hadronic
matrix element of the weak current $(\bar sd)_{V-A}$ can be measured
in the leading decay $K^+\to\pi^0e^+\nu$, the resulting theoretical
expression for $B(\kpn)$ is only a function of the CKM parameters, the
QCD scale $\Lambda_{\overline{MS}}$ and the quark masses $m_t$ and
$m_c$. The long-distance contributions to \kpnn have been found to be
very small: a few percent of the charm contribution to the amplitude at
most, which is safely negligible \cite{reinsehgal:89},
\cite{hagelinlittenberg:89} and \cite{luwise:94}.

Conventionally the branching fraction $B(K^+\to\pi^+\nu\bar\nu)$ is
related to the experimentally well known quantity
$B(K^+\to\pi^0e^+\nu)$ using isospin symmetry. Corrections to this
approximation have recently been studied in \cite{marcianoparsa:95}.
The breaking of isospin is due to quark mass effects and electroweak
radiative corrections.  In the case of $K^+\to\pi^+\nu\bar\nu$ these
effects result in a decrease of the branching ratio by $10\%$. The
corresponding corrections in $K_L\to\pi^0\nu\bar\nu$ lead to a $5.6\%$
reduction of $B(K_L\to\pi^0\nu\bar\nu)$. We have checked the analysis
of \cite{marcianoparsa:95} and agree with their findings.  Once
calculated, the inclusion of these effects is straightforward as they
only amount to an overall factor for the branching ratio and do not
affect the short-distance structure of $K\to\pi\nu\bar\nu$.  We shall
neglect the isospin violating corrections in the following chapters,
where the focus is primarily on the short-distance physics. The effects
are however incorporated in the final prediction quoted in our summary
table in section \ref{sec:summary}.

In the following we shall concentrate on a discussion of $K^+ \to \pi^+
\nu\bar\nu$ within the framework of the standard model. The impact of
various scenarios of new physics on this decay has been considered for
instance in \cite{bigigabbiani:91}.

\subsection{Master Formulae for \kpnn}
\label{sec:Kpnn:MasterKp}
Using the effective hamiltonian (\ref{hkpn}) and summing over the three
neutrino flavors one finds
\begin{equation}\label{bkpn}
B(\kpn)=\kappa_+\cdot\left[\left({\imlt\over\lambda^5}X(x_t)\right)^2+
\left({\relc\over\lambda}P_0(X)+{\relt\over\lambda^5}X(x_t)\right)^2
\right]
\end{equation}
\begin{equation}\label{kapp}
\kappa_+={3\alpha^2 B(K^+\to\pi^0e^+\nu)\over 2\pi^2\sin^4\Theta_W}
 \lambda^8=4.57\cdot 10^{-11}
\end{equation}
where we have used
\begin{equation}\label{alsinbr}
\alpha=\frac{1}{129}\qquad \sin^2\Theta_W=0.23 \qquad
B(K^+\to\pi^0e^+\nu)=4.82\cdot 10^{-2}
\end{equation}
Here $\lambda_i=V^\ast_{is}V_{id}$ with $\lambda_c$ being
real to a very high accuracy. The function $X$ of (\ref{xx})
can also be written as
\begin{equation}\label{xeta}
X(x)=\eta_X\cdot X_0(x) \qquad\quad \eta_X=0.985
\end{equation}
where $\eta_X$ summarizes the NLO corrections discussed in section
\ref{sec:HeffRareKB:kpnn}. With $m_t\equiv m_t(m_t)$ the QCD factor $\eta_X$
is practically independent of $m_t$ and $\Lambda_{\overline{MS}}$.
Next
\begin{equation}\label{p0k}
P_0(X)=\frac{1}{\lambda^4}\left[\frac{2}{3} X^e_{NL}+\frac{1}{3}
 X^\tau_{NL}\right]
\end{equation}
with the numerical values for $X_{NL}^l$ given in table \ref{tab:xnlnum}.
The corresponding values for $P_0(X)$ as a function of
$\Lambda_{\overline{MS}}$ and $m_c\equiv m_c(m_c)$ are collected in
table \ref{tab:P0Kplus}.
We remark that a negligibly small term $\sim(X_{NL}^e-X_{NL}^\tau)^2$
($\sim 0.2\%$ effect on the branching ratio)
has been discarded in formula (\ref{bkpn}).

\begin{table}[htb]
\caption[]{The function $P_0(X)$ for various $\Lms^{(4)}$ and $m_c$.
\label{tab:P0Kplus}}
\begin{center}
\begin{tabular}{|c|c|c|c|}
&\multicolumn{3}{c|}{$P_0(X)$}\\
\hline
$\Lms^{(4)}$ $\backslash$ $m_c$ & $1.25\gev$ & $1.30\gev$ & $1.35\gev$  \\
\hline
$215\mev$ & 0.402 & 0.436 & 0.472 \\
$325\mev$ & 0.366 & 0.400 & 0.435 \\
$435\mev$ & 0.325 & 0.359 & 0.393 
\end{tabular}
\end{center}
\end{table}

Using the improved Wolfenstein parametrization and the approximate
formulae (\ref{2.51}) -- (\ref{2.53}) we can next write
\begin{equation}\label{108}
B(K^{+} \to \pi^{+} \nu \bar\nu) = 4.57 \cdot 10^{-11} A^4 X^2(x_t)
\frac{1}{\sigma} \left[ (\sigma \bar\eta)^2 +
\left(\varrho_0 - \bar\varrho \right)^2 \right]
\end{equation}
where
\begin{equation}\label{109}
\sigma = \left( \frac{1}{1- \frac{\lambda^2}{2}} \right)^2
\end{equation}

The measured value of B($K^{+} \to \pi^{+} \nu \bar\nu$) then
determines  an ellipse in the $(\bar\varrho,\bar\eta)$ plane  centered at
$(\varrho_0,0)$ with \cite{burasetal:94b}
\begin{equation}\label{110}
\varrho_0 = 1 + \frac{P_0(X)}{A^2 X(x_t)}
\end{equation}
and having the squared axes
\begin{equation}\label{110a}
\bar\varrho_1^2 = r^2_0 \qquad \bar\eta_1^2 = \left( \frac{r_0}{\sigma}
\right)^2
\end{equation}
where
\begin{equation}\label{111}
r^2_0 = \frac{1}{A^4 X^2(x_t)} \left[
\frac{\sigma \cdot BR(K^{+} \to \pi^{+} \nu \bar\nu)}{4.57 \cdot 10^{-11}} \right]
\end{equation}
The departure of $\varrho_0$ from unity measures the relative importance
of the internal charm contributions.

The ellipse defined by $r_0$, $\varrho_0$ and $\sigma$ given above
intersects with the circle (\ref{2.94}).  This allows to determine
$\bar\varrho$ and $\bar\eta$  with 
\begin{equation}\label{113}
\bar\varrho = \frac{1}{1-\sigma^2} \left( \varrho_0 - \sqrt{\sigma^2
\varrho_0^2 +(1-\sigma^2)(r_0^2-\sigma^2 R_b^2)} \right) \qquad
\bar\eta = \sqrt{R_b^2 -\bar\varrho^2}
\end{equation}
and consequently
\begin{equation}\label{113aa}
R_t^2 = 1+R_b^2 - 2 \bar\varrho
\end{equation}
where $\bar\eta$ is assumed to be positive.

In the leading order of the Wolfenstein parametrization
\begin{equation}\label{113ab}
\sigma \to 1 \qquad \bar\eta \to \eta \qquad \bar\varrho \to \varrho
\end{equation}
and $B(K^+ \to \pi^+ \nu \bar\nu)$ determines a circle in the
$(\varrho,\eta)$ plane centered at $(\varrho_0,0)$ and having the radius
$r_0$ of (\ref{111}) with $\sigma =1$. Formulae (\ref{113}) and
(\ref{113aa}) then simplify to \cite{buchallaburas:94}
\begin{equation}\label{113a}
R_t^2 = 1 + R_b^2 + \frac{r_0^2 - R_b^2}{\varrho_0} - \varrho_0 \qquad
\varrho = \frac{1}{2} \left( \varrho_0 + \frac{R_b^2 - r_0^2}{\varrho_0}
\right)
\end{equation}
Given $\bar\varrho$ and $\bar\eta$ one can determine $V_{td}$:
\begin{equation}\label{vtdrhoeta}
V_{td}=A \lambda^3(1-\bar\varrho-i\bar\eta)\qquad
|V_{td}|=A \lambda^3 R_t
\end{equation}
Before proceeding to the numerical analysis a few remarks are in
order:
\begin{itemize}
\item
The determination of $|V_{td}|$ and of the unitarity triangle requires
the knowledge of $V_{cb}$ (or $A$) and of $|V_{ub}/V_{cb}|$. Both
values are subject to theoretical uncertainties present in the existing
analyses of tree level decays. Whereas the dependence on
$|V_{ub}/V_{cb}|$ is rather weak, the very strong dependence of
$B(\kpn)$ on $A$ or $V_{cb}$ makes a precise prediction for this
branching ratio difficult at present. We will return to this below.
\item
The dependence of $B(\kpn)$ on $m_t$ is also strong. However $m_t$
should be known already in this decade within $\pm 5\%$ and
consequently the uncertainty in $m_t$ will soon be less serious for
$B(\kpn)$ than the corresponding uncertainty in $V_{cb}$.
\item
Once $\varrho$ and $\eta$ are known precisely from CP asymmetries in
B decays, some of the uncertainties present in (\ref{108}) related
to $|V_{ub}/V_{cb}|$ (but not to $V_{cb}$) will be removed.
\item
A very clean determination of $\sin 2\beta$ without essentially
any dependence on $m_t$ and $V_{cb}$ can be made by combining
$B(\kpn)$ with $B(\klpn)$ discussed below. We will present an
analysis of this type in section \ref{sec:Kpnn:sin2b}.
\end{itemize}

\subsection{Numerical Analysis of \kpnn}
\label{sec:Kpnn:NumericalKp}
\subsubsection{Renormalization Scale Uncertainties}
\label{sec:Kpnn:NumericalKp:RSU}
We will now investigate the uncertainties in $X(x_t)$, $X_{NL}$,
$B(\kpn)$, $|V_{td}|$ and in the determination of the unitarity
triangle related to the choice of the renormalization scales $\mu_t$
and $\mu_c$ (see section \ref{sec:HeffRareKB:kpnn}). To this end we
will fix the remaining parameters as follows
\begin{equation}\label{mcmtnum}
m_c\equiv m_c(m_c)=1.3\gev \qquad m_t\equiv m_t(m_t)=170\gev
\end{equation}
\begin{equation}\label{vcbubnum}
V_{cb}=0.040 \qquad |V_{ub}/V_{cb}|=0.08
\end{equation}
In the case of $B(\kpn)$ we need the values of both $\bar\varrho$
and $\bar\eta$. Therefore in this case we will work with
\begin{equation}\label{rhetnum}
\bar\varrho=0 \qquad\quad  \bar\eta=0.36
\end{equation}
rather than with $|V_{ub}/V_{cb}|$. Finally we will set
$\Lambda_{\overline{MS}}^{(4)}=0.325\gev$ and
$\Lambda_{\overline{MS}}^{(5)}=0.225\gev$ for the charm part and top
part, respectively.
We then vary the scales $\mu_c$ and $\mu_t$, entering $m_c(\mu_c)$
and $m_t(\mu_t)$ respectively, in the ranges
\begin{equation}\label{muctnum}
1\gev\leq\mu_c\leq 3\gev \qquad 100\gev\leq\mu_t\leq 300\gev
\end{equation}

\begin{figure}[hbt]
\vspace{0.10in}
\centerline{
\epsfysize=5in
\rotate[r]{
\epsffile{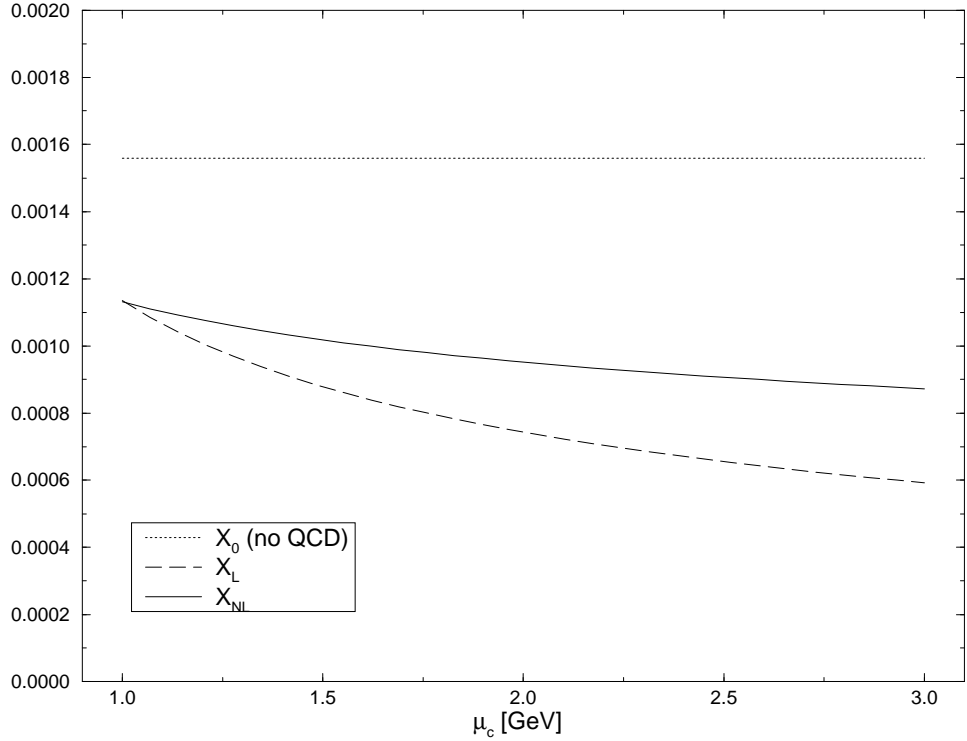}
} }
\vspace{0.08in}
\caption[]{
Charm quark function $X_{NL}$ (for $m_l=0$) compared to the leading-log
result $X_L$ and the case without QCD as functions of $\mu_c$.
\label{fig:kpnunu:XNL}}
\end{figure}

In fig.\ \ref{fig:kpnunu:XNL} we show the charm function $X_{NL}$ (for
$m_l=0$) compared to the leading-log result $X_L$ and the case without
QCD as functions of $\mu_c$. We observe the following features:
\begin{itemize}
\item
The residual slope of $X_{NL}$ is considerably reduced in
comparison to $X_L$, which exhibits a quite substantial dependence
on the unphysical scale $\mu_c$. The variation
of $X$ (defined as $(X(1\gev)-X(3\gev))/X(m_c)$)
is 24.5\% in NLLA compared to 56.6\% in LLA.
\item
The suppression of the
uncorrected function through QCD effects is somewhat less pronounced
in NLLA.
\item
The next-to-leading effects amount to a $\sim 10\%$ correction relative
to $X_L$ at $\mu=m_c$. However the size of this correction strongly
depends on $\mu$ due to the scale ambiguity of the leading order
result. This means that the question of how large the next-to-leading
effects compared to the LLA really are cannot be answered uniquely.
Therefore the relevant result is actually the reduction of the
$\mu$-dependence in NLLA .
\end{itemize}

\begin{figure}[hbt]
\vspace{0.10in}
\centerline{
\epsfysize=5in
\rotate[r]{
\epsffile{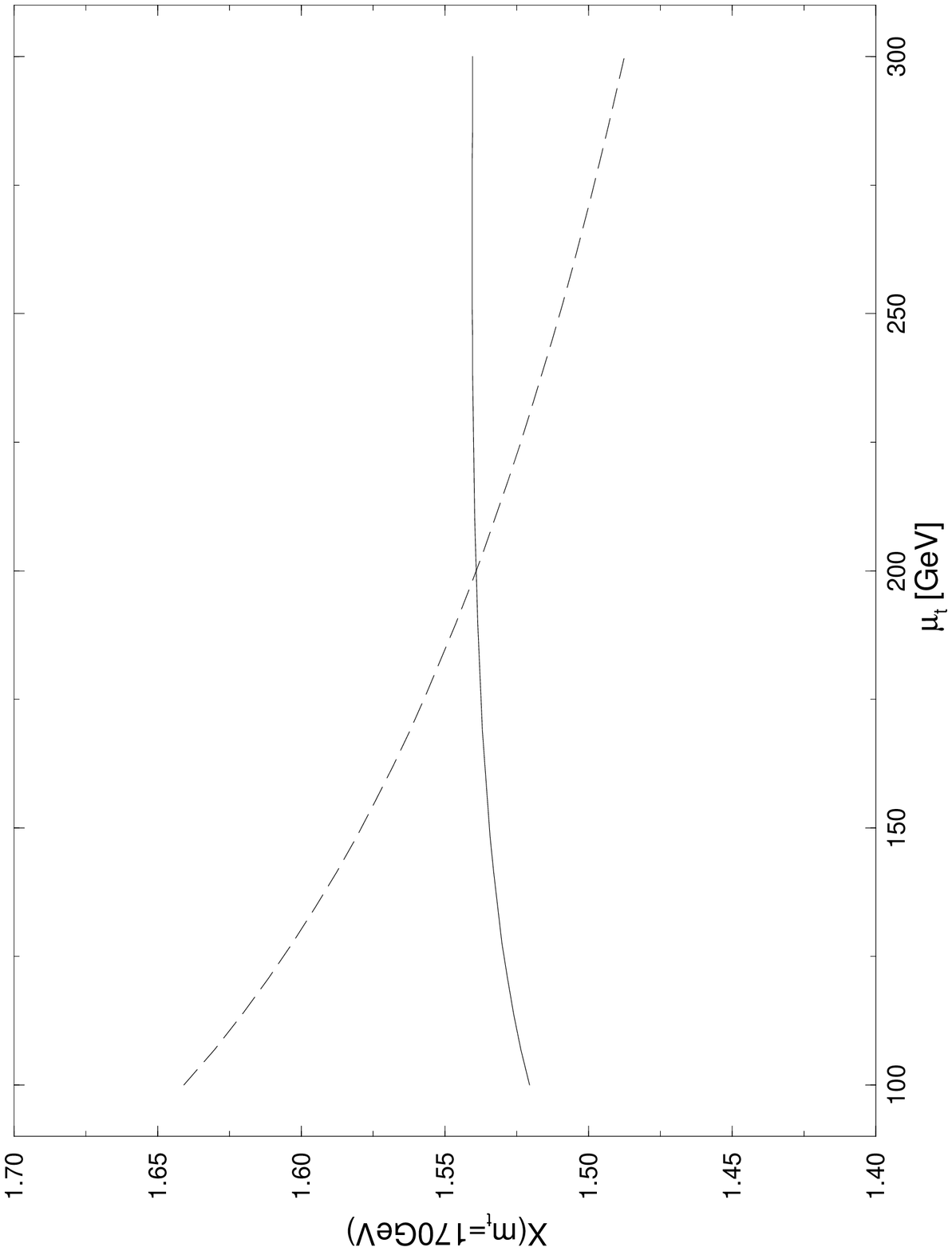}
} }
\vspace{0.08in}
\caption[]{
Top quark function $X(x_t)$ as a function of $\mu_t$ for fixed
$\mt(\mt)=170\gev$ with (solid curve) and without (dashed curve)
$\ord(\as)$ corrections.
\label{fig:kpnunu:X}}
\end{figure}

\noindent
In fig.\ \ref{fig:kpnunu:X} we show the analogous results for the top
function $X(x_t)$ as a function of $\mu_t$. We observe:
\begin{itemize}
\item
Due to $\mu_t\gg\mu_c$ the scale dependences in the top function
are substantially smaller than in the case of charm.
Note in particular how the yet appreciable scale dependence of $X_0$
gets flattened out almost perfectly when the $\ord(\as)$
effects are taken into account. The total variation of $X(x_t)$
with $100\gev\leq\mu_t\leq 300\gev$ is around 1\% in NLLA compared
to 10\% in LLA.
\item
As already stated above after (\ref{xeta}), with the choice
$\mu_t=m_t$ the NLO correction is very small. It is substantially
larger for $\mu_t$ very different from $m_t$.
\end{itemize}
Using (\ref{bkpn}) and varying $\mu_{c, t}$ in the ranges
(\ref{muctnum}) we find that for the above choice of the remaining
parameters the uncertainty in $B(\kpn)$
\begin{equation}\label{varbkpnLO}
0.76\cdot 10^{-10}\leq B(\kpn)\leq 1.20\cdot 10^{-10}
\end{equation}
present in the leading order is reduced to
\begin{equation}\label{varbkpnNLO}
0.88\cdot 10^{-10}\leq B(\kpn)\leq 1.02\cdot 10^{-10}
\end{equation}
after including NLO corrections. Similarly we obtain
\begin{equation}\label{varvtdLO}
8.24\cdot 10^{-3}\leq |V_{td}|\leq 10.97\cdot 10^{-3} \qquad {\rm LLA}
\end{equation}
\begin{equation}\label{varvtdNLO}
9.23\cdot 10^{-3}\leq |V_{td}|\leq 10.10\cdot 10^{-3}  \qquad {\rm NLLA}
\end{equation}
where we have set $B(\kpn)=1\cdot 10^{-10}$. We observe that including
the full next-to-leading corrections reduces the uncertainty in the
determination of $|V_{td}|$ from $\pm 14\%$ (LLA) to $\pm 4.6\%$ (NLLA)
in the present example. The main bulk of this theoretical error stems
from the charm sector. Indeed, keeping $\mu_c=m_c$ fixed and varying
only $\mu_t$, the uncertainties in the determination of $|V_{td}|$
would shrink to $\pm 4.7\%$ (LLA) and $\pm 0.6\%$ (NLLA).
Similar comments apply to $B(\kpn)$ where, as seen in
(\ref{varbkpnLO}) and (\ref{varbkpnNLO}), the theoretical uncertainty
due to $\mu_{c,t}$ is reduced from $\pm 22\%$ (LLA) to $\pm 7\%$ (NLLA).

\begin{figure}[hbt]
\vspace{0.10in}
\centerline{
\epsfysize=5in
\rotate[r]{
\epsffile{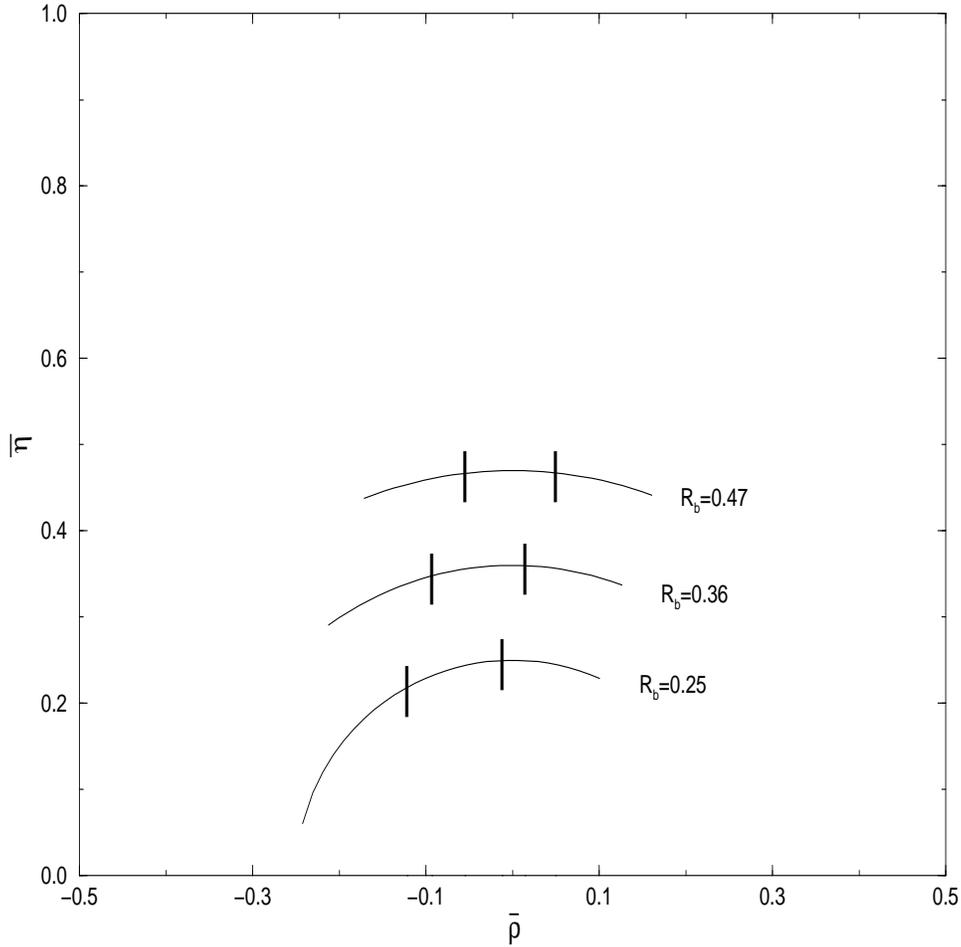}
} }
\vspace{0.08in}
\caption[]{
The theoretical uncertainties in the determination of the unitarity
triangle (UT) in the $(\bar\varrho, \bar\eta)$ plane from
$B(K^+\to\pi^+\nu\bar\nu)$. With fixed input parameters the vertex of
the UT has to lie on a circle around the origin with radius $R_b$.  A
variation of the scales $\mu_c$, $\mu_t$ within  $1\gev \le \mu_c \le 3\gev$
and $100\gev \le \mu_t \le 300\gev$ then yields the indicated ranges in LLA
(full) and NLLA (reduced). We show the cases $R_b=0.25, 0.36, 0.47$.
\label{fig:kpnunu:rhoetabar}}
\end{figure}

Finally in fig.\ \ref{fig:kpnunu:rhoetabar} we show the position of the
point ($\bar\varrho$, $\bar\eta$) which determines the unitarity
triangle.  To this end we have fixed all parameters as stated above
except for $R_b$, for which we have chosen three representative
numbers, $R_b=0.25$, $0.36$, $0.47$. The full and the reduced ranges
represent LLA and NLLA respectively. The impact of the inclusion of NLO
corrections on the accuracy of determining the unitarity triangle is
clearly visible.

\subsubsection{Expectations for $B(\kpn)$}
\label{sec:Kpnn:NumericalKp:EfB}
The purely theoretical uncertainties discussed so far should be
distinguished from the uncertainties coming from the input parameters
such as $m_t$, $V_{cb}$, $|V_{ub}/V_{cb}|$
etc.. As we will see the latter uncertainties are still rather large to
date. Consequently the progress achieved by the NLO calculations
\cite{buchallaburas:94} cannot yet be fully exploited
phenomenologically at present.  However the determination of the
relevant parameters should improve in the future. Once the precision in
the input parameters will have attained the desired level, the gain in
accuracy of the theoretical prediction for \kpnn in NLLA by a factor of
more than 3 compared to the LLA will become very important.

Using our standard set of input parameters specified in appendix
\ref{app:numinput} and the constraints implied by the analysis of
$\varepsilon_K$ and $B_d-\bar B_d$ mixing as described in section
\ref{sec:epsBBUT}, we find for the \kpnn branching fraction
the range
\begin{equation}\label{kpnn1}
0.6\cdot 10^{-10}\leq B(\kpn)\leq 1.5\cdot 10^{-10}
\end{equation}
Eq. (\ref{kpnn1}) represents the current standard model expectation for
$B(\kpn)$ (neglecting small isospin breaking corrections). To obtain
this estimate we have allowed for a variation of the parameters $m_t$,
$|V_{cb}|$, $|V_{ub}/V_{cb}|$, $B_K$, $F^2_B B_B$, $x_d$ within their
uncertainties as summarized in appendix \ref{app:numinput}. The
uncertainties in $m_c$ and $\Lambda_{\overline{MS}}$, on the other
hand, are small in comparison and have been neglected in this context.
The above range would be reduced to \begin{equation}\label{kpnn2}
0.8\cdot 10^{-10}\leq B(\kpn)\leq 1.0\cdot 10^{-10} \end{equation} if
the uncertainties in the input parameters could be decreased as assumed
by our ``future'' scenario in appendix \ref{app:numinput}.

It should be remarked that the $x_d$-constraint, excluding a part of
the second quadrant for the CKM phase $\delta$, plays an essentail role
in obtaining the upper bounds given above, without essentially any
effect on the lower bounds. Without the $x_d$-constraint the upper
bounds in \eqn{kpnn1} and \eqn{kpnn2} are relaxed to $2.3 \cdot
10^{-10}$ and $1.6 \cdot 10^{-10}$, respectively.

\subsection{General Remarks on \klpnn}
            \label{sec:Kpnn:GeneralKL}
The rare decay \klpnn is even cleaner than $\kpn$. It proceeds almost
entirely through direct CP violation \cite{littenberg:89} and
is completely dominated by short-distance loop diagrams with top quark
exchanges. In fact the $m_t$-dependence of $B(\klpn)$ is again
described by $X(x_t)$.  Since the charm contribution can be fully
neglected also the theoretical uncertainties present in \kpnn due to
$m_c$, $\mu_c$ and $\Lambda_{\overline{MS}}$ are absent here. For this
reason \klpnn is very well suited for the determination of CKM
parameters, in particular the Wolfenstein parameter $\eta$.

\subsection{Master Formulae for \klpnn}
\label{sec:Kpnn:MasterKL}
Using the effective hamiltonian (\ref{hxnu}) and summing over three
neutrino flavors one finds
\begin{equation}\label{bklpn}
B(K_L\to\pi^0\nu\bar\nu)=\kappa_L\cdot\left({\imlt\over\lambda^5}X(x_t)
  \right)^2
\end{equation}
\begin{equation}\label{kapl}
\kappa_L=\kappa_+ {\tau(K_L)\over\tau(K^+)}=1.91\cdot 10^{-10}
\end{equation}
with $\kappa_+$ given in (\ref{kapp}). Using the Wolfenstein
parametrization we can rewrite (\ref{bklpn}) as
\begin{equation}\label{bklpnwol1}
B(\klpn)=1.91\cdot 10^{-10} \eta^2 A^4 X^2(x_t)
\end{equation}
or
\begin{equation}\label{bklpnwol2}
B(\klpn)=3.48\cdot 10^{-5} \eta^2 |V_{cb}|^4 X^2(x_t)
\end{equation}
A few remarks are in order:
\begin{itemize}
\item
The determination of $\eta$ using $B(\klpn)$ requires the knowledge
of $V_{cb}$ and $m_t$. The very strong dependence on $V_{cb}$ or $A$
makes a precise prediction for this branching ratio difficult at
present.
\item
It has been pointed out \cite{buras:94b} that the strong
dependence of $B(\klpn)$ on $V_{cb}$, together with the clean nature of
this decay, can be used to determine this element without any hadronic
uncertainties. To this end $\eta$ and $m_t$ have to be known with
sufficient precision in addition to $B(\klpn)$. $\eta$ should be
measured accurately in CP asymmetries in $B$ decays and the value of
$m_t$ known to better than $\pm 5\gev$ from TEVATRON and future LHC
experiments. Inverting (\ref{bklpnwol2}) and using a very accurate
approximation for $X(x_t)$ (valid for $\mt = \overline{m}_{\rm t}(\mt)$)
as given by \eqn{xeta} and \eqn{eq:approxSXYZE}
\begin{equation}\label{xxappr}
X(x_t)=0.65\cdot x_t^{0.575}
\end{equation}
one finds
\begin{equation}\label{vcbklpn}
V_{cb}=39.3\cdot 10^{-3} \sqrt{\frac{0.39}{\eta}}
\left[\frac{170\gev}{m_t}\right]^{0.575}
\left[\frac{B(\klpn)}{3\cdot 10^{-11}}\right]^{1/4}
\end{equation}
We note that the weak dependence of $V_{cb}$ on $B(\klpn)$ allows
to achieve a high precision for this CKM element even when $B(\klpn)$
is known with only relatively moderate accuracy, e.g. 10--15\%.
Needless to say that any measurement of $B(\klpn)$ is extremely
challenging. A numerical analysis of (\ref{vcbklpn}) can be found in
\cite{buras:94b}.
\end{itemize}

\subsection{Numerical Analysis of \klpnn}
\label{sec:Kpnn:NumericalKL}
\subsubsection{Renormalization Scale Uncertainties}
\label{sec:Kpnn:NumericalKL:RSU}
\begin{figure}[hbt]
\vspace{0.10in}
\centerline{
\epsfysize=5in
\rotate[r]{
\epsffile{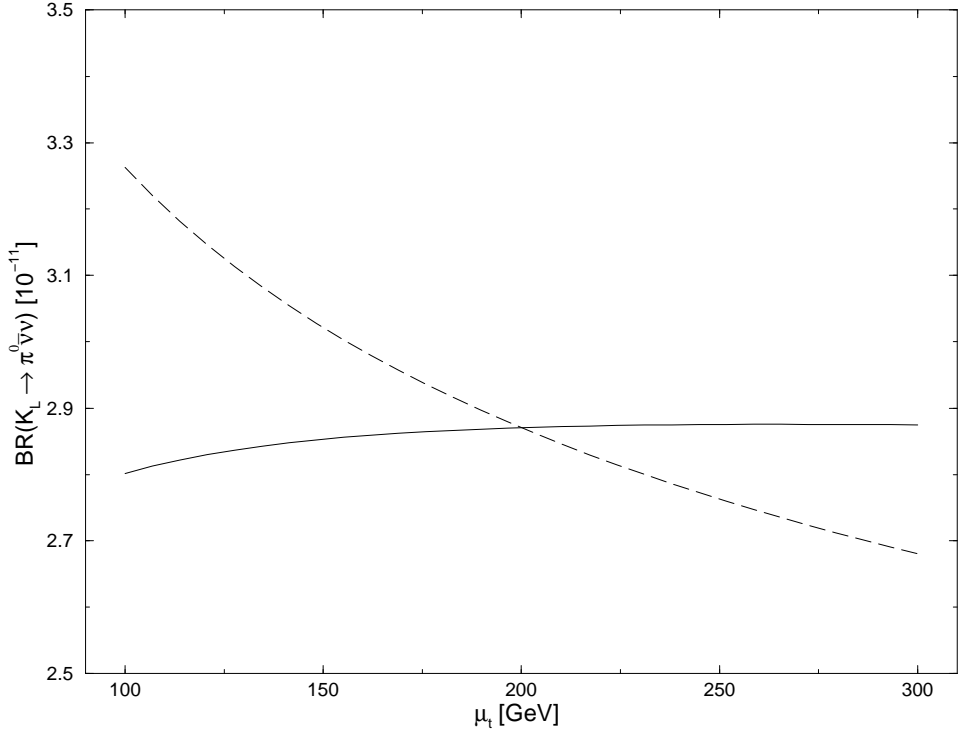}
} }
\vspace{0.08in}
\caption[]{
The $\mu_t$-dependence of $B(K_L \to \pi^0\nu\bar\nu)/10^{-11}$ with
(solid curve) and without (dashed curve) $\ord(\as)$ corrections for
$\mt(\mt)=170\gev$, $|V_{cb}|=0.04$ and $\bar\eta=0.36$.
\label{fig:klpnmut}}
\end{figure}

The scale ambiguities present in the function $X(x_t)$ have already been
discussed in connection with $\kpn$. After the inclusion of NLO
corrections they are so small that they can be neglected for all
practical purposes. Effectively they could also be taken into
account by introducing an additional error $\Delta m_t\leq\pm 1\gev$.
At the level of $B(\klpn)$ the ambiguity in the choice of $\mu_t$ is
reduced from $\pm 10\%$ (LLA) down to $\pm 1\%$ (NLLA), which
considerably increases the predictive power of the theory. Varying
$\mu_t$ according to (\ref{muctnum}) and using the input parameters
of section \ref{sec:Kpnn:NumericalKp} we find that the uncertainty
in $B(\klpn)$
\begin{equation}\label{varbklpnLO}
2.68\cdot 10^{-11}\leq B(\klpn)\leq 3.26\cdot 10^{-11}
\end{equation}
present in the leading order is reduced to
\begin{equation}\label{varbklpnNLO}
2.80\cdot 10^{-11}\leq B(\klpn)\leq 2.88\cdot 10^{-11}
\end{equation}
after including NLO corrections. This means that the theoretical
uncertainty in the determination of $\eta$ amounts to only $\pm 0.7\%$
in NLLA which is safely negligible.
The reduction of the scale ambiguity for $B(\klpn)$ is further 
illustrated in fig.\ \ref{fig:klpnmut}.

\subsubsection{Expectations for $B(\klpn)$}
\label{sec:Kpnn:NumericalKL:EfB}
From an analysis of $B(\klpn)$ similar to the one described for
\kpnn in section \ref{sec:Kpnn:NumericalKp:EfB} we obtain the
standard model expectation
\begin{equation}\label{klpn1}
1.1\cdot 10^{-11}\leq B(\klpn)\leq 5.0\cdot 10^{-11}
\end{equation}
corresponding to present day errors in the relevant input
parameters. This would change into
\begin{equation}\label{klpn2}
2.2\cdot 10^{-11}\leq B(\klpn)\leq 3.6\cdot 10^{-11}
\end{equation}
if the parameter uncertainties would decrease as anticipated by our 
``future'' scenario defined in appendix \ref{app:numinput}.

\subsection{Unitarity Triangle from $K\to\pi\nu\bar\nu$}
\label{sec:Kpnn:Triangle}
The measurement of $B(\kpn)$ and $B(\klpn)$ can determine the
unitarity triangle completely provided $m_t$ and $V_{cb}$ are known.
Using these two branching ratios simultaneously allows to eliminate
$|V_{ub}/V_{cb}|$ from the analysis which removes considerable
uncertainty. Indeed it is evident from (\ref{bkpn}) and
(\ref{bklpn}) that, given $B(\kpn)$ and $B(\klpn)$, one can extract
both $\imlt$ and $\relt$. We get
\begin{equation}\label{imre}
\imlt=\lambda^5{\sqrt{B_2}\over X(x_t)}\qquad
\relt=-\lambda^5{{\relc\over\lambda}P_0(X)+\sqrt{B_1-B_2}\over X(x_t)}
\end{equation}
where we have defined the ``reduced'' branching ratios
\begin{equation}\label{b1b2}
B_1={B(\kpn)\over 4.57\cdot 10^{-11}}\qquad
B_2={B(\klpn)\over 1.91\cdot 10^{-10}}
\end{equation}
Using next the expressions for $\imlt$, $\relt$ and $\relc$ given
in (\ref{2.51}) -- (\ref{2.53}) we find
\begin{equation}\label{rhetb}
\bar\varrho=1+{P_0(X)-\sqrt{\sigma(B_1-B_2)}\over A^2 X(x_t)}\qquad
\bar\eta={\sqrt{B_2}\over\sqrt{\sigma} A^2 X(x_t)}
\end{equation}
with $\sigma$ defined in (\ref{109}). An exact treatment of the CKM
matrix shows that the formulae (\ref{rhetb}) are rather precise
\cite{buchallaburas:94c}. The error in $\bar\eta$ is below 0.1\% and
$\bar\varrho$ may deviate from the exact expression by at most
$\Delta\bar\varrho=0.02$ with essentially negligible error for
$0\leq\bar\varrho\leq 0.25$.
\\
As an illustrative example, let us consider the following scenario.
We assume that the branching ratios are known to within $\pm 10\%$
\begin{equation}\label{bkpkl}
B(\kpn)=(1.0\pm 0.1)\cdot 10^{-10}\qquad
B(\klpn)=(2.5\pm 0.25)\cdot 10^{-11}
\end{equation}
Next we take ($m_i\equiv m_i(m_i)$)
\begin{equation}\label{mtcv}
m_t=(170\pm 5)\gev\quad m_c=(1.30\pm 0.05)\gev\quad
V_{cb}=0.040\pm 0.001
\end{equation}
where the quoted errors are quite reasonable if one keeps in mind
that it will take at least ten years to achieve the accuracy
assumed in (\ref{bkpkl}).
Finally, we use
\begin{equation}\label{lamuc}
\Lms^{(4)}=(200 - 350)\mev \qquad \mu_c=(1-3)\gev
\end{equation}
where $\mu_c$ is the renormalization scale present in the analysis of
the charm contribution. Its variation gives an indication of the
theoretical uncertainty involved in the calculation.  In comparison to
this error we neglect the effect of varying $\mu_W=\ord(M_W)$, the high
energy matching scale at which the W boson is integrated out, as well
as the very small scale dependence of the top quark contribution.  As
reference parameters we use the central values in (\ref{bkpkl}) and
(\ref{mtcv}) and $\Lms^{(4)} = 300\mev$, $\mu_c=m_c$.  The results that
would be obtained in such a scenario for $\bar\eta$, $|V_{td}|$ and
$\bar\varrho$ are collected in table \ref{tab:utkpnn}.

\begin{table}[htb]
\caption[]{$\bar\eta$, $|V_{td}|$ and $\bar\varrho$ determined from
\kpnn and \klpnn for the scenario described in the text together with
the uncertainties related to various parameters.
\label{tab:utkpnn}}
\begin{center}
\begin{tabular}{|c||c||c|c|c|c||c|}
&&$\Delta(BR)$&$\Delta(m_t, V_{cb})$&$\Delta(m_c,\Lms^{(4)})
$&$\Delta(\mu_c)$&$\Delta_{total}$\\ \hline
$\bar\eta$&$0.33$&$\pm 0.02$&$\pm 0.03$&$\pm 0.00$&$\pm 0.00$&
$\pm 0.05$\\ \hline
$|V_{td}|/10^{-3}$&$9.3$&$\pm 0.6$&$\pm 0.6$&$\pm 0.5$&$\pm 0.4$&
$\pm 2.1$\\ \hline
$\bar\varrho$&$0.00$&$\pm 0.08$&$\pm 0.09$&$\pm 0.06$&$\pm 0.04$&
$\pm 0.27$
\end{tabular}
\end{center}
\end{table}

There we have also displayed separately the associated, symmetrized
errors ($\Delta$) coming from the uncertainties in the branching ratios,
$m_t$ and $V_{cb}$, $m_c$ and $\Lms^{(4)}$, $\mu_c$,
as well as the total uncertainty.
\\
We observe that respectable determinations of $\bar\eta$ and
$|V_{td}|$ can be obtained. On the other hand the determination of
$\bar\varrho$ is rather poor. We also note that a sizable part of the
total uncertainty results in each case from the strong dependence of
both branching ratios on $m_t$ and $V_{cb}$. There is however one
important quantity for which the strong dependence of $B(\kpn)$ and
$B(\klpn)$ on $m_t$ and $V_{cb}$ does not matter at all.

\subsection{$\sin 2\beta$ from $K\to\pi\nu\bar\nu$}
\label{sec:Kpnn:sin2b}
Using (\ref{rhetb}) one finds \cite{buchallaburas:94c}
\begin{equation}\label{sin}
r_s=r_s(B_1, B_2)\equiv{1-\bar\varrho\over\bar\eta}=\cot\beta \qquad
\sin 2\beta=\frac{2 r_s}{1+r^2_s}
\end{equation}
with
\begin{equation}\label{cbb}
r_s(B_1, B_2)=\sqrt{\sigma}{\sqrt{\sigma(B_1-B_2)}-P_0(X)\over\sqrt{B_2}}
\end{equation}
Thus within the approximation of (\ref{rhetb}) $\sin 2\beta$ is
independent of $V_{cb}$ (or $A$) and $m_t$. An exact treatment of
the CKM matrix confirms this finding to a high accuracy. The
dependence on $V_{cb}$ and $m_t$ enters only at order
$\ord(\lambda^2)$ and as a numerical analysis shows this
dependence can be fully neglected.
\\
It should be stressed that $\sin 2\beta$ determined this way depends
only on two measurable branching ratios and on the function
$P_0(X)$ which is completely calculable in perturbation theory.
Consequently this determination is free from any hadronic
uncertainties and its accuracy can be estimated with a high degree
of confidence. To this end we use the input given in
(\ref{bkpkl}) -- (\ref{lamuc}) to find
\begin{equation}\label{sin2bnum}
\sin 2\beta=0.60\pm 0.06\pm 0.03\pm 0.02
\end{equation}
where the first error comes from $B(\kpn)$ and $B(\klpn)$, the second
from $m_c$ and $\Lambda_{\overline{MS}}$ and the last one from the
uncertainty due to $\mu_c$. We note that the largest partial
uncertainty results from the branching ratios themselves. It can be
probably reduced with time as is the case with the $\pm 0.03$
uncertainty related to $\Lambda_{\overline{MS}}$ and $m_c$.  Note that
the theoretical uncertainty represented by $\Delta(\mu_c)$, which
ultimately limits the accuracy of the analysis, is small.  This
reflects the clean nature of the $K\to\pi\nu\bar\nu$ decays. However
the small uncertainty of $\pm 0.02$ is only achieved by including
next-to-leading order QCD corrections.  In the leading logarithmic
approximation the corresponding error would amount to $\pm 0.05$,
larger than the one coming from $m_c$ and $\Lambda_{\overline{MS}}$.
\\
The accuracy to which $\sin 2\beta$ can be obtained from
$K\to\pi\nu\bar\nu$ is, in our example, comparable to the one expected
in determining $\sin 2\beta$ from CP asymmetries in B decays prior to
LHC experiments.  In this case $\sin 2\beta$ is determined best by
measuring the time integrated CP violating asymmetry in $B^0_d\to\psi
K_S$ which is given by
\begin{eqnarray}
A_{CP}(\psi K_S) &=& \frac{
 \int^\infty_0\left[\Gamma(B\to\psi K_S)-\Gamma(\bar B\to\psi K_S)\right] dt}
{\int^\infty_0\left[\Gamma(B\to\psi K_S)+\Gamma(\bar B\to\psi K_S)\right] dt}
\nn \\
 &=& -\sin 2\beta {x_d\over 1+x^2_d}
\label{acp}
\end{eqnarray}
where $x_d=\Delta m/\Gamma$ gives the size of $B^0_d-\bar B^0_d$
mixing. Combining (\ref{sin}) and (\ref{acp}) we obtain an
interesting connection between rare K decays and B physics
\begin{equation}\label{kbcon}
{2 r_s(B_1,B_2)\over 1+r^2_s(B_1,B_2)}=-A_{CP}(\psi K_S){1+x^2_d\over x_d}
\end{equation}
which must be satisfied in the Standard Model. We stress that except
for $P_0(X)$ given in table \ref{tab:P0Kplus} all quantities in
(\ref{kbcon}) can be directly measured in experiment and that this
relationship is essentially independent of $m_t$ and $V_{cb}$.

\section{The Decays \klmm and $K^+\to\pi^+\mu^+\mu^-$}
\label{sec:KLmm}
\subsection{General Remarks on \klmm}
\label{sec:KLmm:GeneralKL}
The rare decay \klmm is CP conserving and in addition to its
short-distance part receives important contributions from the
two-photon intermediate state, which are difficult to calculate
reliably \cite{gengng:90}, \cite{belangergeng:91}, \cite{ko:92}.

This latter fact is rather unfortunate because the
short-distance part is, similarly to $\kpn$, free of hadronic
uncertainties and if extracted from the data would give a useful
determination of the Wolfenstein parameter $\varrho$. The separation
of the short-distance from the long-distance piece in the measured
rate is very difficult however.
\\
In spite of all this we will present here the analysis of the short-distance
contribution because on one hand it may turn out to be useful
one day for \klmm and on the other hand it also plays an important
role in a parity violating asymmetry, which can be measured in
$K^+\to\pi^+\mu^+\mu^-$. We will discuss this latter topic later on in
this section.
\\
The analysis of $(\klm)_{SD}$ proceeds in essentially the same
manner as for $\kpn$. The only difference enters through the lepton
line in the box contribution. This change introduces two new
functions $Y_{NL}$ and $Y(x_t)$ for the charm and top
contributions respectively (section \ref{sec:HeffRareKB:klmm}),
which will be discussed in detail below.

\subsection{Master Formulae for $(K_L \to \mu^+ \mu^-)_{\rm SD}$}
\label{sec:KLmm:MasterKL}
Using the effective hamiltonian (\ref{hklm}) and relating
$\langle 0|(\bar sd)_{V-A}|K_L\rangle$ to $B(K^+\to\mu^+\nu)$
we find
\begin{equation}\label{bklm}
B(\klm)_{SD}=\kappa_\mu\left[\frac{\relc}{\lambda}P_0(Y)+
\frac{\relt}{\lambda^5} Y(x_t)\right]^2
\end{equation}
\begin{equation}\label{kapm}
\kappa_\mu=\frac{\alpha^2 B(K^+\to\mu^+\nu)}{\pi^2\sin^4\Theta_W}
\frac{\tau(K_L)}{\tau(K^+)}\lambda^8=1.68\cdot 10^{-9}
\end{equation}
where we have used
\begin{equation}\label{klmpar}
\alpha=\frac{1}{129}\qquad \sin^2\Theta_W=0.23\qquad
B(K^+\to\mu^+\nu)=0.635
\end{equation}
The function $Y(x)$ of (\ref{yy}) can also be written as
\begin{equation}\label{yeta}
Y(x)=\eta_Y\cdot Y_0(x) \qquad\quad \eta_Y=1.026\pm 0.006
\end{equation}
where $\eta_Y$ summarizes the NLO corrections discussed in section
\ref{sec:HeffRareKB:klmm}. With $m_t\equiv m_t(m_t)$ this QCD factor
depends only very weakly on $m_t$. The range in (\ref{yeta})
corresponds to $150\gev\leq m_t\leq 190\gev$. The dependence on
$\Lambda_{\overline{MS}}$ can be neglected. Next
\begin{equation}\label{p0kldef}
P_0(Y)=\frac{Y_{NL}}{\lambda^4}
\end{equation}
with $Y_{NL}$ calculated in section \ref{sec:HeffRareKB:klmm}.
Values for $P_0(Y)$ as a function of $\Lambda_{\overline{MS}}$
and $m_c\equiv m_c(m_c)$ are collected in table \ref{tab:P0KL}.

\begin{table}[htb]
\caption[]{The function $P_0(Y)$ for various $\Lms^{(4)}$ and $m_c$.
\label{tab:P0KL}}
\begin{center}
\begin{tabular}{|c|c|c|c|}
&\multicolumn{3}{c|}{$P_0(Y) $}\\
\hline
$\Lms^{(4)}$ / $m_c$  & $1.25\gev$ & $1.30\gev$ & $1.35\gev$\\
\hline
$215\mev$ & 0.132 & 0.141 & 0.151 \\
$325\mev$ & 0.140 & 0.149 & 0.159 \\
$435\mev$ & 0.145 & 0.156 & 0.166
\end{tabular}
\end{center}
\end{table}

Using the improved Wolfenstein parametrization and the approximate
formulae (\ref{2.51}) -- (\ref{2.53}) we can next write
\begin{equation}\label{bklmnum}
B(\klm)_{SD}=1.68\cdot 10^{-9} A^4 Y^2(x_t)\frac{1}{\sigma}
\left(\bar\varrho_0-\bar\varrho\right)^2
\end{equation}
with
\begin{equation}\label{rhosig}
\bar\varrho_0=1+\frac{P_0(Y)}{A^2 Y(x_t)}\qquad
\sigma=\left(\frac{1}{1-\frac{\lambda^2}{2}}\right)^2
\end{equation}
The "experimental" value of $B(\klm)_{SD}$ determines the value of
$\bar\varrho$ given by
\begin{equation}\label{rhor0}
\bar\varrho=\bar\varrho_0-\bar r_0
\qquad\qquad
\bar r_0^2=\frac{1}{A^4 Y^2(x_t)}\left[
\frac{\sigma B(\klm)_{SD}}{1.68\cdot 10^{-9}}\right]
\end{equation}
Similarly to $r_0$ in the case of $\kpn$, the value of $\bar r_0$
is fully determined by the top contribution which has only a very
weak renormalization scale ambiguity after the inclusion of
$\ord(\as)$ corrections. The main scale ambiguity resides in
$\bar\varrho_0$ whose departure from unity measures the relative
importance of the charm contribution.

\subsection{Numerical Analysis of $(\klm)_{SD}$}
\label{sec:KLmm:NumericalKL}
\subsubsection{Renormalization Scale Uncertainties}
\label{sec:KLmm:NumericalKL:RSU}
We will now investigate the uncertainties in $Y(x_t)$, $Y_{NL}$,
$B(\klm)_{SD}$ and $\bar\varrho$ related to the dependence of
these quantities on the choice of the renormalization scales $\mu_t$
and $\mu_c$. To this end we proceed as in section
\ref{sec:Kpnn:NumericalKp:RSU}. We fix all the remaining parameters
as given in (\ref{mcmtnum}) and (\ref{vcbubnum}) and we vary
$\mu_c$ and $\mu_t$ within the ranges stated in (\ref{muctnum}).

\begin{figure}[hbt]
\vspace{0.10in}
\centerline{
\epsfysize=5in
\rotate[r]{
\epsffile{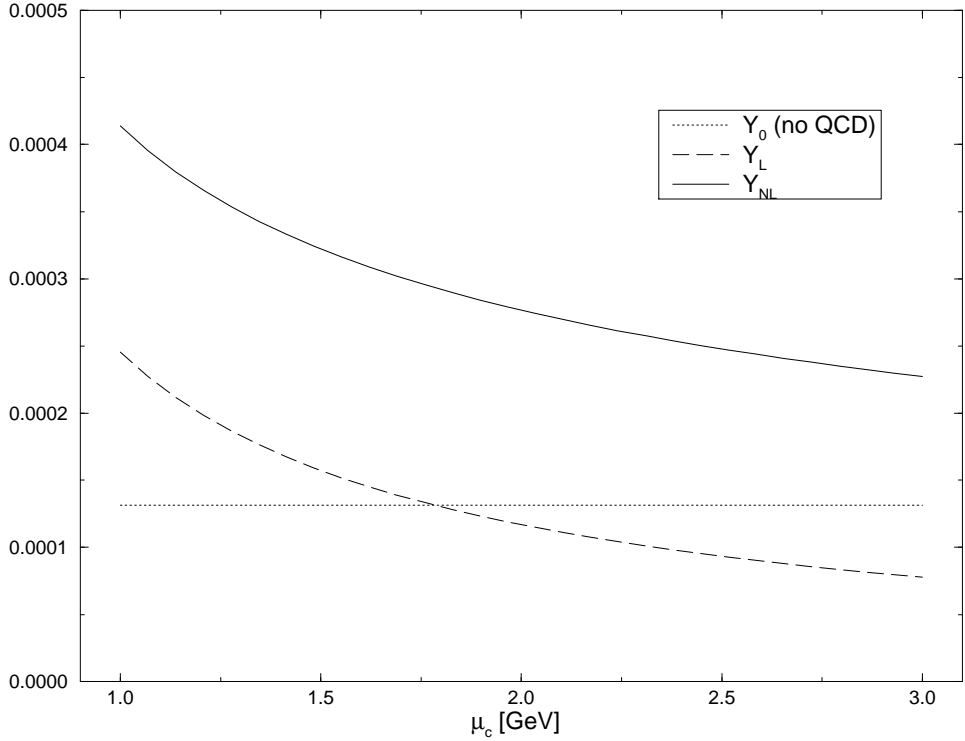}
} }
\vspace{0.08in}
\caption[]{
Charm quark function $Y_{NL}$ compared to the leading-log
result $Y_L$ and the case without QCD as functions of $\mu_c$.
\label{fig:kpmumu:YNL}}
\end{figure}

Fig.\ \ref{fig:kpmumu:YNL} shows the charm function $Y_{NL}$ compared
to the leading-log result $Y_L$ and the case without QCD as a function
of $\mu_c$.  We note the following points:
\begin{itemize}
\item
The residual slope of $Y_{NL}$ is considerably smaller than
in $Y_L$ although still sizable. The variation of $Y$ with $\mu$
defined as $(Y(1\gev)-Y(3\gev))/Y(m_c)$ is 53\% in NLLA compared
to 92\% in LLA.
\item
There is a strong enhancement of $Y_0$ through QCD corrections
in contrast to the suppression found in the case of $X_0$.
\end{itemize}

\begin{figure}[hbt]
\vspace{0.10in}
\centerline{
\epsfysize=5in
\rotate[r]{
\epsffile{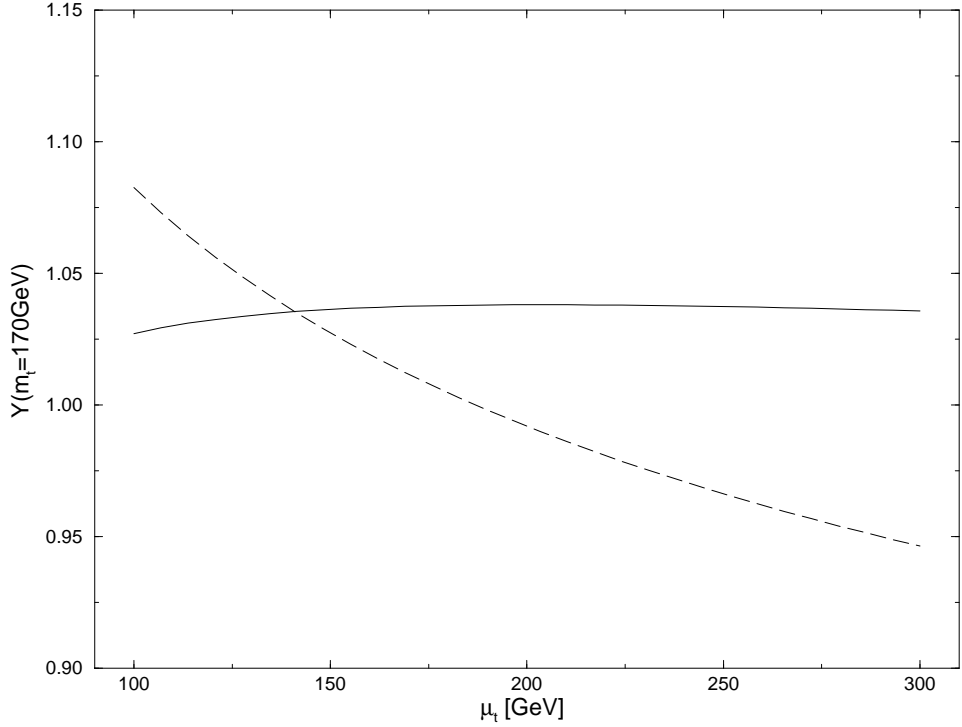}
} }
\vspace{0.08in}
\caption[]{
Top quark function $Y(x_t)$ as a function of $\mu_t$ for fixed
$\mt(\mt)=170\gev$ with (solid curve) and without (dashed curve) $\ord(\as)$
corrections.
\label{fig:kpmumu:Y}}
\end{figure}

In fig.\ \ref{fig:kpmumu:Y} we show the analogous results for $Y(x_t)$
as a function of $\mu_t$. The observed features are similar to the ones
found in the case of $X(x_t)$:
\begin{itemize}
\item
Considerable reduction of the scale uncertainties in NLLA
relative to the LLA with a tiny residual uncertainty after the
inclusion of NLO corrections.
\item
Small NLO correction for the choice $\mu_t=m_t$ as summarized
by $\eta_Y$ in (\ref{yeta}).
\end{itemize}
Using (\ref{bklm}) and varying $\mu_{c,t}$ in the ranges (\ref{muctnum})
we find that for our choice of input parameters the uncertainty in
$B(\klm)_{SD}$
\begin{equation}\label{varbklmLO}
0.816\cdot 10^{-9}\leq B(\klm)_{SD}\leq 1.33\cdot 10^{-9}
\end{equation}
present in the leading order is reduced to
\begin{equation}\label{varbklmNLO}
1.02\cdot 10^{-9}\leq B(\klm)_{SD}\leq 1.25\cdot 10^{-9}
\end{equation}
after including NLO corrections. Here we have assumed $\bar\varrho=0$.
\\
Similarly we find
\begin{equation}\label{varrhbLO}
-0.117\leq \bar\varrho\leq 0.165  \qquad {\rm LLA}
\end{equation}
\begin{equation}\label{varrhbNLO}
0.011\leq \bar\varrho\leq 0.134  \qquad {\rm NLLA}
\end{equation}
where we have set $B(\klm)_{SD}=1\cdot 10^{-9}$.
We observe again a considerable reduction of the theoretical
error when the NLO effects are included in the analyses. Also in this
case the remaining ambiguity is largely dominated by the uncertainty
in the charm sector.

\subsubsection{Expectations for $B(\klm)_{SD}$}
\label{sec:KLmm:NumericalKL:EfB}
We finally quote the standard model expectation for the 
short-distance contribution to the \klmm branching ratio. Using the
analysis of $\varepsilon_K$ and the constraint implied by
$B_d-\bar B_d$ mixing in analogy to the case of \kpnn described in
section \ref{sec:Kpnn:NumericalKp:EfB}, we find
\begin{equation}\label{klmm1}
0.6\cdot 10^{-9}\leq B(\klm)_{SD}\leq 2.0\cdot 10^{-9}
\end{equation}
and
\begin{equation}\label{klmm2}
0.9\cdot 10^{-9}\leq B(\klm)_{SD}\leq 1.2\cdot 10^{-9}
\end{equation}
for present parameter uncertainties and our "future" scenario,
respectively. The relevant sets of input parameters and their errors
are collected in appendix \ref{app:numinput}.  Removing the $x_d$
constraint would increase the upper bounds in \eqn{klmm1} and
\eqn{klmm2} to $3.5 \cdot 10^{-9}$ and $2.2 \cdot 10^{-9}$,
respectively.

\subsection{General Remarks on $K^+\to\pi^+\mu^+\mu^-$}
\label{sec:KLmm:GeneralKp}
Obviously, the short distance effective hamiltonian in \eqn{hklm} also
gives rise to an amplitude for the transition $K^+\to\pi^+\mu^+\mu^-$.
This amplitude, however, is by three orders of magnitude smaller than
the dominant contribution to $K^+\to\pi^+\mu^+\mu^-$ given by the
one-photon exchange diagram \cite{eckeretal:87} and is therefore
negligible in the total decay rate. On the other hand the coupling to
the muon pair is purely vector-like for the one-photon amplitude,
whereas it contains an axial vector part in the case of the SD
contribution mediated by $Z^0$-penguin and W-box diagrams.  Thus, as
was pointed out by \cite{savagewise:90} and discussed in detail in
\cite{luetal:92}, the {\em interference\/} of the one-photon and the SD
contribution, which is odd under parity, generates a parity violating
longitudinal muon polarization asymmetry

\begin{equation}\label{delr}
\Delta_{LR}=\frac{\Gamma_R-\Gamma_L}{\Gamma_R+\Gamma_L}
\end{equation}
in the decay $K^+\to\pi^+\mu^+\mu^-$. Here $\Gamma_R$ ($\Gamma_L$)
denotes the rate of producing a right- (left-) handed $\mu^+$, that is
a $\mu^+$ with spin along (opposite to) its three-momentum direction.
In this way a measurement of the asymmetry $\Delta_{LR}$ could probe
the phenomenologically interesting short distance physics, which is not
visible in the total rate.

The $K^+\to\pi^+\gamma^\ast$ vertex is described by a form factor
$f(s)$ ($s$ being the invariant mass squared of the muon pair), that
determines the one-photon amplitude and hence the total rate of
$K^+\to\pi^+\mu^+\mu^-$, but also enters the asymmetry $\Delta_{LR}$.
This formfactor has been analyzed in detail in \cite{eckeretal:87}
within the framework of chiral perturbation theory. The imaginary part
$\IM f(s)$ turns out to be much smaller than ${\RE} f(s)$ and can
safely be neglected in the calculation of $\Delta_{LR}$. For this
reason $f(s)\approx {\RE} f(s)$, which depends on a constant not fixed
by chiral perturbation theory, may also be directly extracted from
experimental data on $K^+\to\pi^+e^+e^-$ \cite{alliegro:92}, sensitive
to $|f(s)|$. We follow \cite{luetal:92} in adopting this procedure.

The dominance of ${\RE} f(s)$ further implies that $\Delta_{LR}$
actually measures the real part of the short distance amplitude.  As
emphasized in \cite{belangeretal:93}, $\Delta_{LR}$ is therefore
closely related to the short distance part of $K_L\to\mu^+\mu^-$ and
could possibly yield useful information on this contribution, which is
difficult to extract from experimental results on $K_L\to\mu^+\mu^-$.
Like $(K_L\to\mu^+\mu^-)_{SD}$, $\Delta_{LR}$ is in particular a
measure of the Wolfenstein parameter $\varrho$.

The authors of \cite{luetal:92} have also considered potential long
distance contributions to $\Delta_{LR}$ originating from two-photon
exchange amplitudes. Unfortunately these are very difficult to
calculate in a reliable manner. The discussion in \cite{luetal:92}
indicates however, that they are likely to be much smaller than the
short distance contributions considered above. We will focus here on
the short distance part, keeping in mind the uncertainty due to
possible non-negligible long distance corrections.

One should stress that the short distance part by itself, although
calculable in a well defined perturbative framework, is not completely
free from theoretical uncertainty. The natural context to discuss this
issue is a next-to-leading order analysis, which for $\Delta_{LR}$ has
been presented in \cite{buchallaburas:94b}, generalizing the previous
leading log calculations \cite{savagewise:90}, \cite{luetal:92},
\cite{belangeretal:93}. We will summarize the results of
\cite{buchallaburas:94b} below.

We finally mention that other asymmetries in $K^+\to\pi^+\mu^+\mu^-$,
which are odd under time reversal and are also sensitive to short
distance contributions, have been discussed in the literature
\cite{savagewise:90}, \cite{luetal:92}, \cite{agrawaletal:91},
\cite{agrawaletal:92}. They involve both the $\mu^+$ and $\mu^-$
polarizations and are considerably more difficult to measure than
$\Delta_{LR}$. Possibilities for measuring the polarization of muons
from $K^+\to\pi^+\mu^+\mu^-$ in future experiments, based on studying
the angular distribution of $e^\pm$ from muon decay, are described in
\cite{kuno:92}.

\subsection{Master Formulae for $\Delta_{LR}$}
\label{sec:KLmm:MasterDeLR}
The absolute value of the asymmetry $\Delta_{LR}$ can be written as
\begin{equation}\label{drxi}
|\Delta_{LR}|=r\cdot |{\RE}\xi|
\end{equation}
The factor $r$ arises from phase space integrations. It depends only on
the particle masses $m_K$, $m_\pi$ and $m_\mu$, on the form factors of
the matrix element $\langle\pi^+\mid(\bar sd)_{V-A}\mid K^+\rangle$, as
well as on the form factor of the $K^+\to\pi^+\gamma^\ast$ transition,
relevant for the one-photon amplitude. In addition $r$ depends on a
possible cut which may be imposed on $\theta$, the angle between the
three-momenta of the $\mu^-$ and the pion in the rest frame of the
$\mu^+\mu^-$ pair.  Without any cuts one has $r=2.3$ \cite{luetal:92}. If
$\cos\theta$ is restricted to lie in the region $-0.5\leq\cos\theta\leq
1.0 $, this factor is increased to $r=4.1$.  As discussed in
\cite{luetal:92}, such a cut in $\cos\theta$ could be useful since it
enhances $\Delta_{LR}$ by 80\% with only a 22\% decrease in the total
number of events.

${\RE}\xi$ is a function containing the information on the
short distance physics. It depends on CKM parameters, the QCD scale
$\Lambda_{\overline{MS}}$, the quark masses $m_t$ and $m_c$ and is given by
\begin{equation}
\label{rexi}
{\RE}\xi=
\kappa\cdot\left[\frac{\relc}{\lambda}P_0(Y)+\frac{\relt}{\lambda^5}Y(x_t)
\right]            
\end{equation}
\begin{equation}\label{kap}
\kappa=\frac{\lambda^4}{2\pi\sin^2\Theta_W(1-\frac{\lambda^2}{2})}
 =1.66\cdot 10^{-3}   
\end{equation}
Here $\lambda=|V_{us}|=0.22$, $\sin^2\Theta_W=0.23$,
$x_t=m^2_t/M^2_W$, $\lambda_i=V^\ast_{is}V_{id}$ and
\begin{equation}\label{p0ynl}
P_0(Y)=\frac{Y_{NL}}{\lambda^4}
\end{equation}
The functions $Y_{NL}$ and $Y(x_t)$ represent the charm and the top
contribution, respectively. They are to next-to-leading logarithmic
accuracy given in \eqn{ynl} and \eqn{yy} and have already been
discussed in chapter \ref{sec:HeffRareKB:klmm} and in the previous
sections on the phenomenology of $(K_L\to\mu^+\mu^-)_{SD}$. Numerical
values for $P_0(Y)$ can be found in table \ref{tab:P0KL}.
From (\ref{drxi}) and (\ref{rexi}) we can obtain $\relt$ expressed as a
function of $|\Delta_{LR}|$
\begin{equation}\label{rltdlr}
\relt=-\lambda^5\frac{{|\Delta_{LR}|}/{r \kappa}-
  \left(1-\frac{\lambda^2}{2}\right) P_0(Y)}{Y(x_t)}  
\end{equation}
Since $\relt$ is related to the Wolfenstein parameter $\bar\varrho$
(see section \ref{sec:sewm}), one may use (\ref{rltdlr}) to extract
$\bar\varrho$ from a given value of $|\Delta_{LR}|$.

\subsection{Numerical Analysis of $\Delta_{LR}$}
\label{sec:KLmm:NumericalDeLR}
To illustrate the phenomenological implications of the next-to-leading
order calculation, let us consider the following scenario. We assume a
typical value for $\Delta_{LR}$, allowing for an uncertainty of $\pm
10\%$
\begin{equation}\label{delrnum}
\Delta_{LR}=(6.0\pm 0.6)\cdot 10^{-3}
\end{equation}
Here a cut on $\cos\theta$, $-0.5\leq\cos\theta\leq 1.0$,
is understood.
Next we take ($m_i\equiv\bar m_i(m_i)$)
\begin{equation}\label{mtcV}
m_t=(170\pm 5)\gev\quad m_c=(1.30\pm 0.05)\gev\quad
V_{cb}=0.040\pm 0.001  
\end{equation}
\begin{equation}\label{lams}
\Lms^{(4)}=(300 \pm 50)\mev   
\end{equation}
Table \ref{tab:rhoDeLR} shows the central value of $\bar\varrho$ that
is extracted from $\Delta_{LR}$ in our example together with the
uncertainties associated to the relevant input. Combined errors due to
a simultaneous variation of several parameters can be obtained to a
good approximation by simply adding the errors in table \ref{tab:rhoDeLR}.

\begin{table}[htb]
\caption[]{$\bar\varrho$ determined from $\Delta_{LR}$ for the scenario
described in the text together with the uncertainties related to
various input parameters.
\label{tab:rhoDeLR}}
\begin{center}
\begin{tabular}{|c||c||c|c|c|c|c|}
&&$\Delta(\Delta_{LR})$&$\Delta(m_t)$
&$\Delta(V_{cb})$&$\Delta(m_c)$&$\Delta(\Lambda_{\overline{MS}})$
\\ \hline
$\bar\varrho$&$-0.06$&$\pm 0.13$&$\pm 0.05$&$\pm 0.06$&$\pm 0.01$&
$\pm 0.00$
\end{tabular}
\end{center}
\end{table}

These errors should be compared with the purely theoretical uncertainty
of the short distance calculation, estimated by a variation of the
renormalization scales $\mu_c$ and $\mu_t$. Varying these scales as given
in \eqn{muctnum} and keeping all other parameters at their central
values we find
\begin{equation}\label{rhor}
-0.15\leq\bar\varrho\leq -0.03 \qquad {\rm (NLLA)}   
\end{equation}
\begin{equation}\label{rhor0num}
-0.31\leq\bar\varrho\leq 0.02 \qquad {\rm (LLA)}   
\end{equation}
We observe that at NLO the scale ambiguity is reduced by almost a
factor of 3 compared to the leading log approximation. However, even in
the NLLA the remaining uncertainty is still sizable, though moderate in
comparison with the errors in table \ref{tab:rhoDeLR}.  Note that the
remaining error in (\ref{rhor}) is almost completely due to the charm
sector, since the scale uncertainty in the top contribution is
practically eliminated at NLO.
\\
We remark that for definiteness we have incorporated the numerically 
important piece $x_c/2$ in the leading log expression for the
charm function $Y$, although this is strictly speaking a next-to-leading
order term. This procedure corresponds to a central value of
$\bar\varrho=-0.12$ in LLA. Omitting the $x_c/2$ term and employing the
strict leading log result shifts this value to $\bar\varrho=-0.20$.
Within NLLA this ambiguity is avoided in a natural way.
\\
Finally we give the Standard Model expectation for $\Delta_{LR}$,
based on the short distance contribution in (\ref{drxi}), for the
Wolfenstein parameter $\varrho$ in the range
$-0.25\leq\varrho\leq 0.25$, $V_{cb}=0.040\pm 0.004$ and
$m_t=(170\pm 20)\gev$. Including the uncertainties due to
$m_c$, $\Lambda_{\overline{MS}}$, $\mu_c$ and $\mu_t$ and
imposing the cut $-0.5\leq\cos\theta\leq 1$, we find
\begin{equation}\label{dlr1}
3.0\cdot 10^{-3}\leq |\Delta_{LR}|\leq 9.6\cdot 10^{-3}   
\end{equation}
employing next-to-leading order formulae.
Anticipating improvements in $V_{cb}$, $m_t$ and $\varrho$ we also
consider a future scenario in which
$\varrho=0.00\pm 0.02$, $V_{cb}=0.040\pm 0.001$ and
$m_t=(170\pm 5)\gev$. The very precise determination of $\varrho$
used here should be achieved through measuring CP asymmetries
in B decays in the LHC era \cite{buras:94b}. Then (\ref{dlr1}) reduces to
\begin{equation}\label{dlr2}
4.8\cdot 10^{-3}\leq |\Delta_{LR}|\leq 6.6\cdot 10^{-3}   
\end{equation}
\\
One should mention that although the top contribution dominates the
short distance prediction for $|\Delta_{LR}|$, the charm part is still
important and should not be neglected, as done in
\cite{belangeretal:93}.  It is easy to convince oneself that the charm
sector contributes to $\bar\varrho$ the sizable amount
$\Delta\bar\varrho_{charm}\approx 0.2$.  Furthermore, as we have shown
above, the charm part is the dominant source of theoretical uncertainty
in the short distance calculation of $\Delta_{LR}$.

To summarize, we have seen that the scale ambiguity in the
perturbative short distance contribution to $\Delta_{LR}$ can be
considerably reduced by incorporating next-to-leading order QCD
corrections. The corresponding theoretical error in the
determination of $\bar\varrho$ from an anticipated measurement
of $|\Delta_{LR}|$ is then decreased by a factor of 3, in a typical
example. Unfortunately the remaining scale uncertainty is quite
visible even at NLO. In addition there are further uncertainties due to
various input parameters and due to possible long distance effects.
Together this implies that the accuracy to which $\bar\varrho$ can be
extracted from $\Delta_{LR}$ appears to be limited and $\Delta_{LR}$
can not fully compete with the "gold-plated" $K\to\pi\nu\bar\nu$
decay modes. Still, a measurement of $\Delta_{LR}$ might give
interesting constraints on SM parameters, $\bar\varrho$ in particular,
and we feel it is worthwhile to further pursue this interesting
additional possibility.

\section{The Decays $B\to X\nu\bar\nu$ and $B\to\mu^+\mu^-$}
\label{sec:BXnnBmm}
\subsection{General Remarks}
\label{sec:BXnnBmm:General}
The rare decays $B\to X_{s}\nu\bar\nu$, $B\to X_{d}\nu\bar\nu$ and
$B_{s}\to\mu^+\mu^-$, $B_{d}\to\mu^+\mu^-$ are fully dominated by
internal top quark contributions. The relevant effective hamiltonians
are given in (\ref{hxnu}) and (\ref{hyll}) respectively.  Only the top
functions $X(x_t)$ and $Y(x_t)$ enter these expressions and the
uncertainties due to $m_c$ and $\Lambda_{\overline{MS}}$ affecting
\kpnn and \klmm are absent here.  Consequently these two decays are
theoretically very clean. In particular the residual renormalization
scale dependence of the relevant branching ratios, though sizable in
leading order, can essentially be neglected after the inclusion of
next-to-leading order corrections. On the other hand a measurement of
these rare B decays, in particular of $B\to X_{s}\nu\bar\nu$ and $B\to
X_{d}\nu\bar\nu$ , is experimentally very challenging. In addition, as
we will see below, $B(B_{s}\to\mu^+\mu^-)$ and $B(B_{d}\to\mu^+\mu^-)$
is subject to the uncertainties in the values of the B meson decay
constants $F_{B_{s}}$ and $F_{B_{d}}$, which hopefully will be removed
one day.

\subsection{The Decays $B\to X_{s}\nu\bar\nu$ and $B\to X_{d}\nu\bar\nu$}
\label{sec:BXnnBmm:BXnn}
The branching fraction for $B\to X_s\nu\bar\nu$ is given by
\begin{equation}\label{bbxnn}
\frac{B(B\to X_s\nu\bar\nu)}{B(B\to X_c e\bar\nu)}=
\frac{3 \alpha^2}{4\pi^2\sin^4\Theta_W}
\frac{|V_{ts}|^2}{|V_{cb}|^2}\frac{X^2(x_t)}{f(z)}
\frac{\bar\eta}{\kappa(z)}
\end{equation}
Here $f(z)$, $z=m_c/m_b$ is the phase-space factor for $B\to X_c
e\bar\nu$ defined already in \eqn{g} and $\kappa(z)$ is the
corresponding QCD correction \cite{CM:78} given in \eqn{eq:kappaz}. The
factor $\bar\eta$ represents the QCD correction to the matrix element
of the $b\to s\nu\bar\nu$ transition due to virtual and bremsstrahlung
contributions and is given by the well known expression
\begin{equation}\label{etabar}
\bar\eta=\kappa(0)=
1+\frac{2\alpha_s(m_b)}{3\pi}\left(\frac{25}{4}-\pi^2\right)
\approx 0.83
\end{equation}
For the numerical analysis we will use $\Lambda^{(5)}_{QCD}=225\mev$,
(\ref{alsinbr}), $|V_{ts}|=|V_{cb}|$, $m_t(m_t)=170\gev$, $B(B\to X_c
e\bar\nu)=0.104$, $f(z)=0.49$ and $\kappa(z)=0.88$, keeping in mind
the QCD uncertainties in $B\to X_c e\bar\nu$ discussed in section
\ref{sec:InclB}.

Varying $\mu_t$ as in (\ref{muctnum}) we find that the ambiguity
\begin{equation}\label{lobxn}
3.82\cdot 10^{-5}\leq B(B\to X_s\nu\bar\nu)\leq 4.65\cdot 10^{-5}
\end{equation}
present in the leading order is reduced to
\begin{equation}\label{nlobxn}
3.99\cdot 10^{-5}\leq B(B\to X_s\nu\bar\nu)\leq 4.09\cdot 10^{-5}
\end{equation}
after the inclusion of QCD corrections \cite{buchallaburas:93b}.

It should be noted that $B(B \to X_s \nu\bar\nu)$ as given in
\eqn{bbxnn} is in view of $|V_{ts}/V_{cb}|^2 \approx 0.95 \pm 0.03$
essentially independent of the CKM parameters and the main uncertainty
resides in the value of $\mt$. Setting all parameters as given above and
in appendix \ref{app:numinput}, and using \eqn{xxappr} we have
\begin{equation}
B(B \to X_s \nu\bar\nu) = 4.1 \cdot 10^{-5} \,
\frac{|V_{ts}|^2}{|V_{cb}|^2} \,
\left[ \frac{\mt(\mt)}{170\gev} \right]^{2.30} \, .
\label{eq:bxsnnnum}
\end{equation}
In view of a new interest in this decay \cite{grossmanetal:95} we quote
the Standard Model expectation for $B(B \to X_s \nu\bar\nu)$ based on
the input parameters collected in the appendix \ref{app:numinput}. We
find
\begin{equation}
3.1 \cdot 10^{-5} \le B(B \to X_s \nu\bar\nu) \le 4.9 \cdot 10^{-5}
\label{eq:bxsnnnum2}
\end{equation}
for the ``present day'' uncertainties in the input parameters and
\begin{equation}
3.6 \cdot 10^{-5} \le B(B \to X_s \nu\bar\nu) \le 4.2 \cdot 10^{-5}
\label{eq:bxsnnnum3}
\end{equation}
for our ``future'' scenario. 

In the case of $B\to X_d\nu\bar\nu$ one has to replace $V_{ts}$ by
$V_{td}$ which results in a decrease of the branching ratio by
roughly an order of magnitude.

\subsection{The Decays $B_{s}\to\mu^+\mu^-$ and $B_{d}\to\mu^+\mu^-$}
\label{sec:BXnnBmm:Bmm}
\begin{figure}[hbt]
\vspace{0.10in}
\centerline{
\epsfysize=5in
\rotate[r]{
\epsffile{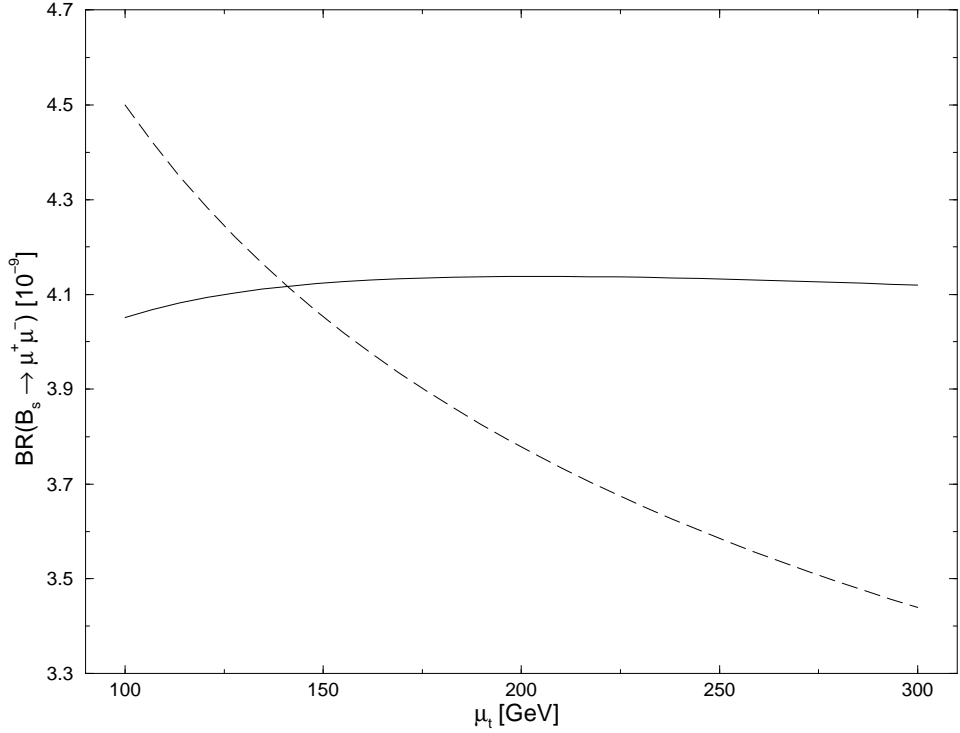}
} }
\vspace{0.08in}
\caption[]{
The $\mu_t$-dependence of $B(B_s \to \mu^+\mu^-) [10^{-9}]$
with (solid curve) and without (dashed curve) $\ord(\as)$
corrections for fixed parameter values as described in the text.
\label{fig:bmmmut}}
\end{figure}

The branching ratio for $B_s\to l^+l^-$ is given by \cite{buchallaburas:93b}
\begin{equation}\label{bbll}
B(B_s\to l^+l^-)=\tau(B_s)\frac{G^2_F}{\pi}
\left(\frac{\alpha}{4\pi\sin^2\Theta_W}\right)^2 F^2_{B_s}m^2_l m_{B_s}
\sqrt{1-4\frac{m^2_l}{m^2_{B_s}}} |V^\ast_{tb}V_{ts}|^2 Y^2(x_t)
\end{equation}
where $B_s$ denotes the flavor eigenstate $(\bar bs)$ and $F_{B_s}$ is
the corresponding decay constant (normalized as $F_\pi=131\mev$). Using
(\ref{alsinbr}), (\ref{yeta}) and \eqn{eq:approxSXYZE} we find in the
case of $B_s\to\mu^+\mu^-$
\begin{equation}\label{bbmmnum}
B(B_s\to\mu^+\mu^-)=4.18\cdot 10^{-9}\left[\frac{\tau(B_s)}{1.6 ps}\right]
\left[\frac{F_{B_s}}{230\mev}\right]^2 
\left[\frac{|V_{ts}|}{0.040}\right]^2 
\left[\frac{m_t(m_t)}{170\gev}\right]^{3.12}
\end{equation}
which approximates the next-to-leading order result.
\\
Taking the central values for $\tau(B_s)$, $F_{B_s}$, $|V_{ts}|$ and
$m_t(m_t)$ and varying $\mu_t$ as in (\ref{muctnum}) we find that the
uncertainty
\begin{equation}\label{lobmm}
3.44\cdot 10^{-9}\leq B(B_s\to\mu^+\mu^-)\leq 4.50\cdot 10^{-9}
\end{equation}
present in the leading order is reduced to
\begin{equation}\label{nlobmm}
4.05\cdot 10^{-9}\leq B(B_s\to\mu^+\mu^-)\leq 4.14\cdot 10^{-9}
\end{equation}
when the QCD corrections are included.
This feature is once more illustrated in fig.\ \ref{fig:bmmmut}.

Finally, we quote the standard model expectation 
for $B(B_s\to\mu^+\mu^-)$ based on the
input parameters collected in the Appendix. We find
\begin{equation}\label{bsmm1}
1.7\cdot 10^{-9}\leq B(B_s\to\mu^+\mu^-)\leq 8.4\cdot 10^{-9}
\end{equation}
using present day uncertainties in the parameters and
$F_{B_s}=230\pm 40\mev$. With reduced errors for the input
quantities, corresponding to our second scenario as defined in 
Appendix \ref{app:numinput}, 
and taking $F_{B_s}=230\pm 10\mev$ this range would
shrink to
\begin{equation}\label{bsmm2}
3.1\cdot 10^{-9}\leq B(B_s\to\mu^+\mu^-)\leq 5.0\cdot 10^{-9}
\end{equation}

For the case of $B_d\to\mu^+\mu^-$ similar formulae hold with obvious
replacements of labels $(s\to d)$. Provided the decay constants
$F_{B_s}$ and $F_{B_d}$ will have been calculated reliably by
non-perturbative methods or measured in leading leptonic decays one
day, the rare processes $B_{s}\to\mu^+\mu^-$ and $B_{d}\to\mu^+\mu^-$
should offer clean determinations of $|V_{ts}|$ and $|V_{td}|$. The
accuracy of the related analysis will profit considerably from the
reduction of theoretical ambiguity achieved through the inclusion of
short-distance QCD effects. In particular $B(B_s\to\mu^+\mu^-)$, which
is expected to be ${\cal O}(4\cdot 10^{-9})$, should be attainable at 
hadronic machines such as HERA-B, Tevatron and LHC.

\section{Summary}
        \label{sec:summary}
In this review we have described in detail the present status of higher
order QCD corrections to weak decays of hadrons. We have emphasized
that during the last years considerable progress has been made in this
field through the calculation of the next--to--leading QCD corrections
to essentially all of the most interesting and important processes.
This effort reduced considerably the theoretical uncertainties in the
relevant formulae and thereby improves the determination of the CKM
parameters to be achieved in future experiments. We have illustrated
this with several examples.

In this review we have concentrated on weak decays in the Standard
Model.  The structure of weak decays in extensions of the Standard
Model will generally be modified. Although we do not expect substantial
effects due to "new physics" in tree level decays, the picture of loop
induced processes, such as rare and CP violating decays, may turn out
to be different from the one presented here. The basic structure of QCD
calculations will remain valid, however. In certain extensions of the
Standard Model, in which no new local operators occur, only the initial
conditions to the renormalization group evolution will have to be
modified. In more complicated extensions additional operators can be
present and in addition to the change of the initial conditions, also
the evolution matrix will have to be generalized.

Yet in order to be able to decide whether modifications of the standard
theory are required by the data, it is essential that the theoretical
calculations within the Standard Model itself reach the necessary
precision.  As far as the {\it short distance}
contributions are concerned, we think that in most cases such a
precision has been already achieved.

Important exceptions are the $b \to s \gamma$ and $b \to s g$
transitions for which the complete NLO corrections are not yet
available.  On the other hand the status of {\it long distance}
contributions represented by the hadronic matrix elements of local
operators or equivalently by various $B_i$ parameters, is much less
satisfactory. This is in particular the case of non--leptonic decays,
where the progress is very slow. Yet without these difficult
non--perturbative calculations it is impossible to give reliable
theoretical predictions for non-leptonic decays even if the Wilson
coefficients of the relevant operators have been calculated with high
precision. Moreover these coefficients have unphysical renormalization
scale and renormalization scheme dependences which can only be canceled
by the corresponding dependences in the hadronic matrix elements. All
efforts should be made to improve the status of non-perturbative
calculations.

The next ten years should be very exciting for the field of weak
decays.  The experimental efforts in several laboratories will provide
many new results for the rare and CP violating decays which will offer
new tests of the Standard Model and possibly signal some "new
physics".  As we have stressed in this review the NLO calculations
presented here will play undoubtedly an important role in these
investigations. Let us just imagine that $B_s^0-\bar B_s^0$ mixing and
the branching ratios for $K^+ \to \pi^+\nu\bar\nu$, $K_L\to \pi^0
\nu\bar\nu$, $B \to X_s \nu \bar\nu$ and $B_s \to \mu^+\mu^-$ have been
measured to an acceptable accuracy.  Having in addition at our disposal
accurate values of $|V_{ub}/V_{cb}|$, $|V_{cb}|$, $m_t$, $F_B$, $B_B$
and $B_K$ as well as respectable results for the angles
$(\alpha,\beta,\gamma)$ from the CP asymmetries in B--decays, we could
really get a great insight into the physics of quark mixing and CP
violation. One should hope that this progress on the experimental side
will be paralleled by the progress in the calculations of hadronic
matrix elements as well as by the calculations of QCD corrections in
potential extensions of the Standard Model.

We would like to end our review with a summary of theoretical
predictions and present experimental results for the rare and CP
violating decays discussed by us. This summary is given in table
\ref{tab:sumtab}.

\begin{table}[htb]
\caption[]{Summary of theoretical predictions and experimental results
for the rare and CP violating processes discussed in this review. The
entry ``input'' indicates that the corresponding measurement is used to
determine or to constrain CKM parameters needed for the calculation of
other decays. For $B(K_L\to\mu^+\mu^-)$ the theoretical value refers
only to the short-distance contribution.  In the case of $B(K_L \to
\pi^0 e^+e^-)$ the SM prediction corresponds to the contribution from
direct CP violation.
The SM predictions for $K^+\to\pi^+\nu\bar\nu$ and
$K_L\to\pi^0\nu\bar\nu$ include the isospin breaking corrections
considered in \cite{marcianoparsa:95}.
\label{tab:sumtab}}
\begin{center}
\begin{tabular}{|c|c|c|l|}
{\bf Quantity}  &  {\bf SM Prediction}  &  {\bf Experiment}  &
\phantom{XXX} {\bf Exp. Reference} \\
\hline \hline
\multicolumn{4}{|c|}{\bf K--Decays}\\
\hline
$|\varepsilon_K|$ & input & $(2.266\pm 0.023)\cdot 10^{-3}$
 & \cite{particledata:94} \\
\hline
$\varepsilon'/\varepsilon$ & $(5.6\pm 7.7)\cdot 10^{-4}$
 & $(15\pm 8)\cdot 10^{-4}$ & \cite{particledata:94} \\
\hline
$B(K_L\to\pi^0e^+e^-)$ & $(4.5 \pm 2.8) \cdot 10^{-12}$
[$\hbox{CP}_{\rm dir}$] & $<4.3\cdot 10^{-9}$
 & \cite{harrisetal:93} )\\
\hline
$B(K^+\to\pi^+\nu\bar\nu)$ & $(1.0\pm 0.4)\cdot 10^{-10}$
 & $<2.4\cdot 10^{-9}$ & \cite{adleretal:95} \\
\hline
$B(K_L\to\pi^0\nu\bar\nu)$ & $(2.9\pm 1.9)\cdot 10^{-11}$
 & $<5.8\cdot 10^{-5}$ & \cite{weaveretal:94} \\
\hline
$B(K_L\to\mu^+\mu^-)$ & $(1.3\pm 0.7)\cdot 10^{-9}$ [SD]
 & $(7.4\pm 0.4)\cdot 10^{-9}$ & \cite{particledata:94} \\
\hline
$|\Delta_{LR}(K^+\to\pi^+\mu^+\mu^-)|$ & $(6\pm 3)\cdot 10^{-3}$
 & --- & ---\\
\hline
\multicolumn{4}{|c|}{\bf B--Decays}\\
\hline
$x_d$ & input & $0.75\pm 0.06$ & \cite{browderhonscheid:95} \\
\hline
$B(B\to X_s\gamma)$ & $(2.8\pm
0.8)\cdot 10^{-4}$
 & $(2.32\pm 0.67)\cdot 10^{-4}$ & \cite{CLEO:94} \\
\hline
$B(B\to X_s\nu\bar\nu)$ & $(4.0\pm 0.9)\cdot 10^{-5}$
 & $< 3.9 \cdot 10^{-4}$ & \cite{grossmanetal:95} \\
\hline
$B(B_s\to\tau^+\tau^-)$ & $(1.1\pm 0.7)\cdot 10^{-6}$
 & --- & ---\\
\hline
$B(B_s\to\mu^+\mu^-)$ & $(5.1\pm 3.3)\cdot 10^{-9}$
 & $<8.4\cdot 10^{-6}$ & \cite{krolletal:95} \\
\hline
$B(B_s\to e^+e^-)$ & $(1.2\pm 0.8)\cdot 10^{-13}$
 & --- & ---\\
\hline
$B(B_d\to\mu^+\mu^-)$ & $\sim 10^{-10}$ & $<1.6\cdot 10^{-6}$ &
\cite{krolletal:95} \\
\hline
$B(B_d\to e^+e^-)$ & $\sim 10^{-14}$ & $<5.9\cdot 10^{-6}$ &
\cite{ammaretal:94} \\
\end{tabular}
\end{center}
\end{table}

Let us hope that the next ten years will bring a further reduction of
uncertainties in the theoretical predictions and will provide us with
accurate measurements of various branching ratios for which, as seen in
table \ref{tab:sumtab}, only  upper bounds are available at present.

\acknowledgements

We would like to thank M.\ Jamin for extensive discussions and
numerical checks in sections \ref{sec:HeffdF1:1010} and
\ref{sec:nloepe}, M.\ M{\"u}nz for providing the updated figures in
sections \ref{sec:Heff:Bsgamma} and \ref{sec:Heff:BXsee:nlo}, and
S.\ Herrlich and U.\ Nierste for valuable comments on section
\ref{sec:HeffKKbar}.
\\
Thanks are also due to B.\ Bardeen, M.\ Beneke, G.\ Burdman,
I.\ Dunietz, A.\ El-Khadra, B.\ Gough, C.\ Greub, C.\ Hill, B.\ Kayser,
W.\ Kilian, A.\ Kronfeld, A.\ Lenz, L.\ Littenberg, M.\ Misiak,
T.\ Onogi, S.\ Parke, J.\ Simone and B.\ Winstein for useful
discussions.  M.E.L.\ acknowledges support by the Deutsche
Forschungsgemeinschaft (DFG) and the hospitality of the SLAC theory
group during parts of this work.

This work has been supported by the German Bundesministerium f\"ur
Bildung und Forschung under contract 06 TM 743, the CEC Science project
SC1-CT91-0729 and DFG contracts La 924/1-1 and ESMEX Li 519/2-1.
\\
Fermilab is operated by Universities Research Association, Inc., 
under contract DE-AC02-76CHO3000 with the United States Department
of Energy.

\appendix

\section{Compilation of Numerical Input Parameters}
        \label{app:numinput}
Below we give for the convenience of the reader a compilation of input
parameters that were used in the numerical parts of this review.

\smallskip

\leftline{\bf Running quark masses:}
\begin{displaymath}
\begin{array}{lclclcl}
\overline{m}_{\rm d}(\mc) &=& 8\mev   &\quad&
\overline{m}_{\rm s}(\mc) &=& (170 \pm 20)\mev \\
\overline{m}_{\rm c}(\mc) &=& 1.3\gev &\quad& \\
\overline{m}_{\rm b}(\mb) &=& 4.4\gev &\quad&
m_{\rm b}^{\rm (pole)}    &=& 4.8\gev
\end{array}
\end{displaymath}

\smallskip

\leftline{\bf Scalar meson masses and decay constants:}
\begin{displaymath}
\begin{array}{lclclcl}
m_\pi     &=& 135\mev &\qquad& F_\pi     &=& 131\mev \\
m_{\rm K} &=& 498\mev &\qquad& F_{\rm K} &=& 160\mev \\
m_{B_d} &=& 5.28\gev &\quad& \tau(B_d) &=& 1.6 \cdot 10^{-12}\,{\rm s} \\
m_{B_s} &=& 5.38\gev &\quad& \tau(B_s) &=& 1.6 \cdot 10^{-12}\,{\rm s}
\end{array}
\end{displaymath}

\smallskip

\leftline{\bf QCD and electroweak parameters:}
\begin{displaymath}
\begin{array}{lclclcl}
\as(\mz)  &=& 0.117 \pm 0.007
&\quad&
\Lms^{(5)}&=& (225 \pm 85)\mev \\
\aem      &=& 1/129 &\quad& \mw &=& 80.2\gev \\
\sin\theta_{\rm W} &=& 0.23 &\quad& &&
\end{array}
\end{displaymath}

\smallskip

\leftline{\bf CKM elements:}
\begin{displaymath}
\begin{array}{lclclcl}
\left|\V{us}\right| &=& 0.22 &\qquad&
\left|\V{ud}\right| &=& 0.975
\end{array}
\end{displaymath}

\smallskip

\leftline{\bf $K$-decays, $K^0-\bar{K}^0$ and $B^0-\bar{B}^0$ mixing:}
\begin{displaymath}
\begin{array}{lclclcl}
\tau(K_{\rm L}) &=& 5.17  \cdot 10^{-8}\,{\rm s} &\quad&
\tau(K^+)       &=& 1.237 \cdot 10^{-8}\,{\rm s} \\
\multicolumn{7}{l}{ BR(\Kpienu) = 0.0482 } \\
|\eps_{\rm K}| &=& (2.266 \pm 0.023) \cdot 10^{-3} &\quad&
\Delta M_{\rm K} &=& 3.51 \cdot 10^{-15}\gev \\
\svs
\RE A_0 &=& 3.33 \cdot 10^{-7}\gev &\quad&
\RE A_2 &=& 1.50 \cdot 10^{-8}\gev \\
\Omega_{\eta\eta'} &=& 
\multicolumn{5}{l}{0.25} \\
\svs
\eta_1 &=& 1.38 &\quad& \eta_2 &=& 0.57 \\
\eta_3 &=& 0.47 &\quad& \eta_B &=& 0.55 \\
\end{array}
\end{displaymath}
The values for $\RE A_{0,2}$ have been obtained from PDG using isospin
analysis.

\smallskip

\leftline{\bf Hadronic matrix element parameters for $\Kpipi$:}
\begin{displaymath}
\begin{array}{lclclcl}
B_{2,LO}^{(1/2)}(\mc)  &=& 5.7 \pm 1.1 & \quad &
B_{2,NDR}^{(1/2)}(\mc) &=& 6.6 \pm 1.0 \\
B_{2,HV}^{(1/2)}(\mc)  &=& 6.2 \pm 1.0 & \quad & & & \\
\end{array}
\mbox{\quad for\ } \Lms^{(4)} = 325\mev
\end{displaymath}

\begin{displaymath}
B_3^{(1/2)} = B_5^{(1/2)} = B_6^{(1/2)} = B_7^{(1/2)} = B_8^{(1/2)} = 
B_7^{(3/2)} = B_8^{(3/2)} = 1
\qquad
\hbox{\rm (central values)}
\end{displaymath}

\goodbreak

In numerical investigations we have for illustrative purposes sometimes
used actual present as well as estimated future errors for various
input parameters. In the table below this is indicated by labels
``present'' and ``future''.

\begin{center}
\begin{tabular}{|c|c|c|c|}
\hline
{\bf Quantity} & {\bf Central} & {\bf Present} & {\bf Future} \\
\hline
$\left|\V{cb} \right|$         & $0.040$ & $\pm 0.003$ & $\pm 0.001$ \\
$\left| \V{ub}/\V{cb} \right|$ & $0.08$  & $\pm 0.02$  & $\pm 0.01$  \\
\svs
$B_{\rm K}$ & $0.75$ & $\pm 0.15$ & $\pm 0.05$ \\
\svs
$\sqrt{B_d} F_{B_{d}}$  & $200\mev$ & $\pm 40\mev$ & $\pm 10\mev$ \\
$x_d$ & $0.75$ & $\pm 0.06$ & $\pm 0.03$ \\
\svs
$\mt$ & $170\gev$ & $\pm 15\gev$ & $\pm 5\gev$ \\
\hline
\end{tabular}
\end{center}

\newpage

\end{document}